\ifdefined\fullversion
\else
\def\fullversion{1}    
\fi

\ifdefined\cameraversion
\else
\def\cameraversion{0}    
\fi

\def\anonymous{1}      

\documentclass[
  reprint,
  aps,
  pra,
  superscriptaddress,
  amsmath,amssymb,
  floatfix,
  nofootinbib,
  a4paper
]{revtex4-2}

\usepackage[utf8]{inputenc}
\usepackage[T1]{fontenc}
\usepackage{svg}
\usepackage{hyperref}
\hypersetup{hidelinks}
\usepackage{verbatim}
\usepackage{tikz}
\usepackage{stmaryrd}
\usetikzlibrary{arrows.meta, calc, fit, positioning, quantikz2, decorations.pathreplacing}
\usepackage[dvipsnames,svgnames]{xcolor}
\definecolor{Coral}{HTML}{D85A30}
\definecolor{Purple}{HTML}{534AB7}
\definecolor{Teal}{HTML}{1D9E75}
\usepackage{xspace}
\usepackage{amsmath} 
\usepackage{amssymb}
\usepackage{bbm}
\usepackage{mathtools}
\usepackage{pifont}
\usepackage{float}
\usepackage{placeins}
\usepackage{etoolbox}
\usepackage[normalem]{ulem}
\usepackage{booktabs}
\usepackage[capitalise,noabbrev]{cleveref}
\usepackage{multibib}
\usepackage{url}
\usepackage{algorithm}
\usepackage{algpseudocode}
\usepackage{paralist}
\usepackage{mathrsfs}
\usepackage{relsize}
\usepackage{stmaryrd}
\usepackage[textsize=small]{todonotes}
\usepackage{to-be-determined}
\usepackage{multirow}
\usepackage[modulo]{lineno}
\usepackage[lambda,n,operators]{cryptocode}
\usepackage{physics}
\usepackage[table-column-type={}]{siunitx}
\usepackage[most]{tcolorbox}
\tcbuselibrary{minted}
\usepackage{minted}
\usepackage{subcaption}
\DeclareCaptionJustification{fulljustified}{\leftskip=0pt\rightskip=0pt\parfillskip=0pt plus 1fil\relax}
\usepackage{adjustbox}
\usepackage{tablefootnote}
\usepackage{anb-colors}

\newtoggle{notes}
\toggletrue{notes} 

\usepackage{amsthm}
\theoremstyle{plain}
\newtheorem{lemma}{Lemma}[section]

\theoremstyle{definition}
\newtheorem{definition}[lemma]{Definition}
\theoremstyle{remark}
\newtheorem{remark}[lemma]{Remark}

\usepackage{etoolbox}

\makeatletter
\let\oldaddcontentsline\addcontentsline
\newcommand{\notoc}{\renewcommand{\addcontentsline}[3]{}}
\newcommand{\dotoc}{\let\addcontentsline\oldaddcontentsline}
\makeatother

\makeatletter

\def\p@paragraph{\thesection\,\thesubsection\,\thesubsubsection\,}
\makeatother

\newcommand{\C}{\mathord{\mathrm{C}}}
\newcommand{\CX}{\mathord{\C X}}
\newcommand{\CZ}{\mathord{\C Z}}
\newcommand{\CCX}{\mathord{\C\C X}}
\newcommand{\CCZ}{\mathord{\C\C Z}}
\newcommand{\CXCXX}{\mathord{\C_X \C_X X}}
\newcommand{\DCZ}{\mathord{\mathrm{D}\CZ}}
\newcommand{\Adder}{\textsc{Adder}}
\newcommand{\Comparison}{\textsc{Comparison}}
\newcommand{\Lookup}{\textsc{Lookup}}
\newcommand{\Phaseup}{\textsc{Phaseup}}
\newcommand{\ComputeExponent}{\textsc{ComputeExponent}}
\newcommand{\ReduceExponent}{\textsc{ReduceExponent}}
\newcommand{\ComputeResidue}{\textsc{ComputeResidue}}
\newcommand{\AccumulateResidue}{\textsc{AccumulateResidue}}
\newcommand{\UnderflowPhasefix}{\textsc{UnderflowPhasefix}}
\newcommand{\Reshuffle}{\textsc{Reshuffle}}
\newcommand{\SWAP}{\mathord{\mathrm{SWAP}}}

\definecolor{ABYellow}{HTML}{FBA821}
\definecolor{ABTeal}{HTML}{00B5C8}
\definecolor{ABSalmon}{HTML}{FB6168}

\colorlet{ABTrack}{ABTeal!80!black}
\colorlet{ABExcept}{ABSalmon!85!black} 
\colorlet{ABSalmonDark}{ABSalmon!85!black} 

\definecolor{abTealDark}{HTML}{00676A}
\definecolor{abInk}{HTML}{010102}
\definecolor{ABInk}{HTML}{010102}
\definecolor{abLine}{HTML}{6B6B6B}
\definecolor{ABLine}{HTML}{6B6B6B}

\definecolor{ABTealDark}{HTML}{00676A}
\definecolor{ABPurple}{HTML}{984F91}
\definecolor{ABPurpleDark}{HTML}{6B3368}

\definecolor{abTeal}{HTML}{00B5C8}
\definecolor{abTealDark}{HTML}{00676A}
\definecolor{abClassical}{HTML}{EEEFF1}  

\definecolor{abYellow}{HTML}{FAC500}      
\definecolor{abTeal}{HTML}{00B5C8}        
\definecolor{abPurple}{HTML}{984F91}      
\definecolor{abPurpleDark}{HTML}{6B3368}  

\definecolor{ABBlack}{HTML}{010102}
\definecolor{ABGray}{HTML}{6B6B6B}

\let\vec\mathbf
\newcommand{\bit}{\ensuremath{\{0,1\}}\xspace}

\newcommand{\prob}[1]{{\Pr}\left[\,{#1}\,\right]}

\newcommand{\NC}{\ensuremath{\mathsf{NC}}\xspace}

\newcommand{\Z}{\mathbb{Z}}
\newcommand{\F}{\mathbb{F}}
\newcommand{\N}{\mathbb{N}}
\newcommand{\R}{\mathbb{R}}

\newcommand{\pk}{\ensuremath{\mathsf{pk}}\xspace}

\newcommand{\mode}{\mathsf{mode}}
\newcommand{\token}{\ensuremath{\mathsf{token}}\xspace}

\makeatletter
\DeclareFontFamily{OMX}{MnSymbolE}{}
\DeclareSymbolFont{MnLargeSymbols}{OMX}{MnSymbolE}{m}{n}
\SetSymbolFont{MnLargeSymbols}{bold}{OMX}{MnSymbolE}{b}{n}
\DeclareFontShape{OMX}{MnSymbolE}{m}{n}{
    <-6>  MnSymbolE5
   <6-7>  MnSymbolE6
   <7-8>  MnSymbolE7
   <8-9>  MnSymbolE8
   <9-10> MnSymbolE9
  <10-12> MnSymbolE10
  <12->   MnSymbolE12
}{}
\DeclareFontShape{OMX}{MnSymbolE}{b}{n}{
    <-6>  MnSymbolE-Bold5
   <6-7>  MnSymbolE-Bold6
   <7-8>  MnSymbolE-Bold7
   <8-9>  MnSymbolE-Bold8
   <9-10> MnSymbolE-Bold9
  <10-12> MnSymbolE-Bold10
  <12->   MnSymbolE-Bold12
}{}

\let\llangle\@undefined
\let\rrangle\@undefined
\DeclareMathDelimiter{\llangle}{\mathopen}%
                     {MnLargeSymbols}{'164}{MnLargeSymbols}{'164}
\DeclareMathDelimiter{\rrangle}{\mathclose}%
                     {MnLargeSymbols}{'171}{MnLargeSymbols}{'171}
\makeatother

\begin{document}

\title{Shor's algorithm compiled on cat qubits: 22\,593 qubits for RSA-2048 in 3.14~days}

\ifnum\anonymous=0
\author{Coauthor Name}
\affiliation{Coauthor University}
\email{mail@mail.com}
\author{Coauthor Name}
\affiliation{Coauthor University}
\email{mail@mail.com}
\else
\author{Axel Pappalardo}
\thanks{These authors contributed equally to this work.}
\affiliation{Alice \& Bob, 49 Bd du Général Martial Valin, 75015 Paris, France}
\author{Dahmun Goudarzi}
\thanks{These authors contributed equally to this work.}
\affiliation{Alice \& Bob, 49 Bd du Général Martial Valin, 75015 Paris, France}
\author{Victor Hertz}
\affiliation{Alice \& Bob, 49 Bd du Général Martial Valin, 75015 Paris, France}
\author{Hugo Jacinto}
\affiliation{Alice \& Bob, 49 Bd du Général Martial Valin, 75015 Paris, France}
\affiliation{Université Paris--Saclay, CEA, CNRS, Institut de Physique Théorique, \num[detect-all]{91191} Gif-sur-Yvette, France}
\author{Larissa Kroell}
\affiliation{Alice \& Bob, 49 Bd du Général Martial Valin, 75015 Paris, France}
\author{André Schrottenloher}
\affiliation{Univ Rennes, Inria, CNRS, IRISA, Rennes, France}
\author{Peter Shanahan}
\affiliation{Alice \& Bob, 49 Bd du Général Martial Valin, 75015 Paris, France}
\affiliation{School of Informatics, University of Edinburgh, United Kingdom}
\author{Élie Gouzien}
\affiliation{Alice \& Bob, 49 Bd du Général Martial Valin, 75015 Paris, France}
\author{Rémi de La Vieuville}
\affiliation{Alice \& Bob, 49 Bd du Général Martial Valin, 75015 Paris, France}
\author{Vivien Londe}
\affiliation{Alice \& Bob, 49 Bd du Général Martial Valin, 75015 Paris, France}
\noaffiliation
\fi

\begin{abstract}
\noindent We present an end-to-end resource estimation of Shor's algorithm on superconducting 
cat qubits, from its mathematical description to the allocation and scheduling of physical 
qubits on a chip layout. After proposing a comprehensive view of the modern versions of Shor's 
algorithm for RSA, ECDLP and FFDLP, we make explicit their logical circuits and the trade-offs 
they enable. We compile the dominant subroutines down to physical-qubit operations on two architectures: one based on repetition cat codes and one based on LDPC cat codes. We design a parity-\textsc{Cnot} gate for cat qubits, extend a multi-target \textsc{Cnot} gate to the LDPC-cat architecture, and adapt an active injection procedure to cat-qubit architectures. Building on these primitives, we adapt to cat-qubit architectures the adder from Gidney’s 2025 resource estimate for factoring RSA-2048. We combine our adder compilation and a coarser resource estimate for the phaseup with the factory designs and lookup compilation from the FeMoco resource estimation of Kroell et al. Together, these compilations lead to chip layouts and schedules for factoring RSA-2048.
 
Assuming a \SI{1}{\micro\second} error-correction cycle time, a \SI{10}{\micro\second} reaction time,
circuit-level phase-flip noise with probability $p_Z = 5 \times 10^{-4}$ and no bit-flips, factoring
RSA-2048, the configuration we compile in full, requires about 23\,000 cat qubits
and 3.14~days. Reusing the operations compiled for RSA-2048, solving ECDLP on secp256k1 requires at most
30\,000 cat qubits and 1 hour, and FFDLP-512, which we propose as an
intermediate benchmark, about 6\,700 cat qubits and 21~minutes. Trading space
for time, RSA-2048 runs in 1.25~hours on about 611\,000 cat qubits. The requirement on no bit-flips 
(which corresponds to a bit-flip time of about 100 years) can be relaxed to a bit-flip time of about 
one hour by using an additional concatenation layer into a distance-3 bit-flip error-correcting code. 
In that case, we estimate that the cost of RSA-2048 increases to about 45\,000 cat qubits.
 
\end{abstract}

\maketitle

\clearpage
\notoc
\newpage

\section{Shor's algorithm, a universal benchmark for quantum computing}

In 1994, Peter Shor introduced a quantum algorithm to factor a large number or compute a discrete logarithm in polynomial time~\cite{Shor1994, Shor97}. The consequences of Shor’s algorithm went far beyond the community of cryptographers, as the three main public-key schemes in use at the time were threatened: \emph{RSA} (factorization, with a classical subexponential complexity), \emph{FFDLP} (discrete logarithm on finite field $\mathbb{Z}/p\mathbb{Z}$ as well as classically subexponential) and \emph{ECDLP} (discrete logarithm on elliptic curves, for which the best known classical attack is exponential). This significant impact on encrypted communications, digital signature, Internet authentication, or more recently crypto-currencies, is one of the reasons the race toward fault-tolerant quantum computing began~\cite{PhysRevLett.127.140503, GouzienCatQubits23,cain2026shors,Babbush26ECC,cryptoeprint:2026/1916}.

\medskip

Since its orginal publication, the study of Shor's algorithm has led to a dedicated research field, yielding substantial gains in terms of number of logical qubits and duration required to run the algorithm. The avenues for tackling RSA and FFDLP converged into a single algorithm and innovations from Ekerå and Håstad~\cite{PQCRYPTO:EkeHas17}, followed by May and Schlieper~\cite{MaySchlieper22}, then by Chevignard et al~\cite{C:CheFouSch25} and Gidney~\cite{gidney2025factor2048bitrsa}, reduced the number of logical qubits from several times the size of the key to half of it. The development of the algorithm for ECDLP is just as dense, with three different paths of optimization emerging in recent publications~\cite{Schrottenloher26PointAdd, LuoEtAl26, CFS26ECDLP}. This work relies on the modern version of these algorithms to solve RSA, FFDLP, and ECDLP (see Appendix~\ref{sec:algo_overview} for a comprehensive presentation and a step-by-step example).

\medskip

Apart from its practical impact, Shor's algorithm applied to RSA is also a perfect candidate for benchmarking and comparing the architectures of quantum computers, for three reasons. First, no known classical algorithm can factor integers of cryptographic size (e.g. RSA keys of more than 2048\,bits) in a reasonable time. A successful run of Shor's algorithm at such sizes is therefore strong evidence that the computation was run on a quantum computer. Second, the problem it solves is simple to state, and the results it provides are easy to verify (e.g. checking that the product of two numbers is equal to a public key). Finally, due to its required depth, the computation cannot succeed without large-scale quantum error correction. In fact, to run the algorithm, billions of logical operations need to be executed without errors, several orders of magnitude beyond the thousands of operations that a machine without error correction can execute.

\medskip

Benchmarking on RSA-2048 and beyond is useful for comparing architectures, but will only become practical in a few years. Thus, we introduce here an additional intermediate benchmark, \emph{FFDLP-512}, defined as an instance of the FFDLP problem for a 512-bit prime group and an 80-bit exponent. Although the cryptanalysis of this problem can be performed classically, it offers three advantages that make it relevant as a first cryptographic milestone.

First, Shor's algorithm for FFDLP is space-efficient. The size of the group targeted in FFDLP-512 makes this intermediate benchmark practical significantly earlier than the RSA-2048 benchmark.

Second, the implementation uses all the blocks needed for RSA-2048, hence it can demonstrate an actual end-to-end execution of Shor's algorithm without shortcuts, contrary to toy problems such as factoring 128-bit numbers where some of the building blocks of the algorithm become irrelevant.

Finally, such Diffie--Hellman parameters have been observed at a large scale on production systems. A scan performed in 2015 by the authors of the Logjam attack~\cite{weakdh15} revealed that 8.4\% of the top 1~million domains were accepting 512-bit groups (``export grade''), with two groups covering 92\% of such vulnerable servers.

\section{Measuring the cost of a quantum computation}\label{sec:metrics}

Formally comparing two descriptions of a quantum computer (with different qubit modalities, architecture and/or error correction scheme) requires agreement on metrics to take into account, among them:
\begin{itemize}
\item Application considered for the benchmark (Shor’s algorithm applied respectively to RSA-2048 challenge number, Bitcoin curve (secp256k1), and FFDLP-512 for an intermediate benchmark),
\item Number of logical qubits (qubits having low error rates, resulting from quantum error correction applied to groups of manufacturable physical qubits with higher error rates), which derives from the algorithm considered,
\item Number of physical qubits (depending on the qubit modality, the error correction architecture and the precise compilation of the algorithm to a specific architecture),
\item Total execution time (trade-offs between number of physical qubits and time are generally possible),
\item Number of shots (average number of executions of the algorithm before success, in particular when one instance fails with constant probability).
\end{itemize}

\medskip

Other metrics measure industrial aspects, with similar importance: design complexity (constraints on the architecture such as connectivity, scalability, availability of enabling technologies), space (footprint), cost of acquisition (bill of materials and installation) and recurring cost (maintenance, electricity), and energy consumption.

\medskip

The two primary metrics used across publications to compare architectures are \emph{number of physical qubits} and \emph{total execution time}. Other metrics are important but are more applicable to economic analysis, or for assessing the plausibility of the architecture.

\medskip

Three rather independent levers enable the improvement of these two main metrics, and explain the rapid evolution of Shor’s algorithm from 1994 to 2026:
\begin{itemize}
\item \emph{The high-level algorithm:} reducing the number of logical qubits, the number of costly gates (typically Toffoli or T gates in most fault-tolerant architectures) or improving success probability;
\item \emph{Hardware architecture:} improving individual physical qubits (faster, more stable, or with a noise profile favorable to error correction), physical operations (1-qubit and multiple-qubits gates), qubit arrangement in the system (connectivity, modularity, routing), and improvements on fabrication and enabling technologies;
\item \emph{Quantum Error Correction (QEC):} better logical-to-physical qubit ratio, faster error correction cycles, higher fidelity of costly gates, etc.
\end{itemize}

\medskip

Each publication improving Shor’s algorithm performance acts on one to three of these levers, defining a triplet: a specific algorithm, targeting a specific hardware platform, compiling the algorithm to the architecture through specific QEC implementation~\cite{Ruiz_2025, Webster2026Pinnacle}. 

\section{Superconducting cat qubits}\label{sec:cat_qubits}

The qubit technology considered in this work is the dissipative cat qubit~\cite{leghtas2013hardware, mirrahimi2014dynamically,leghtas2015confining, Touzard2018} (a variant, the Kerr-cat qubit, is stabilized by a two-photon drive instead~\cite{Grimm2020, Frattini2024, Hajr2024}), which belongs to the family of superconducting qubits, i.e.\ small electronic circuits exhibiting quantum properties. The main difference from transmons (superconducting modality assumed in Gidney--Ekerå~\cite{GidneyEkera21}) is that instead of encoding the state in two levels of energy of an anharmonic oscillator, the cat qubit circuit is engineered so that the state is encoded in superpositions of the two coherent states of opposite phase (``Schrödinger cat states'') of an electromagnetic mode.

\medskip

This specific encoding has very practical consequences. Qubits are typically subject to two kinds of noise, \emph{bit-flip} (an unwanted flip of $\ket{0}$ into $\ket{1}$ and vice versa, as for classical bits) and \emph{phase-flip} (from $\ket{+}$ to $\ket{-}$, changing the phase of the qubit). In practice, the errors affecting a qubit can essentially be described as a combination of both bit-flips and phase-flips, so correcting these two types is enough to protect it. 

For cat qubits, the bit-flip rate decreases exponentially with a tunable parameter (the number of photons in the cat state), while the phase-flip rate increases only linearly with this parameter~\cite{lescanne2020exponential, berdou2023hundred, marquet2024autoparametric, reglade2024quantum, Putterman2025PRX, rousseau2025squeezing}.

This bias between the two types of noise enables simpler error correction schemes (such as a repetition code architecture~\cite{guillaud2019repetition, leregent2023high, chamberland2022building, hann2025hybrid, putterman2025hardware}), with a much lower logical-to-physical qubit ratio, as only phase-flips need to be corrected. More advanced error correction codes, such as LDPC (low-density parity-check) codes, further lower this ratio~\cite{Ruiz_2025}. For the deepest computations, such as Shor's algorithm for RSA-2048, some bit-flip correction beyond the intrinsic protection of cat qubits might be required~\cite{Ruiz_2025}, and architectures like the Elevator code have already been proposed to do this~\cite{shanahan2026elevator}. Lower-depth computations such as FFDLP-512 require a shorter bit-flip time, of about two months instead of about a hundred years for RSA-2048, which nevertheless remains far beyond the bit-flip times measured to date~\cite{berdou2023hundred, reglade2024quantum, rousseau2025squeezing} (Appendix~\ref{ssec:cat_qubits_bitflips}).
In the context of this paper, the noise model of a cat qubit can be summarized with a single parameter $p_Z$ (phase-flip probability), which is mostly driven by the ratio between $\kappa_2$ (two-photon dissipation, which is the effect that stabilizes the cat state) and $\kappa_1$ (one-photon loss, the effect that degrades the cat state) (Appendix~\ref{sssec:cat_qubits_physics}), a critical hardware feature. Bit flips are assumed to be negligible, and Appendix~\ref{sssec:bitflip_assumptions} quantifies the corresponding requirement on the bit-flip time.

\section{Abstraction layers and compilation}\label{sec:layers}

A fault-tolerant quantum algorithm is compiled through several layers of abstraction down to instructions for a specific machine. 

\medskip

The highest level of abstraction is the \emph{Algorithm layer}: in the context of this work, a description of Shor’s algorithm and its sub-routines (modular exponentiation, Quantum Fourier Transform…).  Algorithms at this level are described with block diagrams, pseudo-code and mathematical representation of the transformations applied to registers.\footnote{We recommend reading~\cite{nielsen00} to gain familiarity with qubits, gates and Dirac notation $\ket{\cdot}$ for quantum states.}

\medskip

The algorithm is then compiled to a \emph{Logical layer}. This is the lowest level of abstraction that is still independent of the quantum technology considered, thus enabling hardware-agnostic comparisons between algorithms. Each subroutine is described precisely in terms of logical qubits and logical gates (e.g., \textsc{Cnot} gate, Toffoli gate, ...) or logical operations (e.g., adder, lookup, ...). A basic resource estimation is possible at this level, using approximate metrics such as the logical-to-physical qubit ratio of an architecture. However, this approach misses many optimizations such as parallelism or selecting physical gates for the most appropriate time-qubits trade-off. Logical circuits are formally specified in languages such as Qualtran or Qarton (two Python libraries)~\cite{cryptoeprint:2026/2191,harrigan2024expressing}.

\medskip

The lowest layer of abstraction considered here is the \emph{Physical layer},
which is specific to the targeted quantum hardware. This level specifies the allocation of each physical qubit, or group of physical qubits, to a function
(data logical qubit, routing logical qubit, ancilla, magic state factory),
through a dynamic mapping between logical and physical qubits, possibly with a different error-correcting code depending on the operations applied to a
logical qubit or the duration for which its state must be stored. 
Optimizations of the compilation can go directly from operation specifications (e.g., an adder description) to this physical level, without going through the logical
circuit abstraction. We describe the execution of these operations with space-time slices, representing stabilizer measurements at each time step~\cite{Raussendorf2007,fujii2015quantumcomputationtopologicalcodes,tqec}.

\medskip

With this last layer, we can precisely evaluate the number of physical qubits
used at each time step, the duration of each subroutine, and the number and
type of magic state factories, and deduce the total number of qubits and
duration. This analysis enables the comparison of different hardware
platforms, but also the design of other parts of the system such as the
throughput of the decoders.

\medskip

In this work, we propose architectures where each cat qubit is allocated to a
specific zone, and their use is scheduled at every time step. 

In the repetition-cat architecture, every logical qubit is a repetition code of a single distance $d$ (the size of the code, which sets its level of protection). 

\begin{figure*}[!tp]
    \centering
    \includegraphics[width=\linewidth]{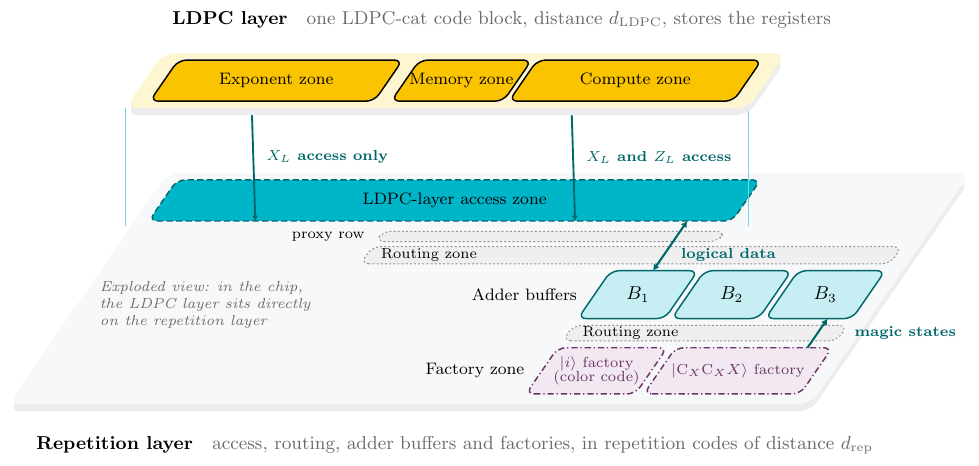}
    \caption{Simplified global layout for the LDPC-cat architecture during the most expensive subroutine of Shor's algorithm
    as presented in Figure~\ref{fig:global_layout_ldpc} of Appendix~\ref{sec:resource_estim}.}
    \label{fig:simplified_global_layout_ldpc}
\end{figure*}

In the LDPC-cat architecture (Figure~\ref{fig:simplified_global_layout_ldpc}),
the chip is a flip chip made of two facing layers. An LDPC layer stores the
registers of the algorithm in a single, dense LDPC-cat code block of distance
$d_{\mathrm{LDPC}}$. A repetition layer underneath gives access to these
registers and holds the routing qubits, some auxiliary qubits and buffers that store the outputs of distillation factories in repetition codes of distance $d_{\mathrm{rep}}$.

\section{Contributions and main results}

This work gives an end-to-end resource estimation of Shor’s algorithm, from its mathematical description to the allocation and scheduling of every physical qubit on a cat-qubit architecture. The contributions follow the structure of the abstraction layers introduced in Section~\ref{sec:layers}.

\medskip

\paragraph*{A reference description of modern Shor.}
Since 1994, improvements to Shor’s algorithm have appeared across many publications, improving either a specific part of the algorithm, or targeting only one cryptographic problem (RSA, ECDLP or FFDLP). Appendix A presents these improvements in a unified way, with examples making it usable as a comprehensive introduction for readers outside of the quantum community.\\
For RSA and FFDLP, four optimizations are presented: reduction to a short discrete logarithm problem (Ekerå--Håstad), shorter input register (May--Schlieper), residue arithmetic and approximate period finding (Chevignard--Fouque--Schrottenloher, then Gidney). We derive the failure budget of the algorithm, and provide the expected number of shots and the algorithmic error rate. We also introduce the notion of noisy hidden subgroup problem in order to formalize the introduction of approximate arithmetic induced by the latest optimizations.\\
An alternative approach to Shor's algorithm exists, introduced by Regev in 2023~\cite{Regev2025Factoring} and further optimized in various works~\cite{ragavan2024space, KahanamokuMeyer2026Parallel,vankirk2026streamingregevsquantumfactoring}. Nonetheless, the current best optimizations of Regev's algorithm are still an order of magnitude less practical than the latest optimizations of Shor's algorithm.
For ECDLP, there is no single best algorithm. We compare gate-optimized, compressed and register-sharing constructions, and identify the resulting Pareto frontier.

\medskip

\paragraph*{An explicit logical implementation.}
Appendix B describes all logical circuits explicitly, and their implementation in the Qarton and Qualtran libraries. The circuit uses classical gates ($X$, $\CX$, $\CCX$), phase gates and measurement-based uncomputation, providing exact logical qubit and Toffoli counts. 
This section enables the extraction, for each algorithm, of the \emph{profile} of its implementation: time and space of each subroutine, share of the overall qubit usage, and percentage of the total runtime. The tables derived for RSA/FFDLP and ECDLP allow anyone, from a set of physical instructions, to compile the algorithm and output the required number of physical qubits and runtime, depending on their architecture. We also propose the first implementation of FFDLP for different sets of parameters in Qarton and Qualtran.

\medskip

\paragraph*{Physical compilation on cat qubits.}
Most of the novelty of this work lies at the Physical layer (Appendix C). Instead of multiplying logical qubit numbers by an average logical-to-physical ratio, we compile the dominant subroutines directly to physical qubits. \\
We design a \emph{parity-\textsc{Cnot} gate} for cat qubits, which runs in $d$ error-correction cycles regardless of the number of controls, extend the \emph{multi-target \textsc{Cnot} gate} to the LDPC-cat architecture, and adapt from Gidney~\cite{gidney2025factor2048bitrsa} a procedure we call \emph{active injection}, which applies a Toffoli gate to parities of qubits in the time of a standard injection.\\
We build on these primitives to redesign the adder, the most frequently called operation. The construction of~\cite{gidney2025factor2048bitrsa} assumes surface code costs, in which $|i\rangle$ states and Hadamard gates are cheap. On a cat-qubit architecture, both are expensive, so we rebuild the adder around auto-\textsc{And} and delayed-$\CZ$ resource states. Simultaneous lattice surgery \cite{Simultaneous2026prep} and pipelined resource-state preparation let the adder keep pace with a single magic-state factory, as long as the code distance stays below the 21 error-correction cycles the factory takes per magic state. For both the repetition-cat and the LDPC-cat architectures, we give the full layout and schedule of the adder, together with the sizing of the factories that feed it.
Combining this adder with the factory designs and the lookup compilation of~\cite{FeMoco2026prep}, and with a coarser estimate for phaseups, we obtain global chip layouts and schedules, and from them a precise count of physical qubits and of the execution time.

\medskip
For ECDLP, we used the operations we optimized for RSA (adder and lookup) and applied the profiling table of Appendix~\ref{sec:logical_impl} to obtain resource estimations. This is not an optimal path as most of the subroutines for ECDLP and its physical layout could be further optimized. Nevertheless, it allows us to derive upper bounds on the number of physical qubits and runtime required for this problem.

\begin{table*}[ht]
\centering
\begin{tabular}{lcccccc}
\hline
Instance & Logical qubits & Toffoli / shot & Physical qubits
         & Time / shot & Shots & Total time \\
\hline
RSA-2048   & 1\,433 & $2^{29.5}$ & 22\,593
           & 7.84 h   & 9.60 & 3.14 days \\
ECDLP-256  & 1\,462 & $2^{25.8}$ & 29\,588
           & 60 min   & 1.03 & 61.8 min \\
FFDLP-512  & 208    & $2^{22.2}$ & 6\,656
           & 2.51 min & 8.44 & 21.2 min \\
\hline
\end{tabular}
\caption{Logical and physical resources for the three target instances
(balanced configuration, LDPC-cat architecture, $p_Z = 5\times10^{-4}$, negligible bit flips, see Appendix~\ref{sssec:bitflip_assumptions}).
Space-time trade-offs are given in Table~\ref{tab:benchmark-matrix}.}
\label{tab:results}
\end{table*}

\medskip
\paragraph*{Resource estimations.}

Table~\ref{tab:results} summarizes the results for the three instances.
We assume an error-correction cycle of \SI{1}{\micro\second}, a decoder
reaction time of \SI{10}{\micro\second}, a phase-flip circuit-level noise model with probability
$p_Z = 5 \times 10^{-4}$, and no bit flips during the computation.
In practice, this requires a bit-flip time of at least about 100 years, for which the bit flips lengthen the runtime by about \SI{11}{\percent}; for a bit-flip time ten times longer, this overhead becomes negligible (Appendix~\ref{sssec:bitflip_assumptions}).
Under these assumptions:
\begin{itemize}
  \item factoring a 2048-bit RSA modulus requires 22\,593 cat qubits and 3.14~days;
  \item solving ECDLP on the Bitcoin curve (secp256k1) requires about 29\,588 cat qubits and 1~hour;
  \item the FFDLP-512 benchmark requires about 6\,656 cat qubits and 21~minutes.
\end{itemize}
The RSA-2048 configuration is the main focus of this work: it is the only one
compiled in full at the physical level (Appendix~C), and its figures are the
most precise. The other results (Appendix~D) are derived from it. The ECDLP figures reuse
the RSA operations and are therefore upper bounds, and the FFDLP-512 figures
reuse the same compiled operations with the parameters of Table~\ref{tab:instances}.

\medskip

Possible space-time trade-offs are presented in Table~\ref{tab:benchmark-matrix}. With more qubits, RSA-2048 runs in about one hour on 610\,530 cat qubits. With fewer qubits, RSA-2048 needs 21\,188 cat qubits over 8 days. These trade-offs are extrapolated from the balanced configuration.
We now discuss how this figure changes if we degrade the physical assumptions or the architecture.
With a phase-flip probability $p_Z = 10^{-3}$ instead of $5\times10^{-4}$, factoring RSA-2048 requires $25\,673$ cat qubits and $3.63$~days (Table~\ref{tab:arch-comparison}).
If the bit-flip time is only of the order of one hour, an additional layer of bit-flip protection is needed, and we estimate that the cost increases to about $45\,000$ cat qubits and at least $3.14$~days (Appendix~\ref{ssec:elevator_code}).
Finally, without LDPC-cat codes, the repetition-cat architecture requires $49\,649$ cat qubits and $2.67$~days at $p_Z = 5\times10^{-4}$ (Appendix~\ref{ssec:physical_resource_count}).

\begin{table*}[ht]
    
    \begin{tabular}{lcccccc}
        \toprule
         & \multicolumn{2}{c}{RSA-2048} & \multicolumn{2}{c}{ECDLP-256} & \multicolumn{2}{c}{FFDLP-512} \\
        \cmidrule(lr){2-3} \cmidrule(lr){4-5} \cmidrule(lr){6-7}
         & Nb cat qubits & Duration & Nb cat qubits & Duration & Nb cat qubits & Duration \\
        \midrule
        Balanced        & 22\,593  & 3.14 days  & 29\,588 & 61.8 minutes & 6\,656  & 21.2 minutes \\
        Space-efficient & 21\,188  & 8.08 days  & 28\,183  & 146 minutes  & 5\,251 & 58.5 minutes \\
        Time-efficient  & 610\,530 & 1.25 hours & -  & -  & 66\,560 & 2.6 minutes \\
        \bottomrule
    \end{tabular}%
    \caption{Trade-off between physical qubit count and runtime for the three targeted problems,
    on the LDPC-cat architecture.}
    \label{tab:benchmark-matrix}
\end{table*}

\section{Conclusion}

\begin{figure*}[t]
  \centering
  \includegraphics[width=\textwidth]{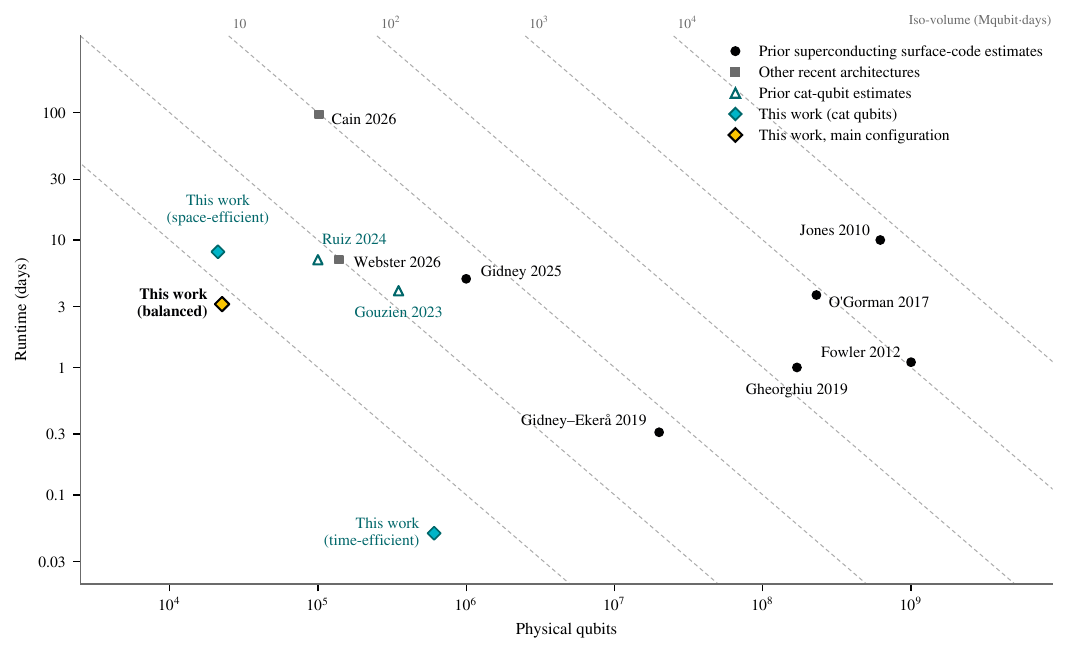}
  \caption{%
    Physical qubit count versus runtime for factoring RSA-2048, on logarithmic
    axes. Dashed diagonals are lines of constant spacetime volume
    (qubits $\times$ days), labelled in megaqubit$\cdot$days along the top axis.
  }
  \label{fig:rsa2048-tradeoff}
\end{figure*}

Figure~\ref{fig:rsa2048-tradeoff} presents a comparison of the results of this work with previous publications since 2010. We observe that in fifteen years,  progress on algorithms, error correction and hardware architecture cut the space-time volume by 5 orders of magnitude, from $10^4$ Mqubit.day for initial surface-code estimations to $7 \times 10^{-2}$ Mqubit.day.
With different hardware hypotheses, this work offers a 40x reduction of the physical qubit count compared to~\cite{gidney2025factor2048bitrsa}, for a shorter runtime. Hence, these results position cat qubits as both a hardware-efficient and time-efficient platform.

\medskip

From a practical standpoint, the reduction of the size of a quantum computer running Shor’s algorithm for RSA-2048 increases the need for a fast transition to post-quantum cryptography. Our results push for an immediate migration effort, relying on hybrid schemes during the transition phase.

Results on ECDLP have practical consequences for the security of Bitcoin cryptocurrency. Space-efficient configuration is a threat to exposed public keys, such as P2PK outputs or reused addresses~\cite{Babbush26ECC}.

\medskip

The FFDLP-512 milestone proposed in this work enables the validation of the end-to-end architecture with far fewer qubits. We recommend adopting FFDLP-512 and RSA-2048 as common benchmarks for all fault-tolerant architectures.

\medskip

All the algorithms and subroutines discussed here are still an active object of research, and leave room for improvements over the coming years. However, we want to highlight the importance of also studying the primitives that will replace RSA and elliptic curves. The trust we place in them for post-quantum cryptography relies on the lack of efficient quantum cryptanalysis. Thus, the same level of effort should be put toward the study of algorithms with potential application to these new primitives. Maturity of quantum cryptanalysis is a precondition for trust in the cryptography intended to resist it.

\section{Acknowledgments and author contributions}

The authors would like to thank André Chailloux and Antoine Joux for their careful review of preliminary versions of this work; Timo Achard for his expertise in Qualtran; and Christelle Yammine for her attention to style coherence and accessibility. We also thank Jérémie Guillaud, Christophe Vuillot and Wes Roberts for their careful proofreading of the manuscript and Laurent Prost for insightful discussions on Shor's algorithm.

PS acknowledges EPSRC CDT grant EP/Y035097/1. A.S. acknowledges support by the French Agence Nationale de la Recherche through the France 2030 program under grant agreement No. ANR-22-PETQ-0008 PQ-TLS. A.S.'s work was done in part while visiting the Simons Institute for the Theory of Computing, supported by a grant from the National Science Foundation.

\medskip

DG wrote the algorithm overview in Appendix~\ref{sec:algo_overview} and contributed to Appendices~\ref{sec:logical_impl} and~\ref{sec:variant_and_tradeoff}. AP wrote the logical implementation in Appendix~\ref{sec:logical_impl} for RSA/FFDLP and parts of the resource estimation in Appendices~\ref{sec:resource_estim} and~\ref{sec:variant_and_tradeoff}. VH wrote the \Adder{} implementation in ~\ref{sec:adder} and contributed to the rest of the resource estimation in Appendix~\ref{sec:resource_estim}. HJ and LK contributed to the resource estimation in Appendix~\ref{sec:resource_estim}. AS wrote the logical implementation in Appendix~\ref{sec:logical_impl} for ECDLP and wrote the different Qarton implementations. PS estimated the Elevator code overhead in Appendix~\ref{ssec:elevator_code}. EG contributed to the algorithm overview and its logical implementation and supervised preliminary work on injection gadgets. AP and DG wrote the Qualtran implementation. RLV managed the project and wrote the main text. VL supervised and contributed to the logical implementation and the resource estimation in Appendices~\ref{sec:logical_impl} and~\ref{sec:resource_estim}. 

\medskip

Claude Opus~5.5 (Anthropic)~\cite{anthropic2026claudeopus55} was used to review the text of this work, our Qualtran implementations, and to polish the style of the figures. The authors checked all of its outputs and take full responsibility for the content of this work.

\dotoc
\appendix
    \clearpage
    \tableofcontents
    \newpage
\section{Deep Dive into Shor's Algorithm}\label{sec:algo_overview}
\setcounter{subsection}{-1}

Shor's algorithm~\cite{Shor97} is the central quantum threat to classical
public-key cryptography. It solves the integer factoring problem and the
discrete logarithm problem in polynomial time on a quantum computer. This breaks RSA, Diffie--Hellman, and elliptic-curve-based schemes. This section develops both
the original algorithm and its modern descendants for factoring:
Ekerå--Håstad~\cite{PQCRYPTO:EkeHas17}, May--Schlieper~\cite{MaySchlieper22}, and
CFS/Gidney~\cite{C:CheFouSch25, gidney2025factor2048bitrsa}. Together, the most recent
algorithms reduce the logical qubit count of attacking RSA from several times the key size
to asymptotically half of it. For RSA-2048 this means about $1400$ logical
qubits~\cite[Table~5]{gidney2025factor2048bitrsa}. We begin with the factoring instance,
which serves as a high-level sketch of the algorithm.

\subsection{High-level Overview}

We start by informally explaining how Shor's algorithm works to recover, 
from an RSA modulus $N$, the two factors $p$ and $q$.

\paragraph{Classical pre-processing.} One first picks a random integer $1 < g < N$
and computes $b = \gcd(g, N)$ using the Euclidean algorithm. If $b \neq 1$, then $b$ is
already a non-trivial factor of $N$, and the algorithm terminates immediately. This case
is rare in practice: it would require $g$ to be a multiple of $p$ or $q$, which happens
only with negligible probability $O(\frac{1}{p}+\frac{1}{q})$. If $b = 1$, then  $g \in
\mathbb{Z}_N^*$, meaning $g$ is invertible modulo $N$. In fact, since $\gcd(g,N)=1$,
Bézout's identity gives two integers $u,v$ such that $ug + vN = 1$. Reducing mod $N$
yields $ug \equiv 1 \pmod N$, confirming that $g$ is invertible. 
Invertibility decides whether $r$ exists at all. The powers $g^1, g^2, g^3, \ldots$ 
can only take finitely many values modulo $N$. Two of them must eventually repeat: 
$g^i \equiv g^j \pmod N$ for some $i < j$. Multiplying both sides by $g^{-i}$ gives 
$g^{j-i} \equiv 1 \pmod N$. So $r := j-i$ is a positive integer with $g^r \equiv 1 \pmod N$.

The algorithm proceeds to find the multiplicative order $r$ of $g$ modulo $N$, 
i.e.\ the smallest integer $r > 0$ such that $g^r \equiv 1 \pmod{N}$.

\paragraph{Classical post-processing.} 
At this stage, the order $r$ is known thanks to a quantum oracle (see hereafter).
We now need to recover $p$ and $q$, the two non-trivial factors of $N$. 
Assuming two conditions hold, this can be done classically with probability $\ge \frac12$. 
Specifically, suppose that (i) $r$ is even and (ii) $g^{r/2}\not\equiv -1 \bmod N$.
Then $g^r-1$ can be factored as
\begin{equation}\label{eq:factorn}
    g^r - 1 = (g^{r/2}-1)(g^{r/2}+1).
\end{equation}
By definition of the order, together with the two assumptions, we obtain:
\begin{itemize}[leftmargin=2em]
    \item[(a)] $N\mid g^r-1 $
    \item[(b)] $g^{r/2} \not\equiv \pm1\bmod N$
\end{itemize}
Therefore, $N$ divides the product of Eq.~\ref{eq:factorn} but neither factor. As a result,
$\gcd(g^{r/2}-1,\,N)$ and $\gcd(g^{r/2}+1,\,N)$ are both non-trivial factors of $N$.
Conditions (i) and (ii) hold with probability at least $\frac12$ for a random $g$\footnote{%
  Let $p,q$ be distinct odd primes. By the Chinese remainder theorem a uniform
  $g\in\mathbb{Z}_N^{*}$ corresponds to an independent uniform pair
  $(g_p,g_q)\in\mathbb{Z}_p^{*}\times\mathbb{Z}_q^{*}$, with
  $r=\operatorname{lcm}(\operatorname{ord}_p g_p,\operatorname{ord}_q g_q)$. Let $c_p,c_q$
  be the exponents of $2$ in $\operatorname{ord}_p g_p$ and $\operatorname{ord}_q g_q$.
  The two failure modes are exactly $c_p=c_q$. The order $r$ is odd iff $c_p=c_q=0$,
  and $g^{r/2}\equiv-1$ iff $c_p=c_q\ge 1$. It therefore suffices to show
  $\Pr[c_p=c_q]\le\tfrac12$. Write $p-1=2^{e_p}m_p$ with $m_p$ odd, and write
  $g_p=\gamma^k$ for a generator $\gamma$ of the cyclic group $\mathbb{Z}_p^{*}$, with $k$
  uniform in $\{0,\dots,p-2\}$. Then $\operatorname{ord}_p g_p=(p-1)/\gcd(k,p-1)$, so
  $c_p=e_p$ exactly when $k$ is odd. This happens with probability $\tfrac12$, hence
  $\Pr[c_p=c]\le\tfrac12$ for every $c$. Since $g_p$ and $g_q$ are independent,
  $\Pr[c_p=c_q]=\sum_{c}\Pr[c_q=c]\Pr[c_p=c]\le\tfrac12\sum_{c}\Pr[c_q=c]=\tfrac12$.
  The bound is tight, for example when $p\equiv q\equiv 3\pmod 4$. Drawing $1<g<N$
  excludes $g=1$, which fails, so it can only raise the success probability.
  See~\cite[Theorem~A4.13]{nielsen00} for $N$ with more prime factors.
}. 
Hence, only a constant number of repetitions, with independently chosen $g$, suffices to recover 
$p$ and $q$ successfully.

\begin{figure*}[ht]
  \centering
  \includegraphics[width=1\linewidth]{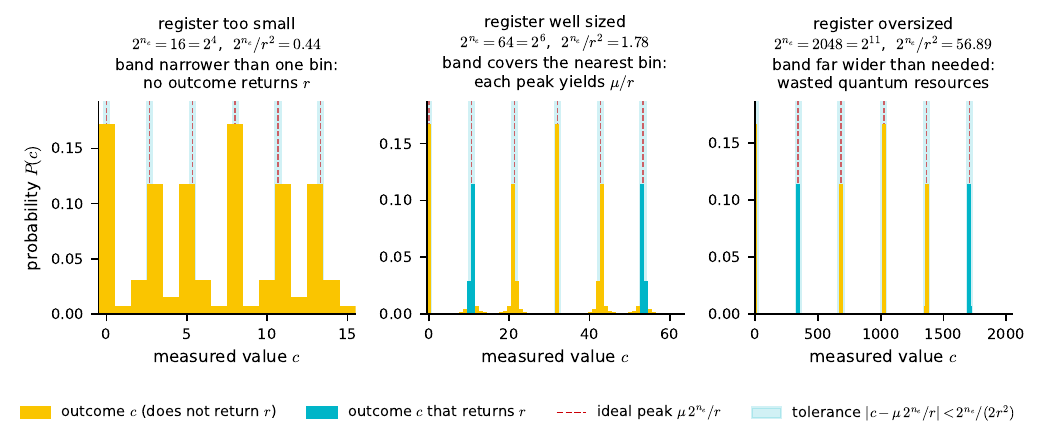}
    \caption{Register size sets peak resolution, for period $r=6$. Each panel shows the
    probability $P(c)$ of measuring the outcome $c$ on the $n_e$-qubit exponent register
    after the inverse QFT. Dashes mark the ideal peaks $\mu\,2^{n_e}/r$, with
    $\mu\in\{0,\dots,r-1\}$. Continued fractions return $\mu/r$ whenever $c$ falls in the
    shaded band $|c-\mu\,2^{n_e}/r| < 2^{n_e}/(2r^2)$ around a peak, and teal bars are the
    outcomes that return $r$. Left: $2^{n_e}=16<r^{2}$, so each band is narrower than one
    bin and no outcome returns $r$. The two bins inside a band, $c=0$ and $c=8$, belong to
    $\mu=0$ and $\mu=3$, which share a factor with $r$ and return only a divisor of it.
    Middle: $2^{n_e}\gtrsim r^{2}$, so the band always covers the nearest bin. Right: the
    register is far larger than needed.}
  \label{fig:tfq-resolution}
\end{figure*}

\paragraph{Processing: quantum period finding.} Only one step can be accelerated
quantumly. It is the central one, and it computes the order $r$. Classically this is expensive. The best
known methods take time growing sub-exponentially in the number of bits of $N$. Let $f\colon e \mapsto g^e \bmod N$ 
be the function that evaluates all the exponentiation of $g$ modulo $N$. $f$ is a periodic function of period $r$, 
and recovering $r$ from queries to $f$ is an expensive process.
A quantum computer instead evaluates $f$ on a superposition of all inputs at once. 
Because $f$ is periodic with period $r$, applying
the quantum Fourier transform (QFT) concentrates the amplitude on frequencies that are
multiples of $1/r$. Section~\ref{sec:shor} explains this in detail. Measuring then returns such a multiple, from which $r$ can often be recovered. Running 
the circuit and measuring the input register yields an integer $c \approx \mu 2^{n_e}/r$.
Here $n_e$ is the input register size, and
$\mu$ is sampled by the measurement. The order $r$ is recovered only when
$\gcd(\mu,r)=1$, which happens with probability $\varphi(r)/r$. Here $\varphi$ is
Euler's totient function, which counts the positive integers up to its argument that are
coprime to it. Since $\varphi(r)/r = \Omega(1/\log\log r)$, $O(\log\log r)$ repetitions
of the circuit suffice in expectation (Section~\ref{sec:shor}).

The register holding the input exponents must be large enough for this to work. A QFT over
a register of $n_e$ qubits resolves a frequency to a precision of about $1/2^{n_e}$.
Separating neighboring multiples of $1/r$ would only need $2^{n_e}\gtrsim r$. Recovering
$r$ needs more. The measured fraction $c/2^{n_e}$ must lie within $1/(2r^{2})$ of
$\mu/r$, so that continued fractions return $\mu/r$ in lowest terms
(Section~\ref{sec:shor}). This requires $2^{n_e}\gtrsim r^{2}$, i.e.\ about $2n$ qubits
for an $n$-bit modulus. Between $r$ and $r^{2}$ the peaks are still separated, but
continued fractions can no longer identify $r$. Beyond $r^{2}$ the arithmetic on the
register only grows more costly for no benefit. We depict the register size impact in
Figure~\ref{fig:tfq-resolution}. Crucially, the whole routine (one evaluation of $f$ in
superposition, one QFT, and one measurement), repeated $O(\log\log r)$ times in
expectation, uses
only $\operatorname{poly}(n)$ operations. This is the superpolynomial speedup at the heart of
Shor's algorithm: order finding, classically intractable, becomes efficient with a quantum
algorithm.

\medskip\noindent\emph{A concrete run.} Take $N = 77$. Suppose the pre-processing draws $g
= 18$. Then $\gcd(18,77)=1$, so we proceed to find the order. Computing successive powers
modulo $77$, 
\begin{align*}
  18^1 = 18,\quad 18^2 = 16,\quad 18^3 = 57,\ \ldots,\ \\ 18^{30}\equiv 1 \pmod{77},
\end{align*}
the first return to $1$ happens at $r = 30$, so the order of $18$ is $r = 30$. This $r$ is
even, and $18^{r/2} = 18^{15} \equiv 43 \pmod{77}$, which is neither $1$ nor $-1 \equiv 76$.
Post-processing then gives
\begin{align*}
  \gcd(18^{15}-1,\,77) = \gcd(42,77) = 7,
  \\
  \gcd(18^{15}+1,\,77) = \gcd(44,77) = 11,
\end{align*}
and indeed $77 = 7 \times 11$. Not every base works. Had the pre-processing drawn
$g = 15$ instead, its powers are
\begin{align*}
  15^1 = 15,\quad 15^2 = 71,\quad 15^3 = 64,\\ 15^4 = 36,\quad 15^5 \equiv 1 \pmod{77},
\end{align*}
so the order is $r = 5$, which is \emph{odd}. The identity $g^r - 1 =
(g^{r/2}-1)(g^{r/2}+1)$ requires an integer exponent $r/2$, so it does not apply
here. This run therefore yields no factor. One simply discards $g = 15$ and draws a fresh
base. This is the ``probability $\ge \tfrac12$'' from the post-processing above, so a
handful of tries succeeds with overwhelming probability.

\medskip

\medskip\noindent\emph{Why a period looks like a hidden subgroup.} The quantum period-finding
step is the heart of the algorithm. The structure the quantum step actually exploits is best 
seen through \emph{cosets}. Take the integers $\mathbb{Z}$ under addition and the subgroup 
$H = 2\mathbb{Z}$ of even numbers. Translating $H$ by a fixed element $a$ gives the set 
$a + H = \{a + h : h \in H\}$, called a coset. Here there are exactly two: $0 + H$, the even 
numbers, and $1 + H$, the odd numbers. They \emph{partition} $\mathbb{Z}$: every integer lies 
in exactly one, and two integers land in the same coset precisely when they differ by an element
of $H$, i.e.\ by an even number.

Order finding has the same shape. Fix $g$ and consider the function $f(x) = g^{x} \bmod N$
on the exponents $x \in \mathbb{Z}$. Because $g^{r} \equiv 1$, we have $f(x + r) = f(x)$:
the function repeats with period $r$. Playing the role of $H$ is the subgroup
$r\mathbb{Z} = \{0, \pm r, \pm 2r, \ldots\}$ of exponents that leave $f$ unchanged, and its
cosets $a + r\mathbb{Z}$ are the sets of exponents sharing a common value of $f$. Two
properties, which are what the quantum step relies on, hold:
\begin{itemize}
  \item $f$ is \emph{constant on each coset}: if $x$ and $x'$ differ by a multiple of $r$,
    then $f(x) = f(x')$.
  \item $f$ \emph{separates distinct cosets}: if $x$ and $x'$ lie in different cosets, then
    $f(x) \neq f(x')$.
\end{itemize}
Concretely, for $g = 18$, $N = 77$ above, $r = 30$, so the exponents split into $30$ cosets
$\{0 + 30\mathbb{Z},\ 1 + 30\mathbb{Z},\ \ldots,\ 29 + 30\mathbb{Z}\}$. The coset $1 +
30\mathbb{Z}$, for instance, is exactly the set of exponents $x$ with $18^{x}\equiv 18
\pmod{77}$. The values $x = 1, 31, 61, \ldots$ all give $18$, and no exponent outside this
coset does. Recovering the period $r$ is therefore the same task as recovering the hidden
subgroup $H$. The function available is constant on the cosets of $H$ and distinct across
them. This
is the \emph{Hidden Subgroup Problem}, and stating it precisely is what the next
subsection does. 

\medskip

In his seminal paper, Shor~\cite{Shor97} also provides a simple extension of this
algorithm to the discrete logarithm problem. The extension is natural: both factoring and
the discrete logarithm reduce to the same underlying computational primitive, finding a
hidden subgroup of an abelian group, and the quantum subroutine is structurally identical
in both cases. The only differences lie in the classical pre- and post-processing. We make
both reductions explicit in the following subsection.

\subsection{Hidden Subgroup Problem}\label{sec:hsp}

Let us first start by defining what a coset is and the notion of constant and separating
for a coset.

\begin{definition}[Coset]\label{def:coset}
  Let $G$ be an abelian group and $H$ a subgroup of $G$. The \emph{coset} of $H$
  associated with an element $g \in G$ is the set
  \[
    gH = \{ gh : h \in H \}.
  \]
  When $G$ is finite, the cosets of $H$ partition $G$ into $|G|/|H|$ subsets, each of size $|H|$.
\end{definition}

All groups in this section are abelian and written additively, namely $\mathbb{Z}$,
$\mathbb{Z}_N$, and products of such groups. For them, the coset $gH$ of
Definition~\ref{def:coset} is written $g+H = \{g+h : h \in H\}$ instead. This is
exactly the notation already used above, where the cosets of $H=2\mathbb{Z}$ in
$\mathbb{Z}$ were written $a+H$.
 
\begin{definition}[Constant and separating on cosets]\label{def:const-sep}
  Let $G$ be a group and $S$ a set and $H$ a subgroup of $G$. A function $f : G \to S$ is said to be 
  \emph{constant and separating on the cosets of $H$ } if it takes the same value 
  on every element of a given coset, and distinct values on distinct cosets. Formally,
  \[
    f(g_1) = f(g_2) \iff g_1 H = g_2 H.
  \]
  In other words, $f$ perfectly labels the cosets of $H$: knowing $f(g)$ tells you
  exactly which coset $g$ belongs to, but nothing more about $g$ itself.
\end{definition}

The \emph{Hidden Subgroup Problem} (HSP)~\cite{Kitaev95,BonehLipton95} asks the following
question. We are given a group and a function on it. The function is constant and separating
on the cosets of some unknown hidden subgroup. The task is to determine that subgroup by
querying the function. It
is a generalization of well-known problems such as factoring or discrete logarithm. More
precisely, the problem is stated as follows: 

\begin{definition}[Hidden Subgroup Problem~\cite{HRT03}]\label{def:hsp}
  Let $f\colon G \rightarrow S$ be an efficiently computable function from a group $G$
  to a set $S$, where $G$ is finite or finitely generated abelian. Suppose $f$ is constant
  and separating on the cosets of some subgroup $H$. Determine the subgroup $H$.
\end{definition}

The definition of~\cite{HRT03} is stated for finite groups. We also allow finitely
generated abelian groups such as $\mathbb{Z}$, which order finding requires.

We now instantiate Definition~\ref{def:hsp} on the three computational problems
that this paper studies: integer factoring, the finite-field discrete logarithm
problem (FFDLP), and the elliptic-curve discrete logarithm problem
(ECDLP). All three reduce to finding a hidden subgroup of an abelian
group, and all three use a group exponentiation as the HSP oracle. They differ in three ways. 
First, the group order may or may not be known in advance.
Second, the element order compares differently with the group order. Third, they need
different numbers of exponent registers, or registers of different sizes. 

\subsubsection*{Factoring.} Let $N = pq$ be an $n$-bit RSA integer with $p,q$ unknown odd
primes of similar size. The goal is to determine $p$ and $q$.

Pick a random $g\in\mathbb{Z}_N^{\ast}$ and set $f(x)=g^{x}\bmod N$. Its period is the
multiplicative order $r=\mathrm{ord}_{\mathbb{Z}_N^{\ast}}(g)$. The function $f$ separates
cosets of the hidden subgroup $H=\langle r\rangle=r\mathbb{Z}=\{0,\pm r,\pm 2r,\dots\}\le
\mathbb{Z}$ (where $H\le G$ means $H$ is a subgroup of $G$):  
\begin{equation}\label{eq:orderfinding-coset}
    \begin{split}
      f(x_1)=f(x_2)& \iff g^{x_1}\equiv g^{x_2}\!\!\pmod N \\ & \iff r\mid(x_1-x_2)\\ & \iff x_1+H=x_2+H.
    \end{split}
\end{equation}
Order finding is therefore the hidden subgroup problem for $(G,H)=(\mathbb{Z},\,r
\mathbb{Z})$. Recovering $H$ is equivalent to recovering its generator $r$. In the
modern Ekerå--Håstad variant, however, the algorithm no longer performs order finding. It
reduces factoring to a short discrete logarithm in $\mathbb{Z}_N^*$, a group of unknown
order~\cite[Sect.~5.2]{PQCRYPTO:EkeHas17}.

This HSP lives on the infinite group $\mathbb{Z}$, while the quantum circuit can only ever
query $f$ on a finite register. Lemma~\ref{lem:orderfinding-hsp} shows that the finite register preserves which inputs collide under $f$; its size controls an approximation error. The factoring algorithm is summarized as follows: 

\begin{description}
  \item[\textbf{Given}] An $n$-bit RSA modulus $N=pq$ and a random $g\in\mathbb{Z}_N^{\ast}$. A
    $2n$-qubit exponent register, so that $2^{2n}\ge N^{2}$. A quantum black box that
    evaluates the function $f:\ket{x}\ket{y}\mapsto\ket{x}\ket{(g^{x}\bmod N)\oplus y}$.
  \item[\textbf{Promise}] $f$ is constant and separating on the cosets of
    $H=r\mathbb{Z}\le\mathbb{Z}$, where $r=\mathrm{ord}_{\mathbb{Z}_N^{\ast}}(g)$
    (Lemma~\ref{lem:orderfinding-hsp}). 
  \item[\textbf{Output}] The order $r$, in time $O(n^3)$, using at most $O(n)$ qubits\footnote{The
      time-space complexity from the seminal Shor
      paper~\cite{Shor97}.}. The factors $p$ and $q$ are then recovered classically via
    $\gcd(g^{r/2} \pm 1, N)$. 
\end{description}

\paragraph*{Finite register for factoring.} The register size $2^{2n}\ge N^{2}>r^{2}$ for 
the factoring case is chosen for purely computational reasons
rather than structural ones. It guarantees $\approx 2^{2n}/r>r$ periods inside the
register, so the resulting phase peaks are sharp. It also gives the approximation bound
$1/(2\cdot 2^{2n})\le 1/(2r^{2})$ that continued-fraction recovery needs (Section~\ref{sec:shor}).  
Since $r$ itself is unknown before the algorithm runs, the checkable condition $2^{2n}\ge N^{2}$ is used in its place. 

\begin{lemma}[The finite register preserves coset collisions]\label{lem:orderfinding-hsp}
Restrict $f$ to the register $X=\{0,1,\dots,2^{2n}-1\}$. Two values $x_1,x_2\in X$
satisfy $f(x_1)=f(x_2)$ if and only if $x_1+H=x_2+H$ in $\mathbb{Z}$. The collision
classes $X\cap(a+H)$ are the cosets of a subgroup of $\mathbb{Z}_{2^{2n}}$ if and only if
$r\mid 2^{2n}$.
\end{lemma}

\noindent\emph{Proof.} The first claim is Equation~\eqref{eq:orderfinding-coset} applied
to $x_1,x_2\in X$. Restricting the domain removes pairs but changes no equivalence between
the pairs that remain. For the second claim, write $2^{2n}=kr+s$ with $0\le s<r$. Since
$r<N\le 2^{2n}$, we have $k\ge1$. The class of $a\in\{0,\dots,r-1\}$ has $k+1$ elements if
$a<s$ and $k$ elements otherwise. The cosets of a subgroup all have the same size, so
$s=0$ is necessary. Conversely, if $r\mid 2^{2n}$, the classes are exactly the cosets of
$r\mathbb{Z}_{2^{2n}}$.

The register therefore preserves which inputs collide, whatever its size. What its size
controls is the Fourier side. When $r\nmid 2^{2n}$, the QFT over $\mathbb{Z}_{2^{2n}}$
concentrates the amplitude near the multiples $\mu\,2^{2n}/r$ rather than on them. This is
why the measured frequency only approximates $\mu/r$, why the continued-fraction step of
Section~\ref{sec:shor} is needed, and why the register must satisfy
$2^{2n}\gtrsim r^{2}$ (Figure~\ref{fig:tfq-resolution}).

\subsubsection*{Discrete logarithm.} Let $p$ be a prime, $g$ a generator of
$\mathbb{Z}_p^*$, and $x = g^d \bmod p$ a given group element. The goal is to compute the
discrete logarithm $d = \log_g x \in \mathbb{Z}_{p-1}$.

Let $G = \mathbb{Z}_{p-1} \times \mathbb{Z}_{p-1}$ and define
\[
  f : G \to \mathbb{Z}_p^*, \qquad f(a,b) = g^a x^{-b} \bmod p = g^{a - bd} \bmod p.
\]
The function $f$ separates cosets of the hidden subgroup
\begin{align*}
  H &= \bigl\{(a,b) \in \mathbb{Z}_{p-1}^2 : a \equiv bd \pmod{p-1}\bigr\} \\
    &= \langle (d, 1) \rangle \leq G.
\end{align*}
This holds because $f(a_1, b_1) = f(a_2, b_2)$ if and only if $a_1 - b_1 d \equiv a_2 - b_2 d
\pmod{p-1}$, i.e.\ $(a_1,b_1)$ and $(a_2,b_2)$ lie in the same coset of $H$.
Determining $H$ is equivalent to determining the generator $(d,1)$, which yields $d$
directly. The overall process is summarized as follows: 

\begin{description}
  \item[\textbf{Given}] The group $G = \mathbb{Z}_{p-1} \times \mathbb{Z}_{p-1}$
    for a prime $p$, with generator $g \in \mathbb{Z}_p^*$. A target element
    $x = g^d \bmod p$. A quantum black box that evaluates the function
    $f : |a\rangle|b\rangle|y\rangle \to |a\rangle|b\rangle|g^a x^{-b}
    \bmod p \oplus y\rangle$.
  \item[\textbf{Promise}] $f$ is constant and separating on the cosets of
    $H=\langle(d,1)\rangle\le G$, as shown above.
  \item[\textbf{Output}] The discrete logarithm $d = \log_g x$, in time
    $O(n^3)$, and using at most $O(n)$
    qubits\footnote{Here again, the time-space complexity comes
      from~\cite{Shor97}.}. 
\end{description}

In contrast to order finding, the discrete logarithm is already a hidden subgroup problem
on a \emph{finite} group, $G=\mathbb{Z}_{p-1}^2$. Its hidden subgroup $H=\langle(d,1)\rangle$ is a genuine
subgroup of $G$, so no passage to $\mathbb{Z}$ is required (cf.\
Lemma~\ref{lem:orderfinding-hsp}). Because the order $p-1$ is known, the exponent
registers could in principle be matched to $\mathbb{Z}_{p-1}$. This requires a Fourier
transform modulo $p-1$, which is not the power-of-two transform of standard circuits.
Practical circuits instead use power-of-two registers of about the bit length of $p$, as
Shor already did for general $p$~\cite[Sect.~7]{Shor97}. In factoring, by contrast, the 
unknown order $r$ forces the over-provisioned register $2^{2n}\ge N^{2}$. 

In the following, we distinguish two cases of the discrete logarithm problem, because the
optimizations of Shor's algorithm developed separately for each. They are the finite field discrete logarithm problem 
($\mathtt{FFDLP}$) and the elliptic curve discrete logarithm problem ($\mathtt{ECDLP}$).
What distinguishes $\mathtt{ECDLP}$ from $\mathtt{FFDLP}$ is not the hidden
subgroup but the cost of evaluating $f$, and we take it up in Section~\ref{sec:ecdlp}.

\subsection{The Noisy Hidden Subgroup Problem}

Definition~\ref{def:hsp} is an idealized promise: $f$ is \emph{exactly} constant on
cosets of $H$ and \emph{exactly} distinct across distinct cosets. This is the right
level of abstraction for stating what Shor's algorithm solves in principle, but it is
not, in general, the object that a physical circuit actually queries. Realizing $f$ on
a finite or truncated register can perturb its values away from the exact promise
(Section~\ref{sec:oracle-rns}). Any measurement-based read-out built on
top of $f$ can occasionally fail on a query (Sections~\ref{sec:input-register}
and~\ref{sec:output-register}). Both effects recur throughout the rest of this
section. The finite register used for order finding preserves the coset collisions
\emph{exactly} (Lemma~\ref{lem:orderfinding-hsp}). The perturbations therefore come from the successive
optimizations of the oracle, namely smaller registers, compressed outputs and approximate
arithmetic. Each introduces a controlled perturbation, either a bounded drift in the value
of $f$ or a read-out that can occasionally fail. Rather than handling each perturbation with a separate ad hoc
argument, we capture the two ways the exact HSP promise can be relaxed in a single
definition. Each later perturbation can then be located on the same two axes.

We introduce two parameters to make this relaxation precise. The first, that we denote $\epsilon$, bounds how far a
single query's output can drift from the value the exact promise would have given. It is a
property of the oracle alone. The second, that we denote $\delta$, bounds how often the algorithm's
own read-out, whatever measurement and post-processing it performs on top of $f$, is
allowed to fail. The two behave differently. $\epsilon$ does not depend on how the oracle
is used, while $\delta$ depends entirely on the choice of read-out built on top of it. The same noisy 
oracle can be turned into a reliable or an unreliable instance of the HSP depending only on that choice. 
This device is the organizing lens that we introduce for comparing variants, not a standard notion. Related
robustness analyses exist, for instance the approximate-periodicity arguments
of~\cite{C:CheFouSch25} and~\cite[Sect.~2.2]{gidney2025factor2048bitrsa}.

The per-shot failure rate $\delta$ splits into two kinds with very different costs, and we
track them separately.
 
\begin{definition}[Read-out: abort and corruption]\label{def:sep-proc}
A \emph{read-out} $R$ is the measurement, classical post-processing and acceptance test
applied to a single invocation of an oracle $\mathcal{O}$ computing $f$. One shot returns
a verdict $R\in\mathcal{Y}\cup\{\bot\}$, where $\bot$ is a \emph{discard} symbol returned
when the acceptance test fails. The verdict is random over the shot's measurement
outcomes, together with any auxiliary randomness the read-out draws. The post-processing
reads each $y\in\mathcal{Y}$ as a \emph{constraint} on the hidden subgroup, that is, as a
set $C_y$ of subgroups of $G$ compatible with $y$. The read-out must state $C_y$,
including any tolerance it allows. A verdict $y$ is \emph{$H$-consistent} if $H\in C_y$,
that is, if the constraint it is read as is true. Let
$\mathcal{Y}_H=\{y\in\mathcal{Y}: H\in C_y\}$. A shot is a \emph{success} if
$R\in\mathcal{Y}_H$, an \emph{abort} if $R=\bot$, and a \emph{corruption} if
$R\in\mathcal{Y}\setminus\mathcal{Y}_H$. The two failure probabilities and their sum,
\begin{align*}
  \delta_{\bot}=\Pr[R=\bot],\qquad
  \delta_{\times}=\Pr[R\in\mathcal{Y}\setminus\mathcal{Y}_H],\\
  \delta(R,\mathcal{O})=\delta_{\bot}+\delta_{\times}=\Pr[R\notin\mathcal{Y}_H],
\end{align*}
are taken as worst case over the instances of the problem, that is, over $H$ and over the
oracles that hide it. The subscripts name the verdict each rate tracks: $\bot$ for the
discard symbol itself, and $\times$ for a wrong-but-accepted answer.
\end{definition}

An abort is repaid by taking more shots. A corruption cannot be repaid that way, since the
post-processing cannot tell it from a success, so it must instead be made rare or caught
by a later check on a whole attempt (Lemma~\ref{lem:compose}). A success need not be
informative. A true constraint can be weak, and the number of successes the
post-processing needs is a separate quantity. The rates above are absolute. When a
mechanism is added to a baseline read-out, we report the rates it adds.

\begin{definition}[$(\epsilon,\delta)$-noisy HSP]\label{def:noisy-hsp}
Let $G$ be a finite group, $H\le G$, and $S$ a set carrying a metric $d_S$. Let
$\mathcal{O}$ be an oracle computing $f\colon G\to S$, read by a read-out $R$
(Definition~\ref{def:sep-proc}). For $\epsilon\ge 0$ and $\delta\in[0,1)$,
$\mathcal{O}$ is an \emph{$(\epsilon,\delta)$-noisy instance of the HSP for $H$ under
$R$} if
\begin{enumerate}
  \item[(i)] there is a function $f_0\colon G\to S$, constant and separating on the
    cosets of $H$ in the sense of Definition~\ref{def:const-sep}, with 
    $d_S\bigl(f(g),f_0(g)\bigr)\le\epsilon/2$ for every $g\in G$;
  \item[(ii)] $\delta(R,\mathcal{O})\le\delta$.
\end{enumerate}
\end{definition}
 
The factor $\tfrac12$ in~(i) is a normalization. At $\epsilon=0$ the oracle meets the
exact promise of Definition~\ref{def:hsp}, and $\delta$ then measures the read-out alone.
The metric $d_S$ matters only when $\epsilon>0$. For the residue-arithmetic oracle,
$S=\mathbb{Z}_N$ and $d_S(u,v)=\min_{k\in\mathbb{Z}}|u-v-kN|/N$, the normalized modular
distance of Equation~\eqref{eq:deviation}.
 
\begin{remark}[When corruption needs value drift]\label{rem:corruption-epsilon}
An exact oracle ($\epsilon=0$) can still corrupt. Whether it does depends on the
read-out. Two situations rule it out. With an exact Fourier transform over $G$, every
outcome lies in the subgroup $H^{\perp}$ orthogonal to $H$, so the constraint
``$y\in K^{\perp}$'' is always true of $K=H$. With an acceptance test that verifies the
constraint, such as the check $g^{r'}\equiv 1$ in order finding, a false constraint
becomes an abort instead. Outside these two situations an exact oracle can corrupt. The
bad pairs of the Ekerå--Håstad read-out are an example
(Section~\ref{sec:input-register}). Value drift adds a second source of corruption. The
same $\epsilon$-drifting oracle turns its drift into corruptions ($\delta_{\times}\to 1$)
under a full-precision measurement, yet the masked read-out of
Section~\ref{sec:oracle-rns} absorbs most of that drift into aborts. This is why $\delta$
is stated relative to $R$. Among the instances of Table~\ref{tab:four-relaxation}, only
the residue-arithmetic oracle has $\epsilon>0$.
\end{remark}

\begin{remark}[$G$ need not contain a literal subgroup]
Conditions~(i) and~(ii) of Definition~\ref{def:noisy-hsp} only ever use the coset
partition of $G$ induced by $H$. Condition~(i) uses it through the exact $H$-periodicity of $f_0$.
Condition~(ii) uses it through the $H$-consistent set $\mathcal{Y}_H$. That set, in turn, is
fixed by the same partition. Neither needs a literal subgroup $H\le G$, so any $G$
realizing the partition exactly will do, whether or not an actual subgroup sits inside it.
\end{remark}

\subsubsection*{Four relaxations of the exact promise.}

The remainder of this section develops the factoring case, starting with one baseline
algorithm and three successive improvements. Section~\ref{sec:shor} builds Shor's full
order-finding algorithm: a large exponent register, a full-width output register, and
exact modular arithmetic. 
Three sections that follow each reduce one of
these costs. Section~\ref{sec:input-register} shrinks the exponent register, so that a
single measurement no longer determines the answer and it is recovered from repeated
sampling instead. Section~\ref{sec:output-register} compresses the output register,
hashing the oracle's output down toward a single qubit. Section~\ref{sec:oracle-rns} makes
the oracle's arithmetic space-efficient, computing the modular exponentiation
approximately in a residue number system rather than exactly. None of the three reductions is free. Each relaxes the exact promise of
Definition~\ref{def:hsp} in a controlled way. It lets either the oracle's values drift or
the read-out occasionally fail. These are the
two parameters $\epsilon$ and $\delta$ of
Definition~\ref{def:noisy-hsp}. Table~\ref{tab:four-relaxation} places the baseline and
the three optimizations on them, and points to the section that derives each. The
elliptic-curve problem of Section~\ref{sec:ecdlp} reuses these relaxations only in part
and adds one the factoring problem never needed. Because it has no single best algorithm,
its landscape is mapped separately there.

\begin{table*}[t]
\centering
\renewcommand{\arraystretch}{1.35}
\setlength{\tabcolsep}{6pt}

\begin{tabular}{@{}p{2.8cm} c c c p{3.6cm} p{3.0cm}@{}}
\toprule
\textbf{Instance} & $\boldsymbol{\epsilon}$
 & \multicolumn{2}{c}{$\boldsymbol{\delta=\delta_{\bot}+\delta_{\times}}$}
 & \textbf{Read-out} $\boldsymbol{R}$ & \textbf{In \S\ref{sec:full-modern}} \\
\cmidrule(lr){3-4}
 & & $\delta_{\bot}$ {\footnotesize\itshape abort}
   & $\delta_{\times}$ {\footnotesize\itshape corrupt} & & \\
\midrule
Finite register\newline(order finding, baseline)
 & $0$ & $\le 1-\tfrac{4}{\pi^{2}}\tfrac{\varphi(r)}{r}$ & $0$
 & measure the exponent register, continued fractions, accept $r'$ iff $g^{r'}\equiv 1$
 & no. Superseded by the short-DLP reduction and the register of \S\ref{sec:input-register} \\
\addlinespace
Ekerå--Håstad good pair\newline(\S\ref{sec:input-register})
 & $0$ & $0$ & $\le 7/8$
 & measure $(j,k)$, accept every pair. A bad pair cannot be detected per shot, so it
   corrupts
 & register shape only. Read-out replaced by graded post-processing~\cite{Eke20} \\
\addlinespace
May--Schlieper hash\newline(\S\ref{sec:output-register})
 & $0$ & $\le 2^{-t}$ & $0$
 & draw $h\sim\mathcal{H}_t$, measure $h\circ f$, group by equality
 & Indirectly: masking is used instead \\
\addlinespace
RNS $\tilde V$ + masking\newline(\S\ref{sec:oracle-rns})
 & $O(|P|\,\ell\,2^{-\kappa})$ & $\approx S$ & $\approx \epsilon/S$
 & draw mask, add $\tilde V$, measure (mask width $SM$)
 & Yes \\
\bottomrule
\end{tabular}
\caption{The four $(\epsilon,\delta)$-noisy HSP relaxations studied in this section,
  with the per-shot failure resolved into aborts $\delta_{\bot}$ (a wasted shot) and
  corruptions $\delta_{\times}$ (a wrong constraint accepted). Each row lists the failure
  that mechanism introduces when taken on its own. The rows are not stages of one
  pipeline, and some of them replace one another. The last column states which of them
  the RSA-2048 pipeline of Section~\ref{sec:full-modern} uses. Corruption occurs in two
  rows. In the Ekerå--Håstad row an exact oracle corrupts through bad pairs, which no
  per-shot test can detect. In the RNS row the oracle values themselves drift
  ($\epsilon>0$). The other two rows can fail only by aborting
  (Remark~\ref{rem:corruption-epsilon}). The RNS optimum
  $S=\sqrt{\epsilon}$ is the point that balances $\delta_{\bot}$ against
  $\delta_{\times}$, giving $\delta\approx 2\sqrt{\epsilon}$. Both RNS rates are heuristic estimates
(Section~\ref{sec:oracle-rns}).}

\label{tab:four-relaxation}

\end{table*}

\subsection{Shor's Algorithm}\label{sec:shor}

We first present the algorithm in its one-dimensional form, order finding for the
factoring problem. This is the classical baseline and the clearest setting for the quantum
core. The modern optimizations instead treat factoring in two dimensions, through a
reduction to a short discrete logarithm that we give in Section~\ref{sec:short-dlp},
immediately afterward.

\begin{figure*}[tp]
  \centering
  \def\rad{3pt}           
  \def\labelband{0.55cm}  
  \def\blockw{1.9cm}
  \tikzset{
    abgroup/.style={draw=abTealDark, fill=abTeal!12, dashed, line width=0.7pt,
                    rounded corners=\rad, inner xsep=3pt},
    abgate/.style={draw=abInk, fill=white, rounded corners=\rad},
    abhighlight/.style={draw=abInk, fill=abYellow, line width=1.2pt, rounded corners=\rad},
    abclassical/.style={draw=abPurpleDark, fill=abPurple!18, line width=1pt,
                        rounded corners=\rad},
  }
  \providecommand{\grouplabel}[2][abTealDark]{{\small\color{#1}#2}}
  \begin{tikzpicture}[
      flow/.style={->, >=stealth, draw=abLine, line width=1pt, shorten >=2pt, shorten <=2pt},
    ]
    \node[inner sep=0pt] (circuit) {
      \begin{quantikz}[row sep={0.65cm,between origins}, column sep=0.8cm]
        \lstick{$\ket{0}$} & \gate[style=abgate]{H}
          \gategroup[4,steps=1,style=abgroup,background]{\grouplabel{State preparation}} & &
        \gate[wires=9, style=abhighlight][2cm]{O_f}
          \gategroup[9,steps=1,style=abgroup,background]{\grouplabel[abInk]{\bfseries Oracle evaluation}}
        & \gate[wires=4, style=abgate][3cm]{\mathrm{QFT}^{-1}}
          \gategroup[4,steps=2,style=abgroup,background]{\grouplabel{Quantum Fourier transform}}
        & \meter[style=abgate]{} & \setwiretype{n} \\
        \lstick{$\ket{0}$} & \gate[style=abgate]{H} & & & & \meter[style=abgate]{} & \setwiretype{n} \\
        \lstick{$\vdots$}  & \lstick{$\vdots$} & & & & \meter[style=abgate]{} & \setwiretype{n} \\
        \lstick{$\ket{0}$} & \gate[style=abgate]{H} & & & & \meter[style=abgate]{} & \setwiretype{n} \\
        \setwiretype{n} & \\
        \lstick{$\ket{0}$} & & & & \meter[style=abgate]{} & \setwiretype{n} & \\
        \lstick{$\ket{0}$} & & & & \meter[style=abgate]{} & \setwiretype{n} & \\
        \lstick{$\vdots$}  & & & & \meter[style=abgate]{} & \setwiretype{n} & \\
        \lstick{$\ket{1}$} & & & & \meter[style=abgate]{} & \setwiretype{n} &
      \end{quantikz}
    };
    \coordinate (wtop) at ([yshift=-\labelband]circuit.north);
    \coordinate (wbot) at ([yshift=0.15cm]circuit.south);
    \node[abclassical, inner sep=0pt, fit={([xshift=-0.6cm-\blockw]circuit.west |- wtop)
                                         ([xshift=-0.6cm]circuit.west |- wbot)}] (pre) {};
    \node at (pre.center) [font=\small\bfseries, text=abPurpleDark, align=center] {Pre-\\processing};
    \node[abclassical, inner sep=0pt, fit={([xshift=-0.15cm]circuit.east |- wtop)
                                         ([xshift=-0.15cm+\blockw]circuit.east |- wbot)}] (post) {};
    \node at (post.center) [font=\small\bfseries, text=abPurpleDark, align=center] {Post-\\processing};
    \draw[flow] (pre.east) -- (pre.east -| circuit.west);
    \draw[flow] ([xshift=-0.75cm]circuit.east |- post.west) -- (post.west);

    \node[below=0.35cm of circuit.south, anchor=north, font=\small, text=abInk] (legend) {%
      \tikz[baseline=-0.6ex]\draw[abclassical] (0,-0.13) rectangle (0.5,0.13);~Classical processing
      \hspace{1.2em}%
      \tikz[baseline=-0.6ex]\draw[abgroup] (0,-0.13) rectangle (0.5,0.13);~Quantum stage
      \hspace{1.2em}%
      \tikz[baseline=-0.6ex]\draw[abhighlight] (0,-0.13) rectangle (0.5,0.13);~Dominant cost};
  \end{tikzpicture}
    \caption{Textbook Shor's algorithm, end to end. The exponent register $\psi_1$
    (top, $n_e=2n$ qubits) is put in uniform superposition by Hadamard gates. The
    accumulator $\psi_2$ (bottom, $n$ qubits) starts in the integer $1$, that is
    $\ket{0^{n-1}1}$. The oracle $O_f$ maps $\ket{e}\ket{1}$ to $\ket{e}\ket{g^{e}\bmod N}$.
    The inverse quantum Fourier transform is then applied to $\psi_1$, which is measured.
    Measuring $\psi_2$ is optional. By the principle of deferred measurement it does not
    change the outcome distribution of $\psi_1$, and it is drawn here because it simplifies
    the analysis. Classical pre- and post-processing choose $g$ and recover $r$ from the
    measured value.
    }
  \label{fig:ms_fullshor}

  \vspace{1.5em}

  \tikzset{
    abmult/.style={draw=abTealDark, fill=abTeal!35, line width=1pt,
                   rounded corners=\rad, text=abInk, inner xsep=6pt},
  }
  \begin{quantikz}[row sep={0.9cm,between origins}, column sep=0.7cm]
    \\
    \lstick{$\ket{e_0}$}       & & \ctrl{4} & & ~\ldots~ & & \\
    \lstick{$\ket{e_1}$}       & & & \ctrl{3} & ~\ldots~ & & \\
    \lstick{$\vdots$}          & & & & ~\ldots~ & & \\
    \lstick{$\ket{e_{n_e-1}}$} & & & & ~\ldots~ & \ctrl{1} & \\
    \lstick{$\ket{1}$} & \qwbundle{n}
      & \gate[style=abmult]{\times g^{2^0} \bmod N}
      & \gate[style=abmult]{\times g^{2^1} \bmod N}
      & ~\ldots~
      & \gate[style=abmult]{\times g^{2^{n_e-1}} \bmod N}
      & \rstick{$g^e \bmod N$} \\
  \end{quantikz}
  \caption{Quantum circuit of the oracle function evaluation. Each of the controlled blocks
  has the same computational cost, showcasing one of the advantages of Shor's algorithm: each constant
  has equal cost since the constant values $g^{2^i} \bmod N$ are precomputed classically.
  }
  \label{fig:ms_oracleF}
\end{figure*}

Shor's algorithm is a three-stage algorithm, state preparation, oracle evaluation, and
quantum Fourier transform, as depicted in Figure~\ref{fig:ms_fullshor}. It is surrounded
by two classical components: pre-treatment and post-treatment. Its principle is the
following: it performs superposition evaluations of the oracle function $f$ to query it on
all its possible inputs, namely all the exponentiations $f \colon e \rightarrow g^e \bmod
N$ for some randomly selected $g$. It then finds the smallest period of $f$, namely its
order, using the quantum Fourier transform and some classical post-processing. 

\subsubsection*{State preparation.}

The first stage of the algorithm is to prepare two quantum registers, $\psi_1$ and
$\psi_2$. Register $\psi_1$ holds the superposition of all $2^{n_e}$ possible exponents,
where $n_e$ is the width of the exponent register. Register $\psi_2$ holds the accumulator
used during the computation of the oracle $O_f$, namely an $n$-qubit register initialized
at $1$, for $n = \log N$. For order finding, $\psi_1$ is the $2n$-qubit register of
Lemma~\ref{lem:orderfinding-hsp}, so $n_e = 2n = O(\log N)$. After state preparation,
$\psi_1$ and $\psi_2$ hold the following values:
\begin{align}
  &\psi_1\colon \ket{0^{n_e}} \xrightarrow{H^{\otimes n_e}}
    \frac{1}{\sqrt{2^{n_e}}}\sum_{x=0}^{2^{n_e} -1}\ket{x}\\ 
  &\psi_2\colon \ket{0^{n-1}1} \xrightarrow{I^{\otimes n}} \ket{0^{n-1}1}
\end{align}

As shown in Figure~\ref{fig:ms_fullshor}, the two registers are easily and efficiently
initialized. $\psi_1$ is put into uniform superposition using a Hadamard layer, and
$\psi_2$ is set to the basis state $\ket{0^{n-1}1}$.

\subsubsection*{Oracle evaluation.} The second stage evaluates the modular exponentiation
$g^{e} \bmod N$ and accumulates the result into the second register $\psi_2$. The exponent
$e$ is encoded in the first register $\psi_1$, and the base $g$ is a fixed classical
constant. The high-level process is depicted in Figure~\ref{fig:ms_oracleF}. It is
structurally identical to the controlled stage of quantum phase estimation of a unitary
$U$, which maps $\ket{j}\ket{u} \mapsto \ket{j}\,U^{j}\ket{u}$ (see Section~5.2 of the
Nielsen and Chuang~\cite{nielsen00}). Writing the exponent in binary, $e = \sum_i 2^i
e_i$, with $e_0$ the least significant bit, the exponentiation can be rewritten as
\begin{equation}
  g^{e} \bmod N = \prod_{i} \bigl(g^{2^{i}}\bigr)^{e_i} \bmod N,
\end{equation}
so the oracle reduces to a sequence of modular multiplications by the constants $g^{2^{i}}
\bmod N$, each controlled by the corresponding exponent qubit $e_i$. Throughout this paper, 
integers are written in this little-endian form, and circuits draw bit $0$ on the top wire.

In a generic phase-estimation setting, the controlled powers $U^{2^{i}}$ are obtained by
repeated application of $U$, so applying $U^{2^{i}}$ costs on the order of $2^{i}$
applications of $U$. The total cost is then dominated by the deepest power, and grows
geometrically. Shor's algorithm avoids this blow-up because the relevant operator is
multiplication in $(\mathbb{Z}/N\mathbb{Z})^\ast$. Each constant $g^{2^{i}} \bmod N$ is
precomputed on a classical computer, and the corresponding controlled multiplication has
the same cost regardless of $i$. Specifically, multiplying an $n$-bit register by a fixed
$n$-bit constant modulo $N$ has a cost that does not depend on the value of that
constant. Consequently an $n_e$-qubit exponent register, spanning $2^{n_e}$ values, is
handled by only $n_e$ controlled multiplications of uniform cost, rather than by
explicitly forming $U^{2^{i}}$ through repeated squaring. This is what makes the modular
exponentiation and hence the whole phase-estimation core of Shor's algorithm efficient.

\subsubsection*{Quantum Fourier transform.}

The third stage applies the inverse quantum Fourier transform to the exponent register
$\psi_1$ and measures it, extracting the period $r$ from the periodicity created by the
oracle. To understand why this works, think of the
oracle not as a black box computing $g^e \bmod N$. It is better seen as the repeated
application of a unitary operator, whose spectral structure encodes $r$ directly.

\begin{figure*}[tp]
\centering
\begin{tikzpicture}[>={Stealth[length=5pt]}, font=\footnotesize]
 
  \begin{scope}[shift={(0,0)}]
    \def\R{1.6}
    \def\rr{8}
 
    \draw[abLine!55, line width=0.5pt] (-\R-0.2,0) -- (\R+0.2,0)
      node[below right, abInk, inner sep=1pt, yshift=-7pt, xshift=1pt]{$\mathrm{Re}$};
    \draw[abLine!55, line width=0.5pt] (0,-\R-0.2) -- (0,\R+0.2)
      node[left, abInk, inner sep=1pt, xshift=-9pt, yshift=-2pt]{$\mathrm{Im}$};
 
    \draw[abLine, line width=0.75pt] (0,0) circle (\R);
 
    \draw[abLine, line width=0.75pt] (0.5,0) arc (0:45:0.5);
    \node[abInk] at (22:0.85) {$\tfrac{2\pi}{r}$};
    \draw[abLine, dashed, line width=0.6pt] (0,0) -- (45:\R);
    \draw[abLine, line width=0.6pt] (0,0) -- (0:\R);
 
    \foreach \d/\lab in {1/{\omega}, 2/{\omega^2}, 3/{\omega^3},
                         4/{\omega^4}, 5/{\omega^5}, 6/{\omega^6}, 7/{\omega^7}} {
      \filldraw[fill=ABTeal, draw=abTealDark, line width=0.6pt] (\d*360/\rr:\R) circle (2pt);
      \node[abTealDark] at (\d*360/\rr:\R+0.3) {$\lab$};
    }

    \filldraw[fill=abYellow, draw=abInk, line width=0.9pt] (0:\R) circle (2.8pt);
    \node[right, abInk] at (0:\R+0.12) {$\lambda_0 = 1$};
 
    \node[align=center, abInk] at (0.2,-\R-0.85)
      {Eigenvalues $\lambda_{\mu} = \omega^{\mu}$,\\ $\omega = e^{2\pi i/r}$\ (here $r=8$)};
  \end{scope}
 
  \draw[abLine!55, dashed, line width=0.5pt] (2.95,-2.95) -- (2.95,2.25);
 
  \begin{scope}[shift={(4.6,0)}]
    \tikzset{ket/.style={draw=abTealDark, fill=ABTeal!14, rounded corners=2.5pt,
                         line width=0.75pt, minimum width=2.2cm, minimum height=0.55cm}}
 
    \node[align=center, abInk] at (0,1.95) {Uniform sum\\ of eigenvectors};
 
    \node[ket] (v0) at (0,1.2)  {$\ket{v_0}$};
    \node[ket] (v1) at (0,0.5)  {$\ket{v_1}$};
    \node[abInk] at (0, 0.05)   {$\vdots$};
    \node[ket] (vr) at (0,-0.55) {$\ket{v_{r-1}}$};
 
    \node[abInk] at (0,-1.3) {$\tfrac{1}{\sqrt{r}}\sum_{\mu}\ket{v_\mu}\;=$};
 
    \node[draw=abInk, fill=abYellow, rounded corners=2.5pt, line width=1.1pt,
          minimum width=2.2cm, minimum height=0.75cm, align=center, text=abInk]
          (one) at (0,-2.15) {\normalsize $\ket{1}$};
 
    \node[align=center, abInk] at (0,-2.98)
      {all phases cancel\\ except on $\ket{1}$};
  \end{scope}
 
\end{tikzpicture}
\caption{Left: the eigenvalues $\lambda_\mu=\omega^\mu$ of $U_g$, the $r$-th roots of
unity (here $r=8$). Right: the corresponding eigenvectors $\ket{v_\mu}$ of
Equation~\eqref{eq:amplitude_interferences} sum to $\ket{1}$,
$\tfrac{1}{\sqrt r}\sum_{\mu}\ket{v_\mu}=\ket{1}$
(Equation~\eqref{eq:uniform-sum}). This is an exact identity. In the sum, the phases
cancel on every basis state except $\ket{1}$.}
\label{fig:ms_eigenvalues}
\end{figure*}
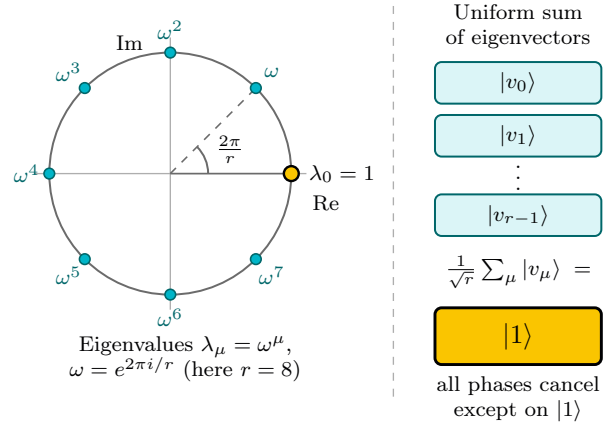

\paragraph{The eigenvalue decomposition.} Define the operator $U_g : \ket{x} \mapsto
\ket{gx \bmod N}$ acting on the output register $\psi_2$. Its action on the output
register is multiplication by $g$ modulo $N$. Since the order of $g$ is $r$, applying
$U_g$ exactly $r$ times returns every state to itself:
\begin{equation}
  U_g^{\,r}\ket{x} = \ket{g^r x \bmod N} = \ket{x},
  \quad\text{so}\quad U_g^{\,r} = I.\footnote{Note that this holds on the full
    computational basis $\{0,\dots,N-1\}$, not only on $\mathbb{Z}_N^*$. Since $g^r \equiv
    1 \pmod{N}$, we have $N \mid x(g^r - 1)$ for every integer $x$. Hence $g^r x \equiv x
    \pmod{N}$, regardless of whether $\gcd(x,N)=1$.} 
\end{equation}
The polynomial $X^r - 1$ therefore annihilates $U_g$. So the eigenvalues of $U_g$
are among the roots of $X^r - 1$, namely the $r$-th roots of unity. Writing $\omega = e^{2\pi i/r}$,
the eigenvalues $\lambda$ and their eigenvectors $v$ are:
\begin{equation}\label{eq:amplitude_interferences}
  \begin{aligned}
  &U_g\ket{v_{\mu}} = \lambda_{\mu}\ket{v_{\mu}},
  \quad \lambda_{\mu} = \omega^{\mu},\\
  &\ket{v_{\mu}} = \frac{1}{\sqrt{r}}\sum_{k=0}^{r-1}\omega^{-{\mu}k}\ket{g^k \bmod N},
   {\mu} \in \{0,\dots,r-1\},
  \end{aligned}
\end{equation}
as illustrated in Figure~\ref{fig:ms_eigenvalues}. These $r$ vectors diagonalize $U_g$
on the $r$-dimensional subspace spanned by the orbit $\{\ket{g^k \bmod N}:0\le k<r\}$ of
$\ket 1$. Being among the roots of $X^r-1$ bounds how many distinct eigenvalues
$U_g$ can have. Since $U_g^r=I$ on the whole $N$-dimensional space (footnote above),
$U_g$ is diagonalizable everywhere. Its remaining $N-r$ eigenvectors simply reuse
these same $r$ values with higher multiplicity, rather than requiring new ones. $U_g$, 
being a permutation of $\{0,\dots,N-1\}$, maps every orbit strictly into itself. 
Starting at $\ket 1$ confines the state, for the whole computation, to the one
$r$-dimensional orbit it starts in, and the eigenvectors of every other orbit never
acquire any amplitude. The phase of the $\mu$-th eigenvalue is $\mu/r$: it encodes $r$ 
directly. The task of period finding therefore reduces to estimating that phase. Preparing any 
specific eigenvector $\ket{v_{\mu}}$ would require knowing $r$, the very quantity we are trying
to find. The key observation that breaks this circularity is the identity
\begin{equation}\label{eq:uniform-sum}
\begin{aligned}
  \frac{1}{\sqrt{r}}\sum_{\mu=0}^{r-1}\ket{v_{\mu}}
  &= \frac{1}{r}\sum_{\mu=0}^{r-1}\sum_{k=0}^{r-1}\omega^{-\mu k}\ket{g^k \bmod N}\\
  &= \frac{1}{r}\sum_{k=0}^{r-1}\Bigl(\sum_{\mu=0}^{r-1}\omega^{-\mu k}\Bigr)\ket{g^k \bmod N}
   = \ket{1}.
\end{aligned}
\end{equation}
The first equality inserts the definition of $\ket{v_\mu}$. The second groups the terms by
basis state. The inner sum adds up the $r$-th roots of unity $\omega^{-k\mu}$. It equals
$r$ for $k=0$ and vanishes for every other $k$, so all basis states cancel except
$\ket{g^0 \bmod N}=\ket{1}$.
The state $\ket{1}$, preparable by setting the output register to the
integer $1$, is already a uniform superposition over these $r$ eigenvectors. By
linearity, running phase estimation on $\ket{1}$ runs it simultaneously on all
eigenvectors at once, without ever having to construct any individual $\ket{v_{\mu}}$.

\paragraph{Running phase estimation on all eigenvectors at once.} The oracle applies
controlled powers of $U_g$: since $\ket{g^e \bmod N} = U_g^{\,e}\ket{1}$, it acts as
$\ket{e}\ket{1}\mapsto\ket{e}\,U_g^{\,e}\ket{1}$. Feeding in $\ket{1}=\frac{1}{\sqrt r}\sum_{\mu}\ket{v_{\mu}}$
and using $U_g^{\,e}\ket{v_{\mu}}=\omega^{\mu e}\ket{v_{\mu}}$,
\begin{equation}
\begin{aligned}
  \frac{1}{\sqrt{2^{n_e}}}\sum_{e=0}^{2^{n_e}-1}\ket{e}\ket{g^{e}\bmod N}
  &=\frac{1}{\sqrt{2^{n_e}}}\sum_{e=0}^{2^{n_e}-1}\ket{e}\,U_g^{\,e}\ket{1}\\
  =\frac{1}{\sqrt r}\sum_{{\mu}=0}^{r-1}\ket{\phi_{\mu}}\ket{v_{\mu}},
  \end{aligned}
\end{equation}
with $\ket{\phi_{\mu}}:=\frac{1}{\sqrt{2^{n_e}}}\sum_{e=0}^{2^{n_e}-1}e^{2\pi i \mu e/r}\ket{e}$.
In each branch ${\mu}$, this single entangled state correlates the eigenvector
$\ket{v_{\mu}}$ in $\psi_2$ with a state $\ket{\phi_{\mu}}$ in $\psi_1$. The amplitude of
$\ket{\phi_{\mu}}$ has the same magnitude on every $\ket e$. Its phase $2\pi\mu e/r$
advances by a fixed step each time $e$ increases by one, a \emph{phase ramp} of frequency
$\mu/r$. The equal weights
$1/\sqrt r$ are inherited directly from $\ket{1}=\frac{1}{\sqrt r}\sum_{\mu}\ket{v_{\mu}}$: every 
eigenphase $\mu/r$ is present with the same amplitude. This is the concrete meaning of 
``running phase estimation on $\ket{1}$ probes all eigenvectors at once'': one run of the circuit 
realizes all $r$ phase-estimation instances $\ket{\phi_{\mu}}$ in superposition.

By the principle of deferred measurement (Nielsen and Chuang~\cite{nielsen00}, \S4.4), one may 
measure $\psi_2$ first, the implicit measurement of Figure~\ref{fig:ms_fullshor}. 
Obtaining a value $g^{e_0}$ leaves $\psi_2$ in the computational-basis state 
$\ket{g^{e_0}}=\frac{1}{\sqrt r}\sum_{\mu=0}^{r-1}\omega^{\mu e_0}\ket{v_{\mu}}$, 
an equal-weight combination of \emph{all} $r$ eigenvectors, not a single one. 
Correspondingly, $\psi_1$ collapses to the periodic superposition
\[
  \ket{\psi_1}\;\longrightarrow\;\frac{1}{\sqrt A}\sum_{k=0}^{A-1}\ket{e_0+kr},
  \qquad
  A=\Bigl\lceil\tfrac{2^{n_e}-e_0}{r}\Bigr\rceil\approx\frac{2^{n_e}}{r},
\]
over the $A\approx 2^{n_e}/r$ exponents consistent with it. Measuring $\psi_2$ therefore does not 
single out one eigenvector. The inverse QFT resolves this periodic superposition of $A$ equally spaced 
spikes into the same $r$ frequency peaks as the equal-weight eigenvector sum above. We follow the
eigenvector form in what follows. 

\paragraph{Phase estimation read-out.} The inverse QFT of size $2^{n_e}$ resolves each
phase ramp $\ket{\phi_{\mu}}$ into a sharp frequency peak. With the phase-estimation
convention $\mathrm{QFT}^{-1}_{2^{n_e}}\ket{x}=\frac{1}{\sqrt{2^{n_e}}}\sum_c e^{-2\pi i xc/2^{n_e}}\ket{c}$,%
\footnote{The standard phase-estimation convention (Nielsen and Chuang~\cite{nielsen00}, \S5.2),
under which a frequency-$\mu/r$ ramp is resolved to a peak at $c\approx \mu\,2^{n_e}/r$. It is the
conjugate of the opposite sign choice and has no direct effect on the recovery of $r$.}
\begin{equation*}\label{eq:qft-output}
  \mathrm{QFT}^{-1}_{2^{n_e}}\ket{\phi_{\mu}}
  =\frac{1}{2^{n_e}}\sum_{c=0}^{2^{n_e}-1}
    \Bigl(\sum_{e=0}^{2^{n_e}-1}e^{2\pi i e\,(\mu/r-c/2^{n_e})}\Bigr)\ket{c},
\end{equation*}
a finite geometric series in $e$. When $\mu/r-c/2^{n_e}$ is close to an integer, its
terms add constructively and the sum reaches magnitude near $2^{n_e}$, so the coefficient
of $\ket{c}$ approaches $1$ in modulus. Elsewhere the terms largely cancel, which produces
a sharp peak at $c\approx \frac{\mu\,2^{n_e}}{r}$\footnote{Explicitly, the
sum evaluates to the Dirichlet kernel $\bigl|\sin\!\bigl(\pi 2^{n_e}(\mu/r-c/2^{n_e})\bigr)
\big/\sin\!\bigl(\pi(\mu/r-c/2^{n_e})\bigr)\bigr|\big/2^{n_e}$, but only its peak location
and sharpness are used below.}. Because the QFT acts on $\psi_1$ alone, it transforms
every branch at once,  
\begin{equation}
  \frac{1}{\sqrt r}\sum_{\mu=0}^{r-1}\ket{\phi_{\mu}}\ket{v_{\mu}}
  \xrightarrow{\mathrm{QFT}^{-1}\otimes I}
  \frac{1}{\sqrt r}\sum_{\mu=0}^{r-1}\bigl(\mathrm{QFT}^{-1}_{2^{n_e}}\ket{\phi_{\mu}}\bigr)\ket{v_{\mu}},
\end{equation}
producing $r$ peaks, one per eigenphase, each still weighted $1/r$. Measuring $\psi_1$ (the
explicit measurement of Figure~\ref{fig:ms_fullshor}) therefore returns an integer $c$ with
\begin{equation}
  \frac{c}{2^{n_e}}\approx\frac{\mu}{r},
  \qquad \mu\in\{0,\dots,r-1\}\ \text{uniform}.
\end{equation}
Since $c$ is the nearest integer to $\frac{\mu\,2^{n_e}}{r}$, the error satisfies
$\lvert \frac{c}{2^{n_e}}-\frac{\mu}{r}\rvert\le 2^{-n_e-1}$. The standard
phase-estimation accuracy bound (Nielsen and Chuang~\cite{nielsen00}, \S5.2) gives each
peak probability at least $4/\pi^2\approx 0.405$, so with constant probability the
measured $c$ is a best approximation to $\mu/r$. The uniformity of $\mu$ is a property of
the spectrum, not of any measurement randomness: the $r$ peaks carry equal weight
precisely because $\ket{1}=\frac{1}{\sqrt r}\sum_{\mu}\ket{v_{\mu}}$ weights the $r$ eigenvectors equally. 

\begin{figure*}[tp]
  \centering
  \def\rad{3pt}
  \tikzset{
    abgate/.style={draw=abInk, fill=white, rounded corners=\rad},
    abrot/.style={draw=abTealDark, fill=abTeal!35, line width=1pt,
                  rounded corners=\rad, text=abInk},
  }
  \begin{quantikz}[row sep={0.9cm,between origins}, column sep=0.30cm]
  \lstick{$\ket{e_0}$} & & & & & & & & ~\ldots~ & \gate[style=abrot]{R_{n_e}^\dagger} & \gate[style=abrot]{R_{n_e-1}^\dagger}
    & ~\ldots~ & & \gate[style=abrot]{R_2^\dagger}\vcw{1} & \gate[style=abgate]{H} & \meter[style=abgate]{} & \rstick{$c_{n_e-1}$} \\
  \lstick{$\ket{e_1}$} & & & & & & & & ~\ldots~ & & & ~\ldots~ & \gate[style=abgate]{H}
    & \meter[style=abgate]{} & \setwiretype{c} & & \rstick{$c_{n_e-2}$} \\
  \lstick{$\vdots$} & & & & & & & & ~\ldots~ & & & ~\ldots~ & & & & & \\
  \lstick{$\ket{e_{n_e-3}}$} & & & & \gate[style=abrot]{R_3^\dagger} & \gate[style=abrot]{R_2^\dagger}\vcw{1} & \gate[style=abgate]{H}
    & \meter[style=abgate]{} & ~\ldots~\setwiretype{c} & & & ~\ldots~ & & & & & \rstick{$c_2$} \\
  \lstick{$\ket{e_{n_e-2}}$} & & \gate[style=abrot]{R_2^\dagger}\vcw{1} & \gate[style=abgate]{H} & & \meter[style=abgate]{}
    & \setwiretype{c} & & ~\ldots~ & & \ctrl[vertical wire=c]{-4} & ~\ldots~ & & & & & \rstick{$c_1$} \\
  \lstick{$\ket{e_{n_e-1}}$} & \gate[style=abgate]{H} & \meter[style=abgate]{} & \setwiretype{c}
    & \ctrl[vertical wire=c]{-2} & & & & ~\ldots~ & \ctrl[vertical wire=c]{-5} & & ~\ldots~ & & & & & \rstick{$c_0$}
\end{quantikz}
    \caption{Semi-classical inverse QFT on the exponent register $\psi_1$. The qubits are measured one at a time,
      from the bottom up. The qubit $e_{n_e-1}$, which controlled the highest power
      $g^{2^{n_e-1}}$, is measured first and gives the least significant bit $c_0$ of the
      outcome $c=\sum_i 2^i c_i$. In general, $e_{n_e-1-i}$ gives $c_i$. Each later qubit is
      first corrected by rotations $R_k^\dagger$, classically controlled on the outcomes already
      measured, and then measured. The rotations on one qubit are drawn separately for clarity.
      In practice they merge into a single rotation, so each qubit needs one Hadamard, one
      rotation and one measurement.}
  \label{fig:ms_qft}
\end{figure*}
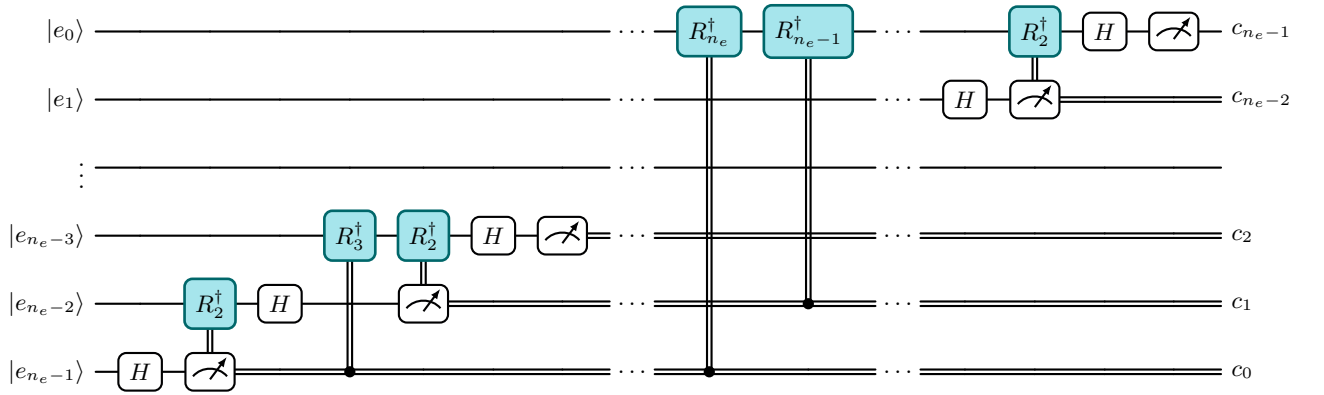

\paragraph{Semi-classical implementation.} Because each qubit of $\psi_1$ is measured
immediately after its Hadamard gate, the QFT can be implemented in its semi-classical
form~\cite{PhysRevLett.76.3228}. The controlled phase rotations
$R_k^{\dagger}\ket{b} = e^{-2\pi i b/2^k}\ket{b}$, for $b\in\{0,1\}$, which the inverse
transform requires on the not-yet-measured qubits, are replaced by rotations
\emph{classically} controlled on the already-measured outcomes. This fixes the order of
the measurements. The qubit $e_{n_e-1}$, which controlled the highest power
$g^{2^{n_e-1}}$, is measured first, and its outcome is the least significant bit $c_0$ of
$c$. The others follow in decreasing order of the power they
controlled~\cite{PhysRevLett.76.3228}, so that $e_{n_e-1-i}$ yields $c_i$. This order is
compatible with the oracle. The controlled multiplications of Figure~\ref{fig:ms_oracleF}
commute, since multiplication modulo $N$ is commutative, so they can be applied in the
order in which the semi-classical transform measures their control qubits. Successive
classically controlled rotations on the same qubit merge into a single rotation, leaving
one Hadamard, one phase gate, and one measurement per qubit. This defers to the classical
world everything that need not be quantum, exactly as the oracle deferred the powers
$g^{2^i} \bmod N$ to classical precomputation. It reduces the QFT to a cost linear in
$n_e$, negligible next to the exponentiation. The resulting circuit is shown in
Figure~\ref{fig:ms_qft}. 

\paragraph{Recycling the exponent register.}\label{par:recycling}
The semi-classical form also saves space. Each qubit of $\psi_1$ is needed only to
control its own multiplications. It is then rotated, measured, and never used again.
Since the controlled multiplications commute, they can be performed in the order in which
the semi-classical transform measures the qubits. The three stages can therefore be
interleaved. One prepares a single control qubit with a Hadamard, applies the controlled
multiplication it steers, applies the classically controlled rotation and the final
Hadamard, measures the qubit, and resets it for the next exponent
bit~\cite{MoscaEkert99,ParkerPlenio00,Zalka98}. With windowing, a block of $w_e$ control
qubits is recycled in the same way. The exponent register then never exists in full. Only
one window of it is alive at any time, whatever the value of $n_e$. The cost analysis
below keeps the three stages separate for clarity, and states the qubit count both with
and without recycling.

The classical recovery of $r$ from $c/2^{n_e}$ follows.

\subsubsection*{Classical post-processing}

The classical step that follows the measurement depends on the modulus of the quantum
Fourier transform. When the order of the group is known, the transform can in principle be
taken modulo that order, and it is then exact. This needs a quantum Fourier transform for
an arbitrary modulus~\cite{MoscaZalka04}, which is more involved than the power-of-two
transform and is not what implementations use. In practice the transform is taken modulo
a power of two, whether the order is known or not, and it only approximates the ideal one.
The register must then be large compared with the order. We describe the exact case
first, because it shows the structure of the outcome most clearly.

\paragraph{Known order: exact transform (idealization).}
In the discrete-logarithm problem, the order $p-1$ of $\mathbb{Z}_p^*$ is known. Assume
an exact quantum Fourier transform over $G = \mathbb{Z}_{p-1} \times \mathbb{Z}_{p-1}$ is
available~\cite{MoscaZalka04}. The measurement then yields a pair $(j,k)$ uniformly distributed
over the subgroup orthogonal to $H = \langle (d,1) \rangle$, that is, over the solutions of
\begin{equation}
  dj + k \equiv 0 \pmod{p-1},
\end{equation}
and every other outcome has amplitude zero. Whenever $j$ is invertible modulo $p-1$, the
discrete logarithm follows directly as $d = -k\,j^{-1} \bmod (p-1)$, with no
approximation. Since $j$ is uniform over $\mathbb{Z}_{p-1}$, this happens with
probability $\varphi(p-1)/(p-1) = \Omega(1/\log\log p)$~\cite{hardywright}.
Shor treats this exact case only when $p-1$ is smooth, so that the transform modulo $p-1$
is easy to build~\cite[Sect.~6]{Shor97}. In that case the Pohlig--Hellman algorithm
already solves the problem classically. For general $p$ he uses a power-of-two register
$q$ with $p\le q<2p$~\cite[Sect.~7]{Shor97}. The outcome then only lies close to the line
$dj+k\equiv 0 \pmod{p-1}$, and recovering $d$ needs an approximate analysis of the same
kind as for order finding below. Every implementation we cite follows this power-of-two
route, including those of Sections~\ref{sec:input-register} and~\ref{sec:ecdlp}.

\paragraph{Unknown order: transform modulo a power of two.}
In order finding, the order $r$ is precisely the unknown, so the transform cannot be
matched to it. It is instead taken over $\mathbb{Z}_q$ with $q = 2^{n_e}$. Since $r$
does not in general divide $q$, the off-peak amplitudes no longer cancel exactly. The
outcome $c$ concentrates near the multiples $q\mu/r$, with some spread. Choosing
$q \geq N^2 > r^2$, phase estimation gives, with probability at least $4/\pi^2$,
\begin{equation}
  \left|\frac{c}{q} - \frac{\mu}{r}\right| \leq \frac{1}{2q} < \frac{1}{2r^2}.
\end{equation}
The measurement reveals $c$ and $q$, but neither $\mu$ nor $r$ individually. They are
recovered with a classical result on continued fractions due to
Legendre~\cite{legendre1830}. If a real number $x$ satisfies
$|x - \mu/r| < 1/(2r^2)$, then $\mu/r$, in lowest terms, is a convergent of the
continued fraction expansion of $x$. Moreover, two distinct fractions with denominators
below $N$ differ by more than $1/N^2 \geq 1/q$. Thus $\mu/r$ is the only such fraction
within $1/(2q)$ of $c/q$, and it is the last convergent of $c/q$ whose denominator is
below $N$.

It therefore suffices to expand $c/q$ as a continued fraction,
\begin{equation}
  \frac{c}{q}
  = \cfrac{1}{a_1 + \cfrac{1}{a_2 + \cfrac{1}{a_3 + \cdots}}}.
\end{equation}
One then computes its successive convergents. This yields $\mu/r$ in lowest terms, whose
denominator is the order $r$ provided $\gcd(\mu, r) = 1$. By Mertens' third
theorem~\cite{hardywright}, the proportion of integers in $\{0, \dots, r-1\}$ coprime
to $r$ satisfies $\varphi(r)/r = \Omega(1/\log\log r)$. A uniformly random $\mu$ is
therefore coprime to $r$ with probability $\Omega(1/\log\log r)$, and
$O(\log\log r)$ repetitions suffice in expectation.

\subsubsection*{Efficient implementation of the oracle.}

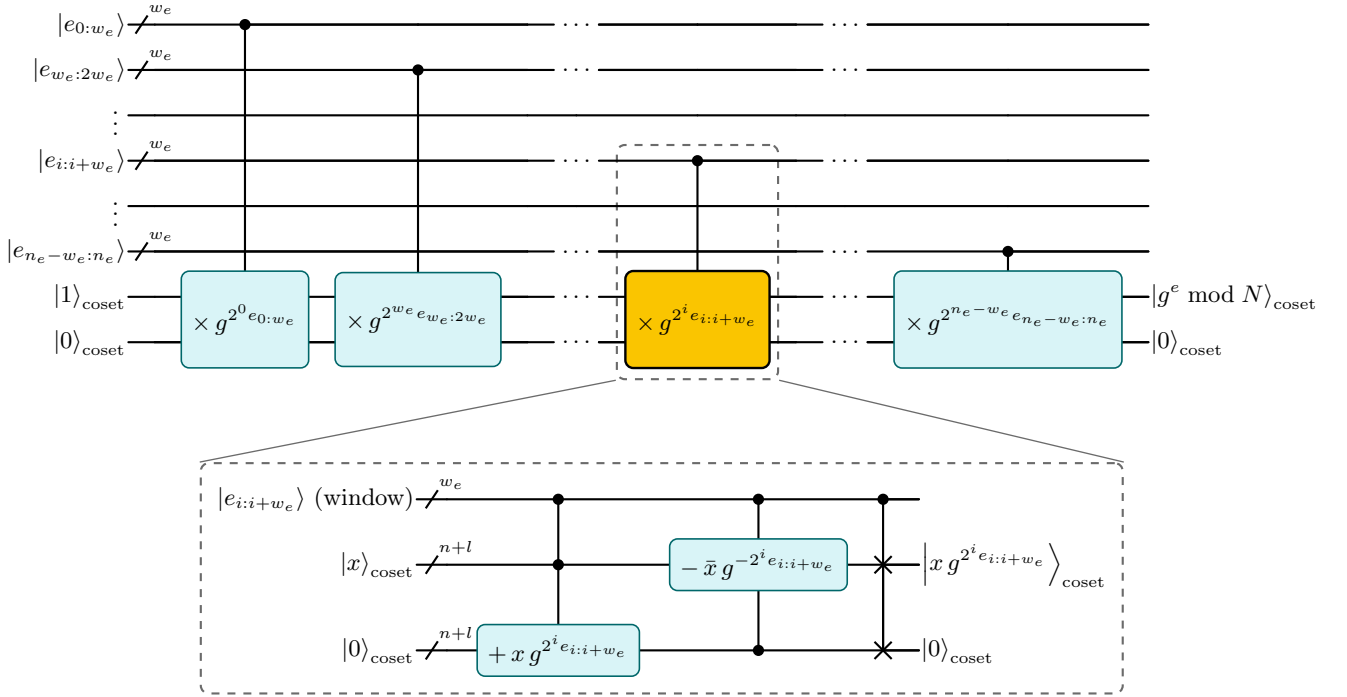
\begin{figure*}[tp]
\centering
\begin{tikzpicture}[remember picture,
  abgate/.style={fill=ABTeal!15, draw=ABTealDark, line width=0.6pt, rounded corners=3pt},
  abhl/.style={fill=abYellow, draw=ABBlack, line width=0.9pt, rounded corners=3pt},
  abzoom/.style={dashed, draw=ABGray, line width=0.8pt, rounded corners=3pt, inner xsep=3pt}]
\node[inner sep=0pt] (topcircuit) {%
\begin{quantikz}[remember picture, column sep=0.35cm, row sep={0.6cm,between origins}]
  \lstick{$\ket{e_{0:w_e}}$} & \qwbundle{w_e}
    & \ctrl{6} & \qw & \qw & \ \ldots\ & \qw & \qw & \ \ldots\ & \qw & \qw \\
  \lstick{$\ket{e_{w_e:2w_e}}$} & \qwbundle{w_e}
    & \qw & \ctrl{5} & \qw & \ \ldots\ & \qw & \qw & \ \ldots\ & \qw & \qw \\
  \lstick{$\vdots\;$} &
    & & & & & & & & & \\
  \lstick{$\ket{e_{i:i+w_e}}$} & \qwbundle{w_e}
    & \qw & \qw & \qw & \ \ldots\
    & \ctrl{3}\gategroup[5,steps=1,style=abzoom,background]{}
    & \qw & \ \ldots\ & \qw & \qw \\
  \lstick{$\vdots\;$} &
    & & & & & & & & & \\
  \lstick{$\ket{e_{n_e-w_e:n_e}}$} & \qwbundle{w_e}
    & \qw & \qw & \qw & \ \ldots\ & \qw & \qw & \ \ldots\ & \ctrl{1} & \qw \\
  \lstick{$\ket{1}_{\mathrm{coset}}$} &
    & \gate[2,style=abgate]{\times\, g^{2^{0}e_{0:w_e}}}
    & \gate[2,style=abgate]{\times\, g^{2^{w_e}e_{w_e:2w_e}}}
    & \qw & \ \ldots\
    & \gate[2,style=abhl]{\times\, g^{2^{i}e_{i:i+w_e}}}
    & \qw & \ \ldots\
    & \gate[2,style=abgate]{\times\, g^{2^{n_e-w_e}e_{n_e-w_e:n_e}}}
    & \qw \rstick{$\ket{g^e \bmod N}_{\mathrm{coset}}$} \\
  \lstick{$\ket{0}_{\mathrm{coset}}$} &
    & \qw & \qw & \qw & \ \ldots\ & \qw & \qw & \ \ldots\ & \qw & \qw \rstick{$\ket{0}_{\mathrm{coset}}$}
\end{quantikz}};
\node[inner sep=0pt, below=1.3cm of topcircuit] (botcircuit) {%
\begin{quantikz}[column sep=0.4cm, row sep=0.45cm]
  \lstick{$\ket{e_{i:i+w_e}}$ (window)} & \qwbundle{w_e}
    & \ctrl{2} & \ctrl{1} & \ctrl{2} & \qw \\
  \lstick{$\ket{x}_{\mathrm{coset}}$} & \qwbundle{n+l}
    & \control{} & \gate[style=abgate]{\mathbin{-}\, \bar{x}\,g^{-2^ie_{i:i+w_e}}}
    & \swap{1} & \qw
      \rstick{$\ket{x\,g^{2^i e_{i:i+w_e}}}_{\mathrm{coset}}$} \\
  \lstick{$\ket{0}_{\mathrm{coset}}$} & \qwbundle{n+l}
    & \gate[style=abgate]{\mathbin{+}\, x\,g^{2^ie_{i:i+w_e}}} & \ctrl{-1}
    & \targX{} & \qw
      \rstick{$\ket{0}_{\mathrm{coset}}$}
\end{quantikz}};
\node[abzoom, fit=(botcircuit), inner xsep=6pt, inner ysep=6pt] (zoomframe) {};
\draw[ABGray, line width=0.5pt]
  (ggroup-4-7.south west) -- (zoomframe.north west)
  (ggroup-4-7.south east) -- (zoomframe.north east);
\end{tikzpicture}
\caption{Hierarchical decomposition of the modular exponentiation. Top: the
exponentiation $g^e \bmod N$ is a chain of $\lceil n_e/w_e\rceil$ controlled modular
multiplications on the coset-represented accumulator. Bottom: one controlled
multiplication, realized out-of-place as described in the text above. Each
product-addition is itself expanded into windowed modular additions $+T_k \bmod N$ via
table lookups, as in~\cite{PhysRevLett.127.140503}. Each wire marked $w_e$ is a window of
$w_e$ exponent qubits. A dot on a window means that its value selects the multiplier
applied, not a control on a single qubit. All arithmetic is modulo $N$, and the
$\bmod N$ is omitted inside the boxes.}
\label{fig:modexp_alg}
\end{figure*}

The modular exponentiation $|e\rangle|1\rangle \mapsto |e\rangle|g^e \bmod
N\rangle$ is built up through a hierarchy of nested decompositions and improved
representations. Exponentiation is reduced to multiplications (as seen in
Figure~\ref{fig:ms_oracleF}), and multiplications to product-additions. Modular additions
use the coset representation~\cite{Zalka06,Gidney19}, and the arithmetic is windowed for
efficiency~\cite{10.1007/978-3-030-44223-1_23,Gidney19windowed}. The overall implementation is detailed at great length
in~\cite{PhysRevLett.127.140503} and is depicted in Figure~\ref{fig:modexp_alg}. We give
hereafter a brief overview of the full computation of the modular exponentiation using
these different techniques.

\paragraph{From exponentiation to multiplications.} Writing $e$ in binary form
$e_{n_e-1}\dots e_1 e_0$, the exponentiation factors as a product of controlled multiplications,
\begin{equation}
  g^{e} = \prod_{i=0}^{n_e-1} \left[g^{2^{i}}\right]^{e_i},
\end{equation}
each factor $\left[g^{2^i}\right]^{e_i}$ being applied to the accumulator register conditioned
on the exponent bit $e_i$. This yields $n_e$ modular multiplications. 

\paragraph{From multiplications to product-additions.} Each multiplication $x \mapsto x
\cdot g^{2^i e_i} \bmod N$ is realized with two controlled product-additions and one
controlled swap, following the out-of-place scheme $\ket{y}\ket{z} \mapsto \ket{y}\ket{z +
  y\gamma \bmod N}$. Expanding the product over the bits $y_j$ of $y$,
\begin{equation}
  y \cdot \gamma = \sum_{j=0}^{n-1} \left[\gamma 2^j\right] y_j,
\end{equation}
turns each product-addition into a sequence of $n$ controlled modular additions. 
Concretely, the procedure has three steps. First, the product-addition writes $\bar{x} =
x\,g^{2^i}$ into an ancilla register. Second, it subtracts $\bar{x}\,g^{-2^i} = x$ back
out of the input register, leaving it at $0$. Third, a controlled swap exchanges the two
registers, so that the product $x\,g^{2^i e_i}$ ends up in the accumulator with the
ancilla reset to $\ket{0}$. 

\paragraph{Coset representation for the modular additions.} Rather than implementing each modular addition with the usual
add, compare, correct, uncompute sequence, the target register is stored in the
\emph{coset representation}~\cite{Zalka06,Gidney19}. The basic idea is to extend
the $n$-qubit register holding $z$ with $l$ additional qubits and to encode the
integer $z$ into the uniform superposition over $2^l$ coset shifts,
\begin{equation}
  \ket{z}_{\mathrm{coset}}
  = \frac{1}{\sqrt{2^{l}}} \sum_{k=0}^{2^{l}-1} \ket{z + kN}.
\end{equation}
The shift by $N$ maps the superposition to itself except for the two end terms ($k=0$
lost, $k=2^{l}$ gained). This invariance implies that adding $\gamma$ modulo $N$ is well
approximated by a plain addition modulo $2^{n+l}$,
\begin{equation}
\begin{aligned}
  \frac{1}{\sqrt{2^{l}}} \sum_{k=0}^{2^{l}-1} \ket{(z + \gamma + kN) \bmod 2^{n+l}}\\
  \;\approx\;
  \frac{1}{\sqrt{2^{l}}} \sum_{k=0}^{2^{l}-1} \ket{(z + \gamma \bmod N) + kN},
  \end{aligned}
\end{equation}
where the approximation fails only at the boundary of the superposition. Write
$z'=(z+\gamma)\bmod N$, so that $z+\gamma$ equals either $z'$ or $z'+N$. The plain
addition then maps the $2^{l}$ shifts $k\in\{0,\dots,2^{l}-1\}$ of $z$ either to the same
shifts of $z'$ or to the shifts $k+1$. An addition therefore never scrambles the encoding.
It only slides the window of shifts by at most one position. After $n_{\mathrm{add}}$
additions the window has slid by at most $n_{\mathrm{add}}$ positions. It then shares at
least $2^{l}-n_{\mathrm{add}}$ of its $2^{l}$ terms with the ideal coset state, so the
overlap of the two states is at least $1-n_{\mathrm{add}}\,2^{-l}$. The trace distance
between the actual and the ideal state is then at most $\sqrt{2\,n_{\mathrm{add}}\,2^{-l}}$.
The exponentiation performs $n_{\mathrm{add}}=O(n_en)$ additions, so keeping this distance
below a fixed constant requires $l=\log_2 n_{\mathrm{add}}+O(1)=\Theta(\log n)$. The error
falls exponentially in $l$, which is why a logarithmic padding suffices. A rigorous bound is 
given in~\cite{Gidney19}, and its application to the full exponentiation, including the uncomputation 
within each multiplication, in~\cite{GidneyEkera21}.

\paragraph{Windowed arithmetic.} Finally, the exponent bits are grouped into windows of
$w_e$ bits and the multiplicand bits into windows of $w_l$
bits~\cite{Gidney19windowed}. Each window value indexes a classically precomputed constant
$T_{k} = 2^{j} k_2\, g^{2^{i} k_1} \bmod N$, with $k=(k_1,k_2)$, $k_1 = e_{i:i+w_e}$ and
$k_2 = x_{j:j+w_l}$. Here $i$ indexes the exponent window and $j$ the multiplicand
window, and $y_{a:b}=\lfloor y/2^{a}\rfloor \bmod 2^{b-a}$ denotes the integer formed by
bits $a$ to $b-1$ of $y$. It is loaded into an ancilla via a quantum table lookup
$\ket{k}\ket{0}\mapsto\ket{k}\ket{T_k}$, added unconditionally to the accumulator, and
then unloaded. This divides the number of multiplications by $w_e$ and the number of
additions per multiplication by $w_l$, reducing the leading cost of the exponentiation
from $O(n_en^2)$ to $O(n_e(n+l)^2 / (w_e w_l))$ Toffoli
gates~\cite{PhysRevLett.127.140503}\footnote{This is taking the assumption that the lookup 
table can be implemented at a similar cost as an addition, which is true with proper windowing 
sizes and fan-out of cost $O(1)$.}. 

The implementation of the exponentiation oracle illustrates one of the key considerations
when realizing complex arithmetic operations on a quantum computer. The goal is to defer
as much of the computation as possible to the classical world, reducing the time-space
cost of the quantum part.

\subsubsection*{Overall cost.} We now summarize the space-time requirements of the
three stages, in terms of logical qubits and Toffoli gates, as functions of the modulus size $n =\log N$ and the exponent size $n_e$. Please note that this cost 
is not linked to up-to-date results on Shor where algorithmic optimizations have been made
but to the seminal paper, where we only augmented it with arithmetic tricks. The aim is to understand how and where the different improvements of the modern Shor algorithm 
make their impact.

\paragraph{State preparation.} The first stage acts with Hadamard gates only: a
layer of $n_e$ Hadamards on $\psi_1$ and none on $\psi_2$. It uses no Toffoli gates
and requires the two registers to be allocated, namely the $n_e$-qubit exponent
register and the $(n+l)$-qubit accumulator. With recycling, only $w_e$ qubits of the
exponent register are allocated at a time, and their Hadamards are applied just before
the window is used. Its cost is therefore negligible,
contributing $O(n_e)$ Clifford gates and no non-Clifford cost.

\paragraph{Oracle evaluation.} The second stage dominates the entire algorithm. The
exponentiation is a chain of $n_e$ controlled modular multiplications. Each multiplication
is two product-additions over $(n+l)$-qubit registers, and each product-addition is a
sequence of modular additions. Windowing the exponent into blocks of $w_e$ bits reduces the number of
multiplications by a factor $w_e$. Windowing the multiplicand into blocks of $w_l$ bits
reduces the number of additions per multiplication by a factor $w_l$. This leaves a
leading Toffoli cost of 
\begin{equation}
  T_{\mathrm{oracle}} = O\!\left(\frac{n_e\,(n+l)^2}{w_e\,w_l}\right),
\end{equation}
as derived in~\cite{PhysRevLett.127.140503}. The space requirement is set by the two
registers and the lookup ancilla. The input register $\ket{x}_{\mathrm{coset}}$ and the
accumulator each use $n+l$ qubits. The table lookup for a window of $w_e + w_l$ address
bits temporarily allocates an $n$-qubit ancilla to hold $T_k$. The stage therefore
uses 
\begin{equation}
  Q_{\mathrm{oracle}} = 2(n+l) + O(n)
\end{equation}
logical qubits. The coset padding is chosen as $l = O(\log n)$, which keeps the
accumulated error of all $O(n_e\,n)$ additions below a fixed target.

\paragraph{Quantum Fourier transform.} The third stage acts only on the $n_e$-qubit
exponent register. In its semi-classical form each qubit contributes one Hadamard,
one classically controlled phase rotation, and one measurement, so the transform
costs $O(n_e)$ single-qubit rotations, negligible but non-Clifford, and allocates no
qubits beyond the exponent register already present. Its cost is thus negligible
next to the oracle.

\paragraph{Total.} Non-Clifford cost is incurred almost entirely by the oracle, so
the full algorithm requires
\begin{equation}
  T_{\mathrm{total}} = O\!\left(\frac{n_e\,(n+l)^2}{w_e\,w_l}\right)
  \quad\text{Toffoli gates},
\end{equation}
and, when the exponent register is held in full,
\begin{equation}
  Q_{\mathrm{total}} = n_e + 2(n+l) + O(n) = O(n)
  \quad\text{logical qubits},
\end{equation}
using $l = O(\log n)$. With recycling, the $n_e$ term drops to $w_e$, the size of one
window, so $Q_{\mathrm{total}} = w_e + 2(n+l) + O(n)$ whatever the value of $n_e$. 
In the plain (unwindowed) regime $w_e = w_l = 1$ and with $l =
O(\log n)$, the Toffoli count reaches $O(n_e\,n^2)$. For order finding the exponent
is taken as $n_e = 2n$, recovering the familiar $O(n^3)$ scaling of Shor's algorithm. 

We now have a clear view of Shor's algorithm and how it solves the HSP for factoring. The
modern, state-of-the-art algorithm lowers its cost significantly by composing three
results: the input-register reduction of Ekerå--Håstad~\cite{PQCRYPTO:EkeHas17}, the
output-register reduction of May--Schlieper~\cite{MaySchlieper22}, and the oracle
compression of Chevignard, Fouque and Schrottenloher~\cite{C:CheFouSch25}. All three
operate in two dimensions. The first step is the reduction that puts factoring there, to
which we now turn.

\subsection{From Factoring to Short Discrete Logarithm Problem}\label{sec:short-dlp}

The order-finding algorithm of the previous section works in one dimension. The
optimizations that follow all operate in two dimensions, and they reach factoring only
through the reduction of factoring to a short discrete logarithm developed here. For
factoring, we rely on the Ekerå--Håstad period-finding algorithm~\cite{PQCRYPTO:EkeHas17},
a recent variant of Shor's algorithm. The core observation of their paper is the
following. The multiplicative group of integers modulo $N$, $\mathbb{Z}^{*}_N$, has order $\phi(N)  =
(p-1)(q-1)$. By definition of the Euler totient, every element
$g\in\mathbb{Z}^{*}_N$ satisfies 
\begin{equation}
  g^{\phi(N)}\equiv 1 \bmod N
\end{equation}

Since $N = p\cdot q$, we have:

\begin{equation}
  \phi(N) = (p-1)(q-1) = pq - p - q + 1 = N - (p+q - 1)
\end{equation}
Therefore:
\begin{equation}
  g^{N-1} = g^{\phi(N)}\cdot g^{(p+q-2)}\equiv g^{(p+q-2)}\bmod N 
\end{equation}

Since $p$ and $q$ are distinct odd primes, $p-1$ and $q-1$ are both even, so
$\gcd(p-1,q-1)$ is even. Hence $\gcd(p-1,q-1)/2$ is an integer. We use the Carmichael function
$\lambda(N) = \mathrm{lcm}(p-1,q-1) = (p-1)(q-1)/\gcd(p-1,q-1)$. Then 
\begin{equation}
\begin{aligned}
  \frac{\phi(N)/2}{\lambda(N)}
  &= \frac{(p-1)(q-1)/2}{(p-1)(q-1)/\gcd(p-1,q-1)}\\
  &= \frac{\gcd(p-1,q-1)}{2} \in \mathbb{Z},
  \end{aligned}
\end{equation}
so $\lambda(N) \mid \phi(N)/2$, and hence $g^{\phi(N)/2} \equiv 1 \pmod{N}$ for every $g
\in \mathbb{Z}_N^*$. 
Let $x := g^{(N-1)/2} \bmod N$ and $d := (p+q-2)/2$. Since $(N-1)/2 = \phi(N)/2 + d$ and
$g^{\phi(N)/2} \equiv 1 \pmod{N}$, we obtain
\begin{equation}
  x = g^{(N-1)/2} = g^{\phi(N)/2} \cdot g^d \equiv g^d \pmod{N},
\end{equation}
so $x = g^d \bmod N$. Since $p, q \approx 2^{n/2}$, we have $d \approx 2^{n/2}$, so $d$ is
small compared to $N$. What matters is its size relative to the order $r$ of $g$, not to the order $\phi(N)$ of
the whole group, because the algorithm only ever sees the cyclic group generated by $g$.
We call $d$ a \emph{short} discrete logarithm if it is much smaller than $r$. The precise
condition the algorithm needs is stated in Section~\ref{sec:input-register}. For a random
$g$ it holds with overwhelming probability, because $r$ is then within a small factor of
$\phi(N)=\Theta(2^{n})$.

This reformulation allows us to express the problem of period finding for factoring
into the search of a short discrete log $d$. More specifically, $x$ can be easily
classically computed, as it is the exponentiation of a group element $g$ to the power
$(N-1)/2$ which is known. Once $d$ is recovered using Shor's algorithm, one can recover $p$
and $q$ by solving the quadratic equation $N = q(2d+2-q)$. Unlike the ordinary DLP case
where $G =\mathbb{Z}_{p-1}^2$ is built from the known group order $p-1$, the group order
$\phi(N)$ is unknown here (factoring $N$ is the whole point). The registers are therefore
built from a public upper bound on $d$'s bit-length rather than the exact group
order. This distinction is precisely why the two-register Ekerå--Håstad construction of
Section~\ref{sec:input-register} looks different from the ordinary DLP circuit, where each
register is sized directly from the known order $p-1$, up to the usual binary
padding. Please note that with this reduction, Shor's algorithm for factoring is now in
the two-dimensional set-up, namely we look for the generator $(d,1)$ instead of the order
$r$.

We can summarize the reduction from factoring to a short discrete-log problem as follows:
\begin{equation}
  \underbrace{x = g^{(N-1)/2} \bmod N}_{\text{classically computable}}
  \;\Rightarrow\;
  \underbrace{d = \log_g x}_{\substack{\text{short DLP} \\ d \text{ of } \approx n/2 \text{ bits}}}
  \;\Rightarrow\;
  \underbrace{p + q = 2d + 2}_{\substack{\text{factor } N \text{ via} \\ q(2d+2-q) = N}}
\end{equation}

We can now use the same algorithm to tackle the factoring problem and the
discrete logarithm problem.

\paragraph*{On the periodicity of the oracle $f$ for two-dimensional factoring.}
The function $f(a,b) = g^{a-bd} \bmod N$ takes the same value on two pairs $(a_1,b_1)$ and
$(a_2,b_2)$ if and only if $g^{a_1-b_1 d} \equiv g^{a_2-b_2 d} \pmod{N}$. Since $g$ has
multiplicative order $r$, this holds if and only if
\begin{equation}
  a_1 - b_1 d \equiv a_2 - b_2 d \pmod{r}.
\end{equation}
The oracle $f$ is
therefore periodic in the $(a,b)$-plane in two independent directions. First, shifting $a$
by $r$ with $b$ fixed leaves $f$ unchanged: 
\begin{equation}
  f(a+r,\,b) = g^{a+r-bd} = g^{a-bd}\cdot g^r = g^{a-bd} = f(a,b),
\end{equation}
so $f$ has period $r$ in the $a$-direction. Second, shifting $(a,b)$ by $(d,1)$
also leaves $f$ unchanged:
\begin{equation}
  f(a+d,\,b+1) = g^{(a+d)-(b+1)d} = g^{a-bd} = f(a,b),
\end{equation}
so $f$ is also periodic along the direction $(d,1)$ in the $(a,b)$-plane. The first
periodicity comes from $g^{r}=1$. The second comes from $x=g^{d}$, that is, from $d$
itself. Together they generate the period lattice
\begin{equation}
\begin{aligned}
  H &= \bigl\{(a,b)\in\mathbb{Z}^{2} : a\equiv bd \pmod r\bigr\}\\
    &= \bigl\langle (r,0),\,(d,1)\bigr\rangle \le \mathbb{Z}^{2},
\end{aligned}
\end{equation}
and $f$ is constant and separating on its cosets. This $H$ plays the role of the hidden
subgroup of the discrete-logarithm instance of Section~\ref{sec:hsp}, with $\mathbb{Z}^{2}$
in place of $\mathbb{Z}_{p-1}^{2}$ because the order $r$ is unknown. Its image in
$\mathbb{Z}_r\times\mathbb{Z}_r$ is the cyclic subgroup $\langle(d,1)\rangle$ of order $r$.
Recovering $H$ determines $d$ only modulo $r$, since $(d,1)$ and $(d+r,1)$ generate the
same lattice. The reduction therefore needs $d<r$, which the hypothesis of
Section~\ref{sec:input-register} guarantees.

\subsection{Reducing the Input Register Size}\label{sec:input-register}

The exponent register has two costs, and which one this section reduces depends on the
oracle. With recycling (Section~\ref{par:recycling}), the number of qubits held live at
any one time is independent of how long the exponent is. What shrinks below is then the
length itself, and with it the number of windows the recycling loop has to step through,
and so the number of controlled multiplications the oracle performs. In that setting the
reductions of this section are depth optimizations. Recycling is not always available.
The residue-arithmetic oracle of~\cite{C:CheFouSch25,gidney2025factor2048bitrsa} is
incompatible with it~\cite[Sect.~1]{gidney2025factor2048bitrsa}, so the whole exponent
register must be held for the entire run. There the reductions of this section save
qubits. This is the regime of the RSA-2048 pipeline of Section~\ref{sec:full-modern}.

As we have seen, the Ekerå--Håstad reduction recasts factoring as the search for the short
logarithm $d=(p+q-2)/2$. Since $p,q \approx 2^{n/2}$, we have $d \approx 2^{n/2}$, so $d$
has bit-length $m \approx n/2$, roughly half the bit-length of the group order $\phi(N)
\approx 2^n$. The algorithm of~\cite{PQCRYPTO:EkeHas17} exploits this by trading a single run for
several runs combined through classical lattice reduction. Its two exponent registers
together hold $n_e=\left(\tfrac12+\tfrac1s\right)n$ bits, for an integer parameter
$s\ge 2$ (the hypothesis below excludes $s=1$ for RSA moduli). This is $n$ bits at $s=2$,
half of Shor's $2n$. As $s$ grows it approaches $n/2$ bits, a factor of four below Shor,
at the price of more runs.

\paragraph{Two index registers and a shorter exponent.} The algorithm uses
two exponent registers rather than the single one of order finding. They have
$\tau_{s} + m$ and $\tau_{s}$ qubits, with $\tau_{s} \approx m/s$. Write $g$ for the chosen generator and $x = g^{d}$ for the target. The
quantum stage prepares a uniform superposition over both registers and evaluates
\begin{equation}
\begin{aligned}
  \ket{a}\ket{b}\ket{0}
  \;\longmapsto\;
  &\ket{a}\ket{b}\ket{\,g^{a} x^{-b} \bmod N} \\
  &= \ket{a}\ket{b}\ket{\,g^{a - bd} \bmod N},
  \end{aligned}
\end{equation}
with $0 \le a < 2^{\tau_{s}+m}$ and $0 \le b < 2^{\tau_{s}}$. Since $m \approx n/2$ once the
short-DLP recasting has been applied and $\tau_{s} \approx m/s$, the first register has
$\tau_{s} + m \approx \left(\tfrac{1}{2} + \tfrac{1}{2s}\right)n$ qubits and the second
$\tau_{s}\approx n/(2s)$. Increasing $s$ shrinks $\tau_{s}$, and
therefore the register, at the price of requiring more runs. Specifically, this approach
trades the exact single-shot recovery of Shor's QFT for a noisier measurement that still constrains 
$d$. To fully recover $d$ from
those noisy linear constraints, one now requires multiple measurements to have a suitable,
solvable system. Counting both registers, the exponent has
$n_e=(\tau_{s}+m)+\tau_{s}=\left(\tfrac12+\tfrac1s\right)n$ bits.

\paragraph{The hypothesis on the order of $g$.} Let $m$ be the smallest integer with
$0<d<2^{m}$, so $m\approx n/2$ here. Ekerå and Håstad require the order $r$ of $g$ to
satisfy
\begin{equation}\label{eq:eh-order}
  r \;\ge\; 2^{\tau_{s}+m} + 2^{\tau_{s}}\,d
\end{equation}
\cite[Theorem~1]{PQCRYPTO:EkeHas17}. It guarantees that the exponent $a-bd$ is never
reduced modulo $r$ for any $(a,b)$ in the registers, and the analysis below relies on
this. Without it, the algorithm only determines $d \bmod r$. The order $r$ need not be
known. For factoring, write $r=\phi(N)/t$. Since $d<2^{m}$, it suffices that
$r\ge 2^{\tau_{s}+m+1}$, and with $\phi(N)\approx 2^{n}$ this amounts to a bound of
roughly $2^{n-\tau_{s}-m}$ on the index $t$. For random $p$, $q$ and a random $g$ the
index is small, and Ekerå and Håstad show that the bound holds with overwhelming
probability for every $s>1$~\cite[Sect.~5.2.4]{PQCRYPTO:EkeHas17}. It fails for $s=1$,
since $2^{\tau_{s}+m}$ then reaches about $2^{n}$, which exceeds $N>r$.

\paragraph{Where the QFT peaks in the two-dimensional setting.}

\begin{figure*}[tp]
  \centering
  \includegraphics[width=1\linewidth]{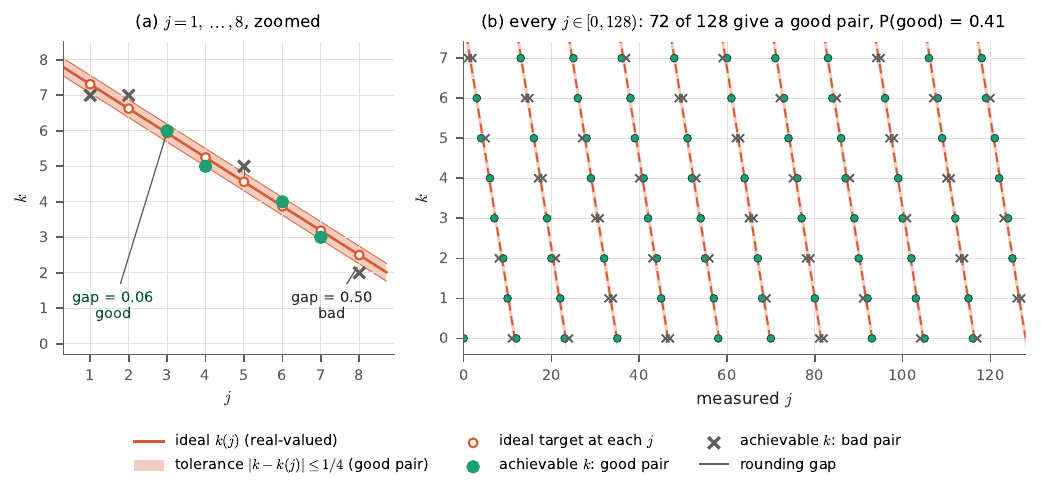}
    \caption{Illustration of the good-pair condition~\eqref{eq:good-pair} of
    Section~\ref{sec:good_pair}, for a toy instance with $\tau_{s}=3$, $m=4$ and $d=11$, so
    that $2^{\tau_{s}+m}=128$ and $k$ takes $2^{\tau_{s}}=8$ values.
    Hypothesis~\eqref{eq:eh-order} requires $r\ge 2^{\tau_{s}+m}+2^{\tau_{s}}d=216$, which
    holds for instance for $g=2$ in $\mathbb{Z}_{227}^{*}$, of order $r=226$. The
    condition itself does not depend on $r$. For each $j$, the curve gives the real-valued
    $k(j)=-\tfrac{d}{2^m}j \bmod 2^{\tau_{s}}$ that would make $dj+2^mk$ an exact multiple of
    $2^{\tau_{s}+m}$. Condition~\eqref{eq:good-pair} is equivalent to
    $|k-k(j)|\le 2^{m-2}/2^{m}=1/4$, the shaded band, so a pair is good when its integer
    $k$ falls inside it. \textbf{(a)} The first $8$ values of $j$, zoomed in. Filled
    circles are good pairs, crosses bad ones, and the thin line is the rounding gap
    between $k(j)$ and the nearest integer. \textbf{(b)} Every $j\in[0,2^{\tau_{s}+m})$,
    each wrap of the curve back to $k=2^{\tau_{s}}$ starting a new period. For $72$ of the
    $128$ values of $j$, the nearest integer $k$ gives a good pair.}
  \label{fig:jk-good-pair}
\end{figure*}

In the one-dimensional order finding of Section~\ref{sec:shor}, one register produced one
measured frequency, sharply peaked at a near-multiple of $2^{n_e}/r$, from which continued
fractions recovered $r$. The two-dimensional algorithm behaves differently. Under
hypothesis~\eqref{eq:eh-order} the registers hold less than one period of $f$, so no
value of $j$ is singled out on its own. All the information about $d$ sits in a single
approximate alignment between $j$ and $k$, which we now derive
following~\cite[Sect.~4.3]{PQCRYPTO:EkeHas17}.

After the oracle, the joint state is
\begin{equation}
  |\psi\rangle = \frac{1}{\sqrt{2^{\tau_{s}+m} \cdot 2^{\tau_{s}}}}
  \sum_{a=0}^{2^{\tau_{s}+m}-1}\sum_{b=0}^{2^{\tau_{s}}-1}
  |a\rangle|b\rangle|g^{a-bd} \bmod N\rangle.
\end{equation}
Write $e=a-bd$ for the exponent. For $(a,b)$ in the registers, $e$ ranges over an interval
of fewer than $r$ integers, by~\eqref{eq:eh-order}. Distinct values of $e$ therefore give
distinct group elements $g^{e}$, so the output register determines $e$ exactly, with no
reduction modulo $r$. Applying $\mathrm{QFT}^{-1}_{2^{\tau_{s}+m}}$ to the first register
and $\mathrm{QFT}^{-1}_{2^{\tau_{s}}}$ to the second, the state $|j,k,g^{e}\rangle$ is
observed with probability
\begin{equation}\label{eq:2d-amplitude}
\begin{aligned}
  P(j,k,e) = \frac{1}{2^{2(2\tau_{s}+m)}}
  \Bigl|\sum_{b\in B_e}
  &e^{2\pi i\,((e+bd)j + 2^{m}bk)/2^{\tau_{s}+m}}\Bigr|^{2},
\end{aligned}
\end{equation}
where $B_e=\{0\le b<2^{\tau_{s}} : 0\le e+bd<2^{\tau_{s}+m}\}$ collects the pairs with
$a=e+bd$ in range. Write $T_e=|B_e|$. The two phase factors have been combined into one,
since $k/2^{\tau_{s}} = k\cdot 2^m/2^{\tau_{s}+m}$. Extracting the factor
$e^{2\pi i\,ej/2^{\tau_{s}+m}}$, which has modulus one, leaves a sum over $b$ alone,
\begin{equation}
  P(j,k,e) = \frac{1}{2^{2(2\tau_{s}+m)}}
  \Bigl|\sum_{b\in B_e}
  e^{2\pi i\, b\,\{dj+2^{m}k\}_{2^{\tau_{s}+m}}/2^{\tau_{s}+m}}\Bigr|^{2}.
\end{equation}
Here $\{\,\cdot\,\}_{2^{\tau_{s}+m}}$ denotes the centered residue in
$[-2^{\tau_{s}+m-1}, 2^{\tau_{s}+m-1})$. Replacing $dj+2^mk$ by its residue shifts each
phase by a multiple of $2\pi$ and changes nothing.

This is a geometric series in $b$ with per-step phase
$2\pi\{dj+2^m k\}_{2^{\tau_{s}+m}}/2^{\tau_{s}+m}$. It is large only when that phase is
small, that is when the integer $dj + 2^m k$ is close to a multiple of $2^{\tau_{s}+m}$.
This is a linear relation between the unknown $d$ and the measured $j,k$. Because $2^m k$
takes only values that are multiples of $2^m$, the frequency $k$ aligns with the
high-order part of $dj$. The remainder $dj \bmod 2^m$ is the part of the relation that
constrains $d$. This alignment is only ever approximate, and it is the source of the
noise in the two-dimensional outcome.
\begin{figure*}[t]
\centering
\def\rad{3pt}
\tikzset{
  abwire/.style={draw=abInk},
  abgate/.style={draw=abInk, fill=white, line width=0.7pt, rounded corners=\rad},
  abexp/.style={draw=abTealDark, fill=abTeal!35, line width=1pt, rounded corners=\rad},
  abreg/.style={draw=abTealDark, dashed, line width=0.7pt, rounded corners=\rad},
  abctrl/.style={circle, fill=abInk, inner sep=0pt, minimum size=3.2pt},
  abbrace/.style={draw=abInk, decorate, decoration={brace,amplitude=4pt,mirror}, line width=0.6pt},
  wire/.style={abwire, line width=0.6pt, ->, >={Stealth[length=3.5pt,width=3pt]}},
  meter/.pic={%
    \draw[abInk,line width=0.5pt] (-0.16,-0.06) arc (180:0:0.16);
    \draw[abInk,-{Stealth[length=2.5pt]},line width=0.5pt] (0,-0.06) -- (0.13,0.17);},
}
\begin{subfigure}[t]{0.48\linewidth}
\centering
\begin{tikzpicture}[x=1cm, y=1cm, font=\footnotesize]
  \draw[abreg] (0.30,1.85) rectangle (6.10,3.90);
  \foreach \i in {0,...,7}{\draw[wire] (0,{2.0+0.25*\i}) -- (6.25,{2.0+0.25*\i});}
  \draw[abreg] (0.30,0.10) rectangle (6.10,1.15);
  \foreach \i in {0,...,3}{\draw[wire] (0,{0.25+0.25*\i}) -- (6.25,{0.25+0.25*\i});}
  \draw[abgate] (0.45,1.92) rectangle (0.95,3.83); \node at (0.70,2.875){$H$};
  \draw[abexp]  (1.20,0.15) rectangle (3.45,1.10); \node at (2.325,0.625){$\times g^{e}$};
  \foreach \i in {0,...,7}{%
    \pgfmathsetmacro\xc{1.38+0.27*(7-\i)}
    \draw[abwire, line width=0.5pt] (\xc,{2.0+0.25*\i}) -- (\xc,1.10);
    \node[abctrl] at (\xc,{2.0+0.25*\i}){};}
  \draw[abgate] (4.00,1.92) rectangle (4.85,3.83); \node at (4.425,2.875){QFT};
  \draw[abgate] (5.15,1.92) rectangle (5.85,3.83); \pic at (5.50,2.875){meter};
  \node[anchor=east] at (-0.05,2.875){$\ket{0}$};
  \node[anchor=east] at (-0.05,0.625){$\ket{1}$};
  \node[anchor=west] at (6.28,2.875){$\ket{j}$\;\;$2n$};
  \node[anchor=west] at (6.28,0.625){$\ket{g^{e}}$\;\;$n$};
  \draw[abbrace] (1.20,-0.05) -- (3.45,-0.05) node[midway,below=4pt]{$2n$ multiplications};
\end{tikzpicture}
\caption{Shor~\cite{Shor97}}
\label{fig:register-shor}
\end{subfigure}
\hfill
\begin{subfigure}[t]{0.48\linewidth}
\centering
\begin{tikzpicture}[x=1cm, y=1cm, font=\footnotesize]
  \draw[abreg] (0.30,3.10) rectangle (6.10,3.90);
  \foreach \i in {0,...,2}{\draw[wire] (0,{3.25+0.25*\i}) -- (6.25,{3.25+0.25*\i});}
  \draw[abreg] (0.30,2.52) rectangle (6.10,2.98);
  \draw[wire] (0,2.75) -- (6.25,2.75);
  \draw[abreg] (0.30,0.10) rectangle (6.10,1.15);
  \foreach \i in {0,...,3}{\draw[wire] (0,{0.25+0.25*\i}) -- (6.25,{0.25+0.25*\i});}
  \draw[abgate] (0.45,3.15) rectangle (0.95,3.85); \node at (0.70,3.50){$H$};
  \draw[abgate] (0.45,2.56) rectangle (0.95,2.94); \node[font=\scriptsize] at (0.70,2.75){$H$};
  \draw[abexp] (1.20,0.15) rectangle (2.45,1.10); \node at (1.825,0.625){$\times g^{a}$};
  \foreach \i in {0,...,2}{%
    \pgfmathsetmacro\xc{1.45+0.30*(2-\i)}
    \draw[abwire, line width=0.5pt] (\xc,{3.25+0.25*\i}) -- (\xc,1.10);
    \node[abctrl] at (\xc,{3.25+0.25*\i}){};}
  \draw[abexp] (2.75,0.15) rectangle (3.45,1.10); \node at (3.10,0.625){$\times x^{-b}$};
  \draw[abwire, line width=0.5pt] (3.10,2.75) -- (3.10,1.10); \node[abctrl] at (3.10,2.75){};
  \draw[abgate] (4.00,3.15) rectangle (4.85,3.85); \node at (4.425,3.50){QFT};
  \draw[abgate] (4.00,2.56) rectangle (4.85,2.94); \node[font=\scriptsize] at (4.425,2.75){QFT};
  \draw[abgate] (5.15,3.15) rectangle (5.85,3.85); \pic at (5.50,3.50){meter};
  \draw[abgate] (5.15,2.56) rectangle (5.85,2.94); \pic[scale=0.7] at (5.50,2.71){meter};
  \node[anchor=east] at (-0.05,3.50){$\ket{0}$};
  \node[anchor=east] at (-0.05,2.75){$\ket{0}$};
  \node[anchor=east] at (-0.05,0.625){$\ket{1}$};
  \node[anchor=west] at (6.28,3.50){$\ket{j}$\;\;$\tfrac{3}{4}n$};
  \node[anchor=west] at (6.28,2.75){$\ket{k}$\;\;$\tfrac{1}{4}n$};
  \node[anchor=west] at (6.28,0.625){$\ket{g^{a}x^{-b}}$\;\;$n$};
  \draw[abbrace] (1.20,-0.05) -- (2.45,-0.05) node[midway,below=4pt]{$\tfrac34 n$};
  \draw[abbrace] (2.75,-0.05) -- (3.45,-0.05) node[midway,below=4pt]{$\tfrac14 n$};
\end{tikzpicture}
\caption{Ekerå--Håstad~\cite{PQCRYPTO:EkeHas17}, drawn for $s=2$}
\label{fig:register-eh}
\end{subfigure}
\caption{Comparison of the input registers, drawn to a common scale in which one wire
stands for $n/4$ qubits. (a): Shor's original circuit uses a single $2n$-qubit index
register evaluating $g^{e}$. (b): the Ekerå--Håstad circuit uses two index registers of
$n/2+n/(2s)$ and $n/(2s)$ qubits, evaluating $g^{a}x^{-b}$ in superposition, shown here
for $s=2$. The output register holds $n$ qubits in both. Measuring it is optional
(Figure~\ref{fig:ms_fullshor}) and is not drawn. All arithmetic is modulo $N$, and the
$\bmod N$ is omitted inside the boxes.
}
\label{fig:register-comparison}
\end{figure*}
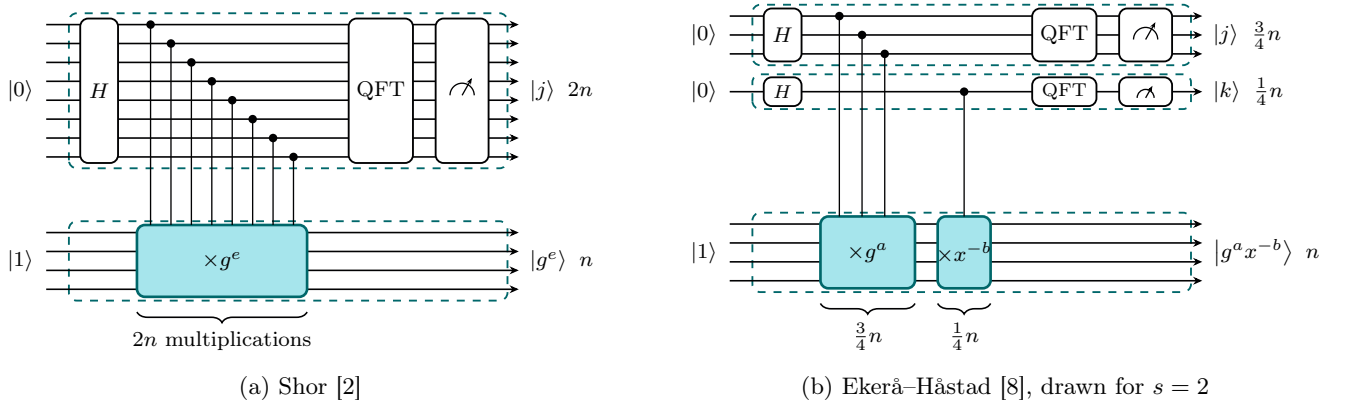

\paragraph{The good pair condition.}\label{sec:good_pair}
That approximate alignment has a precise geometric meaning. The $T_e$ phasors summed
in~\eqref{eq:2d-amplitude} spread across an arc of the unit circle whose width grows with
$|\{dj+2^mk\}_{2^{\tau_{s}+m}}|$. When the spread is too large they cancel, giving a
low-probability outcome that carries no usable information about $d$. Ekerå and
Håstad~\cite{PQCRYPTO:EkeHas17} show (Claim~1) that when every phase angle satisfies
$|\theta_i| \le \pi/4$, the squared norm of a sum of $T$ phasors is at least $T^{2}/2$,
half of its maximal value. This is their $\pm\pi/4$ constructive-interference condition.
Shifting the summation index to $b-2^{\tau_{s}-1}$ only multiplies the sum by a phase, and
centers the angles around zero.

Call a pair $(j,k)$ \emph{good} when
\begin{equation}\label{eq:good-pair}
  \bigl|\{\, dj + 2^{m} k \,\}_{2^{\tau_{s}+m}}\bigr| \le 2^{m-2}.
\end{equation}
Since $|b-2^{\tau_{s}-1}|\le 2^{\tau_{s}-1}$, this bound caps every centered angle at
$2^{\tau_{s}-1}\cdot 2\pi\cdot 2^{m-2}/2^{\tau_{s}+m}=\pi/4$. The probability of observing
a given good pair together with a given $e$ is therefore at least
$T_e^{2}/(2\cdot 2^{2(2\tau_{s}+m)})$. The values $T_e$ sum to $2^{2\tau_{s}+m}$ over
fewer than $2^{\tau_{s}+m+1}$ values of $e$, so the Cauchy--Schwarz inequality gives
$\sum_e T_e^{2}\ge 2^{3\tau_{s}+m-1}$. Summing over $e$, each good pair is observed with
probability at least $2^{-m-\tau_{s}-2}$~\cite[Lemma~2]{PQCRYPTO:EkeHas17}. At least
$2^{\tau_{s}+m-1}$ values of $j$ admit a good $k$~\cite[Lemma~1]{PQCRYPTO:EkeHas17}, so a
single shot yields a good pair with probability at least $1/8$. A good pair is a usable
linear constraint on $d$. Its $j$ acts as a known random multiplier, and its $k$ records
the high-order bits of $dj$, fixing where $dj$ wraps modulo $2^m$. 
Figure~\ref{fig:jk-good-pair} illustrates the condition on a toy instance.

The pair $(j,k)$ is computed exactly, so $\epsilon=0$. The read-out of Ekerå and Håstad
cannot tell a good pair from a bad one, since the test~\eqref{eq:good-pair} involves the
unknown $d$. Every pair is therefore accepted and read as the
constraint~\eqref{eq:good-pair}. For a bad pair that constraint is false, so under
Definition~\ref{def:sep-proc} a bad pair is a corruption, not an abort. This is a
$(0,\delta)$-noisy HSP instance with $\delta_{\bot}=0$ and
$\delta=\delta_{\times}\le 7/8$. The corruptions are caught one level up, on a whole
attempt. Ekerå and Håstad run $8s$ shots, apply the lattice step below to every subset of
$s$ pairs, and accept a candidate only if it satisfies $x=g^{d}$~\cite[Sect.~4.5]{PQCRYPTO:EkeHas17}. The check is exact, so a wrong $d$ is never output.
A corrupted subset only costs another subset, or another attempt.

\paragraph{Lattice-based recovery of $d$.} A single good pair does not determine $d$. The
short logarithm is recovered by combining $s$ of them, $(j_1,k_1),\dots,(j_s,k_s)$,
through lattice reduction. Consider the $(s{+}1)$-dimensional integer lattice $\Lambda$
generated by the rows of
\begin{equation}
  \begin{bmatrix}
    j_1 & j_2 & \cdots & j_s & 1 \\
    2^{\tau_{s}+m} & 0 & \cdots & 0 & 0 \\
    0 & 2^{\tau_{s}+m} & \cdots & 0 & 0 \\
    \vdots & & \ddots & & \vdots \\
    0 & 0 & \cdots & 2^{\tau_{s}+m} & 0
  \end{bmatrix},
\end{equation}
together with the target vector
\begin{equation}
  \vec{v} = \bigl(\{-2^{m} k_1\}_{2^{\tau_{s}+m}},\,\dots,\,\{-2^{m} k_s\}_{2^{\tau_{s}+m}},\,0\bigr).
\end{equation}
By construction, the lattice contains the vector
\begin{equation}
  \vec{u} = \bigl(\{dj_1\}_{2^{\tau_{s}+m}} + m_1 2^{\tau_{s}+m},\,\dots,\,\{dj_s\}_{2^{\tau_{s}+m}} + m_s 2^{\tau_{s}+m},\, d\bigr),
\end{equation}
where $m_1,\dots, m_s$ are the unknown wrap counts, whose last component is exactly $d$, 
and Ekerå and Håstad derive that the good-pair
condition guarantees $|\vec{u} - \vec{v}| < \sqrt{s/4 + 1}\cdot 2^{m}$. Recovering $d$
therefore amounts to a bounded-distance search. One enumerates the lattice points within
this radius of $\vec{v}$ and returns the last component that satisfies $x = g^{d}$. Since
$\Lambda$ has determinant $\approx 2^{m(s+1)}$ in dimension $s+1$, its shortest vector has
norm $\approx 2^{m}$. One shows that, with high probability, only a constant number of
lattice points (depending on $s$) fall within the search radius. The search is thus
efficient, and the whole post-processing runs in classical polynomial time.

\medskip

This is the same design philosophy seen throughout the arithmetic computations of Shor's
algorithm, shifting work from the quantum to the classical world. Here the single-shot
continued-fraction recovery of the order-finding case is replaced by a multi-shot lattice
computation. It tolerates a smaller input register, at the cost of a constant number of
additional runs and a polynomial-time classical lattice search. 

An illustration of the Ekerå--Håstad circuit that allows the reduction of the input
register is depicted in Figure~\ref{fig:register-comparison}.

\subsection{Reducing the Output Register Size}\label{sec:output-register}

The previous optimization shrank the input register but left the output register
untouched. It still stores the full $n$-bit value $g^{a}x^{-b} \bmod N$, which for RSA
dominates the qubit budget. Counting only the registers that are measured, Ekerå--Håstad
uses $\bigl(\tfrac12+\tfrac1s\bigr)n$ input qubits and $n$ output qubits, which
approaches $\tfrac32 n$ as $s$ grows. Both counts leave out the workspace of the
arithmetic, which Section~\ref{sec:shor} counts separately. May and
Schlieper~\cite{MaySchlieper22} observed that the output register is far larger than
period finding needs, and can be compressed to as few as $t=1$ qubit. Their result holds
in an \emph{oracle model}. It assumes a circuit $U_{h\circ f}$ that writes the hash
$h(f(a,b))$ directly into $t$ qubits. Computing $f$ first and hashing the result
afterwards is not enough, since that needs the $n$ qubits hashing was meant to
save~\cite[Sect.~1]{MaySchlieper22}. How to build $U_{h\circ f}$ was left open, and
Section~\ref{sec:oracle-rns} describes how it was later solved.

\begin{figure*}[ht]
\centering
\begin{tikzpicture}[>=Stealth, font=\small,
  lbl/.style={font=\scriptsize, align=center},
  rc/.style={rounded corners=1.5mm},
  reg/.style={rc, draw=black!70, line width=0.75pt},
  frame/.style={rc, draw=black!70, line width=0.75pt},
  flow/.style={->, ABLine, line width=0.9pt},
  sep/.style={black!20, line width=0.5pt},
  cA/.style={fill=ABTeal,        draw=ABTealDark,   line width=0.4pt},
  cB/.style={fill=ABPurple,      draw=ABPurpleDark, line width=0.4pt},
  cC/.style={fill=black!12,      draw=black!50,     line width=0.4pt},
  cD/.style={fill=black!30,      draw=black!55,     line width=0.4pt},
  cE/.style={fill=ABPurple!30,   draw=ABPurpleDark, line width=0.4pt},
  callout/.style={rc, fill=abYellow, draw=ABInk, line width=0.75pt,
                  text=ABInk, inner xsep=5pt, inner ysep=3pt, font=\small},
]
\def\BW{7.0}\def\NC{15}\def\CW{0.4667}\def\PH{1.5}
\newcommand{\abcells}[2]{%
  \begin{scope}\clip[rc] (0,#1) rectangle (\BW,{#1+0.5});
  \foreach \i/\c in {#2}{\path[\c] ({\i*\CW},#1) rectangle ({(\i+1)*\CW},{#1+0.5});}
  \end{scope}\draw[reg] (0,#1) rectangle (\BW,{#1+0.5});}
\newcommand{\aballcells}[1]{\abcells{#1}{0/cA,1/cB,2/cC,3/cD,4/cE,5/cA,6/cB,7/cC,8/cD,9/cE,10/cA,11/cB,12/cC,13/cD,14/cE}}
\newcommand{\abpeak}[3]{\filldraw[fill=#3,draw=#3Dark,line width=0.5pt] (#1-0.13,0) rectangle (#1+0.13,{#2*\PH});}
\newcommand{\abghost}[3]{\draw[ABLine,dash pattern=on 1.6pt off 1.2pt,line width=0.6pt] (#1-0.13,{#3*\PH}) -- (#1-0.13,{#2*\PH}) -- (#1+0.13,{#2*\PH}) -- (#1+0.13,{#3*\PH});}
\def\pk{0.45,1.80,3.15,4.50,5.85}
 
\begin{scope}
  \node[font=\normalsize\bfseries] at (\BW/2,0.95) {Exact output};
  \node[lbl] at (\BW/2,0.55) {measure the value $f(x)$};
  \node[lbl, anchor=south] at (\BW/2,-0.42) {output values $f(x)$, one cell per input $x$};
  \aballcells{-1.0}
  \draw[flow] (\BW/2,-1.05) -- (\BW/2,-1.55) node[midway,right=2pt,lbl,black]{measure: teal};
  \abcells{-2.1}{0/cA,5/cA,10/cA}
  \node[lbl] at (\BW/2,-2.32) {input register: one coset $x_0,\,x_0+r,\,x_0+2r$};
  \draw[flow] (\BW/2,-2.5) -- (\BW/2,-3.0) node[midway,right=2pt,lbl,black]{QFT};
  \begin{scope}[shift={(0,-4.75)}]
    \draw[frame] (0,0) rectangle (\BW,\PH);
    \abpeak{0.45}{0.30}{ABSalmon}
    \foreach \x in {1.80,3.15,4.50,5.85}{\abpeak{\x}{0.30}{ABTeal}}
    \node[lbl, text=ABSalmonDark] at (0.45,-0.22) {$y{=}0$};
    \node[callout] at (\BW/2,-0.72) {$p(y)=1/r$ at every multiple of $1/r$};
  \end{scope}
\end{scope}
 
\draw[sep] (7.75,1.15) -- (7.75,-6.0);
 
\begin{scope}[shift={(8.5,0)}]
  \node[font=\normalsize\bfseries] at (\BW/2,0.95) {Hashed output};
  \node[lbl] at (\BW/2,0.55) {measure the $t$-bit hash $h(f(x))$, here $t=1$};
  \node[lbl, anchor=south] at (\BW/2,-0.42) {teal, purple and light purple share one hash value};
  \aballcells{-1.0}
  \draw[flow] (\BW/2,-1.05) -- (\BW/2,-1.55) node[midway,right=2pt,lbl,black]{measure: that hash value};
  \abcells{-2.1}{0/cA,1/cB,4/cE,5/cA,6/cB,9/cE,10/cA,11/cB,14/cE}
  \node[lbl] at (\BW/2,-2.32) {input register: three cosets at once, each still of period $r$};
  \draw[flow] (\BW/2,-2.5) -- (\BW/2,-3.0) node[midway,right=2pt,lbl,black]{QFT};
  \begin{scope}[shift={(0,-4.75)}]
    \draw[frame] (0,0) rectangle (\BW,\PH);
    \foreach \x in {1.80,3.15,4.50,5.85}{\abpeak{\x}{0.15}{ABTeal}\abghost{\x}{0.30}{0.15}}
    \abpeak{0.45}{0.90}{ABSalmon}
    \node[lbl, text=ABSalmonDark] at (0.45,-0.22) {$y{=}0$};
    \node[callout] at (\BW/2,-0.72) {$p_h(y)=(1-2^{-t})\,p(y)$ for $y\neq0$};
  \end{scope}
\end{scope}
\end{tikzpicture}

\caption{How hashing the output creates collisions (Eq.~\eqref{eq:p_h0}), for a toy
function of period $r=5$ on $15$ inputs, in one dimension as in Section~\ref{sec:shor}.
The same picture applies to each coset of the two-dimensional lattice of
Section~\ref{sec:input-register}. In the top bars, each cell is one input $x$ and its color
is the value $f(x)$, so colors repeat every $r$ cells. Left: measuring $f(x)$ collapses the
input register to the one coset of the measured value, and the QFT gives equal peaks
$p(y)=1/r$ at the multiples of $1/r$, including the uninformative $y=0$. Right: a $t$-bit
hash ($t=1$ shown) gives the same value to about $r/2^{t}$ cosets, here three, so the
input register collapses to all of them at once. Each is still of period $r$, so the peaks
stay at the same positions. Here $p_h(y)$ is the outcome probability with the hashed
oracle, averaged over the random choice of $h$. Each informative peak is scaled by
$1-2^{-t}$ (dashed: exact height), and the removed weight goes to $y=0$. A large $t$
recovers exact Shor, while $t=1$ doubles the number of runs and saves $n$ output qubits.}
\label{fig:compression-collisions}
\end{figure*}
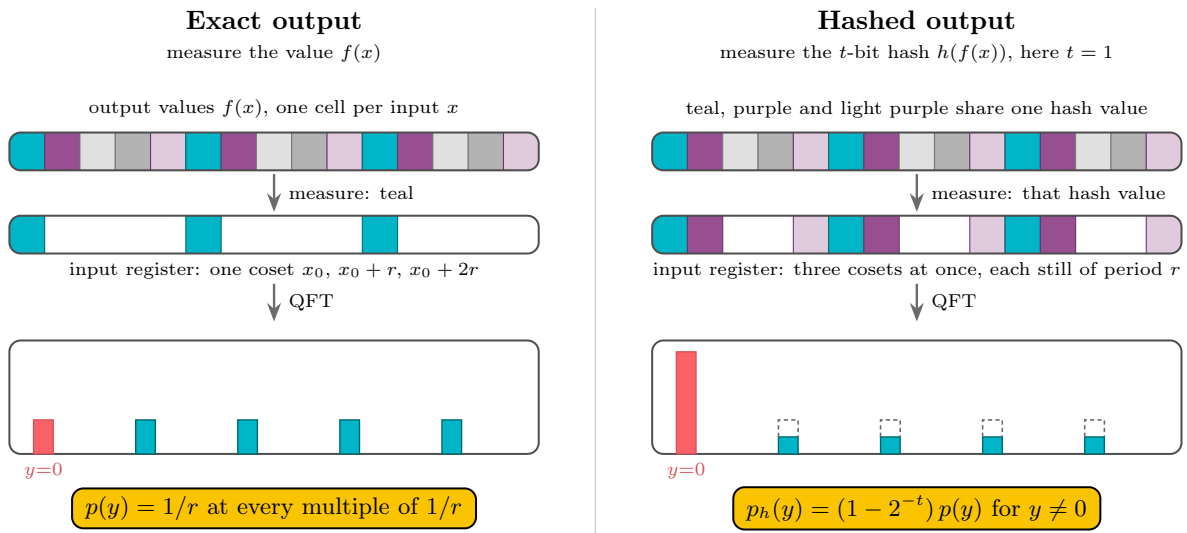

\paragraph{Compression is periodicity-preserving.} The key observation is that period
finding never uses the value of $f(a,b) = g^{a}x^{-b} \bmod N$ itself. It only uses the
fact that $f$ is periodic. Composing $f$ with any function $h$ preserves that periodicity,
since $f(a,b) = f(a+d,b+1)$ immediately gives 
\begin{equation}
  h\bigl(f(a,b)\bigr) = h\bigl(f(a+d,b+1)\bigr), ~\text{ for all } a,b.
\end{equation}
In particular one may take $h$ to hash the output all the way down to a single bit. Doing
so introduces many collisions $h(f(a,b)) = h(f(a',b'))$ between distinct cosets of $H$,
that is, with $(a-a')-(b-b')d\not\equiv 0 \pmod r$. Remarkably, these collisions do not corrupt the period
signal. They only add weight to the useless outcome.

\paragraph{May--Schlieper main result.} Let $\mathcal{H}_t = \{\, h : \mathbb{Z}_N \to
\{0,1\}^{t}\,\}$ be a \emph{universal} family of hash functions in the sense of May and
Schlieper~\cite[Def.~1]{MaySchlieper22}, meaning that any two distinct inputs collide
with probability \emph{exactly} $2^{-t}$:
\begin{equation}
  \Pr_{h \in \mathcal{H}_t}[h(z_1) = h(z_2)] = 2^{-t}
  \qquad \text{for all } z_1 \neq z_2.
\end{equation}
Suppose we replace the oracle $f$ by its hashed version $h \circ f$, drawing a fresh $h$
uniformly from $\mathcal{H}_t$ at each run. Write $p(y)$ for the probability of measuring
the frequency $y$ with the original $f$, and $p_h(y)$ for the probability with
$h \circ f$, taken over both the measurement and the random choice of $h$. May and
Schlieper prove that
\begin{equation}
  p_h(y) = \bigl(1 - 2^{-t}\bigr)\, p(y)
  \qquad \text{for every } y \neq 0
\end{equation}
\cite[Thm.~7]{MaySchlieper22}, provided the circuit meets their cancellation
criterion~\cite[Eq.~(14)]{MaySchlieper22}. Write the state before measurement as
$\sum_{y}\sum_{z} w_{y,z}\ket{y}\ket{z}$, with $z$ ranging over the values of $f$. The
criterion asks that $\sum_{z} w_{y,z}=0$ for every $y\neq 0$. It holds for Shor's circuit
and for its Ekerå--Håstad variant. The sum $\sum_{z} w_{y,z}$ is the amplitude the circuit
would give $y$ if $f$ were constant, and a QFT of a uniform superposition puts all of that
amplitude on $y=0$. Every informative outcome $y\neq 0$ therefore keeps its probability
up to the same factor $1-2^{-t}$, on average over $h$. Since both distributions sum to
one, the removed probability reappears on the outcome $y=0$:
\begin{equation}\label{eq:p_h0}
  p_h(0) = p(0) + 2^{-t}\!\sum_{y \neq 0} p(y).
\end{equation}
On average over $h$, collisions move weight from every informative outcome to the bin
$y=0$ at the same rate $2^{-t}$. For a fixed $h$ the rate depends on
$y$~\cite[Eq.~(13) and Fig.~11b]{MaySchlieper22}. This is depicted in Figure~\ref{fig:compression-collisions}.

\paragraph{Consequences.} Two conclusions follow directly. First, the classical
post-processing is unchanged. Averaged over the random choice of $h$, the outcomes
$y\neq 0$ keep their relative probabilities, so the lattice-based recovery of the
previous section still applies. Second, the cost is only a constant factor in runs. With
a $t$-bit hash a nonzero outcome appears with probability $(1-2^{-t})(1-p(0))$, so the
expected number of measurements grows by a factor $1/(1-2^{-t})$. For the extreme choice
$t = 1$ this is a factor of two, that is, twice as many shots in exchange for discarding
all but one output qubit. Applied to Ekerå--Håstad in the oracle model, this lowers the
measured registers for factoring $n$-bit RSA from about $\tfrac32 n$ to
$\bigl(\tfrac{1}{2}+o(1)\bigr)n$ qubits as $s$ grows, workspace excluded.

The oracle $f$ is exact ($\epsilon=0$). The read-out $R$ draws a fresh hash $h$ from
$\mathcal{H}_t$ for each run and measures $h\circ f$. Hashing adds no corruption. Under
Definition~\ref{def:sep-proc}, $\delta$ is a probability over the whole shot, including
the draw of $h$. What matters is therefore the outcome distribution averaged over $h$. By
Theorem~7 of~\cite{MaySchlieper22}, that distribution, conditioned on $y\neq 0$, is
exactly the one without hashing, so the conditional corruption rate is unchanged. For an
exact transform there is a stronger reason, which holds for every $h$. Two elements of
the same coset of $H$ share the value of $f$, hence of $h\circ f$, so $h\circ f$ is
constant on the cosets of $H$. Its spectrum therefore stays on the subgroup orthogonal to
$H$, and no outcome can carry a false constraint. For a fixed $h$ and a power-of-two
transform no such per-$h$ statement is available, since the rescaling then depends on
$y$. What hashing adds is aborts. The abort rate of Definition~\ref{def:sep-proc} is
$\delta_{\bot}=p_h(0)=p(0)+2^{-t}\bigl(1-p(0)\bigr)$, which is at least $2^{-t}$. The
read-out without hashing already discards $y=0$ with probability $p(0)$, so the abort
rate the hash adds is $2^{-t}\bigl(1-p(0)\bigr)\le 2^{-t}$. This added rate is the entry
of Table~\ref{tab:four-relaxation}.

\begin{figure*}[tp]
\centering
\def\rad{3pt}
\tikzset{
  abwire/.style={draw=abInk},
  abgate/.style={draw=abInk, fill=white, line width=0.7pt, rounded corners=\rad},
  abexp/.style={draw=abTealDark, fill=abTeal!35, line width=1pt, rounded corners=\rad},
  abreg/.style={draw=abTealDark, dashed, line width=0.7pt, rounded corners=\rad},
  abhash/.style={draw=abInk, fill=abYellow, line width=1.2pt, rounded corners=\rad},
  abhashreg/.style={draw=abInk, fill=abYellow!20, densely dashed, line width=0.9pt,
                    rounded corners=\rad},
  abctrl/.style={circle, fill=abInk, inner sep=0pt},
  abbrace/.style={draw=abInk, decorate, decoration={brace,amplitude=5pt,mirror}, line width=0.6pt},
}
 
\begin{subfigure}[b]{0.48\linewidth}
\centering
\resizebox{\linewidth}{!}{%
\begin{tikzpicture}[
  x=1cm,y=1cm,
  wire/.style={abwire,line width=0.6pt},
  ctrl/.style={abctrl,minimum size=3.4pt},
  >={Stealth[length=4pt,width=4pt]}
]
 
\draw[abreg] (0.35,4.90) rectangle (8.15,6.80);
\draw[abreg] (0.35,3.72) rectangle (8.15,4.42);
\draw[abreg] (0.35,1.35) rectangle (8.15,3.25);
 
\foreach \i in {0,...,6}{%
  \pgfmathsetmacro\yy{5.1+0.25*\i}
  \draw[wire,->] (0,\yy) -- (8.5,\yy);
}
 
\draw[wire,->] (0,3.95) -- (8.5,3.95);
\draw[wire,->] (0,4.20) -- (8.5,4.20);
 
\draw[wire,->] (0,3.00) -- (8.5,3.00);
\draw[wire,->] (0,2.65) -- (8.5,2.65);
\draw[wire,->] (0,2.30) -- (8.5,2.30);
\draw[wire,->] (0,1.95) -- (8.5,1.95);
\draw[wire,->] (0,1.60) -- (8.5,1.60);
 
\draw[abgate] (0.60,4.95) rectangle (1.50,6.75); \node at (1.05,5.85){$H$};
\draw[abgate] (0.60,3.80) rectangle (1.30,4.35); \node at (0.95,4.075){$H$};
 
\draw[abexp] (1.70,1.40) rectangle (3.00,3.20); \node at (2.35,2.30){$\times g^{a}$};
\draw[abexp] (4.00,1.40) rectangle (5.00,3.20); \node at (4.50,2.30){$\times x^{-b}$};
 
\draw[abgate] (6.00,4.95) rectangle (6.90,6.75); \node[font=\small] at (6.45,5.85){QFT};
\draw[abgate] (6.00,3.80) rectangle (6.70,4.35); \node[font=\scriptsize] at (6.35,4.075){QFT};
 
\draw[abgate] (7.20,4.95) rectangle (7.90,6.75);
\draw[abInk,line width=0.5pt] (7.37,5.78) arc (180:0:0.18);
\draw[abInk,-{Stealth[length=3pt]},line width=0.5pt] (7.55,5.78) -- (7.70,6.02);
\draw[abgate] (7.20,3.80) rectangle (7.90,4.35);
\draw[abInk,line width=0.5pt] (7.44,4.02) arc (180:0:0.11);
\draw[abInk,-{Stealth[length=2.5pt]},line width=0.5pt] (7.55,4.02) -- (7.65,4.17);
 
\foreach \i in {0,...,6}{%
  \pgfmathsetmacro\yy{5.1+0.25*\i}
  \pgfmathsetmacro\xc{1.9+0.16*(6-\i)}
  \draw[wire] (\xc,\yy) -- (\xc,3.20);
  \node[ctrl] at (\xc,\yy){};
}
 
\draw[wire] (4.45,4.20) -- (4.45,3.20);
\draw[wire] (4.60,3.95) -- (4.60,3.20);
\node[ctrl] at (4.45,4.20){};
\node[ctrl] at (4.60,3.95){};
 
\node[anchor=east] at (-0.10,5.850){$\lvert 0\rangle$};
\node[anchor=east] at (-0.10,4.075){$\lvert 0\rangle$};
\node[anchor=east] at (-0.10,2.300){$\lvert 1\rangle$};
 
\node[rotate=90] at (10.25,5.850){$n/2+n/(2s)$};
\node at (9.15,5.850){$\lvert j\rangle$};
\node[rotate=90] at (10.25,4.075){$n/(2s)$};
\node at (9.15,4.075){$\lvert k\rangle$};
\node[rotate=90] at (10.25,2.300){$n$ qubits};
\node[anchor=west] at (8.55,2.300){$\lvert g^{a}x^{-b}\rangle$};
 
\draw[abbrace] (1.70,1.20) -- (3.00,1.20) node[midway,below=6pt]{$n/2+n/(2s)$};
\draw[abbrace] (4.00,1.20) -- (5.00,1.20) node[midway,below=6pt]{$n/(2s)$};
 
\end{tikzpicture}%
}
\caption{Ekerå--Håstad~\cite{PQCRYPTO:EkeHas17}}
\label{fig:ms-transition-eh}
\end{subfigure}
\hfill
\begin{subfigure}[b]{0.48\linewidth}
\centering
\resizebox{\linewidth}{!}{%
\begin{tikzpicture}[
  x=1cm,y=1cm,
  wire/.style={abwire,line width=0.6pt},
  ctrl/.style={abctrl,minimum size=3.4pt},
  >={Stealth[length=4pt,width=4pt]}
]
\useasboundingbox (-1.00,0.45) rectangle (10.80,6.95);
\draw[abreg]     (0.35,4.90) rectangle (8.15,6.80);
\draw[abreg]     (0.35,3.72) rectangle (8.15,4.42);
\draw[abreg]     (0.35,1.80) rectangle (8.15,2.85);

\foreach \i in {0,...,6}{%
  \pgfmathsetmacro\yy{5.1+0.25*\i}
  \draw[wire,->] (0,\yy) -- (8.5,\yy);
}
\draw[wire,->] (0,3.95) -- (8.5,3.95);
\draw[wire,->] (0,4.20) -- (8.5,4.20);

\draw[wire,->] (0,2.65) -- (8.5,2.65);
\draw[wire,->] (0,2.35) -- (8.5,2.35);
\draw[wire,->] (0,2.05) -- (8.5,2.05);

\draw[abgate] (0.60,4.95) rectangle (1.50,6.75); \node at (1.05,5.85){$H$};
\draw[abgate] (0.60,3.80) rectangle (1.30,4.35); \node at (0.95,4.075){$H$};

\foreach \i in {0,...,6}{%
  \pgfmathsetmacro\yy{5.1+0.25*\i}
  \pgfmathsetmacro\xc{2.1+0.22*(6-\i)}
  \draw[wire] (\xc,\yy) -- (\xc,2.85);
  \node[ctrl] at (\xc,\yy){};
}
\draw[wire] (4.00,4.20) -- (4.00,2.85);
\draw[wire] (4.20,3.95) -- (4.20,2.85);
\node[ctrl] at (4.00,4.20){};
\node[ctrl] at (4.20,3.95){};

\draw[abhash] (1.80,1.85) rectangle (4.60,2.85);
\node[text=abInk] at (3.20,2.35){$U_{h\circ f}$};

\draw[abgate] (6.00,4.95) rectangle (6.90,6.75); \node[font=\small] at (6.45,5.85){QFT};
\draw[abgate] (6.00,3.80) rectangle (6.70,4.35); \node[font=\scriptsize] at (6.35,4.075){QFT};

\draw[abgate] (7.20,4.95) rectangle (7.90,6.75);
\draw[abInk,line width=0.5pt] (7.37,5.78) arc (180:0:0.18);
\draw[abInk,-{Stealth[length=3pt]},line width=0.5pt] (7.55,5.78) -- (7.70,6.02);
\draw[abgate] (7.20,3.80) rectangle (7.90,4.35);
\draw[abInk,line width=0.5pt] (7.44,4.02) arc (180:0:0.11);
\draw[abInk,-{Stealth[length=2.5pt]},line width=0.5pt] (7.55,4.02) -- (7.65,4.17);
\draw[abgate] (7.20,1.85) rectangle (7.90,2.80);
\draw[abInk,line width=0.5pt] (7.37,2.27) arc (180:0:0.18);
\draw[abInk,-{Stealth[length=2.5pt]},line width=0.5pt] (7.55,2.27) -- (7.70,2.55);

\node[anchor=east] at (-0.10,5.850){$\lvert 0\rangle$};
\node[anchor=east] at (-0.10,4.075){$\lvert 0\rangle$};
\node[anchor=east] at (-0.10,2.350){$\lvert 0\rangle$};

\node[rotate=90] at (10.55,5.850){$n/2+n/(2s)$};
\node at (9.15,5.850){$\lvert j\rangle$};
\node[rotate=90] at (10.55,4.075){$n/(2s)$};
\node at (9.15,4.075){$\lvert k\rangle$};
\node[rotate=90] at (10.55,2.35){$t$};
\node[anchor=west] at (8.55,2.35){$\lvert h(g^{a}x^{-b})\rangle$};
\end{tikzpicture}%
}
\caption{May--Schlieper, oracle model~\cite{MaySchlieper22}}
\label{fig:ms-transition-ms}
\end{subfigure}
\caption{Output-register compression, drawn to a common scale in which one wire
stands for $n/4$ qubits. (a): in Ekerå--Håstad the measured output
register holds the full $n$-bit value $g^{a}x^{-b}\bmod N$. (b): in the oracle model of
May and Schlieper, a single circuit $U_{h\circ f}$ writes $h(g^{a}x^{-b}\bmod N)$ into $t\ll n$
measured qubits, at the cost of a factor $1/(1-2^{-t})$ in the number of runs. All arithmetic is modulo $N$, and the
$\bmod N$ is omitted inside the boxes.
}
\label{fig:ms-transition}
\end{figure*}
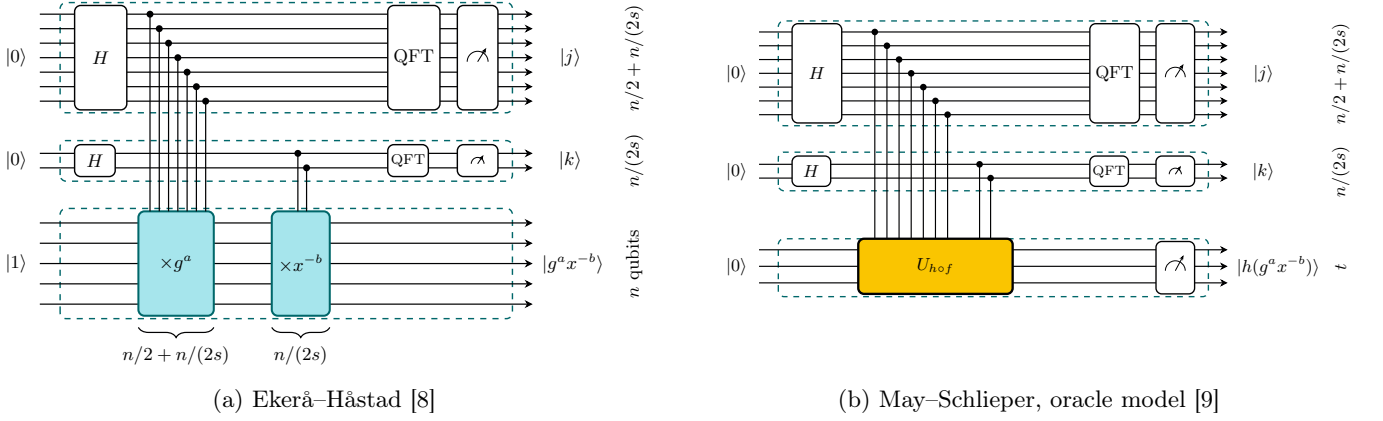

The effect of this compression on the output register is illustrated in
Figure~\ref{fig:ms-transition}, contrasting the full $n$-qubit measured register of
Ekerå--Håstad with the $t$-qubit hashed register of May-Schlieper. 

\paragraph{The remaining obstacle.} As noted at the start of this section, this bound holds
only in the oracle model, which assumes black-box access to $U_{h\circ f}$. Computing
$g^a x^{-b} \bmod N$ first and hashing the result afterwards is not enough. That would
reintroduce the $n$ output qubits one is trying to eliminate. Computing $h(g^a
x^{-b} \bmod N)$ directly, without ever materializing the $n$-bit intermediate value, is
left open by May and Schlieper and is precisely the problem solved by the
residue-arithmetic construction of the next section. 

\subsection{Making the Oracle Function Space Efficient}\label{sec:oracle-rns}

The May-Schlieper compression leaves one obstacle: computing $h(g^a x^{-b} \bmod
N)$ without ever forming the $n$-bit value $g^a x^{-b} \bmod N$ in an intermediate
register. Chevignard, Fouque and Schrottenloher~\cite{C:CheFouSch25} solved it,
and Gidney~\cite{gidney2025factor2048bitrsa} later improved the
construction. The key idea is to carry out the entire modular exponentiation in a
\emph{residue number system} (RNS), where every intermediate value fits in
$O(\log\log N)$ qubits. Only the most significant bits of the result are then extracted,
through an approximate accumulation.

As in the description of Shor's algorithm in Section~\ref{sec:shor}, we present the
residue arithmetic, and the approximate period finding that follows it, for a
one-dimensional exponent $e$ and period $r$. This keeps the notation light. Nothing in this construction
or the next depends on the dimension. Every statement below applies unchanged to the
two-register Ekerå--Håstad setting, with $e$ replaced by the pair $(a,b)$ and
$H=\langle r\rangle$ replaced by the lattice $H=\langle (r,0),(d,1)\rangle$ of
Section~\ref{sec:short-dlp}. This is the setting actually
used once the reductions of Sections~\ref{sec:input-register}
and~\ref{sec:output-register} are combined with the oracle below. 

\paragraph{Residue arithmetic.} Fix a set $P$ of $\ell$-bit primes and let
$L = \prod_{p \in P} p$. The residue number system represents an integer $z < L$ by
its residues $(z \bmod p)_{p \in P}$, and performs additions and multiplications
independently modulo each prime, reconstructing the result through the Chinese
remainder theorem (CRT). By the prime number theorem, the primes can be taken of size
$\ell = \Theta(\log\log L)$ to have enough primes to supply $P$: there are $\sim
2^{\ell}/\ell\ln 2$ primes of size $\ell$, forcing $2^{\ell}\ge n_en$. Each residue
occupies only $O(\log\log N)$ qubits. 
The modular exponentiation
\begin{equation}
\begin{aligned}
  V = g^{e} \bmod N &= \left(\prod_{k=0}^{n_e-1} M_k^{e_k}\right) \bmod
  N,\\
  &\text{ with } M_k = g^{2^k} \bmod N,\footnote{Recall that $n_e$ is the number of exponent
    bits. In the two-registers setting $g^a x^{-b}$ this counts both exponentiations, $n_e
    = (\tau_{s}+m)+\tau_{s}$.}
\end{aligned}
\end{equation}
is not reduced modulo $N$ during the quantum computation. In a residue number system,
a reduction modulo $N$ is expensive, because it would require reconstructing the value (or
comparing it against $N$) inside the circuit. That defeats the purpose of the RNS. Its additions and multiplications are cheap precisely because they act componentwise modulo
the fixed primes $p\in P$. The circuit therefore accumulates the unreduced product
$\prod_{k} M_k^{e_k}$ modulo $L$ and performs a single reduction modulo $N$ only at the
end. This is safe provided the accumulator never wraps modulo $L$. Each $M_k < N$, and at
most $n_e$ factors are multiplied, so the unreduced product stays below $N^{n_e}$. Hence,
as in~\cite{C:CheFouSch25}, one requires
\begin{equation}
  L \geq N^{n_e}.
\end{equation}
This guarantees that the product never exceeds $L$. Write $\bar V=\prod_{k} M_k^{e_k}$ for this 
unreduced product, so that $V=\bar V \bmod N$.
The circuit holds $\bar V$ through its residues $r_j=\bar V \bmod p_j$. Since $\bar V<L$,
the Chinese remainder theorem recovers it from them, using the CRT contribution factors
$u_{p_j} = (L/p_j)\cdot ((L/p_j)^{-1}\bmod p_j)$. Writing $r_{j,k}$ for the $k$-th bit of
$r_j$, with $j$ ranging over $0,\dots,|P|-1$ and $k$ over $0,\dots,\ell-1$, this gives the
exact identity~\cite[Eq.~(13)]{gidney2025factor2048bitrsa}
\begin{equation*}\label{eq:rns-exact}
\begin{aligned}
  V &= \Bigl(\sum_{j} r_j\, u_{p_j}\Bigr) \bmod L \bmod N\\
    &= \Bigl(\sum_{j}\sum_{k} r_{j,k}\, u_{p_j}\, 2^{k}\Bigr) \bmod L \bmod N.
\end{aligned}
\end{equation*}
This turns a product modulo $N$ into a sum, which is what makes the subsequent
approximation stable. An error introduced early is no longer amplified by later
multiplications. The coefficients $u_{p_j}$ satisfy $u_{p_j} \equiv 1 \pmod{p_j}$ and
$u_{p_j} \equiv 0 \pmod{p_i}$ for $i \neq j$, by definition of the modular inverse. Hence
$\sum_i r_i u_{p_i} \equiv r_j \pmod{p_j}$ for each $j$, which recovers $\bar V \bmod p_j$
from the residues. This is precisely the CRT reconstruction.

The circuit cannot afford the outer reduction modulo $L$ either. It therefore reduces
each term separately and accumulates modulo $N$ only,
\begin{equation}\label{eq:rns-approx}
  V \;\approx\; \Bigl(\sum_{j}\sum_{k} r_{j,k}\,
  \bigl[(u_{p_j}2^{k}) \bmod L \bmod N\bigr]\Bigr) \bmod N.
\end{equation}
This is not an identity. After its own reduction modulo $L$, each term is below $L$, so
the sum of the reduced terms wraps modulo $L$ fewer than $A=|P|\cdot\ell$ times. Each wrap
that the right-hand side ignores shifts it by $L \bmod N$. The two sides of
\eqref{eq:rns-approx} therefore differ by a modular distance of at most
$A\,\Delta_N(L)$~\cite[Eq.~(19)]{gidney2025factor2048bitrsa}. Here
$\Delta_N(z) = \min_{k \in \mathbb{Z}} |z - kN|$ denotes the distance from $z$ to the
nearest multiple of $N$, i.e.\ the size of the centered residue
$\{z\}_N \in [-N/2, N/2)$. To keep this error at the truncation scale, a second condition
on $L$ is required,
\begin{equation}\label{eq:dev}
  \Delta_N(L) \;<\; N\,2^{-\kappa},
\end{equation}
where $\kappa$ denotes the number of top bits kept (see hereafter). The primes in $P$ are
ours to choose, and Gidney finds a suitable $L$ by random search over prime sets, under a
stated conjecture~\cite[Assumption~1]{gidney2025factor2048bitrsa}.

\paragraph{Extracting the most significant bits.} Rather than compute $V$ in full,
the algorithm keeps only its top $\kappa$ bits. Additions are truncated: to add an offset
$c$ one instead adds $(c \bmod N) \gg \bar\kappa$ with $\bar\kappa = \log N - \kappa$, and
the accumulator is operated modulo $N \gg \bar\kappa$. Each truncated addition drops the
low $\bar\kappa$ bits, perturbing the result by an absolute modular distance of at most
$O(2^{\bar\kappa}) = O(N\cdot 2^{-\kappa})$, equivalently, by a fractional deviation
$O(2^{-\kappa})$ of the modulus. Since deviations accumulate linearly over the
$|P|\cdot\ell$ additions, and the wrap error of~\eqref{eq:rns-approx} is below
$A\,N2^{-\kappa}$ by~\eqref{eq:dev}, the total deviation of the approximation
$\tilde V \approx V \gg \bar\kappa$ is
\begin{equation}\label{eq:deviation}
  \Delta_N\!\bigl(V - (\tilde V \ll \bar\kappa)\bigr) \leq \frac{\epsilon}{2}\, N,
  \qquad
  \epsilon = O\!\left(|P|\cdot\ell\cdot 2^{-\kappa}\right).
\end{equation}
The accumulator holds only the top $\kappa$ bits of the result, so its raw value is about
$2^{\bar\kappa}$ times smaller than $V$. To compare the two on the same scale, we write
\begin{equation}\label{eq:def_hat_v}
  \hat V(e) := \tilde V(e) \ll \bar\kappa = 2^{\bar\kappa}\,\tilde V(e)
\end{equation}
for the accumulator content scaled back to the size of $N$. Equation~\eqref{eq:deviation}
then reads $\Delta_N\bigl(V(e)-\hat V(e)\bigr)\le\tfrac{\epsilon}{2}N$.

Equation~\eqref{eq:deviation} is stated with the same $\epsilon/2$ normalization as
condition~(i) of Definition~\ref{def:noisy-hsp}. So $\hat V$ instantiates that
condition with exactly this $\epsilon$, and no rescaling is needed. The bound is only informative
for $\epsilon<1$: since $\Delta_N$ never exceeds $N/2$, any claimed bound of
$\tfrac{\epsilon}{2}N\ge N/2$ holds. The choice of $\kappa$ in the success-rate analysis
keeps $\epsilon$ far below this threshold, driving it to an arbitrarily small constant,
not merely below $1$. The approximation reduces the modular exponentiation to a
$\kappa$-bit dot product against classically precomputed constants. These constants are
$C_{j,k} =\bigl[((u_{p_j}2^k) \bmod L \bmod N) \gg \bar\kappa\bigr] \bmod (N \gg \bar\kappa)$, and the dot product reads
\begin{equation}\label{eq:approximate_modular_exponentiation}
  \tilde V = \left(\sum_{j=0}^{|P|-1} \sum_{k=0}^{\ell-1} r_{j,k}\, C_{j,k}\right)
             \bmod (N \gg \bar\kappa).
\end{equation}
           
Every quantum register involved, apart from the input register itself, has size
$O(\log\log N)$: the $n$-qubit intermediate value has been eliminated. 
Figure~\ref{fig:crt-rns} summarizes the three steps on a toy example: residues,
lane-wise arithmetic, and keeping only the top bits. The output is then
hidden by superposition masking rather than by an explicit hash. The approximate value
$\tilde V$ is added directly into the mask register, which is immediately measured. As a
result, no $n$-qubit intermediate state ever accumulates, and no reverse exponentiation is
needed to uncompute it. The two techniques share the same shape, fresh randomness drawn
each run, then a measurement that reveals only part of the value, but not the same
guarantee. May-Schlieper's $h$ acts on the exact value $f(a,b)$, so its row of
Table~\ref{tab:four-relaxation} has $\epsilon=0$ and can only waste a shot, never corrupt
one. Masking instead acts on $\tilde V$, which already carries the truncation drift
$\epsilon$ from above, so it can corrupt a shot too. $\kappa$ and the mask width $S$ are
accordingly two separate parameters, with $\kappa$ fixing the drift $\epsilon$ that $S$
then has to be chosen large enough to absorb, made precise in the success-rate paragraph
below. We explain this technique in more detail below.  

\begin{figure*}[tp]
\centering
\def\rad{3pt}
\tikzset{
  abwire/.style={draw=abInk},
  abgate/.style={draw=abInk, fill=white, line width=0.7pt, rounded corners=\rad},
  abexp/.style={draw=abTealDark, fill=abTeal!35, line width=1pt, rounded corners=\rad},
  abreg/.style={draw=abTealDark, dashed, line width=0.7pt, rounded corners=\rad},
  abhash/.style={draw=abInk, fill=abYellow, line width=1.2pt, rounded corners=\rad},
  abhashreg/.style={draw=abInk, fill=abYellow!20, densely dashed, line width=0.9pt,
                    rounded corners=\rad},
  abresreg/.style={draw=abTealDark, fill=abTeal!12, densely dashed, line width=0.9pt,
                   rounded corners=\rad},
  abacc/.style={draw=abInk, line width=0.7pt, rounded corners=\rad},
  abctrl/.style={circle, fill=abInk, inner sep=0pt},
  abbrace/.style={draw=abInk, decorate, decoration={brace,amplitude=5pt,mirror}, line width=0.6pt},
}
 
\begin{subfigure}{\linewidth}
\centering
\begin{tikzpicture}[
  x=1cm,y=1cm,
  wire/.style={abwire,line width=0.6pt},
  cwire/.style={abwire,line width=0.45pt,double,double distance=1.1pt},
  ctrl/.style={abctrl,minimum size=3.4pt},
  >={Stealth[length=4pt,width=4pt]}
]
\draw[abreg]     (0.35,4.90) rectangle (10.62,6.80);
\draw[abreg]     (0.35,3.72) rectangle (10.62,4.42);
\draw[abhashreg] (-0.80,1.35) rectangle (9.20,3.25);   
\draw[abreg]     (0.35,-0.10) rectangle (10.62,0.95);

\foreach \i in {0,...,6}{%
  \pgfmathsetmacro\yy{5.1+0.25*\i}
  \draw[wire] (0,\yy) -- (9.80,\yy);
  \draw[cwire] (10.50,\yy) -- (11.10,\yy);
}
\foreach \yy in {3.95,4.20}{%
  \draw[wire] (0,\yy) -- (9.80,\yy);
  \draw[cwire] (10.50,\yy) -- (11.10,\yy);
}
\foreach \yy in {3.00,2.65,2.30,1.95,1.60}{\draw[wire] (0,\yy) -- (8.45,\yy);}
\foreach \yy in {0.15,0.45,0.75}{%
  \draw[wire] (0,\yy) -- (9.80,\yy);
  \draw[cwire] (10.50,\yy) -- (11.10,\yy);
}

\draw[abgate] (0.60,4.95) rectangle (1.50,6.75); \node at (1.05,5.85){$H$};
\draw[abgate] (0.60,3.80) rectangle (1.50,4.35); \node at (1.05,4.075){$H$};

\draw[abexp]  (1.70,1.40) rectangle (3.00,3.20); \node at (2.35,2.30){$\times g^{a}$};
\draw[abexp]  (3.30,1.40) rectangle (4.30,3.20); \node at (3.80,2.30){$\times x^{-b}$};
\draw[abhash] (4.60,1.40) rectangle (5.40,3.20); \node[text=abInk] at (5.00,2.30){$h$};
\draw[abexp]  (5.70,1.40) rectangle (6.70,3.20); \node at (6.20,2.30){$\times x^{b}$};
\draw[abexp]  (7.00,1.40) rectangle (8.30,3.20); \node at (7.65,2.30){$\times g^{-a}$};

\draw[abgate] (8.60,4.95) rectangle (9.50,6.75); \node[font=\small] at (9.05,5.85){QFT};
\draw[abgate] (8.60,3.80) rectangle (9.50,4.35); \node[font=\small] at (9.05,4.075){QFT};

\foreach \ylo/\yhi in {4.95/6.75, 3.80/4.35, -0.05/0.90}{%
  \pgfmathsetmacro\ym{(\ylo+\yhi)/2}
  \draw[abgate] (9.80,\ylo) rectangle (10.50,\yhi);
  \draw[abInk,line width=0.5pt] (9.98,\ym-0.08) arc (180:0:0.17);
  \draw[abInk,-{Stealth[length=2.5pt]},line width=0.5pt] (10.15,\ym-0.08) -- (10.29,\ym+0.15);
}

\foreach \i in {0,...,6}{%
  \pgfmathsetmacro\yy{5.1+0.25*\i}
  \pgfmathsetmacro\xc{1.9+0.16*(6-\i)}
  \pgfmathsetmacro\xd{7.2+0.16*(6-\i)}
  \draw[wire] (\xc,\yy) -- (\xc,3.20); \node[ctrl] at (\xc,\yy){};
  \draw[wire] (\xd,\yy) -- (\xd,3.20); \node[ctrl] at (\xd,\yy){};
}
\foreach \xb in {3.75,6.15}{%
  \draw[wire] (\xb,4.20) -- (\xb,3.20); \node[ctrl] at (\xb,4.20){};
  \draw[wire] (\xb+0.15,3.95) -- (\xb+0.15,3.20); \node[ctrl] at (\xb+0.15,3.95){};
}
\foreach \yy in {0.15,0.45,0.75}{
  \draw[wire] (5.00,1.40) -- (5.00,\yy);
  \draw[abInk,line width=0.7pt] (5.00,\yy) circle (0.11);
  \draw[abInk,line width=0.7pt] (4.89,\yy) -- (5.11,\yy);
  \draw[abInk,line width=0.7pt] (5.00,\yy-0.11) -- (5.00,\yy+0.11);
}

\node[anchor=east] at (-0.10,5.850){$\lvert 0\rangle$};
\node[anchor=east] at (-0.10,4.075){$\lvert 0\rangle$};
\node at (-0.40,2.300){$\lvert 1\rangle$};
\node[anchor=east] at (-0.10,0.450){$\lvert 0\rangle$};
\node at (8.82,2.300){$\lvert 1\rangle$};

\node[anchor=west] at (11.15,5.850){$j$};
\node[anchor=west] at (11.15,4.075){$k$};
\node[anchor=west] at (11.15,0.45){$h(g^{a}x^{-b})$};
\node[rotate=90] at (13.15,5.850){$n/2+n/(2s)$};
\node[rotate=90] at (13.15,4.075){\scriptsize$n/(2s)$};
\node[font=\bfseries] at (13.15,2.30){$n$ at peak};
\node[rotate=90] at (13.15,0.45){$t$};
\end{tikzpicture}%
\caption{May--Schlieper, computed directly~\cite{MaySchlieper22}}
\label{fig:oracle-ms}
\end{subfigure}

\par\medskip

\begin{subfigure}{\linewidth}
\centering
\begin{tikzpicture}[
  x=1cm,y=1cm,
  wire/.style={abwire,line width=0.6pt},
  cwire/.style={abwire,line width=0.45pt,double,double distance=1.1pt},
  ctrl/.style={abctrl,minimum size=3.0pt},
  >={Stealth[length=4pt,width=4pt]}
]
\def\xm{15.55}   
\def\xc{16.25}   
\def\xe{16.85}   
\def\Vhw{0.6}
\def\xcut{7.30}  

\draw[abresreg] (-0.80,1.95) rectangle (14.90,3.25);

\foreach \yy in {5.10,5.35,5.60,5.85,6.10,6.35,6.60, 3.95,4.20, 0.15,0.45,0.75}{%
  \draw[wire] (0,\yy) -- (\xcut-0.30,\yy);
  \draw[wire] (\xcut+0.30,\yy) -- (\xm,\yy);
  \draw[cwire] (\xc,\yy) -- (\xe,\yy);
  \node[inner sep=0pt, font=\scriptsize] at (\xcut,\yy) {$\cdots$};
}
\foreach \yy in {2.30,2.65,3.00}{%
  \draw[wire] (0,\yy) -- (\xcut-0.30,\yy);
  \draw[wire] (\xcut+0.30,\yy) -- (14.35,\yy);
  \node[inner sep=0pt, font=\scriptsize] at (\xcut,\yy) {$\cdots$};
}

\draw[abgate] (0.60,4.95) rectangle (1.50,6.75); \node at (1.05,5.85){$H$};
\draw[abgate] (0.60,3.80) rectangle (1.50,4.35); \node at (1.05,4.075){$H$};
\draw[abgate] (\xm-1.20,4.95) rectangle (\xm-0.30,6.75); \node[font=\small] at (\xm-0.75,5.85){QFT};
\draw[abgate] (\xm-1.20,3.80) rectangle (\xm-0.30,4.35); \node[font=\small] at (\xm-0.75,4.075){QFT};
\foreach \ylo/\yhi in {4.95/6.75, 3.80/4.35, -0.05/0.90}{%
  \pgfmathsetmacro\ymid{(\ylo+\yhi)/2}
  \draw[abgate] (\xm,\ylo) rectangle (\xc,\yhi);
  \draw[abInk,line width=0.5pt] (\xm+0.18,\ymid-0.08) arc (180:0:0.17);
  \draw[abInk,-{Stealth[length=2.5pt]},line width=0.5pt] (\xm+0.35,\ymid-0.08) -- (\xm+0.49,\ymid+0.15);
}

\foreach \xx/\lab in {2.6/{r_0}, 5.0/{r_1}, 8.6/{r_j}, 11.0/{r_{|P|-1}}}{
  \draw[abexp] (\xx-\Vhw,2.05) rectangle (\xx+\Vhw,3.20);
  \node[text=abInk] at (\xx,2.625){$\lab$};
  \foreach \i in {0,...,6}{%
    \pgfmathsetmacro\yy{5.1+0.25*\i}
    \pgfmathsetmacro\xk{\xx-0.50+0.08*(6-\i)}
    \draw[wire] (\xk,\yy) -- (\xk,3.20); \node[ctrl] at (\xk,\yy){};
  }
  \draw[wire] (\xx+0.18,4.20) -- (\xx+0.18,3.20); \node[ctrl] at (\xx+0.18,4.20){};
  \draw[wire] (\xx+0.30,3.95) -- (\xx+0.30,3.20); \node[ctrl] at (\xx+0.30,3.95){};
}
\draw[abexp] (13.05,2.05) rectangle (14.15,3.20);
\node[text=abInk] at (13.60,2.625){$r_{|P|-1}^{\dagger}$};
\foreach \i in {0,...,6}{%
  \pgfmathsetmacro\yy{5.1+0.25*\i}
  \pgfmathsetmacro\xk{13.15+0.08*(6-\i)}
  \draw[wire] (\xk,\yy) -- (\xk,3.20); \node[ctrl] at (\xk,\yy){};
}
\draw[wire] (13.85,4.20) -- (13.85,3.20); \node[ctrl] at (13.85,4.20){};
\draw[wire] (13.97,3.95) -- (13.97,3.20); \node[ctrl] at (13.97,3.95){};

\foreach \cx/\lab in {3.80/{C_{0,k}}, 6.20/{C_{1,k}}, 9.80/{C_{j,k}}, 12.25/{C_{|P|-1,k}}}{
  \foreach \yy in {2.30,2.65,3.00}{\node[ctrl] at (\cx,\yy){};}
  \draw[wire] (\cx,3.00) -- (\cx,0.95);
  \draw[abacc, fill=white] (\cx-0.72,-0.02) rectangle (\cx+0.72,0.95);
  \node[text=abInk, font=\footnotesize] at (\cx,0.465){$+\lab$};
}

\node[anchor=east] at (-0.10,5.850){$\lvert 0\rangle$};
\node[anchor=east] at (-0.10,4.075){$\lvert 0\rangle$};
\node at (-0.40,2.650){$\lvert 1\rangle$};
\node[anchor=east] at (-0.10,0.450){$\lvert \sigma\rangle$};
\node[anchor=west] at (\xe+0.05,5.850){$j$};
\node[anchor=west] at (\xe+0.05,4.075){$k$};
\node[anchor=west] at (\xe+0.05,0.45){$\tilde V$};
\node[rotate=90] at (\xe+0.95,5.850){$n/2+n/(2s)$};
\node[rotate=90] at (\xe+0.95,4.075){\scriptsize$n/(2s)$};
\node[rotate=90] at (\xe+0.95,0.42){\scriptsize$\kappa{+}1$};
\node at (14.62,2.650){$\lvert 1\rangle$};
\node[rotate=90] at (\xe+0.95,2.650){\scriptsize$\ell$};
\end{tikzpicture}
\caption{Gidney~\cite{gidney2025factor2048bitrsa}}
\label{fig:oracle-gidney}
\end{subfigure}
\caption{Making the oracle space-efficient, both panels at the same scale. All arithmetic
is modulo $N$, and the $\bmod N$ is omitted inside the boxes. (a) Computing the
May--Schlieper oracle directly forms the full $n$-qubit value $g^{a}x^{-b}\bmod N$ in the
highlighted register, hashes it, and uncomputes it, so $n$ qubits are still needed at
peak. (b) The residue-number-system construction
of~\cite{C:CheFouSch25,gidney2025factor2048bitrsa} never forms that value. The residues
$r_j$ live in a single $\ell=O(\log n)$-qubit register. From the second box on, each box
uncomputes the previous residue while computing the next one
(Figure~\ref{fig:residue-formation}), and the last box uncomputes $r_{|P|-1}$. A box
$+C_{j,k}$ on the accumulator adds the constant $C_{j,k}$ of
Equation~\eqref{eq:approximate_modular_exponentiation}, controlled by the bit $r_{j,k}$
of the residue $r_j$. The $(\kappa{+}1)$-qubit accumulator starts in the mask state
$\ket{\sigma}$ of Section~\ref{sec:approx-period} and is measured at the end. Dots mark the primes omitted between $r_1$
and $r_j$.}
\label{fig:oracle-transition}
\end{figure*}
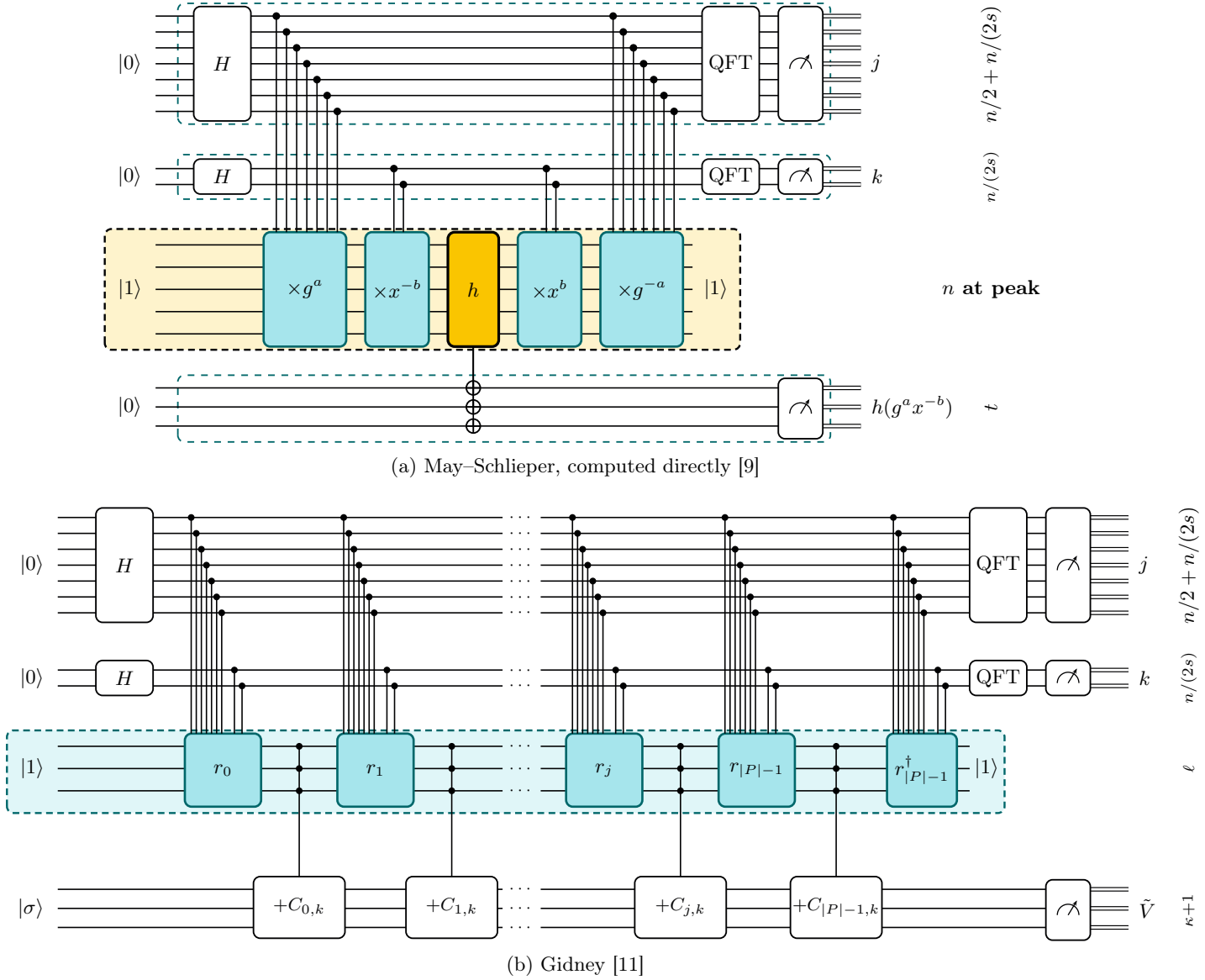

\begin{figure*}[tp]
\centering
\begin{tikzpicture}[font=\footnotesize, x=1cm, y=1cm,
  >={Stealth[length=3.2pt,width=2.8pt]},
  box/.style={draw=abLine, fill=white, rounded corners=3pt, line width=0.75pt,
              inner sep=3pt, align=center, text=abInk},
  res/.style={box, draw=ABTealDark, fill=ABTeal!10},
  never/.style={box, dashed, draw=abLine},
  keep/.style={box, draw=abInk, fill=abYellow!55, line width=1.1pt},
  inreg/.style={box, draw=abInk, line width=0.9pt},
  arr/.style={->, line width=0.75pt, draw=abLine},
  narr/.style={->, line width=0.75pt, draw=abLine, dashed},
  proj/.style={-{Triangle[open,length=5pt,width=5pt]}, line width=0.9pt, draw=abInk, densely dotted},
  xmark/.style={line width=0.6pt, draw=abLine},
  lab/.style={font=\scriptsize, text=abInk, inner sep=1.5pt},
  note/.style={font=\scriptsize\itshape, text=abLine, align=center, inner sep=1pt},
  head/.style={font=\footnotesize, text=abInk, anchor=north west, align=left, inner sep=0pt},
  sep/.style={draw=abLine, line width=0.3pt},
]
\newcommand{\abcross}[1]{\draw[xmark] (#1.south west)--(#1.north east); \draw[xmark] (#1.north west)--(#1.south east);}
 
\node[head] at (0,0) {\textbf{1.} Chinese remainder theorem: a number is its residues};
\node[box, minimum width=1.5cm, minimum height=0.6cm] (V) at (1.0,-1.35) {$V = 41$};
\node[res, minimum width=2.3cm] (r0) at (4.6,-0.75) {$41 \bmod 3 = 2$};
\node[res, minimum width=2.3cm] (r1) at (4.6,-1.35) {$41 \bmod 5 = 1$};
\node[res, minimum width=2.3cm] (r2) at (4.6,-1.95) {$41 \bmod 7 = 6$};
\draw[arr] (V.east) -- (r0.west); \draw[arr] (V.east) -- (r1.west); \draw[arr] (V.east) -- (r2.west);
\node[lab, above=1pt] at (2.35,-0.95) {$\bmod p$};
\node[note, anchor=west, text width=2.1cm] at ($(r1.east)+(0.15,0)$)
  {each residue fits in $\ell$ bits, here $\ell=3$};
\draw[arr] (r2.south) .. controls +(0,-0.75) and +(0,-0.75) .. (V.south);
\node[lab, align=center] at (3.6,-2.95)
  {back, with the CRT: $V=(2\cdot 70+1\cdot 21+6\cdot 15)\bmod 105=41$};
\draw[sep] (8.75,0.1) -- (8.75,-3.3);
 
\begin{scope}[xshift=9.0cm]
\node[head] at (0,0) {\textbf{2.} Arithmetic runs lane by lane: the diagram commutes};
\node[never, minimum width=2.2cm] (a) at (1.25,-0.85) {$41 \times 2$};
\node[never, minimum width=1.5cm] (b) at (6.6,-0.85) {$82$};
\node[res, minimum width=2.2cm] (c) at (1.25,-2.35) {$(2,1,6)\times(2,2,2)$};
\node[res, minimum width=1.5cm] (d) at (6.6,-2.35) {$(1,2,5)$};
\draw[narr] (a) -- node[lab, above, align=center]{$n$-bit multiplication\\never performed} (b);
\draw[arr] (a) -- node[lab, left]{$\bmod p$} (c);
\draw[arr] (b) -- node[lab, right]{$\bmod p$} (d);
\draw[arr] (c) -- node[lab, below=3pt, align=center]{three $\ell$-bit multiplications\\$(2{\cdot}2 \bmod 3,\ 1{\cdot}2 \bmod 5,\ 6{\cdot}2 \bmod 7)$} (d);
\node[note] at (3.9,-1.6) {$82<L=105$: no reduction needed};
\end{scope}
\draw[sep] (0,-3.45) -- (17.5,-3.45);
 
\begin{scope}[yshift=-3.75cm]
\node[head] at (0,0) {\textbf{3.} Keep only the top bits: the diagram commutes approximately};
\node[res, minimum width=1.6cm] (R) at (1.0,-1.0) {$(1,2,5)$};
\node[never, minimum width=2.8cm] (W) at (6.3,-1.0) {$V = 82 = 1010010_2$};
\node[keep, minimum width=1.6cm] (T) at (1.0,-2.7) {$\tilde V \approx 101_2$};
\node[never, minimum width=2.8cm] (E) at (6.3,-2.7) {top $\kappa=3$ bits: $101_2$};
\abcross{W}
\draw[narr] (R) -- node[lab, above]{CRT, never performed} (W);
\draw[narr] (W) -- node[lab, right]{truncate} (E);
\draw[arr] (R) -- node[lab, right, align=left]{accumulate $\sum_{j,k} r_{j,k}C_{j,k}$,\\low bits dropped} (T);
\draw[double, double distance=1.2pt, abLine, line width=0.4pt] (T) -- node[lab, fill=white]{$\approx$} (E);
\node[note, text width=4.8cm] at (3.65,-3.45) {error $O(|P|\,\ell\,2^{-\kappa})$, with $\kappa=O(\log n)$};

\begin{scope}[xshift=10.2cm, yshift=-0.9cm, every node/.style={font=\scriptsize, text=abInk}]
  \node[anchor=west, font=\scriptsize\bfseries] at (-0.05,0) {Legend};
  \node[res, minimum width=0.45cm, minimum height=0.26cm, inner sep=0pt] at (0.22,-0.42) {};
  \node[anchor=west] at (0.55,-0.42) {$\ell$-bit residues};
  \node[keep, minimum width=0.45cm, minimum height=0.26cm, inner sep=0pt] at (0.22,-0.82) {};
  \node[anchor=west] at (0.55,-0.82) {top $\kappa$ bits, kept};
  \node[never, minimum width=0.45cm, minimum height=0.26cm, inner sep=0pt] at (0.22,-1.22) {};
  \node[anchor=west] at (0.55,-1.22) {never formed};
  \draw[narr] (0,-1.62) -- (0.45,-1.62); \node[anchor=west] at (0.55,-1.62) {never performed};
\end{scope}
\end{scope}
\end{tikzpicture}

\caption{How the oracle stays space-efficient, on a toy residue number system with
$P=\{3,5,7\}$ and $L=105$. \textbf{(1)} By the Chinese remainder theorem, a number is its
residues and the CRT coefficients $u=(70,21,15)$ rebuild it. \textbf{(2)}
Arithmetic acts on each residue independently. Multiplying and then reducing modulo each
prime gives the same residues as reducing first and multiplying lane by lane, so the
$n$-bit multiplication is never performed. The product $82$ stays below $L$, so no
reduction modulo $L$ arises. \textbf{(3)} The full value $V$ is never assembled. The
circuit accumulates the residue bits against precomputed constants and keeps only the top
$\kappa=O(\log n)$ bits, which agree with the top bits of $V$ up to the truncation error.
Measuring this accumulator projects the exponent register onto the exponents consistent
with the outcome, and period finding then reads $r$ from the exponent register. The final
reduction modulo $N$ is left out of the toy. Chevignard, Fouque and Schrottenloher hash
the value to its lower bits, while Gidney keeps its top bits, as shown.}
\label{fig:crt-rns}
\end{figure*}
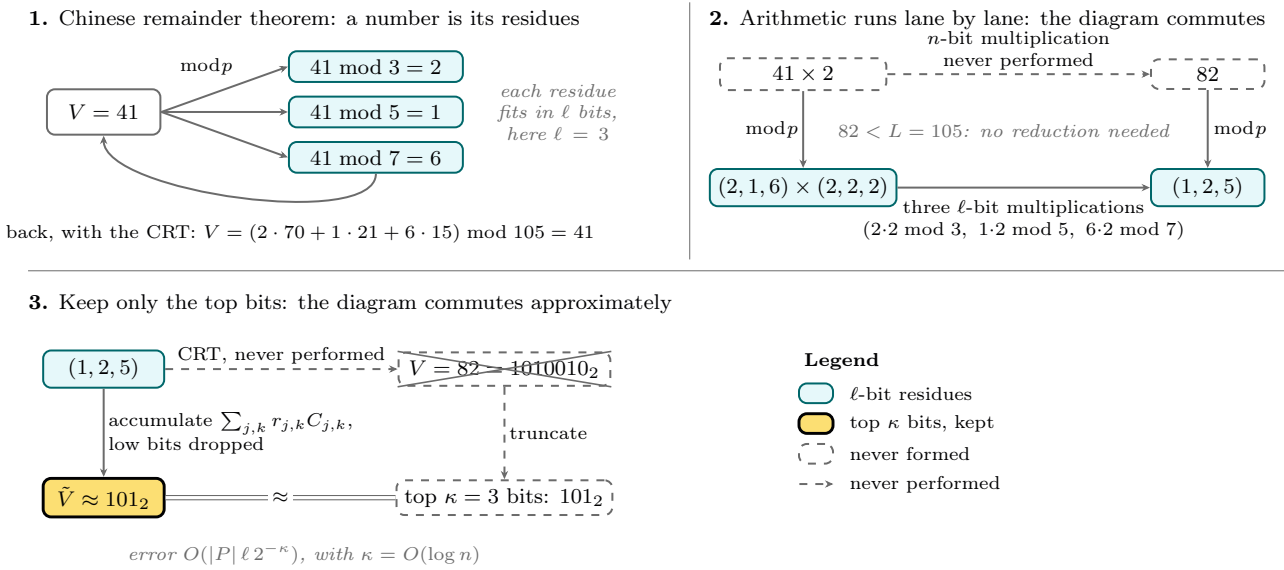

\paragraph{The prime basis and its parameters.} The residue system is where the two
principal cost parameters of the algorithm are set, and they act on the two axes
independently. First, the \emph{size} of the primes fixes the qubit count: each
residue is $\ell = \Theta(\log\log N)$ qubits wide, so the space cost of the
arithmetic is logarithmic in the size of $N$. Second, the \emph{number} of primes
fixes the amount of computation. The size promise $L = \prod_{p\in P} p \geq N^{n_e}$
combined with $\ell = \Theta(\log\log N)$ forces, via the prime number theorem,
\begin{equation}
  |P| = \Theta\!\left(\frac{\log L}{\log\log L}\right)
      = \Theta\!\left(\frac{n_e\,n}{\ell}\right),
\end{equation}
and each prime demands its own controlled exponentiation to form the residue $r_j$.
The number of these exponentiations therefore scales with $|P|$. Anything that lowers
$n_e$, the number of input qubits, is beneficial. A smaller $n_e$ shrinks the exponent, and with it $L$, $|P|$ and the
number of exponentiations. This is precisely
why combining the RNS with the Ekerå--Håstad reduction and with windowing is so
effective. The Ekerå--Håstad reduction lowers $n_e$ to $n_e = \tfrac{n}{2} + 2\lceil
\tfrac{n}{2s}\rceil \approx \bigl(\tfrac12+\tfrac1s\bigr)n$. This is also why increasing the parameter $s$ yields a computation that is
both smaller and faster~\cite{gidney2025factor2048bitrsa}. 

This residue-based approach has a clear space-time trade-off. Rather than
holding the full $n$-qubit intermediate value in memory, the algorithm reuses a
single $\ell$-qubit register to compute one residue at a time. Each contribution is
accumulated into the mask, and the residue is uncomputed before moving on to the next
prime. This
lowers the working qubit count from $O(\log N)$ to $O(\log\log N)$, but at the price of
more gate operations. The value must now be reconstructed from $|P| = \Theta(n_en/\ell)$
sequential residue computations, each a small exponentiation in its own right. The
construction therefore trades gates for space. Gidney estimates that the original
construction of~\cite{C:CheFouSch25} pays roughly a thousand times more Toffoli gates for
a six-fold reduction in space~\cite[Sect.~1]{gidney2025factor2048bitrsa}. The
optimizations below reduce that overhead by more than a factor of $100$.

\paragraph{Computing the residues and arithmetic optimizations.} It remains to
describe how each residue $r_j = \bar V \bmod p_j$ is actually formed on the small
$\ell$-qubit register, and here Gidney~\cite{gidney2025factor2048bitrsa} introduces
several optimizations on top of the residue arithmetic of~\cite{C:CheFouSch25}. The naive
route would compute $r_j$ as a product of $n_e$ controlled modular multiplications, one
per exponent qubit, exactly as in the schoolbook exponentiation of Shor's algorithm. This
costs $O(n_e)$ modular multiplications per prime. The first optimization, which was already 
proposed by the authors of~\cite{C:CheFouSch25} replaces this
product by a sum of discrete logarithms. Fixing a generator $g_{p_j}$ of
$\mathbb{Z}_{p_j}^*$, each precomputed factor $M_k = g^{2^k}\bmod N$, reduced modulo
$p_j$, can be written as $(M_k \bmod p_j) = g_{p_j}^{D_{p_j,k}}$ for a classically known
discrete log $D_{p_j,k}$. This requires $M_k\not\equiv 0 \pmod{p_j}$, since $0$ has no
discrete logarithm. Gidney enforces it in the choice of the prime set. Before searching
for $P$, he excludes every prime that divides one of the multipliers the circuit
uses~\cite[Eq.~(49)]{gidney2025factor2048bitrsa}. The prime set therefore depends on $g$
and on the windowing parameters, and is chosen per shot. The construction
of~\cite{C:CheFouSch25} instead handles these cases separately. Under this hypothesis,
\begin{equation}\label{eq:reduceexponent}
\begin{aligned}
  r_j &= \Bigl(\prod_{k=0}^{n_e-1} M_k^{\,e_k}\Bigr)\bmod p_j = \bar V \bmod p_j\\
    &= g_{p_j}^{\,S_j \bmod (p_j-1)}\bmod p_j,\\
\end{aligned}
\end{equation}
with $\quad S_j = \sum_{k=0}^{n_e-1} e_k\, D_{p_j,k}$.
Here $S_j$ is the integer sum, of at most about $\ell+\log_2 n_e$ bits. It is reduced
modulo the order $p_j-1$ of $g_{p_j}$ only just before the
exponentiation~\cite[Eqs.~(50)--(51)]{gidney2025factor2048bitrsa}. Reducing each $M_k$
modulo $N$ before reducing modulo $p_j$ is essential. It is the integer $\bar V$, whose
residue modulo $N$ is $g^{e}\bmod N$, that the CRT reconstructs, not the residue of the
true power $g^{e}$, which would give an unrelated value.

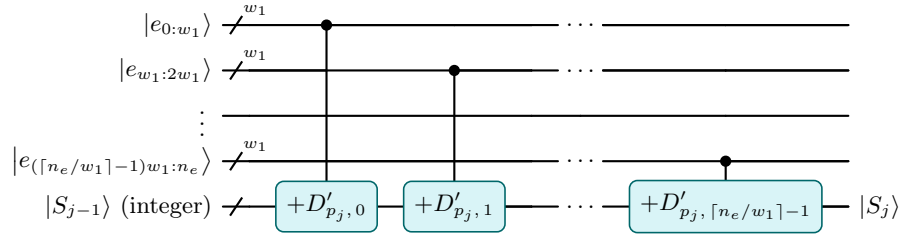
\begin{figure*}[tp]
\centering
\tikzset{abgate/.style={fill=ABTeal!15, draw=ABTealDark, line width=0.6pt, rounded corners=3pt}}

\begin{quantikz}[column sep=0.35cm, row sep={0.6cm,between origins}]
  \lstick{$\ket{e_{0:w_1}}$} & \qwbundle{w_1}
    & \ctrl{4} & \qw & \qw & \ \ldots\ & \qw & \qw \\
  \lstick{$\ket{e_{w_1:2w_1}}$} & \qwbundle{w_1}
    & \qw & \ctrl{3} & \qw & \ \ldots\ & \qw & \qw \\
  \lstick{$\vdots\;$} &
    & & & & & & \\
  \lstick{$\ket{e_{(\lceil n_e/w_1\rceil-1)w_1:n_e}}$} & \qwbundle{w_1}
    & \qw & \qw & \qw & \ \ldots\ & \ctrl{1} & \qw \\
  \lstick{$\ket{S_{j-1}}$ (integer)} & \qwbundle{}
    & \gate[style=abgate]{+D'_{p_j,\,0}}
    & \gate[style=abgate]{+D'_{p_j,\,1}}
    & \qw & \ \ldots\
    & \gate[style=abgate]{+D'_{p_j,\,\lceil n_e/w_1\rceil-1}}
    & \qw \rstick{$\ket{S_{j}}$}
\end{quantikz}

\caption{Accumulation of the exponent $S_j$ for a single prime $p_j$. Here
$S_j=\sum_{k} e_k D_{p_j,k}$, where $D_{p_j,k}$ is the classically precomputed discrete
logarithm of $M_k \bmod p_j$ to the base $g_{p_j}$
(Equation~\eqref{eq:reduceexponent}). It is held, unreduced, on $\ell+\log n_e$ qubits. The exponent register is read in
windows of $w_1$ qubits, indexed by $k$. Each gate $+D'_{p_j,k}$ adds the table entry
$\sum_{i<w_1} e_{kw_1+i}\bigl(D_{p_j,kw_1+i}-D_{p_{j-1},kw_1+i}\bigr)$ selected by window
$k$, so a single pass of $\lceil n_e/w_1\rceil$ additions, instead of $n_e$, turns
$S_{j-1}$ into $S_j$. It merges the uncomputation of the previous prime with the
computation of the current one. $S_j$ is then reduced modulo $p_j-1$, and a windowed
modular exponentiation computes $r_j=g_{p_j}^{S_j\bmod (p_j-1)}\bmod p_j$ into the
$\ell$-qubit residue register of Figure~\ref{fig:oracle-gidney}.}
\label{fig:residue-formation}
\end{figure*}

Computing $r_j$ thus reduces to accumulating the exponent $S_j$, a sum of at most $n_e$
classical constants controlled on the exponent qubits, followed by a single modular
exponentiation $g_{p_j}^{S_j \bmod (p_j-1)} \bmod p_j$, as depicted in
Figure~\ref{fig:residue-formation}. The number of multiplications drops from $O(n_e)$ to
$\operatorname{len} p_j = O(\log\log N)$.

The remaining optimizations reduce the cost of accumulating $S_j$. Windowing the
accumulation over blocks of $w_1$ exponent qubits replaces individual controlled additions
by table lookups. A window of $w_1$ qubits addresses a precomputed table holding the
combined discrete-log contribution $\sum_{k} e_k D_{{p_j},k}$ for that block. The inner loop then runs $\lceil n_e/w_1 \rceil$ times instead of $n_e$, at the cost of a lookup of address 
size $w_1$ per iteration. Here $w_1$ is a windowing parameter specific to the residue-accumulation step,
independent of the windowing parameters $w_e$ and $w_l$ used for the main modular
exponentiation. It controls the granularity of the discrete-log lookup table within each
residue computation. As with the Ekerå--Håstad parameter $s$, increasing $w_1$ trades a
lookup table $2^{w_1}$ against fewer additions, reducing gate count per residue. The residues are formed one prime at a time on a single reused register
(Figure~\ref{fig:oracle-transition}). As a result, the uncomputation of $S_{j-1}$ can be merged
with the computation of $S_{j}$ by tabulating the differences
$D'_{p_j,k} = D_{p_j,k}-D_{p_{j-1},k}$, halving the number of accumulation passes. This
works because $S_{j-1}$ and $S_j$ are integers, so that
$S_j=S_{j-1}+\sum_k e_k D'_{p_j,k}$. Reduced modulo $p_{j-1}-1$ and $p_j-1$, they could
not be related this way, since the two moduli differ. Together these optimizations 
reduce the Toffoli count by more than a factor of $100$ relative to~\cite{C:CheFouSch25}.

\subsection{Approximate Period Finding}\label{sec:approx-period}

The exact modular exponentiation $V(e)=g^{e}\bmod N$ satisfies $V(e+kr)=V(e)$ for every $e$ and every integer $k$, since $g^{r}\equiv 1 \bmod N$.
The approximate oracle $\hat{V}$ (see Equation~\eqref{eq:def_hat_v}) computed by the residue arithmetic described above is only approximately periodic.
By Equation~\eqref{eq:deviation}, each $\hat{V}(e+kr)$ is within $\tfrac{\epsilon}{2}N$ of the exact value $V(e+kr)$.
The triangle inequality gives:
\begin{equation}\label{eq:orbit-uniform-bound}
	\begin{aligned}
		\Delta_N\!\bigl(\hat{V}(e+kr)-\hat{V}(e)\bigr)
		&\le \Delta_N\!\bigl(\hat{V}(e+kr)-V(e+kr)\bigr)\\
		&\quad+\Delta_N\!\bigl(V(e+kr)-V(e)\bigr)\\
		&\quad+\Delta_N\!\bigl(V(e)-\hat{V}(e)\bigr).
	\end{aligned}
\end{equation}
The first and last terms are at most $\tfrac{\epsilon}{2}N$ and the middle term is $0$, so the sum is at most $\epsilon N$:
\begin{equation}\label{eq:approx-periodic}
\begin{gathered}
  \Delta_N\!\bigl(\hat{V}(e+kr) - \hat{V}(e)\bigr) \leq \epsilon\, N\\
  \text{for all } e,k\in\mathbb{Z} \text{ with } e,\ e+kr\in[0,2^{n_e}).
\end{gathered}
\end{equation}

The algorithm measures the accumulator holding $\tilde{V}$ and then applies a quantum Fourier transform to the exponent register.
The transform recovers $r$ if, after the first measurement, the exponent register is close to a state whose amplitudes are constant on each coset $e+r\Z$.
Measuring the full accumulator and obtaining $V^\ast$ keeps the exponents $e$ with $\tilde{V}(e)=V^\ast$.
Because of the drift, this set contains some elements of a coset and misses others.
The resulting state is not periodic, and the Fourier transform cannot recover $r$.
There are two ways around this.

\paragraph{Measure only the most significant bits.}
Write $M=N\gg\bar\kappa=\lfloor N/2^{\bar\kappa}\rfloor$ for the modulus of the accumulator (see Equation~\eqref{eq:approximate_modular_exponentiation}). 
Since $\bar\kappa=\log N-\kappa$, with $\log N$ the bit length of $N$, it satisfies $2^{\kappa-1}\le M<2^{\kappa}$.
A natural fix is to measure only the top $t$ bits of the accumulator.
The drift changes these bits only when the value lies within $\tfrac{\epsilon}{2}M$ of a boundary between two of the $2^{t}$ buckets.
This concerns a fraction of about $\epsilon\,2^{t}$ of the cosets.
Since the read-out is partial, the support of the exponent register collapses to the union of the roughly $r/2^{t}$ cosets whose values share the measured top bits.
This resembles the hashed oracle of Section~\ref{sec:output-register}, with a fixed hash instead of a random one.
The unmeasured low bits, however, remain entangled with the exponent register. 
Tracing them out leaves the exponent register in the same mixed state as measuring them, which is the failing full read-out above.
Clearing them requires running the exponentiation in reverse, which roughly doubles the dominant cost~\cite{gidney2025factor2048bitrsa}.
Superposition masking avoids this reverse pass~\cite{gidney2025factor2048bitrsa}.

\paragraph{Superposition masking.}
Superposition masking~\cite{Zalka06,Gidney19,jaques2020offloadingquantumcomputationsuperposition} 
reveals no more than a top-bit read-out with $2^{t}=1/S$, yet it measures every qubit of the 
accumulator and so leaves nothing to uncompute.
The key idea is to never let the measurement reveal the low bits of $\tilde{V}$.
The $(\kappa+1)$-qubit accumulator that receives the truncated additions of Equation~\eqref{eq:approximate_modular_exponentiation} is simply not initialized to $\ket{0}$.
It is initialized to the uniform superposition
\begin{equation*}\label{eq:mask-state}
  \ket{\sigma} = \frac{1}{\sqrt{W}}\sum_{s=0}^{W-1}\ket{s},
  \qquad W=2^{b},
\end{equation*}
where $S=W/M\in(0,1)$ is the \emph{fraction of the value range} $[0,M)$ covered by the mask.
It is optimized below, then rounded so that $W$ stays a power of $2$.
Note that $S$ is a fraction of the values, not of the qubits.
A uniform superposition over $W=2^{b}$ consecutive values is prepared by one Hadamard on each of the $b=\log_2 W=\log_2(SM)\approx\kappa-\log_2(1/S)$ lowest qubits of the accumulator.
Halving $S$ removes one qubit from the superposition, and the register keeps its $\kappa+1$ qubits whatever $S$ is.
Note that the mask is directly applied to the accumulation register.
Placing it in a separate register would require an extra adder to combine it with $\tilde{V}$, and that register would remain entangled with the output.

Initializing the accumulator to $\ket{\sigma}$ only differs from the standard register initialization by those Hadamards.

Every truncated addition then acts on the masked register.
Each one is a cyclic shift of the basis states modulo $M$, and its circuit depends only on the width of the register, not on its contents.
A superposition of $W$ consecutive values shifted by a constant stays a superposition of $W$ consecutive values.
At the end of the accumulation procedure, after the last prime, the joint state is therefore
\begin{equation*}
  \frac{1}{\sqrt{2^{n_e}W}}\sum_{e}\ket{e}\otimes\sum_{s=0}^{W-1}
  \ket{\bigl(s+\tilde{V}(e)\bigr)\bmod M},
\end{equation*}
and the whole accumulator is measured, returning some $V^\ast$.
Since addition commutes, starting from $\ket{\sigma}$ is equivalent to adding a uniform $s$ at the end.

In the units of $\hat{V}$, that is after multiplying by $2^{\bar\kappa}$, the mask spans $SM\,2^{\bar\kappa}\approx SN$ values, as presented in~\cite{gidney2025factor2048bitrsa}.
Physically it still lives inside the $\kappa+1$ qubits of the accumulator.
A register holding $SN$ values at full precision would need $\log_2(SN)\approx n$ qubits and defeat the whole construction.
Three observations explain why masking works.

For an integer $a$, write
\begin{equation*}
  \ket{B_a}=\frac{1}{\sqrt{W}}\sum_{s=0}^{W-1}\ket{(a+s)\bmod M}
\end{equation*}
for the mask state shifted by $a$, so that $\ket{\sigma}=\ket{B_0}$.
In this notation, the state displayed above is $\ket{\tilde\psi}=2^{-n_e/2}\sum_{e}\ket{e}\ket{B_{\tilde{V}(e)}}$.
We compare it with the ideal state $\ket{\psi}$, defined in the same way with $\tilde{V}(e)$ replaced by $V(e)\gg\bar\kappa$.
Since $V$ is exactly periodic, the accumulator part of $\ket{\psi}$ depends on $e$ only through its coset $e+r\Z$.

\emph{Observation (1): the drift costs an infidelity of at most $\epsilon/S$.}
For each $e$, the boxes $\ket{B_{\tilde{V}(e)}}$ and $\ket{B_{V(e)\gg\bar\kappa}}$ are offset by an integer $\eta_e$, 
the representative of $\tilde{V}(e)-(V(e)\gg\bar\kappa)$ modulo $M$ in $[-M/2,M/2)$.
Two boxes of width $W\le M/2$ offset by $\eta$, with $\abs{\eta}\le W$, share $W-\abs{\eta}$ values, so
\begin{equation*}\label{eq:box-overlap}
  \braket{B_a}{B_{a+\eta}}=1-\frac{\abs{\eta}}{W}.
\end{equation*}
Without a mask, $W=1$ and an offset of one unit already makes the two boxes orthogonal.
This is why the full measurement fails.
With the mask, condition~(i) of Definition~\ref{def:noisy-hsp} gives, after division by $2^{\bar\kappa}$, 
$\abs{\eta_e}\le\tfrac{\epsilon}{2}M$ up to two units of rounding, one from the floor in $V(e)\gg\bar\kappa$ and one from $N\neq M\,2^{\bar\kappa}$.
Hence $\abs{\eta_e}/W\le\epsilon/(2S)$.
The branches $\ket{e}$ are orthogonal and equally weighted, so $\braket{\psi}{\tilde\psi}$ is the average of these overlaps over $e$.
When $\epsilon\le 2S$, this gives
\begin{equation*}
  1-\abs{\braket{\psi}{\tilde\psi}}^{2}\;\le\;1-\Bigl(1-\frac{\epsilon}{2S}\Bigr)^{2}\;\le\;\frac{\epsilon}{S}.
\end{equation*}
The rounding units add a term of order $1/W$, which is negligible for the parameters below.
Following~\cite{gidney2025factor2048bitrsa}, we charge this infidelity as a per-shot failure probability and continue with the ideal state $\ket{\psi}$.%
\footnote{Charging infidelity directly as a failure probability is a heuristic.
The rigorous route goes through the trace distance, $\sqrt{1-\abs{\braket{\psi}{\tilde\psi}}^2}\le\sqrt{\epsilon/S}$, which is larger.
The linear charge is the one used in~\cite{gidney2025factor2048bitrsa} to choose the parameters.}
This charge is a bookkeeping device, not a description of individual shots.
For a fixed outcome $V^\ast$, a coset is cut, with some of its exponents kept and others discarded, only when $V^\ast$ falls within the drift of its box edges.
The outcomes that cut a given coset form at most a fraction $2\epsilon$ of $\Z_M$.
Each exponent is kept by $W$ outcomes, and its coset is cut for at most $2\epsilon M$ of them.
Since $\Pr[v]$ is proportional to $\abs{K_v}$, the cut cosets hold on average at most a fraction $2\epsilon M/W=2\epsilon/S$ of the kept exponents, without any assumption on the coset values.
For the real parameters the cosets far outnumber $1/\epsilon$, so cut cosets appear in essentially every shot.
This weight counts every kept exponent of a cut coset, so it bounds the affected exponents rather than the infidelity.
The linear charge books this weight as if a fraction $\epsilon/S$ of the shots were lost.

\emph{Observation (2): the ideal read-out keeps about a fraction $S$ of the exponents.}
Measuring the accumulator of $\ket{\psi}$ returns a value $V^\ast$ and keeps the exponents whose ideal box contains it,
\begin{equation*}\label{eq:mask-collapse}
  \ket{\psi_2}\;\propto\;\sum_{e}\ket{e}\cdot
  \mathbf{1}\!\left[\,\bigl(V^\ast-(V(e)\gg\bar\kappa)\bigr)\bmod M\in[0,W)\,\right].
\end{equation*}
Here $\mathbf{1}[\cdot]$ equals $1$ when its condition holds and $0$ otherwise. The condition depends on $e$ only through $V(e)$, so each coset of $H=r\Z$ is kept or discarded as a whole, and the kept set stays periodic.
A coset is kept when its value lies among the $W$ values ending at $V^\ast$ modulo $M$, a fraction $S$ of $\Z_M$.
Write $K_v$ for the set of exponents kept on outcome $v$, so that $\Pr[v]=\abs{K_v}/(2^{n_e}W)$ and $\sum_{v\in\Z_M}\abs{K_v}=2^{n_e}W$.
Write $y$ for the outcome of the QFT.
The post-measurement state is uniform on $K_v$, so $\Pr[y=0\mid v]=\abs{K_v}/2^{n_e}$ exactly, for any kept set.
Averaging over $v$ and applying the Cauchy--Schwarz inequality gives
\begin{equation*}
  \Pr[y=0]=\sum_{v\in\Z_M}\frac{\abs{K_v}^2}{2^{2n_e}W}=S\bigl(1+\mathrm{CV}^2\bigr)\ge S,
\end{equation*}
where $\mathrm{CV}$ is the coefficient of variation of $\abs{K_v}$ over uniform $v\in\Z_M$.
The same computation holds for $\ket{\tilde\psi}$, with $\tilde{V}(e)$ in place of $V(e)\gg\bar\kappa$.
The zero frequency therefore costs at least $S$.
It costs about $S$ when the values $V(e)\gg\bar\kappa$ are equidistributed in $\Z_M$ at the scale $W$, so that $\abs{K_v}$ hardly depends on $v$.
This equidistribution is the assumption behind $\delta_\bot\approx S$.
The read-out resembles the hashed oracle of Section~\ref{sec:output-register}, with $S$ in the role of the collision rate $2^{-t}$.
The analogy is not exact.
The window test is not a universal hash in the sense of~\cite[Def.~1]{MaySchlieper22}, since two values collide with a probability that depends on their distance.
The guarantee of~\cite[Thm.~7]{MaySchlieper22} therefore does not apply to the informative peaks.
For those peaks, Gidney models the kept set as random~\cite[Eqs.~(37)--(42)]{gidney2025factor2048bitrsa}.
In this model they share the remaining weight with unchanged relative weights~\cite[Eq.~(42)]{gidney2025factor2048bitrsa}, which underlies his cost estimate~\cite[Assumption~2]{gidney2025factor2048bitrsa}.
His numerical experiments support the model for small $N$ but do not prove it.

\begin{figure*}[tp]
	\centering

	\def\maskToyData{0/1.60/1.69, 1/2.15/2.07, 2/0.70/0.76, 3/1.60/1.50, 4/2.15/2.27,
		5/0.70/0.63, 6/1.60/1.80, 7/2.15/2.24, 8/0.70/0.55}%
	\def\maskDx{0.55}%
	\def\maskW{1.0}%
	\def\maskM{3.5}%
	\newcommand{\maskToyAxes}{%
		\draw[maskaxis] (0,0) -- (9*\maskDx+0.3,0);
		\draw[maskaxis] (0,0) -- (0,\maskM+0.4);
		\draw[ABInk, line width=0.6pt] (-0.08,\maskM) -- (0.08,\maskM);
		\node[masklbl, anchor=east] at (-0.1,0) {$0$};
		\node[masklbl, anchor=east] at (-0.1,\maskM) {$M$};
	}
	\newcommand{\maskToyBars}{%
		\foreach \toyE/\toyV/\toyVt in \maskToyData {
			\pgfmathsetmacro{\toyX}{\toyE*\maskDx+0.5}
			\draw[maskbar] (\toyX-0.08,0) rectangle (\toyX+0.08,\toyVt);
		}%
	}
	\newcommand{\maskToyDrift}{%
		\foreach \toyE/\toyV/\toyVt in \maskToyData {
			\pgfmathsetmacro{\toyX}{\toyE*\maskDx+0.5}
			\draw[maskdrift] (\toyX-0.08,\toyV) rectangle (\toyX+0.08,\toyVt);
		}%
	}
	\newcommand{\maskToyBoxes}{%
		\foreach \toyE/\toyV/\toyVt in \maskToyData {
			\pgfmathsetmacro{\toyX}{\toyE*\maskDx+0.5}
			\pgfmathsetmacro{\toyLo}{max(\toyV,\toyVt)}
			\pgfmathsetmacro{\toyHi}{min(\toyV,\toyVt)+\maskW}
			\fill[ABTeal!25] (\toyX-0.18,\toyVt) rectangle (\toyX+0.18,\toyVt+\maskW);
			\fill[maskstrip] (\toyX-0.18,\toyVt) rectangle (\toyX+0.18,\toyLo);
			\fill[maskstrip] (\toyX-0.18,\toyHi) rectangle (\toyX+0.18,\toyVt+\maskW);
			\draw[maskbox] (\toyX-0.18,\toyVt) rectangle (\toyX+0.18,\toyVt+\maskW);
		}%
	}
	\newcommand{\maskToyOutcome}[2]{%
		\draw[maskvstar] (0,#1) -- (9*\maskDx+0.1,#1);
		\node[masklbl, anchor=east] at (-0.1,#1) {$V^\ast$};
		\foreach \toyE in {#2} {\node[maskkept] at (\toyE*\maskDx+0.5,0) {};}%
	}
	\newcommand{\maskToyTick}[3][-0.15]{%
		\node[anchor=north, font=\scriptsize, inner sep=1pt] at (#2*\maskDx+0.5,#1) {#3};%
	}
	\resizebox{\linewidth}{!}{%
		\begin{tikzpicture}[>=Stealth, font=\small]
			\def\rc{3pt}
			\tikzset{
				maskaxis/.style  = {->, draw=ABInk, line width=0.6pt},
				masklbl/.style   = {font=\footnotesize, inner sep=1pt},
				maskbar/.style   = {fill=abClassical, draw=ABLine, line width=0.5pt},
				maskdrift/.style = {fill=ABSalmon!40, draw=ABSalmonDark, line width=0.5pt,
				                    densely dashed},
				maskbox/.style   = {draw=ABTeal!80!black, line width=0.6pt},
				maskstrip/.style = {fill=ABSalmon!55},
				maskvstar/.style = {draw=ABInk, line width=0.7pt, densely dashed},
				maskkept/.style  = {circle, inner sep=0pt, minimum size=5pt, fill=abYellow,
				                    draw=ABInk, line width=0.6pt},
				masktitle/.style = {anchor=south, align=center},
				verdict/.style   = {rounded corners=\rc, minimum width=5.0cm,
				                    minimum height=0.95cm, align=center, inner sep=3pt},
				fail/.style      = {verdict, draw=ABLine, line width=0.6pt, dashed, fill=white},
				win/.style       = {verdict, draw=ABInk, line width=1.2pt, fill=abYellow},
			}

			\maskToyAxes
			\node[masklbl, anchor=south west] at (-0.2,\maskM+0.45) {accumulator value};
			\node[masklbl, anchor=north] at (9*\maskDx+0.3,-0.15) {exponent};
			\maskToyBars
			\maskToyDrift
			\maskToyOutcome{1.69}{0}
			\maskToyTick{0}{$e$}
			\maskToyTick{3}{$e{+}r$}
			\maskToyTick{6}{$e{+}2r$}
			\node[masktitle] at (2.6,\maskM+0.95)
				{\textbf{(a) No mask}};
			\node[fail] at (2.6,-1.35)
				{\textbf{Not periodic}\\[1pt]{\footnotesize kept: $e$ only}};

			\begin{scope}[shift={(6.6,0)}]
				\maskToyAxes
				\maskToyBars
				\maskToyBoxes
				\maskToyOutcome{1.93}{0,3,6}
				\maskToyTick{0}{$e$}
				\maskToyTick{3}{$e{+}r$}
				\maskToyTick{6}{$e{+}2r$}
				\draw[decorate, decoration={brace, amplitude=3pt}, draw=ABTeal!80!black,
				      line width=0.6pt]
					(8*\maskDx+0.76,1.55) -- (8*\maskDx+0.76,0.55)
					node[midway, right=3pt, masklbl, text=ABTeal!60!black] {$W$};
				\node[masktitle] at (2.6,\maskM+0.95)
					{\textbf{(b) Mask}};
				\node[win] at (2.6,-1.35)
					{\textbf{Periodic}\\[1pt]{\footnotesize kept: $e$, $e{+}r$, $e{+}2r$}};
			\end{scope}

			\begin{scope}[shift={(13.2,0)}]
				\maskToyAxes
				\maskToyBars
				\maskToyBoxes
				\maskToyOutcome{2.39}{0,1,3,4,6,7}
				\maskToyTick{0}{$e$}
				\maskToyTick[-0.42]{1}{$e'$}
				\maskToyTick{3}{$e{+}r$}
				\maskToyTick[-0.42]{4}{$e'{+}r$}
				\maskToyTick{6}{$e{+}2r$}
				\maskToyTick[-0.42]{7}{$e'{+}2r$}
				\node[masktitle] at (2.6,\maskM+0.95)
					{\textbf{(c) Mask, another outcome}};
				\node[win] at (2.6,-1.35)
					{\textbf{Periodic}\\[1pt]{\footnotesize kept: cosets $e{+}r\mathbb{Z}$ and $e'{+}r\mathbb{Z}$}};
			\end{scope}

			\begin{scope}[every node/.style={font=\footnotesize, anchor=west}]
				\draw[maskbar] (0,-2.49) rectangle (0.45,-2.3);
				\node at (0.55,-2.4) {computed value $\tilde{V}(e)$};
				\draw[maskdrift] (6.6,-2.49) rectangle (7.05,-2.3);
				\node at (7.15,-2.4) {drift $\tilde{V}(e)-(V(e)\gg\bar\kappa)$};
				\fill[ABTeal!25] (13.2,-2.49) rectangle (13.65,-2.3);
				\draw[maskbox] (13.2,-2.49) rectangle (13.65,-2.3);
				\node at (13.75,-2.4) {values $\tilde{V}(e)+s$, $0\le s<W$};
				\fill[ABTeal!25] (0,-3.04) rectangle (0.45,-2.85);
				\fill[maskstrip] (0,-3.04) rectangle (0.45,-2.97);
				\draw[maskbox] (0,-3.04) rectangle (0.45,-2.85);
				\node at (0.55,-2.95) {part of the box moved by the drift};
				\node[maskkept, anchor=center] at (6.83,-2.95) {};
				\node at (7.15,-2.95) {kept exponent};
				\draw[maskvstar] (13.2,-2.95) -- (13.65,-2.95);
				\node at (13.75,-2.95) {measured value $V^\ast$};
			\end{scope}
		\end{tikzpicture}%
	}
	\caption{How superposition masking restores periodicity.
	Bars are the computed values $\tilde{V}(e)$ of the accumulator.
	The exact value $V(e)\gg\bar\kappa$, in the same units, is constant on each coset $e+r\mathbb{Z}$, and $\tilde{V}(e)$ is not.
	\textbf{(a)}~Without mask, only the exponents with $\tilde{V}(e)=V^\ast$ are kept.
	The drift separates $e$ from $e+r$ and $e+2r$, so the kept set is not periodic.
	\textbf{(b)}~With the mask, when the input register holds $e$, the accumulator holds the $W=SM$ consecutive values $\tilde{V}(e)+s$, $0\le s<W$ (teal box).
	Without drift, all boxes of a coset would start at the same value $V(e)\gg\bar\kappa$.
	The drift shifts each box, and the salmon strip is the part of the box shifted out of that common position.
	An outcome within the drift of the box edges keeps only part of the coset.
	For the real parameters, every outcome cuts some cosets in this way, and they hold at most about a fraction $2\epsilon/S$ of the kept exponents (observation~(1)).
	\textbf{(c)}~Same state as in (b), another outcome.
	The boxes of the coset of $e'$ also contain $V^\ast$, so this coset is kept too.
	The kept set is a union of cosets and still has period $r$.
	For the real parameters, every outcome keeps about a fraction $S$ of all cosets whole, as here, and cuts a fraction at most about $2\epsilon$ of them as in~(b). 	
    The zero frequency of the QFT then has weight about $S$ (observation~(2)).
	}
	\label{fig:masking-boxes}
\end{figure*}
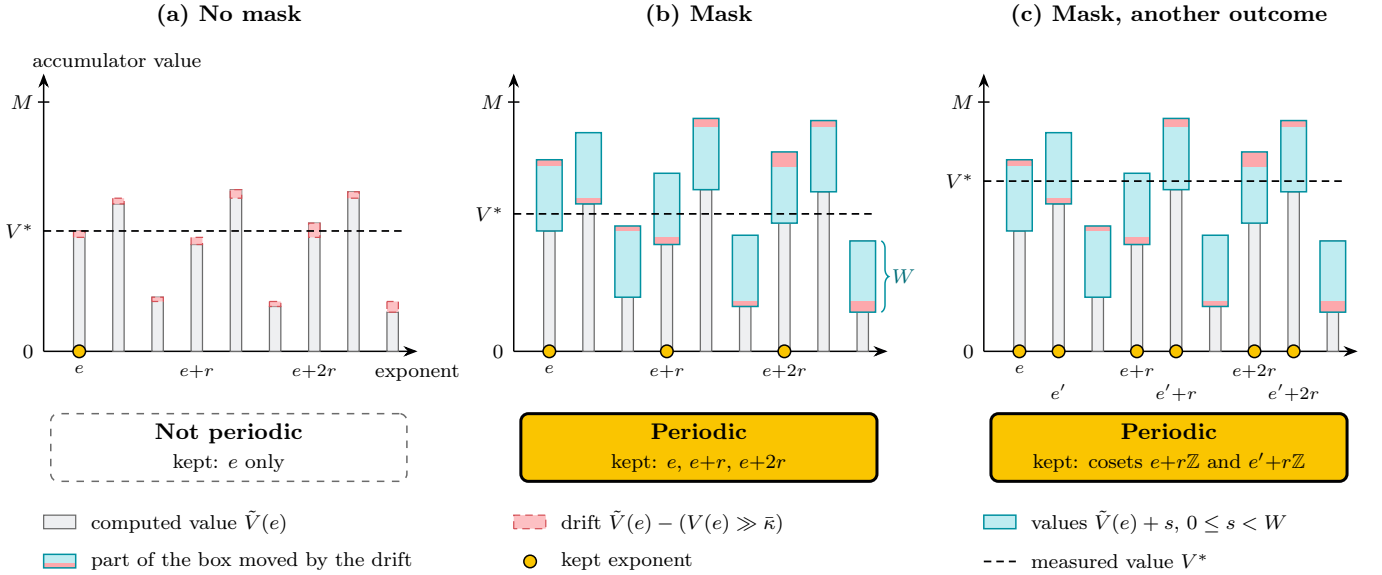

\emph{Observation (3): no garbage is left to uncompute.}
All $\kappa+1$ qubits of the accumulator are measured, so no part of it remains entangled with the exponent register, and the exponentiation need not be run backwards.
This holds for any full measurement, including the unmasked one that fails above.
What the mask adds is that this full measurement is harmless.
Given $e$, the outcome $V^\ast=(s+\tilde{V}(e))\bmod M$ is uniform over $W$ consecutive values.
It therefore carries at most $\log_2(M/W)=\log_2(1/S)$ bits of information about $\tilde{V}(e)$, and essentially nothing about the low bits that hold the drift.
Observations~(1) and~(2) quantify what this costs.
Compare with the top-bit read-out for $2^{t}=1/S$.
Both reveal the same amount of information, and in both the drift cuts a fraction of order $\epsilon\,2^{t}=\epsilon/S$ of the kept cosets in every shot.
Masking differs in two ways.
Its bucket boundaries sit at a random offset instead of fixed positions, and it leaves no unmeasured register behind.
Figure~\ref{fig:masking-boxes} illustrates the three cases on a toy example.

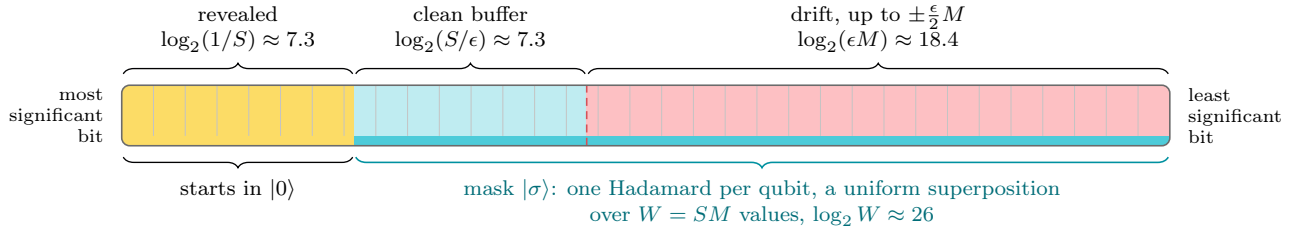
\begin{figure*}[tp]
\centering
\begin{tikzpicture}[x=0.42cm, y=1cm, font=\footnotesize]
  \def\rc{3pt}
  \begin{scope}
    \clip[rounded corners=\rc] (0,0) rectangle (33,0.8);
    \fill[abYellow!60]  (0,0)     rectangle (7.32,0.8);
    \fill[ABTeal!25]    (7.32,0)  rectangle (14.64,0.8);
    \fill[ABSalmon!40]  (14.64,0) rectangle (33,0.8);
    \fill[ABTeal!70]    (7.32,0)  rectangle (33,0.13);
  \end{scope}
  \foreach \i in {1,...,32}{\draw[ABLine!40, line width=0.2pt] (\i,0.13) -- (\i,0.8);}
  \draw[ABSalmonDark, line width=0.6pt, densely dashed] (14.64,0) -- (14.64,0.8);
  \draw[ABLine, line width=0.6pt, rounded corners=\rc] (0,0) rectangle (33,0.8);
  \node[anchor=east, align=right, font=\scriptsize] at (-0.3,0.4) {most\\significant\\bit};
  \node[anchor=west, align=left, font=\scriptsize] at (33.3,0.4) {least\\significant\\bit};
  \draw[decorate, decoration={brace, amplitude=4pt}, draw=ABInk, line width=0.5pt]
    (0.05,0.95) -- (7.27,0.95) node[midway, above=5pt, align=center]
    {revealed\\ $\log_2(1/S)\approx 7.3$};
  \draw[decorate, decoration={brace, amplitude=4pt}, draw=ABInk, line width=0.5pt]
    (7.37,0.95) -- (14.59,0.95) node[midway, above=5pt, align=center]
    {clean buffer\\ $\log_2(S/\epsilon)\approx 7.3$};
  \draw[decorate, decoration={brace, amplitude=4pt}, draw=ABInk, line width=0.5pt]
    (14.69,0.95) -- (32.95,0.95) node[midway, above=5pt, align=center]
    {drift, up to $\pm\tfrac{\epsilon}{2}M$\\ $\log_2(\epsilon M)\approx 18.4$};
  \draw[decorate, decoration={brace, amplitude=4pt, mirror},
        draw=ABTeal!80!black, line width=0.6pt]
    (7.37,-0.15) -- (32.95,-0.15) node[midway, below=5pt, align=center,
    text=ABTeal!60!black]
    {mask $\ket{\sigma}$: one Hadamard per qubit, a uniform superposition\\
     over $W=SM$ values, $\log_2 W\approx 26$};
  \draw[decorate, decoration={brace, amplitude=4pt, mirror}, draw=ABInk, line width=0.5pt]
    (0.05,-0.15) -- (7.27,-0.15) node[midway, below=5pt, align=center]
    {starts in $\ket{0}$};
\end{tikzpicture}
\caption{Bit layout of the $\kappa=33$-bit masked accumulator for RSA-2048, with the most significant bit on the left (one extra wrap-around qubit not shown).
It shows, bit by bit, the accumulator of Figure~\ref{fig:oracle-gidney}.
All qubits are measured at the end.
Because of the mask, the measured value is uniform over $W$ values for every exponent, so the measurement reveals at most $\log_2(1/S)$ bits of information about $\tilde{V}$.
The clean buffer between those bits and the drift sets the corruption rate: its width $\log_2(S/\epsilon)$ gives the heuristic rate $\epsilon/S$.
Widths are in bits, derived from the failure rate of~\cite[Table~5]{gidney2025factor2048bitrsa} with $S=\sqrt\epsilon$ and $M=2^{\kappa}$, before $W$ is rounded to a power of two.}
\label{fig:accumulator-anatomy}
\end{figure*}

\paragraph{Anatomy of the accumulator.}
The three observations assign a role to each of the $\kappa$ value bits of the accumulator.
The extra wrap-around qubit plays no role here.
From the least significant bit upwards, the accumulator splits into three layers.
\begin{itemize}
  \item The drift occupies the low $\log_2(\epsilon M)$ bits. The computed value deviates from the ideal one by at most $\tfrac{\epsilon}{2}M$ in either direction, a band of total width $\epsilon M$.
  \item The clean buffer occupies the next $\log_2(S/\epsilon)$ bits. Together with the drift band it forms the $\log_2(SM)$ bits covered by the mask. The ratio $S/\epsilon$ of the mask width to the drift width is the inverse of the corruption rate $\epsilon/S$ of observation~(1).
  \item The top $\log_2(1/S)$ bits are what the measurement effectively reveals.
\end{itemize}
At the optimum $S=\sqrt\epsilon$ derived below, the buffer and the revealed part have the same width $\log_2(1/S)$.
For RSA-2048 with Gidney's parameters ($\kappa=33$, called $f$ in~\cite{gidney2025factor2048bitrsa}, and $P_{\mathrm{deviant}}\approx\SI{1.25}{\percent}$, the per-shot failure rate defined in the success-rate paragraph below), this gives $S\approx\SI{0.6}{\percent}$, about $7$ revealed bits, a buffer of about $7$ bits and a drift of about $18$ bits (Figure~\ref{fig:accumulator-anatomy}).%
\footnote{Back-of-envelope values, taking $\kappa=33$ and $P_{\mathrm{deviant}}$ from the $n=2048$ row of Table~5 of~\cite{gidney2025factor2048bitrsa} as inputs.
We take this failure rate as given and do not recompute it from the circuit of~\cite{gidney2025factor2048bitrsa}.
The values above are those that our bound~\eqref{eq:Sopt} needs to reach it, obtained by inverting $P_{\mathrm{deviant}}=2\sqrt\epsilon$.
The actual mask width $W$ is a power of two, so $S$ is close to $\sqrt\epsilon$ but not exactly equal to it.}

\begin{remark}[Back to two dimensions]
The argument above carries over to the two-dimensional setting of Section~\ref{sec:short-dlp}.
Replace $e$ by $(a,b)$ and the coset $e+r\Z$ by a coset of the lattice $H$.
The bound~\eqref{eq:approx-periodic}, the box overlap and the fact that the masked read-out keeps or discards whole cosets use only that $V$ is constant on cosets, so they hold in any dimension.
The zero-frequency identity of observation~(2) holds in any dimension as well.
The equidistribution behind $\delta_\bot\approx S$ remains an assumption, as in one dimension.
Gidney presents his analysis for one-dimensional period finding and states that it is meant for the Ekerå--Håstad variant~\cite[Sect.~2]{gidney2025factor2048bitrsa}.
In two dimensions, the relative weights of the informative peaks are therefore a modelling assumption as well.
This two-dimensional setting is the one used by the full algorithm of Section~\ref{sec:full-modern}.
\end{remark}

\paragraph{Success rate.}
In the language of Definition~\ref{def:noisy-hsp}, $\hat{V}$ satisfies condition~(i) with the $\epsilon$ of Equation~\eqref{eq:deviation}, and masking is the read-out $R$.
We do not establish condition~(ii) with the worst case over instances that $\delta_\bot$ and $\delta_\times$ require.
The only worst-case statements are $\delta_\times\le\sqrt{\epsilon/S}$ and $\delta_\bot\ge S$, and the rates below are heuristic estimates for the instances at hand.
Let $P_{\mathrm{deviant}}=\delta_\bot+\delta_\times$ be the probability that this read-out spoils a shot.
The two terms move in opposite directions as $S$ varies.
A shot aborts when the QFT returns the zero frequency, which carries no information and is discarded.
By observation~(2), this happens with probability $\delta_\bot\ge S$, with near equality when the coset values are equidistributed at the scale of the mask.
This near equality is an assumption, not a proven bound.
A shot is corrupted when the drift becomes visible.
By observation~(1), this happens with probability $\delta_\times\approx\epsilon/S$.
This is also a heuristic, because it charges an infidelity as a probability (see the footnote to observation~(1)).
Counting all of this mass as corruption is conservative, since part of it may land on the zero frequency and be discarded.
Adding the two rates gives
\begin{equation}\label{eq:Sopt}
\begin{aligned}
  P_{\mathrm{deviant}}&\lesssim S+\epsilon/S,\\
  \min_{S}\bigl(S+\epsilon/S\bigr)&=2\sqrt{\epsilon}\quad\text{at } S=\sqrt{\epsilon}.
\end{aligned}
\end{equation}
At this optimum the abort and corruption rates are equal, and $P_{\mathrm{deviant}}\lesssim 2\sqrt\epsilon$.
With the rigorous charge $\delta_\times\le\sqrt{\epsilon/S}$ instead, and still $\delta_\bot\approx S$, the optimum moves to $S=(\epsilon/4)^{1/3}$.
It then gives $P_{\mathrm{deviant}}\lesssim 3(\epsilon/4)^{1/3}\approx 1.9\,\epsilon^{1/3}$.

Equation~\eqref{eq:Sopt} also fixes the truncation length $\kappa$.
Write the drift of Equation~\eqref{eq:deviation} as $\epsilon=c\,A_{\mathrm{acc}}\,2^{-\kappa}$, where $A_{\mathrm{acc}}$ is the number of truncated additions into the accumulator and one truncated addition shifts the value by at most $\tfrac{c}{2}\,2^{-\kappa}N$.
Setting $2\sqrt\epsilon=P_{\mathrm{deviant}}$ and solving for $\kappa$ gives
\begin{equation}\label{eq:kappa-budget}
  \kappa \;=\; \underbrace{\log_2(cA_{\mathrm{acc}})}_{\text{drift}}
  \;+\; \underbrace{2\log_2\!\bigl(2/P_{\mathrm{deviant}}\bigr)}_{\text{buffer}+\text{revealed}}.
\end{equation}
The first term pays for the drift.
Without windowing, $A_{\mathrm{acc}}=|P|\,\ell=\Theta(n_e n)$, so this term is $\log_2(n_e n)+O(1)=O(\log\log N)$.
The second term equals $\log_2(1/\epsilon)$, the combined width of the buffer and the revealed bits.
It depends only on the target failure rate and grows by two bits each time $P_{\mathrm{deviant}}$ is halved.
For RSA-2048 at $P_{\mathrm{deviant}}=\SI{1.25}{\percent}$, it is about $14.6$ bits.
Gidney's parameter search gives $\kappa=33$ for this target~\cite{gidney2025factor2048bitrsa}, so his windowed circuit spends the remaining $\approx 18$ bits on drift.
We infer this value from $\kappa=33$ and do not derive it from the windowing parameters.
Under the rigorous charge, the second term becomes $3\log_2(3/P_{\mathrm{deviant}})-2\approx 21.7$ bits at the same target, and $\kappa$ would grow to about $40$.

\subsection{Assembling Modern Shor}\label{sec:full-modern}

The previous subsections built the modern algorithm one relaxation at a time. This
subsection makes one choice per stage, assembles the choices into a single pipeline,
and instantiates it for an $n=2048$-bit RSA modulus. The pipeline uses the
short-discrete-log reduction, the Ekerå--Håstad register shape with Ekerå's graded
post-processing~\cite{Eke20}, and the residue-arithmetic oracle with masking. The
$(\kappa{+}1)$-qubit accumulator of that oracle already compresses the output, so the
hash of Section~\ref{sec:output-register} is not used. Table~\ref{tab:four-relaxation}
marks these choices. We follow the run from the choice of $g$
to the recovered factors $p$ and $q$, and we state, at each stage, the failure probability
already derived above. Composing them gives one number: the \emph{algorithmic error rate}
of the full pipeline. This rate is algorithmic. It comes from the structure of
the algorithm itself, from the randomness of $g$, from the statistics of the measured
frequency pairs, and from the noise budget of the space-efficient oracle. Two of its inputs
are heuristic: the masking rates $\delta_\bot\approx S$ and $\delta_\times\approx\epsilon/S$
(Section~\ref{sec:oracle-rns}), and the $\approx 1~\%$ failure floor of Ekerå's
post-processing, which is measured by simulation~\cite{Eke20}. It says nothing
about the physical qubits that execute the algorithm. A real run must also survive the
physical error rate of the hardware, which is the subject of a separate part of this work.
Table~\ref{tab:full-algo-budget} summarizes the failure budget derived below, before we
walk through each row in turn.

At this level of detail we keep visible only the parameters that set the algorithmic error
rate: the lattice parameter $s$ of Section~\ref{sec:input-register}, the truncation length
$\kappa$, and the mask width $S$ of Section~\ref{sec:oracle-rns}. The windowing parameters
$w_e$ and $w_l$, the lookup table sizes, and the exact set of RNS primes $P$ mainly set the
gate count and the qubit count of the resulting circuit. They are not neutral for the
success probability, however. Windowing changes the number $A$ of truncated additions, and
the prime set fixes $\Delta_N(L)$. At fixed $\kappa$, both change the drift $\epsilon$
through Equations~\eqref{eq:dev} and~\eqref{eq:kappa-budget}, and hence
$P_{\mathrm{deviant}}$. We take them as fixed by Gidney's parameter search, which sets
$\kappa=33$ for the resulting drift. Their concrete values are given in
Appendix~\ref{sec:logical_impl} and used for the resource estimate in
Appendix~\ref{sec:resource_estim}.

\begin{table*}[t]
\centering
\renewcommand{\arraystretch}{1.3}
\setlength{\tabcolsep}{5pt}
\begin{tabular}{@{}p{4.0cm} p{3.6cm} c@{}}
\toprule
\textbf{Stage} & \textbf{Failure} & \textbf{Rate} \\
\midrule
Pick $g$ (once per run) & $\gcd(g,N)\neq 1$: a factor found directly (early success) &
$O(2^{-1024})$ \\
\addlinespace
RNS oracle + masking (per shot) & mask abort & $\approx S$ \\
\addlinespace
RNS oracle + masking (per shot) & corruption & $\approx \epsilon/S$ \\
\addlinespace
Lattice recovery of $d$ (once, from enough consistent shots) & post-processing fails
despite enough consistent shots~\cite{Eke20} & $\approx 1~\%$ \\
\addlinespace
Check $q(2d+2-q)=N$ (once per attempt) & (catches remaining corruption) & --- \\
\bottomrule
\end{tabular}
\caption{Failure budget of the RSA-2048 pipeline, one row per stage. Shot counts throughout
this section already use Ekerå's graded post-processing~\cite{Eke20} (see below), not a
hard good/bad filter on the measured frequency pair. The last row is not a failure. It is
the classical check that turns every remaining corruption into a caught and retried
attempt rather than a wrong output.}
\label{tab:full-algo-budget}
\end{table*}

\paragraph{Classical pre-processing.} Fix the $2048$-bit RSA modulus $N=pq$. Draw a uniformly random
$g\in\{2,\dots,N-2\}$ and compute $\gcd(g,N)$ with the Euclidean algorithm. If the gcd is
not $1$, it is $p$ or $q$, and the factorization is already found. This early success
happens with probability $O(2^{-1024})$ for $1024$-bit primes, so it plays no role in the
budget, and we assume $\gcd(g,N)=1$ from here on. Compute classically the value
\begin{equation}
  x = g^{(N-1)/2} \bmod N.
\end{equation}
By the reduction of Section~\ref{sec:short-dlp}, $x = g^d \bmod N$ for the short logarithm
$d=(p+q-2)/2$, of bit-length $m\approx n/2 = 1024$. Recovering $p$ and $q$ is now
equivalent to recovering $d$. This step is deterministic. It introduces no failure
probability of its own.

\paragraph{Fixing the register-shape parameters for $n=2048$.} One of the three
register-shape parameters is a genuine trade-off and we keep it symbolic. Recall from
Section~\ref{sec:input-register} that the lattice parameter $s\ge 2$ sets the exponent
register length $n_e = \left(\tfrac12+\tfrac1s\right)n$. For $n=2048$ this ranges from
$n_e=2048$ at $s=2$ down toward $1024$ for large $s$, trading qubits for shots. The other two
are not free in the same sense. Once a target per-shot failure rate
$P_{\mathrm{deviant}}$ is chosen, the mask width is fixed at its optimum
$S=\sqrt\epsilon$ by Equation~\eqref{eq:Sopt}, and the truncation length $\kappa$
follows from Equation~\eqref{eq:kappa-budget}. That equation splits $\kappa$ into two
parts. The drift part, $\log_2(cA)$, counts the truncated additions into the
accumulator. Without windowing it is $\log_2(n_e n)\approx 21$ to $22$ for $n=2048$,
and windowing lowers it. The precision part,
$2\log_2(2/P_{\mathrm{deviant}})$, buys the clean buffer and the revealed bits of
Figure~\ref{fig:accumulator-anatomy}. It is not a hidden constant: it grows by two
bits each time the target failure rate is halved. Gidney's optimized parameter
search~\cite{gidney2025factor2048bitrsa} for $n=2048$ (with $s=8$, so $n_e=1280$)
lands on $\kappa=33$ bits for a mask-and-truncation failure rate
$P_{\mathrm{deviant}}\approx1.25~\%$. We take $\kappa=33$ as an input. Of it,
$2\log_2(160)\approx 14.6$ bits are the precision part, so the windowed drift term accounts
for the remaining $\approx 18$ bits, down from $21$ to $22$ without windowing. 

\paragraph{One shot of the quantum core.} With $g$, the target $x$, and $(s,\kappa,S)$
fixed, the algorithm runs the following quantum core, and repeats it.
\begin{enumerate}
  \item \emph{State preparation.} Two registers of size $\tau_{s}+m$ and $\tau_{s}$, with
    $\tau_{s}\approx m/s$, are put in a uniform superposition by Hadamards. 
  \item \emph{Oracle evaluation.} The residue-arithmetic oracle of
    Section~\ref{sec:oracle-rns} computes the top $\kappa$ bits $\tilde V$ of
    $g^ax^{-b}\bmod N$, so that $\hat V\approx g^ax^{-b}\bmod N$, and adds them into the
    $(\kappa{+}1)$-qubit accumulator, initialized to a mask of width $\lceil SM\rceil$.
    The whole accumulator is then measured.
  \item \emph{Fourier transform and measurement.} A semi-classical QFT on the two exponent
    registers returns a pair $(j,k)$.
\end{enumerate}
Whether a shot ends up contributing to the recovery of $d$ depends on how $(j,k)$ relates
to the hidden logarithm, and on how much noise the oracle introduced. Ekerå and Håstad's
original analysis of the first dependency singles out a \emph{good} pair, one meeting the
constructive-interference condition of Equation~\eqref{eq:good-pair}, and shows that this
happens with probability at least $1/8$. Good pairs cannot be recognized one at a time, so
their read-out runs about $8s$ shots and tries every subset of $s$ pairs against the check
$x=g^{d}$. Ekerå's later post-processing~\cite{Eke20} refines this. Rather than a hard
good/bad split, it grades every possible outcome and combines the graded evidence directly
in the lattice step below, so that a pair the simple analysis would count as bad can still
contribute. Under this read-out a single pair asserts nothing that can be false. A bad pair
is therefore neither an abort nor a corruption, and its cost moves into the success
probability $q_{\mathrm{pp}}$ of the lattice step (Lemma~\ref{lem:compose}). This is the
post-processing this section uses from here on, and it needs on average only about $s+1$
shots, rather than the roughly $8s$ of the subset search.

\begin{figure*}[t]
\centering
\begin{tikzpicture}[font=\footnotesize, x=1cm, y=1cm,
  >={Stealth[length=3.2pt,width=2.8pt]},
  boxst/.style={draw, rounded corners=3pt, execute at begin node={\hyphenpenalty=10000}, align=center, text=abInk, fill=white,
    line width=0.75pt, text width=3.1cm, minimum height=0.95cm, inner sep=3pt},
  start/.style={boxst, draw=abLine},
  main/.style={boxst, draw=ABTealDark, fill=ABTeal!10},
  result/.style={boxst, draw=abInk, fill=abYellow, line width=1.1pt},
  abortbox/.style={boxst, densely dotted, draw=ABExcept, fill=ABSalmon!12, line width=0.9pt},
  corrupt/.style={boxst, dashed, draw=ABPurpleDark, fill=ABPurple!10, line width=0.9pt,
    text width=3.4cm},
  lbl/.style={font=\scriptsize, align=left, inner sep=1pt},
  path/.style={->, line width=0.9pt, draw=ABTealDark},
]
  \node[start]  (shot) at (1.75,0)  {One shot: measure $(j,k)$ and $V^*$};
  \node[main]   (mask) at (6.15,0)  {Kept by the mask?};
  \node[main]   (corr) at (10.55,0) {Not corrupted?};
  \node[result] (use)  at (14.95,0) {Passed to Ekerå's post-processing};
 
  \node[abortbox] (discard) at (6.15,-2.0) {Discarded (abort)};
  \node[corrupt]  (caught)  at (10.55,-2.0)
    {Wrong constraint, kept:\\ caught by \mbox{$q(2d{+}2{-}q)=N$},\\ at the cost of a subset\\ or an attempt};
 
  \draw[path] (shot) -- (mask);
  \draw[path] (mask) -- (corr);
  \draw[path] (corr) -- (use);
 
  \draw[->, line width=0.9pt, densely dotted, draw=ABExcept] (mask.south) -- (discard.north)
    node[lbl, text=abInk, midway, right=2pt] {no, $\le S$};
  \draw[->, line width=0.9pt, dashed, draw=ABPurpleDark] (corr.south) -- (caught.north)
    node[lbl, text=ABPurpleDark, midway, right=2pt] {no, $\approx \epsilon/S$};
 
  \node[lbl, text=ABTealDark, below=4pt of use] {prob.\ $\ge 1-P_{\mathrm{deviant}}$};
  \begin{scope}[shift={(13.35,-1.35)}, every node/.style={font=\scriptsize, text=abInk}]
    \node[anchor=west, font=\scriptsize\bfseries] at (-0.05,0) {Legend};
    \draw[path] (0,-0.45) -- (0.6,-0.45);
    \node[anchor=west] at (0.7,-0.45) {surviving path};
    \draw[->, line width=0.9pt, densely dotted, draw=ABExcept] (0,-0.9) -- (0.6,-0.9);
    \node[anchor=west] at (0.7,-0.9) {discarded outright};
    \draw[->, line width=0.9pt, dashed, draw=ABPurpleDark] (0,-1.35) -- (0.6,-1.35);
    \node[anchor=west] at (0.7,-1.35) {kept but wrong};
  \end{scope}
\end{tikzpicture}
\caption{Per-shot failure budget contributed by the RNS oracle and its masked read-out.
The teal path survives and is handed to Ekerå's graded post-processing~\cite{Eke20}. On
the salmon path the shot is discarded outright. On the dashed purple path the shot is kept
but wrong. This corruption is not detected here but is caught downstream by the
factorization check, at the cost 
of another subset of shots or a fresh attempt (Lemma~\ref{lem:compose}).}
\label{fig:shot-funnel}
\end{figure*}
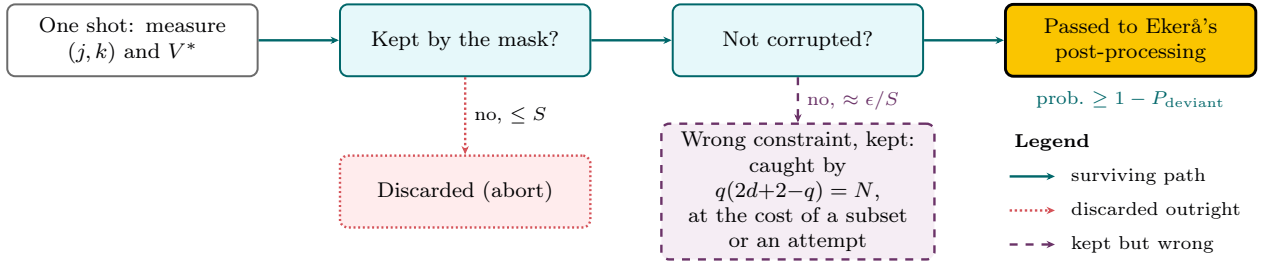

Given Ekerå's post-processing, a shot can still be wasted or corrupted, but only through
the oracle's own noise, from Section~\ref{sec:oracle-rns}. The masked measurement can abort
on its own, with probability about $S$, when the shot collapses onto the uninformative
zero-frequency outcome. It can also corrupt, with probability about $\epsilon/S$ (a heuristic estimate,
Section~\ref{sec:oracle-rns}),
returning a constraint built from the wrong value of $\tilde V$ without any visible sign
that it is wrong. Together these two contributions give
$P_{\mathrm{deviant}}\lesssim S+\epsilon/S$, which the optimal choice $S=\sqrt\epsilon$
brings down to $2\sqrt\epsilon$, as in
Equation~\eqref{eq:Sopt}. Figure~\ref{fig:shot-funnel} traces this per-shot budget: every
shot reaches Ekerå's post-processing unless the oracle's own noise wastes or corrupts it
first. 

\paragraph{Classical post-processing.} Shots are collected until Ekerå's graded
post-processing~\cite{Eke20} can support a single consistent value of $d$, which for
RSA-2048 needs on average close to $s+1$ shots (made precise in the algorithmic error rate
paragraph below). The lattice-based recovery of Section~\ref{sec:input-register} then
returns a candidate $d$: a bounded-distance decoding search in a lattice of small, constant
dimension, which succeeds with probability around $99~\%$ once given enough consistent
shots~\cite{Eke20}. The candidate $d$ gives $p+q=2d+2$, and $p,q$ are the two roots of
$q(2d+2-q)=N$. This last equation, together with the condition $1<q<N$, is also the check that the
whole run succeeded. It holds if and only if $d$ was correct. Its roots $q$ and $2d+2-q$
are then integers whose product is $N$. Since $N=pq$ with $p$ and $q$ prime, the only
factorizations of $N$ into two positive integers are $\{p,q\}$ and $\{1,N\}$, and the
condition $1<q<N$ excludes the second. The roots are therefore $p$ and $q$, so
$2d+2=p+q$, which fixes $d$. The excluded case would need $d=(N-1)/2$, which the shortness
of $d$ rules out anyway.

\begin{figure*}[t]
\centering
\begin{tikzpicture}[
  >={Stealth[length=4.5pt,width=3.6pt]}, font=\footnotesize, text=abInk,
  pbox/.style={draw, rounded corners=3pt, align=center, minimum width=2.0cm,
    minimum height=1.15cm,  inner xsep=5pt, inner ysep=3pt, line width=0.6pt},
  cstep/.style={pbox, draw=ABTealDark, fill=ABTeal!14},
  qstep/.style={pbox, draw=ABPurpleDark, fill=ABPurple!14, line width=1.1pt},
  answer/.style={pbox, draw=abInk, fill=abYellow, line width=1.1pt},
  flow/.style={->, abLine, line width=0.7pt, shorten >=1.5pt, shorten <=1.5pt},
  retry/.style={->, anbSalmonDeep, dashed, line width=0.7pt, shorten >=1.5pt, shorten <=1.5pt},
  rlabel/.style={font=\scriptsize, text=anbSalmonDeep, align=center},
  legtext/.style={font=\scriptsize, anchor=west, inner sep=2pt},
]
  \node[cstep]  (g)    at (0,0)     {Pick $g$, reduce\\ to short DLP};
  \node[qstep]  (core) [right=0.62cm of g] {Quantum core:\\ one shot (Figure~\ref{fig:shot-funnel})};
  \node[cstep]  (coll) [right=0.62cm of core] {Collected enough\\ consistent shots?};
  \node[cstep]  (latt) [right=0.42cm of coll] {Lattice reduction\\ $\to$ candidate $d$};
  \node[cstep]  (chk)  [right=0.42cm of latt] {Check\\ $q(2d{+}2{-}q)=N$};
  \node[answer] (out)  [right=0.42cm of chk] {Output $p,\,q$};
 
  \node[draw=ABPurpleDark, densely dotted, rounded corners=3pt, line width=0.6pt,
        fit=(core), inner xsep=4pt, inner ysep=4pt] (corebox) {};
 
  \draw[flow] (g)    -- (core);
  \draw[flow] (core) -- (coll);
  \draw[flow] (coll) -- (latt);
  \draw[flow] (latt) -- (chk);
  \draw[flow] (chk)  -- (out);
 
  \draw[retry] (coll.north) .. controls +(0,0.95) and +(0,0.95) .. (corebox.north)
    node[pos=0.5, above=1pt, rlabel] {not yet: run another shot};
  \draw[retry] (chk.north) .. controls +(0,0.95) and +(0,0.95) .. (latt.north)
    node[pos=0.5, above=1pt, rlabel] {check fails: another\\ subset of the shots};
  \draw[retry] (chk.south) .. controls +(0,-1.2) and +(0,-1.2) .. (g.south)
    node[pos=0.5, below=1pt, rlabel] {subsets exhausted: restart with a fresh $g$};
 
  \path ($(g.west)!0.5!(out.east)$) ++(-5.85,-2.5) coordinate (leg);
  \begin{scope}[shift={(leg)}]
    \node[draw=ABTealDark, fill=ABTeal!14, rounded corners=1.5pt, line width=0.6pt,
          minimum width=0.45cm, minimum height=0.28cm] (l1) at (0,0) {};
    \node[legtext] at (l1.east) {classical step};
    \node[draw=ABPurpleDark, fill=ABPurple!14, rounded corners=1.5pt, line width=1.1pt,
          minimum width=0.45cm, minimum height=0.28cm] (l2) at (2.55,0) {};
    \node[legtext] at (l2.east) {quantum core (dotted: repeated)};
    \draw[retry, shorten >=0pt, shorten <=0pt] (7.25,0) -- (7.95,0);
    \node[legtext] at (7.95,0) {retry loop};
    \node[draw=abInk, fill=abYellow, rounded corners=1.5pt, line width=1.1pt,
          minimum width=0.45cm, minimum height=0.28cm] (l4) at (9.95,0) {};
    \node[legtext] at (l4.east) {accepted output};
  \end{scope}
\end{tikzpicture}
\caption{The full RSA-2048 pipeline. Teal: classical steps. Purple: the quantum core,
repeated until enough consistent shots, in Ekerå's sense~\cite{Eke20}, are collected.
Salmon, dashed: the three retry loops, for an insufficient shot count, for a candidate $d$
that fails the final check (retry the lattice step on another subset of the shots), and
for exhausted subsets (restart with a fresh $g$). Yellow: the accepted output, reached only once
$pq=N$ has been verified.
}
\label{fig:full-pipeline}
\end{figure*}
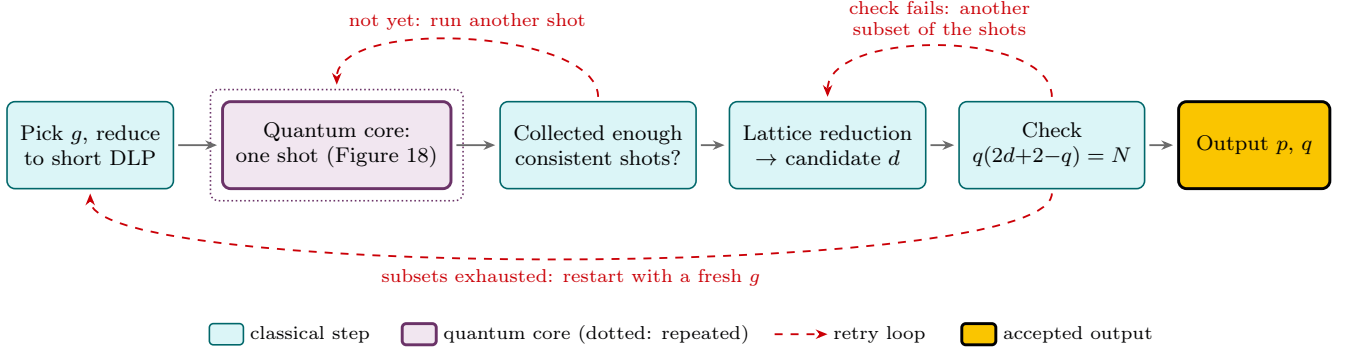

This last point deserves emphasis. A corrupted shot does not make the algorithm output a
wrong factorization. It makes the lattice step fail its own consistency check, since a
corrupted constraint is inconsistent with the other good ones. Corruption costs a subset search or a whole
attempt rather than a single shot (Lemma~\ref{lem:compose}), but like an abort it never
produces a wrong and undetected answer. Factoring
is self-verifying. $p$ and $q$ are only ever accepted once $pq=N$ has been checked. In practice, a run with too many corrupted shots is not simply
discarded either~\cite{gidney2025factor2048bitrsa}. One can first retry the lattice step
over different subsets of the collected shots. Only if that also fails does one restart
with a fresh $g$. For the person running the algorithm, every failure mode in
this section is the same kind of event. It wastes an attempt or a subset, which is then
retried. It never yields a silently wrong RSA key. Figure~\ref{fig:full-pipeline} places this retry loop around the full
pipeline, from $g$ to $p,q$.

\paragraph{Expected number of shots.} The algorithm does not fix a shot count in advance.
It draws shots one at a time. It stops as soon as Ekerå's
post-processing recovers a $d$ that passes the check $q(2d+2-q)=N$. Let $R$ be the number
of shots this takes in one attempt, that is, with a single fixed $g$. This is a
negative-binomial quantity, the number of draws needed to accumulate the $s+1$ consistent
shots the lattice step requires. Ekerå's post-processing already absorbs the
frequency-pair statistics, so the per-shot bookkeeping reduces to the RNS oracle's own
noise. A fraction $P_{\mathrm{deviant}}\le 2\sqrt\epsilon$ of shots is wasted or corrupted
by the masked, approximate oracle (Equation~\eqref{eq:Sopt}), which inflates the count by
$1/(1-P_{\mathrm{deviant}})$. A further factor $0.99$ covers the residual chance that the
post-processing fails even on a clean sample~\cite{Eke20}. Collecting the $s+1$ shots the
lattice step needs then costs, on average,
\begin{equation}\label{eq:Rshots}
  \mathbb E[R] \;\approx\; \frac{s+1}{(1-P_{\mathrm{deviant}})\cdot 0.99}.
\end{equation}
For RSA-2048, with $s=8$ and $\kappa=33$ and $P_{\mathrm{deviant}}\approx\SI{1.25}{\percent}$, 
both taken from~\cite{gidney2025factor2048bitrsa}, Equation~\eqref{eq:Rshots} gives $\mathbb
E[R]\approx9.2$~\cite{gidney2025factor2048bitrsa}. The baseline is the $s+1$ shots, and
the RNS and masking noise is a percent-level correction on top. The value $s+1$ is itself
the asymptotic count, and the finite-$n$ figures we quote are read from Ekerå's
tables~\cite{Eke20}.

Equation~\eqref{eq:Rshots} is the estimate of Gidney~\cite{gidney2025factor2048bitrsa}.
It rests on his Assumption~2 and charges a corrupted shot like an aborted one, as a
single lost shot. A corruption can in fact cost a whole attempt.
Lemma~\ref{lem:compose} below charges it that way and gives a heuristic worst case of
$\approx 10$ shots. 

This is also where the choice of masking over hashing, made when we fixed the pipeline
earlier, pays off concretely. With a $t$-bit hash, the output compression of
Section~\ref{sec:output-register} sends a fraction $2^{-t}$ of the informative shots to the
zero outcome, a factor $1/(1-2^{-t})$ in shots. The penalty shrinks as $t$ grows, but only
by spending output qubits. Masking instead reaches $P_{\mathrm{deviant}}\approx 1.25~\%$
inside the $(\kappa{+}1)$-qubit accumulator it needs anyway. The implementation
of~\cite{C:CheFouSch25} uses the $1$-bit extreme, where zeros appear in half of all runs.
For RSA-2048 it takes $s=17$ and $\mu=20$ measurement results, following Ekerå's tables,
and therefore needs $2\mu=40$ runs on average, against the $9.2$ shots of
Equation~\eqref{eq:Rshots} at $s=8$. The two counts differ in $s$ as well as in the
read-out, so the ratio is not due to masking alone.

\paragraph{A Las Vegas procedure.} The check $q(2d+2-q)=N$ makes the whole pipeline
self-verifying. A run never returns a wrong $(p,q)$. A corrupted shot, or a batch that
fails post-processing, makes the check fail, and the algorithm responds by retrying the
lattice step over subsets of the collected shots, or by restarting with a fresh
$g$~\cite{gidney2025factor2048bitrsa}. The algorithm is thus Las Vegas. Its only failure
mode is exhausting a chosen shot or restart budget before the check passes. We define
$P_{\mathrm{alg}}$ as the probability of that event. Independent fresh-$g$
attempts each succeed with a constant probability $p_{\mathrm{ok}}$ close to $1$, so after
$k$ such restarts,
\begin{equation*}\label{eq:Palg}
  P_{\mathrm{alg}} \;\le\; (1-p_{\mathrm{ok}})^{\,k},
\end{equation*}
with each restart costing about $\mathbb E[R]$ shots. Lemma~\ref{lem:compose} below bounds
$1-p_{\mathrm{ok}}$ by about $7~\%$ for RSA-2048. This is a heuristic worst case that
ignores subset retry, which can only lower it. A budget of a few times $\mathbb E[R]$
therefore drives $P_{\mathrm{alg}}$ far below any target. At $R=\mathbb E[R]$ the
algorithm sits near the median of its own shot distribution, where failure is of order one
half, so smallness comes only from running a constant factor past the expected count,
never from the expected count itself. One modeling caveat applies to both equations. They
treat the per-shot deviant rate as independent across shots. It is not exactly
independent. Within a single run $g$ is fixed, so the mask-abort rate carries a common,
$g$-dependent skew~\cite{gidney2025factor2048bitrsa}, and the corruption part is a
coherent error bounded through trace distance rather than a classical
coin~\cite{C:CheFouSch25}. This correlation affects the spread of the shot count, not its
mean or the Las Vegas guarantee. Both~\cite{gidney2025factor2048bitrsa} and~\cite{C:CheFouSch25} validate
the mean directly by simulation.  

Lemma~\ref{lem:compose} makes $p_{\mathrm{ok}}$ precise and states how aborts and
corruptions combine across shots, attempts and restarts.

\begin{lemma}[Aborts and corruptions across shots and attempts]\label{lem:compose}
Fix one attempt, that is, one value of $g$. Assume that each shot aborts with
probability at most $\delta_{\bot}$ and corrupts with probability at most
$\delta_{\times}$, conditionally on all earlier shots. Assume that the lattice step
consumes $s+1$ kept shots and, when none of them is corrupted, returns the correct $d$
with probability at least $q_{\mathrm{pp}}$. This last assumption is heuristic. It rests
on the simulations of Ekerå~\cite{Eke20}. Then the following hold.
\begin{enumerate}
  \item The attempt outputs either the correct pair $(p,q)$ or nothing. It never
    outputs a wrong factorization.
  \item (Heuristic.) The attempt succeeds with probability
    $p_{\mathrm{ok}}\ge q_{\mathrm{pp}}-(s+1)\,\delta_{\times}/(1-\delta_{\bot})$.
  \item If aborts are independent across shots, the attempt draws on average
    $(s+1)/(1-\delta_{\bot})$ shots.
  \item If restarts use independent fresh values of $g$, then after $k$ attempts
    $P_{\mathrm{alg}}\le(1-p_{\mathrm{ok}})^{k}$. The expected number of shots over the
    whole run is at most $(s+1)/\bigl((1-\delta_{\bot})\,p_{\mathrm{ok}}\bigr)$.
\end{enumerate}
\end{lemma}

\noindent\emph{Proof sketch.} Item~1 holds because the check $q(2d+2-q)=N$ with $1<q<N$
is satisfied only by the correct $d$. For item~2, a kept shot is corrupted with
probability at most $\delta_{\times}/(1-\delta_{\bot})$ given the earlier shots. A union
bound over the $s+1$ kept shots and over the failure of the lattice step on clean input
gives the claim. Item~3 is the mean of
a negative binomial count. Item~4 follows from the independence of the attempts and from
Wald's identity.

For RSA-2048, the optimum $S=\sqrt\epsilon$ splits $P_{\mathrm{deviant}}\approx1.25~\%$
into $\delta_{\bot}\approx\delta_{\times}\approx0.63~\%$. This uses the heuristic charge
$\delta_{\times}\approx\epsilon/S$ of Section~\ref{sec:approx-period}. With $s+1=9$ and
$q_{\mathrm{pp}}=0.99$, Lemma~\ref{lem:compose} then gives $p_{\mathrm{ok}}\ge0.93$ and
at most about $9.7$ expected shots. The rigorous charge is the trace distance,
$\delta_{\times}\le\sqrt{\epsilon/S}\approx7.9~\%$ at the same parameters. It gives only
$p_{\mathrm{ok}}\ge0.27$ and at most about $33$ expected shots, before subset retry
(Remark~\ref{rem:subset-retry}).

\begin{remark}[Subset retry]\label{rem:subset-retry}
Gidney~\cite{gidney2025factor2048bitrsa} does not end an attempt at the first failed
check. He takes a few extra shots and runs the post-processing on all
$\binom{t}{s+1}$ subsets of the $t$ shots collected so far. He restarts with a fresh $g$
only when $t$ grows too large for this search. A single corruption therefore no longer
ends the attempt. Lemma~\ref{lem:compose} ignores this recovery, so its bound on the
expected shot count is a worst case.
\end{remark}

\paragraph{Composition with the physical error rate.}\label{par:composition-physical-error} $P_{\mathrm{alg}}$ is internal to
the algorithm's randomness. It does not depend on the hardware. Running the pipeline on a
physical device adds a second error source. That source enters in the same way, not as a new
term. A logical error during a shot is also silent. It too surfaces only as a
post-processing or check failure~\cite{gidney2025factor2048bitrsa}. It is therefore caught
and retried by the very mechanism above, so it inflates the expected shot count rather
than creating a wrong-answer probability. The full cost of a real run is then captured by
two multipliers on the $s+1$ baseline. The first is the algorithmic
$1/(1-P_{\mathrm{deviant}})$ of Equation~\eqref{eq:Rshots}. The second is a physical
factor set by the per-shot logical error rate,
treated in the logical error budget of Section~\ref{sec:C7-budget}. $P_{\mathrm{alg}}$ is the part of that
budget that no amount of qubit quality can remove. It comes from the structure of the
algorithm, not from the device that runs it. 

\subsection{FFDLP Algorithm}

The finite field discrete logarithm problem (FFDLP) is a discrete logarithm problem in a
cyclic subgroup of $\mathbb{F}_p^*$. In safe-prime groups like those
of~\cite{HC98,KK03}, the subgroup has prime order $r=(p-1)/2$. The logarithm itself,
however, is usually much shorter than $p$. This is an implementation choice, not a
property of the group. Standards size the secret exponent to the target security level,
for instance about $220$ bits for the $2048$-bit group of~\cite{KK03}. The hardness of
the problem is therefore parametrized \emph{both} by the bit-size $n$ of $p$ and by the
bit-size $m$ of the discrete logarithm.

This is an important difference with RSA instances, which are reduced to an $n/2$-bit
discrete logarithm in an $n$-bit group. Here $m \ll n$ is possible. The exponent only has
to resist generic attacks on a logarithm known to lie in an interval of size $2^m$, such as
Pollard's kangaroo method or baby-step giant-step restricted to that interval, which take
$O(2^{m/2})$ group operations~\cite{Pollard78,vOW99}. The (generalized) Number Field
Sieve in $\mathbb{F}_p^*$ has subexponential complexity in $n$ and does not depend on $m$.

This parametrization modifies the asymptotic and practical complexities of the RNS-based
truncated exponentiation. With a logarithm of $m$ bits, the number of qubits becomes
$m + o(n)$~\cite{C:CheFouSch25} (instead of $n/2 + o(n)$), and the gate count becomes
$\widetilde{O}(m^2 n)$ (instead of $O(n^3)$). This is because both the number of RNS
residues (of order $mn$, up to logarithmic factors) and the time to compute such a residue
(of order $m$, up to logarithmic factors) depend on $m$, and have been reduced with respect
to a factoring instance of comparable security level.

Apart from these different sizes, the circuit for a single shot is unchanged.

\begin{table*}[ht]
\centering
\begin{tabular}{|l|r|r|r|r|r|r|r|r|r|r|}
\hline
 Instance                     & $n$    & $d$  & $\kappa$  & $s$ & Window & Integer  & RNS primes & \multicolumn{2}{c}{Prime bits} & Dlog acc.\\
 &&&&& bit-size &&& Min & Max & bit-size \\
\hline
FFDLP-512    &  512 &   80 &  26 &   7 &   7 &     7680 &      591 &    14 &   14 &    18  \\
FFDLP-1024   & 1024 &  160 &  26 &   7 &   7 &    30720 &     2012 &    16 &   16 &    21  \\
FFDLP-1536   & 1536 &  180 &  28 &   7 &   6 &    59904 &     3974 &    15 &   16 &    22  \\
FFDLP-2048   & 2048 &  224 &  28 &   7 &   6 &   100352 &     6291 &    16 &   17 &    23  \\
RSA-1024    & 1024 &  512 &  28 &   8 &   6 &   109568 &     6330 &    18 &   18 &    25  \\
RSA-2048    & 2048 & 1024 &  33 &   8 &   6 &   438272 &    22658 &    20 &   20 &    28  \\
\hline
\end{tabular}
\caption{Instance parameters for RSA and FFDLP. $n$ is the bit-size of the ambient group (RSA key size or ambient prime for FFDLP), $d$ is the bit-size of the discrete logarithm, $\kappa$ is the number of bits of the accumulator register (named $f$ in~\cite{gidney2025factor2048bitrsa}), $s$ determines the input register size; ``integer bit-size'' is the bit-size of integers represented by the RNS; ``RNS primes'' is the number of primes in the RNS; ``min prime bits'' and ``max prime bits'' are the minimal and maximal bit-size of primes in the RNS (we start from a given bit-size, and take all possible smallest primes). The window bit-size of the primes are selected by minimizing the $\CCX$ gate count. First the window size, which has more impact, then the bit-size of primes. Both have only an impact on the first decimal of the $\log_2$, and the exact implementation of the residue circuit matters here.}
\label{tab:instances}
\end{table*}

\subsection{Parameter Selection for RSA and FFDLP}

While the logical qubit count is mostly determined by the input register sizes, and interacts with the post-processing parameters (such as the number of shots in a run), the logical gate count is mostly determined by the parameters of the problem: the bit-size of $N$ (or $p$) and the bit-size of the discrete logarithm determine the number of RNS primes and their maximal size. It also depends on the windowing parameter in the exponent register: the larger the window size, the smaller the values in the integer domain.

For several choices of bit-size, exponent size and window size, we compute an upper bound on the total gate count. We do the following approximations:
\begin{itemize}
    \item The RNS depends on the ambient group, the exponent size, and the precomputed windows (as we have to avoid any prime which divides one of the integers in the windows). We take a random prime for the ambient group (as this does not change the sizes), but compute the windows genuinely. We take primes of minimal bit-size 12. The input register size is computed as $m + 2\lceil m/s\rceil$, and padded to a multiple of the window size. 
    \item The truncation length $\kappa$ is taken from~\cite[Table~5]{gidney2025factor2048bitrsa} for the RSA rows. The FFDLP rows have fewer RNS primes,
    hence less drift, and we use smaller values there. We report the resulting failure rate
    $P_{\mathrm{deviant}}$ in Table~\ref{tab:instances}. For the FFDLP rows it is estimated
    with Equation~\eqref{eq:kappa-budget}, scaling the drift term $\log_2(cA)$ of the
    RSA-2048 row in proportion to $A=|P|\ell$. This estimate is rough. Applied to RSA-1024
    at $\kappa=28$ it gives about $3.6\%$, against the $2.87\%$ reported
    in~\cite{gidney2025factor2048bitrsa}. The value of $\kappa$ does not change the $\CCX$
    count of the residue circuits reported in the table, but it sets the width of the
    $|P|\ell$ additions into the accumulator and the qubit count.
    \item The $\CCX$ gate count and qubit count of a residue computation circuit increase with the RNS prime bit-size. We just take the largest primes in the RNS to upper bound the cost, and multiply by the number of primes in the RNS. Note that a residue computation circuit actually merges the uncomputation of a residue with the computation of the next one.
    \item The window size can be freely chosen to minimize the cost. We try window sizes between 6 and 8, and observe variations of a factor at most $2^{0.2}$ in the gate count; in the end a window size 6 seems most appropriate, except for small instances.
\end{itemize}

The overall instance parameters for different instances of RSA and FFDLP are depicted in Table~\ref{tab:instances}. Note that the residue circuit is computed explicitly using our Qarton code.

\subsection{The Elliptic-Curve Instance}\label{sec:ecdlp}

Section~\ref{sec:hsp} showed that $\mathtt{ECDLP}$ is formally the same hidden subgroup
problem as $\mathtt{FFDLP}$. The group is $G = \mathbb{Z}_r^2$, the subgroup is
$H = \langle (d,1)\rangle$, and the function is $f(a,b) = [a]P + [b](-Q)$, evaluated
using elliptic curve point additions. Throughout this section $n$ denotes the bit-length of the curve's field or order, the role the RSA modulus bit-length played in the sections above. The two
are different numbers even at comparable classical security (a 256-bit curve against a
3072-bit RSA modulus), so comparing figures across the two halves should not carry over a
specific value of $n$. In the factoring case, the literature seems to have converged to a single best algorithm, which is the one presented in the preceding sections. On the contrary, the elliptic-curve case has multiple competing trade-offs between qubit and gate count, which we detail in this section. We start by making explicit some differences with respect to the discrete logarithm instances previously seen in factoring and $\mathtt{FFDLP}$.

\emph{The order is known and prime.} For the NIST and SECG curves the (prime) order $r$ of
the base point is a published parameter, so the exponent registers can be sized to
about $\log_2 r$ bits from the start. They remain power-of-two registers, and the Fourier
transform on them is semi-classical and approximate, as described below. Two simplifications follow. 
There is no analogue of the
over-provisioned register that the unknown order forced on order finding in
Section~\ref{sec:shor}, where it required $2^{2n}\ge N^{2}$.

\emph{The logarithm is general, not short.} The Ekerå--Håstad
analysis~\cite{PQCRYPTO:EkeHas17} needs the unknown logarithm to be \emph{short} for its lattice
recovery to succeed. Concretely, in the notation of Section~\ref{sec:input-register}, this is
hypothesis~\eqref{eq:eh-order}, $r \ge 2^{\tau_s+m}+2^{\tau_s}d$. Ekerå gives the sharp form
of this condition, with $2^{\tau_s}-1$ in place of $2^{\tau_s}$~\cite[Sect.~1.4]{Ekera26ShortDLP}.
With $\tau\approx m/s$ it allows $d$ up to about
$r^{s/(s+1)}$, that is $d \lesssim \sqrt r$ for $s=1$. For factoring,
the reduction of Section~\ref{sec:short-dlp} arranged exactly such a short logarithm. Here
$d$ is a full-length exponent, uniform on $[0,r)$. That reduction therefore does not apply, and
its heuristic variant degrades as well~\cite{Ekera19GeneralDLP}. The input-register saving
is still available, but through a different route, Ekerå's algorithm for \emph{general}
logarithms, which trades runs against register size rather than exploiting shortness. We
explore this saving in more details hereafter.

\emph{The group law does not reduce to multiplication.} Evaluating
$f$ means adding points on the curve. The ordinary \emph{affine} addition law represents a
point by its coordinates $(x,y)$, and it contains a division in $\mathbb{F}_q$. For $P_1
\neq P_2$ the slope is $\lambda = (y_2-y_1)/(x_2-x_1)$, giving $x_3 = \lambda^2 - x_1 -
x_2$ and $y_3 = \lambda(x_1-x_3) - y_1$. Division becomes the most expensive sub-circuit, and the primary target for trade-offs. The alternative is to use \emph{projective coordinates}, in which addition formulas do not contain divisions. These are only used in~\cite{CFS26ECDLP}.

\begin{figure*}[ht]
\centering
\definecolor{ABTeal}{HTML}{00B5C8}     
\definecolor{ABTealDark}{HTML}{00676A} 
\definecolor{ABInk}{HTML}{010102}      
\definecolor{ABLine}{HTML}{6B6B6B}     
\resizebox{\linewidth}{!}{%
\begin{tikzpicture}[
    >=Stealth, font=\small, text=ABInk,
    box/.style={draw=ABTealDark, rounded corners=5pt, line width=0.75pt,
                minimum width=6.0cm, minimum height=1.3cm,
                align=center, inner xsep=10pt, inner ysep=4pt},
    neutral/.style={box, draw=ABLine, fill=black!6},
    hdr/.style={box, fill=ABTeal!22, line width=1.15pt},
    mech/.style={box, fill=white},
    res/.style={box, fill=ABTeal!10},
    arr/.style={->, line width=0.8pt, draw=ABLine},
    bus/.style={line width=0.8pt, draw=ABLine},
  ]
 
  \def\cL{0}   \def\cM{6.4}  \def\cR{12.8}
  \def\yBase{6.8} \def\yHdr{4.2} \def\yMid{2.1} \def\yRes{0.0} \def\yDepth{-2.0}
  \def\yBus{5.45}
 
  \node[neutral, minimum width=6.6cm]
    (base) at (\cM,\yBase)
    {\textbf{Shor's Algorithm for ECDLP}\\[1pt]{\scriptsize Proos--Zalka, Häner et al.}};
 
  \node[hdr] (h1) at (\cL,\yHdr)
    {\textbf{Optimize the gates}\\[1pt]{\scriptsize Babbush et al.\ 2026 / Schrottenloher
        2026 }};
  \node[hdr] (h2) at (\cM,\yHdr)
    {\textbf{Shrink the register}\\[1pt]{\scriptsize Luo et al.\ 2026}};
  \node[hdr] (h3) at (\cR,\yHdr)
    {\textbf{Avoid the inversion}\\[1pt]{\scriptsize Chevignard et al.\ 2026}};

  \node[mech] (m1) at (\cL,\yMid)
    {Dialog-split EEA\\[1pt]{\scriptsize approximate arithmetic}};
  \node[mech] (m2) at (\cM,\yMid)
    {Length registers\\[1pt]{\scriptsize exact inversion}};
  \node[mech] (m3) at (\cR,\yMid)
    {Projective $+$ RNS\\[1pt]{\scriptsize one Legendre bit}};

  \node[res] (r1) at (\cL,\yRes)
    {\textbf{1208 qubits}\\[1pt]{\scriptsize $2^{26.11}$ Toffoli, 1 run}};
  \node[res] (r2) at (\cM,\yRes)
    {\textbf{835 qubits}\\[1pt]{\scriptsize $2^{30.88}$ Toffoli, 1 run}};
  \node[res] (r3) at (\cR,\yRes)
    {\textbf{1193 qubits}\\[1pt]{\scriptsize $2^{38.98}$ Toffoli, 22 runs}};

  \node[neutral, fit={(r1.west |- 0,\yDepth) (r3.east |- 0,\yDepth)},
        inner sep=0pt, minimum height=1.3cm] (depthframe) {};
  \node[align=center] (depth) at (depthframe.center)
    {\textbf{Fourth axis: circuit depth}\\[1pt]{\scriptsize Kim et al.\ 2026, lowest depth given ancillae}};
 
  \draw[bus] (base.south) -- (\cM,\yBus);
  \draw[bus] (\cL,\yBus) -- (\cR,\yBus);
  \draw[arr] (\cL,\yBus) -- (h1.north);
  \draw[arr] (\cM,\yBus) -- (h2.north);
  \draw[arr] (\cR,\yBus) -- (h3.north);

  \draw[arr] (h1.south) -- (m1.north);  \draw[arr] (m1.south) -- (r1.north);
  \draw[arr] (h2.south) -- (m2.north);  \draw[arr] (m2.south) -- (r2.north);
  \draw[arr] (h3.south) -- (m3.north);  \draw[arr] (m3.south) -- (r3.north);
 
\end{tikzpicture}%
}

\caption{The elliptic-curve frontier, oriented by the resource each direction pushes:
\emph{optimize the gates}, \emph{shrink the register}, \emph{avoid the inversion}, and a
fourth axis, circuit depth. Unlike the factoring pipeline, no algorithm dominate for all metrics. 
The figures are current $n=256$ estimates. 
}
\label{fig:ecdlp-axes}
\end{figure*}
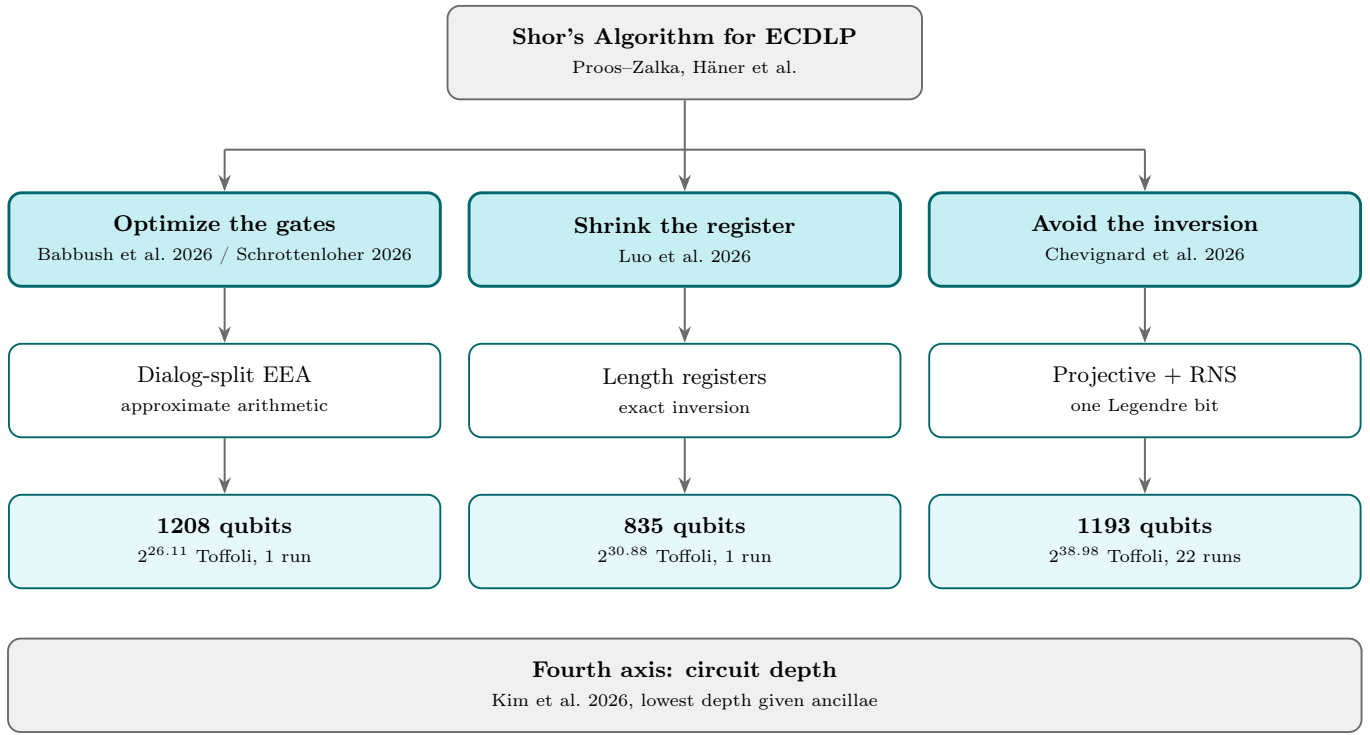

We will therefore refer in the following to two families. The
affine family keeps the point in $(x,y)$ coordinates and computes
inversions. Its two members, reaching the current best trade-offs,
use two different reversible
implementations of the extended Euclidean algorithm (EEA). 
The projective family has a single member~\cite{CFS26ECDLP}, which avoids inversions and reuses some
of the structure of the factoring case: the output-register compression of
Section~\ref{sec:output-register} with the residue arithmetic of
Section~\ref{sec:oracle-rns}. While it was the most space-efficient at the time of publication, it is now beaten by~\cite{LuoEtAl26} in both space and gate count, as we will detail in the rest of this section. A fourth resource, circuit \emph{depth} (the
number of sequential steps needed once independent gates are parallelized, rather than
their raw total), is pushed by a separate line of work again. We set it aside until the
comparison at the end. Figure~\ref{fig:ecdlp-axes} maps the three space/gate directions
and the depth axis.

Figure~\ref{fig:ecdlp-axes} maps the resource trade-off. Table~\ref{tab:ecdlp-relaxation}
maps the same three directions onto the $(\epsilon,\delta)$ axes of
Definition~\ref{def:noisy-hsp}, alongside the factoring relaxations of
Table~\ref{tab:four-relaxation}. As in that table, these three rows are alternatives, not
a pipeline. Two of the three rows never abort. They either succeed exactly or corrupt. The third
compresses the output to a single Legendre bit, which is the May--Schlieper move of
Section~\ref{sec:output-register} with $t=1$. About half of its runs therefore land on the
discarded zero outcome, which is why its run count doubles.
By the identity of Section~\ref{sec:approx-period}, this abort rate is at least $1/2$, with near equality when the hash is balanced.

\begin{table*}[t]
\centering
\renewcommand{\arraystretch}{1.35}
\setlength{\tabcolsep}{6pt}

\begin{tabular}{@{}p{3.4cm} c c c p{4.0cm}@{}}
\toprule
\textbf{Direction} & $\boldsymbol{\epsilon}$
 & \multicolumn{2}{c}{$\boldsymbol{\delta=\delta_{\bot}+\delta_{\times}}$}
 & \textbf{Read-out} $\boldsymbol{R}$ \\
\cmidrule(lr){3-4}
 & & $\delta_{\bot}$ {\footnotesize\itshape abort}
   & $\delta_{\times}$ {\footnotesize\itshape corrupt} & \\
\midrule
Optimize the gates
  & $0$ & $0$ & $\lesssim 2^{-6.9}$
  & measure the accumulated point (rare circuit failure corrupts silently, empirical rate,
   $95\%$ bound) \\
\addlinespace
Shrink the register
 & $0$ & $0$ & $\approx 2^{-w}$
 & exact arithmetic. Corrupts only when the first window is $0$ \\
\addlinespace
Avoid the inversion
  & $0$ & $\approx 1/2$ & $\lesssim 0.01$
  & measure $1$ Legendre bit, a $1$-bit hash (about half the runs abort). Exact except on
   a sparse set of inputs, where it corrupts \\
\bottomrule
\end{tabular}
\caption{The three $\mathtt{ECDLP}$ directions of this section, placed on the same
  $(\epsilon,\delta)$ axes as Table~\ref{tab:four-relaxation}. All three oracles are exact
  ($\epsilon=0$). The first and last rows fail on a sparse set of inputs, which counts as
  corruption $\delta_\times$ rather than as a drift $\epsilon$. The last row also aborts on
  about half of its runs, because it compresses the output to a single bit.}
\label{tab:ecdlp-relaxation}
\end{table*}

\paragraph{Recycling the input register.} This is not a trick specific to elliptic curves: it
is the same semi-classical-QFT recycling already used for the factoring baseline of Section~\ref{sec:shor}, 
applied here to two registers instead of one. The two exponent registers, holding $a$ and
$b$, never exist in full at any one time. The $w$ qubits of one window of the exponents
enter the circuit together, as a single control block. They control one windowed point
addition and are then measured, before the next window is prepared. Because the quantum
Fourier transform at read-out can be performed
\emph{semi-classically}~\cite{PhysRevLett.76.3228}, each measurement outcome is a
classical value that only sets the phase of the gates still to come. Nothing therefore
needs to be kept coherent once it has been read. A measured window of control qubits can
thus be reset and reused for the next window~\cite{MoscaEkert99,ParkerPlenio00,Zalka98}. Carried through, a single recycled
$w$-qubit window does the work of both exponent registers, saving the $\approx 2n$ qubits
they would otherwise occupy~\cite{LuoEtAl26}. This is why the $(a,b)$ registers cost
essentially nothing beyond the single $w$-qubit window already totaled elsewhere, and why
the space figures of the recycling constructions count only the \emph{point} registers, the
workspace of the arithmetic on the
curve~\cite{ProosZalka03,RNSL17,10.1007/978-3-030-44223-1_23,Litinski23,Babbush26ECC,%
  Schrottenloher26PointAdd,LuoEtAl26}. The compressed construction of~\cite{CFS26ECDLP} does
not recycle (Remark~\ref{rem:recycling-tension}), so its figure includes the index
registers.

The input-register optimization of Section~\ref{sec:input-register} therefore buys no
space on the elliptic curve. What it can still buy is fewer point additions per run, which lowers both the gate count and the depth.
For factoring, the picture depends on the oracle. With recycling, the Ekerå--Håstad
reduction of Section~\ref{sec:input-register} likewise buys fewer modular multiplications
per run, not space. With the residue-arithmetic oracle, which cannot recycle, it saves
qubits, and this is how the modern factoring pipeline uses it.
Ekerå's algorithm for general logarithms with tradeoffs~\cite{Ekera21Tradeoffs} realizes
this. With a tradeoff factor $s \ge 1$ and $\ell=\lceil m/s\rceil$, its two exponent
registers hold $m+\ell$ and $\ell$ bits, so it evaluates about $(1+2/s)m$ point additions 
per run instead of Shor's $2m$, fewer as soon as $s>2$. The price is recovering $d$ from several runs by lattice
reduction rather than from one. Slightly more than $s$ runs typically reach $99\%$
success~\cite{Ekera21Tradeoffs}. Shor's original algorithm, with no tradeoff, behaves
differently. In Ekerå's analysis of it, a single run succeeds only with probability between
about $60\%$ and $82\%$, and this is a feature of the exact algorithm rather than of any
approximation. Pushing this past $99\%$ costs either a few extra point additions plus one
bounded classical search, or two bounded classical searches in post-processing with no
extra point additions~\cite{Ekera19GeneralDLP}.\footnote{The heuristic $60$--$82\%$ range
  and the two routes past $99\%$ are from Ekerå's own analysis~\cite[Abstract]{Ekera19GeneralDLP}.
  The range reflects the group order $r$ and search-space size, and the extra operations
  sharpen the QFT peak in the same way the register size does.}

\begin{remark}[Recycling and compression are exclusive]\label{rem:recycling-tension}
The free input register and the compressed oracle do not combine for free. Recycling the
control qubit (above) requires measuring each control as the oracle runs. The
residue-arithmetic construction of Section~\ref{sec:oracle-rns}, by contrast, compresses
the \emph{work} register but needs its control registers intact, so it leaves them
unrecycled~\cite[Sect.~1]{gidney2025factor2048bitrsa}. The tension is concrete in the compressed
$\mathtt{ECDLP}$ construction~\cite{CFS26ECDLP}: it does not recycle its index registers,
and spends its mid-circuit measurements instead on a \emph{pebbling} schedule. A pebble
game is a bookkeeping scheme. For each intermediate result of a reversible circuit,
it records whether that result is kept in memory or freed and recomputed later. This
trades space against gate count. \emph{Spooky} pebbling~\cite{Gidney19spooky,Kornerup25} is the
variant that frees a result's qubits early by \emph{measuring} it, and repairs the phase
error this leaves behind at a later point. Here the results being pebbled are the nodes of
a binary tree. Its leaves are the windowed point additions and its root is the accumulated
projective point. Each node stores a residue triple in the RNS of
Section~\ref{sec:oracle-rns}. The tree shape comes from how the point accumulation is
structured. The affine
constructions~\cite{Babbush26ECC,Schrottenloher26PointAdd,LuoEtAl26} make the opposite
choice: they recycle the control qubit and use measurement only inside the arithmetic. A
construction can thus collapse the input register or compress the output register. Obtaining both at once is
open. This is the fault line along which the frontier of
Table~\ref{tab:ecdlp-costs} splits.
\end{remark}

\paragraph{From scalar multiplication to point additions.} The oracle stage decomposes
exactly as the modular exponentiation of Section~\ref{sec:shor} did, one level at a
time. Writing both exponents in windows of $w$ bits,
$a = \sum_{k} 2^{kw} a_k$ and $b = \sum_{k} 2^{kw} b_k$ with $0 \le k < \lceil n/w \rceil$,
the point multiplication factors as
\begin{equation*}\label{eq:point-mult-windowed}
    [a]P + [b](-Q) = \sum_{k} \bigl[a_k\bigr]\bigl([2^{kw}]P\bigr)
                 + \sum_{k} \bigl[b_k\bigr]\bigl([2^{kw}](-Q)\bigr).
\end{equation*}
Collecting the $K = 2\lceil n/w \rceil$ window values in a single sequence
$(i_0,\dots,i_{K-1}) = (a_0,\dots,a_{K/2-1},\,b_0,\dots,b_{K/2-1})$, the oracle reduces
to $K \approx 2n/w$ \emph{windowed point additions}
\begin{equation}\label{eq:windowed-point-add}
  \ket{i_k}\ket{R} \;\longmapsto\; \ket{i_k}\ket{R + P^{(k)}_{i_k}},
\end{equation}
where the table of window $k$, $(P^{(k)}_0,\dots,P^{(k)}_{2^w-1})$, holds the
classically precomputed multiples $P^{(k)}_j = [j]\bigl([2^{kw}]P\bigr)$ for $k < K/2$
and $P^{(k)}_j = [j]\bigl([2^{(k-K/2)w}](-Q)\bigr)$ otherwise. The addend is loaded by a
table lookup at a cost of $2^w$ Toffoli gates and unloaded almost for free by
measurement-based uncomputation~\cite{Babbush18}. This is the same design philosophy
as before: the multiples $[2^{kw}]P$ and $[2^{kw}](-Q)$ play the role of the constants
$g^{2^i} \bmod N$, they are computed classically, and every window costs the same
regardless of $k$.

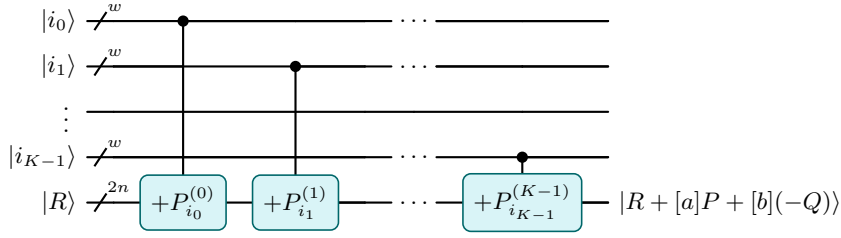
\begin{figure*}[tp]
  \centering
  \tikzset{abgate/.style={fill=ABTeal!15, draw=ABTealDark, line width=0.6pt, rounded corners=3pt}}
  \begin{quantikz}[column sep=0.35cm, row sep={0.6cm,between origins}]
    \lstick{$\ket{i_0}$}
      & \qwbundle{w} & \ctrl{4} & \qw & \qw & \ \ldots\ & \qw & \qw \\
    \lstick{$\ket{i_1}$}
      & \qwbundle{w} & \qw & \ctrl{3} & \qw & \ \ldots\ & \qw & \qw \\
    \lstick{$\vdots\;$}
      & & & & & & & \\
    \lstick{$\ket{i_{K-1}}$}
      & \qwbundle{w} & \qw & \qw & \qw & \ \ldots\ & \ctrl{1} & \qw \\
    \lstick{$\ket{R}$}
      & \qwbundle{2n}
      & \gate[style=abgate]{+P^{(0)}_{i_0}}
      & \gate[style=abgate]{+P^{(1)}_{i_1}}
      & \qw & \ \ldots\
      & \gate[style=abgate]{+P^{(K-1)}_{i_{K-1}}}
      & \qw \rstick{$\ket{R+[a]P+[b](-Q)}$}
  \end{quantikz}
  \caption{Windowed point-accumulation stage, in coherent form: block $k$ adds to the
    accumulator $R$ the table point $P^{(k)}_{i_k}$ selected by the window register
    $\ket{i_k}$, one instance of Eq.~\eqref{eq:windowed-point-add} per window. In the
    implementation, a single $w$-qubit window register is recycled through the
    semi-classical Fourier transform.
    }
  \label{fig:ecdlp-oracleF}
\end{figure*}

Figure~\ref{fig:ecdlp-oracleF} shows the resulting accumulation stage. Block $k$ reads
its addend from table $k$, indexed by the window register $\ket{i_k}$, and adds it to the
shared accumulator $R$ with the point-addition circuit of
Fig.~\ref{fig:ec-point-addition}. The figure is drawn in the coherent form, with all $K$
window registers present, that is $2n$ exponent qubits. In practice, the inverse quantum Fourier transform on the exponents is performed semi-classically~\cite{PhysRevLett.76.3228}.
The windows of each exponent are processed from the most significant one down, since each measured window sets the phase corrections of the less significant ones.
Once a block is done, its $w$ qubits receive classically controlled phase rotations, are measured, and are re-prepared for the next window. A single $w$-qubit
register is therefore reused throughout, and the logical qubit counts of this section
include only $w$ exponent qubits.

The count $K$ is itself only the naive baseline. Following Litinski~\cite{Litinski23},
the very first addition, into the starting point $\mathcal{O}$, is just a lookup of the table point itself. This also avoids feeding the point at infinity to the affine
addition circuit, which does not accept it as input, except when $i_0=0$, where the looked-up point is itself $\mathcal{O}$. The last three additions can be
folded into extra classical post-processing, using the single-run, bounded-search mode
of Ekerå's general-logarithm algorithm~\cite{Ekera19GeneralDLP}, instead of being run
quantumly. For $\mathtt{secp256k1}$ with $w = 16$, this brings $K = 32$ down to the
$N_A = K - 4 = 28$ windowed additions actually
needed~\cite{Babbush26ECC,Schrottenloher26PointAdd,LuoEtAl26}. The addend is no longer a
classical constant but a point read from the table, so the number of lookups per window
depends on how much free space the surrounding circuit has left. The gate-oriented
constructions can afford three such lookups~\cite{Schrottenloher26PointAdd}. The
space-optimal one, with only a single free $n$-qubit register to work with, reports
five~\cite{LuoEtAl26}, at $5\cdot 2^{w}$ rather than $3\cdot 2^{w}$ Toffoli gates per
window.

\begin{figure*}[ht]
\centering
\resizebox{0.92\linewidth}{!}{%
\begin{tikzpicture}[
    >={Stealth[length=5pt, width=4pt]}, font=\small,
    box/.style={rounded corners=3pt, minimum width=3.8cm, minimum height=1.3cm,
                align=center, inner sep=3pt},
    databox/.style={box, fill=gray!10, draw=ABGray, dashed, line width=0.6pt},
    opbox/.style={box, fill=ABTeal!15, draw=ABTealDark, line width=0.6pt},
    hlbox/.style={box, fill=abYellow, draw=ABBlack, line width=0.9pt},
    arr/.style={->, line width=0.8pt, draw=ABGray},
  ]
  \def\dx{4.5}
  \def\yA{3.0} \def\yB{-0.2}
 
  \node[databox] (in1) at (0,4.2) {Accumulator\\[1pt]$(x_R,y_R)$};
  \node[databox] (in2) at (0,1.8) {Table point\\[1pt]$(x_i,y_i)$};
 
  \node[opbox] (sub) at (\dx,\yA)   {Subtract\\[1pt]$\Delta x = x_i - x_R$\\[-1pt]$\Delta y = y_i - y_R$};
  \node[hlbox] (inv) at (2*\dx,\yA) {Invert\\[1pt]$\Delta x \leftarrow \Delta x^{-1}$};
  \node[opbox] (mul) at (3*\dx,\yA) {Multiply\\[1pt]$\lambda = \Delta y \cdot \Delta x$};
 
  \node[opbox]   (cx)  at (3*\dx,\yB) {Combine\\[1pt]$x_{R'} = \lambda^2 - x_R - x_i$};
  \node[opbox]   (cy)  at (2*\dx,\yB) {Combine\\[1pt]$y_{R'} = \lambda(x_R - x_{R'}) - y_R$};
  \node[opbox]   (unc) at (\dx,\yB)   {Uncompute\\[1pt]$\Delta x,\ \Delta y,\ \lambda,\ (x_i,y_i)$};
  \node[databox] (out) at (0,\yB)     {Accumulator\\[1pt]$(x_{R'},y_{R'})$};
 
  \draw[arr] (in1.east) -- ++(0.35,0) |- ([yshift=5pt]sub.west);
  \draw[arr] (in2.east) -- ++(0.35,0) |- ([yshift=-5pt]sub.west);
  \draw[arr] (sub) -- (inv);
  \draw[arr] (inv) -- (mul);
  \draw[arr] (mul) -- (cx);
  \draw[arr] (cx) -- (cy);
  \draw[arr] (cy) -- (unc);
  \draw[arr] (unc) -- (out);
\end{tikzpicture}%
}
\caption{Textbook view of one windowed point addition $R \mapsto R + P_i$, following the
  affine addition law of Eq.~\eqref{eq:affine-add-law}. The table point is loaded by a lookup
  ($2^w$ Toffoli gates). The subtractions cost $O(n)$, while the inversion, the
  multiplications and the squaring all cost $O(n^2)$. In the circuits considered here,
  \emph{invert} and \emph{multiply} are fused into a single in-place division
  $\Delta y \leftarrow \Delta y/\Delta x$ that never materializes $\Delta x^{-1}$. The slope $\lambda$ is not uncomputed separately. It is
  overwritten by a second in-place multiplication, $\lambda(x_i - x_{R'}) = y_{R'} + y_i$
  (Fig.~\ref{fig:ec-point-addition}). These two in-place multiplications account for
  most of the Toffoli count of the addition.
  }
\label{fig:point-add-generic}
\end{figure*}
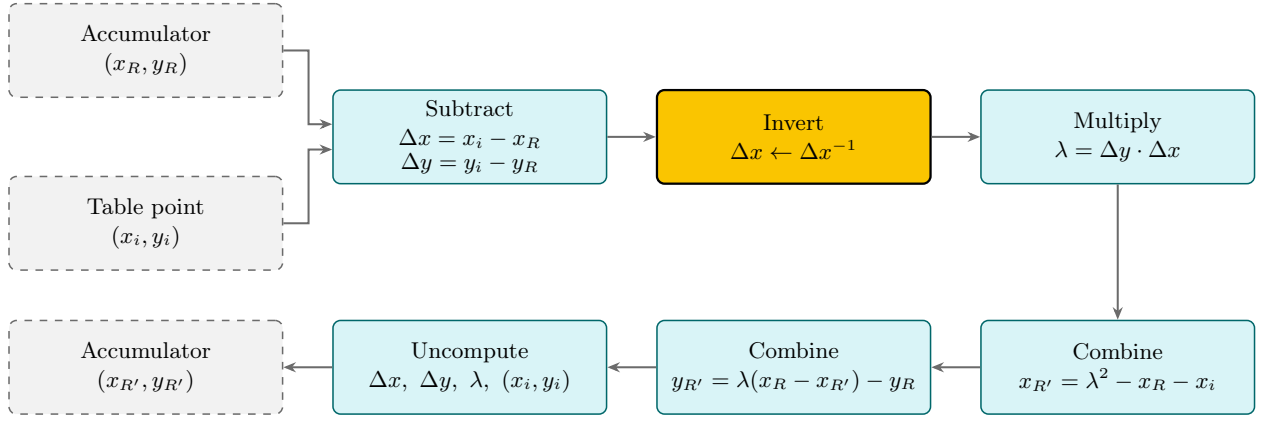

\paragraph{From point additions to field arithmetic.} A single controlled multiplication
in Section~\ref{sec:shor} still had to be built out of modular arithmetic. A single
windowed point addition $R \mapsto R+P_i$ is no different. It still has to be built out of
the affine addition law. Write the running accumulator as $R=(x_R,y_R)$ and the freshly loaded table
point as $P_i=(x_i,y_i)$. In the generic case, the law reads 
\begin{equation}\label{eq:affine-add-law}
\begin{aligned}
  \lambda &= \frac{y_i - y_R}{x_i - x_R},\\
  x_{R'} &= \lambda^2 - x_R - x_i,\\
  y_{R'} &= \lambda(x_R - x_{R'}) - y_R,
\end{aligned}
\end{equation}
where $R'=(x_{R'},y_{R'})$ is the value written back into the accumulator. ``Generic''
means $R\neq\pm P_i$ and neither is $\mathcal{O}$. Doubling and the point at infinity use
different formulas, but they occur with negligible probability once the accumulator is
seeded at a nonzero point rather than at $\mathcal{O}$ itself. Both are therefore neglected here, as elsewhere in this literature.
The first-addition lookup described above seeds the accumulator at $P^{(0)}_{i_0}$, which is itself $\mathcal{O}$ whenever $i_0=0$.
That branch has weight $2^{-w}$ per run, so it must be handled explicitly or counted as corruption. The case $i=0$ is
handled as the only special case~\cite{GouzienCatQubits23,10.1007/978-3-030-44223-1_23,Schrottenloher26PointAdd}. 
This last case is not a matter of negligible probability like the other two. The window
$(P_0,\dots,P_{2^w-1})$ always sets $P_0=\mathcal{O}$ by construction, so whenever a
window's value happens to be $0$ the addend is certainly $\mathcal{O}$. The circuit must
therefore explicitly recognize this and leave $R$ unchanged, rather than run the generic
formulas above. Equation~\eqref{eq:affine-add-law} fixes the registers a reversible
circuit must hold. Besides the $2n$-qubit accumulator $(x_R,y_R)$, which already holds the
point accumulated so far, the circuit needs the table point $(x_i,y_i)$ just loaded by
Eq.~\eqref{eq:windowed-point-add} and a workspace register for $\lambda$. Three more
$n$-qubit registers are needed, which together with the accumulator and the $w$-qubit
window index bring the qubit count in play to $2n+3n+w=5n+w$. Realizing the map $R\mapsto
R'$ reversibly then means executing, in order,

\begin{align*}
\textbf{subtract:} \quad & \Delta x \leftarrow x_i - x_R, \\
& \Delta y \leftarrow y_i - y_R \\
\textbf{invert:} \quad & \Delta x \leftarrow \Delta x^{-1} \\
\textbf{multiply:} \quad & \lambda \leftarrow \Delta y \cdot \Delta x \\
\textbf{square, combine:} \quad & x_{R'} \leftarrow \lambda^2 - x_R - x_i \\
\textbf{combine, multiply,} \quad & \\
\textbf{combine:} \quad & y_{R'} \leftarrow \lambda(x_R - x_{R'}) - y_R \\
\textbf{uncompute:} \quad & \text{release } \Delta x,\, \Delta y,\, \lambda,\, (x_i, y_i)
\end{align*}

and Fig.~\ref{fig:point-add-generic} lays the same sequence out as a data-flow diagram.
(In practice \textbf{invert} and \textbf{multiply} are usually fused into
a single divide primitive~\cite{GouzienCatQubits23,10.1007/978-3-030-44223-1_23}.
They are kept apart here to show exactly where the inversion enters.) The uncompute step
is not free, but it is cheap to route. The line of slope $\lambda$ through $R$ and $P_i$
also passes through $(x_{R'},-y_{R'})$, the unreflected third intersection point that $R'$
is the negative of. Since $R$ is overwritten in place by $R'$, $\lambda$ is recovered from
$R'$ and the table point $P_i$, which can be loaded again by lookup, as
$\lambda = (y_{R'}+y_i)/(x_i-x_{R'})$. It is uncomputed by recomputing it this way rather
than by keeping a separate garbage copy~\cite{GouzienCatQubits23}. Of the five arithmetic steps, one is
qualitatively different from the rest. Subtraction costs $O(n)$ gates, while
multiplication, squaring and inversion all cost $O(n^2)$, as in the caption of
Fig.~\ref{fig:point-add-generic}. What sets \textbf{invert} apart is its constant. It
is the one place this sequence asks for a modular inversion rather than a multiplication,
and as the next paragraph makes precise, that constant makes it the dominant cost.

Gouzien et al.\ report a total of $9n+w+4$ logical qubits~\cite{GouzienCatQubits23} for
the unoptimized construction. The total counts this layer's registers together with the
workspace of the extended Euclidean algorithm, described in the next paragraph. It is of
the same order as the
$9n+2\lceil\log_2 n\rceil+10$ already quoted for Roetteler et al.\ in
Table~\ref{tab:ecdlp-costs}. This is the baseline every direction below cuts down from,
rather than a new number in its own right.

\paragraph{The inversion bottleneck.} Of the operations in
Fig.~\ref{fig:point-add-generic}, both the gate count and the qubit count of the whole
construction are dominated by \textbf{invert}. It is realized through a modular inversion
computed by a reversible extended Euclidean algorithm (EEA). Although inversion and the
multiplications elsewhere in the same figure are both $O(n^2)$, the constant is
significantly larger for inversion~\cite{RNSL17,10.1007/978-3-030-44223-1_23,Litinski23}. 
This is the structural difference with factoring. There, the innermost primitive was a modular \emph{addition}. The direct,
large-register pipeline made it cheap with the coset representation of
Section~\ref{sec:shor}. The modern construction does so by simply working modulo small
primes in the residue system of Section~\ref{sec:oracle-rns}. Here it is a modular \emph{inversion}, and no comparable
technique makes it free. 

\paragraph{How the inversion is actually computed: Kaliski's algorithm.}\label{par:kaliski} The reversible
EEA invoked above is, concretely, a \emph{binary} extended Euclidean algorithm due to
Kaliski~\cite{Kaliski95}, chosen because the ordinary
Euclidean algorithm is awkward to make reversible. Reversible dividers exist, but inside
the Euclidean loop each quotient has a data-dependent length, so a fixed circuit would
have to provision every step for the longest possible quotient. Kaliski's algorithm replaces 
division by repeated halving, so every branch is selected by parity tests and one comparison, and every step does the same 
fixed amount of work regardless of the input. We denote its four registers, $(u,v,w,s)$, throughout. 

The four registers are initialized to $(q,x,0,1)$ for computing $x^{-1}\bmod q$. The
algorithm then applies the following \emph{round} operation:

\begin{tabular}{@{}l@{\quad}l@{}}
\textbf{if} $u$ even: & $u \leftarrow u/2$, \quad $s \leftarrow 2s$ \\
\textbf{elif} $v$ even: & $v \leftarrow v/2$, \quad $w \leftarrow 2w$ \\
\textbf{elif} $u > v$: & $u \leftarrow (u-v)/2$, \quad $w \leftarrow w+s$, \quad $s \leftarrow 2s$ \\
\textbf{else}: & $v \leftarrow (v-u)/2$, \quad $s \leftarrow w+s$, \quad $w \leftarrow 2w$
\end{tabular}

Each round is one of four mutually exclusive branches, selected by the parities of $u,v$
and a comparison. Each branch is a shift, an addition, or a conditional subtraction, so
no division ever appears. Run classically, the loop stops as soon as $v=0$. A
quantum circuit cannot stop early on a data-dependent condition, so instead it runs a fixed
$2n$ rounds for an $n$-bit modulus, an upper bound on how many rounds any input can need.
The rounds after $v=0$ are not automatically no-ops. With $v=0$ the branch ``$v$ even''
still fires and doubles $w$. Reversible implementations therefore add a termination flag,
set once $v=0$, that disables every later round, and a counter that records the number $k$
of active rounds~\cite{RNSL17}. At termination $w \equiv -x^{-1}2^{k} \pmod q$, with
$n \le k \le 2n$ depending on the input~\cite{Kaliski95}. At this point $0<w<2q$~\cite{GouzienCatQubits23}, 
so $w$ is first reduced, $w\leftarrow w-q$ if $w\ge q$, and then negated, $w\leftarrow q-w$, 
giving $w\equiv x^{-1}2^k\pmod q$ with $0<w<q$. Only then does a second pass strip
the extra factors of two. It runs $2n$ further rounds of ``if $w$ even,
$w \leftarrow w/2$, else $w \leftarrow (w+q)/2$'', each controlled on the counter, so that
exactly $k$ of them act and the rest are disabled. This leaves $w = x^{-1} \bmod q$.
Both passes are therefore circuits of fixed, input-independent length, built entirely from
halvers, comparators, one conditional subtraction, a single negation, controlled adders, and the flag and counter logic,
which is what made Kaliski's algorithm the basis of the earlier reversible constructions~\cite{RNSL17,10.1007/978-3-030-44223-1_23,GouzienCatQubits23}.
Gouzien et al.\ need neither the flag nor the second pass. They let the idle rounds keep doubling $w$, so that phase~1 always ends with $w\equiv -x^{-1}2^{2n}\pmod q$, and absorb the fixed factor $2^{2n}$ into the Montgomery representation~\cite{GouzienCatQubits23}.
Each of the three directions below pays for, avoids, or reorganizes exactly this inversion, and the two affine ones replace Kaliski's algorithm, by a different binary algorithm and by the quotient-based Euclidean algorithm respectively.

As an example, computing $3^{-1}\bmod 11$ starts at $(u,v,w,s)=(11,3,0,1)$ and runs:
\begin{center}
\begin{tabular}{@{}c|c|c|c|c@{}}
round & branch taken & $u$ & $v$ & $(w,s)$ \\
\hline
1 & $u>v$    & $4$ & $3$ & $(1,2)$ \\
2 & $u$ even & $2$ & $3$ & $(1,4)$ \\
3 & $u$ even & $1$ & $3$ & $(1,8)$ \\
4 & else     & $1$ & $1$ & $(2,9)$ \\
5 & else     & $1$ & $0$ & $(4,11)$
\end{tabular}
\end{center}
Round 5 reaches $v=0$, so phase 1 halts after $k=5$ rounds with $w=4$. Since
$3^{-1}\bmod 11=4$, the natural guess is $w\equiv
x^{-1}2^k=4\cdot2^5\equiv7\pmod{11}$. Instead $w=4$, and $4\equiv-7\pmod{11}$. This sign
flip is not specific to this example: the recurrence as stated always computes
$w\equiv-x^{-1}2^{k}\pmod q$, not $w\equiv x^{-1}2^k\pmod q$, and running the halving
pass below on the unfixed $w$ only carries the wrong sign through to the end rather than
correcting it. The fix is a conditional subtraction of $q$ followed by a single negation, $w\leftarrow q-w$, applied once phase~1 halts and before phase~2 begins.
Here $w=4<11$, so the subtraction does nothing, and $w\leftarrow 11-4=7$, which is now $\equiv x^{-1}2^k\pmod{11}$ as required.
The subtraction is not idle in general. For $x=5$ and $q=7$, phase~1 ends with $w=8\ge q$.  Phase 2 then strips the $k=5$ powers of two by halving
modulo $q$:

\begin{center}
\begin{tabular}{@{}c|c|c|c@{}}
round & branch taken & $w$ \\
\hline
1 & else     & $9$ \\
2 & else     & $10$ \\
3 & $w$ even & $5$ \\
4 & else     & $8$ \\
5 & $w$ even & $4$
\end{tabular}
\end{center}
After exactly $k$ rounds, $w=4=3^{-1}\bmod 11$, as required. A real circuit cannot know
$k$ in advance. It runs both passes for their full length of $2n$ rounds, and the flag and
the counter disable the rounds past the true stopping point, here $2n-k=3$ rounds in each
pass. Without the flag, round~6 would take the branch ``$v$ even'' and double $w$ from
$(4,11)$ to $(8,11)$.

\subsection{Optimize the Gates: Splitting the EEA} 

\begin{figure*}[htbp]
  \centering
  \resizebox{\linewidth}{!}{%
\begin{tikzpicture}[>={Stealth[length=4pt, width=3.5pt]}, font=\small,
  gatebox/.style={draw=ABTealDark, fill=ABTeal!15, rounded corners=3pt,
                  line width=0.6pt, minimum width=1.95cm, minimum height=1.0cm,
                  align=center, font=\scriptsize, inner sep=2pt},
  idlebox/.style={draw=ABGray, fill=white, dashed, rounded corners=3pt,
                  line width=0.6pt, minimum width=1.95cm, minimum height=1.0cm,
                  align=center, font=\scriptsize, inner sep=2pt, text=ABGray},
  gslot/.style={draw=ABBlack, fill=abYellow, rounded corners=3pt,
                line width=0.9pt, minimum width=1.95cm, minimum height=0.7cm,
                align=center, font=\scriptsize},
  gidle/.style={draw=ABGray, fill=white, dashed, rounded corners=3pt,
                line width=0.6pt, minimum width=1.95cm, minimum height=0.7cm,
                align=center, font=\scriptsize, text=ABGray},
  wire/.style={line width=0.9pt, draw=black!70},
  darr/.style={dashed, line width=0.6pt, draw=ABGray},
  ctrldot/.style={circle, fill=ABBlack, inner sep=1.3pt},
  lbl/.style={font=\scriptsize, align=center},
  vlbl/.style={font=\scriptsize, align=center, text=ABGray},
  rowtag/.style={font=\bfseries\small, align=right, text=ABBlack, anchor=east},
]
\def\dx{2.3}
\def\yWtop{6.6} \def\yWbot{5.9} \def\yG{4.1} \def\yRtop{2.4} \def\yRbot{1.7}
\pgfmathsetmacro\xend{8*\dx}
 
\node[rowtag] at (-1.3,{(\yWtop+\yWbot)/2}) {Pass 1};
\node[rowtag] at (-1.3,\yG) {dialog $g$};
\node[rowtag] at (-1.3,{(\yRtop+\yRbot)/2}) {Pass 2};

\foreach \k in {1,...,8}{\node[lbl, text=ABGray] at ({(\k-0.5)*\dx},{\yWtop+0.85}) {round \k};}
 
\foreach \yy/\nm in {\yWtop/u, \yWbot/v, \yRtop/w, \yRbot/s}{
  \draw[wire] (-0.9,\yy) -- (\xend+0.9,\yy);
  \node[lbl, anchor=east] at (-1.0+0.05,\yy) {$\nm$};
}
 
\foreach \k/\val in {0/{(11,3)},1/{(4,3)},2/{(2,3)},3/{(1,3)},4/{(1,1)},5/{(1,0)},6/{(1,0)},7/{(1,0)},8/{(1,0)}}{
  \node[vlbl] at ({\k*\dx},\yWtop+0.32) {$\val$};}
\foreach \k/\val in {0/{(0,1)},1/{(1,2)},2/{(1,4)},3/{(1,8)},4/{(2,9)},5/{(4,11)},6/{(4,11)},7/{(4,11)},8/{(4,11)}}{
  \node[vlbl] at ({\k*\dx},\yRbot-0.38) {$\val$};}
 
\foreach \k/\pa/\gg/\pb in {
  1/{$u{>}v$\\$u\leftarrow(u{-}v)/2$}/{$u{>}v$}/{$w\leftarrow w{+}s$\\$s\leftarrow 2s$},
  2/{$u$ even\\$u\leftarrow u/2$}/{$u$ even}/{$s\leftarrow 2s$},
  3/{$u$ even\\$u\leftarrow u/2$}/{$u$ even}/{$s\leftarrow 2s$},
  4/{else\\$v\leftarrow(v{-}u)/2$}/{else}/{$s\leftarrow s{+}w$\\$w\leftarrow 2w$},
  5/{else\\$v\leftarrow(v{-}u)/2$}/{else}/{$s\leftarrow s{+}w$\\$w\leftarrow 2w$}}{
  \pgfmathsetmacro\xc{(\k-0.5)*\dx}
  \node[gatebox] (A\k) at (\xc,{(\yWtop+\yWbot)/2}) {\pa};
  \node[gslot] (G\k) at (\xc,\yG) {\gg};
  \node[gatebox] (B\k) at (\xc,{(\yRtop+\yRbot)/2}) {\pb};
}

\foreach \k in {6,7,8}{
  \pgfmathsetmacro\xc{(\k-0.5)*\dx}
  \node[idlebox] (A\k) at (\xc,{(\yWtop+\yWbot)/2}) {idle\\(flag set)};
  \node[gidle] (G\k) at (\xc,\yG) {idle};
  \node[idlebox] (B\k) at (\xc,{(\yRtop+\yRbot)/2}) {idle};
}
\foreach \k in {1,...,8}{
  \draw[darr, ->] (A\k.south) -- (G\k.north);
  \draw[darr] (G\k.south) -- (B\k.north);
  \node[ctrldot] at (B\k.north) {};
}

\draw[ABGray, line width=0.8pt, rounded corners=3pt, dashed]
  ({0.5*\dx-1.1},{\yG-0.5}) rectangle ({7.5*\dx+1.1},{\yG+0.5});
 
\begin{scope}[shift={(0,0.35)}, every node/.style={font=\scriptsize, align=left, anchor=north west}]
  \node[font=\scriptsize\bfseries] at (-1.6,0) {Each round\\takes one of\\four branches};
  \foreach \i/\c/\up/\dn in {
    0/{1. $u$ even}/{$u\leftarrow u/2$}/{$s\leftarrow 2s$},
    1/{2. else, $v$ even}/{$v\leftarrow v/2$}/{$w\leftarrow 2w$},
    2/{3. else, $u>v$}/{$u\leftarrow(u{-}v)/2$}/{$w\leftarrow w{+}s,\ s\leftarrow 2s$},
    3/{4. else}/{$v\leftarrow(v{-}u)/2$}/{$s\leftarrow s{+}w,\ w\leftarrow 2w$}}{
    \node at ({0.9+\i*4.3},0) {\textbf{\c}\\pass 1: \up\\pass 2: \dn};
  }
\end{scope}
\end{tikzpicture}%

  }
  \caption{Unrolled trace of the dialog-split extended Euclidean algorithm (EEA), applied to
Kaliski's algorithm on the worked example of Section~\ref{par:kaliski} ($x=3$, $q=11$).
Each round takes one of the four branches listed below the trace. Pass~1 updates $(u,v)$
and records only the branch taken, in the dialog $g$ (dashed frame). Pass~2 replays $g$ on
$(w,s)$ and never revisits $(u,v)$. Both passes run the fixed $2n=8$ rounds of a reversible
circuit. Once $v=0$, after round~5, the termination flag makes the remaining rounds idle.
Pass~2 ends at $(4,11)$, Kaliski's phase-1 output, where the direct computation applies the
negation $w\leftarrow q-w$. That negation and the halving phase are unchanged by the
split and are not shown. The variant of~\cite{Schrottenloher26PointAdd} records different
bits, starts $(r,s)$ at $(1,0)$, and ends at $(0,x^{-1}\bmod q)$ directly.}
  \label{fig:dialog-split-eea}
\end{figure*}
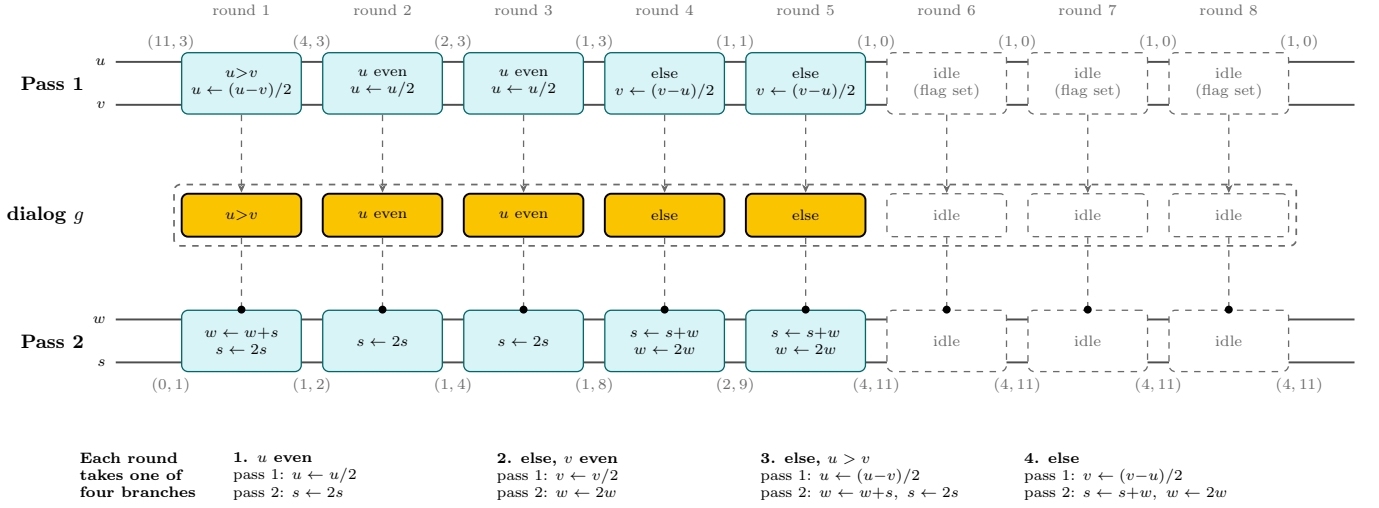

The first affine direction
reorganizes the inversion itself, borrowing an idea from the reversible Reed--Solomon
decoders of~\cite{Khattar25DQI}. Decoding a Reed--Solomon code also reduces to an EEA
computation (finding an error-locator polynomial via a polynomial Bézout identity), and~\cite{Khattar25DQI} needed a fully reversible, low-qubit version of it for an unrelated
quantum algorithm. The ordinary extended Euclidean algorithm carries the Bézout
coefficients alongside the running remainders, and it is keeping those coefficients
coherent that makes it expensive to run reversibly. The trick is to split it into two
cheaper passes that never store the coefficients at all. The \emph{Euclidean pass} runs
the plain, non-extended algorithm on $(u,v)=(q,x)$ and records only which branch it takes
at each iteration, as a string of \emph{garbage} bits $g$. In the variant
of~\cite{Schrottenloher26PointAdd}, each iteration records two bits, $b_0 = v \bmod 2$ and
$b_0\wedge b_1$ with $b_1=[u>v]$. This is a different binary Euclidean algorithm from
Kaliski's, with its own Bézout coefficients, which we denote $(r,s)$. The \emph{reconstruction pass}
then replays those recorded choices to rebuild the inverse. (The record $g$ is
what is called a \emph{dialog}). Three properties make the split
profitable:  
\begin{enumerate}
  \item[(i)] \emph{Space.} The Euclidean pass runs $\approx 1.413n$ iterations and emits
    on average $1.5$ garbage bits per iteration, so $g$ is about $1.413n \times 1.5 =
    2.12n$ bits long after the pass. The exact length of the garbage depends on two factors: 
    the exact number of iterations, and the exact number of garbage bits emitted per iteration. 
    The standard deviation of the iteration count around the average is of order $O(\sqrt n)$, 
    and one must actually run $1.413n + O(\sqrt{n})$ iterations to succeed with constant probability,
    where the constant in the $O$ determines the chance of success. The second factor is removed 
    in~\cite{Schrottenloher26PointAdd} by an \emph{encoding-decoding} method which packs the bits 
    created by three iterations into 5 bits deterministically. The register thus has $2.355n + O(\sqrt{n})$ qubits.

    \item[(ii)] \emph{In-place multiplication for free.} Starting the reconstruction pass
    from $(r,s)=(1,0)$ ends it at $(0,\, x^{-1} \bmod q)$. The Bézout updates are linear
    operations controlled by the recorded bits, so starting from $(r,s) = (y,0)$ instead
    ends it at $(0,\, yx^{-1} \bmod q)$. The inversion and one modular multiplication come
    out of a single circuit~\cite{Khattar25DQI,Schrottenloher26PointAdd}. Reading the
    recorded bits in the opposite order yields $(0,\, yx \bmod q)$ instead, which is how
    the in-place multiplications of the point addition are
    implemented~\cite[Algs.~3--4]{Schrottenloher26PointAdd}.
    
  \item[(iii)] \emph{No expensive reverse pass.} When a reversible binary EEA is run, the 
  entire algorithm (including updates on $(u,v)$ and $(r,s)$) runs once forwards and once 
  backwards. In the new variant, the updates on $(r,s)$ only need to be computed once, in 
  the forward pass. Only the updates on $(u,v)$ (conversion into dialog, and then back to 
  the initial value) are run twice, removing the cost of the extensive reverse pass on $(r,s)$.
\end{enumerate}
The qubit count is maximal during the reconstruction step, since it must hold the dialog record $g$ together
with the two $n$-bit modular integers it combines. With the ideal $g$ this is $2.12n + 2n
= 4.12n$ qubits. With the fixed packed register actually implemented, it is $2.355n + 2n
\approx 4.36n$~\cite{Schrottenloher26PointAdd}. Figure~\ref{fig:dialog-split-eea} illustrates
the same split on Kaliski's algorithm, whose
four-branch rounds can be recorded and replayed in the same way. Its invariant differs from
that of~\cite{Schrottenloher26PointAdd}. Replaying from Kaliski's initial $(w,s)=(0,1)$
reproduces his phase-1 output rather than $(0,x^{-1})$, so the figure shows the split
itself, not item~(ii).

A last factor of improvement on this direction comes from allowing the point addition to
be inexact. All comparisons inside the Euclidean pass and inside the modular adders are
performed on the top $40$ to $50$ bits only. The number of EEA iterations is fixed at
$1.413n + c_{\mathrm{iter}}\sqrt n$ with $c_{\mathrm{iter}} \approx 2.4$ rather than being
data-dependent. The margin $c_{\mathrm{iter}}\sqrt{n}$ is a tail bound, not a tuning
parameter. The number of iterations this algorithm needs on a random input is empirically
normal, with mean $1.413n$ and standard deviation $\approx 0.6\sqrt{n}$. The constant
$c_{\mathrm{iter}}\approx 2.4$ is four such standard deviations, chosen so that an input
needing more than the fixed length is a four-sigma event, about $3\times10^{-5}$ per addition under the normal approximation.
That is small, but of the same order as the measured failure bound below.
A fixed length is needed for the reason met with Kaliski's algorithm above, since a reversible circuit cannot stop on a data-dependent condition, but Kaliski's $2n$ rounds are a worst-case guarantee whereas this length is a tail bound. When $q$ is a pseudo-Mersenne prime $2^{k}-f$ with $f\ll
2^{k}$, as for $\mathtt{secp256k1}$, a controlled subtraction of $q$ further degenerates
into erasing the most significant bit and adding the small constant $f$ into the low
bits. The composite circuit succeeded on all $10\,000$ random inputs tested
in~\cite{Schrottenloher26PointAdd}, which reports this as a failure probability of at most
$2^{-13.3}$, the reciprocal of the number of trials. This is an empirical observation, not
a bound. Zero failures in $10^{4}$ trials give a $95\%$ upper confidence bound of about
$3\times10^{-4}\approx 2^{-11.7}$ per addition (the rule of three). Comparable rates are
reported in~\cite{Babbush26ECC}.
 
It is tempting to record this failure rate as an $\epsilon$, but it is not one. Recall
that Definition~\ref{def:noisy-hsp} splits imperfection along two axes: $\epsilon$ bounds
how far the oracle's values may drift from an exactly periodic function, while $\delta$
bounds the probability that the read-out $R$ fails, split into an abort rate $\delta_\bot$
and a corruption rate $\delta_\times$. Condition~(i), the one that fixes $\epsilon$, is a
deterministic, worst-case statement: $d_S(f(g),f_0(g)) \le \epsilon/2$ for \emph{every} $g
\in G$. An approximate point addition does not drift: on almost every input it returns the
correct point, and on a sparse subset it returns an unrelated one.
Curve points carry no modular distance, and with the discrete metric condition~(i) fails for every $\epsilon<2$, that is, for every non-trivial $\epsilon$.
The failure must therefore be accounted on the other axis, as a corruption $\delta_{\times}$ of the read-out. The
distinction is not cosmetic: an $\epsilon$-drift is a property of the oracle that no
read-out can remove, and masking converts it into aborts and corruptions at rates about $S$ and $\epsilon/S$ (Section~\ref{sec:approx-period}),  
whereas a sparse circuit failure is suppressed by making the circuit better and admits no read-out remedy at all.
With $28$ windowed additions per run~\cite{Babbush26ECC,Schrottenloher26PointAdd}, a union bound on the $95\%$ confidence bound puts the failing inputs at a fraction $p\lesssim 28\cdot 3\times10^{-4}\approx 2^{-6.9}$ of the superposition, per run.
Charging this weight as a corruption rate, $\delta_\times\lesssim p$, is the heuristic of Section~\ref{sec:approx-period}.
The failing branches lower the overlap with the ideal state to no less than $1-p$, so the infidelity is at most $2p\approx 2^{-5.9}$, and the rigorous trace-distance bound is only $\delta_\times\le\sqrt{2p}\approx 0.13$.
The point estimate of $p$ is lower, but it rests on zero observed failures. Please note that register sharing 
is also used in this EEA variant: the bits of $g$ share the same register as the bits of
$(u,v)$ during dialog computation and uncomputation.

\subsection{Shrink the Register: Sharing the EEA Exactly}\label{sec:shrink-register}

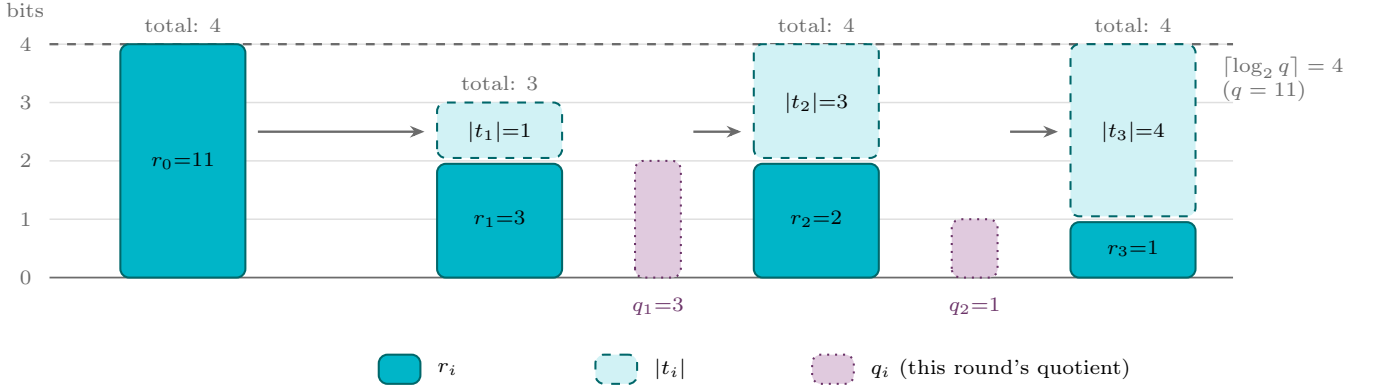
\begin{figure*}[ht]
  \centering
  \resizebox{\linewidth}{!}{%
\begin{tikzpicture}[>={Stealth[length=5pt, width=4pt]}, font=\small,
  rfield/.style={draw=ABTealDark, fill=ABTeal, line width=0.7pt, rounded corners=3pt},
  tfield/.style={draw=ABTealDark, fill=ABTeal!18, line width=0.7pt, dashed, rounded corners=3pt},
  qfield/.style={draw=ABPurpleDark, fill=ABPurple!30, line width=0.7pt, dotted, rounded corners=3pt},
  arr/.style={->, line width=0.8pt, draw=ABGray},
  lbl/.style={font=\scriptsize, align=center},
  axlbl/.style={font=\scriptsize, align=center, text=ABGray},
  inlbl/.style={font=\scriptsize, align=center, text=ABBlack},
  qlbl/.style={font=\scriptsize, align=center, text=ABPurpleDark},
  tot/.style={font=\scriptsize, align=center, text=ABGray},
  rowtag/.style={font=\bfseries\small, align=left, text=ABBlack},
  U/.style={y=0.7cm},
]
 
\def\xA{0} \def\xB{3.8} \def\xC{7.6} \def\xD{11.4}
\def\barw{1.5}
\def\qw{0.55}
\def\gap{0.05}     
\def\yArrow{1.75}
 
\begin{scope}[U]
\foreach \b in {1,2,3,4} {
  \draw[gray!25, line width=0.5pt] (-1.6,\b) -- (12.6,\b);
  \node[axlbl, anchor=east] at (-1.7,\b) {$\b$};
}
\draw[ABGray, line width=0.6pt] (-1.6,0) -- (12.6,0);  
\node[axlbl, anchor=east] at (-1.7,0) {$0$};

\node[axlbl, anchor=south east] at (-1.55,4.3) {bits};
\draw[dashed, ABGray, line width=0.8pt] (-1.6,4) -- (12.6,4);
\node[lbl, anchor=north west, align=left, text=ABGray] at (12.35,3.95) {$\lceil\log_2 q\rceil = 4$\\($q=11$)};
\end{scope}
 
\newcommand{\regbar}[6]{%
  \begin{scope}[U]
    \draw[rfield] (#1-\barw/2,0) rectangle (#1+\barw/2,{#2-\gap*(#3>0)});
    \ifnum#3>0
      \draw[tfield] (#1-\barw/2,{#2+\gap}) rectangle (#1+\barw/2,#2+#3);
      \node[inlbl] at (#1,{#2+#3/2}) {#5};
    \fi
    \node[inlbl] at (#1,{#2/2}) {#4};
    \node[tot, anchor=south] at (#1,{#2+#3+0.05}) {total: $#6$};
  \end{scope}}
 
\regbar{\xA}{4}{0}{$r_0{=}11$}{}{4}
\regbar{\xB}{2}{1}{$r_1{=}3$}{$|t_1|{=}1$}{3}
\regbar{\xC}{2}{2}{$r_2{=}2$}{$|t_2|{=}3$}{4}
\regbar{\xD}{1}{3}{$r_3{=}1$}{$|t_3|{=}4$}{4}
 
\newcommand{\qbar}[3]{%
  \begin{scope}[U]
    \draw[qfield] ({#1-\qw/2},0) rectangle ({#1+\qw/2},#2);
  \end{scope}
  \node[qlbl] at ({#1},-0.35) {#3};}
 
\qbar{(\xB+\xC)/2}{2}{$q_1{=}3$}
\qbar{(\xC+\xD)/2}{1}{$q_2{=}1$}

\draw[arr] (\xA+\barw/2+0.15,\yArrow) -- (\xB-\barw/2-0.15,\yArrow);
\foreach \xl/\xr in {\xB/\xC, \xC/\xD}
  \draw[arr] ({(\xl+\xr)/2+\qw/2+0.15},\yArrow) -- (\xr-\barw/2-0.15,\yArrow);
  
\node[rfield, minimum width=0.5cm, minimum height=0.35cm] (legr) at (2.6,-1.1) {};
\node[lbl, anchor=west] at (legr.east) {\;$r_i$};
\node[tfield, minimum width=0.5cm, minimum height=0.35cm] (legt) at (5.2,-1.1) {};
\node[lbl, anchor=west] at (legt.east) {\;$|t_i|$};
\node[qfield, minimum width=0.5cm, minimum height=0.35cm] (legq) at (7.8,-1.1) {};
\node[lbl, anchor=west] at (legq.east) {\;$q_i$ (this round's quotient)};
 
\end{tikzpicture}%
  }
  \caption{Register sharing in the quotient-based EEA of Section~\ref{sec:shrink-register},
on $q=11$ and $x=3$, with $r_i$, $t_i$ and $q_i$ as defined there. Bar heights are
bit-lengths. As $r_i$ shrinks, $|t_i|$ grows into the qubits it frees. By
invariant~\eqref{eq:pz-invariant}, $r_i|t_i|<q$, so the two bit-lengths sum to at most
$\lceil\log_2 q\rceil+1$. The tag above each column shows that here they never exceed $4$.
Each quotient $q_i=\lfloor r_{i-1}/r_i\rfloor$ is drawn just before the column it
produces. The last one, $q_3=2$, gives $r_4=0$ and ends the algorithm. The figure shows
bit-lengths only, not the layout of the triple $(r_{i-1},t_i,q_i)$ in the register
of~\cite{LuoEtAl26}.}
  \label{fig:shrink-register}
\end{figure*}

The second affine direction
returns to an original observation of Proos and Zalka~\cite{ProosZalka03}. Along the EEA,
the sequence $\{r_i\}$ decreases while the paired B\'ezout coefficient $\{t_i\}$
increases, and they satisfy
\begin{equation}\label{eq:pz-invariant}
  r_{i-1}t_i + r_i t_{i-1} = q.
\end{equation}
The two may therefore share a register.\footnote{Here
  $r_i,t_i,q_i$ follow Luo et al.'s own notation and are unrelated to the curve order $r$
  used throughout Section~\ref{sec:ecdlp} or to the modulus $q$ of Kaliski's algorithm
  above. We have $r_0:=q,\,r_1:=x$, and each step divides $r_{i-1}$ by $r_i$ to get the
  quotient $q_i=\lfloor r_{i-1}/r_i\rfloor$ and the next remainder
  $r_{i+1}=r_{i-1}-q_ir_i$. In parallel, $t_0:=0,\,t_1:=1$ and $t_{i+1}=t_{i-1}+q_it_i$. When $r_k=1$, the inverse is then $(-1)^{k-1}t_k \bmod q$ (a sign correction analogous
  to Kaliski's final negation above).} Luo et al.~\cite{LuoEtAl26} push this to a single
$(n+2)$-qubit register holding the \emph{triple} $(r_{i-1}, t_i, q_i)$. The register also holds the intermediate states that arise while
computing $q_i$ and updating $r_{i-1},t_{i-1}$. A
second $(n+2)$-qubit register holds the complementary pair $(r_i, t_{i-1})$, written
$(r',t')$ below to distinguish it notationally from the first register. Three ingredients make it
explicit and exact: 
\begin{enumerate}
  \item[(i)] \emph{Length registers.} Four auxiliary registers of $\lfloor\log_2
    n\rfloor+1$ qubits each hold the field boundaries $\ell_t,\ell_q,\ell_{r'}$ and the shift counter
    $\ell_s$. The shift counter is not a field length. It is the number of positions by
    which the second register is currently circularly shifted relative to the first, which
    keeps the two registers' fields aligned during arithmetic between them. Together they let one circuit serve inputs whose field lengths differ in
    superposition.
  \item[(ii)] \emph{Location-controlled arithmetic.} Additions, subtractions, swaps and
    length updates act only on the positions selected by those registers. This is realized by
    unary iteration~\cite{Babbush18} over a \emph{step-dependent active window}
    $[k(T),K(T)]$. Here $T$ denotes the step index, a global counter over all steps of the
    algorithm (different inputs $x$ take different numbers of steps through the
    same fixed schedule). The bounds $k(T)$ and $K(T)$ are derived analytically over every
    possible input. They specify which register positions can possibly be touched at step $T$.
    Bounding the total number of steps means bounding how many iterations the classical
    Euclidean algorithm itself can take on an $n$-bit input, a question solved by Lam\'e's
    theorem. The slowest case occurs when every quotient along the way equals $1$, forcing
    the remainders to shrink as slowly as possible. This happens exactly when they form a
    run of consecutive Fibonacci numbers. Consecutive Fibonacci numbers grow by the golden
    ratio $\varphi = \tfrac{1+\sqrt5}{2}$ at each step, so the worst case takes $c =
    1/\log_2\varphi \approx 1.44$ quotient steps per bit. Each iteration $i$
    of the extended Euclidean algorithm runs through four phases, and each phase takes
    one circuit step per bit of the quotient $q_i$. The iteration therefore costs
    $4(\lfloor\log_2 q_i\rfloor+1)$ steps, and the total is
    $N_{\mathrm{step}}=4\sum_i(\lfloor\log_2 q_i\rfloor+1)$~\cite[App.~A.1]{LuoEtAl26}. Luo et al.\ prove
    $4n\le N_{\mathrm{step}}\le 4\lceil cn\rceil$. The lower bound holds because the steps count quotient
    bits, not quotients. An input that finishes in few iterations has large quotients,
    and their bit lengths still add up to about the bit length $n$ of the modulus. The
    upper bound is the Fibonacci worst case above, where every quotient equals $1$. At
    most $4\lceil cn\rceil\approx 5.76n$ steps of $O(n)$ gates are therefore needed.
  \item[(iii)] \emph{Exact allocation.} Where Proos and Zalka used a uniform worst-case
    allocation, incurring $O(\sqrt n)$ overhead and truncating rare outliers at a cost in
    fidelity, this scheme guarantees a deterministic register bound for every input,
    reducing the overhead to $O(\log n)$ while preserving exact correctness.
\end{enumerate}
The inversion then fits in $2n + 6\lfloor\log_2 n\rfloor + 19$ qubits and $217n^2 +
O(n\log_2 n)$ Toffoli gates.\footnote{These figures are from version~2 of~\cite{LuoEtAl26} (arXiv:2604.02311v2,
  9~August 2026). Version~1 (2~April 2026) reported $3n+4\lfloor\log_2 n\rfloor+O(1)$
  qubits and $204n^2\log_2 n+O(n^2)$ Toffoli gates for the same inversion.} 
  We illustrate this technique in Fig.~\ref{fig:shrink-register}.
  
This direction reaches $3n + O(\log n)$ for the whole point addition, rather than the $4.36n$ of the dialog split, by
one further trick worth stating because it is a second and quite different use of
mid-circuit measurement. Besides the two $n$-qubit registers $R_x,R_y$ holding $x,y$, the
in-place division $\ket{x}\ket{y}\ket{0}_A\ket{0}_S \mapsto
\ket{x}\ket{y/x}\ket{0}_A\ket{0}_S$ uses an $n$-qubit workspace register $A$ and a small
$O(\log n)$-qubit register $S$. The forward EEA runs on $R_x$ using $A$ purely as scratch
space, in the same sense as the $(u,v)$ wires of Fig.~\ref{fig:dialog-split-eea}:
overwritten and discarded, never the thing that has to be undone later. Concretely, $A$
returns to $\ket 0$ once $x^{-1}$ has been written into $R_x$. The leftover state that cannot
yet be uncomputed is parked in $S$, playing the role $g$ played there. $R_y$ is untouched
throughout and still holds the original $y$. A separate modular multiplication
then writes $y/x$ into $A$, and only at this point does $A$ become dirty. The resolution
is that the original $y$, still sitting in $R_y$, has become redundant. It is recoverable from $x$ and $y/x$. One therefore measures $R_y$ in the $X$ basis and resets
it. This yields a uniformly random bitstring $b \in \{0,1\}^n$ and leaves only the phase
$(-1)^{b\cdot y}$. $R_y$ is now clean and serves as workspace for the inverse EEA. The inverse EEA uses it
to clear $S$ and recover $x$ in $R_x$. The phase is removed in three steps. First,
$y$ is recomputed into $R_y$ from $x$ and $y/x$. Second, the correction
$Z^{b} = \bigotimes_{k} Z^{b_k}$ is applied, that is, a Pauli $Z$ gate on each qubit $k$
with $b_k=1$. Third, the recomputation is uncomputed. Swapping $R_y$ and $A$
then completes the map~\cite{LuoEtAl26}. It is exact, which is why it contributes no
relaxation to the $(\epsilon,\delta)$ accounting at all.

\subsection{Avoid the Inversion: Projective Coordinates and a One-Bit Output} 

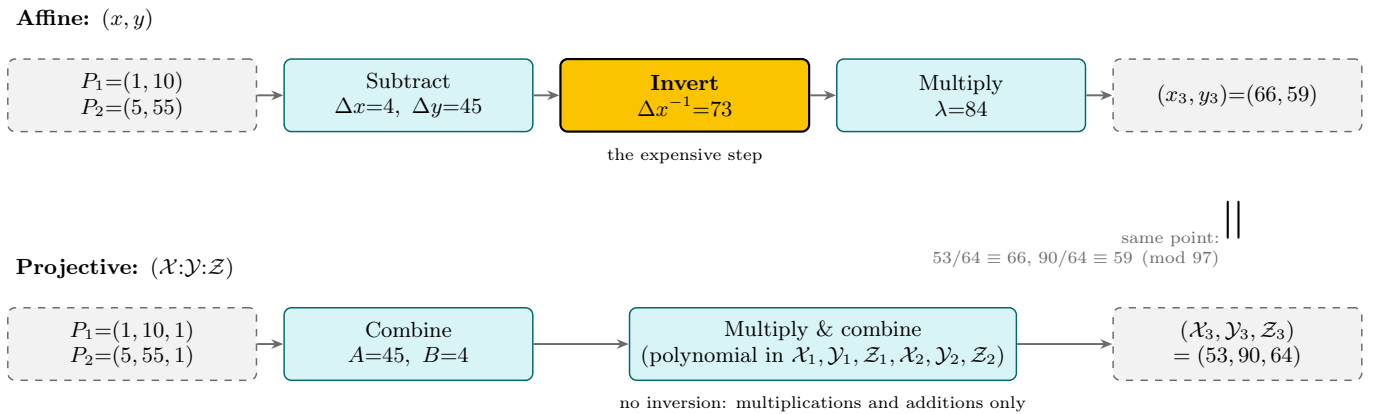
\begin{figure*}[ht]
  \centering
  \resizebox{\linewidth}{!}{%
\begin{tikzpicture}[>={Stealth[length=5pt, width=4pt]}, font=\small,
  box/.style={rounded corners=3pt, minimum width=3.6cm, minimum height=1.05cm,
              align=center, inner sep=3pt},
  databox/.style={box, fill=gray!10, draw=ABGray, dashed, line width=0.6pt},
  opbox/.style={box, fill=ABTeal!15, draw=ABTealDark, line width=0.6pt},
  hlbox/.style={box, fill=abYellow, draw=ABBlack, line width=0.9pt},
  arr/.style={->, line width=0.8pt, draw=ABGray},
  lbl/.style={font=\scriptsize, align=center},
  rowtag/.style={font=\bfseries\small, align=left, text=ABBlack},
  ]
 
\def\yAff{4.2}
\def\yProj{0.6}

\node[rowtag, anchor=west] at (-1.8,\yAff+1.1) {Affine: $(x,y)$};
 
\node[databox] (a_in)  at (0,\yAff)    {$P_1{=}(1,10)$\\$P_2{=}(5,55)$};
\node[opbox]   (a_sub) at (4.0,\yAff)  {Subtract\\$\Delta x{=}4,\ \Delta y{=}45$};
\node[hlbox]   (a_inv) at (8.0,\yAff)  {\textbf{Invert}\\$\Delta x^{-1}{=}73$};
\node[opbox]   (a_mul) at (12.0,\yAff) {Multiply\\$\lambda{=}84$};
\node[databox] (a_out) at (16.0,\yAff) {$(x_3,y_3){=}(66,59)$};
 
\draw[arr] (a_in)  -- (a_sub);
\draw[arr] (a_sub) -- (a_inv);
\draw[arr] (a_inv) -- (a_mul);
\draw[arr] (a_mul) -- (a_out);
 
\node[lbl, text=ABBlack, below=3pt of a_inv] {the expensive step};
 
\node[rowtag, anchor=west] at (-1.8,\yProj+1.1) {Projective: $(\mathcal{X}{:}\mathcal{Y}{:}\mathcal{Z})$};
 
\node[databox] (p_in)  at (0,\yProj)    {$P_1{=}(1,10,1)$\\$P_2{=}(5,55,1)$};
\node[opbox]   (p_ab)  at (4.0,\yProj)  {Combine\\$A{=}45,\ B{=}4$};
\node[opbox, minimum width=5.6cm] (p_mul) at (10.0,\yProj)
  {Multiply \& combine\\(polynomial in $\mathcal{X}_1,\mathcal{Y}_1,\mathcal{Z}_1,\mathcal{X}_2,\mathcal{Y}_2,\mathcal{Z}_2$)};
\node[databox] (p_out) at (16.0,\yProj) {$(\mathcal{X}_3,\mathcal{Y}_3,\mathcal{Z}_3)$\\$=(53,90,64)$};
 
\draw[arr] (p_in)  -- (p_ab);
\draw[arr] (p_ab)  -- (p_mul);
\draw[arr] (p_mul) -- (p_out);
 
\node[lbl, text=ABBlack, below=3pt of p_mul] {no inversion: multiplications and additions only};
 
\node[rotate=90, font=\Huge, text=ABBlack] (same) at ($(a_out.south)!0.5!(p_out.north)$) {$=$};
\node[lbl, text=ABGray, left=4pt of same, align=right]
  {same point:\\$53/64\equiv 66$,\ $90/64\equiv 59 \pmod{97}$};
 
\end{tikzpicture}%
  }
  \caption{The same point addition in affine and in projective coordinates, on the curve
    $y^2=x^3-3x+5$ over $\mathbb{F}_{97}$, with $P_1=(1,10)$ and $P_2=(5,55)$. The affine
    law needs the inverse of $\Delta x$. The projective law needs only multiplications and
    additions, and its output $(\mathcal{X}_3:\mathcal{Y}_3:\mathcal{Z}_3)$ decodes to the
    same affine point $(66,59)$. The projective row shows the structure of the computation.
    The construction of~\cite{CFS26ECDLP} uses the complete addition formulas
    of~\cite{RenesCostelloBatina16,Pornin24}.}
  \label{fig:avoid-inversion}
\end{figure*}

The third
direction is the construction of Chevignard, Fouque and
Schrottenloher~\cite{CFS26ECDLP}. It is the only one that leaves affine coordinates behind. In doing so, it is also the only
one that reuses the factoring optimizations. It fuses
the output-register compression of Section~\ref{sec:output-register} with the residue
arithmetic of Section~\ref{sec:oracle-rns}. Two observations drive it. First, the
inversion exists only because affine coordinates represent a point by $(x,y) = (\mathcal{X}/\mathcal{Z}, \mathcal{Y}/\mathcal{Z})$
for some $\mathcal{Z}\ne0$. Reconstructing $x,y$ from $\mathcal{X,Y,Z}$ is exactly where the division
above came from. \emph{Projective} coordinates instead keep the triple $(\mathcal{X,Y,Z})$ itself,
without ever dividing. The chord-and-tangent addition law then rewrites into polynomial
formulas in $\mathcal{X,Y,Z}$ that use only multiplications and additions in
$\mathbb{F}_q$~\cite{RenesCostelloBatina16,Pornin24}. The point can thus be advanced with
modular multiplications alone. Those multiplications are then accumulated in a residue
number system, exactly as the unreduced product $\prod_k M_k^{e_k}$ was in
Section~\ref{sec:oracle-rns}, so every intermediate value occupies only $O(\log\log q)$
qubits. Second, reading the answer out would normally mean converting the projective point
back to affine form, which reintroduces a division (recovering $x = \mathcal{X}/\mathcal{Z}$). This is
sidestepped by never forming the point at all, and instead outputting a \emph{single bit},
the Legendre symbol of the affine $x$-coordinate. It needs no inversion. With
$x=\mathcal{X}/\mathcal{Z}$, multiplicativity gives
$\bigl(\tfrac{\mathcal{X}\mathcal{Z}}{q}\bigr)=\bigl(\tfrac{x\mathcal{Z}^{2}}{q}\bigr)
=\bigl(\tfrac{x}{q}\bigr)$, so the bit is computed from $\mathcal{X}\mathcal{Z}\bmod q$ and
is the same for every projective representative of the point~\cite{CFS26ECDLP}. The difference between the affine and projective
coordinate computations for point addition is depicted in
Fig.~\ref{fig:avoid-inversion}.

Compressing the output to one bit is exactly the May--Schlieper move of
Section~\ref{sec:output-register}, and it is justified here by three facts, which we take
in turn. First, May--Schlieper~\cite{MaySchlieper22} guarantees that period finding
survives compression to $t$ bits at the cost of a factor $1/(1-2^{-t})$ in the number of
runs, so a single output bit ($t=1$) merely doubles the run count. This doubling applies
to the run count Ekerå's general-logarithm tradeoff~\cite{Ekera19GeneralDLP} already needs
for its own lattice-based post-processing, not to a single-shot baseline. For $n=256$,
that tradeoff already targets $11$ measurements to reach $99\%$ post-processing success,
and doubling gives exactly the $22$ runs quoted below, matching the pattern at every curve
size~\cite{CFS26ECDLP}. Second, that guarantee was stated for a hash drawn at random from a universal family,
which the Legendre symbol is not. Chevignard et al.\ randomize it by adding a random
multiple $uP$ to the point before taking the symbol, and treat the resulting family as
close enough to universal under an explicit heuristic~\cite[Heuristic~1]{CFS26ECDLP}.
Results on the balance of character sums over multiples of a point support that
heuristic~\cite{JaoJetchevVenkatesan07,FarashahiShparlinski12}, but do not prove it. The
hashing step is therefore heuristic, unlike the May--Schlieper theorem it borrows. Third, the Legendre symbol
must itself be computed reversibly. The Legendre symbol is the special case of the
\emph{Jacobi} symbol where the lower argument is prime, so a binary Jacobi-symbol
algorithm~\cite{ShallitSorenson93} computes it directly, with no adaptation needed, and
its structure and garbage accounting mirror the Euclidean passes above. Exceptional cases
in the projective addition law are handled with complete
formulas~\cite{RenesCostelloBatina16,Pornin24}. The residue reconstruction here is not the truncated one of
Section~\ref{sec:oracle-rns}. A Legendre symbol needs the exact residue
$\mathcal{X}\mathcal{Z}\bmod q$, not its top bits, and an arbitrarily small error can flip
it. The reconstruction therefore uses explicit Chinese remaindering, with its precision set
so that it is exact except with negligible probability over the
input~\cite[Lemma~2]{CFS26ECDLP}. The Jacobi-symbol circuit's iteration count is similarly
fixed so that it succeeds with probability at least
$0.99$~\cite[Heuristics~3--4, Lemma~4]{CFS26ECDLP}. By the criterion applied to the
approximate point addition above, a computation that is exact except on a sparse set of
inputs has $\epsilon=0$, and its failures count as corruption $\delta_\times$. Masking has
no role here. The output is already a single bit, and there is no value drift for it to
absorb.

The resulting algorithm uses $3.12n + o(n)$ qubits at the price of raising the gate count
from $O(n^3)$ to $\widetilde{O}(n^4)$. For $n = 256$ the estimate is $1193$ logical
qubits\footnote{We quote the full version of~\cite{CFS26ECDLP} (IACR ePrint 2026/280), which
accompanies the EUROCRYPT 2026 paper. Its abstract and Table~8 agree on $1193$ qubits,
$22$ independent runs, and $2^{38.98}$ Toffoli gates for $n=256$.}
and $22$ independent runs of $2^{38.98}$ Toffoli gates each~\cite{CFS26ECDLP}.

\subsection{Trade-offs for ECDLP} 

\begin{table*}[ht]
\centering
\renewcommand{\arraystretch}{1.3}
\setlength{\tabcolsep}{6pt}
\begin{tabular}{@{}p{4.1cm} c c c@{}}
\toprule
\textbf{Work} & \textbf{Qubits ($n$)} & \textbf{Qubits ($256$)} & \textbf{Toffoli} \\
\midrule
Proos--Zalka~\cite{ProosZalka03}        & $5n + o(n)$ & --- & --- \\
Roetteler et al.~\cite{RNSL17}          & $9n + 2\lceil\log_2 n\rceil + 10$ & $2330$ & $\approx 2^{36.9}$ \\
Häner et al.~\cite{10.1007/978-3-030-44223-1_23} & $8n + 10.2\log_2 n + O(1)$ & $2124$ & $\approx 2^{30}$ \\
Litinski~\cite{Litinski23}              & --- & $\approx 2350$ $^{a}$ & $2^{27.57}$ \\ 
Babbush et al.~\cite{Babbush26ECC}      & --- & $1191$ / $1441$ & $2^{26.27}$ /
                                                                  $2^{25.94}$ \\
Schrottenloher~\cite{Schrottenloher26PointAdd} & $4.36n + O(\sqrt n)$ & $1208$ / $1462$ & $2^{26.11}$ / $2^{25.78}$ \\
\addlinespace
Chevignard et al.~\cite{CFS26ECDLP}     & $3.12n + o(n)$ & $1193$ & $2^{38.98}$ \\
\addlinespace
Luo et al.~\cite{LuoEtAl26}             & $3n + 6\lfloor\log_2 n\rfloor + O(1)$ & $835$ & $2^{30.88}$ \\
\bottomrule
\end{tabular}
\caption{Logical-level cost of solving $\mathtt{ECDLP}$ for $n=256$ (curve
  $\mathtt{secp256k1}$ for~\cite{Babbush26ECC,Schrottenloher26PointAdd,LuoEtAl26}, P-256
  otherwise). Qubit and Toffoli counts are for the whole algorithm, per run, including the
  windowed point additions. For~\cite{Babbush26ECC,Schrottenloher26PointAdd} they are taken
  from~\cite[Table~2]{Schrottenloher26PointAdd}. The asymptotic formulas may count a single
  point addition only. Two-figure entries are the space- and gate-optimized variants of the
  same construction. The last two rows are the \emph{avoid the inversion} and \emph{shrink
  the register} directions of Section~\ref{sec:ecdlp}. A point addition is called
  \emph{exact} here if it uses no approximate comparisons and no truncated iteration count,
  so that it is correct on every input apart from the exceptional cases of the addition
  law, which the affine constructions neglect and the complete formulas of~\cite{CFS26ECDLP} handle. The point additions
  of~\cite{RNSL17,10.1007/978-3-030-44223-1_23,LuoEtAl26} are exact. Those
  of~\cite{Babbush26ECC,Schrottenloher26PointAdd} truncate the Euclidean iteration count
  and fail on a small, empirically measured fraction of inputs. Run counts, circuit depth,
  and physical-level estimates are omitted (discussed above). $^{a}$Litinski's paper does not report 
  a comparable logical-qubit count, since it uses an
  active-volume cost model. The figure $\sim\!2350$ is Schrottenloher's own comparison,
  ``about twice as much qubits'' as the space-optimized figures of Babbush et al.\ and of
  his own work~\cite{Schrottenloher26PointAdd}, not a number from Litinski~\cite{Litinski23}
  itself.}
\label{tab:ecdlp-costs}
\end{table*}

No single algorithm dominates for ECDLP. Space, gate count and depth are pushed by three different 
lines of work, the depth axis being held by the low-depth point-addition circuits of Kim et al.~\cite{KimJang26}. 
More importantly, the strategy that dominates for RSA does \emph{not} survive here. The
compressed construction of~\cite{CFS26ECDLP} was, on publication, the most space-efficient
polynomial-time $\mathtt{ECDLP}$ algorithm known. It has since been dominated on
\emph{both} axes at once by the register-sharing construction, using $835$ rather than
$1193$ logical qubits \emph{and} $2^{30.88}$ rather than $2^{38.98}$ Toffoli
gates~\cite{LuoEtAl26} (Table~\ref{tab:ecdlp-costs}). Asymptotically the same holds: $3n +
O(\log n)$ qubits with $1008\, n^3/\log_2 n + O(n^2)$ gates against $3.12n + o(n)$ with
$\widetilde{O}(n^4)$. 

The reason is visible in the preceding paragraphs. In factoring, residue arithmetic
eliminates the register that dominated the count, the $n$-bit exponentiation accumulator,
and nothing cheaper was available. In $\mathtt{ECDLP}$ the dominant register is the EEA
workspace of the inversion, and \emph{two} reorganizations of the EEA remove it while
staying inside $O(n^3)$. Compression, by contrast, buys the same space by a much more
expensive route, and additionally forfeits the free input register of
Remark~\ref{rem:recycling-tension}. The Pareto frontier now has five points, all on the affine branch. Ordered by qubit
count, they are~\cite{LuoEtAl26} ($835$ qubits, $2^{30.88}$ Toffoli gates), then the
space- and gate-optimized variants of~\cite{Babbush26ECC} and
of~\cite{Schrottenloher26PointAdd}, which trade qubits for gates down to $1462$ qubits and
$2^{25.78}$ Toffoli gates. This comparison mixes curves. The figures
of~\cite{Babbush26ECC,Schrottenloher26PointAdd} are for $\mathtt{secp256k1}$, whose
special prime makes the arithmetic cheaper. For a generic prime, Schrottenloher's variants
cost $2^{26.66}$ and $2^{26.33}$ Toffoli gates instead~\cite[Table~2]{Schrottenloher26PointAdd}. For context, the earlier affine estimates that
these improve upon are Proos--Zalka's $5n + o(n)$ qubits~\cite{ProosZalka03}, and the
point-addition figures of Roetteler et al.~\cite{RNSL17} and Häner et
al.~\cite{10.1007/978-3-030-44223-1_23} (see Table~\ref{tab:ecdlp-costs}).

Two things are worth flagging about Table~\ref{tab:ecdlp-costs}. Its concrete qubit counts
are each taken from the source's own detailed per-curve estimate, never obtained by
substituting $n=256$ into the asymptotic formula quoted in the same row. For Chevignard et
al., that substitution would give $3.12\times256\approx799$, well short of their reported
$1193$, because the suppressed $o(n)$ term is not yet small at this size. Litinski's own
headline figure of $50$ million Toffoli gates assumes an active-volume architecture with batching
across many keys. The table instead reports $2^{27.57}$, the plain single-run Toffoli
count of the same underlying circuit as given in~\cite{Schrottenloher26PointAdd}, for
comparability with the other rows. Circuit depth is a further axis the table omits. See
Kim et al.~\cite{KimJang26} for the lowest-depth point-addition circuits, and~\cite{GouzienCatQubits23,Babbush26ECC} for physical-level cat-qubit and surface-code
estimates. These last are architecture-specific and not comparable across rows. 



    \clearpage
\section{Logical Implementation of Shor's Algorithm}\label{sec:logical_impl}

This appendix turns the algorithms of Appendix~\ref{sec:algo_overview} into explicit logical circuits. 
Each algorithm is first written as classical pseudo-code, then compiled into reversible arithmetic built from a few primitives, namely adders, comparators, table lookups and their measurement-based uncomputation. We focus on the part that dominates the cost of each attack. 
For factoring and FFDLP, this is the approximate modular exponentiation of Section~\ref{sec:full-modern}, detailed in Section~\ref{ssec:approx-modexp}. 
For ECDLP, it is the elliptic curve point addition and, within it, the modular inversion, detailed in Section~\ref{ssec:ecdlp-logical}. 
Every subroutine that produces intermediate values needs a matching uncomputation, so that its work registers return to $\ket{0}$ before the final measurement. 
We therefore describe each subroutine together with its adjoint, when they are not  the direct inverse function. 
The semiclassical quantum Fourier transform is implemented as in~\cite[App.~A.5 and Fig.~15]{gidney2025factor2048bitrsa}. 
We do not detail it further, since its cost is negligible next to that of the oracle.

\paragraph{Software Implementations in Qualtran and Qarton.}
All exponentiation and compressed exponentiation circuits considered in this paper, whether in the RSA, FFDLP or ECDLP cases, are classical arithmetic circuits which, on a computational basis state, produce a computational basis state. Furthermore they can be entirely decomposed into purely classical operations ($X$, $\CX$, $\CCX$) as well as phase operations ($Z$, $\CZ$, $\CCZ$) and classically controlled operations following from measurement-based uncomputation. Additionally, the analysis can be reduced to sub-circuits with modest sizes, either a single residue computation for RSA and FFDLP, or a single elliptic curve point addition for ECDLP. It then becomes natural to entirely implement these circuits (i.e., construct them explicitly) and to assert their correctness by simulating them on classical inputs. The combination of classical operations and phase operations, with classical controls, produces the so-called \emph{kickmix} circuit class~\cite{Babbush26ECC}, which can be efficiently simulated.

For RSA and FFDLP, Chevignard et al.~\cite{C:CheFouSch25} provided partial implementations in Qiskit~\cite{CFS24code}. For ECDLP, Schrottenloher~\cite{Schrottenloher26PointAdd} provided an implementation in Qarton. For their space-efficient version, Luo et al.~\cite{LuoEtAl26} provided a Qiskit implementation. The new implementations we provide use Qualtran~\cite{harrigan2024expressing} and Qarton~\cite{cryptoeprint:2026/2191,qarton}. Both libraries offer a high-level programming interface that abstracts the exact memory layout of a quantum circuit away from the programmer. In Qarton, the circuit is then constructed explicitly. Qualtran can likewise compile a program down to an explicit circuit, but its resource estimates are typically computed hierarchically from the program’s call graph without doing so. The resulting qubit and gate count estimates are precise, but hold only in the logical circuit model, without taking into account routing, distance between logical qubits, etc. More precisely, we give:
\begin{itemize}
    \item A Qarton implementation of the RSA and FFDLP compressed-exponentiation routine of~\cite{C:CheFouSch25}, with the improvements by Gidney~\cite{gidney2025factor2048bitrsa}. The exact gate and qubit counts obtained are similar, but differ slightly due to low-level optimization choices.
    \item A Qualtran implementation of the RSA and FFDLP of the modern algorithm as described in~\ref{sec:full-modern}.
\end{itemize}
and use existing implementation in Qarton for the ECDLP case.

\subsection{Approximate Modular Exponentiation for Factoring}\label{ssec:approx-modexp}

\begin{table*}[ht]
    \centering
        \begin{tabular}{@{}lll@{\hspace{2em}}lll@{}}
        \toprule
        Symbol & Definition & Role & Symbol & Definition & Role \\
        \midrule
        $N$ & integer to factor & classical & $g_{p_i}$ & generator of $\mathbb{Z}_{p_i}^*$ & classical \\
        $g$ & base of the exponentiation & classical & $M_k$ & $g^{2^k} \bmod N$ & classical \\
        $e$ & exponent, in uniform superposition & quantum & $D_{p_i,k}$ & discrete log of $M_k \bmod p_i$ & classical \\
        $V$ & exponentiation modulo $N$ & quantum & $r_i$ & residue for the prime $p_i$ & quantum \\
        $n_e$ & size of the exponent register & classical & $S_i$ & logarithm of $r_i$ & quantum \\
        $p_i$ & $i$-th prime of the RNS & classical & $C_{j,k}$ & truncated CRT factors & classical \\
        $|P|$ & number of primes in the RNS & classical & $\text{Acc}_i$ & partial sum over the first $i$ primes & quantum \\
        $\ell$ & bit length of an RNS prime & parameter & $w_1$ & window, discrete-log computation & parameter \\
        $\kappa$ & size of the accumulation register & parameter & $w_{3a}$ & window, exponentiation outer loop & parameter \\
        $\bar\kappa$ & bits removed by the truncation & parameter & $w_{3b}$ & window, exponentiation inner loop & parameter \\
        & & & $w_4$ & window, accumulation & parameter \\
        \bottomrule
    \end{tabular}
    \caption{All parameters used in the modern Shor algorithm presented in Section \ref{sec:full-modern}.}
    \label{tab:parameter_definition}
\end{table*}

This section provides a detailed low-level description of the implementation of Shor's oracle function for the factoring problem.
We follow the implementation presented in Section~\ref{sec:full-modern} which is described here in detail for completeness.
More precisely, this section focuses on the oracle that computes the approximate modular exponentiation whose high-level circuit can be seen in Figure~\ref{fig:residue-formation}.
For the sake of clarity, Table~\ref{tab:parameter_definition} recalls all the parameters used in the modern Shor algorithm whose notation we rely on, throughout this appendix. 
In order to properly implement this algorithm at the logical level, we decompose the oracle function, namely the computation of the approximate modular exponentiation, into four subroutines. 
Each of these subroutines is then decomposed into arithmetic blocks (adders, lookup tables, etc.), from which we can compute the cost of the overall computation. 
An overview of the circuit, decomposed into 4 main subroutines, is presented in Figure~\ref{fig:full_shor_circuit}.
In the next paragraphs, we detail the subroutines:
\begin{itemize}
    \item \textbf{\ComputeExponent}: computes $S_i$, the logarithm of the residue $r_i$. The
    subroutine combines the uncomputation of $S_{i-1}$ with the computation of $S_i$. The
    last call to the subroutine (outside of the loop) is used to uncompute the last
    exponent, returning the register to $\ket{0}$.
    \item \textbf{\ReduceExponent}: computes the modular reduction for the exponent output of
    \ComputeExponent. It also produces the quotient register ($\log(n_e)$ qubits,
    Section~\ref{par:quotient-register}), which $\ReduceExponent^\dagger$ needs to
    undo the reduction.
    \item \textbf{\ComputeResidue}: computes the residue with a small modular exponentiation.
    \item \textbf{\AccumulateResidue}: adds the residue to the accumulation register to
    build the approximate modular exponentiation $\tilde V$. All necessary phase
    corrections are included in the subroutine.
\end{itemize}
Algorithm~\ref{alg:approx-modexp} lists them in the order in which the oracle calls them,
with $D_{p_{-1},k}=0$ for the first prime. Two subroutines are not unitary on their own, so their adjoints need their own definition. 
\ComputeResidue{} measures some of its work qubits in the $X$ basis and defers the
resulting phase corrections. Its adjoint $\ComputeResidue^\dagger$ is therefore a
separate circuit, a left inverse that satisfies $\ComputeResidue^\dagger \circ \ComputeResidue{} = \mathrm{id}$. 
\ReduceExponent{} maps $S_i$ to $S_i \bmod (p_i-1)$, which on its own loses information, since many values of $S_i$ share the same remainder. 
The circuit makes the reduction reversible by also producing the quotient register (Section~\ref{par:quotient-register}). 
$\ReduceExponent^\dagger$ is therefore not the adjoint of the reduction alone. 
It inverts the whole circuit, and it needs the quotient register, kept until then, to restore $S_i$.  

\begin{figure}[htbp]
\begin{algorithm}[H]
\caption{Approximate modular exponentiation oracle.}\label{alg:approx-modexp}
\begin{algorithmic}[1]
\Statex \textbf{Input:} exponent register $\ket{e}$, accumulator $A$ on $\kappa+1$ qubits,
  prepared in the mask state $\ket{\sigma}$
\Statex \textbf{Output:} $\ket{e}$, and $A$ shifted by
  $\tilde V(e)\approx (g^{e}\bmod N)\gg\bar\kappa$
\State $S \gets 0$ \Comment{$(\ell+\log n_e)$-qubit register}
\For{$j = 0$ to $|P|-1$}
  \State $S \gets S + \sum_k e_k\,(D_{p_j,k}-D_{p_{j-1},k})$
    \Comment{\ComputeExponent, $S_{j-1}\mapsto S_j$}
  \State $(S, a) \gets \bigl(S \bmod (p_j-1),\ \lfloor S/(p_j-1)\rfloor\bigr)$
    \Comment{\ReduceExponent}
  \State $r \gets g_{p_j}^{S} \bmod p_j$ \Comment{\ComputeResidue, $\ell$-qubit register}
  \State $A \gets A + \sum_k r_k\, C_{j,k} \bmod (N\gg\bar\kappa)$
    \Comment{\AccumulateResidue}
  \State uncompute $r$ \Comment{\ComputeResidue$^\dagger$}
  \State $S \gets S + a\,(p_j-1)$, uncompute $a$ \Comment{\ReduceExponent$^\dagger$}
\EndFor
\State $S \gets S - \sum_k e_k D_{p_{|P|-1},k}$ \Comment{\ComputeExponent, $S\gets 0$}
\State apply the deferred phase corrections on $\ket{e}$ \Comment{\Phaseup}
\end{algorithmic}
\end{algorithm}
\end{figure}

Each subroutine is decomposed into arithmetic operations, which serve as the main building blocks for the entire circuit.
There are three main operations:
\begin{itemize}
    \item \textbf{\Adder}: addition, subtraction, and comparison operations rely on circuits that are very similar in structure and cost. When two quantum registers are added together, we denote it a Q-Q \Adder{} (quantum-quantum adder), and when one of the two operands is classical, a Q-C \Adder{} (quantum-classical adder). A similar notation is used for comparison operations, namely Q-Q and Q-C \Comparison. A subtraction is the adjoint of an addition and has the same cost. Every addition is performed modulo $2^n$, where $n$ is the size of the input number.
    \item \textbf{\Lookup}: given an \emph{address register}, loads into a \emph{target register} a value precomputed classically and associated to each address value. Its cost is directly correlated to the two input parameters. The address size determines the Toffoli cost and impacts the qubit cost. The target register size only impacts the qubit cost.
    \item \textbf{\Phaseup}: applies to an address register a phase given by a classical table. Combined with measurement-based uncomputation~\cite{Berry_2019}, it uncomputes \Lookup{} operations efficiently.\footnote{Uncomputation of multiple \Lookup{}s that share the same address register can be merged into a single \Phaseup{} as long as the address register is only used as control qubits before the \Phaseup.}
\end{itemize}
Table~\ref{tab:operation_count_recap} gives the operation counts of every subroutine, and
Table~\ref{tab:logical_qubit_count} its qubit counts.

\begin{figure*}[tp]
    \centering
    \includegraphics[width=1\linewidth]{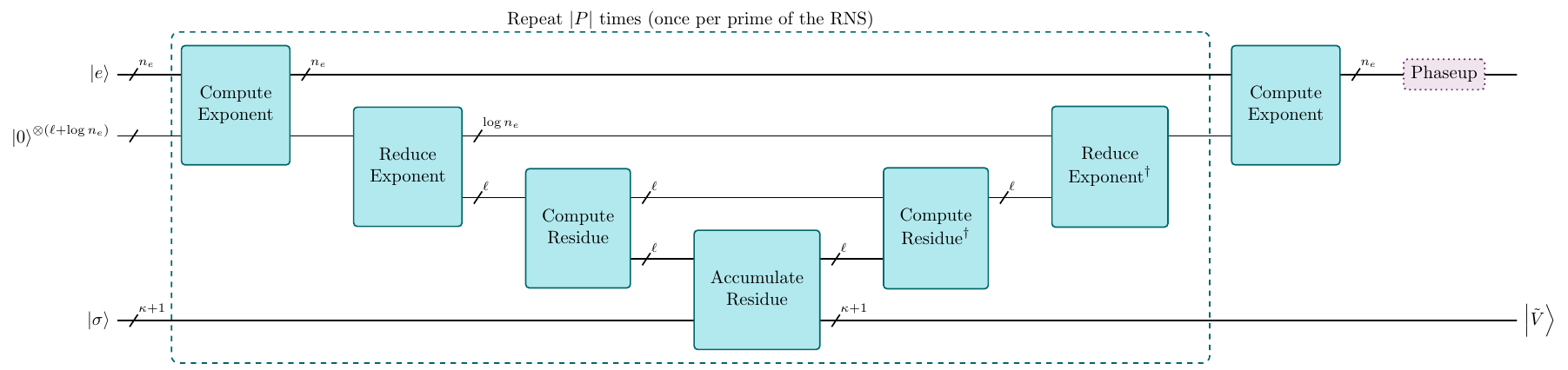}
    \caption{Overview of the oracle of Shor's algorithm at the logical level. It shifts the
    accumulator, prepared in the mask state $\ket{\sigma}$, by
    $\tilde V(e)=\sum_{j}\sum_{k} r_{j,k}C_{j,k} \bmod (N\gg\bar\kappa)$, the top $\kappa$
    bits of $g^{e}\bmod N$ up to the truncation error
    (Equation~\eqref{eq:approximate_modular_exponentiation}). In iteration $j$, for the prime
    $p_j$, \ComputeExponent{} forms $S_j=\sum_k e_kD_{p_j,k}$, \ReduceExponent{} replaces it by
    $S_j\bmod(p_j-1)$ and a quotient register, \ComputeResidue{} forms
    $r_j=g_{p_j}^{S_j\bmod(p_j-1)}\bmod p_j$, and \AccumulateResidue{} adds
    $\sum_k r_{j,k}C_{j,k}$. The two adjoints then erase $r_j$ and restore $S_j$. After the
    loop, a last \ComputeExponent{} erases $S_{|P|-1}$, and a \Phaseup{} applies the deferred phase
    corrections on $\ket{e}$. The subroutines are detailed in Section~\ref{ssec:approx-modexp}
    and listed in order in Algorithm~\ref{alg:approx-modexp}. 
    This is the analog of Figure~\ref{fig:oracle-gidney}, restricted to the oracle.}
    \label{fig:full_shor_circuit}
\end{figure*}

We now detail each of the subroutines, starting with the computation of the logarithm of the residue.

\subsubsection{Computing the logarithm of the residue}

The goal of \ComputeExponent{} is, given the classically known discrete logarithms $D_{p_i,k}$, and the exponent $e$, to compute $S_i$ which is defined as
\begin{equation}
    S_i = \sum_{k=0}^{n_e-1} D_{p_i,k} e_k,
\end{equation}

Since,
\begin{equation}
    r_i = g_{p_i}^{\sum_{k=0}^{n_e-1}D_{p_i,k} e_k} \mod p_i = g_{p_i}^{S_i} \mod p_i,
\end{equation}
each $S_i$ is the exponent used to compute the residue $r_i$.
Following the implementation of~\cite{gidney2025factor2048bitrsa}, we apply three optimizations for the computation of $S_i$.

First, as explained in Appendix~\ref{sec:algo_overview}, windowing is used to reduce the number of iterations on the oracle's exponent $e$ from $n_e$ to $\lceil \frac{n_e}{w_1}\rceil$.
Each window of $w_1$ qubits is the address register of a \Lookup{} and adds a $2^{w_1}-w_1-1$ Toffoli gate cost.\\
Second, the computation of $S_i$ can be merged with the uncomputation of $S_{i-1}$.
Indeed, since the addition that computes $S_i$ and the subtraction that uncomputes $S_{i-1}$ depend on the same address register and windowed exponent $e_{k:k+w_1}$, the two operations can be merged into the same loading operation. 
This means $S_i$ can be expressed as follows,
\begin{equation}
    S_i = S_{i-1} + \sum_{j=0}^{\lceil\frac{n_e}{w_1}\rceil-1} \text{T}_1^{(i,j)}[e_{jw_1:(j+1)w_1}],
\end{equation}
where
\begin{equation}
\text{T}_1^{(i,j)}[e_{jw_1:(j+1)w_1}] \coloneqq \sum_{k=0}^{w_1-1} (D_{p_i,jw_1+k} - D_{p_{i-1,jw_1+k}}) e_{jw_1+k}
\end{equation}
is the value loaded by the \Lookup{} operation.

\begin{figure}[htbp]
    \centering
    \includegraphics[width=\columnwidth]{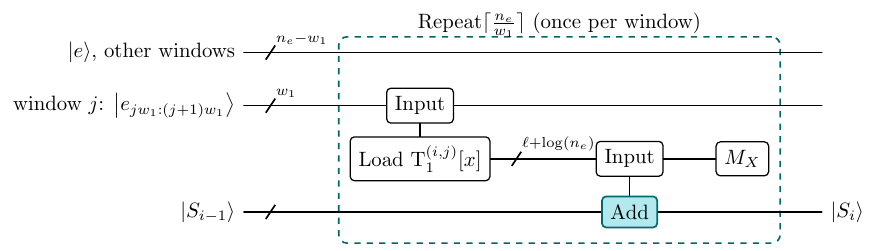}
    \caption{\ComputeExponent: one pass of the loop that turns $S_{i-1}$ into $S_i$, where
    $S_i=\sum_{k} e_k D_{p_i,k}$ is held on $\ell+\log n_e$ qubits and $D_{p_i,k}$ is the
    classically precomputed discrete logarithm of $M_k=g^{2^k}\bmod N$ modulo $p_i$
    (Table~\ref{tab:parameter_definition}). The loop runs over the $\lceil n_e/w_1\rceil$
    windows of the exponent. In iteration $j$, window $j$ is the address $x$ of a \Lookup,
    which loads
    $\text{T}_1^{(i,j)}[x]=\sum_{k=0}^{w_1-1} x_k\,\bigl(D_{p_i,jw_1+k}-D_{p_{i-1},jw_1+k}\bigr)$.
    The loaded value is added into $S$ and then measured, and the other windows are untouched.
    Adding these differences uncomputes $S_{i-1}$ and computes $S_i$ in the same pass.}
    \label{fig:offsetdlog}
\end{figure}

The third optimization is to delay and merge \Phaseup{}s associated to measurement-based uncomputation of the \Lookup{}s.
Once the value loaded into a register by the \Lookup{} operation has been used, this register is measured in the $X$ basis.
Usually, a \Phaseup{} operation would need to be applied immediately to apply phase corrections to the address register.
However, in this context, the address register that contains the exponent is only used as control throughout the entire computation. 
Thus every phase correction on the exponent register can be pushed to the end of the computation and combined into a single \Phaseup{} operation for each window of $w_1$ qubits.
The overall circuit for \ComputeExponent{} with the three optimizations is depicted in Figure~\ref{fig:offsetdlog}. Its total cost is given in Table~\ref{tab:computing_dlog_operation_count}.

\begin{table}[htbp]
    \centering
    \begin{tabular}{|c|c|c|}
        \hline
        Calls & Operation & Operation size \\
        \hline
        \multirow{2}{*}{$(|P|+1)\lceil\frac{n_e}{w_1}\rceil$} & Q-Q \Adder{} & $\ell + \log(n_e)$ \\
        \cline{2-3} 
            & \Lookup{} & address $= w_1$, target $=\ell+\log(n_e)$ \\
        \hline
        $\lceil\frac{n_e}{w_1}\rceil$ & \Phaseup{} & address $= w_1$ \\
        \hline
    \end{tabular}
    \caption{Enumeration of operations used by \ComputeExponent{}. The \Phaseup{}s of all \Lookup{}s are deferred to the end of the computation and merged into one \Phaseup{} per window of $w_1$ exponent qubits (last row).
     }
    \label{tab:computing_dlog_operation_count}
\end{table}

\subsubsection{Reducing the exponent}\label{sssec:dlog_modular_reduction}

\ComputeExponent{} leaves $S_i$ on $\ell+\log(n_e)$ qubits, but \ComputeResidue{} only needs it
modulo $p_i-1$, the order of the generator $g_{p_i}$. \ReduceExponent{} therefore replaces
$S_i$ by its remainder $S_i \bmod (p_i-1)$, which lies in $[0,p_i-1)$ and fits on $\ell$
qubits. This shortens the exponent that \ComputeResidue{} has to process, and so lowers its
cost. The reduction has to be reversible, because $\ReduceExponent^\dagger$ must later
restore $S_i$. \ReduceExponent{} is therefore a reversible Euclidean division of $S_i$ by
$p_i-1$. It keeps the remainder in the $\ell$ least significant qubits of the register, and
the quotient in the $\log(n_e)$ most significant ones.

\begin{figure}[htbp]
    \centering
    \begin{subfigure}[c]{\columnwidth}
        \centering
        \includegraphics[width=\linewidth]{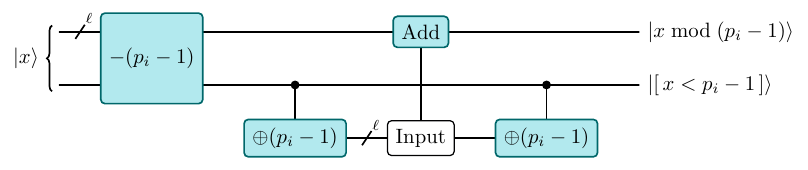}
    \end{subfigure}
    \par\medskip
    \begin{subfigure}[c]{\columnwidth}
        \centering
        \includegraphics[width=\linewidth]{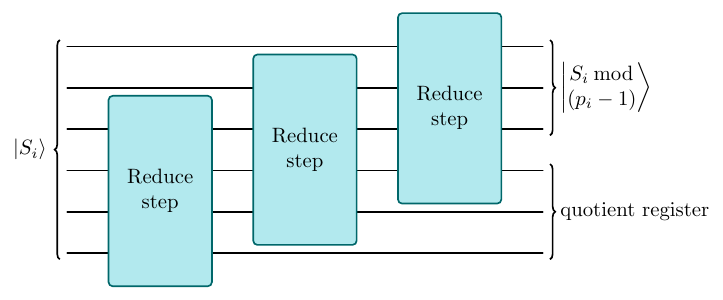}
    \end{subfigure}
    \caption{\ReduceExponent. Top: one reduction step, the construction of
    Lemma~\ref{lemma:efficient_modular_reduction} with $b=p_i-1$. For an input
    $0\le x<2(p_i-1)$ on $\ell+1$ qubits, it leaves $x\bmod (p_i-1)$ on the low $\ell$ qubits
    and the bit $[x<p_i-1]$, the $\mathbbm{1}_{x<b}$ of the lemma, on the top qubit. Bottom:
    \ReduceExponent{} for $\ell=\log n_e=3$, as $\log n_e$ such steps, each on an
    $(\ell+1)$-qubit window that moves down by one qubit
    (Section~\ref{sssec:dlog_modular_reduction}). The top qubits left behind by the steps form
    the quotient register, which is kept to uncompute \ReduceExponent.}
    \label{fig:compress_dlog}
\end{figure}

\paragraph{Long division, one bit at a time.}
\ReduceExponent{} follows schoolbook long division in binary. An $(\ell+1)$-qubit window moves
down the register of $S_i$, one position per step, starting from the most significant end.
At each step the window holds twice the current remainder plus the next bit of $S_i$. A
single conditional subtraction of $p_i-1$ brings it back below $p_i-1$, and the window then
moves down by one position. After $\log(n_e)$ steps the window has reached the bottom of
the register, and its $\ell$ low qubits hold $S_i \bmod (p_i-1)$, as shown in
Figure~\ref{fig:compress_dlog}.

Each step is valid because its input stays below $2(p_i-1)$. The discrete logarithms
satisfy $D_{p_i,k}\le p_i-2$, so $S_i \le n_e(p_i-2) < 2^{\log(n_e)}(p_i-1)$, with
$\log(n_e)=\lceil\log_2 n_e\rceil$. The $\ell$ most significant qubits of $S_i$ therefore
already hold a value below $p_i-1$. By induction, each window holds twice a remainder below
$p_i-1$ plus one bit, so its value is below $2(p_i-1)$.

\paragraph{One reduction step.}
A step must reduce an input $x$ with $0\le x<2b$, where $b=p_i-1$, by subtracting $b$ at
most once. Lemma~\ref{lemma:efficient_modular_reduction} does this without any controlled
adder.

\begin{lemma}\label{lemma:efficient_modular_reduction}
    Let $b$ be an $n$-bit integer, $2^{n-1} \leq b \leq 2^n-1$, and let
    $0 \leq x \leq 2b-1$ be stored on $n+1$ qubits. The following sequence uses an $n$-qubit
    auxiliary register initialized to $\ket{0}$. It leaves $x \bmod b$ in the $n$ low qubits
    of the working register and the bit $\mathbbm{1}_{x<b}$ in its top qubit, and it returns
    the auxiliary register to $\ket{0}$.
\begin{equation} \label{eq:modred_protocol}
    \begin{aligned}
        \ket{x} & \xmapsto{\mathmakebox[\widthof{$\textnormal{\Adder}^\dagger \textnormal{Q-C}$}][c]{\textnormal{\Adder}^\dagger \textnormal{Q-C}}}  \ket{x-b} \\
                & \xmapsto{\mathmakebox[\widthof{$\textnormal{\Adder}^\dagger \textnormal{Q-C}$}][c]{\CX(n)}}  \ket{\mathbbm{1}_{x<b}}\ket{y_t = x-b + 2^n\mathbbm{1}_{x<b}}\ket{0 \oplus b \mathbbm{1}_{x<b}} \\
                & \xmapsto{\mathmakebox[\widthof{$\textnormal{\Adder}^\dagger \textnormal{Q-C}$}][c]{\textnormal{\Adder{} Q-Q}}}  \ket{\mathbbm{1}_{x<b}}\ket{y_t + b \mathbbm{1}_{x<b}}\ket{b\mathbbm{1}_{x<b}} \\
                & \xmapsto{\mathmakebox[\widthof{$\textnormal{\Adder}^\dagger \textnormal{Q-C}$}][c]{\CX(n)}} \ket{\mathbbm{1}_{x<b}}\ket{y_t + (b- 2^n) \mathbbm{1}_{x<b}},
    \end{aligned}
\end{equation}
where $\mathbbm{1}_{x<b} = 1$ if $x<b$, and $0$ otherwise.
\end{lemma}

\begin{proof}
    Since $0 \leq x \leq 2b-1$, there are exactly two mutually exclusive cases, either
    $0\leq x<b$, or $b \leq x \leq 2b-1$. \\
    \underline{If $x<b$:} then $\mathbbm{1}_{x<b} = 1$, and the working register contains
    $x - b + 2^n + (b-2^n) = x = x \bmod b$. \\
    \underline{If $x\geq b$:} then $\mathbbm{1}_{x<b} = 0$, and the working register
    contains $x - b = x \bmod b$.
\end{proof}

Here $x-b$ is stored in two's complement on $n+1$ qubits, so
$\ket{x-b}=\ket{\mathbbm{1}_{x<b}}\ket{y_t}$. Its top qubit is the sign bit, and its $n$
low qubits hold $y_t$.

The direct implementation would first compute $\mathbbm{1}_{x<b}$ with a Q-C comparison,
then use it to control a Q-C adder$^\dagger$ that subtracts $b$. That controlled Q-C
adder$^\dagger$ costs $2n + O(1)$ Toffoli gates. The lemma avoids it. It always subtracts
$b$, with an uncontrolled Q-C adder$^\dagger$ whose top qubit records the underflow. The
underflow qubit then controls a multi-target \textsc{Cnot} gate ($\CX(n)$), which loads $b$
into the auxiliary register if $x<b$, and $0$ otherwise. A Q-Q adder adds the auxiliary
register back, which undoes the subtraction exactly when $x<b$. A second $\CX(n)$ clears the
auxiliary register. This saves $n$ Toffoli gates, at the cost of the $n$-qubit auxiliary
register, which is free at this point of the circuit anyway. The same construction, with
small variations, implements the modular additions of \ComputeResidue{} and
\AccumulateResidue.

\paragraph{The quotient register.}\label{par:quotient-register}
In each step, the top qubit of the window ends in the state $\ket{\mathbbm{1}_{x<b}}$ and
is not measured. It then leaves the window, while the next lower qubit of $S_i$ enters it.
We call the $\log(n_e)$ qubits produced in this way the \emph{quotient qubits}, and the
register they form the \emph{quotient register}. The quotient qubit produced at step $k$
holds the complement of bit $\log(n_e)-k-1$ of the quotient $a$, where
$S_i \bmod (p_i-1) = S_i - a(p_i-1)$. The quotient register therefore stores $a$ bitwise
complemented, which is exactly the information the reversible division must retain. It
occupies the $\log(n_e)$ most significant qubits of the $(\ell+\log(n_e))$-qubit input
register, the $\ell$ least significant ones holding $S_i \bmod (p_i-1)$. It stays allocated
during \ComputeResidue, \AccumulateResidue{} and $\ComputeResidue^\dagger$, until
$\ReduceExponent^\dagger$ runs the same steps in reverse and uses it to restore $S_i$.
All additions and subtractions in \ReduceExponent{} act on registers of the same size. Their
number is given in Table~\ref{tab:dlog_modular_reduction_operation_count}.

\begin{table}[htbp]
    \centering
    \begin{tabular}{|c|c|c|}
        \hline
        Calls & Operation & Operation size \\
        \hline
        \multirow{2}{*}{$|P|\log(n_e)$} & Q-Q \Adder{} & $\ell$ \\
        \cline{2-3} 
            & Q-C \Adder{} (subtraction) & $\ell+1$ \\
        \hline
    \end{tabular}
    \caption{Enumeration of operations used for \ReduceExponent{}.}
    \label{tab:dlog_modular_reduction_operation_count}
\end{table}

\subsubsection{Computing the residue and its uncomputation}

\ComputeResidue{} turns the reduced exponent into the residue
$r_i = g_{p_i}^{S_i \bmod (p_i-1)} \bmod p_i$. It is a small modular exponentiation on
$\ell$-qubit registers, built from windowed table lookups and modular additions in the
style of Lemma~\ref{lemma:efficient_modular_reduction}. Its adjoint
$\ComputeResidue^\dagger$ is not obtained by running \ComputeResidue{} backwards.
\ComputeResidue{} measures its auxiliary registers in the $X$ basis and defers the resulting
phase corrections, so $\ComputeResidue^\dagger$ is a circuit of its own. It erases
$r_i$ and applies all the deferred corrections at once. We first describe \ComputeResidue,
then $\ComputeResidue^\dagger$, and finally the phase corrections they share.

\begin{figure*}[tp]
    \centering
    \includegraphics[width=1\linewidth]{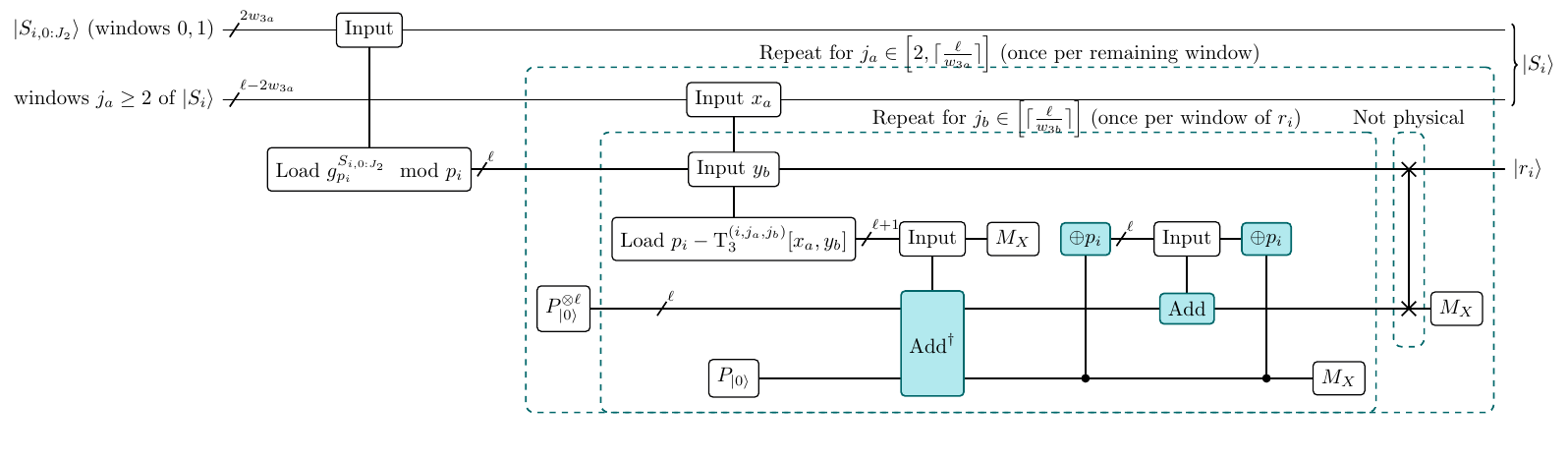}
    \caption{\ComputeResidue, a windowed modular exponentiation
    $r_i=g_{p_i}^{S_i\bmod(p_i-1)}\bmod p_i$, built like the exponentiation of
    Figure~\ref{fig:modexp_alg}. A first lookup loads the product of the first two windows,
    $g_{p_i}^{S_{i,0:J_2}}\bmod p_i$. Each outer iteration $j_a$ then multiplies the partial
    product by the power of $g_{p_i}$ selected by $x_a$, window $j_a$ of $S_i$. The
    multiplication is out of place, each inner iteration $j_b$ adds the table entry
    $\text{T}_3^{(i,j_a,j_b)}[x_a,y_b]$, where $y_b$ is window $j_b$ of the partial product,
    with the modular addition of Lemma~\ref{lemma:efficient_modular_reduction}. The swap at
    the end of each outer iteration only exchanges the roles of the two registers, in
    software. The phase corrections of the measurements, including the \UnderflowPhasefix{},
    are deferred to \ComputeResidue$^\dagger$ (Figure~\ref{fig:unmodexp}).}
    \label{fig:modexp}
\end{figure*}

\paragraph{\ComputeResidue.}
\ComputeResidue{} takes as input the reduced exponent $S_i \bmod (p_i-1)$, which we write
$S_i$ for the rest of this subsection, and outputs
\begin{equation}
    r_i = g_{p_i}^{S_i} \bmod p_i.
\end{equation}
Following~\cite{Gidney19windowed}, it splits $S_i$ into $\lceil \ell/w_{3a}\rceil$ windows
of $w_{3a}$ bits, with offsets $J_{j_a}=j_aw_{3a}$, and writes
\begin{equation}
    r_i = \prod_{j_a=0}^{\lceil \ell/w_{3a} \rceil-1}
    g_{p_i}^{S_{i,J_{j_a}:J_{j_a + 1}}2^{J_{j_a}}} \bmod p_i.
\end{equation}
Let $\alpha_{j_a} = g_{p_i}^{S_{i,0:J_{j_a}}} \bmod p_i$ be the partial product of the first $j_a$ windows. 
Each factor multiplies the partial product by a constant that depends on one window of $S_i$, so that $\alpha_{j_a+1} = \alpha_{j_a}\, g_{p_i}^{S_{i,J_{j_a}:J_{j_a+1}}2^{J_{j_a}}} \bmod p_i$.

This multiplication is itself windowed. 
The partial product $\alpha_{j_a}$ is split into $\lceil \ell/w_{3b}\rceil$ windows of $w_{3b}$ bits, with offsets $J_{j_b}=j_bw_{3b}$.
Write $x_a := S_{i,J_{j_a}:J_{j_a + 1}} \in \{0, \dots, 2^{w_{3a}}-1\}$ for the current window of the exponent, and $y_b := \alpha_{j_a,J_{j_b}:J_{j_b + 1}} \in \{0, \dots, 2^{w_{3b}}-1\}$ for a window of the partial product. 
The multiplication then becomes a sum of table entries,
\begin{equation}\label{eq:double_window}
\begin{aligned}
    \alpha_{j_a+1} &= \sum_{j_b=0}^{\lceil \ell/w_{3b} \rceil-1}
      \text{T}_3^{(i,j_a,j_b)}[x_a,y_b] \bmod p_i, \\
    \text{T}_3^{(i,j_a,j_b)}[x_a,y_b] &:= y_b\, 2^{J_{j_b}}\, g_{p_i}^{x_a 2^{J_{j_a}}}
      \bmod p_i.
\end{aligned}
\end{equation}
The table is precomputed classically for every address pair $(x_a,y_b)$. The inner loop
over $j_b$ therefore computes $\alpha_{j_a+1}$ out of place, with one \Lookup{} and one
modular addition per window. The outer loop over $j_a$ chains these multiplications into
the exponentiation.

\paragraph{The modular addition.}
Each step of the inner loop adds a looked-up value to the helper register modulo $p_i$,
with a variant of Lemma~\ref{lemma:efficient_modular_reduction}. Write $T$ for
$\text{T}_3^{(i,j_a,j_b)}[x_a,y_b]$. The \Lookup{} loads $p_i - T$ rather than $T$, and a
Q-Q adder$^\dagger$ replaces the Q-C adder$^\dagger$ of the lemma,
\begin{equation*}
    \ket{p_i-T}\ket{h} \longmapsto \ket{p_i-T}\ket{h-(p_i-T)}.
\end{equation*}
This single subtraction both adds $T$ and performs the subtraction of $p_i$ that the
modular reduction needs. When it underflows, the last three steps of
Equation~\eqref{eq:modred_protocol} add $p_i$ back. The loaded register is then measured in the
$X$ basis, and its phase correction is deferred to $\ComputeResidue^\dagger$.

\paragraph{Initialization.}
The outer loop starts from $\alpha_2 = g_{p_i}^{S_{i,0:J_2}} \bmod p_i$, loaded directly by
a \Lookup{} of address size $2w_{3a}$. This skips the first two multiplications, by
$g_{p_i}^{S_{i,0:J_1}}$ and by $g_{p_i}^{S_{i,J_1:J_2}2^{J_1}}$. When $w_{3a}=w_{3b}$, as
in our resource estimates, this \Lookup{} has the same address size as those of the inner
loop, namely $w_{3a}+w_{3b}$. One iteration of the outer loop then reads
\begin{equation}\label{eq:modexp_protocol}
    \begin{aligned}
        \ket{S_i}\ket{\alpha_{j_a}}\ket{0}^{\otimes \ell} & \xmapsto{\text{mod mul}}
        \ket{S_i}\ket{\alpha_{j_a}}\ket{\alpha_{j_a+1}} \\
        & \xmapsto{\text{role swap} + M_X} \ket{S_i}\ket{\alpha_{j_a+1}}.
    \end{aligned}
\end{equation}
Figure~\ref{fig:modexp} shows the circuit, and
Table~\ref{tab:modexp_operation_count} counts its operations.

\begin{table*}[tp]
    \centering
    \begin{tabular}{|c|c|c|}
        \hline
        Calls & Operation & Operation size \\
        \hline
        $|P|$ & \Lookup{} (initialization) & address $= 2w_{3a}$, target $=\ell$ \\
        \hline
        \multirow{3}{*}{$|P|\left( \lceil \frac{\ell}{w_{3a}} \rceil -2 \right)\lceil \frac{\ell}{w_{3b}} \rceil$} & Q-Q \Adder{} (subtraction) & $\ell+1$ \\
        \cline{2-3}
            & Q-Q \Adder{} & $\ell$ \\
            \cline{2-3}
            & \Lookup{} & address $= w_{3a} + w_{3b}$, target $=\ell+1$ \\
        \hline
    \end{tabular}
    \caption{Operations used by \ComputeResidue. For resource estimation, we choose
    $w_{3a} = w_{3b} = w_3$.
     }
    \label{tab:modexp_operation_count}
\end{table*}

\begin{figure*}[tp]
    \centering
    \includegraphics[width=\linewidth]{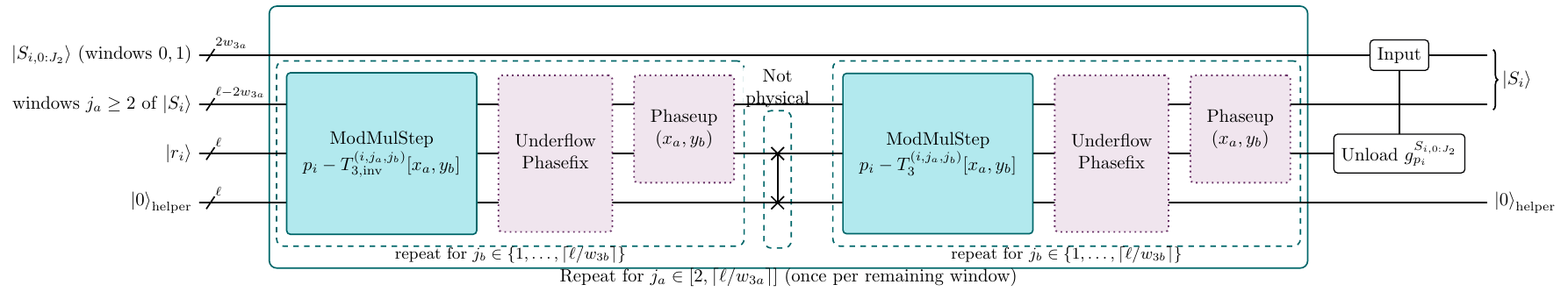}
    \caption{\ComputeResidue$^\dagger$. It erases the residue $r_i$ computed by \ComputeResidue{}
    (Figure~\ref{fig:modexp}) and applies the phase corrections that \ComputeResidue{} deferred.
    It runs the outer loop of Figure~\ref{fig:modexp} again, over the windows $x_a$ of $S_i$.
    Each \textsc{ModMulStep} box is one pass of the inner loop there, that is, one step of the
    out-of-place multiplication shown at the bottom of Figure~\ref{fig:modexp_alg}. It loads
    $p_i-T_{3,\text{inv}}^{(i,j_a,j_b)}[x_a,y_b]$ in the first inner loop and
    $p_i-T_{3}^{(i,j_a,j_b)}[x_a,y_b]$ in the second, where $y_b$ is window $j_b$ of the
    partial product. Each pass is followed by the underflow phase correction of
    Figure~\ref{fig:underflowphasefix} and by a \Phaseup{} addressed by $(x_a,y_b)$. A \Phaseup{}
    applies the phase $(-1)^{m\cdot T[x]}$ to the address $x$ of a lookup whose output $T[x]$
    was measured in the $X$ basis with outcomes $m$, which completes its measurement-based
    uncomputation~\cite{Babbush18}. The swap between the two inner loops only exchanges the
    roles of the two registers, in software. After the outer loop, the initial value
    $g_{p_i}^{S_{i,0:J_2}}$ is unloaded by measurement-based uncomputation, with a \Phaseup{}
    addressed by the first two windows of $S_i$.}
    \label{fig:unmodexp}
\end{figure*}

\paragraph{$\ComputeResidue^\dagger$.}
$\ComputeResidue^\dagger$ erases the residue,
\begin{equation}\label{eq:unmodexp}
    \ComputeResidue^\dagger \ket{S_i}\ket{r_i} = \ket{S_i}\ket{0},
\end{equation}
and applies all the phase corrections that \ComputeResidue{} deferred. It runs the outer loop
backwards. Each iteration turns $\alpha_{j_a+1}$ back into $\alpha_{j_a}$ with two inner
loops, built from the same step as the inner loop of \ComputeResidue{}
(Figure~\ref{fig:unmodexp}). The first inner loop computes $\alpha_{j_a}$ out of place, from
$\alpha_{j_a+1}$, with the table of inverse factors
\begin{equation*}
    T_{3,\text{inv}}^{(i,j_a,j_b)}[x_a,y_b] := y_b\, 2^{J_{j_b}}\, g_{p_i}^{-x_a 2^{J_{j_a}}}
    \bmod p_i.
\end{equation*}
The second inner loop uses $\alpha_{j_a}$ to clear the register that held $\alpha_{j_a+1}$,
\begin{equation}
\begin{aligned}
    \ket{S_i} \ket{\alpha_{j_a+1}} \ket{0}
    &\xmapsto{\text{1st inner loop}}
    \ket{S_i} \ket{\alpha_{j_a+1}} \ket{\alpha_{j_a}} \\
    &\xmapsto{\text{2nd inner loop}}
    \ket{S_i} \ket{\alpha_{j_a+1} - \alpha_{j_a}\, g_{p_i}^{x_a 2^{J_{j_a}}}}
    \ket{\alpha_{j_a}} \\
    &\qquad = \ket{S_i}\ket{0}\ket{\alpha_{j_a}},
\end{aligned}
\end{equation}
since $\alpha_{j_a} = \alpha_{j_a+1}\, g_{p_i}^{-x_a 2^{J_{j_a}}} \bmod p_i$. A role swap
then makes $\alpha_{j_a}$ the input of the next iteration. After the last iteration, the
register holds $\alpha_2$, which is unloaded by measurement-based uncomputation and a \Phaseup{} addressed by the $2w_{3a}$ low
bits of $S_i$, mirroring the initialization of \ComputeResidue. Equation~\eqref{eq:unmodexp}
then holds. Table~\ref{tab:unmodexp_operation_count} counts the operations.

\begin{table*}[tp]
    \centering
    \begin{tabular}{|c|c|c|}
        \hline
        \multicolumn{3}{|c|}{\textbf{$\ComputeResidue^\dagger$}} \\
        \hline
        Calls & Operation & Operation size\\
        \hline
        $|P|$ & \Phaseup{} (final unloading) & address $= 2w_{3a}$ \\
        \hline
        \multirow{4}{*}{$2|P|\left( \lceil \frac{\ell}{w_{3a}} \rceil -2 \right)\lceil \frac{\ell}{w_{3b}} \rceil$} & Q-Q \Adder{} (subtraction) & $\ell+1$ \\
            \cline{2-3}
            & Q-Q \Adder{} & $\ell$ \\
            \cline{2-3}
            & \Lookup{} & address $= w_{3a} + w_{3b}$, target $=\ell+1$ \\
            \cline{2-3}
            & \Phaseup{} & $w_{3a} + w_{3b}$ \\
        \hline
        $|P|\left( \lceil \frac{\ell}{w_{3a}} \rceil -2 \right)\lceil \frac{\ell}{w_{3b}} \rceil$ & \UnderflowPhasefix{} & address $= w_{3a} + w_{3b}$, comparison $=\ell$ \\[0.5ex]
        \hline\hline
        \multicolumn{3}{|c|}{\textbf{\UnderflowPhasefix}} \\
        \hline
        Calls & Operation & Operation size\\
        \hline
        \multirow{2}{*}{$|P|\left( \lceil \frac{\ell}{w_{3a}} \rceil -2 \right)\lceil \frac{\ell}{w_{3b}} \rceil$} & \Lookup{} & address $= w_{3a} + w_{3b}$, target $=\ell$ \\
            \cline{2-3}
            & Q-Q \Comparison{} & $\ell$ \\
        \hline
    \end{tabular}
    \caption{Operations used by $\ComputeResidue^\dagger$ and by \UnderflowPhasefix.
    For resource estimation, we choose $w_{3a} = w_{3b} = w_3$.
     }
    \label{tab:unmodexp_operation_count}
\end{table*}

\paragraph{Phase corrections.}
Each step measures two kinds of qubits in the $X$ basis: the register loaded by the
\Lookup, and the underflow qubit of the modular addition. Both measurements leave phases
that must be corrected.

The phases left by the measurement of the loaded registers are corrected by \Phaseup{} operations. The \Phaseup{} of the first inner
loop of $\ComputeResidue^\dagger$ corrects the phases of the loaded register. That
register is measured twice, once in this loop and once in the inner loop of
\ComputeResidue. The two corrections are added classically, and their sum is applied once.
The same \Phaseup{} also corrects the \Lookup{} inside \UnderflowPhasefix. The \Phaseup{} of the
second inner loop does the same for its own \Lookup{} and \UnderflowPhasefix.

\begin{figure}[htbp]
    \centering
    \includegraphics[width=\columnwidth]{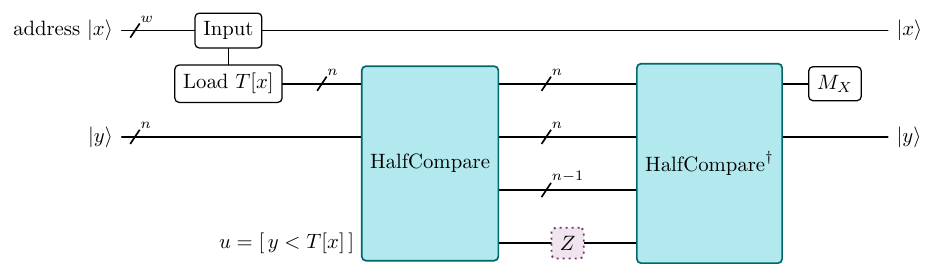}
    \caption{\UnderflowPhasefix, the phase correction
    $\ket{x}\ket{y}\mapsto(-1)^{[\,y<T[x]\,]}\ket{x}\ket{y}$, applied when the $X$-basis
    measurement of the underflow qubit of a modular addition $y+T[x]$ returns $1$. A
    lookup loads $T[x]$. \textsc{HalfCompare} is the first half of the subtraction $y-T[x]$, whose
    last carry is the comparison bit $u=[\,y<T[x]\,]$. A $Z$ gate on $u$ applies the phase,
    and \textsc{HalfCompare}$^\dagger$ erases the carries. The loaded value is then measured in the $X$
    basis. Its phase correction acts on the same address $x$, so it is merged into the \Phaseup{}
    that follows in Figure~\ref{fig:unmodexp}. The lookup is drawn inside the block because
    the block is only used in that figure.}
    \label{fig:underflowphasefix}
\end{figure}

The underflow qubit is corrected by \UnderflowPhasefix, which needs the value of that qubit.
Let $x$ be the input of the step. It is either $0$ or the output of the previous step, so
$0 \leq x < p$, with $p=p_i$. Let $p-T$ be the loaded value, with $0 \leq p-T < p$. The
underflow qubit $u$ is the sign bit of
\begin{equation}
    x - (p-T) = x + T - p.
\end{equation}
Computing the comparison $u = \mathbbm{1}_{x<p-T}$ directly is not possible, because $x$
is gone. The helper register now holds $x+T$ if $u=1$, and $x+T-p$ if $u=0$. Since
$x \geq 0$, the first case gives $x + T \geq T$, and since $x < p$, the second gives
$x+T-p < T$. Comparing the helper register with $T$ therefore recovers $u$.
\UnderflowPhasefix{} (Figure~\ref{fig:underflowphasefix}) loads $T$, performs this comparison,
and corrects the phase if needed. Unlike the other phase corrections, it cannot be delayed
and merged, because the helper register is overwritten at each iteration.

One \UnderflowPhasefix{} per pair of steps can be saved. The second inner loop of
$\ComputeResidue^\dagger$ is identical to the inner loop of \ComputeResidue. 
This loop performs the same operations and loads the same table $p_i-T_3$. 
Its first pass also has the same input as the last pass of the inner loop of \ComputeResidue, namely the last partial product. 
The two underflow qubits therefore carry the same value, and their phase corrections can be merged. 
Writing $\text{meas}_c$ and $\text{meas}_{u2}$ for the two $X$-basis outcomes, \UnderflowPhasefix{} is applied only if $\text{meas}_c \oplus \text{meas}_{u2} = 1$, which happens with probability $1/2$.

\subsubsection{Accumulating the residue}

\AccumulateResidue{} adds the contribution of the residue $r_i$ to the accumulator, so that
after the last prime the accumulator holds $\tilde V$
(Equation~\eqref{eq:approximate_modular_exponentiation}). It is the only subroutine that
writes to the accumulator. It reuses the modular addition of \ComputeResidue, on a
$(\kappa+1)$-qubit register instead of an $(\ell+1)$-qubit one.

\begin{figure*}[tp]
    \centering
    \includegraphics[width=1\linewidth]{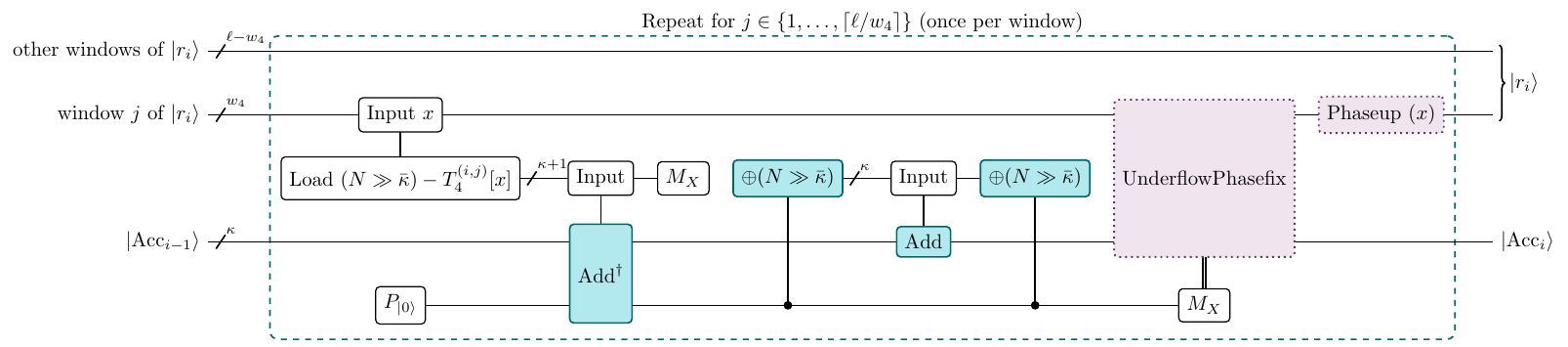}
    \caption{\AccumulateResidue, the multiply-accumulate
    $\mathrm{Acc}_i=\mathrm{Acc}_{i-1}+\sum_k r_{i,k}C_{i,k}\bmod (N\gg\bar\kappa)$, with the
    constants $C_{i,k}$ of Equation~\eqref{eq:approximate_modular_exponentiation}. It
    multiplies the bit vector of $r_i$ by classical constants, window by window. In iteration
    $j$, window $j$ of $r_i$ is the address $x$ of a lookup that loads
    $(N\gg\bar\kappa)-T_4^{(i,j)}[x]$, with
    $T_4^{(i,j)}[x]=\sum_{k=0}^{w_4-1}x_kC_{i,jw_4+k}$. Subtracting this value adds
    $T_4^{(i,j)}[x]$ modulo $N\gg\bar\kappa$. The loaded value is then measured in the $X$
    basis, and the \Phaseup{} addressed by $x$ applies the matching phase correction, as in
    Figure~\ref{fig:unmodexp}. \UnderflowPhasefix{} (Figure~\ref{fig:underflowphasefix})
    corrects the phase left by measuring the underflow qubit of the modular addition.}
    \label{fig:accumulation} 
\end{figure*}

The subroutine takes as input the current residue $\ket{r_i}$ and the accumulated value $\ket{\text{Acc}_{i-1}}$, with $\text{Acc}_i = \sum_{j=0}^{i}\sum_{k=0}^{\ell-1}r_{j,k}C_{j,k}$ the partial sum for the $i$ first primes of the residue number system.
It performs an addition modulo $(N \gg \bar \kappa)$ and outputs $\ket{r_i}$ and $\ket{\text{Acc}_{i}}$. Recall that $(N \gg \bar \kappa) = \lceil \frac{N}{2^{\bar \kappa}} \rceil$ is obtained from the unsigned integer binary representation of  $N$ by deleting its $\bar \kappa$ least significant bits. 
The circuit is presented in Figure~\ref{fig:accumulation}, and is very similar to one inner loop of $\ComputeResidue^\dagger$ detailed before in Figure \ref{fig:unmodexp}, albeit using different tables and operations on registers of different sizes.
The number of operations required for the accumulation is given in Table \ref{tab:accumulation_operation_count}.

\begin{table*}[tp]
    \centering
    \begin{tabular}{|c|c|c|}
        \hline
        \multicolumn{3}{|c|}{\textbf{\AccumulateResidue}} \\
        \hline
        Calls & Operation & Operation size\\
        \hline
        \multirow{4}{*}{$|P|\lceil \frac{\ell}{w_{4}} \rceil$} & Q-Q \Adder{} (subtraction) & $\kappa+1$ \\
            \cline{2-3} 
            & Q-Q \Adder{} & $\kappa$ \\
            \cline{2-3}
            & \Lookup{} & address $= w_4$, target $=\kappa+1$ \\
            \cline{2-3}
            & \Phaseup{} & $w_4$ \\
        \hline
        $\frac{|P|}{2}\lceil \frac{\ell}{w_{4}} \rceil$ & \UnderflowPhasefix{} & address $= w_4$, comparison $=\kappa$ \\[0.5ex]
        \hline\hline
        \multicolumn{3}{|c|}{\textbf{\UnderflowPhasefix}} \\
        \hline
        Calls & Operation & Operation size\\
        \hline
        \multirow{2}{*}{$\frac{|P|}{2}\lceil \frac{\ell}{w_{4}} \rceil$} & \Lookup{} & address $= w_4$, target $=\kappa$ \\
            \cline{2-3} 
            & Q-Q \Comparison{} & $\kappa$ \\
        \hline
    \end{tabular}
    \caption{Enumeration of operations used by \AccumulateResidue{} and the \UnderflowPhasefix.
     }
    \label{tab:accumulation_operation_count}
\end{table*}

\subsubsection{Logical cost breakdown}

This subsection collects the operation and qubit counts of all subroutines, for use in the
resource estimate of Section~\ref{sec:resource_estim}. For additions, the size given is the
size of the input number.

For \Lookup{} operations, the values given are the address register and target register sizes.
For \Phaseup{} operations, the value given is the address register size, which is the only value needed as the target register is only measured out in the $X$ basis before starting the \Phaseup{} operation.
For \Comparison{} operations, the circuit is a subtraction and uses the last carry bit as the comparison bit, thus a comparison of size $n$ is equivalent in cost to an addition of size $n+1$.
Additions, subtractions and comparisons are referred to and accounted for as \Adder{} operations as their circuit and costs are very similar.
Concerning the qubit cost, the logical qubit count at the different steps of the circuit is shown in Table~\ref{tab:logical_qubit_count}.
Observe that the total cost consists mostly of \Adder{} operations, \Lookup{} operations and, to a lesser extent, \Phaseup{} operations.
Therefore, it is important to optimize their cost and implementation at the physical level, which is the focus of Section \ref{sec:resource_estim}.

\begin{table*}[tp]
    \centering
    \begin{tabular}{|c|c|c|c|c|c|c|c|}
        \hline
        Subroutine & Calls & Register size & Address size & \begin{tabular}[c]{@{}c@{}}Q-Q\\\Adder\end{tabular} & \begin{tabular}[c]{@{}c@{}}Q-C\\\Adder\end{tabular} & \Lookup{} & \Phaseup{}  \\
        \hline\hline
        \ComputeExponent{} & $(|P|+1)\lceil\frac{n_e}{w_1}\rceil$ & $\ell + \log(n_e)$ & $w_1$ & 1 & 0 & 1 & 0  \\
        \hline
        \ReduceExponent{} & $|P|\log(n_e)$ & $\ell + 1$ & - & 1 & 1 & 0 & 0  \\
        \hline
        \ComputeResidue{} (initialization) & $|P|$ & $\ell$ & $2w_{3}$ & 0 & 0 & 1 & 0  \\
        \ComputeResidue{} (outer loop) & $|P|\left( \lceil \frac{\ell}{w_{3}} \rceil -2 \right)\lceil \frac{\ell}{w_{3}} \rceil$ & $\ell + 1$ &$2w_3$ & 2 & 0 & 1 & 0  \\
        \hline
        \AccumulateResidue{} & $|P|\lceil \frac{\ell}{w_{4}} \rceil$ & $\kappa + 1$ & $w_4$ & 2.5 & 0 & 1.5 & 1  \\
        \hline
        \ComputeResidue$^\dagger$ (outer loop) & $2|P|\left( \lceil \frac{\ell}{w_{3}} \rceil -2 \right)\lceil \frac{\ell}{w_{3}} \rceil$ & $\ell + 1$ & $2w_3$ & 2.5 & 0 & 1.5 & 1  \\
        \ComputeResidue$^\dagger$ (unloading) & $|P|$ & $\ell$ & $2w_{3}$ & 0 & 0 & 0 & 1  \\
        \hline
        \ReduceExponent$^\dagger$ & $|P|\log(n_e)$ & $\ell + 1$ & - & 1 & 1 & 0 & 0  \\
        \hline
        Last \Phaseup{} (\ComputeExponent) & $\lceil\frac{n_e}{w_1}\rceil$ & - & $w_1$ & 0 & 0 & 0 & 1  \\
        \hline
    \end{tabular}
    \caption{Enumeration of operations used by each subroutine of the algorithm. "Register size" refers to the biggest register involved in an addition or as the target of a \Lookup{} in that subroutine. \Comparison{}s are counted as Q-Q \Adder{}s. The 0.5 comes from the \UnderflowPhasefix{} which is applied or not $50~\%$ of the time (depends on an $M_X$ measurement). The last row gives the $\lceil\frac{n_e}{w_1}\rceil$ \Phaseup{}s that apply, at the end of the computation, the phase corrections deferred by \ComputeExponent. These \Phaseup{}s act on no \Adder{} or \Lookup{} target, hence the empty register size.
     }
    \label{tab:operation_count_recap}
\end{table*}

\begin{table*}[tp]
    \centering
    \begin{tabular}{|c|c|c|c|c|}
        \hline
        Subroutine & Calls & Operation & Subroutine scope & Out of scope   \\
        \hline\hline
        \multirow{3}{*}{\ComputeExponent} & \multirow{3}{*}{$(|P|+1)\lceil\frac{n_e}{w_1}\rceil$} & \Lookup$(w_1,\ell+\log(n_e))$ & $n_e + \ell+\log(n_e) - w_1$ & \multirow{3}{*}{$\kappa$} \\
            \cline{3-4}
            & & Q-Q \Adder$(\ell +\log(n_e))$ & $n_e$ & \\
            \cline{3-4}
            & & $M_X(\ell +\log(n_e))$ & $n_e + \ell + \log(n_e)$ & \\
        \hline
        \multirow{4}{*}{\ReduceExponent} & \multirow{4}{*}{$|P|\log(n_e)$} & Q-C \Adder$(\ell+1)$ & $\log(n_e) - 1$ & \multirow{4}{*}{$n_e + \kappa$} \\
            \cline{3-4}
            & & $\CX(\ell)$ & $\ell + \log(n_e) - 1$ & \\
            \cline{3-4}
            & & Q-Q \Adder$(\ell)$ & $\log(n_e)$ & \\
            \cline{3-4}
            & & $\CX(\ell)$ & $\ell + \log(n_e) - 1$ & \\
        \hline
        \multirow{10}{*}{\ComputeResidue} & $|P|$ & \Lookup$(2w_3,\ell)$ (loading) & $\ell - 2w_3$ & \multirow{10}{*}{$n_e + \log(n_e) + \kappa$} \\
        \cline{2-4}
        & $|P|\left( \lceil \frac{\ell}{w_{3}} \rceil -2 \right)$ & $P_{\ket{0}}(\ell)$ & $2\ell$ & \\
        \cline{2-4}
            & \multirow{7}{*}{$|P|\left( \lceil \frac{\ell}{w_{3}} \rceil -2 \right)\lceil \frac{\ell}{w_{3}} \rceil$} & \Lookup$(2w_3,\ell+1)$ & $3\ell - 2w_3 + 1$ & \\
            \cline{3-4}
            & & Q-Q \Adder$^\dagger(\ell+1)$ & $2\ell$ & \\
            \cline{3-4}
            & & $M_X(\ell+1)$ & $3\ell + 1$ & \\
            \cline{3-4}
            & & $\CX(\ell)$ & $3\ell$ & \\
            \cline{3-4}
            & & Q-Q \Adder$(\ell)$ & $2\ell + 1$ & \\
            \cline{3-4}
            & & $\CX(\ell)$ & $3\ell$ & \\
            \cline{3-4}
            & & $M_X(1)$ & $3\ell$ & \\
        \cline{2-4}
        & $|P|\left( \lceil \frac{\ell}{w_{3}} \rceil -2 \right)$ & $M_X(\ell)$ & $2\ell$ & \\
        \hline
        \multirow{10}{*}{\AccumulateResidue} & \multirow{6}{*}{$|P|\lceil \frac{\ell}{w_{4}} \rceil$} & \Lookup$(w_4,\kappa+1)$ & $\ell - w_4 + \kappa + 1$ & \multirow{10}{*}{$n_e + \log(n_e) + \ell$} \\
            \cline{3-4}
            & & Q-Q \Adder$^\dagger(\kappa+1)$ & $\ell$ & \\
            \cline{3-4}
            & & $M_X(\kappa+1)$ & $\ell + \kappa + 1$ & \\
            \cline{3-4}
            & & $\CX(\kappa)$ & $\ell + \kappa$ & \\
            \cline{3-4}
            & & Q-Q \Adder$(\kappa)$ & $\ell + 1$ & \\
            \cline{3-4}
            & & $\CX(\kappa)$ & $\ell + \kappa$ & \\
            \cline{2-4}
            & \multirow{3}{*}{$\frac{|P|}{2}\lceil \frac{\ell}{w_{4}} \rceil$} & \Lookup$(w_4,\kappa)$ & $\ell - w_4 + \kappa$ & \\
            \cline{3-4}
            & & Q-Q \Comparison$(\kappa)$ & $\ell$ & \\
            \cline{3-4}
            & & $M_X(\kappa)$ & $\ell + \kappa$ & \\
            \cline{2-4}
            & $|P|\lceil \frac{\ell}{w_{4}} \rceil$ & \Phaseup$(w_4)$ & $\ell - w_4 + \kappa$ & \\
        \hline
        \multirow{11}{*}{\ComputeResidue$^\dagger$} & \multirow{6}{*}{$2|P|\left( \lceil \frac{\ell}{w_{3}} \rceil -2 \right)\lceil \frac{\ell}{w_{3}} \rceil$} & \Lookup$(2w_3,\ell+1)$ & $3\ell - 2w_3 + 1$ & \multirow{11}{*}{$n_e + \log(n_e) + \kappa$} \\
            \cline{3-4}
            & & Q-Q \Adder$^\dagger(\ell+1)$ & $2\ell$ & \\
            \cline{3-4}
            & & $M_X(\ell+1)$ & $3\ell + 1$ & \\
            \cline{3-4}
            & & $\CX(\ell)$ & $3\ell$ & \\
            \cline{3-4}
            & & Q-Q \Adder$(\ell)$ & $2\ell + 1$ & \\
            \cline{3-4}
            & & $\CX(\ell)$ & $3\ell$ & \\
            \cline{2-4}
            & \multirow{3}{*}{$|P|\left( \lceil \frac{\ell}{w_{3}} \rceil -2 \right)\lceil \frac{\ell}{w_{3}} \rceil$} & \Lookup$(2w_3,\ell)$ & $3\ell - 2w_3$ & \\
            \cline{3-4}
            & & Q-Q \Comparison$(\ell)$ & $2\ell$ & \\
            \cline{3-4}
            & & $M_X(\ell)$ & $3\ell$ & \\
            \cline{2-4}
            & $2|P|\left( \lceil \frac{\ell}{w_{3}} \rceil -2 \right)\lceil \frac{\ell}{w_{3}} \rceil$ & \Phaseup$(2w_3)$ & $3\ell - 2w_3$ & \\
            \cline{2-4}
            & $|P|$ & \Phaseup$(2w_3)$ (unload) & $\ell$ & \\
        \hline
        \multirow{4}{*}{\ReduceExponent$^\dagger$} & \multirow{4}{*}{$|P|\log(n_e)$} & Q-C \Adder$(\ell+1)$ & $\log(n_e) - 1$ & \multirow{4}{*}{$n_e + \kappa$} \\
            \cline{3-4}
            & & $\CX(\ell)$ & $\ell + \log(n_e) - 1$ & \\
            \cline{3-4}
            & & Q-Q \Adder$(\ell)$ & $\log(n_e)$ & \\
            \cline{3-4}
            & & $\CX(\ell)$ & $\ell + \log(n_e) - 1$ & \\
        \hline
        Last \Phaseup{} & $\lceil\frac{n_e}{w_1}\rceil$ & \Phaseup$(w_1)$ & $n_e-w_1$ & $\kappa$ \\
        \hline
    \end{tabular}
    \caption{Symbolic logical qubit count at each point of the approximate modular exponentiation circuit, and number of calls for each operation.
    "Out of scope" corresponds to the number of logical qubits not involved in the subroutine, while "subroutine scope" corresponds to the number of logical qubits involved elsewhere in the subroutine but not in the current operation.
    $\CX(x)$ is a multi-target \textsc{Cnot} gate with $x$ target logical qubits, $M_X(x)$ is the measurement in the $X$ basis of $x$ logical qubits and $P_{\ket{0}}(x)$ is the preparation of $x$ logical qubits in the $\ket{0}$ state.
    In \ComputeResidue, \AccumulateResidue{} and $\ComputeResidue^\dagger$, the $\log(n_e)$ term of the out-of-scope count is the quotient register.
     }
    \label{tab:logical_qubit_count}
\end{table*}

\newpage
\subsection{Arithmetic circuits for the two ECDLP variants}\label{ssec:ecdlp-logical}

\begin{table*}[t]
\centering
\small
\begin{tabular}{@{}llll@{}}
\toprule
Symbol & Definition & Role & secp256k1 \\
\midrule
$q$            & prime defining the base field, $q = 2^u - f$              & classical  & $2^{256}-4294968273$ \\
$f$            & pseudo-Mersenne defect $2^u - q$                          & classical  & $2^{32}+977$ \\
$\lambda_f$    & bit-size of $f$                                           & classical  & $33$ \\
$n$            & bit-size of $q$                                           & classical  & $256$ \\
$P,\,Q$        & base point and target point, $Q = [k]P$                   & classical  & --- \\
$k$            & discrete logarithm of $Q$ in base $P$                     & output     & --- \\
$w$            & window length of the point addition                       & parameter  & $16$ \\
$N_A$          & number of windowed point additions, $2\lceil n/w\rceil-4$ & classical  & $28$ \\
$i$            & window index (address register)                           & quantum    & --- \\
$c$            & flag $(i \neq 0)$                                         & quantum    & --- \\
$(x_1,y_1)$    & coordinates of the accumulator point                      & quantum    & --- \\
$(x_2,y_2)$    & coordinates of the loaded point $P_i$                     & quantum, from a table & --- \\
$\lambda$      & slope $(y_1-y_2)(x_1-x_2)^{-1} \bmod q$                   & quantum    & --- \\
$u,\,v$        & registers of the Euclidean algorithm                      & quantum    & --- \\
$b_0,\,b_0\&b_1$ & decision bits of one Euclidean iteration                & quantum    & --- \\
$g$            & garbage (``dialog'') bit-vector                           & quantum    & --- \\
$r,\,s$        & registers of the B\'ezout replay                          & quantum    & --- \\
$c_{\rm iter}$ & padding-iteration constant                                & parameter  & $2.4$ \\
$c_{\rm pad}$  & register-padding constant                                 & parameter  & $2.3$ \\
$N_{\rm it}$   & gcd iterations, $3\lceil (1.413n + c_{\rm iter}\sqrt{n})/3\rceil$ & parameter & $402$ \\
$G$            & garbage register, $5N_{\rm it}/3 \simeq 2.355n$           & classical  & $670$ \\
$p$            & register padding, $\lceil c_{\rm pad}\sqrt{n}\rceil$      & parameter  & $37$ \\
$B$            & shared $u/v/$garbage block, $G + 2p + 2\,\textsc{step}$   & classical  & $748$ \\
$A$            & ancilla budget of the hybrid adder, $n+5-(B-G)$           & classical  & $183$ \\
$\nu_j$        & width of $u,v$ at iteration $j$                           & classical  & $n \to 10$ \\
$\gamma_j$     & garbage written after $j$ iterations, $5\lceil j/3\rceil$ & classical  & $0 \to 670$ \\
$\tau$         & width of the truncated gcd comparison, $\textsc{truncate}+p$ & parameter & $77$ \\
$\mu_R$        & MSBs of the replay comparisons                            & parameter  & $40$ \\
$\varphi_R$    & LSBs of the replay constant additions, $\mu_R+\lambda_f$  & parameter  & $73$ \\
$\mu_S$        & MSBs of the square-subtract comparisons                   & parameter  & $50$ \\
$\varphi_S$    & LSBs of the square-subtract constant additions, $\mu_S+\lambda_f$ & parameter & $83$ \\
$N_Q$          & replay iterations handling $x+y=q$                        & parameter  & $50$ \\
$a$            & residual $O(1)$ ancillae at the peak                      & classical  & $\simeq 10$ \\
\bottomrule
\end{tabular}
\caption{Parameters of the elliptic-curve discrete logarithm circuit, with
the values used for \texttt{secp256k1}.}
\label{tab:ecdlp-params}
\end{table*}

\subsubsection{High-level point addition}

Recall that elliptic curve points are represented in affine coordinates $(x,y)$ over the prime field $\mathbb{F}_q$. This representation misses the point at infinity, but with very high probability the operations will not produce it, so that the point addition remains correct for most cases. Likewise, the doubling case is omitted from the addition formulas.

Addition by a constant point $(x_2,y_2)$ is thus performed by Algorithm~\ref{algo:point-add}, like in~\cite{ProosZalka03,GouzienCatQubits23}.

\begin{figure}[htb]
\begin{algorithm}[H]
\caption{Constant point addition on an ordinary prime-field curve, in affine coordinates.}\label{algo:point-add}
\begin{algorithmic}[1]
\Statex \textbf{Input:} Registers $x_1, y_1$ (coordinates of the accumulator point)
\Statex \textbf{Output:} Registers are modified and contain the sum of $(x_2,y_2)$ with $(x_1, y_1)$
\State $x_1 \leftarrow x_1- x_2$
\State $y_1 \leftarrow y_1- y_2$
\State $x_1, y_1 \leftarrow x_1, y_1 \cdot x_1^{-1}$
\State $x_1 \leftarrow x_1 + 3x_2$
\State $x_1 \leftarrow x_1 - y_1^2$
\State $x_1, y_1 \leftarrow x_1, y_1 \cdot x_1$
\State $x_1 \leftarrow -x_1$
\State $y_1 \leftarrow y_1 - y_2$
\State $x_1 \leftarrow x_1 + x_2$
\end{algorithmic}
\end{algorithm}
\end{figure}

\begin{figure}[htb]
\begin{algorithm}[H]
\caption{Dialog computation.}\label{algo:dialog}
\begin{algorithmic}[1]
\Statex \textbf{Input:} $x$
\Statex \textbf{Output:} Dialog vector $\vec{g}$

\State $(u,v) := (q,x)$
\For{$i$ in range($N_{it}$)}
\State $b_0 := v \bmod{2}$
\State $b_1 := u > v$
\State Store $b_0 \& b_1$ and $b_0$ in $\vec{g}$
\If{$b_0 \& b_1$}
\State $u,v = v,u$
\EndIf
\If{$b_0$}
\State $v -= u$ \Comment{Integer subtraction}
\EndIf
\State $v = v/2$ \Comment{Built-in by swapping qubit indices}
\EndFor \Comment{At the end, $v = 0$ and $u = 1$}
\end{algorithmic}
\end{algorithm}
\end{figure}

\begin{figure}[htb]
\begin{algorithm}[H]
\caption{Replay of the dialog}\label{algo:replay}
\begin{algorithmic}[1]
\Statex \textbf{Input:} Integer $y$, dialog $\vec{g}$
\Statex \textbf{Output:} Replaces $y$ by $y x^{-1} \bmod{q}$

\State $r,s \leftarrow y, 0$
\For{$i$ in reversed(range($N_{it}$))}
\State Obtain $b_0$ and $b_0 \& b_1$ for iteration $i$ from $\vec{g}$
\State $s \leftarrow 2s \bmod{q}$
\If{$b_0$}
\State $s \leftarrow s + r \bmod{q}$
\EndIf
\If{$b_0 \& b_1$}
\State $r, s \leftarrow s, r$
\EndIf
\EndFor
\State $(r,s) = (0, y x^{-1} \bmod{q})$
\end{algorithmic}
\end{algorithm}
\end{figure}

\begin{figure}[tb]
  \centering
  \includegraphics[width=\columnwidth]{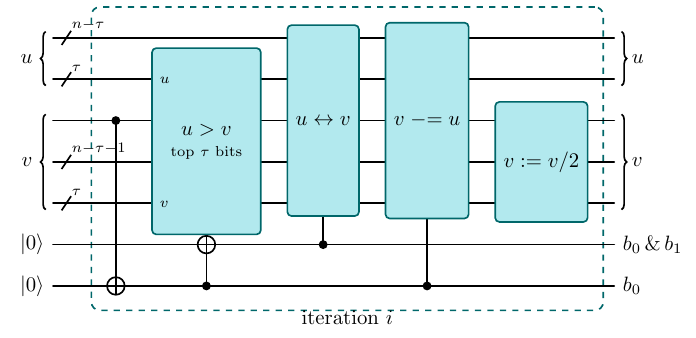}
  \caption{One iteration $i$ of the dialog construction, pass~1 of
    Algorithm~\ref{algo:dialog}.}\label{fig:dialog}
\end{figure}

\begin{figure}[tb]
  \centering
  \includegraphics[width=\columnwidth]{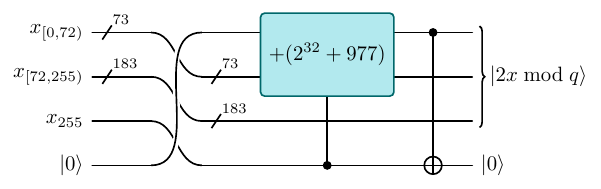}
  \caption{Optimized modular doubling for $q = 2^{256} - (2^{32}+977)$ with 40 bits of padding.}\label{fig:mod_double}
\end{figure}

\begin{figure*}[t]
  \centering
  \includegraphics[width=\linewidth]{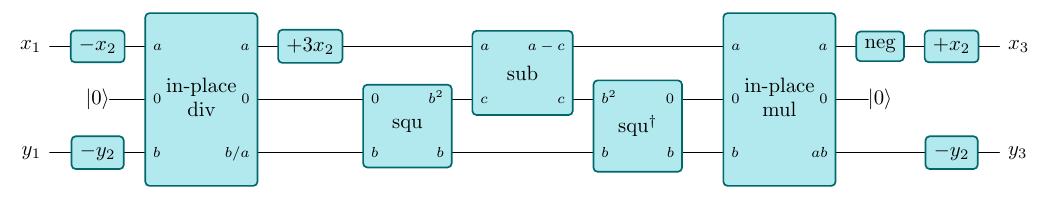}
  \caption{Arithmetic circuit for an elliptic-curve point addition
    (Algorithm~\ref{algo:point-add}). The small labels give the register contents before and
    after each block. Colors are as in Figure~\ref{fig:full_shor_circuit}.
    }\label{fig:ec-point-addition}
\end{figure*}

In this algorithm, the constant $(x_2,y_2)$ is not hard-coded into the circuit: it is loaded in memory using three \Lookup{}s. The first one loads $(x_2,y_2)$ at the beginning of the algorithm, the last one loads $(x_2,y_2)$ at the end, and the middle one loads $3x_2$ only. In the case of~\cite{Babbush26ECC,Schrottenloher26PointAdd}, there are enough ancilla qubits available at the time of the \Lookup{}s to neglect their effect on the total qubit count. This is not the case in~\cite{LuoEtAl26}, but in the latter, one can look up each coordinate individually instead with only a minor impact on the total gate count.

The implementation of modular subtraction and negation also has a minor impact on the gate count, so in the following we focus on in-place modular multiplication ($x_1, y_1 \leftarrow x_1, y_1 \cdot x_1$) and out-of-place squaring ($x_1 \leftarrow x_1 - y_1^2$). Both use the same ancilla qubits, as shown in~\autoref{fig:ec-point-addition}.

\subsubsection{Out-of-place squaring}

Following~\cite{Schrottenloher26PointAdd}, out-of-place squaring is implemented by a sequence of modular doubling and modular adders.

\subsubsection{In-place modular inverse using dialogs}

Modular inverse using dialogs is a desynchronized variant of the binary extended GCD. A fixed number of iterations $N_{it} = 1.413 n + O(\sqrt{n})$ is selected, where the choice of the constant in the $O$ determines the success probability of the algorithm. On input $(x,y)$, the algorithm runs in three phases, which are implemented by three circuits: 1. computation of the dialog (Algorithm~\ref{algo:dialog}, transforms $x$ in-place into the \emph{dialog} bit-vector $\mathbf{g}$); 2.  \emph{replay} of the dialog (Algorithm~\ref{algo:replay}, transforms $y$ in-place into $y x^{-1} \bmod{q}$); 3. uncomputation of the dialog (reverse of the first step). Note that~\cite{Schrottenloher26PointAdd} does not use the term ``dialog'' but we keep it from~\cite{Khattar25DQI}, since this is the same method.

The computation of the dialog follows a binary GCD algorithm (not extended), in which the input $x$ is consumed and the dialog bit-vector $\vec{g}$ is produced. Each iteration records two Boolean values $(b_0, b_0 \& b_1)$ which indicate which controlled operations were performed.

As described in~\cite{Schrottenloher26PointAdd}, an important optimization of the logical circuit is that the registers $u,v$ are truncated throughout the iterations. Their most significant bits are freed and reused as storage for $g$. This optimizes the total qubit count and the gate count. The (controlled) comparison used to compute $b_0 \& b_1$ is only performed on the $\tau$ most significant bits of $u$ and $v$. 

\begin{figure*}[t]
  \centering
  \includegraphics[width=0.75\linewidth]{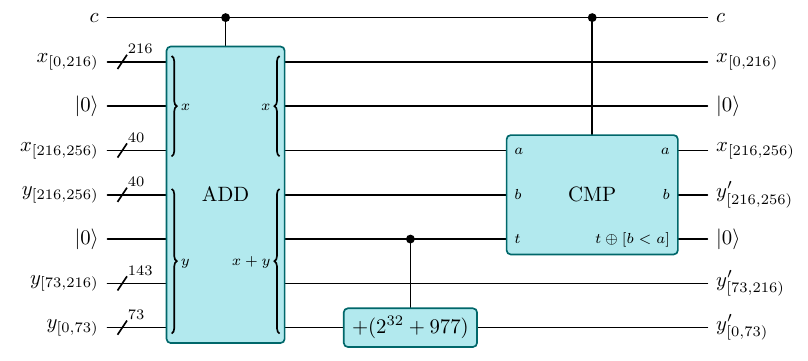}
  \caption{Optimized controlled modular addition for $q = 2^{256} - (2^{32}+977)$ with 40 bits of padding. The two input ancillae $\ket{0}$ are where the MSBs of the ADD output are written (the first one remains 0, the second one becomes 1 if there is an overflow). The circuit computes $y' = y + c\cdot x \bmod q$.
  }\label{fig:mod_cadd}
\end{figure*}

\begin{figure*}[t]
  \centering
  \includegraphics[width=0.9\linewidth]{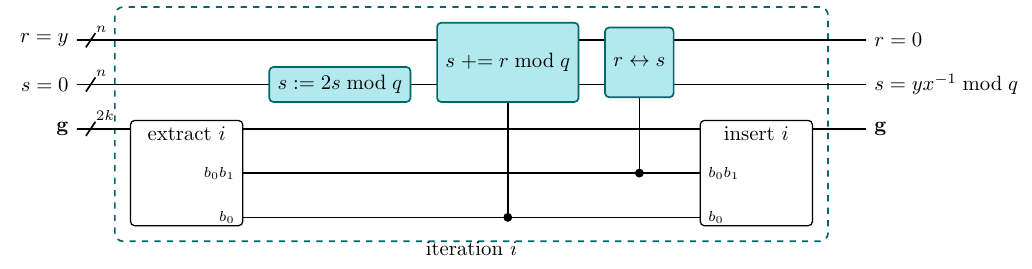}
  \caption{One step $i$ of the dialog replay, pass~2 of Algorithm~\ref{algo:replay}.
    Extract~$i$ moves the two bits recorded at iteration~$i$, $b_0$ and $b_0\wedge b_1$
    (written $b_0b_1$), out of the dialog $\mathbf{g}$, and insert~$i$ puts them back, so the
    replay leaves $\mathbf{g}$ unchanged. The steps run from the last iteration to the first.
    Starting from $(r,s)=(y,0)$, the replay ends at $(0,\,yx^{-1}\bmod q)$. The dialog is
    erased afterwards by running pass~1 (Figure~\ref{fig:dialog}) backwards, which also
    restores $(u,v)$.}\label{fig:dialog_apply}
\end{figure*}

The replay step uses the dialog bits to reconstruct the Bézout coefficients from the extended GCD, asynchronously (Figure~\ref{fig:dialog_apply}). In this algorithm, operations need to be performed modulo $q$.

In both algorithms, the storage of $(b_0, b_0\&b_1)$ into $\mathbf{g}$ should be optimized, in order to minimize space (eventually the storage of $\mathbf{g}$ during the replay is the dominating space of the circuit). \cite{Schrottenloher26PointAdd} does this with a \emph{compression}-\emph{decompression} circuit which packs three iterations into 5 bits. The gate count of this circuit is negligible throughout the algorithm.

\subsubsection{Additive modular operations}

While integer operations (subtraction and comparison) are immediately implemented, modular operations require another level of optimization. For these optimizations, we follow~\cite{Schrottenloher26PointAdd} again. Some of them are specific to the pseudo-Mersenne prime $q = 2^{256} - (2^{32}+977)$ used in the secp256k1 standard curve. These optimizations add some arithmetic errors in the circuit, which are also bounded away by choosing appropriate amounts of \emph{padding}. A padding of $\tau$ bits ensures that, assuming uniformly random inputs, an operation fails with probability $2^{-\tau}$.

\paragraph{Modular Doubling.}
The circuit for modular doubling, in the case of $q = 2^{256} - (2^{32}+977)$ and using 40 bits of padding, is represented in Figure~\ref{fig:mod_double}. The value of $x \bmod{q}$ is first shifted by one bit, and the modular reduction is performed if the most significant bit is 1. It is approximated by simply adding $(2^{32}+977)$ to the least significant bits (this number has 33 bits, and the window is expanded by 40 bits to allow for carry propagation). The most significant bit is erased by XORing the least significant bit (which is 1 if and only if a modular reduction occurred).

\paragraph{Controlled Modular Addition.}
The controlled modular adder for $q = 2^{256} - (2^{32}+977)$ follows the same principle and can also use 40 bits of padding; it is represented in Figure~\ref{fig:mod_cadd}. On this figure the bits of the inputs $x$ and $y$ have been reordered for clarity. A controlled adder on 257 bits is first performed, then the MSB of $y$ controls the modular reduction (performed by adding $2^{32} + 977$ to the least significant bits of $y$). A comparison between $x$ and $y$, which uses only their $40$ MSBs, erases the MSB of $y$.

This circuit works only if the inputs $x,y$ are taken uniformly at random. A slightly more costly variant of this circuit is used in~\cite{Schrottenloher26PointAdd} for the 50 first iterations of the replay step, in which the input may satisfy $x + y = q$ with high probability (this corresponds to ``empty'' dialog steps, needed since the dialog runs for a fixed number of iterations regardless of its input).

\paragraph{Constant addition.}
The additions by the constant $(2^{32}+977)$ are performed using Gidney's  low-space constant adder~\cite{gidney2025classical}.

\subsubsection{Logical cost breakdown}

Tables~\ref{tab:ecdlp-qubits-a} and~\ref{tab:ecdlp-qubits-b} give, for each step of the
windowed point addition, the number of calls over a whole run of Shor's algorithm and the
logical qubit count, in the same format as Table~\ref{tab:logical_qubit_count} for
factoring.

\begin{table*}[tp]
\centering
\footnotesize
\begin{tabular}{|l|c|c|c|c|}
\hline
Subroutine & Calls & Operation & Subroutine scope & Out of scope \\
\hline
$c \leftarrow (i \neq 0)$ & $4N_A$ & $\mathrm{MC}X(w)$ & $2n$ & $0$ \\
\hline
\multirow{6}{*}{$-x_2,\ -y_2$}
 & $N_A$  & \Lookup($w, 2n$)                & $2n$              & \multirow{6}{*}{$0$} \\
 & $2N_A$ & Q-Q \Adder$^{\dagger}(n{+}1)$ & $2n+w-2$          & \\
 & $2N_A$ & Q-C \Adder($\varphi$)           & $4n+w-\varphi$    & \\
 & $2N_A$ & Q-Q \Comparison($\mu$)          & $4n+w-2\mu$       & \\
 & $N_A$  & $M_X(2n)$                      & $2n+w$            & \\
 & $N_A$  & \Phaseup($w$)                   & $2n$              & \\
\hline
\multirow{15}{*}{\begin{tabular}{@{}l@{}}inplace div\\ inplace mul\end{tabular}}
 & $4N_A$                & $P_{|0\rangle}(B-2n)$                                & $3n+2$                    & \multirow{15}{*}{$w$} \\
 & $4N_A N_{\rm it}$     & $\CX(1)$                                              & $B+n$                     & \\
 & $4N_A N_{\rm it}$     & Q-Q \Comparison($\tau$) (ctrl $b_0$)                  & $B+n-2\tau$               & \\
 & $4N_A N_{\rm it}$     & $\C\SWAP(\nu_j)$ (ctrl $b_0\&b_1$)                     & $B+n+1-2\nu_j$            & \\
 & $4N_A N_{\rm it}$     & Q-Q \Adder$^{\dagger}(\nu_j)$ (ctrl $b_0$)          & $B+n+1-2\nu_j$            & \\
 & $4N_A N_{\rm it}$     & \textsc{Absorber}                                             & $B+n-5$                   & \\
\cline{2-4}
 & $2N_A$                & $P_{|0\rangle}(n)$                                   & $G+n+a$                   & \\
 & $4N_A N_{\rm it}$     & \textsc{Swapper}                                              & $G+2n+a-8$                & \\
 & $2N_A N_{\rm it}$     & Q-C \Adder($\varphi_R$) (ctrl)                        & $G+2n+a-\varphi_R-1$      & \\
 & $2N_A N_{\rm it}$     & Q-Q \Adder$(n{+}1)$ (ctrl $b_0$)                  & $G+a-3$                   & \\
 & $2N_A N_{\rm it}$     & Q-C \Adder($\varphi_R$) (ctrl)                        & $G+2n+a-\varphi_R-1$      & \\
 & $2N_A N_{\rm it}$     & Q-Q \Comparison($\mu_R$) (ctrl $b_0$)                 & $G+2n+a-2\mu_R-2$         & \\
 & $2N_A N_{\rm it}$     & $\C\SWAP(n)$ (ctrl $b_0\&b_1$)                         & $G+a-1$                   & \\
 & $4N_A N_Q$            & $\mathrm{MC}X(\mu_R)$                                         & $G+2n+a-\mu_R-1$          & \\
 & $2N_A N_Q$            & $\CX(n)$                                              & $G+n+a-1$                 & \\
\hline
\end{tabular}
\caption{Symbolic logical qubit count at each point of the windowed elliptic
curve point addition circuit (first half, continued in
Table~\ref{tab:ecdlp-qubits-b}), and number of calls for each operation over a whole run of Shor's algorithm.
}
\label{tab:ecdlp-qubits-a}
\end{table*}

\begin{table*}[tp]
\centering
\footnotesize
\begin{tabular}{|l|c|c|c|c|}
\hline
Subroutine & Calls & Operation & Subroutine scope & Out of scope \\
\hline

\multirow{6}{*}{$+3x_2$}
 & $N_A$ & \Lookup($w, n$)            & $n$              & \multirow{6}{*}{$n$} \\
 & $N_A$ & Q-Q \Adder$(n{+}1)$    & $w-2$            & \\
 & $N_A$ & Q-C \Adder($\varphi$)      & $2n+w-\varphi$   & \\
 & $N_A$ & Q-Q \Comparison($\mu$)     & $2n+w-2\mu$      & \\
 & $N_A$ & $M_X(n)$                  & $n+w$            & \\
 & $N_A$ & \Phaseup($w$)              & $n$              & \\
\hline

\multirow{6}{*}{\begin{tabular}{@{}l@{}}square-subtract\\ (ctrl $c$)\end{tabular}}
 & $2N_A(n{-}1)$ & Q-C \Adder($\varphi_S$) (ctrl)              & $2n+1-\varphi_S$   & \multirow{6}{*}{$w$} \\
 & $2N_A(n{-}1)$ & $\CX(1)$                                    & $2n$               & \\
 & $2N_A n$      & \textsc{And} (ctrl $c$)                             & $2n-1$             & \\
 & $N_A n$       & Q-Q \Adder$(n{+}1)$ (ctrl)              & $1$                & \\
 & $N_A n$       & Q-C \Adder($\varphi_S$) (ctrl)              & $2n+2-\varphi_S$   & \\
 & $N_A n$       & Q-Q \Comparison($\mu_S$) (ctrl)             & $2n+1-2\mu_S$      & \\
\hline

\multirow{2}{*}{neg (ctrl $c$)}
 & $N_A$ & $\CX(n)$ (ctrl $c$)              & $0$          & \multirow{2}{*}{$n+w$} \\
 & $N_A$ & Q-C \Adder($\varphi$) (ctrl $c$) & $n-\varphi$  & \\
\hline

\multirow{6}{*}{$+x_2,\ -y_2$}
 & $N_A$  & \Lookup($w, 2n$)                & $2n$              & \multirow{6}{*}{$0$} \\
 & $2N_A$ & Q-Q \Adder$(n{+}1)$         & $2n+w-2$          & \\
 & $2N_A$ & Q-C \Adder($\varphi$)           & $4n+w-\varphi$    & \\
 & $2N_A$ & Q-Q \Comparison($\mu$)          & $4n+w-2\mu$       & \\
 & $N_A$  & $M_X(2n)$                      & $2n+w$            & \\
 & $N_A$  & \Phaseup($w$)                   & $2n$              & \\
\hline

\multirow{2}{*}{Window recycling}
 & $N_A$ & $M_X(w)$            & $2n$ & \multirow{2}{*}{$0$} \\
 & $N_A$ & $P_{|0\rangle}(w)$  & $2n$ & \\
\hline
\end{tabular}
\caption{Symbolic logical qubit count at each point of the windowed elliptic
curve point addition circuit (second half, continuing
Table~\ref{tab:ecdlp-qubits-a})}.
\label{tab:ecdlp-qubits-b}
\end{table*}
    \clearpage

\section{Resource estimation for RSA-2048}\label{sec:resource_estim}

\subsection{Architecture} \label{ssec:architecture}

\subsubsection{Layout, connectivity and noise} \label{sssec:layout_connectivity}
In the repetition-cat architecture, data qubits lie on a square lattice and ancillae at the centers of the corresponding plaquettes (Figures~\ref{fig:adder_global_layout} and~\ref{fig:adder_routing_overlays}).
Outside the factories, the physical operations are preparation in $\ket{0}$ or $\ket{+}$, \textsc{Cnot} gates between neighboring qubits, and $X$ or $Z$ measurements.
Each ancilla is coupled to the four data qubits at the corners of its plaquette, such that every qubit has degree four connectivity.
Transversal \textsc{Cnot} gates between repetition codes laid out in adjacent columns require \textsc{Cnot} gates between horizontally neighboring data qubits, which are not coupled.
We emulate them with the degree-four connectivity.
Each \textsc{Cnot} gate between two horizontally neighboring data qubits is mediated by an ancilla adjacent to both. The ancilla is prepared in $\ket{0}$, receives a \textsc{Cnot} gate from the control, applies a \textsc{Cnot} gate onto the target, and is measured in the $X$ basis. The outcome is absorbed as a $Z$ correction on the control in the Pauli frame.
This sequence has the structure, and hence the duration, of a stabilizer round, and each single fault still causes at most one phase flip per repetition code.
We assume that all physical qubits are identical, such that any qubit of this lattice can serve as a data qubit or an ancilla.

The LDPC-cat architecture~\cite{Ruiz_2025} is a flip chip of two facing layers (Figure~\ref{fig:adder_ldpc_multitarget_cnot_example}).
The \emph{repetition layer} has the lattice and connectivity above.
The \emph{LDPC layer} holds the data qubits of an LDPC-cat code (Section~\ref{sssec:error_correcting_codes}) on the same square lattice, and one ancilla per check, coupled to the at most four data qubits of its check, which fit within a $3\times3$ grid of neighboring data qubits.
Each data qubit of the LDPC layer is coupled to at most four ancillae, and to the facing qubit of the repetition layer.
The two layers do not correspond to two monolithic chips.
To scale to the required number of qubits, the device would be split into chiplets, coupled either directly across separate dies~\cite{Gold_2021} or through cryogenic microwave links~\cite{Magnard_2020,Sundaresan_2026}.
Such links are noisier than on-chip couplings and could therefore add an overhead, which recent works~\cite{Jacinto_2026, Ramette_2023} suggest can remain small given a sufficient rate of shared Bell pairs and a remote link infidelity of the order of a percent.
The partitioning into chiplets and the cost of inter-chip links are beyond the scope of this work.

We denote the durations of a transversal \textsc{Cnot} gate, local initialization, destructive measurement, and one stabilizer round by $T_{\textsc{Cnot}}$, $T_{\mathrm{init}}$, $T_{\mathrm{meas}}$, and $T_{\mathrm{round}}$, respectively.
One stabilizer round defines an error-correction cycle, $T_{\mathrm{round}} = 1$~cycle $=$ \SI{1}{\micro\second}.
A logical measurement outcome is available for feedforward after the decoder reaction time $T_{\mathrm{react}}$.
We take $T_{\textsc{Cnot}} = T_{\mathrm{init}} = T_{\mathrm{meas}} = 1$ cycle and $T_{\mathrm{react}} = 10$ cycles.
These durations hold in both layers, including for transversal inter-layer \textsc{Cnot} gates.
Durations are rounded up to whole cycles.
Moreover, we assume a noise model neglecting bit flips.
At the main point of Section~\ref{ssec:physical_resource_count} (code distance $15$), one RSA-2048 shot involves about $1.2\times10^{15}$ physical qubit-cycles ($49\,649$ qubits $\times$ $2.49\times10^{10}$ error-correction cycles), or a bit-flip probability well below $10^{-15}$ per qubit and cycle.
We treat this as a working hypothesis (Section~\ref{sec:cat_qubits}), which we quantify in Section~\ref{sssec:bitflip_assumptions} and mitigate in Section~\ref{ssec:elevator_code}.
Phase flips follow the generic circuit-level noise model of~\cite[Supplementary Table~1(b)]{Ruiz_2025}, with a single parameter $p_Z$. 
Each initialization in $\ket{+}$ is followed by a $Z$ error with probability $p_Z$, and each $X$-basis measurement outcome is flipped with probability $p_Z$. 
Each \textsc{Cnot} gate is followed by $Z_c$, $Z_t$ or $Z_cZ_t$ with probability $p_Z/3$ each, and every qubit idle during an operation undergoes a $Z$ error with probability $p_Z$.
Phase flips have no effect on preparations in $\ket{0}$ or on $Z$-basis measurements.
In the following, we consider two phase-flip probabilities, $p_Z = 10^{-3}$ and $p_Z = 5\times10^{-4}$, and unless stated otherwise, use the latter.

\subsubsection{Zones} \label{sssec:zones_factories}
The chip is divided into zones whose number and arrangement depend on the subroutine and on the architecture: \emph{compute}, \emph{buffer}, \emph{routing} and \emph{factory} zones for the adder (Figure~\ref{fig:adder_global_layout}), and further \emph{exponent} and \emph{memory} zones in the complete layouts (Figures~\ref{fig:global_layout} and~\ref{fig:global_layout_ldpc}).
A zone is a region of the repetition layer, extended on the LDPC-cat architecture by the facing region of the LDPC layer (Section~\ref{sssec:error_correcting_codes}).
The states $\ket{\CXCXX}$ are produced by a two-level factory with a shared first-level production stage~\cite{FeMoco2026prep} and $\ket{i}$ states from color code factories~\cite{FeMoco2026prep,Bombin_2006,Ruiz_2026} that store their produced states alternately into two output repetition codes.
Their parameters are given in Table~\ref{tab:adder_reference_factories}.

\subsubsection{Error correcting codes} \label{sssec:error_correcting_codes}
In the repetition-cat architecture, logical qubits are stored in repetition codes of distance $d$, each made of $d$ data qubits and $d-1$ ancillae ($2d-1$ physical qubits).
The stabilizers of each code are $X_iX_{i+1}$, its codewords are $\ket{\pm}_L \coloneqq \ket{\pm}^{\otimes d}$, and its logical operators are $X_L = X_1$, which has weight one, and $Z_L = Z^{\otimes d}$, which has weight $d$.
The logical preparation $\ket{+}_L$ of a repetition code is performed at the physical level, by initializing its data qubits in $\ket{+}$ in $T_{\mathrm{init}}$.
The preparation $\ket{0}_L$ also initializes the data qubits, in $\ket{0}$, in $T_{\mathrm{init}}$.
Its first stabilizer outcomes are random and only define a Pauli frame.
Establishing it takes about $d$ rounds, which must precede any use of an $X$-measurement outcome of the code, or of a code coupled to it by transversal \textsc{Cnot} gates.
When a following operation provides these rounds, the preparation adds no time to them.
Otherwise, it lasts $d$ error-correction cycles.
On the repetition-cat architecture, in the color-changed convention (Section~\ref{ssec:colour_changed}), every $\ket{0}_L$ preparation is followed by a lattice surgery (with one harmless exception in the auto-\textsc{And} preparation, Section~\ref{sssec:resource_states}), so none lasts $d$.

In the LDPC-cat architecture, the LDPC layer stores logical qubits in an open-boundary LDPC-cat code of distance $d_{\mathrm{LDPC}}$ (Table~\ref{tab:adder_ldpc_codes}).
The region of the repetition layer facing it, the \emph{LDPC layer access zone}, provides access to both logical operators of every logical qubit of the LDPC layer.
The operator $X_L$ is supported on a single \emph{tip} qubit, which lies in the bottom rows of the code.
The facing qubit of the repetition layer serves as a \emph{leaf} ancilla and measures the $X$ parity of the tip and of a \emph{proxy} data qubit, through which $X_L$ is then accessible.
The operator $Z_L$, supported on at least $d_{\mathrm{LDPC}}$ qubits from the tip upward, is accessed through transversal inter-layer \textsc{Cnot} gates onto a repetition code covering its support.
In this architecture, when necessary, we distinguish the distance $d_{\mathrm{rep}}$ of the repetition codes of the repetition layer from the distance $d_{\mathrm{LDPC}}$ of the LDPC-cat code.
As is common in the literature~\cite{fowler2019lowoverheadquantumcomputation}, we count the duration of each lattice surgery step as
\begin{equation}\label{eq:surgery_duration}
    T_{\mathrm{surg}}=d\,T_{\mathrm{round}}.
\end{equation}
A lattice surgery confined to the repetition layer lasts $T_{\mathrm{surg}}$ of Equation~\eqref{eq:surgery_duration} with $d=d_{\mathrm{rep}}$, whereas a lattice surgery involving the LDPC layer conservatively lasts
\begin{equation}\label{eq:Tsurg-ldpc}\csname ltx@label\endcsname{eq:ldpc-dstar}    T^{(\mathrm{LDPC})}_{\mathrm{surg}}=d^{\star}\,T_{\mathrm{round}},\qquad d^{\star}=\max(d_{\mathrm{rep}},d_{\mathrm{LDPC}}),
\end{equation}
such that its timelike distance is at least the spatial distance of every code that takes part in it.
In particular, an access to $X_L$ through its leaf ancilla, or to $Z_L$ through a covering repetition code that receives the transversal inter-layer \textsc{Cnot} gates during its first stabilizer round, lasts $T^{(\mathrm{LDPC})}_{\mathrm{surg}}$.

\subsection{Lattice-surgery primitives} \label{ssec:gate_gadgets}

\subsubsection{Parity-\textsc{Cnot} gate on cat qubits} \label{sssec:parity_controlled_not}

Given $N\ge 1$ and $x_1,\ldots,x_N,y\in\{0,1\}$, the parity-\textsc{Cnot} gate is defined by the following operation:
\begin{equation}
\begin{aligned}
    &\mathrm{PC}X(N)\,\ket{x_1\ldots x_N}\ket{y}\\
    &\qquad=\ket{x_1\ldots x_N}\,\ket{x_1\oplus\cdots\oplus x_N \oplus y }.
\end{aligned}
\end{equation}

\begin{figure*}[ht]
    \centering
    \textbf{(a) Parity-\textsc{Cnot} gate}\par\smallskip
    \begin{subfigure}[b]{0.4\linewidth}
        \centering
        \raisebox{-0.5\height}{\includegraphics[scale=0.9,page=1]{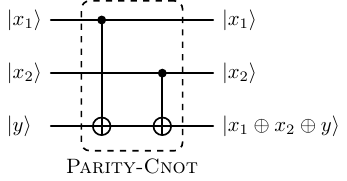}}
    \end{subfigure}
    \hfill
    \Large$=$\normalsize
    \hfill
    \begin{subfigure}[b]{0.5\linewidth}
        \centering
        \raisebox{-0.5\height}{\includegraphics[scale=0.9,page=2]{assets/Parity_controlled_NOT/parity_cnot_circuit.pdf}}
    \end{subfigure}

    \par\smallskip
    \includegraphics[width=\linewidth]{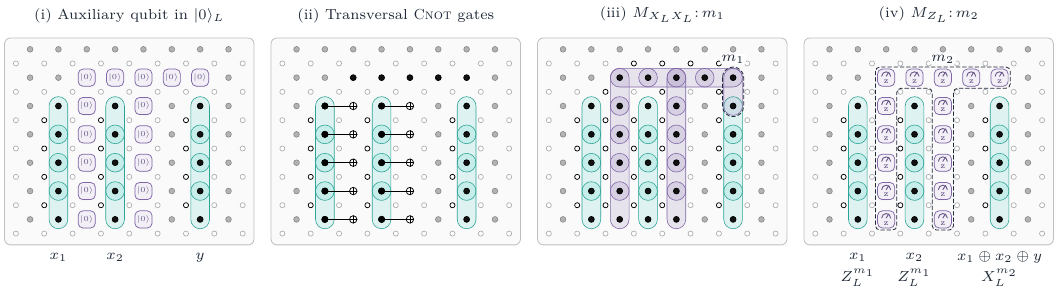}
    \par\bigskip
    \textbf{(b) Multi-target \textsc{Cnot} gate}\par\smallskip
    \begin{subfigure}[b]{0.4\linewidth}
        \centering
        \raisebox{-0.5\height}{\includegraphics[scale=0.9,page=1]{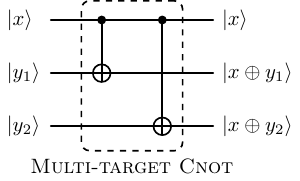}}
    \end{subfigure}
    \hfill
    \Large$=$\normalsize
    \hfill
    \begin{subfigure}[b]{0.5\linewidth}
        \centering
        \raisebox{-0.5\height}{\includegraphics[scale=0.9,page=2]{assets/multitarget_cnot/multitarget_cnot_circuit.pdf}}
    \end{subfigure}

    \par\smallskip
    \includegraphics[width=\linewidth]{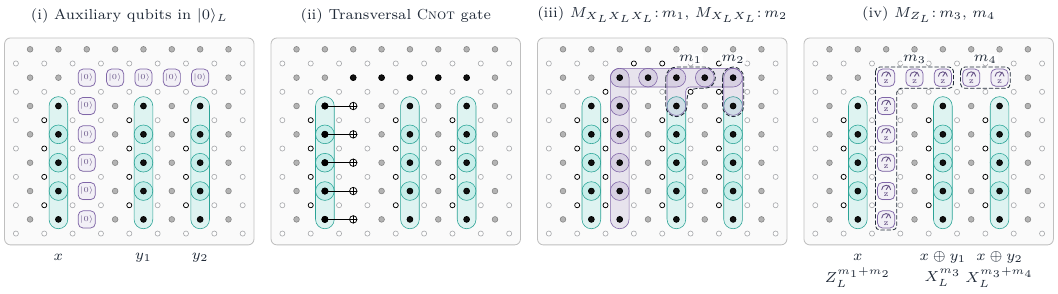}
    \par\medskip
    \includegraphics[width=\linewidth]{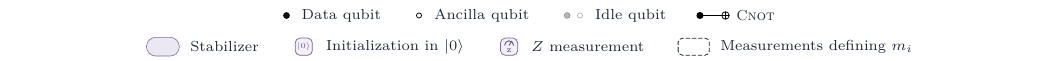}

   \caption{Implementation of the parity-\textsc{Cnot} gate (a) and of the multi-target \textsc{Cnot} gate (b) on the repetition-cat architecture. Top of each panel: circuit identity used to teleport the logical gate. Bottom: physical mapping. 
   (i)~The auxiliary qubits are prepared in $\ket{0}_L$ (Section~\ref{sssec:error_correcting_codes}). In practice, each data qubit (in purple online) is initialized in $\ket{0}$. Since their stabilizers commute with the transversal \textsc{Cnot} gates of (ii), they are measured for the first time in (iii), to save time. 
   (ii)~Transversal \textsc{Cnot} gates from each control to the neighboring branch of the auxiliary qubit. 
   (iii)~A lattice surgery merge measures $X_LX_L$ (a), or $X_LX_LX_L$ and $X_LX_L$ (b), on the auxiliary qubits and the targets, $m_1$ (a), or $m_1$ and $m_2$ (b), being the outcome of the corresponding dashed check. In (b), the last two routing qubits form the second auxiliary qubit.
   (iv)~The auxiliary qubits are measured in the $Z$ basis and Pauli corrections are applied in the Pauli frame.}
     \label{fig:cnot_gadgets_on_cat}
\end{figure*}

The repetition-cat implementation, shown in Figure~\ref{fig:cnot_gadgets_on_cat}(a), is adapted from the long-range \textsc{Cnot} scheme of~\cite{GouzienCatQubits23}.
\textsc{Cnot} gates applied from $x_1,\ldots,x_N$ onto an auxiliary qubit prepared in $\ket{0}_L$ collect the parity $x_1\oplus\cdots\oplus x_N$, and an $XX$ parity measurement between the auxiliary qubit and the target, followed by a destructive $Z$ measurement of the auxiliary qubit, completes the parity-\textsc{Cnot} gate, up to Pauli corrections.
This operation is performed in approximately one lattice surgery step, or $T_{\mathrm{surg}}$ cycles (Equation~\eqref{eq:surgery_duration}), independently of $N$.

The gate layout is asymmetric. 
It accesses each control through the logical $Z_L$ representative, supported on all the $d$ data qubits, and the target through the logical $X_L$ representative, supported on any single data qubit.
The auxiliary qubit is therefore encoded in a tree repetition code with one branch along each control, which reaches the target at a single qubit (Figure~\ref{fig:cnot_gadgets_on_cat}(a)).
Transversal \textsc{Cnot} gates from the $d$ data qubits of a repetition code onto any $d$ neighboring data qubits of a tree repetition code implement a logical \textsc{Cnot} gate in the same direction.
As such, the $N$ \textsc{Cnot} gates act on disjoint qubits and are applied in a single step.
Additional controls therefore cost about $d$ routing data qubits each, but no additional time.

The scheme generalizes to the LDPC-cat architecture~\cite{FeMoco2026prep}, where the auxiliary qubit is encoded in a snake repetition code covering, in the repetition layer, the support of the product of the LDPC logical representatives $Z_L^{x_1},\ldots,Z_L^{x_N}$ of the controls.
The \textsc{Cnot} gates are then transversal inter-layer gates, applied in a single step, and the snake is merged with the target through the $XX$ measurement, then measured out in the $Z$ basis.

\subsubsection{Multi-target \textsc{Cnot} gate on cat qubits} \label{sssec:multitarget_cnot}

Similarly, given $x,y_1,\ldots,y_N\in\{0,1\}$, the multi-target \textsc{Cnot} gate implements the following operation:
\begin{equation}
\begin{aligned}
    &\CX(N)\,\ket{x}\ket{y_1\ldots y_N}\\
    &\qquad=\ket{x}\,\ket{x\oplus y_1, \ldots, x\oplus y_N}.
\end{aligned}
\end{equation}
By the asymmetry of the repetition code, the multi-target \textsc{Cnot} gate, presented in~\cite{GouzienCatQubits23}, is implemented differently from the parity-\textsc{Cnot} gate.
A \textsc{Cnot} gate is applied from $x$ onto an auxiliary qubit prepared in $\ket{0}_L$, and an $XX\cdots X$ parity measurement between the auxiliary qubit and the targets, followed by a destructive $Z$ measurement of the auxiliary qubit, completes the multi-target \textsc{Cnot} gate, up to Pauli corrections.

The gate accesses the control once, through the logical $Z_L$ representative supported on all the associated $d$ data qubits, and each target only through its logical $X_L$ representative supported on any single data qubit.
On the repetition-cat architecture, a single lattice surgery merge measures the parity~\cite{GouzienCatQubits23} (Figure~\ref{fig:cnot_gadgets_on_cat}(b)).
Additional targets therefore cost routing, which scales independently of $d$, but does not require additional time.

\begin{figure*}[t]
    \centering
    \includegraphics[width=\linewidth]{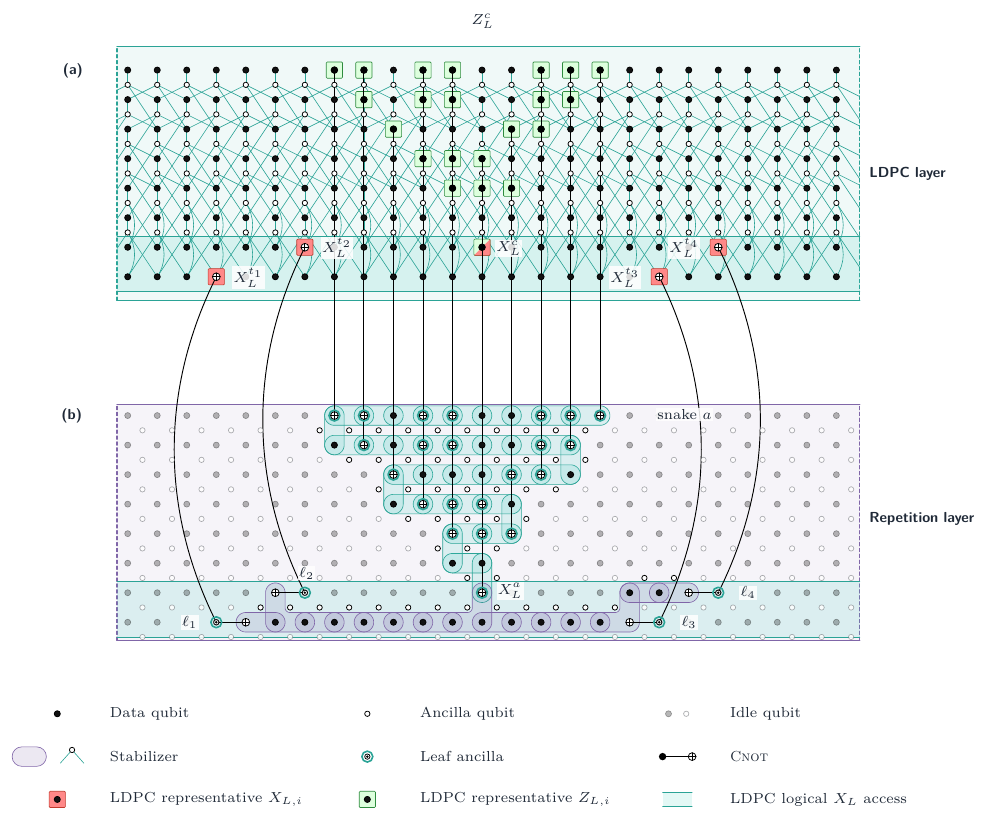}
\caption{Multi-target \textsc{Cnot} gate on the LDPC-cat architecture, from a control $c$ onto targets $t_1,\ldots,t_4$, in one lattice surgery step.
(a)~LDPC layer, with a logical representative $Z_L^c$ of the control and the weight-one logical representatives $X_L^c$ and $X_L^{t_i}$.
(b)~Repetition layer, with a snake repetition code $a$ covering the support of $Z_L^c$, and a fork code jointly measured at $X_L^a$, reaching each target through a leaf ancilla $\ell_i$.
The snake repetition code is prepared in $\ket{0}_L$ and the fork qubits in $\ket{0}$. 
The snake then receives the transversal inter-layer \textsc{Cnot} gates from the support of $Z_L^c$ in the second cycle.
Finally, the lattice surgery merge measures $X_L^aX_L^{t_1}\cdots X_L^{t_4}$, after which the snake and the fork are measured in the $Z$ basis, and the Pauli corrections are applied in the Pauli frame.}
    \label{fig:adder_ldpc_multitarget_cnot_example}
\end{figure*}

\begin{figure*}[t]
    \centering
    \includegraphics[width=0.95\linewidth]{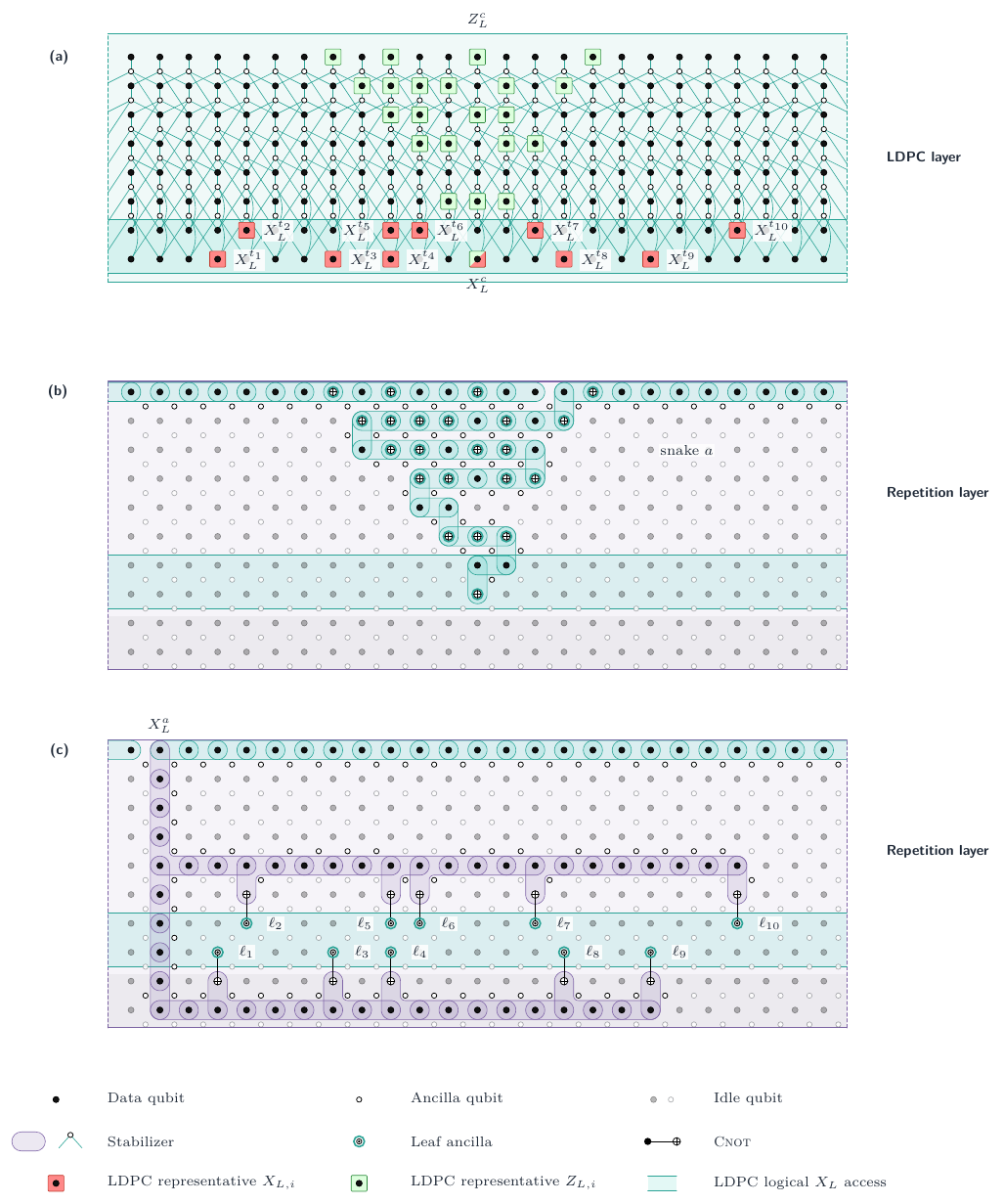}
  \caption{Multi-target \textsc{Cnot} gate on the LDPC-cat architecture, from a control $c$ onto targets $t_1,\ldots,t_{10}$, in approximately two lattice surgery steps and two extra rows in the repetition layer (grey).
(a)~LDPC layer, with logical representatives $Z_L^c$, $X_L^c$ and $X_L^{t_i}$.
(b)~During the first lattice surgery step, the snake $a$, prepared in $\ket{0}_L$, covers the support of $Z_L^c$ and the top row of the repetition layer, the tail, and receives the transversal inter-layer \textsc{Cnot} gates from the support of $Z_L^c$ in the second cycle.
(c)~During the second lattice surgery step, after $Z$ measurements reduce the snake to its tail, the fork code, prepared in $\ket{0}_L$, joins the tail at $X_L^a$ and measures $X_L^aX_L^{t_1}\cdots X_L^{t_{10}}$ through one leaf ancilla $\ell_i$ per target.
The tail and the fork are then measured in the $Z$ basis, and the Pauli corrections are applied in the Pauli frame.}
    \label{fig:adder_ldpc_multitarget_cnot_general_case}
\end{figure*}

The scheme extends to the LDPC-cat architecture, where the auxiliary qubit is encoded in a snake repetition code covering, in the repetition layer, the support of the LDPC logical representative $Z_L^c$ of the control $c$.
The \textsc{Cnot} gate is then a transversal inter-layer gate, and the parity measurement reaches the targets through a fork code (Figures~\ref{fig:adder_ldpc_multitarget_cnot_example} and~\ref{fig:adder_ldpc_multitarget_cnot_general_case}).
If the fork operation can be performed adjacent to the snake, the gate requires approximately a single lattice surgery step (Figure~\ref{fig:adder_ldpc_multitarget_cnot_example}).
Otherwise, the snake is reduced, which requires an additional lattice surgery step and frees space for the fork measurement (Figure~\ref{fig:adder_ldpc_multitarget_cnot_general_case}).

\subsubsection{Active injection} \label{sssec:active_injection}

\begin{figure*}[t]
    \centering
    \begin{subfigure}[b]{0.33\linewidth}
        \centering
        \raisebox{-0.5\height}{\includegraphics[scale=0.8,page=1]{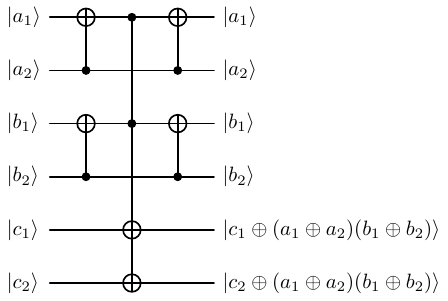}}
    \end{subfigure}
    \hfill
    \Large$=$\normalsize
    \hfill
    \begin{subfigure}[b]{0.62\linewidth}
        \centering
        \raisebox{-0.5\height}{\includegraphics[scale=0.8,page=2]{assets/active_injection/active_injection_circuit.pdf}}
    \end{subfigure}

    \par\medskip
    \includegraphics[width=0.82\linewidth]{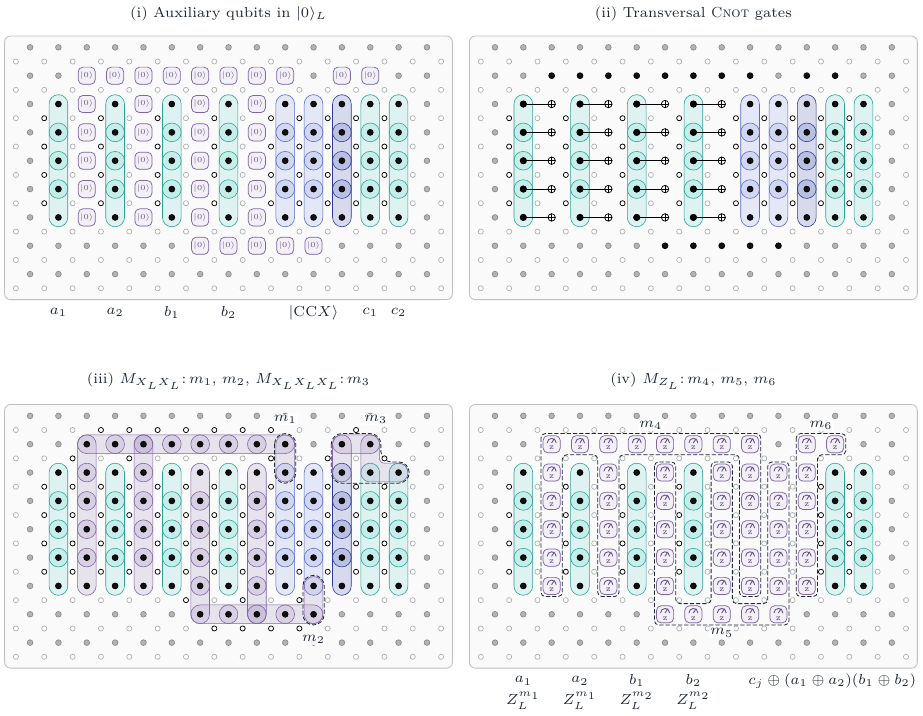}
    \par\medskip
    \includegraphics[width=\linewidth]{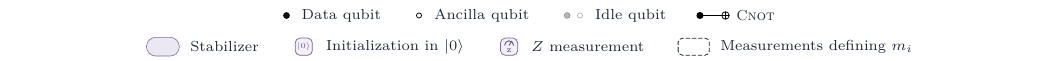}
    \caption{Active injection of $c_j\mapsto c_j\oplus(a_1\oplus a_2)(b_1\oplus b_2)$ for $j=1,2$, with parity-\textsc{Cnot} gates on the control qubits of a $\ket{\CCX}$ state and a multi-target \textsc{Cnot} gate on its target qubit.
    Top: logical operation and its injection, with a parity-\textsc{Cnot} gate onto each control qubit, the multi-target \textsc{Cnot} gate from the target qubit onto $c_1$ and $c_2$ followed by $M_X$ (dashed), implemented as $M_{XXX}$ and $M_Z$, and the Clifford corrections.
    Bottom: repetition-cat layout, with the control qubits of $\ket{\CCX}$ in blue and its target qubit in dark blue.
    (i)~Auxiliary qubits, the routing trees and the $M_{XXX}$ routing, prepared in $\ket{0}_L$.
    (ii)~Transversal \textsc{Cnot} gates onto the routing trees of the parity-\textsc{Cnot} gates.
    (iii)~One merge measures $X_LX_L$ between each tree and its control qubit of $\ket{\CCX}$ ($m_1$, $m_2$, parity-\textsc{Cnot} gates) and $X_LX_LX_L$ on the target qubit of $\ket{\CCX}$, $c_1$ and $c_2$ ($m_3$, multi-target \textsc{Cnot} gate).
    (iv)~$Z$ measurements of each tree and of the $M_{XXX}$ routing, whose outcomes enter those of the $\ket{\CCX}$ qubits they are merged with ($m_4$, $m_5$, $m_6$).}
    \label{fig:active_injection}
\end{figure*}

When a Toffoli gate is performed by injecting $\ket{\CCX}$ or another $\CCZ$-type state (i.e.\ any state obtained from $\ket{\CCZ}$ by applying a Hadamard gate to a subset of its three qubits), each injection \textsc{Cnot} gate can be replaced by a parity-\textsc{Cnot} gate or a multi-target \textsc{Cnot} gate.
Measuring multi-qubit parities is natural in lattice surgery~\cite{Litinski_2019,fowler2019lowoverheadquantumcomputation}. 
We denote this procedure, adapted from~\cite{gidney2025factor2048bitrsa}, by \emph{active injection}.
In the following, $\C_X$ denotes a control in the $X$ basis, so that $\ket{\C_X\C_X Z}\coloneqq H^{\otimes3}\ket{\CCX}$ and $\ket{\CXCXX}\coloneqq H^{\otimes3}\ket{\CCZ}$ are the color-changed counterparts of $\ket{\CCX}$ and $\ket{\CCZ}$.
Let $I,J,K$ be sets of logical qubits with $I,J\neq\emptyset$, and let $x_I$ denote the XOR of the qubits in $I$.
The operations
\begin{equation}
\begin{aligned}
    U_{\CCX}\,\ket{x}\ket{c}&=\ket{x}\,\ket{c\oplus x_I\cdot x_J\oplus x_K},\\
    U_{\CCZ}\,\ket{x}&=(-1)^{x_I\cdot x_J\cdot x_K}\ket{x}\quad(K\neq\emptyset),
\end{aligned}
\label{eq:active_injection}
\end{equation}
where the target $c$ is not among the inputs, can be implemented in the time of a single Toffoli (respectively $\CCZ$) injection by lattice surgery.
The terms $x_I$ and $x_J$ come from replacing the injection \textsc{Cnot} gates onto the control legs of the magic state by parity-\textsc{Cnot} gates (Figure~\ref{fig:active_injection}).
The term $x_K$ comes from adding the qubits of $K$ as extra controls of the \textsc{Cnot} gate from the target leg onto $c$ (for $U_{\CCZ}$, of the parity-\textsc{Cnot} gate onto the third leg).
The sets $I$, $J$ and $K$ may overlap.
For example, the sum bit of Equation~\eqref{eq:adder_sum_identity} is $U_{\CCX}$ with $c=0$, $I=\{a_k,s_k\}$, $J=\{b_k,s_k\}$ and $K=\{a_k,b_k,s_k,a_{k+1},b_{k+1}\}$.

On the target leg, the multi-target \textsc{Cnot} gate simplifies, since the leg is measured in the $X$ basis right after.
Indeed, a multi-target \textsc{Cnot} gate from the leg onto $c_1,\ldots,c_N$ followed by its $X$ measurement amounts to the single parity measurement $X_LX_{c_1}\cdots X_{c_N}$ followed by its $Z$ measurement, up to Pauli corrections on the targets.
Figure~\ref{fig:active_injection} combines a parity-\textsc{Cnot} gate on each control leg with a multi-target \textsc{Cnot} gate on the target leg, implemented as this parity measurement.
The resulting $\ket{\CCX}$ injection computes $c_j\oplus(a_1\oplus a_2)(b_1\oplus b_2)$ for $j=1,2$ in one lattice surgery step, followed by the associated Clifford corrections.

\subsection{Color-changed implementation}
\label{ssec:colour_changed}
Throughout this work, we implement the color-changed (Hadamarized) version of the computation, in which the roles of the $X$ and $Z$ bases are exchanged in every circuit relative to the standard presentation. 
This convention allows a more efficient implementation of the \Adder{} and \Lookup{}
operations, which dominate the cost of the algorithm (see Section~\ref{sec:adder} and Ref.~\cite{FeMoco2026prep}).
The physical code is unchanged, with a logical operator $Z_L$ of weight $d$ and a logical operator $X_L$ of weight one.
Exchanging the bases reverses the direction of every \textsc{Cnot} gates, exchanges parity-\textsc{Cnot} gates with multi-target \textsc{Cnot} gates, and swaps the preparation and measurement bases, such that $\ket{+}_{L}$ preparation and $X$-basis measurement in the standard circuits become $\ket{0}_{L}$ preparation and $Z$-basis measurement. 
Since all circuits are exchanged consistently, the logical operation implemented is unchanged. 
Unless stated otherwise, all circuits and figures of Sections~\ref{sec:adder} to~\ref{ssec:ldpc_cat_layout} are drawn in the color-changed convention.

\subsection{Magic-state factories and cadence}\label{ssec:factories}

The non-Clifford operations of the computation, namely the \Adder, the \Lookup{} and the \Phaseup, consume two types of resource states, $j \in \{\ket{\CXCXX}, \ket{i}\}$, produced by dedicated factories, presented in~\cite{FeMoco2026prep}.
In the following, these operations are decomposed into steps. 
Each step consumes at most one $\ket{\CXCXX}$ state and two $\ket{i}$ states, all loaded simultaneously for consumption in a buffer during a lattice surgery of duration $T_{\mathrm{surg}}$.
Unless stated otherwise, steps consuming factory outputs are applied at least a \emph{cadence} $T_{\mathrm{cad}}$ apart.

\begin{table*}[ht]
    \centering
    \footnotesize
    \renewcommand{\arraystretch}{1.12}
    \begin{tabular}{@{}lccccc@{}}
        \toprule
         & & & & \multicolumn{2}{c}{Output error $p_{Z,\mathrm{out}}$} \\
        \cmidrule(lr){5-6}
        Factory
         & \shortstack{Length                                          \\$l_{\mathrm{fac}}^j$ (columns)}
         & \shortstack{Footprint                                       \\$S_{\mathrm{fac}}^j$ (qubits)}
         & \shortstack{Production                                      \\$T_{\mathrm{prod}}^j$ (cycles)}
         & $p_Z=10^{-3}$ & $p_Z=5\times10^{-4}$ \\
        \midrule
        $\ket{\CXCXX}$
         & $29$ & $1\,566$ & $21$ & $1.48\mathbin{\times}10^{-11}$ & $2.02\mathbin{\times}10^{-13}$ \\
        $\ket{i}$
         & $10.5$ & $153$ & $7$ & $3.1\mathbin{\times}10^{-12}$ & $9.43\mathbin{\times}10^{-13}$ \\
        \bottomrule
    \end{tabular}
    \caption{Factory parameters for $p_Z = 10^{-3}$ and $p_Z = 5\times10^{-4}$.}
    \label{tab:adder_reference_factories}
\end{table*}

The $\ket{\CXCXX}$ factory is composed of two second-level factories of distance $19$ sharing three pairs of first-level factories.
The $\ket{i}$ factory is a distance-$7$ color code with two output repetition codes of distance $19$.
To fit the factory on the lattice of Section~\ref{sssec:layout_connectivity}, three of the six stabilizers along each output repetition code are measured by two ancillae sharing a Bell pair, which adds three qubits per output repetition code to the layout presented in~\cite[Sec.~4.1]{FeMoco2026prep}.
These qubits are included in Table~\ref{tab:adder_reference_factories}, and their small effect on the simulated output error is neglected.
We denote by $l_{\mathrm{fac}}^j$ and $S_{\mathrm{fac}}^j$ the length, in data qubit columns, and the footprint, in physical qubits, of one factory of type $j$, by $T_{\mathrm{prod}}^j$ its mean production time excluding loading, and by $n_{\mathrm{fac}}^j$ the number of such factories.
These parameters are summarized in Table~\ref{tab:adder_reference_factories}.
Each factory produces a state sequentially, in a mean time $T_{\mathrm{prod}}^j$, and alternates between its two output channels.
We assume each channel is occupied from the start of a production to the end of its loading.
Used cyclically and assigning each state of a $\ket{i}$ pair to a dedicated factory, a sufficient condition to sustain the consumption rate is then given by
\begin{equation}\label{eq:factory_counts}
    n_{\mathrm{fac}}^{\ket{\CXCXX}} = \left\lceil \frac{T_{\mathrm{prod}}^{\ket{\CXCXX}}}{T_{\mathrm{cad}}} \right\rceil,
    \qquad
    n_{\mathrm{fac}}^{\ket{i}} = 2\left\lceil \frac{T_{\mathrm{prod}}^{\ket{i}}}{T_{\mathrm{cad}}} \right\rceil,
\end{equation}
where the cadence time is assumed at least as large as the loading time.
In this work, we choose to minimize the factory footprint, with the cadence choice
\begin{equation}\label{eq:cadence}
    T_{\mathrm{cad}} = \max\!\left(T_{\mathrm{surg}},\, T_{\mathrm{prod}}^{\ket{\CXCXX}},\, T_{\mathrm{prod}}^{\ket{i}}\right).
\end{equation}
Using the factories of Table~\ref{tab:adder_reference_factories}, this cadence corresponds to $T_{\mathrm{cad}} = \max(d, 21)$.
Using one $\ket{\CXCXX}$ factory and two $\ket{i}$ factories is then sufficient at every distance.
The $\ket{\CXCXX}$ factory therefore sets the cadence for small distances, while the lattice surgery sets it for large distances.

A step is \emph{cadence-limited} when it is required to be applied at least $T_{\mathrm{cad}}$ cycles after the previous step.
These are the steps of the adder (Section~\ref{sssec:adder_scheduling}), the \textsc{And} gates of a \Lookup{} not performed simultaneously with a parity-\textsc{Cnot} gate, and the \textsc{And} gates of a \Phaseup{} (Section~\ref{ssec:lookup_and_phaseup}).
The steps that are \emph{surgery-limited} last $T_{\mathrm{surg}}$, namely the parity-\textsc{Cnot} gates (Section~\ref{sssec:parity_controlled_not}), the multi-target \textsc{Cnot} gates (Section~\ref{sssec:multitarget_cnot}), the \textsc{And} gates of a \Lookup{} performed simultaneously with a parity-\textsc{Cnot} gate, and the multi-target $\C_X X$ gates of a \Phaseup{} (Section~\ref{ssec:lookup_and_phaseup}).
Note that the multi-target $\C_X X$ gates consume only $\ket{i}$ states and can be applied sequentially every $T_{\mathrm{surg}}$ with the same two $\ket{i}$ factories, since $T_{\mathrm{prod}}^{\ket{i}} \le T_{\mathrm{surg}}$ for the factories of Table~\ref{tab:adder_reference_factories}.

\subsection{\texorpdfstring{\Adder}{Adder}} \label{sec:adder}

In this section, we present an adaptation of the adder introduced in~\cite{gidney2025factor2048bitrsa} to cat-qubit architectures.
Following the convention presented in \cref{ssec:colour_changed}, register states of the adder are written in the $X$ basis, such that we use the shorthand notation $\ket{x}_X \coloneqq H\ket{x}$.
Unless explicitly stated otherwise, all states discussed below are logical states.
Given two integers in binary representation $a,b \in \{0,1\}^n$, the circuit for the addition implements the following operation:
\begin{equation}
    \text{Add}\ket{a}_X\ket{b}_X = \ket{a}_X\ket{a + b \mod 2^n}_X.
\end{equation}
Using a carry $c\in \{0,1\}^n$, this computation can be decomposed into elementary operations by using the relations:
\begin{equation} \label{eq:add_bin_naive}
    \begin{aligned}
        s_k     & = a_k \oplus b_k \oplus c_k              \\
        c_{k+1} & = a_k b_k \oplus b_k c_k \oplus a_k c_k,
    \end{aligned}
\end{equation}
with $s$ the outcome of the adder, $a_k$, $b_k$, $c_k$ and $s_k$ the $k$-th bit of $a$, $b$, $c$ and $s$, respectively.
Alternatively, for an out-of-place addition, the output can be reused directly, without a carry, in order to reduce the number of non-linear operations as follows:
\begin{equation}\label{eq:adder_sum_identity}
    s_{k+1} = ((a_k \oplus s_k) \cdot (b_k \oplus s_k)) \oplus (a_k \oplus b_k \oplus s_k \oplus a_{k+1} \oplus b_{k+1}).
\end{equation}
In practice, this implies that the $k$-th step of the addition can be implemented using a single non-Clifford \textsc{And} gate on a linear combination of inputs, the latter of which can be achieved using Clifford gates.
Uncomputing $b_k$ relies on the same identity, where the non-linear term is implemented with a $\CZ$ gate between $a_{k-1}\oplus s_{k-1}$ and $b_{k-1}\oplus s_{k-1}$. The remaining linear terms, then correspond to a Pauli frame update.

\subsubsection{Logical circuit}
\label{sssec:adder_logical_circuit}
Based on the identity~\eqref{eq:adder_sum_identity}, the logical circuit design is adapted from the construction presented in Appendix~A.2 of~\cite{gidney2025factor2048bitrsa}.
An \textsc{And} gate injected from a $\ket{\CCX}$ state requires two outcome dependent \textsc{Cnot} corrections.
As these corrections depend on measurement outcomes from the \textsc{And} gate injection procedure, further computations must wait for the reaction time of the decoder.
At the cost of additional qubits, it is possible to include all (or part) of these corrections through adaptive measurements on an offline \emph{(semi-)auto} magic state~\cite{gidneyFlexibleLayoutSurface2019}.
Gidney~\cite{gidney2025factor2048bitrsa} optimizes the adder circuit for surface code compilation by using \emph{semi-auto}-\textsc{And} states.
These states allow \textsc{And} gate injections requiring classically controlled phase gates on the control qubits.
Remarkably, in their adder circuit, those phase gate byproducts can be later merged, with no additional cost, into the corrections required for the uncomputation of one of the input registers.
Preparation of this semi-auto-\textsc{And} state requires three $\ket{i}$ states, in addition to a $\ket{\CCZ}$ state and a Hadamard gate.
This implementation is favorable for surface codes, since both the $\ket{i}$ state preparation~\cite{gidneyInplaceAccessSurface2024} and the Hadamard gate are spatially and temporally efficient.

The cost balance is different for the repetition-cat architecture.
Indeed, $\ket{i}$ states are supplied by dedicated color code factories and a Hadamard gate consumes two $\ket{i}$ states, with a third reused as a catalyst~\cite[App.~C.2]{FeMoco2026prep}.
We implement the \textsc{And} gate with an auto-\textsc{And} magic state prepared from a $\ket{\CXCXX}$ state, and the subsequent phase corrections with \emph{delayed-$\CZ$} states $\ket{\DCZ}$.
Each preparation consumes two $\ket{i}$ states and reuses a catalyst, reducing the factory throughput required between consecutive adder steps.

The circuit stores $a$ and $b$ in separate registers and writes the result into a third $s$ register during injection, following the recurrence from Eq.~\eqref{eq:adder_sum_identity}.
We denote the width of each of the three registers by $n$ and index the bits by $k=1,\ldots,n$ from least to most significant, with no incoming carry.
Each logical qubit is denoted by the bit it stores.

The adder performs $n$ injections followed by $n$ corrections in reverse order, $I_1,\ldots,I_n,C_n,\ldots,C_1$.
At the $k$-th step, the injection $I_k$ computes $s_k$ and the correction $C_k$ performs a destructive $Z$ measurement on $b_k$ and applies the corresponding phase correction.
After the corrections, $a$ and $s$ are the remaining quantum registers.
In the context of Section~\ref{sec:logical_impl}, the register $a$ corresponds to the output of a \Lookup, the register $b$ to the previous value of an accumulator, and the register $s$ to the corresponding update of the latter.
As such, uncomputing $b$ leads to an in-place update, while $a$ is cleared afterwards by the measurement-based uncomputation of the \Lookup, whose phases are resolved by a \Phaseup{} (Section~\ref{ssec:lookup_and_phaseup}).

\begin{figure*}[t]
    \centering
    \includegraphics[width=\linewidth]{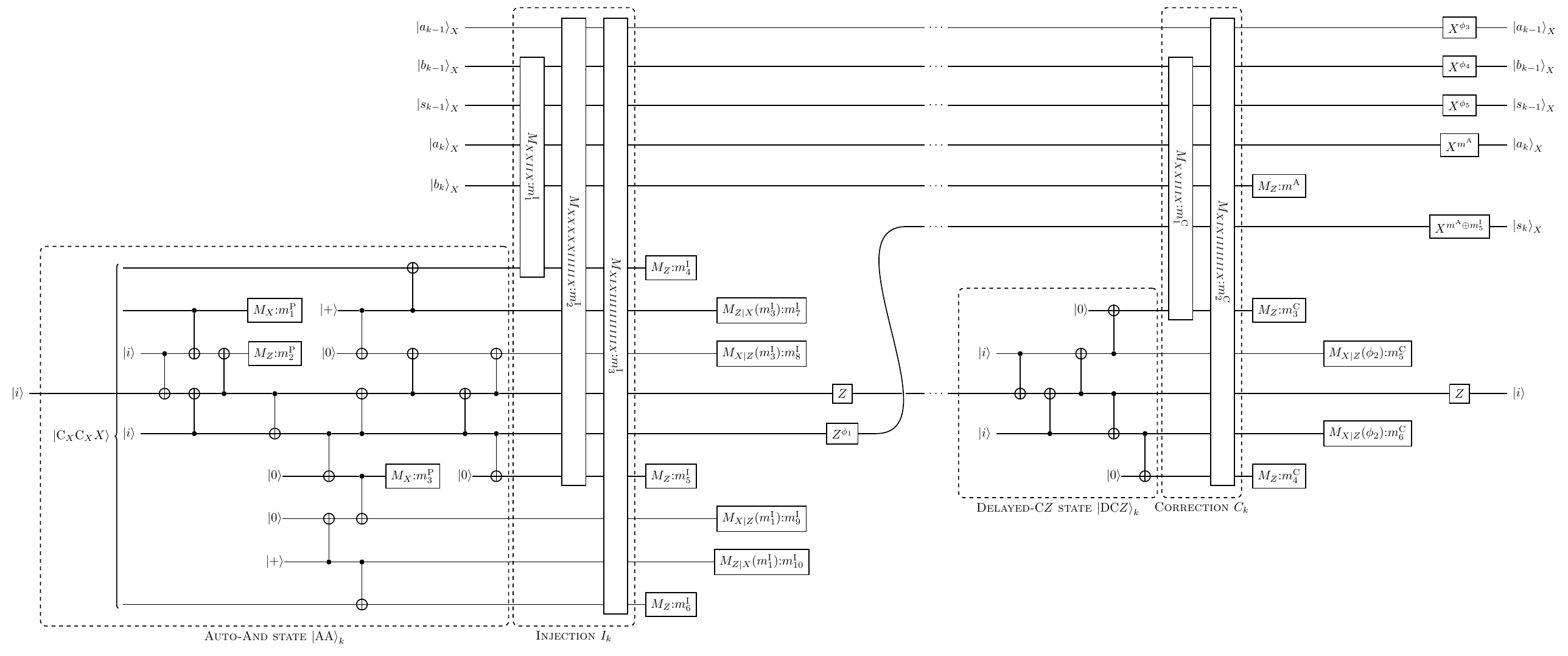}
    \caption{
        Logical circuit for the adder iteration at step $k>1$.}
    \label{fig:logical_adder_circuit}
\end{figure*}

Figure~\ref{fig:logical_adder_circuit} provides the logical circuit implementing Eq.~\eqref{eq:adder_sum_identity} at $k>1$.
The $X$-parity measurements follow the active injection procedure of Section~\ref{sssec:active_injection}.
Section~\ref{sssec:adder_routing} details the implementation of these measurements in a single simultaneous lattice surgery step~\cite{Simultaneous2026prep}.
The recorded measurement outcomes determine the adaptive $X$ or $Z$ measurement bases of the auto-\textsc{And} and delayed-$\CZ$ states.
Tracking the measurement outcomes in the Pauli frame results in the following feedforward Boolean functions:
\begin{equation}\label{eq:adder_pauli_corrections}
\begin{aligned}
\phi_1={}&m_3^{\mathrm I}(m_1^{\mathrm I}\oplus m_8^{\mathrm I}\oplus m_7^{\mathrm I})
 \oplus m_1^{\mathrm I}(m_9^{\mathrm I}\oplus m_{10}^{\mathrm I})\\
 &\oplus m_2^{\mathrm I}\oplus m_2^{\mathrm P}\oplus m_3^{\mathrm P}
 \oplus m_8^{\mathrm I}\oplus m_9^{\mathrm I}\oplus1,\\
\phi_2={}&m^{\mathrm A}\oplus m_5^{\mathrm I}\oplus m_1^{\mathrm P},\\
\phi_3={}&(m_1^{\mathrm C}\oplus m_5^{\mathrm C})(\phi_2\oplus1)
 \oplus m_6^{\mathrm C}\phi_2\oplus m_4^{\mathrm C}\oplus m_5^{\mathrm I}\\
 &\oplus1\oplus m_1^{\mathrm P}\oplus m_6^{\mathrm I}
 \oplus m_{10}^{\mathrm I}(m_1^{\mathrm I}\oplus1)\\
 &\oplus m_1^{\mathrm I}(m_9^{\mathrm I}\oplus1\oplus m_1^{\mathrm P}),\\
\phi_4={}&(m_2^{\mathrm C}\oplus m_6^{\mathrm C})(\phi_2\oplus1)
 \oplus m_5^{\mathrm C}\phi_2\oplus m_3^{\mathrm C}\oplus m_4^{\mathrm I}\\
 &\oplus m_5^{\mathrm I}\oplus1\oplus m_1^{\mathrm P}
 \oplus m_7^{\mathrm I}(m_3^{\mathrm I}\oplus1)\\
 &\oplus m_3^{\mathrm I}(m_8^{\mathrm I}\oplus1\oplus m_1^{\mathrm P}),\\
\phi_5={}&m^{\mathrm A}\oplus\phi_3\oplus\phi_4,
\end{aligned}
\end{equation}
where $m_r^{\mathrm P}$, $m_r^{\mathrm I}$, and $m_r^{\mathrm C}$ denote the preparation, injection, and correction outcomes, respectively, and $m^{\mathrm A}$ is the outcome of the $Z$ measurement on $b_k$.
All outcome arithmetic is modulo two, such that $\oplus$ denotes a XOR operation.

The first injection $I_1$ can be performed by initializing $\ket{s_1}_X=\ket{0}_X$ and computing $s_1=a_1\oplus b_1$ with three transversal \textsc{Cnot} gates $\CX_{b_1 a_1}$, $\CX_{s_1 b_1}$ and $\CX_{b_1 a_1}$.
Alternatively, $s_1$ can be initialized in $\ket{0}_L$ and set to $a_1\oplus b_1$ by an $X$-parity measurement on $a_1$, $b_1$ and $s_1$, with outcome $m$, followed by $Z_{s_1}^{m}$ in the Pauli frame.
The last correction $C_1$ only measures $b_1=a_1\oplus s_1$ in the $Z$ basis and applies $X_{a_1}^{m^{\mathrm A}}X_{s_1}^{m^{\mathrm A}}$ in the Pauli frame.
In both cases, the first and last steps require no resource state.

\subsubsection{Layout on the repetition-cat architecture}
\label{sssec:adder_physical_implementation}

\begin{figure*}[t]
    \centering
    \includegraphics[width=\linewidth]{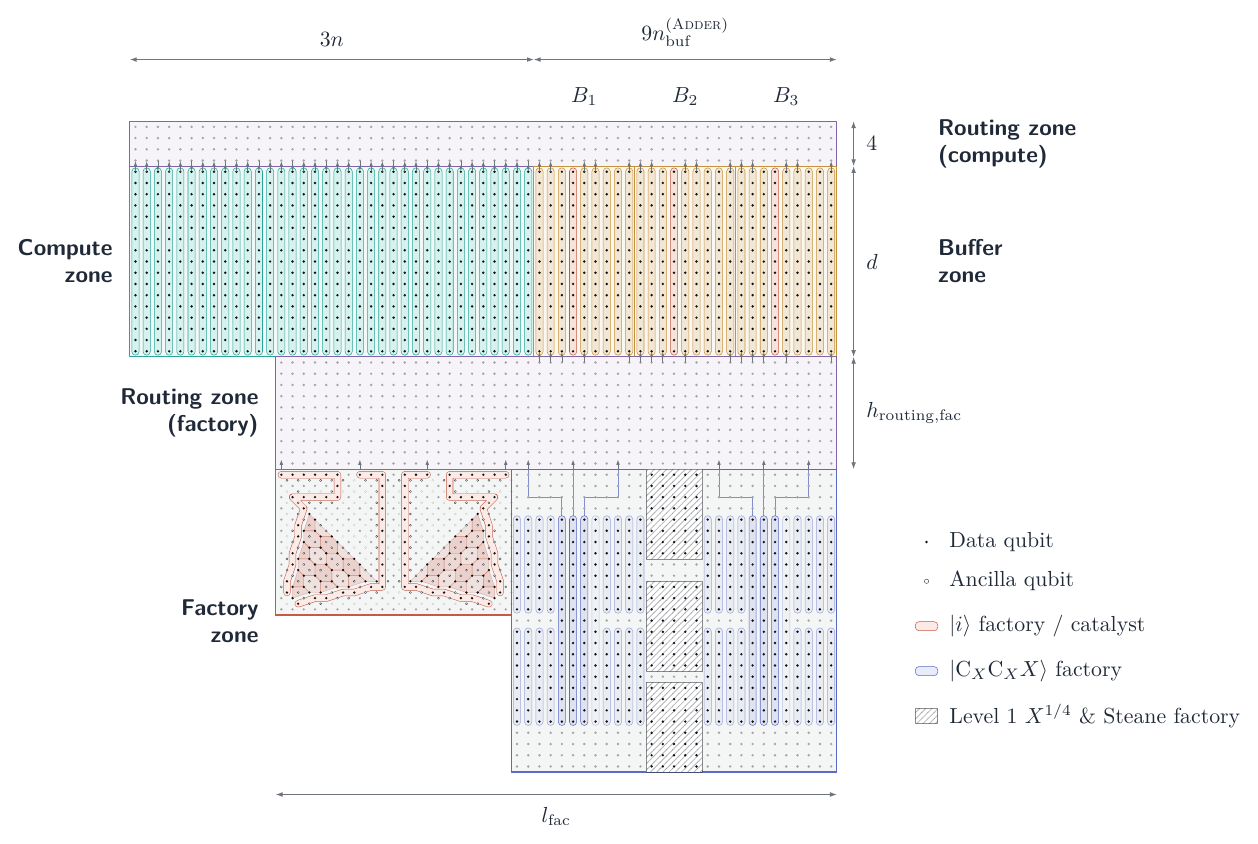}
    \caption{Illustrative layout of the repetition-cat adder at $n=12$, $d=17$ and $T_{\mathrm{cad}} = 21$.
        The compute zone stores the adder inputs and outputs $(a_k,b_k,s_k)$ next to the three buffers $B_1$, $B_2$ and $B_3$.
        The compute routing zone above carries the simultaneous $X$-parity measurements of the injection and correction steps.
        The factory routing zone below carries factory outputs into the buffers.
        The factory zone holds two color code $\ket{i}$ factories and one $\ket{\CXCXX}$ factory, each with two output channels.
        Ancillae are drawn only in the color code factories, whose data and ancilla roles differ from the rest of the layout (Section~\ref{sssec:layout_connectivity}).
        Alternative adder layouts can be obtained by choosing different factory and buffer placements.}
    \label{fig:adder_global_layout}
\end{figure*}

The adder layout on the repetition-cat architecture, shown in Figure~\ref{fig:adder_global_layout}, is organized into five interacting zones.
The \emph{compute zone} stores the registers $a$, $b$ and $s$ of the adder as interleaved repetition codes.
The \emph{buffer zone} is composed of neighboring \emph{buffers}, each comprised of nine adjacent repetition codes of distance $d$, labeled $q_1,\ldots,q_9$.
Each buffer retains an $\ket{i}$ catalyst at $q_4$ and can prepare either resource state.
Delayed-$\CZ$ preparation occupies only $q_2,\ldots,q_6$, leaving the other four columns idle.
A buffer receives two $\ket{i}$ states from the factories for each preparation, together with a single $\ket{\CXCXX}$ state for auto-\textsc{And} preparation, and prepares the corresponding resource state in place.
The number of adder buffers $n_{\mathrm{buf}}^{(\Adder)}$ is discussed in Section~\ref{sssec:adder_scheduling} and the factory counts in Section~\ref{ssec:factories}.
Two distinct \emph{routing zones} interface with the buffer zone. 
The upper compute routing zone spans both the compute and buffer zones, with four rows of data qubits.
This zone supports the injection and correction phases in Figure~\ref{fig:logical_adder_circuit} through simultaneous parity measurements.
The \emph{factory zone} consists of the $\ket{\CXCXX}$ and $\ket{i}$ factories presented in Section~\ref{ssec:factories}.
The lower factory routing zone lies between the factories and the compute and buffer zones, with a height of $h_{\mathrm{routing,fac}}$ rows of data qubits and a length set by the factory placement.
Similarly, this zone is reserved for routing factory outputs into the buffers via simultaneous parity measurements.
In every zone, each row of data qubits is counted with the row of ancillae above it, i.e.\ two physical qubits per data qubit. 
An operation footprint is the sum of its zones. 
As such, the compute zone, the buffers and the upper routing require a core footprint 
\begin{equation}\label{eq:adder_core_count_piecewise}
    S_{\mathrm{core}}^{(\mathrm{rep})}=2(d+4)(3n+9n_{\mathrm{buf}}^{(\Adder)}),
\end{equation}
where $S_{\mathrm{core}}^{(\mathrm{rep})}$ corresponds to the total number of physical qubits for the adder, excluding factories and their associated routing.
The factories are placed side by side along the factory routing zone.
With $n_{\mathrm{fac}}^j$ factories of type $j\in\{\ket{\CXCXX},\ket{i}\}$, the factory zone spans a total length $l_{\mathrm{fac}}$ and requires $S_{\mathrm{fac}}$ physical qubits, given by
\begin{equation}\label{eq:adder_factory_footprint}
\begin{aligned}
    l_{\mathrm{fac}}&=n_{\mathrm{fac}}^{\ket{\CXCXX}}\,l_{\mathrm{fac}}^{\ket{\CXCXX}}
    +n_{\mathrm{fac}}^{\ket{i}}\,l_{\mathrm{fac}}^{\ket{i}},\\
    S_{\mathrm{fac}}&=n_{\mathrm{fac}}^{\ket{\CXCXX}}\,S_{\mathrm{fac}}^{\ket{\CXCXX}}
    +n_{\mathrm{fac}}^{\ket{i}}\,S_{\mathrm{fac}}^{\ket{i}}.
\end{aligned}
\end{equation}
The factory routing zone spans the same length and occupies
\begin{equation}\label{eq:adder_factory_routing_footprint}
    S_{\mathrm{routing,fac}}=2h_{\mathrm{routing,fac}}\,l_{\mathrm{fac}}
\end{equation}
physical qubits.

\subsubsection{Layout on the LDPC-cat architecture}
\label{sssec:adder_physical_implementation_LDPC}

On the LDPC-cat architecture, the buffer, factories and their associated routing zones are those of the repetition-cat layout.
The registers $a$, $b$ and $s$ are stored in a single open-boundary LDPC-cat code~\cite{Ruiz_2025} of height $H$ and width $L$, whose $2L$ logical qubits hold the $3n$ register bits, given $2L\geq 3n$.
The code sits above the compute routing zone, and the buffer row starts under its last five tip columns, as shown in Figure~\ref{fig:adder_ldpc_routing_overlays}.
The $X$-parity measurements of each injection and correction reach the register bits through their proxies, and the $Z$ measurement of $b_k$ through a repetition code covering the support of $Z_L$, as described in Section~\ref{sssec:error_correcting_codes} and shown in Figure~\ref{fig:adder_ldpc_correction_layout}.
The compute zone of the repetition layer covers the footprint of the open-boundary code, including $H$ rows and the proxy row over $p_\ell+L+p_r$ columns, where $p_\ell$ and $p_r$ are the left and right edge paddings of the code. 
Each triple $(a_k,b_k,s_k)$ occupies three consecutive columns, in increasing order of $k$.
Row~1, the bottom row, holds the first $\lceil n/2\rceil$ triples, and Row~2, above it, the remaining ones.
For odd $n$ with $3\lceil n/2\rceil>L_{\min}$, Row~1 holds the first $\lfloor n/2\rfloor$ triples followed by $a_n$ and $b_n$, such that $s_n$ occupies the column of Row~2 above $a_n$.
The proxy of $s_n$ then lies in the first column of the right edge padding.
As such, for an $n$-bit adder, an LDPC code satisfying $L=\max(L_{\min},\lceil 3n/2\rceil)$ suffices.
The LDPC layer holds the $HL+\Delta_{\mathrm{LDPC}}$ data qubits, where $\Delta_{\mathrm{LDPC}}$ counts the qubits added by the open-boundary staircase, and $HL+\Delta_{\mathrm{LDPC}}-2L$ ancillae.
Therefore, the adder on the LDPC-cat architecture, including the compute zone of both layers, the buffer zone and the compute routing zone, requires a core footprint of
\begin{equation}\label{eq:adder_ldpc_core_count}
\begin{aligned}
    S_{\mathrm{core}}^{(\mathrm{LDPC})}&=2(H+1)(p_\ell+L+p_r)+2(H-1)L\\
    &\quad+2\Delta_{\mathrm{LDPC}}+18dn_{\mathrm{buf}}^{(\Adder)}\\
    &\quad+8(L+9n_{\mathrm{buf}}^{(\Adder)}-5).
\end{aligned}
\end{equation}

\subsubsection{Resource states preparation} \label{sssec:resource_states}
\label{sssec:resource_state_preparation}

In the steady state, the adder consumes an auto-\textsc{And} state and a delayed-$\CZ$ state for the injection and correction steps, respectively.
For inputs $A, C$ and a catalyst $K$, all initially in $\ket{i}$, up to Pauli frame updates, the four \textsc{Cnot} gates $\CX_{AK}$, $\CX_{CK}$, $\CX_{KA}$ and $\CX_{KC}$ produce the resource state $\ket{\CZ}=\CZ\ket{+}\ket{+}=\C_X X\ket{0}\ket{0}$.
The auto-\textsc{And} preparation uses this state to teleport a Hadamard gate, and the delayed-$\CZ$ state preparation uses it to inject a $\C_X X$ gate~\cite{FeMoco2026prep}.
Each preparation therefore consumes two $\ket{i}$ states, while the catalyst is supplied once per buffer and per operation. 
The catalyst is seeded during the initialization of the operation and retained for all its preparations. 
Since the buffer zone is reconfigured between operations (Section~\ref{sec:C7-global}), catalysts are not kept from one operation to the next.

\paragraph{Auto-\textsc{And} state.}
The auto-\textsc{And} preparation receives the three $\ket{\CXCXX}$ legs at $q_1,q_2,q_9$, two $\ket{i}$ states at $q_3,q_5$, and the recycled catalyst at $q_4$.
It uses only local \textsc{Cnot} gates, measurements, and initializations (see Figure~\ref{fig:logical_adder_circuit}).
First, the two $\ket{i}$ states and the catalyst form a $\ket{\CZ}$ state, which teleports a Hadamard gate from $q_2$ onto $q_5$.
Then, two Bell pairs are prepared and entangled with the resulting $\ket{\C_X\C_X Z}$ state to implement delayed-choice \textsc{Cnot} corrections.
These entangling operations are achieved through the remaining \textsc{Cnot} gates.
\paragraph{Delayed-$\CZ$ state.}
A delayed-$\CZ$ state is prepared from two $\ket{i}$ states at $q_3,q_5$, the catalyst at $q_4$, and two auxiliary qubits initialized in $\ket{0}$ at $q_2,q_6$.
Assuming that the input states are ready, the preparation consists of six transversal \textsc{Cnot} gates, with independent operations performed in parallel.

\begin{figure*}[!tp]
    \centering
    \begin{subfigure}[t]{0.48\linewidth}
        \centering
        \begin{minipage}[c][6.2cm][c]{\linewidth}
            \centering
            \includegraphics[width=\linewidth,height=6.2cm,keepaspectratio,
                trim=648bp 258bp 662bp 367bp,clip]
            {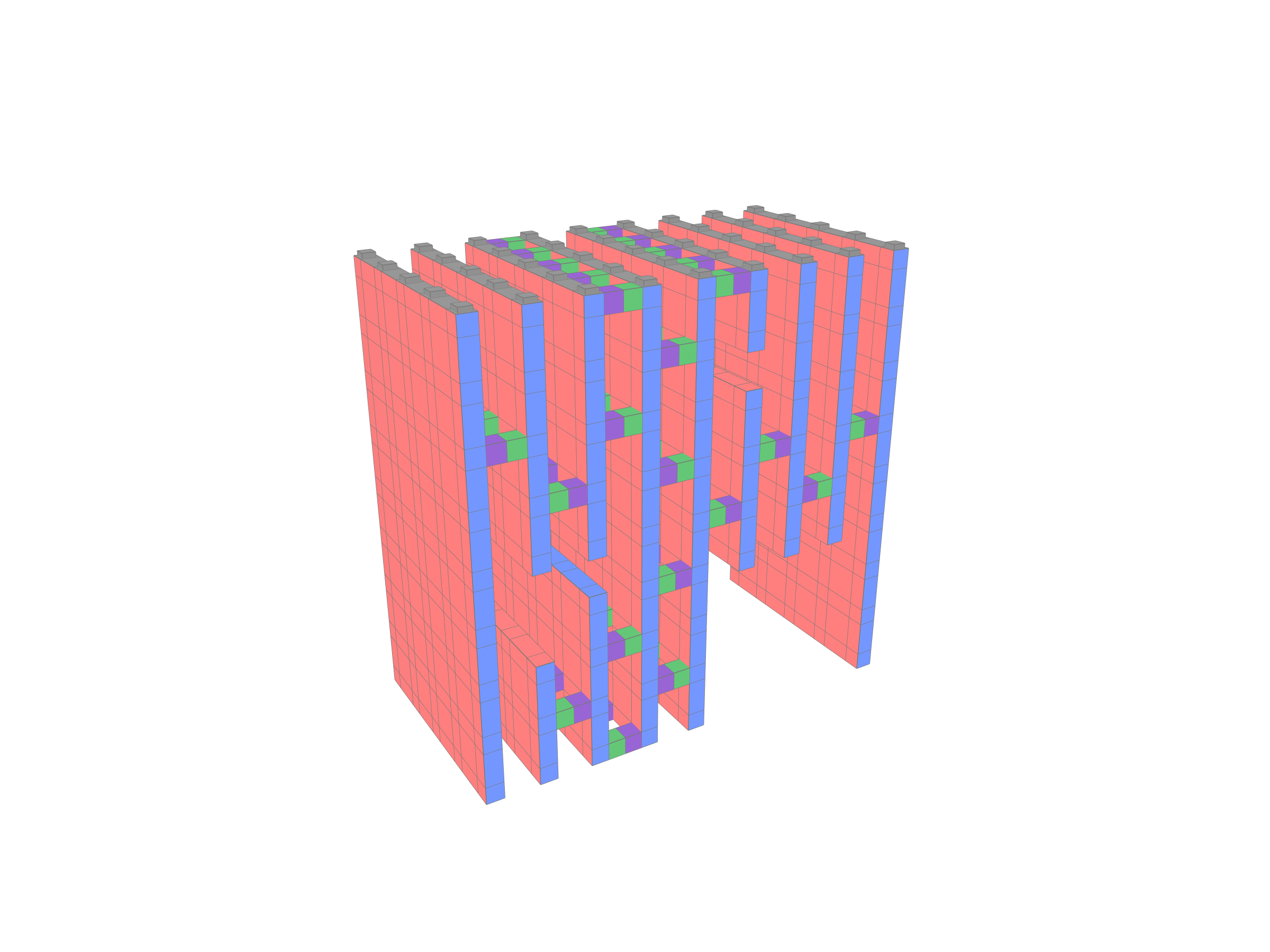}
        \end{minipage}
        \caption{Auto-\textsc{And} state $\ket{\mathrm{AA}}$}
    \end{subfigure}
    \hfill
    \begin{subfigure}[t]{0.48\linewidth}
        \centering
        \begin{minipage}[c][6.2cm][c]{\linewidth}
            \centering
            \includegraphics[width=0.9\linewidth,height=4.2cm,keepaspectratio,
                trim=842bp 516bp 816bp 552bp,clip]
            {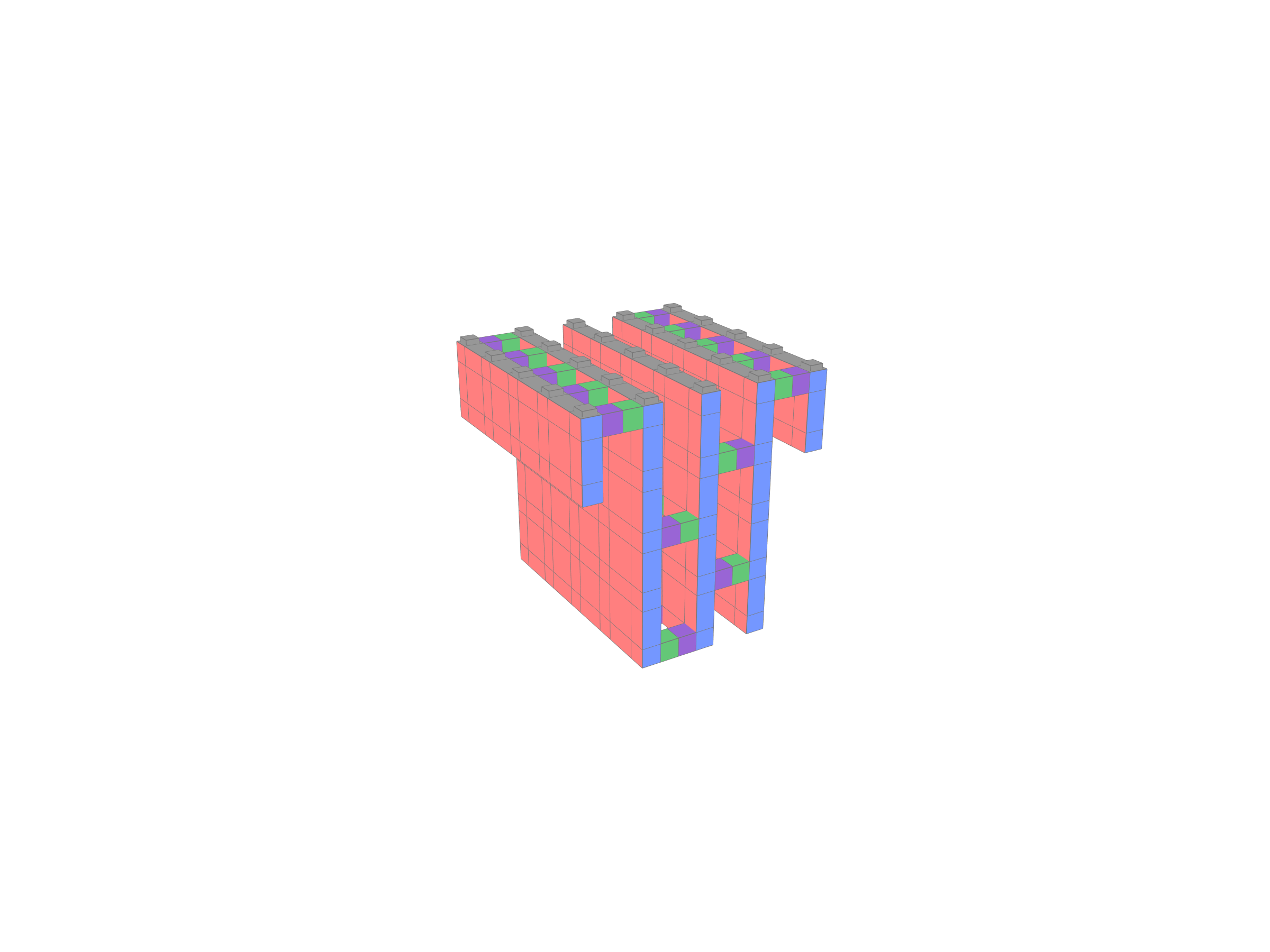}
        \end{minipage}
        \caption{Delayed-$\CZ$ state $\ket{\DCZ}$}
    \end{subfigure}
    \bigskip

    \def\Primitive#1{\raisebox{-0.4\height}{\includegraphics[height=0.55cm,
        trim=143bp 114bp 167bp 186bp,clip]{#1}}}
    {\scriptsize
    \begin{tabular}{@{}cl@{\hspace{3em}}cl@{\hspace{3em}}cl@{\hspace{3em}}cl@{}}
        \Primitive{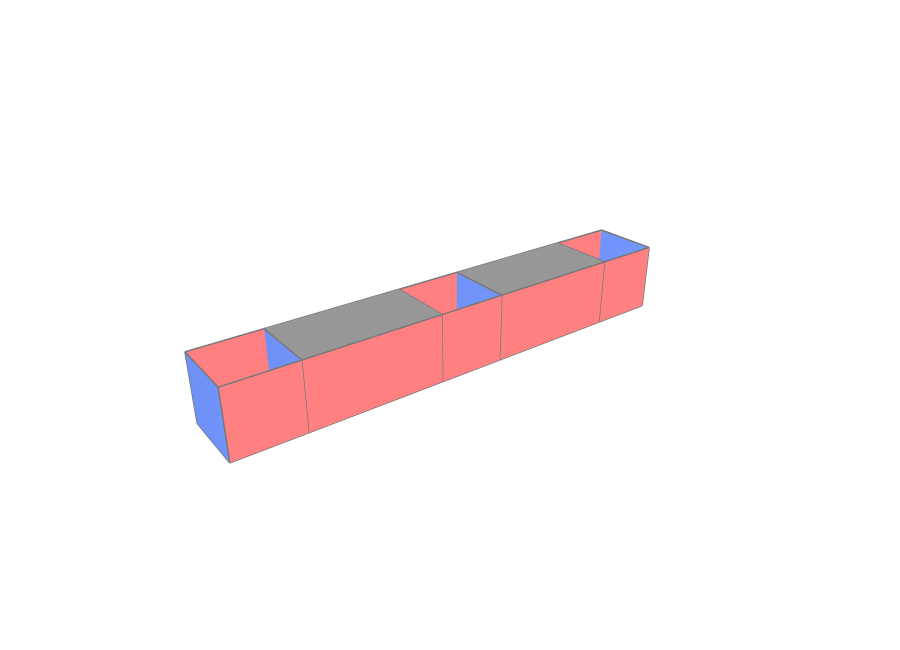} & Logical qubit &
        \Primitive{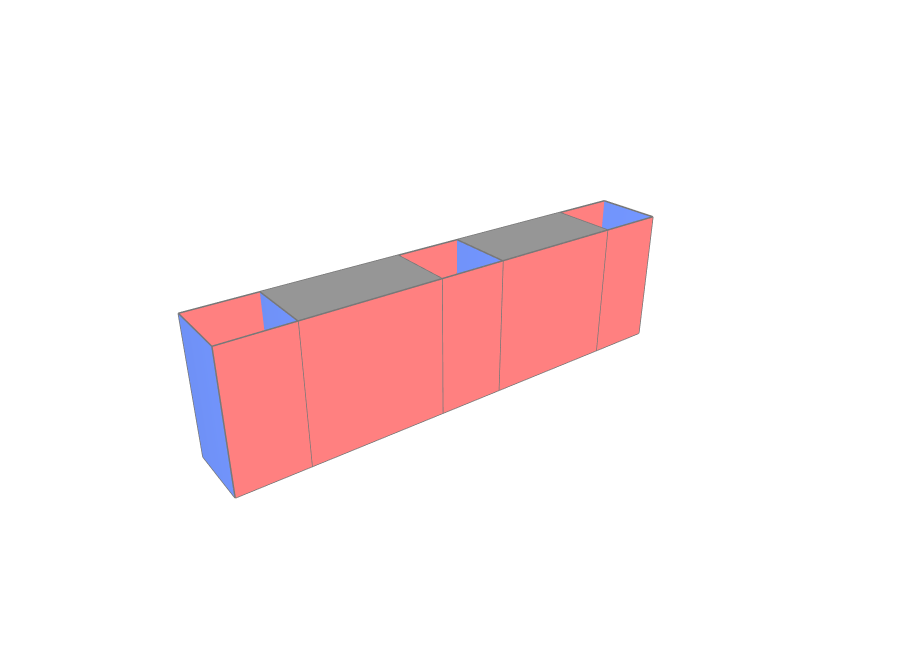} & Idling &
        \Primitive{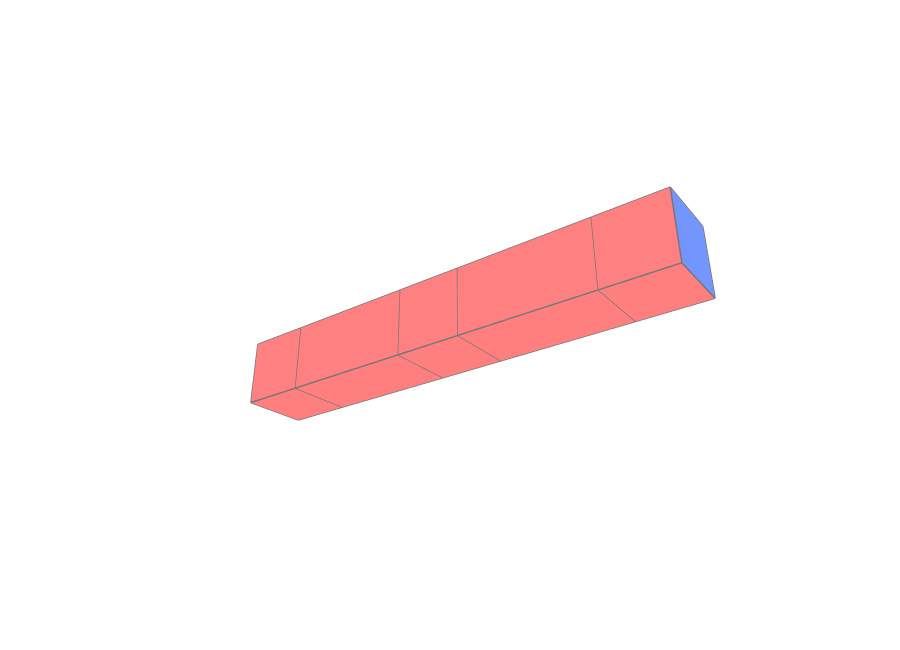} & Preparation $\ket{+}_L$ &
        \Primitive{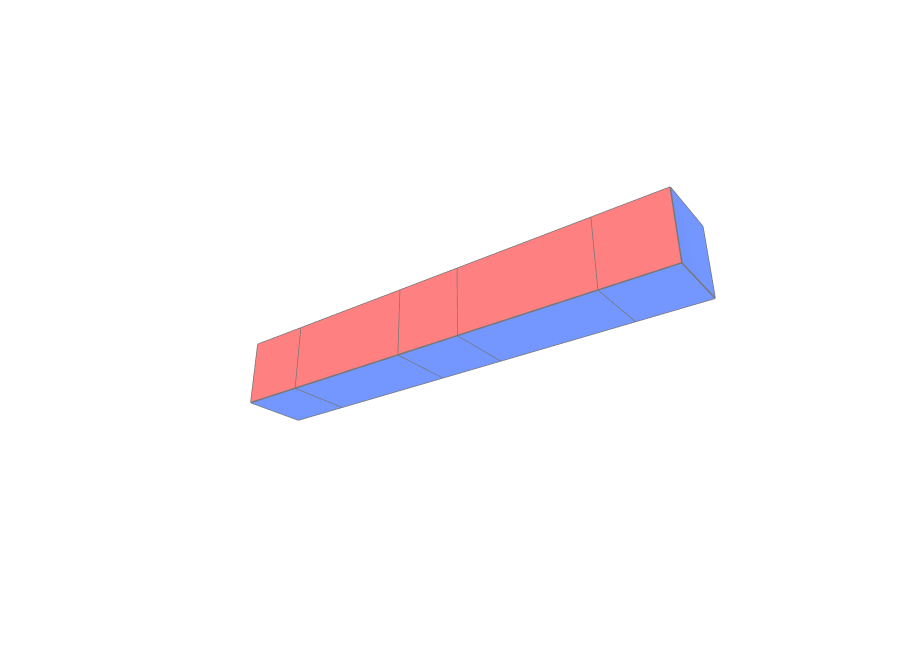} & Preparation $\ket{0}_L$ \\[3.5ex]
        \Primitive{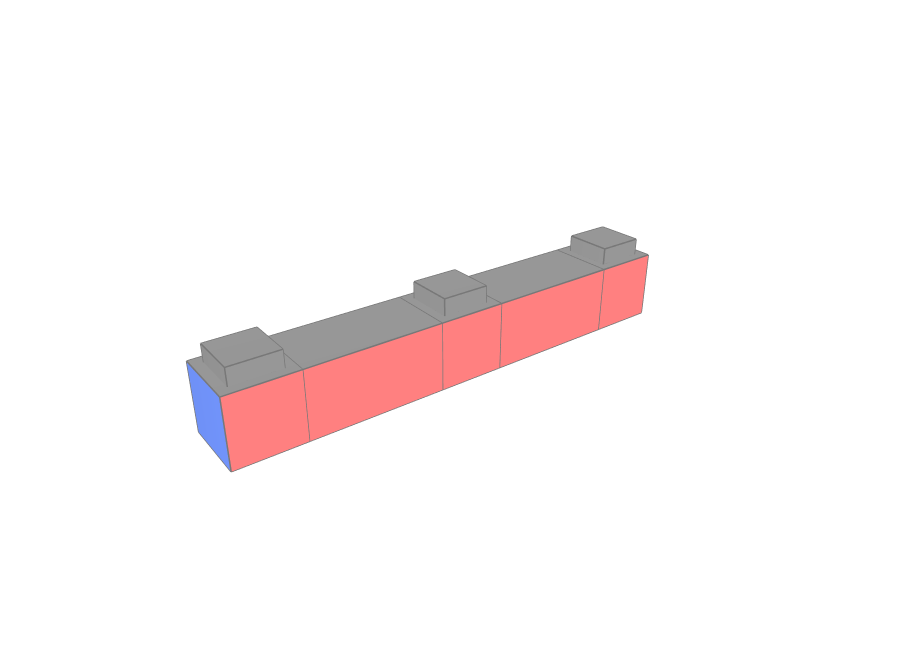} & Logical input/output &
        \Primitive{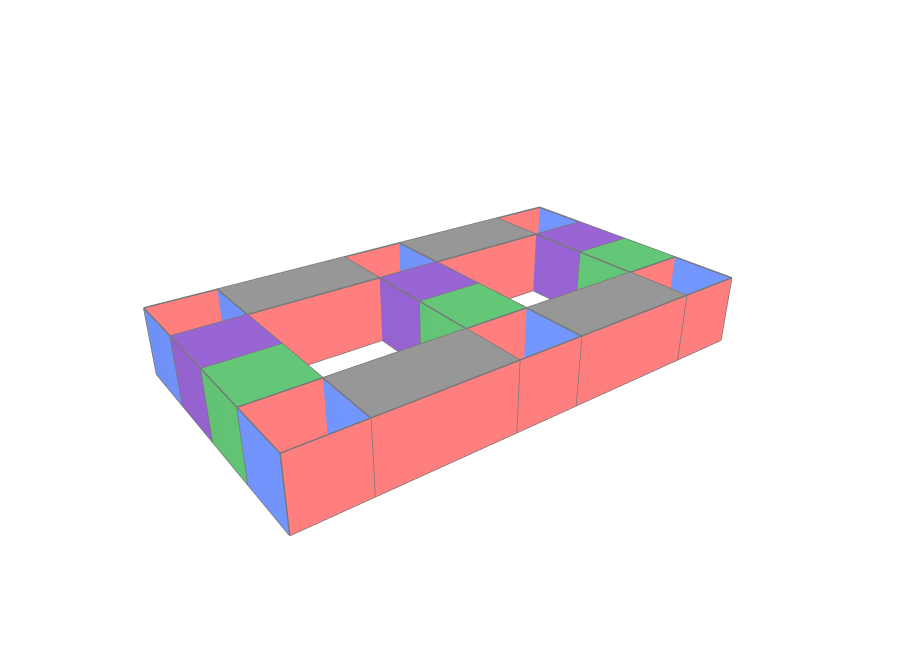} & Transversal \textsc{Cnot} &
        \Primitive{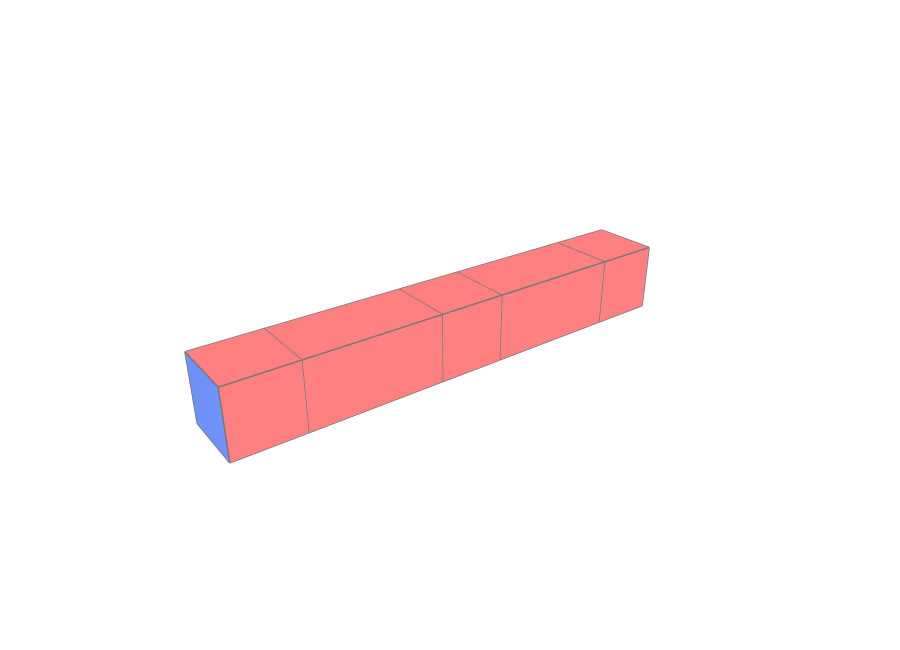} & Measurement $M_{X_L}$ &
        \Primitive{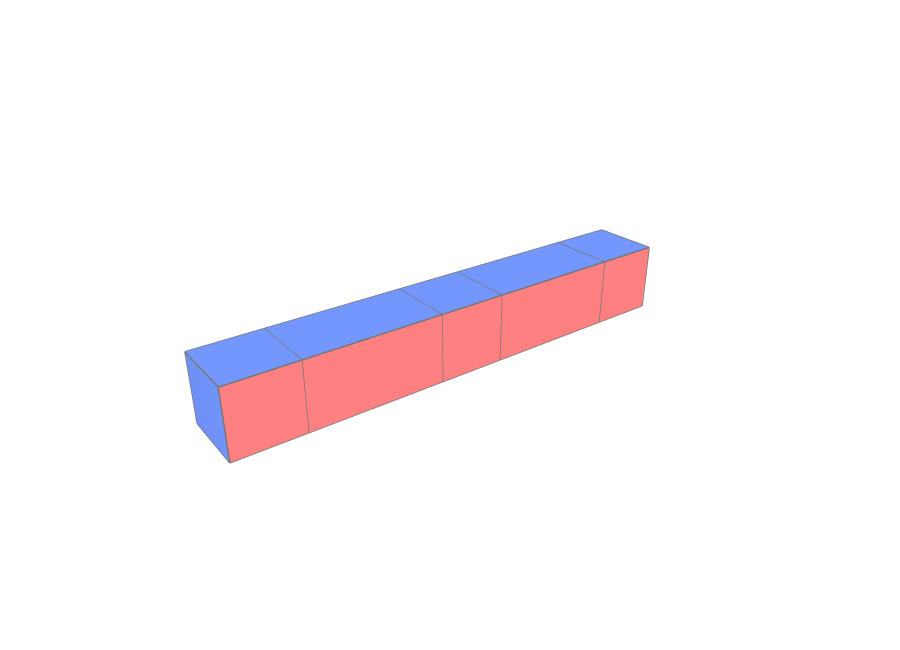} & Measurement $M_{Z_L}$
    \end{tabular}}
    \caption{Spacetime diagrams for the resource preparations at $d=5$.}
    \label{fig:adder_state_preparations}
\end{figure*}

The compiled preparation durations, shown in Figure~\ref{fig:adder_state_preparations}, follow from Section~\ref{sssec:layout_connectivity}, such that $T_{\mathrm{prep}}^{\ket{\mathrm{AA}}}=9$ and $T_{\mathrm{prep}}^{\ket{\DCZ}}=5$ cycles.
The auto-\textsc{And} preparation contains the only $X$ measurement of a $\ket{0}_L$ code made before any lattice surgery on it, $m_3^{\mathrm P}$ on $q_6$ (Section~\ref{sssec:error_correcting_codes}).
This does not lengthen $T_{\mathrm{prep}}^{\ket{\mathrm{AA}}}$, as $I_k$ establishes the Pauli frame of $m_3^{\mathrm P}$ through the codes coupled to $q_6$, and $m_3^{\mathrm P}$ enters only $\phi_1$ in Equation~\eqref{eq:adder_pauli_corrections}, which is evaluated after $I_k$.
We omit Pauli corrections from the timing, as those are performed as a Pauli frame update in classical software with no associated delay on the state preparation.

\paragraph{Adaptive measurements.}
In both cases, the buffer remains occupied during the subsequent lattice surgery, throughout the reaction time and the final adaptive measurement.
The final resource measurements therefore complete $T_{\mathrm{react}}+T_{\mathrm{meas}}$ after the lattice surgery.
\paragraph{Spacetime diagrams.}
The horizontal plane represents the physical layout, while the vertical axis represents time in cycles.
Each cube represents a physical qubit.
Two neighboring cubes are linked by a parallelepiped that represents an $XX$ stabilizer.
Red, blue, and grey faces mark $X_L$ boundaries, $Z_L$ boundaries, and logical interfaces, respectively.
A vertical extension denotes idling through continued syndrome extraction.
Red and blue lower caps prepare $\ket{+}_L$ and $\ket{0}_L$, respectively, while upper caps denote destructive $X_L$ and $Z_L$ measurements, respectively.
A transversal \textsc{Cnot} gate is drawn as a connector with a green control and a purple target between adjacent tubes.
Finally, the Pauli frame corrections are omitted.
These conventions are presented in Figure~\ref{fig:adder_state_preparations}.

\subsubsection{Simultaneous lattice surgery}
\label{sssec:adder_routing}
\begin{figure*}[t]
    \centering
    \begin{subfigure}[t]{0.42\linewidth}
        \centering
        \includegraphics[width=\linewidth]{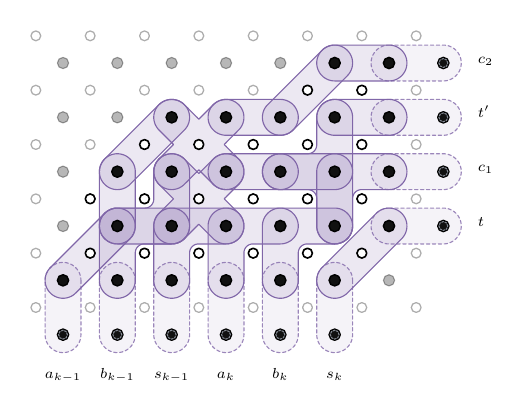}
        \caption{Injection $I_k$ (compute).}
    \end{subfigure}
    \hfill
    \begin{subfigure}[t]{0.5503\linewidth}
        \centering
        \includegraphics[width=\linewidth]{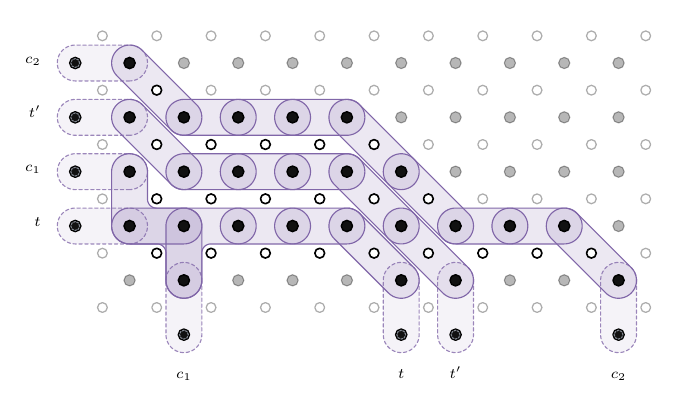}
        \caption{Injection $I_k$ (buffer).}
    \end{subfigure}
    \par\medskip
    \begin{subfigure}[t]{0.42\linewidth}
        \centering
        \begin{minipage}[b][0.7734\linewidth][c]{\linewidth}
            \centering
            \includegraphics[width=0.5864\linewidth]{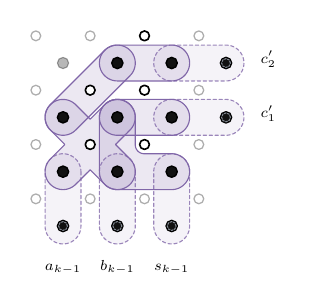}
        \end{minipage}
        \caption{Correction $C_k$ (compute).}
    \end{subfigure}
    \hfill
    \begin{subfigure}[t]{0.5503\linewidth}
        \centering
        \includegraphics[width=0.9211\linewidth]{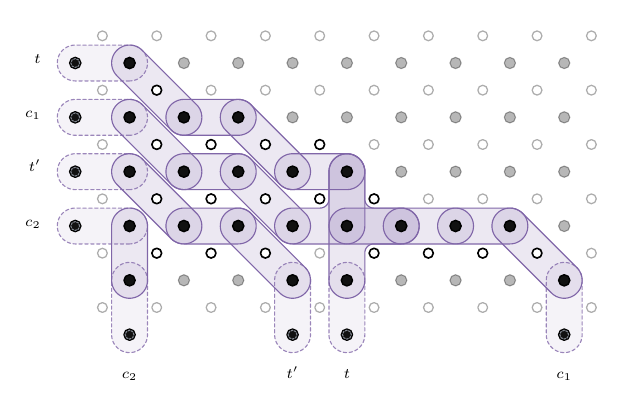}
        \caption{Injection $I_k$ (mirrored buffer).}
    \end{subfigure}
    \par\medskip
    \includegraphics[scale=0.85]{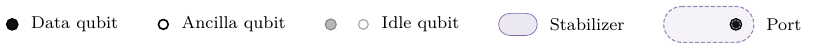}
    \caption{Simultaneous lattice surgery routing primitives of the injection and correction steps.
        Figures (a)--(c) are used on the repetition-cat architecture.
        On the LDPC-cat architecture, (a) and (c) are flipped top to bottom, and (d) replaces (b).
        In the injection, $c_1$ and $c_2$ are the control qubits, while $t$ and $t'$ are $Z$-copies of the target qubit from the auto-\textsc{And} state.
        The qubit $t$ is moved into $s_k$, and $t'$ is involved in an $X$-parity measurement with the compute zone.
        In the correction, $c'_1$ and $c'_2$ are $Z$-copies of the control qubits $c_1$ and $c_2$ from the delayed-$\CZ$ state.}
    \label{fig:adder_routing_primitives}
\end{figure*}

The injection and correction phases in Figure~\ref{fig:logical_adder_circuit} involve three and two $X$-parity measurements, respectively.
A possible implementation protocol consists of sequentially applying for each measurement a merge and split instruction between the repetition codes and the auxiliary qubits involved~\cite{GouzienCatQubits23}.
In the following, we instead use \emph{simultaneous lattice surgery} to perform the measurements of each operation in a single lattice surgery step~\cite{Simultaneous2026prep}.
For $k>1$, the auto-\textsc{And} state consumed during injection $I_k$ is injected into the codes associated with $a_{k-1}$, $b_{k-1}$, $s_{k-1}$, $a_k$, and $b_k$ through their $X_L$ logical representatives.
During this operation, a routing of $q_5$ transfers the result into the slot reserved for $s_k$.
The sum output $t=q_5$ is distinct from its entangled $Z$-copy $t'=q_6$, which participates in one of the three injection measurements.
Similarly, correction $C_k$ for $k>1$ applies the delayed-$\CZ$ state through an $X$-parity measurement on $b_{k-1}$, $s_{k-1}$, and $q_2$, as well as another on $a_{k-1}$, $s_{k-1}$, and $q_6$.
The stabilizer patterns shown in Figure~\ref{fig:adder_routing_primitives} implement these simultaneous measurements and routing, with two complementary injection primitives and one correction primitive.
In each case, the data qubits of the routing zone are initialized in $\ket{0}$, undergo $d$ rounds of stabilizer measurements, and are measured destructively in the $Z$ basis.
The compute routing zone is released after the final $Z$ measurements of each surgery.
The consumed resource state continues to occupy its buffer until its reaction time and adaptive measurements are complete, while the catalyst is recycled for the next state preparation.

The injection primitive facing right connects to the compute codes, and the primitive facing left connects to the prepared buffer codes.
Their horizontal arms are joined in matching order.
On both architectures, we embed these primitives into the four data rows of the compute routing zone, above the buffer zone.

\begin{figure*}[t]
    \centering
    \begin{subfigure}[t]{\linewidth}
        \centering
        \includegraphics[width=\linewidth]{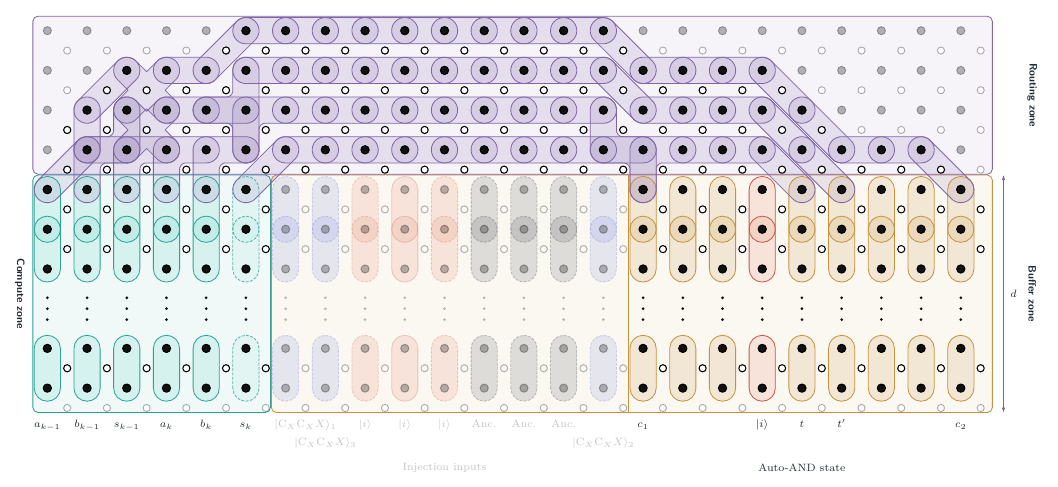}
        \caption{Injection $I_k$.}
    \end{subfigure}
    \par\medskip
    \begin{subfigure}[t]{\linewidth}
        \centering
        \includegraphics[width=\linewidth]{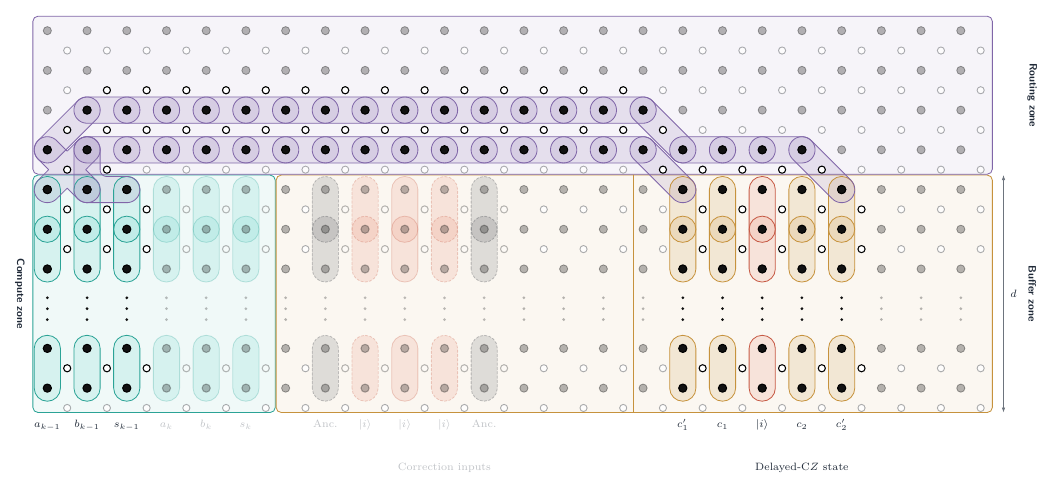}
        \caption{Correction $C_k$.}
    \end{subfigure}
    \par\medskip
    \includegraphics[scale=0.85]{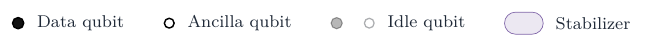}
    \caption{Routing layout for an injection and a correction step ($k > 1$) on the repetition-cat architecture.
        Each figure contains a loading (left) and available (right) buffer.}
    \label{fig:adder_routing_overlays}
\end{figure*}

\begin{figure*}[tp]
    \centering
    \includegraphics[width=\linewidth]{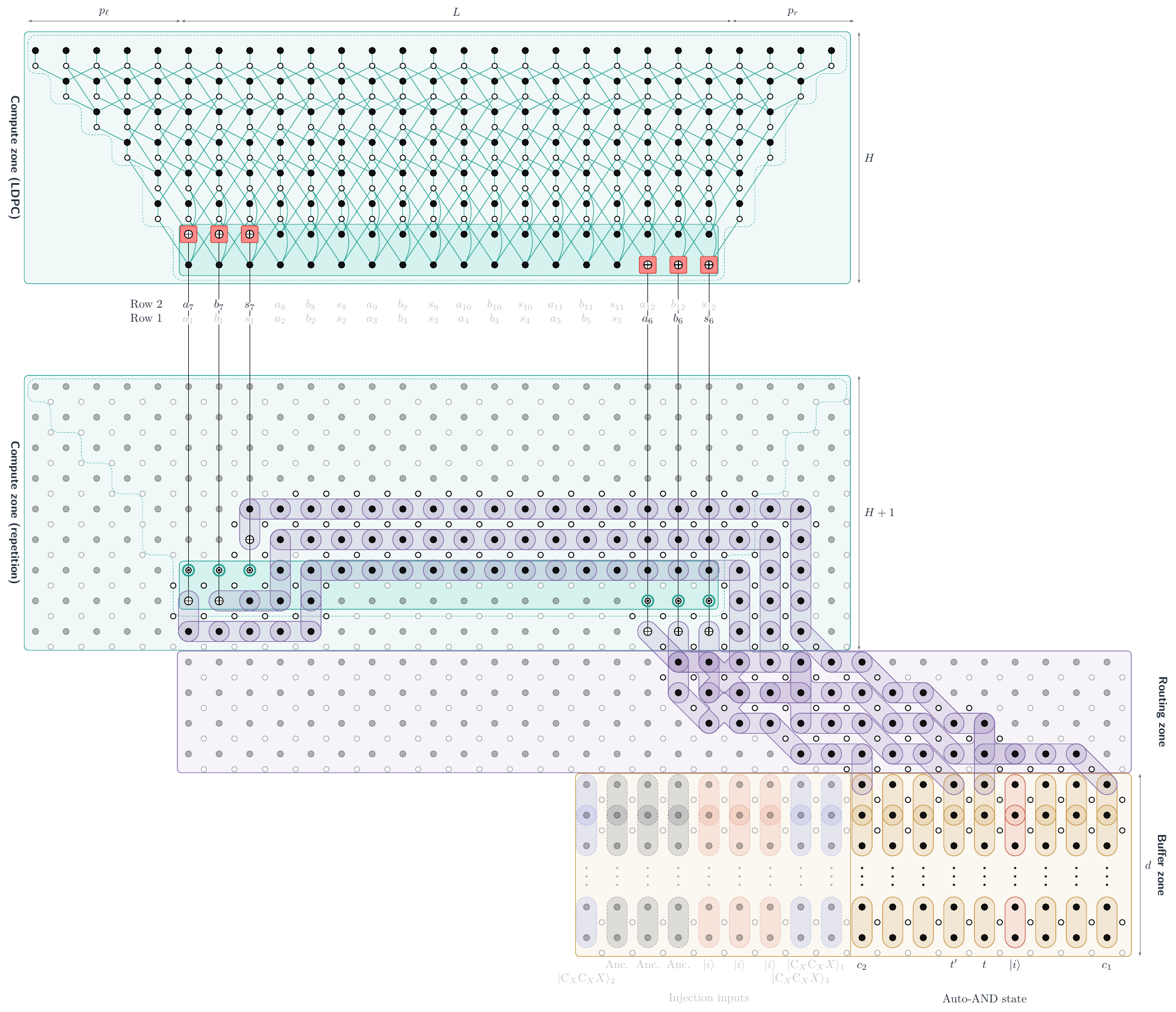}
    \par\medskip
    \includegraphics[scale=0.85]{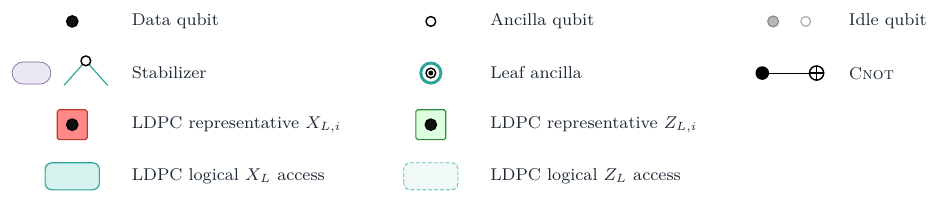}
    \caption{Layout for the injection step $I_7$ of the adder on the LDPC-cat architecture for $n=12$ and $d_{\mathrm{LDPC}}=22$. The compute zone lies above the routing zone and the buffers are mirrored left to right.}
    \label{fig:adder_ldpc_routing_overlays}
\end{figure*}

\begin{figure*}[tp]
    \centering
    \includegraphics[width=\linewidth]{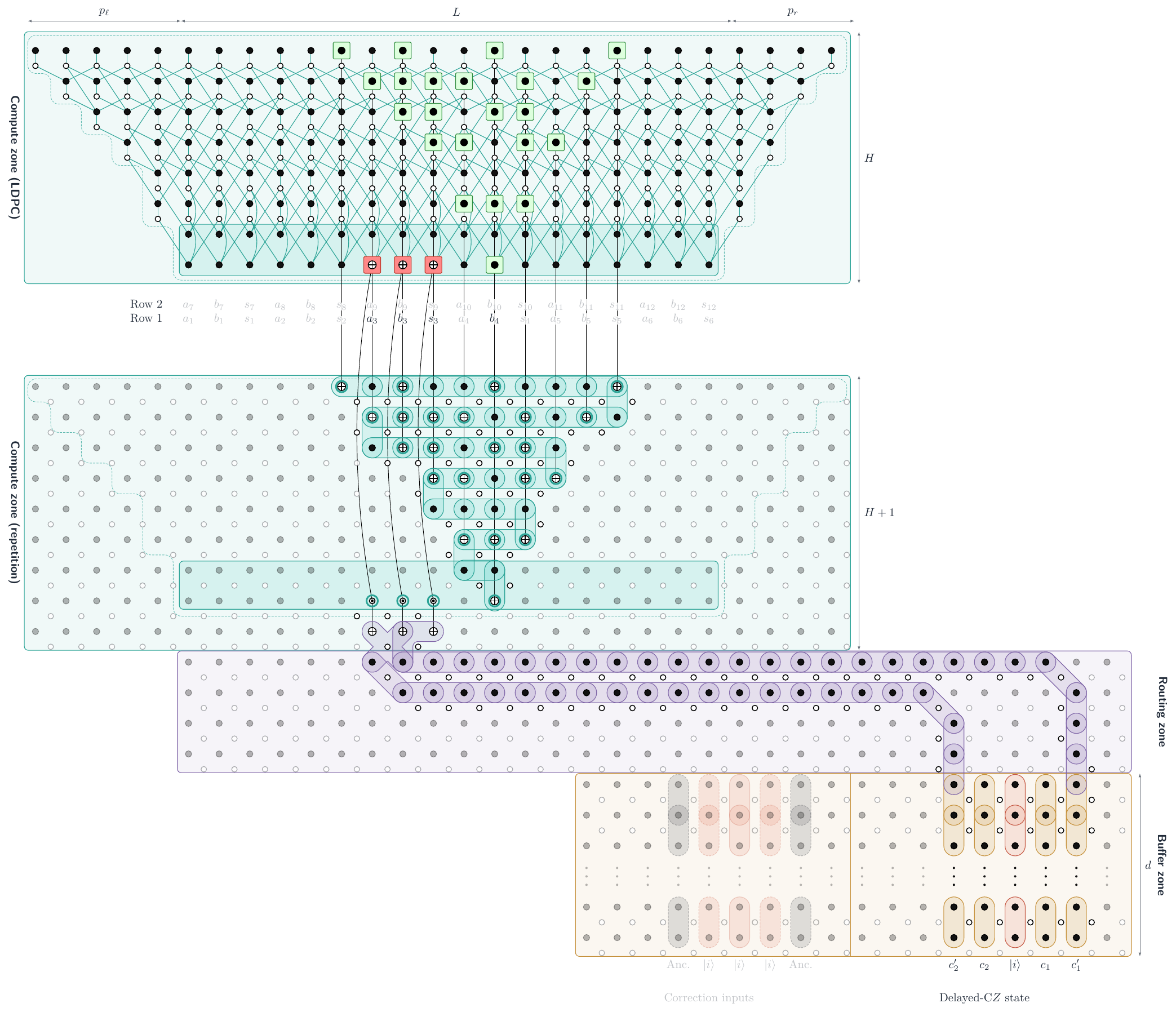}
    \par\medskip
    \includegraphics[scale=0.85]{assets/adder_ldpc/routing_layout_legend.pdf}
    \caption{Layout for the correction step $C_4$ of the adder on the LDPC-cat architecture, similarly to Figure~\ref{fig:adder_ldpc_routing_overlays}.}
    \label{fig:adder_ldpc_correction_layout}
\end{figure*}

Figure~\ref{fig:adder_routing_overlays} shows the injection and correction layouts on the repetition-cat architecture.
Figures~\ref{fig:adder_ldpc_routing_overlays} and~\ref{fig:adder_ldpc_correction_layout} present the same steps on the LDPC-cat architecture, where the compute side uses the primitives of Figure~\ref{fig:adder_routing_primitives}(a) and~(c) flipped top to bottom, and the buffers are mirrored left to right, such that the buffer side of the injection is the mirrored primitive of Figure~\ref{fig:adder_routing_primitives}(d).
For the first buffer, which lies under the compute zone, an injection or correction associated with a compute tuple above can be achieved by combining the compute and buffer primitives into a single routing adjusted to the relative position.
We note that this routing layout, designed for a compute zone on top of the buffers, can also be used for the repetition-cat architecture.

As for the injections and corrections, we allocate $T_{\mathrm{surg}}$ to each simultaneous loading of factory outputs into a buffer, during which the factory output and destination buffer are occupied.
The factory routing zone of height $h_{\mathrm{routing,fac}}$ must accommodate five resource qubits for auto-\textsc{And} preparation, comprising two $\ket{i}$ states and the three inputs of a single $\ket{\CXCXX}$ state, or two fresh $\ket{i}$ states for delayed-$\CZ$ preparation.
As such, each buffer also serves as an $\ket{i}$ and $\ket{\CXCXX}$ state storage area to maximize factory throughput.

\subsubsection{Schedule}
\label{sssec:adder_scheduling}

On the repetition-cat architecture, $I_1$ lasts $T_{\mathrm{init}}+3T_{\textsc{Cnot}}$ and $C_1$ ends with the surgery of $C_2$.
On the LDPC-cat architecture, assuming $s_1$ is initialized in $\ket{0}_L$, $I_1$ measures $X_{a_1}X_{b_1}X_{s_1}$ through the LDPC layer access zone, and $C_1$ measures the $Z_L$ of $b_1$ in its own surgery after that of $C_2$, each lasting $T^{(\mathrm{LDPC})}_{\mathrm{surg}}$.
For $k>1$, the $Z$ measurement of $b_k$ is performed during the surgery of $C_k$ on both architectures.
After $I_1$, the $2(n-1)$ remaining steps start at the cadence $T_{\mathrm{cad}}$ of Equation~\eqref{eq:cadence}, in the order $I_2,\ldots,I_n,C_n,\ldots,C_2$.
Each surgery lasts $T_{\mathrm{surg}}$ while the rest of the slot is idle.
On the LDPC-cat architecture, these surgeries involve the LDPC layer, such that $T^{(\mathrm{LDPC})}_{\mathrm{surg}}$ replaces $T_{\mathrm{surg}}$ in Equation~\eqref{eq:cadence} and in the following.
The sufficient condition $T_{\mathrm{react}}+T_{\mathrm{meas}}\leq T_{\mathrm{cad}}$ allows reaction and adaptive measurement to complete during the next step.
\paragraph{Buffers.}
Let $j\in\{\ket{\mathrm{AA}},\ket{\DCZ}\}$ denote the resource state consumed.
For a step starting at $t$, loading begins at $t-T_{\mathrm{prep}}^j-T_{\mathrm{surg}}$ and preparation begins at $t-T_{\mathrm{prep}}^j$.
A buffer remains reserved for $T_{\mathrm{buf}}$ cycles, through loading, preparation, surgery, and the final reaction and adaptive measurement.
We assign the steps to the $n_{\mathrm{buf}}^{(\Adder)}$ buffers in round-robin order, provided that
\begin{equation}\label{eq:adder_shared_buffer_count}
\begin{aligned}
    T_{\mathrm{buf}}
        &=2T_{\mathrm{surg}}+\max\!\left(T_{\mathrm{prep}}^{\ket{\mathrm{AA}}},T_{\mathrm{prep}}^{\ket{\DCZ}}\right)\\
        &\quad+T_{\mathrm{react}}+T_{\mathrm{meas}},\\
    n_{\mathrm{buf}}^{(\Adder)}
        &=\left\lceil\frac{T_{\mathrm{buf}}}{T_{\mathrm{cad}}}\right\rceil.
\end{aligned}
\end{equation}
Figure~\ref{fig:adder_reference_factory_buffer_timeline} shows the operation order and the complete buffer reservation.

\paragraph{Factories.}
During injections, one $\ket{\CXCXX}$ state is required every $T_{\mathrm{cad}}$, corrections require none, and every preparation requires two $\ket{i}$ states loaded simultaneously.
The loading of an output state in a buffer is executed in parallel with the production of the following state.
The full adder consumes $n-1$ $\ket{\CXCXX}$ states and $4(n-1)$ $\ket{i}$ states, that is, four $\ket{i}$ states per Toffoli gate, two for the auto-\textsc{And} state and two for the delayed-$\CZ$ state.
In addition, it consumes one catalyst seed per buffer, so that an $n$-bit adder consumes $4(n-1) + n_{\mathrm{buf}}^{(\Adder)}$ $\ket{i}$ states in total.
The seeds are produced and loaded during initialization, and each catalyst is retained for the successive preparations of the adder.

\paragraph{Latency.}
The cadence of Equation~\eqref{eq:cadence} minimizes the number of factories.
As the corrections $C_n,\ldots,C_2$ consume no $\ket{\CXCXX}$ state, we note that the correction phase could run at a shorter cadence under the current numbers of factories and buffers, bounded only by the lattice surgery duration, the $\ket{i}$ production time and the buffer reservation time.
For simplicity, we use the cadence $T_{\mathrm{cad}}$ for all steps, such that the latency of an $n$-bit adder is
\begin{equation}\label{eq:adder_current_warm_latency}\csname ltx@label\endcsname{eq:adder_ldpc_latency}
    T_{\Adder}^{(\mathrm{rep})}(n)\simeq T_{\Adder}^{(\mathrm{LDPC})}(n)\simeq 2nT_{\mathrm{cad}}.
\end{equation}

We call this schedule \emph{nominal}, as it assumes that every factory production completes within the associated mean time $T_{\mathrm{prod}}^j$.
A longer production can delay the step consuming its output, and with it all later steps.
This nominal latency is measured from the start of $I_1$ and excludes initialization.

\subsubsection{Summary}
\label{sssec:adder_resources}
Adding the core, factories and their associated routing results in a total footprint of
\begin{equation}\label{eq:adder_current_total_count}
    S_{\mathrm{total}}=S_{\mathrm{core}}+S_{\mathrm{fac}}+S_{\mathrm{routing,fac}},
\end{equation}
with $S_{\mathrm{core}}=S_{\mathrm{core}}^{(\mathrm{rep})}$ or $S_{\mathrm{core}}^{(\mathrm{LDPC})}$ chosen for the repetition-cat and LDPC-cat architectures, respectively.

The estimates use the factory parameters of Table~\ref{tab:adder_reference_factories} and the operation durations of Section~\ref{sssec:layout_connectivity}, such that $T_{\mathrm{buf}}=2d+20$ in Equation~\eqref{eq:adder_shared_buffer_count}.

\begin{figure*}[t]
    \centering
    \resizebox{\linewidth}{!}{\input{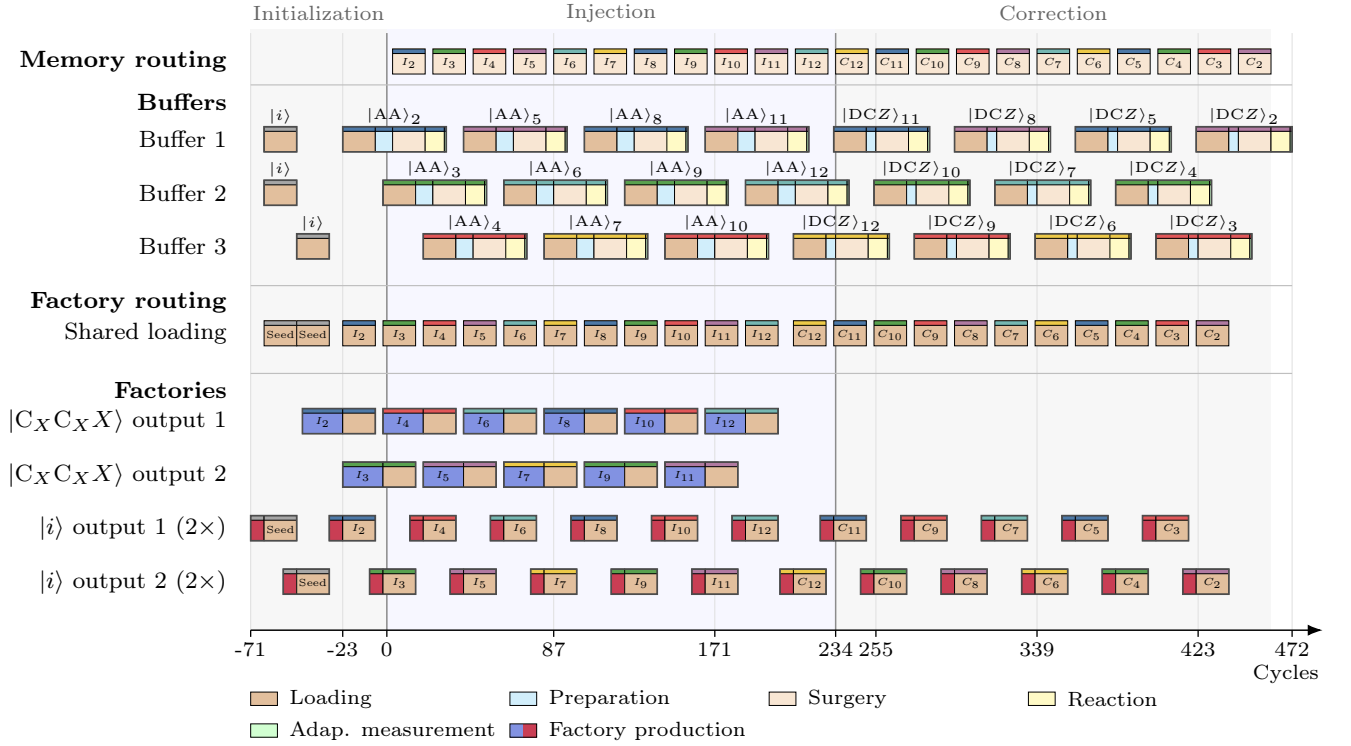}}
    \caption{Complete nominal schedule of a repetition-cat adder at $n=12$, $d=17$ and $T_{\mathrm{cad}} = 21$.}
    \label{fig:adder_reference_factory_buffer_timeline}
    \label{fig:adder_latency_timeline}
\end{figure*}

We choose $h_{\mathrm{routing,fac}}=10$ as a margin of two data rows for each of the five resource qubits in an injection load.

On the LDPC-cat architecture, the registers are stored in the code of Table~\ref{tab:adder_ldpc_codes} with the smallest distance $d_{\mathrm{LDPC}}\geq d_{\mathrm{rep}}$.
Table~\ref{tab:adder_closed_forms} provides the footprint of an $n$-bit adder under various distances, where the choice of $T_{\mathrm{cad}}$ from Equation~\eqref{eq:cadence} gives a latency of about $42n$ cycles at $d_{\mathrm{rep}}=17$, or about $44n$ cycles on the LDPC-cat architecture with $d_{\mathrm{LDPC}}=22$.
Under this cadence choice, one $\ket{\CXCXX}$ factory and two $\ket{i}$ factories suffice at every distance (Section~\ref{ssec:factories}).
The factory overhead, including routing, is $S_{\mathrm{routing,fac}}+S_{\mathrm{fac}}=2872$ qubits.
Figure~\ref{fig:adder_reference_factory_buffer_timeline} illustrates the complete schedule at $n=12$ for this parameter choice.

\begin{table*}[t]
    \centering
    \footnotesize
    \renewcommand{\arraystretch}{1.12}
    \begin{tabular}{@{}lcccccc@{}}
        \toprule
        Code & $H$ & $L_{\min}$ & $p_\ell$ & $p_r$ & $\Delta_{\mathrm{LDPC}}$ & LDPC layer \\
        \midrule
        $[8L+29,2L,22]$ & $8$ & $17$ & $5$ & $4$ & $29$ & $14L+58$ \\
        $[7L+26,2L,16]$ & $7$ & $17$ & $4$ & $5$ & $26$ & $12L+52$ \\
        $[6L+17,2L,12]$ & $6$ & $13$ & $3$ & $4$ & $17$ & $10L+34$ \\
        \bottomrule
    \end{tabular}
    \caption{Open-boundary LDPC-cat codes of height $H$ and width $L$~\cite{Ruiz_2025}, with $L=\max(L_{\min},\lceil 3n/2\rceil)$ for an $n$-bit adder.
    The paddings $p_\ell$ and $p_r$ extend the code to the left and right, $\Delta_{\mathrm{LDPC}}$ counts the qubits added by the open-boundary staircase, and the last column provides the number of data and ancilla qubits in the LDPC layer.}
    \label{tab:adder_ldpc_codes}
\end{table*}

\begin{table*}[t]
    \centering
    \footnotesize
    \renewcommand{\arraystretch}{1.15}
    \begin{tabular}{@{}cccccc@{}}
        \toprule
        & & \multicolumn{2}{c}{Repetition-cat} & \multicolumn{2}{c}{LDPC-cat} \\
        \cmidrule(lr){3-4}\cmidrule(lr){5-6}
        $d_{\mathrm{rep}}$ & $d_{\mathrm{LDPC}}$ & $n_{\mathrm{buf}}^{(\Adder)}$ & $S_{\mathrm{core}}^{(\mathrm{rep})}$
         & $n_{\mathrm{buf}}^{(\Adder)}$ & $S_{\mathrm{core}}^{(\mathrm{LDPC})}$ \\
        \midrule
        $11$ & $12$ & $2$ & $90n+540$   & $3$ & $32L+902$  \\
        $13$ & $16$ & $3$ & $102n+918$  & $3$ & $36L+1074$ \\
        $16$ & $16$ & $3$ & $120n+1080$ & $3$ & $36L+1236$ \\
        $17$ & $22$ & $3$ & $126n+1134$ & $3$ & $40L+1314$ \\
        \bottomrule
    \end{tabular}
    \caption{Footprint without factories, in qubits, of an $n$-bit adder for the cadence of Equation~\eqref{eq:cadence} and the operation durations of Section~\ref{sssec:layout_connectivity}. $L=\max(L_{\min},\lceil 3n/2\rceil)$ is the width of the LDPC code of Table~\ref{tab:adder_ldpc_codes}.}
    \label{tab:adder_closed_forms}
\end{table*}

\subsection{\texorpdfstring{\Lookup{} and \Phaseup}{Lookup and Phaseup}}\label{ssec:lookup_and_phaseup}

\paragraph{\texorpdfstring{\Lookup}{Lookup}}
A \Lookup{} operation, also called a QROM operation (quantum read-only memory), is an operation that loads values from a classical table in a quantum target register using an address register.
Let $w$ be the size of the address register, $n$ the size of the target register and $t$ the classical table that maps each address to an $n$-bit value:
\begin{equation}
    \begin{aligned}
        t:  \{0,1\}^w & \longrightarrow  \{0,1\}^n \\
          k & \longmapsto  t(k) \\
    \end{aligned}
\end{equation}
Let $b\in \{0,1\}^n$ and $(a_k)_{k\in[0,2^w-1]}$ such that $\sum_{k=0}^{2^w-1}|a_k|^2=1$.
The \Lookup{} associated with $t$ acts as:
\begin{equation}
    \mathrm{Lookup}\sum_{k=0}^{2^w-1}a_k\ket{k}\ket{b} = \sum_{k=0}^{2^w-1}a_k\ket{k}\ket{b \oplus t(k)}.
\end{equation}

In this work, we use the cat-qubit implementation of the skew-tree QROM detailed in~\cite{FeMoco2026prep}. We recall the space and time footprints derived in~\cite{FeMoco2026prep} for the reader's convenience. Reference~\cite{FeMoco2026prep} counts data qubits only, whereas Equations~\eqref{eq:lookup_cost_rep} and~\eqref{eq:lookup_cost_ldpc} give physical qubits, i.e.\ twice those counts, and Equation~\eqref{eq:lookup_cost_ldpc} further excludes the LDPC layer, which we count separately (Equation~\eqref{eq:Gamma-ldpc}).
With the repetition-cat architecture, the physical time and space cost of this implementation, for a $w$-qubit address register and an $n$-qubit target register, is
\begin{equation}
    \begin{aligned}
        &T_{\Lookup}^{(\mathrm{rep})}(w,d) = (2^w-1)\,d + (2^{w-1}-w)\,T_{\mathrm{cad}},\\
        &\Gamma_{\Lookup}^{(\mathrm{rep})}(w,n,d,n_{\mathrm{buf}}^{(\Lookup)})\\
        &\quad = 2\bigl[(2w + 3\lceil n/2 \rceil)(d+4) + 6\,n_{\mathrm{buf}}^{(\Lookup)}(d+1)\bigr],
    \end{aligned}
    \label{eq:lookup_cost_rep}
\end{equation}
with $T_{\Lookup}^{(\mathrm{rep})}$ and $\Gamma_{\Lookup}^{(\mathrm{rep})}$ the duration and physical qubit count for the \Lookup, $d$ the distance of the repetition code, $T_{\mathrm{cad}}$ the cadence of Equation~\eqref{eq:cadence} and $n_{\mathrm{buf}}^{(\Lookup)}$ the number of \Lookup{} buffers. In reverse depth-first order, the skew-tree QROM consists of $2^w - 1$ parity-\textsc{Cnot} gates and $2^w - w - 1$ \textsc{And} gates~\cite{FeMoco2026prep}. The $2^{w-1} - 1$ \textsc{And} gates that are directly followed by a parity-\textsc{Cnot} gate are performed simultaneously with it. The two gates use disjoint routing paths. The \textsc{And} gate and the parity-\textsc{Cnot} gate performed simultaneously are then followed by a measurement-based \textsc{And} uncomputation and another parity-\textsc{Cnot} gate, which leaves the factory $2T_{\mathrm{surg}}$ to produce the next $\ket{\CXCXX}$ state. Only the $2^{w-1} - w$ \textsc{And} gates followed by another \textsc{And} gate wait for the cadence. This requires $t_{\ket{\CXCXX}} \le 2T_{\mathrm{surg}}$, i.e.\ $d \ge 11$ with the factories of Table~\ref{tab:adder_reference_factories}, which holds at every distance considered in this work~\cite[Eq.~(4.8)]{FeMoco2026prep}. A \Lookup{} buffer~\cite{FeMoco2026prep} consists of six adjacent repetition-code columns, in which the $\ket{\CXCXX}$ and $\ket{i}$ resource states consumed by the \textsc{And} gates of the \Lookup{} are prepared. It differs from the nine-column adder buffer of Section~\ref{sec:adder}. Three of these columns receive the $\ket{\CXCXX}$ state, and the other three are $\ket{i}$ columns, one of which holds a catalyst, as in the adder buffer (Section~\ref{sssec:resource_states}). Each \textsc{And} gate thus consumes one $\ket{\CXCXX}$ state and two $\ket{i}$ states. Including one catalyst seed per buffer, loaded during the initialization of the \Lookup, a \Lookup{} consumes $2(2^w-w-1) + n_{\mathrm{buf}}^{(\Lookup)}$ $\ket{i}$ states. The auxiliary qubits of the \Lookup{} are uncomputed by measurement-based uncomputation~\cite{Berry_2019}. The resulting phase corrections are $\C_X X$ gates, that is, signs depending only on the address register. They are not applied in the \Lookup{} but folded into the classical table of the \Phaseup{} that resolves its phases, under the same condition as for combining \Lookup{}s: the address register is used only as a control in between. They therefore consume no resource state and add no step to Equation~\ref{eq:lookup_cost_rep}.
The \Lookup{} buffers therefore account for $6n_{\mathrm{buf}}^{(\Lookup)}$ repetition-code columns in Equations~\ref{eq:lookup_cost_rep} and~\ref{eq:lookup_cost_ldpc}. The factor $(d+4)$ of Equation~\eqref{eq:lookup_cost_rep} corresponds to the standalone layout of~\cite{FeMoco2026prep}, with three routing rows between the register and the buffer and one routing row on the other side. In the global layout of Section~\ref{sssec:global_layout_routing}, routing qubits are instead counted in $\Gamma_{\mathrm{global}}$ (Equation~\eqref{eq:Gglobal}), so this factor does not enter $n_{\mathrm{total}}$.
For the LDPC-cat architecture, the time and space cost are 
\begin{equation}
    \begin{aligned}
        &T_{\Lookup}^{(\mathrm{LDPC})}(w) = (2^{w}-1)\,T^{(\mathrm{LDPC})}_{\mathrm{surg}} + (2^{w-1}-w)\,T_{\mathrm{cad}},\\
        &\Gamma_{\Lookup}^{(\mathrm{LDPC})}(w,H,L,d_{\mathrm{rep}},n_{\mathrm{buf}}^{(\Lookup)})\\
        &\quad = 2\bigl[HL + \Delta_H + 2\max(L, 6n_{\mathrm{buf}}^{(\Lookup)}+w)\\
        &\qquad + \max(6n_{\mathrm{buf}}^{(\Lookup)} + w - L, 0)\\
        &\qquad + (6n_{\mathrm{buf}}^{(\Lookup)}+w)(d_{\mathrm{rep}}+1)\bigr],
    \end{aligned}
    \label{eq:lookup_cost_ldpc}
\end{equation}
with $T^{(\mathrm{LDPC})}_{\mathrm{surg}}$ the surgery duration of Equation~\eqref{eq:Tsurg-ldpc}, so that the duration is that of Equation~\eqref{eq:lookup_cost_rep} with $d$ replaced by $T^{(\mathrm{LDPC})}_{\mathrm{surg}}$~\cite[Eq.~(4.12)]{FeMoco2026prep}. Here, $d_{\mathrm{LDPC}}$ is the distance of the LDPC code, and $d_{\mathrm{rep}}$ the distance of the repetition code used for the buffer and the auxiliary qubits. The LDPC code has height $H$ and width $L=\lceil w/2\rceil + \lceil n/2\rceil$~\cite[Eq.~(4.11)]{FeMoco2026prep}, and $\Delta_H$ extra qubits form its triangular edge.
Physical qubits in the LDPC layer are not included in this formula, their count is $2(HL + \Delta_H) -2L$.
All \Lookup{}s in this paper, use $n_{\mathrm{buf}}^{(\Lookup)} = 2$.

\paragraph{\texorpdfstring{\Phaseup}{Phaseup}}
A \Phaseup{} on a $w$-qubit address register applies a phase that depends only on the value of this register, $\sum_k a_k \ket{k}_X \mapsto \sum_k (-1)^{T_k} a_k \ket{k}_X$, where $T : \{0,1\}^w \to \{0,1\}$ is a classically known table.
In this work, \Phaseup{}s resolve the phases left by the measurement-based uncomputation~\cite{Berry_2019} of \Lookup{}s.
Following the color-changed convention introduced in Section~\ref{ssec:colour_changed}, measurement-based uncomputation clears the target register of the \Lookup{} by measuring each target qubit in the $Z$ basis.
This measurement leaves on the address register, whose states are written in the $X$ basis, a sign $(-1)^{T_k}$, where the table $T$ depends on the measurement outcomes.
A \Lookup{} is therefore uncomputed by this measurement followed by a \Phaseup{} that removes this sign.
It does not need to be applied immediately after the measurement: it can be delayed as long as the address register is only used as control qubits between the measurement and the \Phaseup.
Under this condition, the corrections of several \Lookup{}s that share the same address register can be combined into a single \Phaseup.

We use the color-changed version of the \Phaseup{} construction of~\cite[Appendix~A.3]{gidney2025factor2048bitrsa}.
The address register is split into a high half of $h=\lceil w/2\rceil$ qubits and a low half of $l=\lfloor w/2\rfloor$ qubits, each expanded into its power product with \textsc{And} gates. The empty product of each half is omitted, and the single-qubit entries are the address qubits themselves, so only the $2^{h}-h-1$ and $2^{l}-l-1$ entries of weight at least two require an \textsc{And} gate.
The sign is then applied by multi-target $\C_X X$ gates, and the power products are uncomputed by measurement-based uncomputation.
Counting the gates of this construction, a \Phaseup{} on a $w$-qubit address register requires $2^{h}+2^{l}-w-2$ \textsc{And} gates, $2^{l}-1$ multi-target $\C_X X$ gates and $2^{h}+2^{l}-2$ workspace qubits, including the address register.
This agrees with the leading-order count of~\cite{gidney2025factor2048bitrsa} and, for $w=6$, with the 8 Toffoli gates quoted there.
On the repetition-cat architecture, the $2^h + 2^l - w - 2$ \textsc{And} gates are cadence-limited and each of the $2^l - 1$ multi-target $\C_X X$ gates takes one lattice surgery, so that this logical cost translates into the following approximate physical cost:
\begin{equation}
    \begin{aligned}
        T_{\Phaseup}^{(\mathrm{rep})}(w,d) & \simeq \left(2^{h}+2^{l}-w-2\right) T_{\mathrm{cad}} \\
        & \quad + \left(2^{l}-1\right) d, \\
        \Gamma_{\Phaseup}^{(\mathrm{rep})}(w,d,n_{\mathrm{buf}}^{(\Lookup)}) & \simeq 2d\bigl(2^{h}+2^{l}-2 \\
        & \quad + 6\,n_{\mathrm{buf}}^{(\Lookup)}\bigr).
    \end{aligned}
    \label{eq:phaseup_cost}
\end{equation}
with $T_{\Phaseup}^{(\mathrm{rep})}$ and $\Gamma_{\Phaseup}^{(\mathrm{rep})}$ the duration and physical qubit count for the \Phaseup, $d$ the distance of the repetition code, and $n_{\mathrm{buf}}^{(\Lookup)} = 2$ the number of \Lookup{} buffers, each made of six repetition-code columns (see~\cite{FeMoco2026prep}).
The \textsc{And} gates of the \Phaseup{} consume the same resource states as the \textsc{And} gates of a \Lookup. 
Each requires one $\ket{\CXCXX}$ state and two $\ket{i}$ states loaded into a buffer, and is issued at the cadence $T_{\mathrm{cad}}$ of Equation~\ref{eq:cadence}. 
A multi-target $\C_X X$ gate with control $x$ and targets $y_1, \ldots, y_m$ applies the phase $(-1)^{x\cdot(y_1\oplus\cdots\oplus y_m)}$, that is, a single $\C_X X$ gate between $x$ and the $X$-parity of its targets. 
Since $\C_X X$ maps $Z_x$ to $Z_x X_y$, it cannot be implemented with \textsc{Cnot} gates and $X$- and $Z$-basis preparations and measurements alone. 
We apply it with a $\ket{\CZ} = \CZ\ket{+}\ket{+}$ state, prepared in a \Lookup{} buffer from two $\ket{i}$ states and the catalyst with the four \textsc{Cnot} gates of Section~\ref{sssec:resource_states}. 
One lattice surgery that measures the $X$-parity of each $\ket{\CZ}$ qubit with the corresponding data, followed by $Z$ measurements of the two $\ket{\CZ}$ qubits. 
The gate thus consumes no $\ket{\CXCXX}$ state. 
Unlike the corrections of the adder, the $\C_X X$ gates of the \Phaseup{} do not require a delayed-$\CZ$ state: whether each of them is applied is known before its surgery, so no delayed choice, and hence no $Z$-copy, is needed. 
The measurement-based uncomputation of a power product requires, with probability $1/2$, a $\C_X X$ phase correction between its two factors. 
We apply these phase corrections adaptively: once the measurement outcome is known, after the reaction time, a $\ket{\CZ}$ state is used only if the phase correction is needed, and otherwise kept in its buffer for the next phase correction. 
This consumes on average one $\ket{i}$ state per power product. 
Including the two catalyst seeds of the \Lookup{} buffers, a \Phaseup{} on a $w$-qubit address register consumes on average
\begin{equation}
    N_{\ket{i}}^{\Phaseup}(w) = 3\left(2^{h}+2^{l}-w-2\right) + 2\left(2^{l}-1\right) + 2
    \label{eq:phaseup_istates}
\end{equation}
$\ket{i}$ states, that is $40$ for $w = 6$ and $23$ for $w = 5$.
Each \textsc{And} gate occupies a buffer for the same time as an \textsc{And} gate of a \Lookup, and the \textsc{And} gates are issued at the same rate. The \Phaseup{} thus needs the same number of buffers, $n_{\mathrm{buf}}^{(\Lookup)} = 2$, which adds $6 n_{\mathrm{buf}}^{(\Lookup)}$ repetition-code columns to its footprint. A \Phaseup{} never runs concurrently with a \Lookup, so in the global layout it reuses the \Lookup{} buffers and adds no qubits (Section~\ref{sssec:global_layout_routing}).
As everywhere, a repetition-code column costs $2d$ physical qubits, and routing qubits are counted in the global layout. 
Equation~\ref{eq:phaseup_cost} neglects the duration of the measurement-based uncomputation of the power products, including the reaction time and the surgeries of the adaptive phase corrections. 
The equation also assumes that the multi-target $\C_X X$ gates are issued every $d$ cycles, whereas each of them now also occupies a buffer with a $\ket{\CZ}$ state. 
Approximate estimates in space and time are acceptable for \Phaseup{} operations since they are not determinant for global layouts and represent a small fraction of the algorithm's total duration (see Table~\ref{tab:physical_cycle_breakdown}).  
On the LDPC-cat architecture, the address register is stored in the LDPC layer, whereas the power-product auxiliaries and the \Lookup{} buffers are stored in the repetition layer.
The cost of Equation~\eqref{eq:phaseup_cost} then becomes
\begin{equation}
    \begin{aligned}
        &T_{\Phaseup}^{(\mathrm{LDPC})}(w,d_{\mathrm{LDPC}})\\
        &\quad \simeq (2^{h}+2^{l}-w-2)\,T_{\mathrm{cad}} + (2^{l}-1)\,d_{\mathrm{LDPC}},\\
        &\Gamma_{\Phaseup}^{(\mathrm{LDPC})}(w,H,d_{\mathrm{rep}},n_{\mathrm{buf}}^{(\Lookup)})\\
        &\quad \simeq 2\bigl(\lceil w/2\rceil H + \Delta_\mathrm{LDPC}\bigr)\\
        &\qquad + 2d_{\mathrm{rep}}\bigl(2^{h}-h+2^{l}-l-2+6\,n_{\mathrm{buf}}^{(\Lookup)}\bigr),
    \end{aligned}
    \label{eq:phaseup_cost_ldpc}
\end{equation}
The duration is that of Equation~\eqref{eq:phaseup_cost} with $d$ replaced by $d_{\mathrm{LDPC}}$ for the multi-target $\C_X X$ gates.
In the footprint, the first term counts the repetition-layer zone facing the address register, with the same convention as Equation~\eqref{eq:lookup_cost_ldpc}.
Each column of the LDPC-cat code holds two logical qubits, so the $w$ address qubits occupy $\lceil w/2\rceil$ columns of height $H$.
The last term counts the repetition-code columns, each accounted as $2d_{\mathrm{rep}}$ physical qubits, its repetition code with the ancilla row above it.
Compared with Equation~\eqref{eq:phaseup_cost}, the $h$ high and $l$ low address qubits, which are the single-qubit entries of the two power products, are now in the LDPC layer and leave the repetition-code workspace.
This workspace thus reduces to the $2^{h}-h-1$ and $2^{l}-l-1$ entries of weight at least two, that is, one repetition-code column per \textsc{And} gate, plus the $6n_{\mathrm{buf}}^{(\Lookup)}$ columns of the buffers.
As for Equation~\eqref{eq:phaseup_cost}, routing qubits are not included, and the approximations listed above apply.

\subsection{Repetition-cat architecture} \label{ssec:repetition_cat_architecture}

In this section, we estimate the physical resources required to factor a 2048-bit RSA integer on the repetition-cat architecture.
Section~\ref{sssec:global_layout_routing} describes the layout, the duration and the volume of the computation for the algorithmic parameters of Table~\ref{tab:algo_parameter_value} at an arbitrary code distance $d$. Section~\ref{ssec:physical_resource_count} then evaluates the estimate at $d = 13$, $15$ and $17$ for $p_Z = 10^{-3}$ and $p_Z = 5\times10^{-4}$.

\subsubsection{Layout, duration and volume at distance $d$} \label{sssec:global_layout_routing}

\paragraph{Introduction} \label{sec:C7-inputs}
This section estimates the number of physical qubits, the duration and
the space-time volume of one shot of the algorithm of
Section~\ref{sec:logical_impl} on the repetition-cat architecture, as functions of
the code distance $d$. The estimate uses the physical implementation of the
operations of Sections~\ref{sec:adder} and~\ref{ssec:lookup_and_phaseup} and the
algorithmic parameters of Table~\ref{tab:algo_parameter_value}. Following
Table~\ref{tab:logical_qubit_count}, we divide the algorithm into temporal slices, and
for each slice we count the repetition-code columns it uses and its
duration. We assume an error-correction cycle time of
$1\,\mu\mathrm{s}$, a reaction time of $10\,\mu\mathrm{s}$, and no bit
flip during the computation. The factories are those of
Section~\ref{ssec:factories}: one $\ket{\CXCXX}$ factory and two $\ket{i}$
factories, shared by all operations, which run at the cadence
$T_{\mathrm{cad}} = \max(d, 21)$ of Equation~\ref{eq:cadence}. We fix the
number of adder buffers to $n_{\mathrm{buf}}^{(\Adder)} = 3$. This satisfies
Equation~\ref{eq:adder_shared_buffer_count} for every $d$, since
$T_{\mathrm{buf}} = 2d + 20 \le 3\max(d, 21)$, and it is the minimum for
$d \ge 12$. The sufficient condition
$T_{\mathrm{react}} + T_{\mathrm{meas}} = 11 \le T_{\mathrm{cad}}$ also
holds. \Lookup{}s and \Phaseup{}s use
$n_{\mathrm{buf}}^{(\Lookup)} = 2$ resource-state buffers of six
repetition codes each (Section~\ref{ssec:lookup_and_phaseup}). Physical qubits are counted in two parts. Every logical qubit,
resource state or operation slot is held in a repetition-code column:
$d$ data qubits, each counted with the ancilla directly above it, i.e.\ $2d$ physical qubits (Section~\ref{sec:adder}). The active columns of a \Lookup{} are those of its standalone footprint
(Equation~\ref{eq:lookup_cost_rep}), which includes its own routing columns. All other qubits of the chip (except factory qubits) are routing qubits: the routing
rows, the routing column of the compute zone and the factory routing
zone. They are never counted within an operation and instead form the global
routing $\Gamma_{\mathrm{global}}$, which is fixed by the global layout
and present during the whole shot. A slice $s$ therefore costs
$\Gamma_{\mathrm{physical}}(s) = 2d\,n_{\mathrm{col}}(s)$ physical
qubits, with $n_{\mathrm{col}}(s)$ its number of repetition-code columns
(Table~\ref{tab:rep-code-columns}), on top of $\Gamma_{\mathrm{global}}$ and of the
factories. The factories enter only through their footprint
$n_{\mathrm{fac}}$ and the factory routing zone, of height
$h_{\mathrm{routing,fac}}$ and length $l_{\mathrm{fac}}$, whose values
are given in Section~\ref{ssec:physical_resource_count}. The Layout paragraph below states the counting convention and gives
$n_{\mathrm{col}}(s)$ and $\Gamma_{\mathrm{global}}$, the Duration
paragraph the duration of each slice, and the Volume paragraph combines
them into the logical volume of a shot.

\begin{table*}[ht]
    \centering
    \begin{tabular}{|l|p{0.6\textwidth}|r|}
        \hline
        \textbf{Parameter} & \textbf{Meaning} & \textbf{Value} \\
        \hline
        $n$      & Bit size of the RSA modulus $N$. & 2048 \\
        $s$      & Ekerå--Håstad parameter: a larger $s$ shortens the exponent register but requires more runs. & 8 \\
        $n_e$    & Size of the exponent register. & 1280 \\
        $|P|$    & Number of primes in the RNS. & 22\,658 \\
        $\ell$   & Bit length of an RNS prime. & 20 \\
        $\kappa$ & Size of the accumulation register. & 33 \\
        $w_1$    & Window of the discrete-log computation. & 6 \\
        $w_3$    & Window of the exponentiation loops, $w_3 = w_{3a} = w_{3b}$ (Table~\ref{tab:modexp_operation_count}).\footnote{\ReduceExponent{} uses no window, hence there is no $w_2$.} & 3 \\
        $w_4$    & Window of the accumulation. & 5 \\
        \hline
    \end{tabular}
    \caption{Algorithmic parameters of the RSA-2048 instance used in the resource estimates of Sections~\ref{ssec:repetition_cat_architecture} and~\ref{ssec:ldpc_cat_layout}, taken from the RSA-2048 row of Table~\ref{tab:instances}. The exponent register gathers the two Ekerå--Håstad registers, of $(1/2 + 1/(2s))n = 1152$ and $n/(2s) = 128$ qubits, hence $n_e = (1/2 + 1/s)n$ (Section~\ref{sec:input-register}).}
    \label{tab:algo_parameter_value}
\end{table*}

\paragraph{Layout} \label{sec:C7-global} \label{par:accumulation_layout_and_schedule} \label{sec:C7-other-subroutines} \label{sec:C7-reshuffling}
\begin{table*}[h]
    \centering
    \begin{tabular}{|l|l|c|c|c|}
        \hline
        \multirow{2}{*}{\textbf{Subroutine}} & \multirow{2}{*}{\textbf{Operation}} & \multicolumn{3}{c|}{\textbf{Repetition-code columns}} \\
        \cline{3-5}
         & & \textbf{Storage} & \textbf{Active} & \textbf{Total} \\
        \hline
        \multirow{2}{*}{\ComputeExponent} & \Lookup(6,31) & 1338 & 72 & 1410 \\
         & Q-Q \Adder(31) & 1313 & 120 & 1433 \\
        \hline
        \multirow{2}{*}{\ReduceExponent} & Q-C \Adder(21) & 1323 & 90 & 1413 \\
         & Q-Q \Adder(20) & 1324 & 87 & 1411 \\
        \hline
        \multirow{4}{*}{\ComputeResidue} & \Lookup(6,20) (loading) & 1338 & 54 & 1392 \\
         & \Lookup(6,21) & 1378 & 57 & 1435 \\
         & Q-Q \Adder$^\dagger$(21) & 1364 & 90 & 1454 \\
         & Q-Q \Adder(20) & 1365 & 87 & 1452 \\
        \hline
        \multirow{6}{*}{\AccumulateResidue} & \Lookup(5,34) & 1359 & 73 & 1432 \\
         & Q-Q \Adder$^\dagger$(34) & 1331 & 129 & \textbf{1460} \\
         & Q-Q \Adder(33) & 1332 & 126 & 1458 \\
         & \Lookup(5,33) & 1359 & 73 & 1432 \\
         & Q-Q \Comparison(33) & 1331 & 129 & \textbf{1460} \\
         & \Phaseup(5) & 1359 & 22 & 1381 \\
        \hline
        \multirow{7}{*}{\ComputeResidue$^\dagger$} & \Lookup(6,21) & 1378 & 57 & 1435 \\
         & Q-Q \Adder$^\dagger$(21) & 1364 & 90 & 1454 \\
         & Q-Q \Adder(20) & 1365 & 87 & 1452 \\
         & \Lookup(6,20) & 1378 & 54 & 1432 \\
         & Q-Q \Comparison(20) & 1364 & 90 & 1454 \\
         & \Phaseup(6) & 1378 & 26 & 1404 \\
         & \Phaseup(6) (unload) & 1344 & 26 & 1370 \\
        \hline
        Last \Phaseup{} & \Phaseup(6) & 1307 & 26 & 1333 \\
        \hline
    \end{tabular}
    \caption{Breakdown by operations of the number of repetition-code columns for the factorization of an
    RSA-2048 number. Parameters are selected from Table~\ref{tab:algo_parameter_value}. A repetition-code column is a vertical strip of
    the chip that holds one repetition code, counted with one ancilla per data qubit, and costs $2d$ physical qubits. All three
    counts (Storage, Active, Total) are in repetition-code columns. Storage counts the logical qubits held
    outside the operation, one repetition-code column each.
    Active counts the repetition-code columns of the current operation, including its
    buffers: $3m+9n_{\mathrm{buf}}^{(\Adder)}$ with $n_{\mathrm{buf}}^{(\Adder)}=3$ for an adder of size $m$ (Equation~\ref{eq:adder_core_count_piecewise}; a
    comparison of size $m$ is an adder of size $m+1$), $2w+3\lceil m/2\rceil+6n_{\mathrm{buf}}^{(\Lookup)}$ with
    $n_{\mathrm{buf}}^{(\Lookup)}=2$ for a \Lookup$(w,m)$, i.e.\ its standalone footprint (Equation~\ref{eq:lookup_cost_rep}) including its own routing columns, and
    $2^{\lceil w/2\rceil}+2^{\lfloor w/2\rfloor}-2+6n_{\mathrm{buf}}^{(\Lookup)}$ with $n_{\mathrm{buf}}^{(\Lookup)}=2$ for a \Phaseup$(w)$ (Equation~\ref{eq:phaseup_cost}). Total is the sum of Storage and Active. The routing rows and routing zones of the chip
    ($\Gamma_\text{global}$, Section~\ref{sssec:global_layout_routing}) and the factories ($n_\mathrm{fac}=1872$ physical qubits) are not included. The maximum, 1460
    repetition-code columns, is reached by the Q-Q \Adder$^\dagger$(34) and the Q-Q \Comparison(33) of
    \AccumulateResidue, and the global layout of Section~\ref{sssec:global_layout_routing} is sized for this maximum.}
    \label{tab:rep-code-columns}
\end{table*}

\begin{figure*}[htbp]
    \centering
    \includegraphics[width=\linewidth]{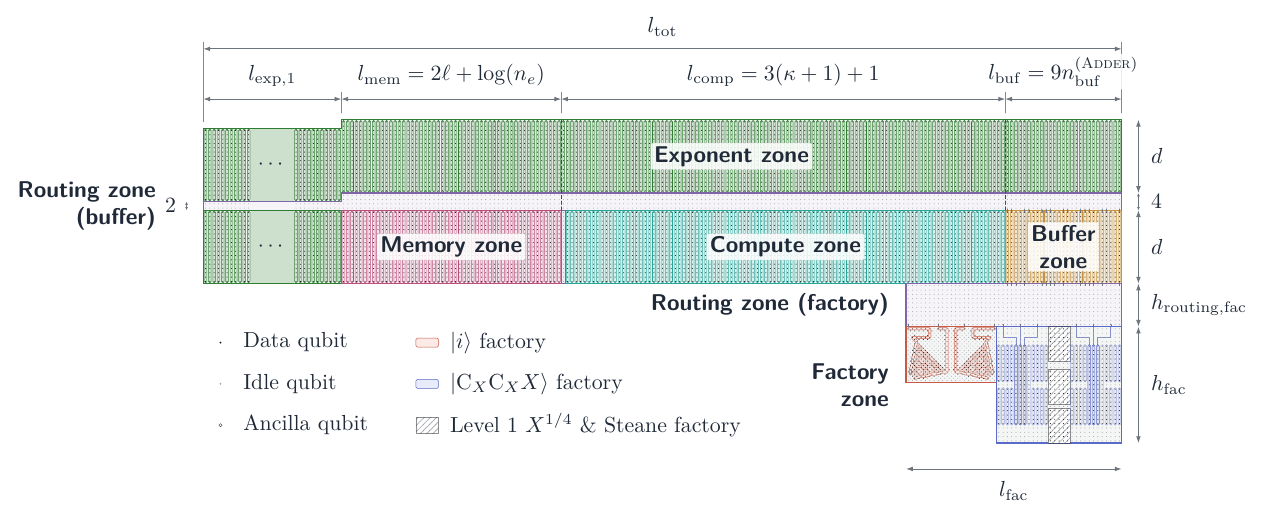}
    \caption{Global layout of the repetition-cat chip for RSA-2048.
    The compute zone holds the $3(\kappa+1)$ interleaved slots of the \Adder{} plus one routing column on its left.}
    \label{fig:global_layout}
\end{figure*}

Figure~\ref{fig:global_layout} shows the global layout. The arrangement of the
zones is set by the most frequent operations, \Lookup{} and \Adder, and the
length of the chip by the widest slice of Table~\ref{tab:rep-code-columns}. The
chip consists of two rectangles stacked vertically, each one repetition
code high, with a routing zone between them that spans the length of
the chip. The top rectangle and the left part of the bottom one store
the exponent. The right end of the bottom rectangle holds the buffer
zone, which stores resource states. The compute zone, which consumes
them, sits directly to its left, and the memory zone, between the bottom
exponent zone and the compute zone, stores the remaining logical qubits.
The routing zone has four rows above the memory, compute and buffer
zones. Above the memory and compute zones, these rows provide the three lanes
and the crossing row used to move registers~\cite{Simultaneous2026prep}, and above the
buffers they carry the injection and correction routing of the adder
(Section~\ref{sssec:adder_routing}). During a \Lookup, the lowest of these rows holds the $\ket{0}_L$ code of the parity-\textsc{Cnot} gates and the three others carry the $X$-parity measurements of the \textsc{And} gates (see below). Two rows suffice between the two exponent
zones, since exponent qubits are only used as \Lookup{} addresses and
controls, so that only their $X_L$ operator needs to be accessible. In
the color-changed implementation (Section~\ref{ssec:colour_changed}), this also holds
for the inverse quantum Fourier transform at the end of the algorithm,
which needs only $X_L$ access to the exponent qubits and a transversal
destructive $Z$-basis measurement. Exponent qubits are therefore stored
side by side, with no routing between them. The compute zone holds the
$3(\kappa+1)$ interleaved slots of the \Adder$^\dagger$ of \AccumulateResidue,
plus one routing column on its left, which gives the
\Lookup{} $Z_L$ access to its leftmost output, hence
$l_{\mathrm{comp}} = 3(\kappa+1)+1$. The memory zone holds two
$\ell$-qubit registers and the quotient register,
$l_{\mathrm{mem}} = 2\ell + \log(n_e)$, and the buffer zone is sized for
the \Adder, $l_{\mathrm{buf}} = 9 n_{\mathrm{buf}}^{(\Adder)}$. The factories sit
below the buffer zone, separated from it by $h_{\mathrm{routing,fac}}$
rows that route resource states to the buffers.

Lengths are counted in repetition-code columns. Both rectangles have the
same length, so that the chip has length
\begin{equation}
  l_{\mathrm{tot}} = \left\lceil
    \frac{l_{\mathrm{buf}} + l_{\mathrm{comp}} + l_{\mathrm{mem}} + n_e}{2}
  \right\rceil ,
  \label{eq:ltot}
\end{equation}
and the bottom exponent zone has length
$l_{\mathrm{exp},1} = l_{\mathrm{tot}} - (l_{\mathrm{buf}} +
l_{\mathrm{comp}} + l_{\mathrm{mem}})$, the top rectangle holding the
remaining exponent qubits. The chip thus offers $2l_{\mathrm{tot}}$
repetition-code columns. Each row of data qubits owns the row of ancillae above it, so that a zone of $h$ rows of data qubits and $l$ columns costs $2hl$ physical qubits, and a repetition-code column $2d$ (Section~\ref{sec:adder}). The global layout adds the zones and subtracts the ancilla rows above exposed zone tops, i.e.\ zone tops with nothing above them, since an ancilla row is only needed between two zones. Here, the only exposed tops are those of the top rectangle, with $l_{\mathrm{tot}}$ ancillae. Both routing zones lie under the top rectangle. The factory routing zone lies under the buffer zone and, for its $l_{\mathrm{fac}} - l_{\mathrm{buf}}$ other columns, under the compute zone, so that it is not exposed either. The global routing consists of the factory routing zone, the
routing zone between the two exponent zones, the routing zone above the
memory, compute and buffer zones, and the routing column of the compute
zone:
\begin{equation}
  \begin{split}
  \Gamma_{\mathrm{global}}
  = {}& 2h_{\mathrm{routing,fac}}\, l_{\mathrm{fac}}
  + 2\times 2\, l_{\mathrm{exp},1} \\
  & + 2\times 4\,(l_{\mathrm{buf}} + l_{\mathrm{comp}} + l_{\mathrm{mem}})
  + 2d .
  \end{split}
  \label{eq:Gglobal}
\end{equation}
The total number of physical qubits is the sum of the zones minus the exposed tops,
\begin{equation}
  \begin{split}
  n_{\mathrm{total}} = {}& 2d\cdot 2l_{\mathrm{tot}} + 2\times 2\, l_{\mathrm{exp},1} \\
  & + 2\times 4\,(l_{\mathrm{buf}} + l_{\mathrm{comp}} + l_{\mathrm{mem}}) \\
  & + 2h_{\mathrm{routing,fac}}\, l_{\mathrm{fac}} - l_{\mathrm{tot}} + n_{\mathrm{fac}} .
  \end{split}
  \label{eq:ntotal}
\end{equation}

For RSA-2048, the parameters of Table~\ref{tab:algo_parameter_value} give
$l_{\mathrm{comp}} = 103$, $l_{\mathrm{mem}} = 51$ and
$l_{\mathrm{buf}} = 27$. The chip has length $l_{\mathrm{tot}} = 731$
and the bottom exponent zone has length $l_{\mathrm{exp},1} = 550$.
Since $l_{\mathrm{buf}} + l_{\mathrm{comp}} + l_{\mathrm{mem}} + n_e =
1461$ is odd, the top rectangle holds the remaining 730 exponent qubits
in 731 columns, so one exponent slot is unused. Equations~\ref{eq:Gglobal}
and~\ref{eq:ntotal} become
\begin{align}
  \Gamma_{\mathrm{global}}(d) &= 2d + 3648 + 2h_{\mathrm{routing,fac}}\, l_{\mathrm{fac}} ,
  \label{eq:Gglobal_d}\\
  n_{\mathrm{total}}(d) &= 2924\, d + 3648
  + 2h_{\mathrm{routing,fac}}\, l_{\mathrm{fac}} - 731 + n_{\mathrm{fac}} ,
  \label{eq:ntotal_d}
\end{align}
where the constant of $\Gamma_{\mathrm{global}}(d)$ collects the routing zone between the exponent zones ($2200$) and the routing zone above the memory,
compute and buffer zones ($1448$), and the routing column gives $2d$. In $n_{\mathrm{total}}(d)$, $3648$ is the routing of $\Gamma_{\mathrm{global}}(d)$ outside the routing column, and $731 = l_{\mathrm{tot}}$ removes the ancillae above the exposed top of the top rectangle.
The column counts $n_{\mathrm{col}}(s)$ of Table~\ref{tab:rep-code-columns} do not
depend on $d$. Their maximum, $1460$, is reached by the
Q-Q \Adder$^\dagger(\kappa+1)$ and the Q-Q \Comparison$(\kappa)$ of
\AccumulateResidue, and equals $2l_{\mathrm{tot}} - 2$. At the peak, every
repetition-code column of the chip is in use except the routing column
and the unused exponent slot. Hence, at every $d$,
\begin{equation}
  n_{\mathrm{total}}(d) = \max_s \Gamma_{\mathrm{physical}}(s)
  + \Gamma_{\mathrm{global}}(d) + 2d + n_{\mathrm{fac}} - l_{\mathrm{tot}} ,
  \label{eq:ntotal_peak}
\end{equation}
the term $2d$ being the unused exponent slot and $l_{\mathrm{tot}}$ the exposed ancillae above the top rectangle.

\begin{figure*}[htbp]
    \centering
        \includegraphics{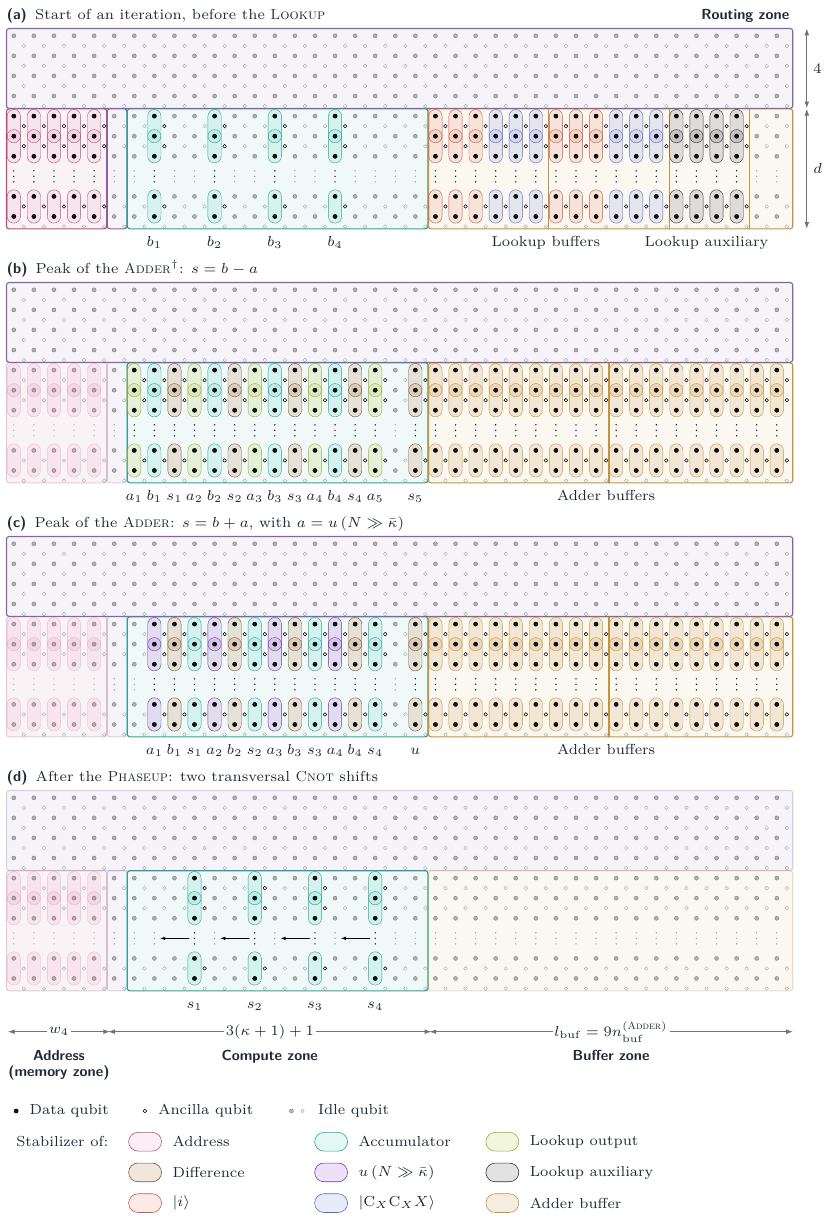}

  \caption{Repetition-cat layouts at four time steps of \AccumulateResidue{} (Figure~\ref{fig:accumulation}), drawn for $\kappa = 4$ and $n_{\mathrm{buf}}^{(\Adder)} = 2$. 
  Columns $a_k$, $b_k$ and $s_k$ are the slots of the \Adder$^\dagger$ in (a) and (b), and of the \Adder, one column to the right, in (c) and (d). 
  Zones that a step does not use are faded.
In (a), $b_{\kappa+1} = 0$ is not allocated (Equation~\ref{eq:adder_sum_identity}).
In (b), between the last injection and the first correction (Figure~\ref{fig:adder_latency_timeline}), the top output bit $s_{\kappa+1}$ is the underflow qubit $u$.
In (c), after the output of the \Lookup{} is measured, $\mathrm{PC}X(\kappa)$ loads $a = u\,(N\gg\bar\kappa)$ into the columns freed by the $b_k$ of (b).
In (d), the shifts move the accumulator two columns left, restoring (a).}
    \label{fig:accumulation_layout}
\end{figure*}

We now check that this layout accommodates every slice. In
\AccumulateResidue, the subroutine with the largest column count
(Figure~\ref{fig:accumulation_layout}), the accumulation qubits sit in the compute zone
with two free columns between consecutive qubits, and the residue
register, whose current $w_4$-bit window is the \Lookup{} address, sits in
the memory zone. The \Lookup{} loads its $k$-th target immediately to the
left of the $k$-th accumulation qubit, and its $(\kappa+1)$-th target
three columns to the right of the $\kappa$-th one. 
The routing column and
the free columns give these targets the $Z_L$ access required by the \Lookup, and the $w_4-1$ auxiliary qubits of its unary
iteration fit in the buffer zone, of which the \Lookup{} buffers use only $12$
positions. Since every adder writes the sum bit $s_k$ in the column to 
the right of its input $b_k$ and consumes $b_k$ by a destructive
$Z$-basis measurement, the \Adder$^\dagger$ and the \Adder{} each shift the
accumulation register by one column. The controls of the parity-\textsc{Cnot} gate $\mathrm{PC}X(\kappa)$ are
prepared in $\ket{+}$ in the columns freed by these measurements, and the comparison
bit of the \UnderflowPhasefix{} is written into the last column of the
compute zone, freed by the $Z$-basis measurement of the underflow qubit. The
\Phaseup{} auxiliary qubits fit in the buffer zone, and two transversal \textsc{Cnot}
shifts restore the initial configuration. Since every adder shifts the register it writes by one column, the same drift occurs in \ComputeExponent, where $S_i$ shifts by one column per iteration, and in the inner loops of \ComputeResidue{} and \ComputeResidue$^\dagger$, where the register written by the adders shifts by two columns per \Adder$^\dagger$/\Adder{} pair. It is compensated in the same way, by transversal \textsc{Cnot} shifts, except at the end of an inner loop, where the register moves anyway because the two $\ell$-qubit registers swap roles. The other subroutines use the
same operations. Outside the exponent zones, each slice must fit in the
$l_{\mathrm{mem}} + l_{\mathrm{comp}} = 154$ columns. In \ComputeResidue{} and
\ComputeResidue$^\dagger$, the widest operations, of size $\ell+1$, need
$3(\ell+1)+1 = 64$ columns next to the $\kappa = 33$ packed accumulation
qubits, i.e.\ $97 \le 103$ columns, while the memory zone holds
$S_i \bmod (p_i - 1)$, the second $\ell$-qubit register and the quotient
register. The Q-Q \Adder$(\ell + \log(n_e))$ of \ComputeExponent{} needs
$94$ columns, so the accumulation register is parked in the memory zone
during \ReduceExponent$^\dagger$, \ComputeExponent{} and \ReduceExponent, next to the
quotient register ($33 + 11 \le 51$). The $27$ positions of the buffer
zone do not depend on the register size (Equation~\ref{eq:adder_shared_buffer_count}) and
serve every adder, and the at most $5$ \Lookup{} and $8$ \Phaseup{} auxiliary
qubits fit in the $15$ positions left free by the $12$ repetition codes of the \Lookup{} buffers.

In the skew-tree QROM of~\cite{FeMoco2026prep}, an \textsc{And} gate and the parity-\textsc{Cnot} gate that follows it act simultaneously on the same auxiliary qubit, the former through its $X$-parity measurement and Hadamard injection, the latter through the $X$-parity measurement that closes its parity. In the standalone layout of~\cite{FeMoco2026prep}, they reach this qubit from opposite ends of its repetition code, the buffer lies below the register and a routing row runs above it. The global layout of Figure~\ref{fig:global_layout} has routing only above the compute zone, but the \Lookup{} buffers and auxiliary qubits sit in the buffer zone, to the right of the compute zone. Both gates then fit in the four-row routing zone above it as follows. First, the $\ket{0}_L$ repetition code of the parity-\textsc{Cnot} gate is prepared in the lowest routing row, from which it reaches the $Z$-access columns of the \Lookup{} targets. Second, the output of an \textsc{And} gate followed by a parity-\textsc{Cnot} gate is not moved to an auxiliary slot. Since the catalytic Hadamard injection is out-of-place, this output lies in one of the $\ket{i}$ columns of the buffer, where the parity-\textsc{Cnot} gate accesses it. It is measured in place by the measurement-based \textsc{And} uncomputation that immediately follows. Third, from left to right, the buffer zone holds the two \Lookup{} buffers next to the compute zone, then the \Lookup{} auxiliary qubits. Within each buffer, the order is the $\ket{i}$ column that receives the \textsc{And} output, the two other $\ket{i}$ columns, the $\ket{\CXCXX}$ qubit consumed by the Hadamard injection, and the two other $\ket{\CXCXX}$ qubits. The $\ket{0}_L$ code then ends at the leftmost column of the active buffer. The qubits accessed by the \textsc{And} gate through the routing zone all lie outside its span: the address register, in the memory zone or in the top rectangle reached from above. The two $\ket{\CXCXX}$ qubits measured against the controls; and the parent auxiliary qubit. The Hadamard injection itself only uses transversal \textsc{Cnot} gates between adjacent buffer columns. The two $X$-parity measurements of the \textsc{And} gate are therefore routed in the three rows above the $\ket{0}_L$ code without crossing it, as in the three routing rows of~\cite[Fig.~12b]{FeMoco2026prep}. The buffer not in use is refilled from below through the factory routing zone.

\begin{figure*}[t]
    \centering
    \includegraphics{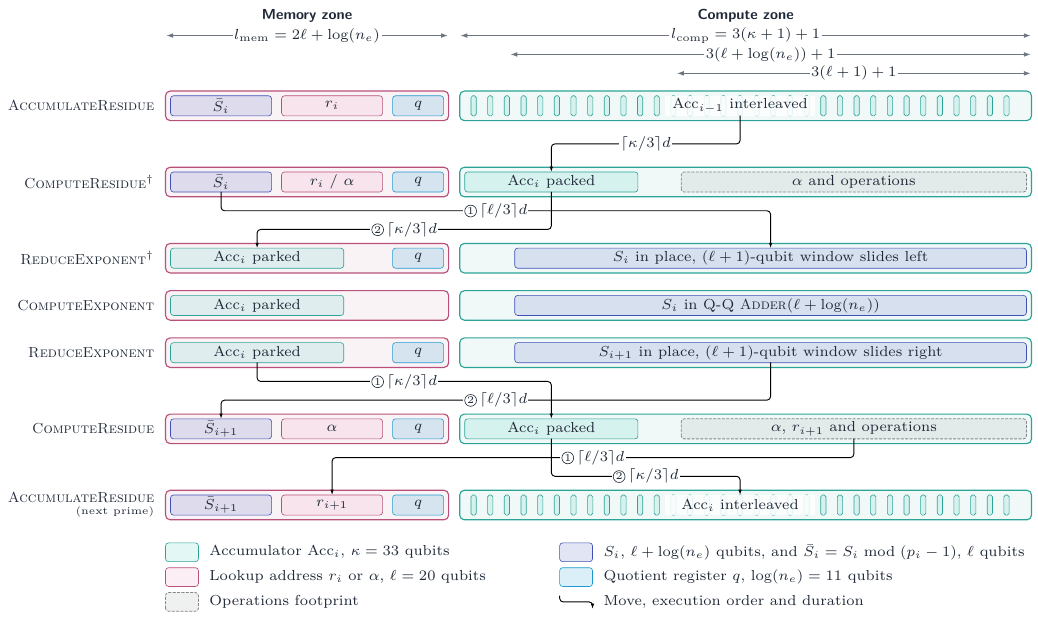}
    \caption{Register movements through the six subroutines of one prime on the repetition-cat chip of Figure~\ref{fig:global_layout} (RSA-2048), drawn to scale with $l_{\mathrm{mem}} = 51$ memory and $l_{\mathrm{comp}} = 103$ compute columns.
    Rows display the layout during each subroutine, starting from \AccumulateResidue, with their input register values. 
    Arrows represent the moves of each transition, their execution order, and the outlined boxes the associated qubit footprints.
    Moves within a subroutine, of the $\ell$-qubit registers in \ComputeResidue{} and \ComputeResidue$^\dagger$ and of the quotient qubits in \ReduceExponent{} and \ReduceExponent$^\dagger$, are not drawn.
    Similarly, the exponent and buffer zones are not drawn, since there are no register movements associated.
    The accumulator is stored in the memory zone during \ReduceExponent$^\dagger$, \ComputeExponent{} and \ReduceExponent, where $S_i$ occupies $3(\ell+\log(n_e))+1 = 94$ of the $l_{\mathrm{comp}} = 103$ compute columns.
    Each lattice surgery move of an $m$-qubit register requires $\lceil m/3 \rceil d$ cycles, such that the moves between the subroutines of a prime have a total duration of $4\lceil \kappa/3 \rceil d + 3\lceil \ell/3 \rceil d = 65d$ cycles.}
    \label{fig:acc_register_path}
\end{figure*}

Between subroutines, at most two registers move (Figure~\ref{fig:acc_register_path}):
the accumulation register, between its interleaved position, the left
end of the compute zone and the memory zone, and one $\ell$-qubit
register that switches between operand in the compute zone and address
or storage in the memory zone. Within subroutines, the underflow qubits
of \ReduceExponent{} and \ReduceExponent$^\dagger$ join or leave the quotient
register, and in \ComputeResidue{} and \ComputeResidue$^\dagger$ the register written by the
adders moves to the memory zone to serve as \Lookup{} address. All moves
use lattice surgery through the four-row routing zone, three qubits at a
time in $d$ cycles, so they need no qubit beyond those of
Equation~\ref{eq:ntotal}. Their duration is bounded in the next
paragraph.

\paragraph{Duration} \label{sec:C7-durations}
With $T_{\mathrm{surg}} = d$, $T_{\mathrm{cad}} = \max(d, 21)$
(Equation~\ref{eq:cadence}), $T_{\mathrm{react}} = 10$, and
$T_{\textsc{Cnot}} = T_{\mathrm{meas}} = 1$ for a transversal \textsc{Cnot} gate
and a transversal destructive measurement (Section~\ref{sssec:layout_connectivity}).
Equation~\ref{eq:adder_current_warm_latency} then gives the
duration of an $n$-bit adder,
\begin{equation}
  T_{\Adder}(n) = 2n\, T_{\mathrm{cad}} ,
  \label{eq:Tadder}
\end{equation}
which also applies to the quantum-classical adder (neglecting the classical value preparation time) and to a comparison of
size $n$, implemented as an adder of size $n+1$. For $d \le 21$, this is
$42n$. Equation~\ref{eq:lookup_cost_rep} gives
\begin{equation}
  T_{\Lookup}^{(\mathrm{rep})}(w, d) = (2^w - 1)\, d + (2^{w-1} - w)\, T_{\mathrm{cad}} ,
  \label{eq:Tlookup}
\end{equation}
that is $63d + 26T_{\mathrm{cad}}$ ($63d + 546$ for $d \le 21$) for the
address size $w_1 = 2w_3 = 6$ and $31d + 11T_{\mathrm{cad}}$
($31d + 231$) for $w_4 = 5$. \Phaseup{}s take $T_{\Phaseup}^{(\mathrm{rep})}(w, d)$
cycles (Equation~\ref{eq:phaseup_cost}), that is $7d + 8T_{\mathrm{cad}}$ for
$w = 6$ and $3d + 5T_{\mathrm{cad}}$ for $w_4 = 5$. Summing over the
numbers of calls $N_{\mathrm{op}}$ of Table~\ref{tab:physical_cycle_breakdown}, which depend
only on Table~\ref{tab:algo_parameter_value}, the duration of one shot is
\begin{align}
  T_{\mathrm{shot}}(d) &= \sum_{\mathrm{op}} N_{\mathrm{op}}\, T_{\mathrm{op}}(d)
  + T_{\mathrm{other}}(d) \nonumber\\
  &= 8.0909\times 10^{8}\, T_{\mathrm{cad}} + 5.2251\times 10^{8}\, d + T_{\mathrm{other}}(d) ,
  \label{eq:Tshot}
\end{align}
where the term in $T_{\mathrm{cad}}$ is the cadence-limited part and $T_{\mathrm{other}}(d)$ collects the slices of
Table~\ref{tab:logical_qubit_count} omitted from Table~\ref{tab:physical_cycle_breakdown}. In the color-changed convention (Section~\ref{ssec:colour_changed}), the $\CX(x)$, $M_X(x)$ and $P_{\ket{0}}(x)$ slices of Table~\ref{tab:logical_qubit_count} become parity-\textsc{Cnot} gates with $x$ controls, $\mathrm{PC}X(x)$, $Z$-basis measurements $M_Z(x)$ and $\ket{+}$ preparations $P_{\ket{+}}(x)$. Their durations
do not depend on the number of qubits involved. A $\mathrm{PC}X(x)$ is a single
lattice surgery (Section~\ref{sssec:parity_controlled_not}), $T_{\mathrm{PC}X} = T_{\mathrm{surg}} = d$, an $M_Z(x)$
is a transversal destructive $Z$-basis measurement, $T_{\mathrm{meas}} = 1$, and a
$P_{\ket{+}}(x)$, a $\ket{+}_L$ preparation of $x$ repetition codes in parallel, takes $T_{\mathrm{prep}} = T_{\mathrm{init}} = 1$ cycle (Section~\ref{sssec:error_correcting_codes}). Summing the
iteration counts of the corresponding rows of Table~\ref{tab:logical_qubit_count},
\begin{align}
  T_{\mathrm{other}}(d) &= N_{\mathrm{PC}X}\, d + N_{\mathrm{prep}}\, T_{\mathrm{init}} + N_{M_Z}\, T_{\mathrm{meas}} \nonumber\\
  &= N_{\mathrm{PC}X}\, d + N_{\mathrm{prep}} + N_{M_Z} ,
  \label{eq:Tother}
\end{align}
with
\begin{align}
  N_{\mathrm{PC}X} &= |P| \Big[ 4\log(n_e)
  + 6\big(\lceil \ell/w_3\rceil - 2\big)\lceil \ell/w_3\rceil \nonumber\\
  &\qquad + 2\lceil \ell/w_4\rceil \Big] , \label{eq:NCX}\\
  N_{M_Z} &= (|P| + 1)\lceil n_e/w_1\rceil \nonumber\\
  &\quad + |P| \Big[ 5\big(\lceil \ell/w_3\rceil - 2\big)\lceil \ell/w_3\rceil \nonumber\\
  &\qquad + \lceil \ell/w_3\rceil - 2 + \tfrac{3}{2}\lceil \ell/w_4\rceil \Big] ,
  \label{eq:NMX}\\
  N_{\mathrm{prep}} &= |P|\big(\lceil \ell/w_3\rceil - 2\big) . \label{eq:Nprep}
\end{align}
For RSA-2048, $N_{\mathrm{PC}X} = 262|P|$, $N_{M_Z} = 400|P| + 214$
and $N_{\mathrm{prep}} = 5|P|$, so that
$T_{\mathrm{other}}(d) \approx 5.94\times 10^{6} d + 9.18\times 10^{6}$.
These slices consume no factory output and are therefore not
cadence-limited. Including them,
$T_{\mathrm{shot}}(d) = 8.0909\times 10^{8}\, T_{\mathrm{cad}}
+ 5.2844\times 10^{8}\, d + 9.18\times 10^{6}$, which reduces to
$5.2844\times 10^{8}\, d + 1.7000\times 10^{10}$ for $d \le 21$.

The register moves of the Layout paragraph are not included in
$T_{\mathrm{shot}}$. A lattice-surgery move of an $m$-qubit register
takes $\lceil m/3\rceil d$ cycles. Between subroutines, the moves of
Figure~\ref{fig:acc_register_path} take $65d$ cycles per prime, i.e.\
$65|P|d \approx 1.5\times 10^{6} d$ cycles, and at most
$12|P|\lceil \kappa/3\rceil d \approx 3.0\times 10^{6} d$ cycles for any
order of the moves. Within subroutines, the
$2|P|\log(n_e) \approx 5.0\times 10^{5}$ single-qubit moves of the
underflow qubits and the at most
$3|P|(\lceil \ell/w_3\rceil - 2) \approx 3.4\times 10^{5}$ moves of
$\ell$-qubit registers in \ComputeResidue{} and \ComputeResidue$^\dagger$ take at most
$\approx 2.9\times 10^{6} d$ cycles. The transversal \textsc{Cnot} shifts that compensate the drift of the adders, about $9\times 10^{6}$ single-column register shifts per shot, take a number of cycles independent of $d$, at most about $3\times 10^{7}$ cycles, i.e.\ less than $0.15\%$ of $T_{\mathrm{shot}}$.
With the schedule of Figure~\ref{fig:acc_register_path}, the moves together stay below $0.5\%$ of $T_{\mathrm{shot}}$ at every $d$ (below $0.6\%$ with the bound for any order of the moves), and we
neglect them. The final inverse quantum Fourier transform is not included in $T_{\mathrm{shot}}$ either. In its semi-classical form (Section~\ref{sec:shor}), each of the $n_e$ exponent qubits receives one rotation, whose angle depends on the outcomes of the qubits already measured, and is then measured by a transversal destructive $Z$-basis measurement. Each rotation is applied by phase kickback, the exponent qubit controls the addition of a classically known $n_{\varphi}$-bit constant into a phase-gradient register of $n_{\varphi}$ qubits~\cite{Gidney18halving}, prepared once before the transform. A rotation therefore costs one Q-Q \Adder$(n_{\varphi})$, i.e.\ $n_{\varphi} - 1$ Toffoli gates, plus the $\ket{+}$ preparation and the parity-\textsc{Cnot} gate that load the constant, one measurement and one reaction time. The transform thus takes $n_e\,[T_{\Adder}(n_{\varphi}) + T_{\mathrm{init}} + d + T_{\mathrm{meas}} + T_{\mathrm{react}}]$ cycles (Equation~\ref{eq:Tadder}). For $n_{\varphi} = 25$, this is $1.38\times 10^{6}$ cycles at $d = 15$, i.e.\ $5\times 10^{-5}$ of $T_{\mathrm{shot}}$, and $3.1\times 10^{4}$ Toffoli gates, i.e.\ less than $5\times 10^{-5}$ of the Toffoli count per shot, and we neglect both, as well as the one-time preparation of the phase-gradient register.

\paragraph{Volume} \label{sec:C7-volume}
We combine the column counts of the Layout paragraph and the durations of
the Duration paragraph into the probability that a shot suffers a
logical error. One physical qubit-cycle corresponds to one physical
qubit being allocated during one error-correction cycle. The physical
volume of a shot is
\begin{equation}
  \begin{split}
  \mathrm{Vol}_{\mathrm{physical}} = \sum_s
  & \big[ 2d\, n_{\mathrm{col}}(s) - n_{e,2} + \Gamma_{\mathrm{global}} \big] \\
  & \times T_{\mathrm{error\,correction}}(s) ,
  \end{split}
  \label{eq:Volphys}
\end{equation}
with $n_{\mathrm{col}}(s)$ the number of repetition-code columns of
slice $s$ (Table~\ref{tab:rep-code-columns}), $n_{e,2} = n_e - l_{\mathrm{exp},1}$ ($730$ for RSA-2048) the number of exponent qubits in the top rectangle, whose ancillae above them are not needed since nothing lies above them (the unused exponent slot is not in $n_{\mathrm{col}}(s)$, so only these $730$ of the $l_{\mathrm{tot}} = 731$ exposed ancillae are removed here), $\Gamma_{\mathrm{global}}$ the global
routing of Equation~\ref{eq:Gglobal}, and
$T_{\mathrm{error\,correction}}(s)$ the duration of the slice in
error-correction cycles. The repetition-code columns of a slice are
those of the logical qubits stored outside the current operation and
those of the operation itself, including its buffers, its unallocated
slots and, for a \Lookup, the routing columns of its standalone footprint
(Equation~\ref{eq:lookup_cost_rep}). No routing row
or routing zone is included, since these routing qubits are all in
$\Gamma_{\mathrm{global}}$, which contributes the same amount to every
slice, whatever the current operation. The factory routing zone is
counted even though the factories are not, since it carries the resource
states after they leave the factories, so its errors are not covered by
the factory output infidelities. Equation~\ref{eq:Volphys}
overestimates the volume, since routing qubits are not all allocated in
every cycle, nor are the unallocated slots of an operation. The physical
volume is translated into an effective logical volume by dividing it by
$2d-1$, the size of the repetition code for which $p_{\mathrm{logical}}$ is defined~\cite{Ruiz_2025} (Section~\ref{sssec:error_correcting_codes}), and by $d$, to
translate error-correction cycles into logical cycles:
\begin{equation}
  \begin{split}
  \mathrm{Vol}^{(\mathrm{eff})}_{\mathrm{logical}}
  &= \frac{\mathrm{Vol}_{\mathrm{physical}}}{d(2d-1)} \\
  &= \frac{1}{d}\sum_s \frac{2d\, n_{\mathrm{col}}(s) - n_{e,2} + \Gamma_{\mathrm{global}}}{2d-1} \\
  &\qquad \times T_{\mathrm{error\,correction}}(s) .
  \end{split}
  \label{eq:Voleff}
\end{equation}
In the effective logical volume, global routing therefore counts as
$\Gamma_{\mathrm{global}}/(2d-1)$ additional effective logical qubits throughout
the shot, whereas the magic-state factories are not counted, since their
logical error rate is accounted for separately. We denote by
$\epsilon_{\mathrm{logical}}$ the probability that at least one logical
error occurs during one shot,
\begin{equation}
  \begin{split}
  \epsilon_{\mathrm{logical}}
  &= 1 - (1 - p_{\mathrm{logical}})^{\mathrm{Vol}^{(\mathrm{eff})}_{\mathrm{logical}}} \\
  &\simeq p_{\mathrm{logical}}\, \mathrm{Vol}^{(\mathrm{eff})}_{\mathrm{logical}} ,
  \end{split}
  \label{eq:eps}
\end{equation}
where $p_{\mathrm{logical}}$ is the logical error rate per logical cycle
of one logical qubit. We also introduce $T_{\mathrm{logical}}$, the
duration of the computation in logical cycles, and
$n^{(\mathrm{eff})}_{\mathrm{logical}}$, the effective number of
logical qubits,
\begin{equation}
  \begin{gathered}
  T_{\mathrm{logical}} = \frac{T_{\mathrm{error\,correction}}}{d} , \\
  n^{(\mathrm{eff})}_{\mathrm{logical}}
  = \frac{\mathrm{Vol}^{(\mathrm{eff})}_{\mathrm{logical}}}{T_{\mathrm{logical}}}
  = \frac{2d\,\bar n_{\mathrm{col}} - n_{e,2} + \Gamma_{\mathrm{global}}}{2d-1} ,
  \end{gathered}
  \label{eq:neff}
\end{equation}
where $T_{\mathrm{error\,correction}} = \sum_s
T_{\mathrm{error\,correction}}(s)$ and $\bar n_{\mathrm{col}}$ is the
time average of $n_{\mathrm{col}}(s)$ over the shot.

For RSA-2048, each operation row $s$ of Tables~\ref{tab:rep-code-columns}
and~\ref{tab:physical_cycle_breakdown} is repeated $N_s$ times and lasts $T_s(d)$ cycles
(Equations~\ref{eq:Tadder}, \ref{eq:Tlookup} and~\ref{eq:phaseup_cost}).
Since $n_{\mathrm{col}}(s)$ does not depend on $d$,
Equation~\ref{eq:Volphys} reads
\begin{equation}
  \mathrm{Vol}_{\mathrm{physical}}(d) = 2d\, X(d)
  + \big[\Gamma_{\mathrm{global}}(d) - n_{e,2}\big]\, T_{\mathrm{shot}}(d) ,
  \label{eq:Volphys_d}
\end{equation}
with
\begin{align}
  X(d) &= \sum_s N_s\, n_{\mathrm{col}}(s)\, T_s(d)
  + n_{\mathrm{col,other}}\, T_{\mathrm{other}}(d) \nonumber\\
  &= 1.1604\times 10^{12}\, T_{\mathrm{cad}} + 7.5024\times 10^{11}\, d \nonumber\\
  &\quad + 1.3398\times 10^{10} ,
  \label{eq:X_d}
\end{align}
$\Gamma_{\mathrm{global}}(d)$ given by Equation~\ref{eq:Gglobal_d} and
$T_{\mathrm{shot}}(d)$ by Equation~\ref{eq:Tshot}. Keeping the factory
routing term $2h_{\mathrm{routing,fac}}\, l_{\mathrm{fac}}$ symbolic,
since its value is set in Section~\ref{ssec:physical_resource_count}, this gives
\begin{align}
  \mathrm{Vol}_{\mathrm{physical}}(d) ={}& 1.502\times 10^{12}\, d^2
  + 2.322\times 10^{12}\, d\, T_{\mathrm{cad}} \nonumber\\
  &+ 1.569\times 10^{12}\, d
  + 2.361\times 10^{12}\, T_{\mathrm{cad}} \nonumber\\
  &+ 2.678\times 10^{10} \nonumber\\
  &+ 2h_{\mathrm{routing,fac}}\, l_{\mathrm{fac}}\, T_{\mathrm{shot}}(d) .
  \label{eq:Volphys_closed}
\end{align}

These coefficients rest on five assumptions. First, the slices of
$T_{\mathrm{other}}(d)$ are assigned the peak column count of the
computation, $n_{\mathrm{col,other}} = 1460$, an upper bound since these
slices are always smaller than the adjacent additions or lookups.
Second, \ReduceExponent$^\dagger$, which applies the operations of \ReduceExponent{}
in reverse, has the same column counts. Third, $n_{\mathrm{buf}}^{(\Adder)} = 3$ at
every $d$ considered, although two adder buffers suffice for $d \le 11$. Fourth, the
register moves are neglected, as in the Duration paragraph.
Fifth, the active columns of a \Lookup{} are those of its standalone footprint (Equation~\ref{eq:lookup_cost_rep}). When a \Lookup{} loads an adder input, its targets sit in the interleaved slots of the adder (Figure~\ref{fig:accumulation_layout}) and each of them uses its own free neighboring column as routing, so the \Lookup{} uses up to 16 more repetition-code columns than counted. These slices stay below the peak of 1460 columns, and the volume is underestimated by about $0.4\%$.

\providecommand{\TBD}[1]{\mbox{\textcolor{red}{[TBD:~#1]}}}
\subsubsection{Evaluation at $p_Z = 10^{-3}$ and $p_Z = 5\times10^{-4}$} \label{ssec:physical_resource_count}

We now evaluate the estimate for RSA-2048 at the distances $d = 13$, $15$ and $17$, and at the phase-flip probabilities $p_Z = 10^{-3}$ and $p_Z = 5\times10^{-4}$. Our main result is at $p_Z = 5\times10^{-4}$ and $d = 15$. As in Section~\ref{sssec:global_layout_routing}, operations are written in the color-changed convention of Section~\ref{ssec:colour_changed}.

\paragraph{Physical qubit count, duration and volume}
We use the factories of Table~\ref{tab:adder_reference_factories}. Their layout and production time do not depend on $p_Z$, only the infidelity of their output states does (paragraph~\ref{sec:C7-budget}). For all distances considered here, $T_{\mathrm{cad}} = \max(d, 21) = 21$ is set by the $\ket{\CXCXX}$ factory, and Equation~\ref{eq:factory_counts} gives one $\ket{\CXCXX}$ factory and two $\ket{i}$ factories. The factory zone has length $l_\text{fac} = w_{\mathrm{fac}}^{\ket{\CXCXX}} + 2 w_{\mathrm{fac}}^{\ket{i}} = 29 + 2\times10.5 = 50$ and contains $n_\text{fac} = 1566 + 2\times153 = 1872$ physical qubits, as in Section~\ref{sssec:adder_resources}.
The factory routing zone has height $h_\text{routing,fac} = 10$, hence $2h_\text{routing,fac}\,l_\text{fac} = 1000$ in Equations~\ref{eq:Gglobal_d} and~\ref{eq:ntotal_d}.

With these values, Equations~\ref{eq:ntotal_d}, \ref{eq:Tshot}, \ref{eq:Volphys_closed} and~\ref{eq:Voleff} give the physical qubit count, the duration of one shot and the effective logical volume listed in Table~\ref{tab:rep_cat_recap} for $d = 13$, $15$ and $17$. At $d = 15$, the circuit is widest at two different times in \AccumulateResidue: the subtraction Q-Q \Adder$^\dagger(\kappa+1)$ and the comparison inside the \UnderflowPhasefix. At those times, the 1460 repetition-code columns of Table~\ref{tab:rep-code-columns} hold $1460\times 2d = 43\,800$ physical qubits, of which $n_e\times 2d = 38\,400$ store the exponent. The global routing $\Gamma_\text{global} = 4678$ of Equation~\ref{eq:Gglobal_d} splits into $1000$ (factory routing zone), $2200$ (routing zone between the two exponent zones), $1448$ (routing zone above the memory, compute and buffer zones) and $30$ (routing column of the compute zone). Adding the factories ($n_\text{fac} = 1872$), the global routing and the unused exponent slot ($2d = 30$), and subtracting the $l_{\mathrm{tot}} = 731$ exposed ancillae gives the $49\,649$ physical qubits of Equation~\ref{eq:ntotal_d}, in agreement with Equation~\ref{eq:ntotal_peak}. One shot takes $T_{\mathrm{shot}} = 2.493\times10^{10}$ cycles, about $6.92$~hours (not accounting for classical precomputation). The moves of registers between and within subroutines stay below $0.5\%$ of this duration at every $d$ (Section~\ref{sssec:global_layout_routing}) and are neglected.

At $d = 15$, $\Gamma_\text{global} = 4678$ amounts to $161$ effective logical qubits, $T_{\text{logical}} = 1.66\times10^{9}$ and $n^{(\text{eff})}_{\text{logical}} = 1615$,
of which $\bar n_{\mathrm{col}} = 1430$. For $d < 21$, the cadence sets about two thirds of the shot length, so that the number of logical cycles $T_{\mathrm{shot}}/d$ grows as $d$ decreases, and $\text{Vol}^{(\text{eff})}_{\text{logical}}$ grows somewhat more slowly than $1/d$. Going from $d = 17$ to $d = 13$ thus saves $21\%$ of the physical qubits but shortens a shot by only $8\%$, while the effective logical volume grows by $24\%$. The logical error probability per shot grows much faster, since $p_{\text{logical}}$ increases by a factor of about $1/(B p_Z)^C$ per step of two in $d$ (Equation~\ref{eq:plogical}). From $d = 17$ to $d = 13$, $\epsilon_{\text{logical}}$ rises from $2.6\times10^{-5}$ to $0.157$ at $p_Z = 5\times10^{-4}$, and from $0.0125$ to almost $1$ at $p_Z = 10^{-3}$ (Table~\ref{tab:rep_cat_recap}). A smaller distance therefore trades space for retries only within a narrow range of $d$ (paragraph~\ref{sec:C7-budget}).

\paragraph{Logical error budget, number of shots and total runtime} \label{sec:C7-budget}
We take the logical error model of the repetition code under generic circuit-level noise from~\cite[Supplementary Table~1(b)]{Ruiz_2025}, in which every operation of the syndrome-extraction circuit has phase-flip probability $p_Z$.
The probability that a logical error occurs on one logical qubit during one logical cycle, i.e.\ $d$ error-correction cycles, is
\begin{equation}
  p_{\text{logical}}(p_Z, d) = A\, d\, (B\, p_Z)^{C \lfloor (d+1)/2 \rfloor} ,
  \label{eq:plogical}
\end{equation}
with $A = 0.12$, $B = 23$ and $C = 0.99$. This fit is obtained from simulations at logical error probabilities above about $10^{-8}$~\cite[Supplementary Fig.~2(b)]{Ruiz_2025}. The values of Table~\ref{tab:rep_cat_recap} extrapolate it to the regime relevant here.
By Equation~\ref{eq:eps}, the probability per shot that a logical error occurs on the data and routing qubits is $\epsilon_{\text{logical}} = 1 - (1 - p_{\text{logical}}(p_Z, d))^{\text{Vol}^{(\text{eff})}_{\text{logical}}}$, with $p_{\text{logical}}(p_Z, d)$ given by Equation~\ref{eq:plogical}. We use this exact form, also given in Table~\ref{tab:rep_cat_recap}, since $\epsilon_{\text{logical}}$ can exceed a few percent at small $d$, where the linear approximation of Equation~\ref{eq:eps} no longer holds.

The output infidelities of the factories of Table~\ref{tab:adder_reference_factories} are $1.48\times10^{-11}$ ($p_Z = 10^{-3}$) and $2.02\times10^{-13}$ ($p_Z = 5\times10^{-4}$) for the $\ket{\CXCXX}$ factory, and $3.1\times10^{-12}$ and $9.43\times10^{-13}$ for the $\ket{i}$ factory, a color code of distance $d_i = 7$.
Counting $n-1$ Toffoli gates per $n$-bit adder (Section~\ref{sssec:adder_scheduling}), a comparison of size $n$ as an adder of size $n+1$, $2^w-w-1$ Toffoli gates per \Lookup{} (Equation~\ref{eq:lookup_cost_rep}) and $2^{\lceil w/2\rceil}+2^{\lfloor w/2\rfloor}-w-2$ per \Phaseup{} (Equation~\ref{eq:phaseup_cost}), the calls of Table~\ref{tab:physical_cycle_breakdown} give $7.56\times10^{8}$ Toffoli gates per shot: $2.81\times10^{8}$ in adders and comparisons ($37\%$), $4.62\times10^{8}$ in \Lookup{}s ($61\%$) and $1.33\times10^{7}$ in \Phaseup{}s ($1.8\%$). The $\ket{\CXCXX}$ factories therefore cause a failure during a shot with probability $\epsilon_{\ket{\CXCXX}} \approx 1.1\%$ at $p_Z = 10^{-3}$ and $1.5\times10^{-4}$ at $p_Z = 5\times10^{-4}$.
Including the catalyst seeds (Section~\ref{sssec:resource_states}), an $n$-bit adder consumes $4(n-1)+n_{\mathrm{buf}}^{(\Adder)}$ $\ket{i}$ states, with $n_{\mathrm{buf}}^{(\Adder)}=3$ (Section~\ref{sssec:adder_scheduling}), a comparison of size $n$ being counted as an adder of size $n+1$. A \Lookup{} on a $w$-qubit address register consumes $2(2^w-w-1)+n_{\mathrm{buf}}^{(\Lookup)}$ $\ket{i}$ states, with $n_{\mathrm{buf}}^{(\Lookup)}=2$ (Section~\ref{ssec:lookup_and_phaseup}), and a \Phaseup{} consumes on average $N_{\ket{i}}^{\Phaseup}(w)$ $\ket{i}$ states (Equation~\ref{eq:phaseup_istates}), that is $40$ for $w=6$ and $23$ for $w=5$. As for the Toffoli count, \Adder$^\dagger$ and Q-C adders are counted as adders of the same size. The calls of Table~\ref{tab:physical_cycle_breakdown} then give $2.17\times10^{9}$ $\ket{i}$ states per shot: $1.16\times10^{9}$ in adders and comparisons ($54\%$), $9.40\times10^{8}$ in \Lookup{}s ($43\%$) and $6.64\times10^{7}$ in \Phaseup{}s ($3.1\%$), catalyst seeds included ($5.46\times10^{7}$ in total). The $\ket{i}$ factories therefore cause a failure during a shot with probability $\epsilon_{\ket{i}} \approx 0.67\%$ at $p_Z = 10^{-3}$ and $0.20\%$ at $p_Z = 5\times10^{-4}$.
These two contributions do not depend on $d$. The per-shot failure probability is $\epsilon = 1 - (1 - \epsilon_{\text{logical}})(1 - \epsilon_{\ket{\CXCXX}})(1 - \epsilon_{\ket{i}})$.

We select the distance with the following criterion: among the distances for which the per-shot failure probability $\epsilon$ is below $10\%$, we choose the one with the fewest physical qubits. On the repetition-cat architecture, $n_{\mathrm{total}}$ grows with $d$ (Equation~\eqref{eq:ntotal_d}), so this is the smallest odd distance with $\epsilon < 10\%$. A shot with a logical error is caught and retried (Section~\ref{par:composition-physical-error}), so $\epsilon$ multiplies the expected number of shots by $1/(1 - \epsilon)$, and the threshold limits this increase to $11\%$. At $p_Z = 5\times10^{-4}$, $d = 13$ gives $\epsilon \approx 16\%$, so we choose $d = 15$, which gives $\epsilon_{\text{logical}} = 0.211\%$ and $\epsilon \approx 0.43\%$. At this point, the data and routing qubits ($0.21\%$) and the $\ket{i}$ factories ($0.20\%$) contribute almost equally to the per-shot failure, while the $\ket{\CXCXX}$ factories ($0.015\%$) are negligible.
At $p_Z = 10^{-3}$, $d = 15$ gives $\epsilon \approx 41\%$, so the same criterion selects $d = 17$, with $\epsilon \approx 3.0\%$. The selected distances are therefore the same for any threshold between $3.0\%$ and $16\%$.
 For $s = 8$, the average number of shots required is $9.2$, as explained in Section~\ref{sec:full-modern}. Accounting for the per-shot failure probability $\epsilon$, it becomes $9.2/(1 - \epsilon) = 9.24$ at $p_Z = 5\times10^{-4}$ and $d = 15$, which brings the total duration to $9.24\times6.92\ \mathrm{h} \approx 64.0$~h $\approx 2.67$~days. At $p_Z = 5\times10^{-4}$, $d = 13$ would save $12\%$ of the physical qubits ($43\,801$ versus $49\,649$) at the cost of $13\%$ more runtime ($3.02$ versus $2.67$ days). The shots are independent, so they can run in parallel on several quantum computers, which reduces the total runtime accordingly.

\begin{table*}[ht]
    \centering
    \begin{tabular}{|c|c|c|c|c|c|c|c|c|}
        \hline
        \multirow{2}{*}{$d$} & \multirow{2}{*}{$n_\text{total}$} & \multicolumn{2}{c|}{$T_{\mathrm{shot}}$} & \multirow{2}{*}{$\text{Vol}^{(\text{eff})}_{\text{logical}}$} & \multicolumn{2}{c|}{$p_Z = 10^{-3}$} & \multicolumn{2}{c|}{$p_Z = 5\times10^{-4}$} \\
        \cline{3-4}\cline{6-9}
         & & error-correction cycles & hours & & $p_{\text{logical}}$ & $\epsilon_{\text{logical}}$ & $p_{\text{logical}}$ & $\epsilon_{\text{logical}}$ \\
        \hline
        13 & $43\,801$ & $2.387\times10^{10}$ & $6.63$ & $3.02\times10^{12}$ & $6.92\times10^{-12}$ & $>99.5\%$ & $5.67\times10^{-14}$ & $15.7\%$ \\
        15 & $49\,649$ & $2.493\times10^{10}$ & $6.92$ & $2.68\times10^{12}$ & $1.91\times10^{-13}$ & $40.0\%$ & $\mathbf{7.87\times10^{-16}}$ & $\mathbf{0.211\%}$ \\
        17 & $55\,497$ & $2.598\times10^{10}$ & $7.22$ & $2.43\times10^{12}$ & $\mathbf{5.16\times10^{-15}}$ & $\mathbf{1.25\%}$ & $1.07\times10^{-17}$ & $2.6\times10^{-3}\%$ \\
        \hline
    \end{tabular}
    \caption{Physical resources and logical error budget of the repetition-cat architecture for RSA-2048, at $p_Z = 10^{-3}$ and $p_Z = 5\times10^{-4}$. $n_\text{total}$ is given by Equation~\ref{eq:ntotal_d}, $T_{\mathrm{shot}}$ by Equation~\ref{eq:Tshot} and $\text{Vol}^{(\text{eff})}_{\text{logical}}$ by Equations~\ref{eq:Volphys_closed} and~\ref{eq:Voleff}, with the factories of Table~\ref{tab:adder_reference_factories}. $p_{\text{logical}}$ is the logical error probability per logical cycle of Equation~\ref{eq:plogical}, and $\epsilon_{\text{logical}}$ the probability of a logical error on the data and routing qubits during one shot (exact form of Equation~\ref{eq:eps}). The entry $>99.5\%$ denotes a probability above $0.995$. The main result is at $p_Z = 5\times10^{-4}$ and $d = 15$.}
    \label{tab:rep_cat_recap}
\end{table*}

\paragraph{Duration breakdown} \label{sec:C7-duration}
Table~\ref{tab:physical_cycle_breakdown} gives the number of cycles of each operation at $d = 15$.
\ComputeExponent{} represents $54\%$ of the total duration. Subroutine shares below include the $\ket{+}$ preparation, $Z$-basis measurement and parity-\textsc{Cnot} slices omitted from Table~\ref{tab:physical_cycle_breakdown}, and therefore slightly exceed the sums of its rows. \ComputeResidue$^\dagger$ is about 3 times longer than \ComputeResidue{} ($30.0\%$ versus $10.5\%$). This is expected since \ComputeResidue$^\dagger$ must properly uncompute the modular exponentiation, which requires two inner loops, each including a phase correction. \ComputeResidue, however, needs a single inner loop without any phase correction. Finally, \ReduceExponent{} (including \ReduceExponent$^\dagger$) and \AccumulateResidue{} represent $3.5\%$ and $1.7\%$ of the total duration respectively. The time spent on addition-like operations and on \Lookup{}s is comparable, accounting respectively for $49.3\%$ and $48.5\%$ of the total duration. Other operations have a much smaller contribution, with \Phaseup{}s representing $1.8\%$ of the duration and the $\mathrm{PC}X$, $M_Z$ and $P_{\ket{+}}$ slices $0.39\%$ combined. The distribution depends only weakly on $d$. The share of the \Lookup{}s goes from $46\%$ at $d = 13$ to $50\%$ at $d = 17$, since a \Lookup{} has a larger surgery-limited part than an adder, whose duration is entirely cadence-limited. We do not account for the reaction time between the $M_Z$ of underflow qubits and the conditional application of an \UnderflowPhasefix, since its contribution to the total duration is negligible.

\begin{table*}[h]
    \centering
    \begin{tabular}{|l|l|c|c|c|c|}
        \hline
        \textbf{Subroutine} & \textbf{Operation} & \textbf{Calls} & \textbf{Cycles/call} & \textbf{Cycles} & \textbf{Share} \\
        \hline

        \multirow{2}{*}{\ComputeExponent}
         & \Lookup(6, 31) & 4849026 & 1491 & $7.23\times10^{9}$ & 29.00\% \\
         & Q-Q \Adder(31)     & 4849026 & 1302 & $6.31\times10^{9}$ & 25.33\% \\
        \hline

        \multirow{2}{*}{\ReduceExponent}
         & Q-C \Adder(21) & 249238 & 882 & $2.20\times10^{8}$ & 0.88\% \\
         & Q-Q \Adder(20)   & 249238 & 840 & $2.09\times10^{8}$ & 0.84\% \\
        \hline

        \multirow{4}{*}{\ComputeResidue}
         & \Lookup(6, 20) (loading) & 22658 & 1491 & $3.38\times10^{7}$ & 0.14\% \\
         & \Lookup(6, 21)         & 793030 & 1491 & $1.18\times10^{9}$ & 4.74\% \\
         & Q-Q \Adder$^\dagger$(21)    & 793030 & 882 & $6.99\times10^{8}$ & 2.81\% \\
         & Q-Q \Adder(20)                & 793030 & 840 & $6.66\times10^{8}$ & 2.67\% \\
        \hline

        \multirow{6}{*}{\AccumulateResidue}
         & \Lookup(5, 34)     & 90632 & 696 & $6.31\times10^{7}$ & 0.25\% \\
         & Q-Q \Adder$^\dagger$(34) & 90632 & 1428 & $1.29\times10^{8}$ & 0.52\% \\
         & Q-Q \Adder(33)           & 90632 & 1386 & $1.26\times10^{8}$ & 0.50\% \\
         & \Lookup(5, 33)       & 45316 & 696 & $3.15\times10^{7}$ & 0.13\% \\
         & Q-Q \Comparison(33)      & 45316 & 1428 & $6.47\times10^{7}$ & 0.26\% \\
         & \Phaseup(5)                & 90632 & 150 & $1.36\times10^{7}$ & 0.05\% \\
        \hline

        \multirow{7}{*}{\ComputeResidue$^\dagger$}
         & \Lookup(6, 21)          & 1586060 & 1491 & $2.36\times10^{9}$ & 9.49\% \\
         & Q-Q \Adder$^\dagger$(21)     & 1586060 & 882 & $1.40\times10^{9}$ & 5.61\% \\
         & Q-Q \Adder(20)                 & 1586060 & 840 & $1.33\times10^{9}$ & 5.34\% \\
         & \Lookup(6, 20)            & 793030 & 1491 & $1.18\times10^{9}$ & 4.74\% \\
         & Q-Q \Comparison(20)            & 793030 & 882 & $6.99\times10^{8}$ & 2.81\% \\
         & \Phaseup(6)                   & 1586060 & 273 & $4.33\times10^{8}$ & 1.74\% \\
         & \Phaseup(6) (unload)          & 22658 & 273 & $6.19\times10^{6}$ & 0.02\% \\
        \hline

        \multirow{2}{*}{\ReduceExponent$^\dagger$}
         & Q-C \Adder(21) & 249238 & 882 & $2.20\times10^{8}$ & 0.88\% \\
         & Q-Q \Adder(20)   & 249238 & 840 & $2.09\times10^{8}$ & 0.84\% \\
        \hline

        Last \Phaseup{} & \Phaseup(6) & 214 & 273 & $5.84\times10^{4}$ & 0.00\% \\
        \hline
    \end{tabular}
    \caption{Breakdown of error-correction cycle counts by operation for the factorization of an RSA-2048 number. All logical data qubits are encoded in a repetition code of distance 15.
    Time slices involving a $\ket{+}$ preparation, a $Z$-basis measurement or a parity-\textsc{Cnot} gate are not reported in this table for clarity, as they have a small contribution of $0.39\%$ to the total duration. Their contribution is still accounted for in the total duration and in the total volume.
    {\it Cycles/call} refers to the number of error-correction cycles required for one call of an operation, while {\it Cycles} gives the total number of error-correction cycles required for all calls of this operation.
    }
    \label{tab:physical_cycle_breakdown}
\end{table*}

\newcommand{\ldpcDrep}{13}
\newcommand{\ldpcDldpc}{16}
\newcommand{\ldpcDstar}{16} 
\newcommand{\ldpcDrepHigh}{15}
\newcommand{\ldpcDldpcHigh}{22}
\newcommand{\ldpcH}{7}
\newcommand{\ldpcDelta}{26}
\newcommand{\ldpcNL}{5045}
\newcommand{\ldpcGammaLDPC}{8656}
\newcommand{\ldpcTprod}{21}
\newcommand{\ldpcTcad}{21}
\newcommand{\ldpcTbuf}{49}
\newcommand{\ldpcNbuf}{3} 
\newcommand{\ldpcNfacI}{two}
\newcommand{\ldpcNfacCCX}{one}
\newcommand{\ldpcNfac}{1872} 
\newcommand{\ldpcLfac}{50}
\newcommand{\ldpcHroutfac}{10}
\newcommand{\ldpcER}{5}
\newcommand{\ldpcCoefH}{2868}
\newcommand{\ldpcCoefDelta}{4}
\newcommand{\ldpcCoefZero}{-1401}
\newcommand{\ldpcPeak}{12774}
\newcommand{\ldpcNtotal}{22577}
\newcommand{\ldpcGammaGlobal}{4043}
\newcommand{\ldpcGammaAddr}{2560}
\newcommand{\ldpcGammaQFT}{5760}
\newcommand{\ldpcExpCost}{12.5}
\newcommand{\ldpcExpReductionShare}{\TBD{exponent share of reduction}}
\newcommand{\ldpcQubitRatio}{\TBD{qubit ratio}} 
\newcommand{\ldpcTshotCycles}{\ensuremath{2.80\times10^{10}}}
\newcommand{\ldpcTshotHours}{7.79} 
\newcommand{\ldpcShareLookup}{44.95\%}
\newcommand{\ldpcShareAdder}{43.63\%}
\newcommand{\ldpcSharePhaseup}{1.66\%}
\newcommand{\ldpcShareMX}{4.77\%}
\newcommand{\ldpcShareReshuffle}{4.65\%}
\newcommand{\ldpcShareOther}{0.35\%} 
\newcommand{\ldpcTMXacc}{160} 
\newcommand{\ldpcMovesBound}{\ensuremath{1.30\times10^{9}}}
\newcommand{\ldpcReshuffleBound}{\ensuremath{1.13\times10^{8}}} 
\newcommand{\ldpcReshuffleShare}{0.40\%}
\newcommand{\ldpcTQFT}{33280} 
\newcommand{\ldpcTimeRatio}{\TBD{time ratio}} 
\newcommand{\ldpcVolRep}{\ensuremath{4.6\times10^{11}}} 
\newcommand{\ldpcVolLDPC}{\ensuremath{2.5\times10^{12}}} 
\newcommand{\ldpcGlobalShare}{76\%}
\newcommand{\ldpcAddrShare}{48\%} 
\newcommand{\ldpcEpsRep}{\ensuremath{2.6\times10^{-2}}} 
\newcommand{\ldpcEpsLDPC}{\ensuremath{9.71\times10^{-3}}} 
\newcommand{\ldpcEpsTot}{\ensuremath{3.5\times10^{-2}}}
\newcommand{\ldpcEpsFac}{\ensuremath{2.2\times10^{-3}}} 
\newcommand{\ldpcShots}{9.6} 
\newcommand{\ldpcRuntimeDays}{3.10} 
\newcommand{\ldpcRuntimeRatio}{\TBD{runtime ratio}}
\newcommand{\repcmpD}{\TBD{repetition-cat $d$}} 
\newcommand{\repcmpNtotal}{\TBD{repetition-cat $n_\mathrm{total}$}} 
\newcommand{\repcmpPeak}{\TBD{repetition-cat peak}}
\newcommand{\repcmpExpCost}{\TBD{repetition-cat exponent cost}} 
\newcommand{\repcmpTshotCycles}{\TBD{repetition-cat $T_\mathrm{shot}$}}
\newcommand{\repcmpShareLookup}{\TBD{repetition-cat \Lookup{} share}} 
\newcommand{\repcmpShareAdder}{\TBD{repetition-cat adder share}} 
\newcommand{\repcmpRuntimeDays}{\TBD{repetition-cat runtime}}

\subsection{LDPC-cat architecture}
\label{sec:ldpc-cat}\label{ssec:ldpc_cat_layout}

This section estimates the physical resources of the computation for
RSA-2048 on the LDPC-cat architecture, following the method of
Section~\ref{ssec:repetition_cat_architecture}. Section~\ref{sec:ldpc-cat-a} describes the
layout, the duration and the volume of one shot at arbitrary distances
$d_{\mathrm{rep}}$ and $d_{\mathrm{LDPC}}$ of the repetition and LDPC layers, and
Section~\ref{sec:ldpc-cat-b} evaluates the estimate at
$p_Z = 5\times10^{-4}$ and $p_Z = 10^{-3}$, where the selected distances
are $(d_{\mathrm{rep}},d_{\mathrm{LDPC}})=(13,16)$ and $(17,22)$, and compares it with the
repetition-cat architecture. As in Section~\ref{ssec:repetition_cat_architecture}, operations are written in
the color-changed convention of Section~\ref{ssec:colour_changed}.

The LDPC-cat architecture differs from the repetition-cat architecture in
two ways. First, the logical data qubits are stored in a single LDPC-cat
code block, in which the operator $Z_L$ of a logical qubit is only
accessible through transversal \textsc{Cnot} gates onto a repetition code of the
repetition layer that covers its support, which must then undergo
$d^{\star}=\max(d_{\mathrm{rep}},d_{\mathrm{LDPC}})$ rounds of stabilizer measurements before it is
read out. Z-basis
measurements and register moves, which take one cycle or are neglected on
the repetition-cat architecture (Section~\ref{sssec:global_layout_routing}), therefore
take $T^{(\mathrm{LDPC})}_{\mathrm{surg}}$ cycles each and are no longer negligible
(Section~\ref{sec:ldpc-cat-zaccess}). Second, the chip has two layers,
with two distances, two logical error models and two volumes
(Section~\ref{sec:ldpc-cat-volume}).

\subsubsection{Layout, duration and volume at distances $d_{\mathrm{rep}}$ and $d_{\mathrm{LDPC}}$}
\label{sec:ldpc-cat-a}

\paragraph{Introduction}
We divide the algorithm into the temporal slices of
Table~\ref{tab:logical_qubit_count}, with the algorithmic parameters of
Table~\ref{tab:algo_parameter_value} and the factories of
Section~\ref{ssec:factories}, and assume, as in
Section~\ref{sssec:global_layout_routing}, an error-correction cycle of
$1~\mu\mathrm{s}$, a reaction time $T_{\mathrm{react}}=10$ cycles and no bit flip
during the computation.

Two distances enter the estimate. The LDPC distance $d_{\mathrm{LDPC}}$ is chosen
among the codes of Table~\ref{tab:adder_ldpc_codes}, which fixes the height $H$
of the block and the number $\Delta_{\mathrm{LDPC}}$ of qubits of its staircases. The
repetition distance $d_{\mathrm{rep}}$ applies to every repetition code of the
repetition layer: the buffers, the factory outputs, the routing codes and
the auxiliary qubits of \Lookup{}s and \Phaseup{}s.

The number of rounds of a lattice surgery sets its timelike distance, which
protects the outcome of the surgery against chains of measurement errors.
Every lattice surgery involving the LDPC layer, such as the Z-basis
measurements and register moves of Section~\ref{sec:ldpc-cat-zaccess},
the surgeries of the adder, the \textsc{And} gates and parity-\textsc{Cnot} gates of
\Lookup{}s, and the \textsc{And} gates and multi-target $\C_X X$ gates of \Phaseup{}s,
lasts $T^{(\mathrm{LDPC})}_{\mathrm{surg}}$ cycles (Equation~\eqref{eq:Tsurg-ldpc}).
Its timelike failure, a flip of the surgery outcome by a chain of
measurement errors, is not a memory error of the codes taking part, and
therefore does not enter the logical error probabilities of
Equation~\eqref{eq:eps-layers}, which apply the memory rates of
Equations~\eqref{eq:plogical} and~\eqref{eq:plogical-ldpc} to the volumes
of Section~\ref{sec:ldpc-cat-volume}. Its probability per surgery is at
most of the order of one logical cycle of the weakest code taking part:
with about $2.3\times10^{9}$ such surgeries per shot and the rates of
Section~\ref{sec:ldpc-cat-b}, it adds of the order of $10^{-4}$ to the
failure probability per shot, which we neglect. A surgery confined to the repetition
layer, such as the loading of factory outputs into the buffers, lasts
$d_{\mathrm{rep}}$ cycles.

The steps that consume factory outputs are surgeries involving the LDPC
layer, and they start
at the cadence of Equation~\eqref{eq:cadence} with $T_{\mathrm{surg}}=T^{(\mathrm{LDPC})}_{\mathrm{surg}}$,
\begin{align}
  T_{\mathrm{cad}} &= \max\!\left(T^{(\mathrm{LDPC})}_{\mathrm{surg}}, T^{\ket{\CXCXX}}_{\mathrm{prod}},
  T^{\ket{i}}_{\mathrm{prod}}\right) \nonumber\\
  &= \max(T^{(\mathrm{LDPC})}_{\mathrm{surg}}, 21),
  \label{eq:Tcad-ldpc}
\end{align}
so that, as in Section~\ref{sssec:global_layout_routing}, the $\ket{\CXCXX}$ factory sets
the cadence $T_{\mathrm{cad}}=21$ for $T^{(\mathrm{LDPC})}_{\mathrm{surg}}\le21$, that is for every pair of
distances considered here with $d_{\mathrm{LDPC}}\le16$, whereas the surgeries
set it for $d_{\mathrm{LDPC}}=22$, where $T^{(\mathrm{LDPC})}_{\mathrm{surg}}=T_{\mathrm{cad}}=22$. In both cases,
Equation~\eqref{eq:factory_counts} gives one $\ket{\CXCXX}$ factory and two $\ket{i}$
factories. Each buffer reservation contains one loading surgery, confined to the repetition layer and lasting $T_{\mathrm{surg}} = d_{\mathrm{rep}}$, and one injection or correction surgery, which involves the LDPC layer and lasts $T^{(\mathrm{LDPC})}_{\mathrm{surg}}$. Replacing one of the two $T_{\mathrm{surg}}$ terms of Equation~\eqref{eq:adder_shared_buffer_count} accordingly, with $\max\bigl(T_{\mathrm{prep}}^{\ket{\mathrm{AA}}}, T_{\mathrm{prep}}^{\ket{\DCZ}}\bigr) = 9$ (Section~\ref{sssec:resource_states}) and $T_{\mathrm{react}} + T_{\mathrm{meas}} = 11$, the reservation time becomes
\begin{align}
  T_{\mathrm{buf}} &= d_{\mathrm{rep}} + T^{(\mathrm{LDPC})}_{\mathrm{surg}}
  + \max\bigl(T_{\mathrm{prep}}^{\ket{\mathrm{AA}}}, T_{\mathrm{prep}}^{\ket{\DCZ}}\bigr) \nonumber\\
  &\quad + T_{\mathrm{react}} + T_{\mathrm{meas}} \nonumber\\
  &= d_{\mathrm{rep}} + T^{(\mathrm{LDPC})}_{\mathrm{surg}} + 20,
  \label{eq:Tbuf-ldpc}
\end{align}
and the number of adder buffers is
$n_{\mathrm{buf}}^{(\Adder)}=\lceil T_{\mathrm{buf}}/T_{\mathrm{cad}}\rceil$. For every pair of
distances considered in Table~\ref{tab:ldpc-scan},
$43\le T_{\mathrm{buf}}\le 61$ and $n_{\mathrm{buf}}^{(\Adder)}=3$, as on the
repetition-cat architecture. \Lookup{}s and
\Phaseup{}s use $n_{\mathrm{buf}}^{(\Lookup)}=2$ resource-state
buffers of six repetition codes each (Section~\ref{ssec:lookup_and_phaseup}).

Physical qubits are counted as in Section~\ref{sssec:global_layout_routing}, with one
difference. In the LDPC layer, each physical qubit lies in the support of
the logical operators of several logical qubits, so that its qubits cannot
be attributed to individual logical qubits. We therefore count the whole
LDPC layer as allocated during the whole shot. On the repetition layer, as
in Section~\ref{sssec:global_layout_routing}, a slice $s$ costs $2d_{\mathrm{rep}}\,n_{\mathrm{col}}(s)$
physical qubits (a repetition-code column counts its $d_{\mathrm{rep}}$ data qubits, each with the ancilla directly above it), with $n_{\mathrm{col}}(s)$ the number of repetition-code
columns allocated by its operation, i.e.\ its buffers and auxiliary qubits
(Buffer zone column of Table~\ref{tab:ldpc-layout}), since its data qubits
are in the LDPC layer. All other qubits of the
repetition layer, except the factories, are never counted within an
operation, including the access zone and the routing zones that an
operation uses: they form the global routing $\Gamma^{(\mathrm{rep})}_{\mathrm{global}}$,
present during the whole shot, and the zone $\Gamma_{\mathrm{QFT}}$, used
only by the final inverse quantum Fourier transform. The Layout paragraph
below gives these quantities, the Access paragraph the cost of reaching
$Z_L$, the Duration paragraph the duration of each slice, and the Volume
paragraph combines them into the logical volumes of the two layers.

\begin{figure*}[t]
  \centering
  \includegraphics[width=\textwidth]{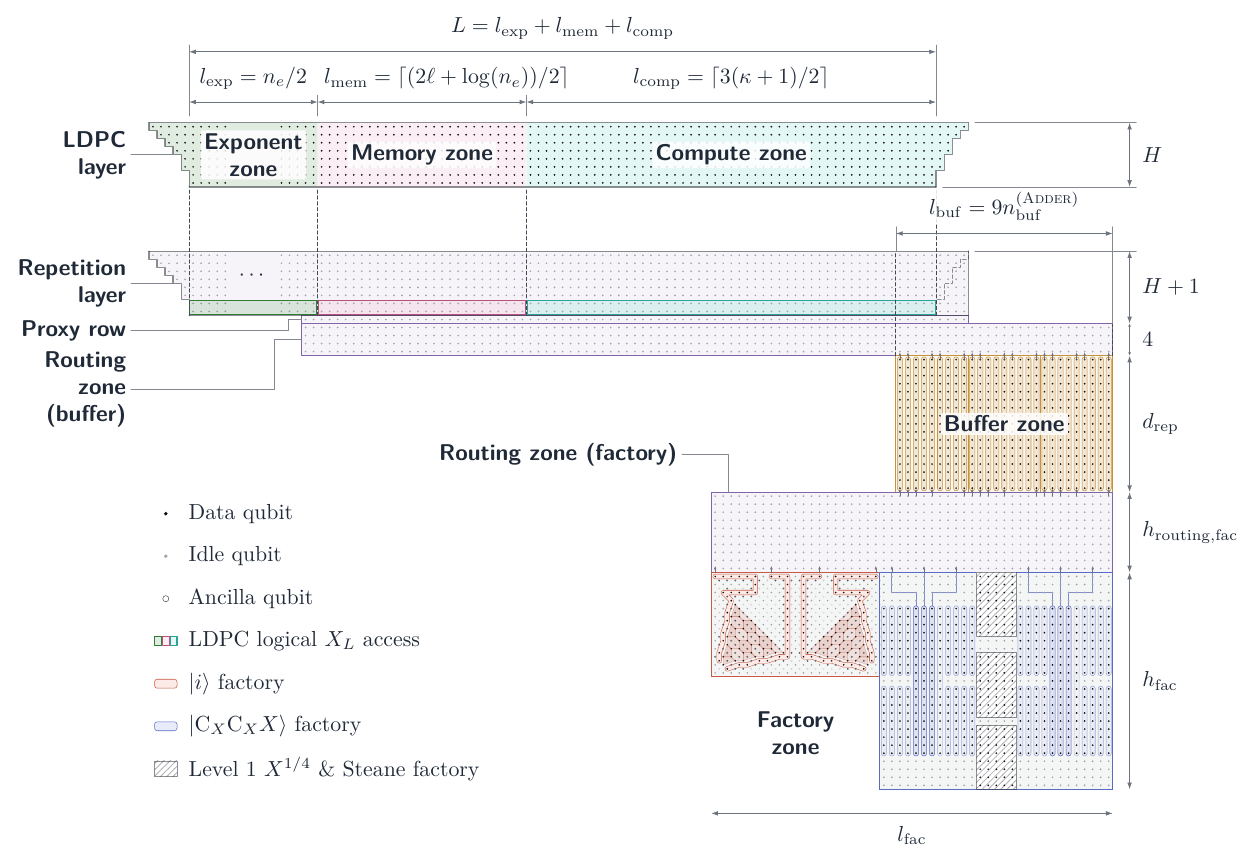}
  \caption{Global layout of the LDPC-cat chip for RSA-2048.}
  \label{fig:ldpc-layout}\label{fig:global_layout_ldpc}
\end{figure*}

\paragraph{Layout}
\label{sec:ldpc-cat-layout}\label{sec:ldpc-global}
The global layout, shown in Figure~\ref{fig:ldpc-layout}, is determined by the requirements of the \AccumulateResidue{} subroutine. The chip is a flip chip of two facing layers
(Section~\ref{sssec:error_correcting_codes}). The LDPC layer holds one open-boundary LDPC-cat
code block~\cite{Ruiz_2025} of $H$ rows and $L$ columns, which stores
all logical data qubits. Its upper rows extend laterally into staircases
of $\Delta_{\mathrm{LDPC}}$ qubits in total, so that the outermost logical qubits keep the
full distance (Table~\ref{tab:adder_ldpc_codes}). The operator $X_L$ of a logical
qubit is the weight-one operator of a tip qubit. The tips form the bottom
two rows of the block, with two logical qubits per column. The operator
$Z_L$ is supported on a set that spreads upward over at most $2H-1$
columns. The block is divided into an exponent zone, a memory zone and a
compute zone of
\begin{gather}
  l_{\mathrm{exp}} = \frac{n_e}{2},\quad
  l_{\mathrm{mem}} = \left\lceil \frac{2\ell + \log(n_e)}{2}\right\rceil,
  \nonumber\\
  l_{\mathrm{comp}} = \left\lceil \frac{3(\kappa+1)}{2}\right\rceil
  \label{eq:ldpc-zones}\csname ltx@label\endcsname{eq:ldpc-zone-lengths}
\end{gather}
columns, so that $L = l_{\mathrm{exp}} + l_{\mathrm{work}}$, with
$l_{\mathrm{work}} = l_{\mathrm{mem}} + l_{\mathrm{comp}}$ the length of
the work zone. The work zone holds the same logical qubits as on the
repetition-cat architecture. The compute zone keeps the $3(\kappa+1)$
interleaved slots of the \Adder$^\dagger$ of \AccumulateResidue, but needs no
routing column, since $Z_L$ access comes from the repetition layer. The
$[n_L, k_L, d_{\mathrm{LDPC}}]$ code has $n_L = HL + \Delta_{\mathrm{LDPC}}$ data qubits, one
ancilla per check and $k_L = 2L$ logical qubits, so the LDPC layer holds
\begin{equation}
  \Gamma_{\mathrm{LDPC}} = 2n_L - k_L = 2(H-1)L + 2\Delta_{\mathrm{LDPC}}
  \label{eq:Gamma-ldpc}\csname ltx@label\endcsname{eq:ldpc-gamma-ldpc}
\end{equation}
physical qubits.

The repetition layer spans the whole length of the LDPC layer and consists
of the following zones, from top to bottom in Figure~\ref{fig:ldpc-layout}.
The \emph{LDPC layer access zone}, of $H$ rows of data qubits facing the
block and its staircases, gives access to both logical
operators of every logical qubit (Section~\ref{sssec:error_correcting_codes}): $X_L$ through a
transversal \textsc{Cnot} gate with the facing tip qubit, and $Z_L$ through a
repetition code covering its support (Section~\ref{sssec:adder_physical_implementation_LDPC}).
Under the work zone, it is completed by the proxy row used by the adder.
The \emph{routing zone (buffer)}, of height $h_{\mathrm{routing,buf}}=4$
and length
$l_{\mathrm{routing,buf}} = l_{\mathrm{work}} + 2 + e_R + 9(n_{\mathrm{buf}}^{(\Adder)}-1)$,
where $e_R$ is the length of the widest point of the right staircase,
carries the routing codes that connect the inputs of each injection or
correction step to the adder buffers (Figure~\ref{fig:adder_ldpc_routing_overlays}).
The \emph{buffer zone} holds $n_{\mathrm{buf}}^{(\Adder)}$ adder buffers of nine
repetition codes of distance $d_{\mathrm{rep}}$, hence
$l_{\mathrm{buf}} = 9n_{\mathrm{buf}}^{(\Adder)}$ columns. The first buffer lies below
the compute zone, with its right edge aligned with the widest row of the
access zone, and the others lie to its right, as required by the routing
of the adder (Figure~\ref{fig:adder_ldpc_routing_overlays}). As on the
repetition-cat architecture, the auxiliary qubits of \Lookup{}s and the
\Phaseup{} workspace are stored in the buffer zone. Finally, as in
Figure~\ref{fig:global_layout}, $h_{\mathrm{routing,fac}}$ rows of factory
routing lie above a \emph{factory zone} of length $l_{\mathrm{fac}}$ that
holds the factories of Section~\ref{ssec:factories}.

Under the exponent zone, the only accesses during the computation are
$X_L$ accesses to at most two control tips at a time, for the \textsc{And}
gates of \Lookup{}s and \Phaseup{}s, which are performed sequentially. They are
routed through the two bottom rows of the access zone, which form the
address-routing allowance $\Gamma_{\mathrm{addr}} = 4l_{\mathrm{exp}}$.
The rest of the access zone under the exponent zone,
\begin{equation}
  \Gamma_{\mathrm{QFT}} = 2H\,l_{\mathrm{exp}} - l_{\mathrm{exp}} - \Gamma_{\mathrm{addr}},
  \label{eq:Gamma-qft}
\end{equation}
i.e.\ its $2H$ qubits per column minus the exposed ancilla row
and the address-routing allowance, is used only by the Z-basis measurements of the final inverse quantum
Fourier transform (Duration paragraph). In the color-changed convention,
these measurements need $Z_L$, which, unlike on the repetition-cat
architecture (Section~\ref{sssec:global_layout_routing}), is not reachable by a
transversal destructive measurement.

Adding the zones of the repetition layer and the factories to the LDPC
layer gives the total number of physical qubits,
\begin{align}
  n^{(\mathrm{LDPC})}_{\mathrm{total}} ={}& \Gamma_{\mathrm{LDPC}}
  + \Big[2H(L + e_R) - (L + e_R) \nonumber\\
  &+ 2(l_{\mathrm{work}}+2) + 2\Delta_{\mathrm{L}}\Big] \nonumber\\
  &+ 2h_{\mathrm{routing,buf}}\,l_{\mathrm{routing,buf}}
  + 2d_{\mathrm{rep}}\, l_{\mathrm{buf}} \nonumber\\
  &+ 2h_{\mathrm{routing,fac}}\,l_{\mathrm{fac}} - \max(l_{\mathrm{fac}} - l_{\mathrm{buf}}, 0) + n_{\mathrm{fac}},
  \label{eq:ntotal-ldpc}\csname ltx@label\endcsname{eq:ntotal_ldpc}
\end{align}
whose terms count, in order, six parts of the chip. 
The first term is the LDPC layer. 
The second term is the access zone, of $2H$ qubits per column over the $L$ columns of the block and the $e_R$ columns of its right end. 
This term subtracts the exposed ancilla row of these $L + e_R$ columns and adds the proxy row under the work zone and the left end of the block. 
The third term is the routing zone (buffer). 
The fourth term is the buffer zone, of $2d_{\mathrm{rep}}$ qubits per column, i.e.\ $d_{\mathrm{rep}}$ data qubits each counted with the ancilla above it, as in Equation~\eqref{eq:adder_ldpc_core_count}. 
The fifth term is the factory routing zone, of $2h_{\mathrm{routing,fac}}$ qubits per column. 
This term subtracts the exposed tops of its $l_{\mathrm{fac}} - l_{\mathrm{buf}}$ columns beyond the buffer zone, which no routing qubit uses. 
The sixth term is the factories. 
At the left end, under the exponent zone, the access zone only faces the $\Delta_{\mathrm{L}}$
qubits of the left staircase. At the right end, the adder compilation uses
the full padded rectangle of Equation~\eqref{eq:adder_ldpc_core_count}, so that the
access zone covers a rectangle of width $e_R$ over its $H$ rows. We
assume that the two staircases have the same number of qubits,
$\Delta_{\mathrm{L}} = \Delta_{\mathrm{LDPC}}/2$.
For RSA-2048, the parameters of Table~\ref{tab:algo_parameter_value} give
$l_{\mathrm{exp}}=640$, $l_{\mathrm{mem}}=26$ and $l_{\mathrm{comp}}=51$,
hence $l_{\mathrm{work}}=77$, $L=717$ and $k_L=1434$. With
$h_{\mathrm{routing,buf}}=4$ and $l_{\mathrm{buf}}=9n_{\mathrm{buf}}^{(\Adder)}$,
Equation~\eqref{eq:ntotal-ldpc} becomes
\begin{align}
  n^{(\mathrm{LDPC})}_{\mathrm{total}} ={}& 2868H + 3\Delta_{\mathrm{LDPC}}
  + (2H+7)\,e_R \nonumber\\
  &+ 18\,n_{\mathrm{buf}}^{(\Adder)} d_{\mathrm{rep}} + 81\,n_{\mathrm{buf}}^{(\Adder)} - 1433 \nonumber\\
  &+ (2h_{\mathrm{routing,fac}}-1)\,l_{\mathrm{fac}} + n_{\mathrm{fac}},
  \label{eq:ntotal-ldpc-num}\csname ltx@label\endcsname{eq:ldpc-ntotal}
\end{align}
for $l_{\mathrm{fac}} \ge 9n_{\mathrm{buf}}^{(\Adder)}$, with $e_R=5$ for the code of Section~\ref{sec:ldpc-cat-b}. The factory
parameters $h_{\mathrm{routing,fac}}$, $l_{\mathrm{fac}}$ and
$n_{\mathrm{fac}}$ are given in the same section.

The qubits of Equation~\eqref{eq:ntotal-ldpc} that belong neither to the
LDPC layer, nor to the repetition-code columns of an operation, nor to the
factories, nor to
$\Gamma_{\mathrm{QFT}}$ form the global routing of the repetition layer,
\begin{align}
  \Gamma^{(\mathrm{rep})}_{\mathrm{global}} ={}&
  2h_{\mathrm{routing,fac}}\,l_{\mathrm{fac}} - \max(l_{\mathrm{fac}} - l_{\mathrm{buf}}, 0) \nonumber\\
  &+ \Gamma_{\mathrm{addr}} + \Gamma_{\mathrm{work}},
  \label{eq:Gamma-global-ldpc}
\end{align}
whose terms are the factory routing zone, the address-routing allowance
and the routing qubits under the work zone,
\begin{align}
  \Gamma_{\mathrm{work}} ={}& 2H(l_{\mathrm{work}} + e_R) - (l_{\mathrm{work}} + e_R)
  \nonumber\\
  &+ 2(l_{\mathrm{work}}+2) + 2\Delta_{\mathrm{L}} \nonumber\\
  &+ 2h_{\mathrm{routing,buf}}\,l_{\mathrm{routing,buf}},
  \label{eq:Gamma-work}
\end{align}
that is the access zone under the work zone and at the right end of the
block, minus their exposed ancilla row, the proxy row, the access zone at
the left end of the block, and the routing zone (buffer). As on the repetition-cat architecture, the global
routing is counted during the whole shot, whatever the current operation.
The largest operation is an adder, whose $9n_{\mathrm{buf}}^{(\Adder)}$ buffer columns
fill the buffer zone, so that Equations~\eqref{eq:ntotal-ldpc}
to~\eqref{eq:Gamma-work} give the identity
\begin{equation}
  \max_s \Gamma_{\mathrm{total}}(s) + \Gamma^{(\mathrm{rep})}_{\mathrm{global}}
  + \Gamma_{\mathrm{QFT}} = n^{(\mathrm{LDPC})}_{\mathrm{total}},
  \label{eq:identity-ldpc}
\end{equation}
where $\Gamma_{\mathrm{total}}(s) = \Gamma_{\mathrm{LDPC}} +
2d_{\mathrm{rep}}\,n_{\mathrm{col}}(s) + n_{\mathrm{fac}}$ is the physical qubit
count of slice $s$. Section~\ref{sec:ldpc-cat-b} evaluates both sides.

\AccumulateResidue{} keeps the arrangement of the repetition-cat layout
(Section~\ref{sssec:global_layout_routing}), with the address register in the memory
zone, left of the compute zone, but the buffers are at the bottom right of
the compute zone (Figure~\ref{fig:ldpc-layout}). The accumulation qubits
occupy the two tip rows of the compute zone, one every three columns in
each row, which leaves room for the \Adder$^\dagger$. The \Lookup{} loads each
qubit of its target register directly left of an accumulation qubit, and
the \Adder$^\dagger$ writes its output in the remaining free slot of each
triple. At its widest point, the \Adder$^\dagger$ therefore uses all
$3(\kappa+1)$ slots of the compute zone and the $n_{\mathrm{buf}}^{(\Adder)}$ adder
buffers. The output qubits of the \Lookup{} are then uncomputed by the
Z-basis measurement $M_Z(\kappa+1,3)$
(Section~\ref{sec:ldpc-cat-zaccess}). The \Adder{} that follows runs on
triples shifted by one slot, so the output of the last bit of the bottom
row lands on the first slot of the top row, at the opposite end of the
compute zone. This long-range output should be routable without
additional qubits, although the routing is not demonstrated here. In the
worst case, one additional compute column is needed, which adds $4H+7$
physical qubits. After the \Adder, the accumulation register sits two slots
to the right of its initial position and moves back with
$\Reshuffle(\kappa,3)$ (Section~\ref{sec:ldpc-cat-zaccess}).

The other subroutines use the same operations, and
Table~\ref{tab:ldpc-layout} checks, as Table~\ref{tab:rep-code-columns} does on
the repetition-cat architecture, that every slice fits in the layout of
Figure~\ref{fig:ldpc-layout}. For each operation, it counts the columns of
the LDPC layer used in the memory and compute zones and the positions of
the buffer zone, with three rules. A register laid out for an $n$-bit
addition, one qubit per triple $(a_k,b_k,s_k)$ in the two tip rows, spans
$\lceil 3n/2\rceil$ columns of the compute zone, whether only one summand
is present, a \Lookup{} has written the second one into the free slots of the
triples, or the adder fills all three slots. A
register stored outside any addition is packed in $\lceil m/2\rceil$
columns for $m$ qubits. Finally, an adder uses the $9n_{\mathrm{buf}}^{(\Adder)}$
positions of its buffers, a \Lookup$(w,\cdot)$ the
$6n_{\mathrm{buf}}^{(\Lookup)}$ positions of its buffers and its
$w-1$ auxiliary qubits, and a \Phaseup$(w)$ the same buffers and its
$2^h-h+2^l-l-2$ power products of weight at least two. These counts depend
only on the parameters of Table~\ref{tab:algo_parameter_value} and on the numbers of
buffers, not on the distances. Their maxima, 26 columns of memory zone, 51
columns of compute zone and 27 buffer positions, are all reached in
\AccumulateResidue, and they set $l_{\mathrm{mem}}$, $l_{\mathrm{comp}}$ and
$l_{\mathrm{buf}}$. With the $l_{\mathrm{exp}} = 640$ columns of the
exponent zone, the block has $L = 717$ columns, which every slice of
\AccumulateResidue{} fills.

In \ComputeResidue{} and \ComputeResidue$^\dagger$, the accumulation register,
packed in 17 columns, sits next to the widest operations, the Q-Q
\Adder$^\dagger(21)$ and the Q-Q \Comparison$(20)$, which take 32 columns
($17+32 = 49 \le 51$), while $S_i \bmod (p_i-1)$, the second $\ell$-qubit
register and the quotient register fill the 26 columns of the memory zone.
The Q-Q \Adder$(31)$ of \ComputeExponent{} takes 47 columns, and $17+47 = 64 >
51$. As on the repetition-cat architecture, the accumulation register is
therefore parked in the memory zone during \ReduceExponent$^\dagger$,
\ComputeExponent{} and \ReduceExponent, next to at most the $\log(n_e)$
quotient qubits ($17+6 = 23 \le 26$ columns). The register moves of
Figure~\ref{fig:acc_register_path} therefore carry over to the LDPC-cat
architecture.
The buffer zone serves every adder, since its $9n_{\mathrm{buf}}^{(\Adder)}$
positions do not depend on the register size, and \Lookup{}s and \Phaseup{}s use
at most 17 and 20 of its 27 positions.

Besides the drift compensation of \AccumulateResidue, registers move within
three subroutines. Each adder shifts its output by one slot, so that the
register held in the compute zone drifts to the right and must move back
before the next iteration: by one slot in \ComputeExponent, and by two
slots per inner-loop iteration in \ComputeResidue{} and
\ComputeResidue$^\dagger$. In \ReduceExponent{} and \ReduceExponent$^\dagger$,
the $(\ell+1)$-qubit window of the reduction slides by one bit per
iteration: the underflow qubit leaves the window and joins the quotient
register in the memory zone, and the next lower bit of $S_i$ enters the
window. Finally, the two $\ell$-qubit registers of \ComputeResidue{} and
\ComputeResidue$^\dagger$ swap roles once per outer-loop iteration, which
requires a physical move between the compute zone and the memory zone.
Between subroutines, the registers follow the moves of
Figure~\ref{fig:acc_register_path}. The durations of all these moves are given in
the Duration paragraph.

\begin{table*}[t]
\centering
\begin{tabular}{llrrrr}
\hline\hline
& & \multicolumn{3}{c}{LDPC-layer columns} & Buffer zone\\
Subroutine & Operation & Memory & Compute & Block & positions\\
\hline
\ComputeExponent{} & \Lookup$(6,31)$ & 17 & 47 & 704 & 17\\
 & Q-Q \Adder$(31)$ & 17 & 47 & 704 & 27\\
\hline
\ReduceExponent{} & Q-C \Adder$(21)$ & 23 & 37 & 700 & 27\\
 & Q-Q \Adder$(20)$ & 23 & 36 & 699 & 27\\
\hline
\ComputeResidue{} & \Lookup$(6,20)$ (loading) & 16 & 47 & 703 & 17\\
 & \Lookup$(6,21)$ & 26 & 49 & 715 & 17\\
 & Q-Q \Adder$^\dagger(21)$ & 26 & 49 & 715 & 27\\
 & Q-Q \Adder$(20)$ & 26 & 47 & 713 & 27\\
\hline
\AccumulateResidue{} & \Lookup$(5,34)$ & 26 & 51 & 717 & 16\\
 & Q-Q \Adder$^\dagger(34)$ & 26 & 51 & 717 & 27\\
 & Q-Q \Adder$(33)$ & 26 & 51 & 717 & 27\\
 & \Lookup$(5,33)$ & 26 & 51 & 717 & 16\\
 & Q-Q \Comparison$(33)$ & 26 & 51 & 717 & 27\\
 & \Phaseup$(5)$ & 26 & 51 & 717 & 17\\
\hline
\ComputeResidue$^\dagger$ & \Lookup$(6,21)$ & 26 & 49 & 715 & 17\\
 & Q-Q \Adder$^\dagger(21)$ & 26 & 49 & 715 & 27\\
 & Q-Q \Adder$(20)$ & 26 & 47 & 713 & 27\\
 & \Lookup$(6,20)$ & 26 & 47 & 713 & 17\\
 & Q-Q \Comparison$(20)$ & 26 & 49 & 715 & 27\\
 & \Phaseup$(6)$ & 26 & 47 & 713 & 20\\
 & \Phaseup$(6)$ (unload) & 16 & 17 & 673 & 20\\
\hline
Last \Phaseup{} & \Phaseup$(6)$ & 17 & 0 & 657 & 20\\
\hline\hline
\end{tabular}
\caption{Layout of each slice on the LDPC-cat architecture for the
factorization of an RSA-2048 number, with the parameters of
Table~\ref{tab:algo_parameter_value}, $n_{\mathrm{buf}}^{(\Adder)} = 3$ and
$n_{\mathrm{buf}}^{(\Lookup)} = 2$. The counts do not depend on the
distances. Memory and Compute are the numbers of columns of the LDPC layer,
two logical qubits each, used in the memory and compute zones. A register
laid out for an $n$-bit addition, one qubit per triple $(a_k,b_k,s_k)$ in
the two tip rows, spans $\lceil 3n/2\rceil$ columns, whether only one
summand is present, a \Lookup{} has written the second one, or the adder
fills all three slots (a comparison of size $m$ is an adder of size
$m+1$). A register stored outside any addition is
packed in $\lceil m/2\rceil$ columns for $m$ qubits. Block adds
the $l_{\mathrm{exp}} = 640$ columns of the exponent zone, used during
the whole computation. Buffer zone positions counts the repetition codes
of the buffer zone used by the operation:
$9n_{\mathrm{buf}}^{(\Adder)}$ for an adder,
$6n_{\mathrm{buf}}^{(\Lookup)} + w - 1$ for a \Lookup$(w,\cdot)$, and
$6n_{\mathrm{buf}}^{(\Lookup)} + 2^h-h+2^l-l-2$ for a \Phaseup$(w)$;
it is the column count $n_{\mathrm{col}}(s)$ of
Equation~\eqref{eq:vol-phys-rep-ldpc}. The maxima, 26 memory columns, 51
compute columns, $L = 717$ columns of the block and 27 buffer positions,
are all reached in \AccumulateResidue{} and fix $l_{\mathrm{mem}}$,
$l_{\mathrm{comp}}$, $L$ and $l_{\mathrm{buf}}$ in the global layout of
Figure~\ref{fig:ldpc-layout}. The table omits the $M_Z$,
$\Reshuffle$, $\mathrm{PC}X$ and $P_{\ket{+}}$ slices, which never
exceed the adjacent \Adder{} or \Lookup{} slices.}
\label{tab:ldpc-layout}
\end{table*}

\paragraph{Access to $Z_L$: Z-basis measurements and register moves}
\label{sec:ldpc-cat-zaccess}
On the repetition-cat architecture, a Z-basis measurement $M_Z(x)$
is a transversal destructive measurement of one cycle
(Section~\ref{sssec:layout_connectivity}), and registers move through the routing
zone at a cost neglected in $T_{\mathrm{shot}}$ (Section~\ref{sssec:global_layout_routing}). In the
LDPC layer, both operations need the operator $Z_L$ of the logical qubits
involved. A Z-basis measurement of a logical qubit of the LDPC layer is
performed by a snake repetition code of the access zone covering
$\mathrm{supp}(Z_L)$: the code is initialized, accesses $Z_L$ during its
first stabilizer round and is measured after $d^{\star}$ rounds
(Section~\ref{sssec:adder_physical_implementation_LDPC}), so that the measurement takes
$T^{(\mathrm{LDPC})}_{\mathrm{surg}}$ cycles instead of one. Two logical qubits whose $Z_L$ supports
overlap cannot be measured simultaneously. Since a $Z_L$ support spans at
most $2H-1$ columns, each holding two logical qubits, measuring a register
of $n$ logical qubits that occupies one column out of $k$ takes at most
\begin{equation}
  T_{M_Z}(n,k) = \min\!\left(2\left\lceil\frac{2H-1}{k}\right\rceil,
  n\right)T^{(\mathrm{LDPC})}_{\mathrm{surg}}
  \label{eq:TMZ}
\end{equation}
cycles, with $T_{M_Z}(n) = T_{M_Z}(n,1)$. The registers
handled by the adders, which occupy one column out of three
(Section~\ref{sssec:adder_physical_implementation_LDPC}), have $k=3$.

A logical qubit moves from a slot $a$ to a slot $b$ of the LDPC layer by
teleportation, with $b$ in a $Z_L$ eigenstate. The $X$-parity
$X_L(a)X_L(b)$ is first measured between the tips of the two slots,
through routing codes of the repetition layer, in $T^{(\mathrm{LDPC})}_{\mathrm{surg}}$ cycles. $Z_L(a)$
is then measured with a snake under the support of $a$ only, as above.
Slot $b$ then holds the state of $a$ up to Pauli corrections, which are
absorbed in the Pauli frame, and slot $a$ is left in a $Z_L$ eigenstate,
ready to receive a later move. The destination slots are always in a $Z_L$
eigenstate, since they are freed either by a Z-basis measurement, such as
the measurement-based uncomputation of the \Lookup{} outputs or the
consumption of an adder input, or by an earlier move. The only access to
$Z_L$ is local, so that the overlap constraint of
Equation~\eqref{eq:TMZ} applies to the source register, while the
long-range part of the move is an $X$-parity, like the other long-range
operations of the architecture. The $X$-parities of several moves run in
parallel on separate lanes, i.e.\ horizontal repetition codes of the
repetition layer. When moving between the memory and compute zones, the
lanes use the access zone, the proxy row and the routing zone (buffer)
under the work zone, which hold about 12 rows of data qubits. Since two
neighboring lanes cannot share a row of ancillae, this gives about
$n_{\mathrm{lane}} = 10$ lanes, the vertical segments to the tips crossing
the lanes as in~\cite{FeMoco2026prep}. A drift compensation moves each qubit by
one or two slots only, so that all its $X$-parities run at once,
$n_{\mathrm{lane}} = n$. Moving an $n$-qubit register that occupies one
column out of $k$ therefore takes
\begin{equation}
  T_{\Reshuffle}(n,k) = T_{M_Z}(n,k) +
  \left\lceil \frac{n}{n_{\mathrm{lane}}}\right\rceil T^{(\mathrm{LDPC})}_{\mathrm{surg}}
  \label{eq:TReshuffle}
\end{equation}
cycles, with $n_{\mathrm{lane}} = n$ for a drift compensation and
$n_{\mathrm{lane}} = 10$ for a move between zones, marked ``swap'' in
Table~\ref{tab:ldpc-cycles}, where these operations appear as
$M_Z(x,k)$ and $\Reshuffle(x,k)$. The second term is an estimate that does not take into account the exact locations of the source and destination slots. An estimate is sufficient because this term is small (Section~\ref{sec:ldpc-cat-b-duration}).

\paragraph{Duration}
\label{sec:ldpc-cat-duration}
All durations are counted in physical error-correction cycles. Lattice
surgeries involving the LDPC layer last $T^{(\mathrm{LDPC})}_{\mathrm{surg}}$ cycles
(Equation~\eqref{eq:Tsurg-ldpc}),
and steps that consume factory outputs run at the cadence $T_{\mathrm{cad}}$ of
Equation~\eqref{eq:Tcad-ldpc}. As in Section~\ref{sssec:global_layout_routing}, Equation~\eqref{eq:adder_ldpc_latency} gives the duration of an $n$-bit adder,
\begin{equation}
  T_{\Adder}^{(\mathrm{LDPC})}(n) = 2n\,T_{\mathrm{cad}},
  \label{eq:T-adder-ldpc}\csname ltx@label\endcsname{eq:ldpc-tadder}
\end{equation}
which also applies to the quantum-classical adder and to a comparison of
size $n$, implemented as an adder of size $n+1$. For $T_{\mathrm{cad}} = 21$, this is $42n$.
In a \Lookup, the \textsc{And} gates and the parity-\textsc{Cnot} gates are surgeries
involving the LDPC layer. Equation~\eqref{eq:lookup_cost_ldpc} then gives
\begin{equation}
  T_{\Lookup}^{(\mathrm{LDPC})}(w) = (2^w-1)\,T^{(\mathrm{LDPC})}_{\mathrm{surg}}
  + (2^{w-1}-w)\,T_{\mathrm{cad}},
  \label{eq:T-lookup-ldpc}
\end{equation}
that is $63T^{(\mathrm{LDPC})}_{\mathrm{surg}} + 26T_{\mathrm{cad}}$ for $w_1 = 2w_3 = 6$ and $31T^{(\mathrm{LDPC})}_{\mathrm{surg}} +
11T_{\mathrm{cad}}$ for $w_4=5$. As in~\cite[Eqs.~(4.8) and (4.12)]{FeMoco2026prep}, this assumes
that the factories produce a $\ket{\CXCXX}$ state within $2T^{(\mathrm{LDPC})}_{\mathrm{surg}}$ cycles,
which holds for every pair of distances considered here
($21\le 2T^{(\mathrm{LDPC})}_{\mathrm{surg}}$, since $T^{(\mathrm{LDPC})}_{\mathrm{surg}}\ge12$).
Equation~\eqref{eq:phaseup_cost_ldpc} gives the duration of a \Phaseup{} on a
$w$-qubit address register, split into $h=\lceil w/2\rceil$ and
$l=\lfloor w/2\rfloor$,
\begin{equation}
  T_{\Phaseup}^{(\mathrm{LDPC})}(w) = (2^h+2^l-w-2)\,T_{\mathrm{cad}}
  + (2^l-1)\,T^{(\mathrm{LDPC})}_{\mathrm{surg}},
  \label{eq:T-phaseup-ldpc}
\end{equation}
since its \textsc{And} gates and multi-target $\C_X X$ gates are surgeries
involving the LDPC layer,
that is $8T_{\mathrm{cad}} + 7T^{(\mathrm{LDPC})}_{\mathrm{surg}}$ for $w=6$ and $5T_{\mathrm{cad}} + 3T^{(\mathrm{LDPC})}_{\mathrm{surg}}$ for $w_4=5$.
The slices of Table~\ref{tab:logical_qubit_count} omitted from
Table~\ref{tab:ldpc-cycles} are, in the color-changed convention, the
parity-\textsc{Cnot} gates $\mathrm{PC}X(x)$ and the preparations $P_{\ket{+}}(x)$.
A $\mathrm{PC}X(x)$ is a single lattice surgery (Section~\ref{sssec:global_layout_routing}), and
takes $T^{(\mathrm{LDPC})}_{\mathrm{PC}X}=T^{(\mathrm{LDPC})}_{\mathrm{surg}}$ cycles. A preparation of logical
qubits of the LDPC layer takes $T^{(\mathrm{LDPC})}_{\mathrm{prep}}=T^{(\mathrm{LDPC})}_{\mathrm{surg}}$
cycles, instead of $T_{\mathrm{init}}=1$ cycle on the repetition-cat
architecture. A logical qubit of the LDPC layer shares its physical qubits
with other logical qubits and cannot be reinitialized at the physical
level. The qubit is therefore prepared in $\ket{+}_L$ by measuring its weight-one $X_L$
through the facing qubit of its tip over $d^{\star}$ rounds, and the sign of the outcome being absorbed in the Pauli frame.
The $M_Z(x,k)$ and $\Reshuffle(x,k)$ slices last
$T_{M_Z}(x,k)$ cycles (Equations~\eqref{eq:TMZ}
and~\eqref{eq:TReshuffle}). The Z-basis measurements are the $N_{M_Z}$
slices of Equation~\eqref{eq:NMX}, which last one cycle on the
repetition-cat architecture and therefore appear here as rows of
Table~\ref{tab:ldpc-cycles}.

Unlike on the repetition-cat architecture, the register moves within
subroutines are included in $T_{\mathrm{shot}}$, as $\Reshuffle$ slices of
Table~\ref{tab:ldpc-cycles}. Each iteration of \ComputeExponent{} adds
$T_{\Reshuffle}(\ell+\log(n_e),3)$, each iteration of
\AccumulateResidue{} adds $T_{\Reshuffle}(\kappa,3)$, and each
iteration of \ReduceExponent{} and \ReduceExponent$^\dagger$ adds
$T_{\Reshuffle}(\ell+1,3) + T_{\Reshuffle}(2,1)$, the
second term moving the underflow qubit and the next bit of $S_i$, that is
$3T^{(\mathrm{LDPC})}_{\mathrm{surg}}$. In \ComputeResidue, the drift of each inner-loop iteration is
compensated by a $\Reshuffle(\ell,3)$. At the end of each
outer-loop iteration, this move is replaced by the move of the newly
computed register to the memory slots freed by the Z-basis measurement of
the former register, a move between zones. In \ComputeResidue$^\dagger$, no
measurement frees the memory slots, so the register they hold first moves
into the compute zone, in the slots freed by the drift of the \Adder, with
one $\Reshuffle(\ell,1)$ per outer-loop iteration. The newly
computed register then moves to the memory slots with a
$\Reshuffle(\ell,3)$, again in place of the drift compensation.
With $N_{\mathrm{out}} = |P|(\lceil \ell/w_3\rceil - 2)$ outer-loop
iterations, $c = \lceil \ell/w_3\rceil$ inner-loop iterations per outer
iteration in \ComputeResidue{} and $2c$ in \ComputeResidue$^\dagger$, and
$T^{\mathrm{swap}}_{\Reshuffle}$ denoting a move between zones, the
moves within \ComputeResidue{} and \ComputeResidue$^\dagger$ take
\begin{align}
  &N_{\mathrm{out}}\Big[(c-1)\,T_{\Reshuffle}(\ell,3)
  + T^{\mathrm{swap}}_{\Reshuffle}(\ell,3)\Big]
  \quad\text{and} \nonumber\\
  &N_{\mathrm{out}}\Big[(2c-1)\,T_{\Reshuffle}(\ell,3)
  + T^{\mathrm{swap}}_{\Reshuffle}(\ell,3) \nonumber\\
  &\qquad + T^{\mathrm{swap}}_{\Reshuffle}(\ell,1)\Big]
  \label{eq:T-moves-cr}
\end{align}
cycles, respectively.

Summing over the numbers of calls $N_{\mathrm{op}}$ of
Table~\ref{tab:ldpc-cycles}, which depend only on
Table~\ref{tab:algo_parameter_value} and are those of Table~\ref{tab:physical_cycle_breakdown}.
as in Equation~\eqref{eq:Tshot}, the duration of one shot is
\begin{equation}
  T_{\mathrm{shot}} = \sum_{\mathrm{op}} N_{\mathrm{op}}\,T_{\mathrm{op}}
  + T_{\mathrm{other}},
  \label{eq:Tshot-ldpc}
\end{equation}
where the sum runs over the operations of Table~\ref{tab:ldpc-cycles},
including the $M_Z$ and $\Reshuffle$ slices, with the
durations of Equations~\eqref{eq:TMZ} to~\eqref{eq:T-moves-cr}, and
$T_{\mathrm{other}}$ collects the parity-\textsc{Cnot} and preparation slices,
\begin{equation}
  T_{\mathrm{other}} = \left(N_{\mathrm{PC}X} + N_{\mathrm{prep}}\right)T^{(\mathrm{LDPC})}_{\mathrm{surg}},
  \label{eq:Tother-ldpc}
\end{equation}
with $N_{\mathrm{PC}X}$ and $N_{\mathrm{prep}}$ given by Equations~\eqref{eq:NCX}
and~\eqref{eq:Nprep}.

As on the repetition-cat architecture, the moves between subroutines are
not included in $T_{\mathrm{shot}}$. They follow Figure~\ref{fig:acc_register_path}: per
prime, the accumulation register moves four times, once from its
interleaved layout ($k=3$) and three times packed ($k=1$), $S_i \bmod
(p_i-1)$ moves twice, packed, and $r_i$ moves once from a layout for an
addition ($k=3$), while \ReduceExponent$^\dagger$, \ComputeExponent{} and
\ReduceExponent{} are entered and left without moves. All these moves are
between zones, so that
\begin{align}
  T_{\mathrm{transit}} = |P|\Big[&3\,T^{\mathrm{swap}}_{\Reshuffle}(\kappa,1)
  + T^{\mathrm{swap}}_{\Reshuffle}(\kappa,3) \nonumber\\
  &+ 2\,T^{\mathrm{swap}}_{\Reshuffle}(\ell,1)
  + T^{\mathrm{swap}}_{\Reshuffle}(\ell,3)\Big]
  \label{eq:T-transit}
\end{align}
cycles, whose share of $T_{\mathrm{shot}}$ is given in Section~\ref{sec:ldpc-cat-b}.
Moves within subroutines occur at almost every iteration, about
$8.4\times10^{6}$ times per shot, and each needs $Z_L$ measurements of its
source qubits (Section~\ref{sec:ldpc-cat-zaccess}), so that they take
about 5\% of $T_{\mathrm{shot}}$ (Section~\ref{sec:ldpc-cat-b}) and are counted.
Moves between subroutines occur only seven times per prime, about
$1.6\times10^{5}$ times per shot, and take 0.21\% of $T_{\mathrm{shot}}$, below
the 0.5\% under which Section~\ref{sssec:global_layout_routing} neglects all register
moves.
The final inverse quantum Fourier transform is not included in $T_{\mathrm{shot}}$
either. As in Section~\ref{sssec:global_layout_routing}, each of the $n_e$ exponent
qubits receives one rotation by phase kickback into a phase-gradient
register of $n_\varphi$ qubits~\cite{Gidney18halving}, and is then
measured in the Z basis, here through the access zone under the exponent
zone, in $T^{(\mathrm{LDPC})}_{\mathrm{surg}}$ cycles. The transform takes
\begin{align}
  T_{\mathrm{QFT}} = n_e\Big[&T_{\Adder}^{(\mathrm{LDPC})}(n_\varphi)
  + T^{(\mathrm{LDPC})}_{\mathrm{prep}} \nonumber\\
  &+ T^{(\mathrm{LDPC})}_{\mathrm{PC}X} + T^{(\mathrm{LDPC})}_{\mathrm{surg}} + T_{\mathrm{react}}\Big]
  \label{eq:T-qft-ldpc}
\end{align}
cycles, and $n_e(n_\varphi-1)$ Toffoli gates.

\paragraph{Volume}
\label{sec:ldpc-cat-volume}
All durations are counted in physical error-correction cycles. The logical
error rate of the repetition code is defined per $d_{\mathrm{rep}}$ cycles
(Equation~\eqref{eq:plogical}, from~\cite[Supplementary Table~1(b)]{Ruiz_2025}),
and that of the LDPC-cat code per $d_{\mathrm{LDPC}}$
cycles~\cite[Eq.~(6.1)]{FeMoco2026prep}. A physical duration $T$ therefore counts
as $T/d_{\mathrm{rep}}$ logical cycles in the repetition layer and as $T/d_{\mathrm{LDPC}}$
logical cycles in the LDPC layer. A repetition code that takes part in a
surgery of $T^{(\mathrm{LDPC})}_{\mathrm{surg}}$ cycles is exposed for $T^{(\mathrm{LDPC})}_{\mathrm{surg}}/d_{\mathrm{rep}}$ logical cycles, and
the volume accounts for this exposure automatically.

For the repetition layer, Equations~\eqref{eq:Volphys}
and~\eqref{eq:Voleff} become
\begin{align}
  \mathrm{Vol}^{(\mathrm{rep})}_{\mathrm{physical}} &= \sum_s
  \Big[2d_{\mathrm{rep}}\,n_{\mathrm{col}}(s) +
  \Gamma^{(\mathrm{rep})}_{\mathrm{global}}\Big] \nonumber\\
  &\qquad\times T_{\mathrm{error\,correction}}(s),
  \label{eq:vol-phys-rep-ldpc}\\
  \mathrm{Vol}^{(\mathrm{eff,rep})}_{\mathrm{logical}} &=
  \frac{\mathrm{Vol}^{(\mathrm{rep})}_{\mathrm{physical}}}{d_{\mathrm{rep}}(2d_{\mathrm{rep}}-1)}.
  \label{eq:vol-eff-rep-ldpc}
\end{align}
For the LDPC layer, counted as allocated during the whole shot,
\begin{equation}
  \mathrm{Vol}^{(\mathrm{LDPC})}_{\mathrm{physical}} =
  \Gamma_{\mathrm{LDPC}}\,T_{\mathrm{shot}}.
  \label{eq:vol-phys-ldpc}
\end{equation}
The LDPC layer encodes $k_L$ logical qubits on $\Gamma_{\mathrm{LDPC}}$
physical qubits, so the factor $2d-1$ of Equation~\eqref{eq:Voleff}
becomes $\Gamma_{\mathrm{LDPC}}/k_L$,
\begin{equation}
  \mathrm{Vol}^{(\mathrm{eff,LDPC})}_{\mathrm{logical}} =
  \frac{k_L\,\mathrm{Vol}^{(\mathrm{LDPC})}_{\mathrm{physical}}}
  {\Gamma_{\mathrm{LDPC}}\,d_{\mathrm{LDPC}}} = k_L\,\frac{T_{\mathrm{shot}}}{d_{\mathrm{LDPC}}},
  \label{eq:vol-eff-ldpc}
\end{equation}
and the effective number of logical qubits of the LDPC layer is $k_L$ by
construction. The logical error rate per logical cycle of one logical
qubit of the repetition layer, $p^{(\mathrm{rep})}_{\mathrm{logical}}$, is
given by Equation~\eqref{eq:plogical} at $d=d_{\mathrm{rep}}$. For the LDPC layer,
we use the fit of~\cite[Eq.~(6.1)]{FeMoco2026prep}, obtained under the same
phase-flip circuit-level noise model,
\begin{equation}
  p^{(\mathrm{LDPC})}_{\mathrm{logical}}(p_Z,d_{\mathrm{LDPC}}) = A\,d_{\mathrm{LDPC}}\,
  (B\,p_Z)^{C\lfloor(d_{\mathrm{LDPC}}+1)/2\rfloor},
  \label{eq:plogical-ldpc}
\end{equation}
with $A = 0.1$, $C = 1$ and $B = 25.8$, $29.8$ and $42.5$ for
$d_{\mathrm{LDPC}} = 12$, 16 and 22~\cite[Table~10]{FeMoco2026prep}. Equation~\eqref{eq:plogical-ldpc} is a per-logical-qubit
rate.
We assume that logical errors occur independently in the two
layers. With $p^{(\mathrm{rep})}_{\mathrm{logical}}$ and
$p^{(\mathrm{LDPC})}_{\mathrm{logical}}$, the probability that at least
one logical error occurs during one shot is, as in
Equation~\eqref{eq:eps},
\begin{equation}
  \epsilon^{(j)}_{\mathrm{logical}} = 1 - \left(1 -
  p^{(j)}_{\mathrm{logical}}\right)^{\mathrm{Vol}^{(\mathrm{eff},j)}_{\mathrm{logical}}},
  \quad j\in\{\mathrm{rep},\mathrm{LDPC}\},
  \label{eq:eps-layers}\csname ltx@label\endcsname{eq:error_rep_layer}
\end{equation}
in each layer, and
\begin{equation}
  \epsilon_{\mathrm{total}} = 1 - \left(1-\epsilon^{(\mathrm{rep})}_{\mathrm{logical}}\right)
  \left(1-\epsilon^{(\mathrm{LDPC})}_{\mathrm{logical}}\right)
  \label{eq:eps-total-ldpc}\csname ltx@label\endcsname{eq:total_error_ldpc}
\end{equation}
in either layer.

As in Section~\ref{sssec:global_layout_routing}, these volumes rest on a few
assumptions. First, the whole LDPC layer is allocated during the whole
shot. The exponent register alone occupies $n_e = 1280$ of the
$k_L = 1434$ slots, so at most 154 slots are unused at any time, and the
LDPC volume is overestimated by at most $k_L/n_e - 1 \approx 12\%$.
Second, the slices omitted from Table~\ref{tab:ldpc-layout}
($M_Z$, $\Reshuffle$, $\mathrm{PC}X$ and $P_{\ket{+}}$)
are assigned the peak column count of the computation,
$n_{\mathrm{col,other}} = 9n_{\mathrm{buf}}^{(\Adder)}$, an upper bound since these
slices are always smaller than the adjacent additions or \Lookup{}s.
Third, as in Equation~\eqref{eq:Gglobal}, the routing qubits are
counted at all times, although an operation only uses part of them. In the
access zone and the routing zone under the work zone, a surgery
uses only the snakes and routing codes it needs.
The allowance $\Gamma_{\mathrm{addr}}$ is only used during the \Lookup{}s of
\ComputeExponent{} and the last \Phaseup, about 27\% of $T_{\mathrm{shot}}$, and each
access uses only the part of its two rows between the control tip and the
end of the exponent zone.
$\Gamma_{\mathrm{QFT}}$, used only during the final transform, enters the
physical count but not the volume. Fourth, \ReduceExponent$^\dagger$ has the
same footprints and durations as \ReduceExponent, and the moves between
subroutines and the final transform are neglected, as in the Duration
paragraph.

\subsubsection{Evaluation at $p_Z = 5\times10^{-4}$ and $p_Z = 10^{-3}$}
\label{sec:ldpc-cat-b}\label{sssec:ldpc-evaluation}

We now evaluate the estimate for RSA-2048 in detail at
$p_Z = 5\times10^{-4}$, the phase-flip probability of the main result of
Section~\ref{ssec:physical_resource_count}, and give its main figures at
$p_Z = 10^{-3}$. Table~\ref{tab:ldpc-scan} gives the logical error budget
at both values of $p_Z$ and for other pairs of distances.

\paragraph{Factories, distances and physical qubit count} \label{sec:ldpc-factories}
We use the factories of Table~\ref{tab:adder_reference_factories}, as in
Section~\ref{ssec:physical_resource_count}, with outputs at distance $d_{\mathrm{rep}}$. We choose
$(d_{\mathrm{rep}},d_{\mathrm{LDPC}}) = (13,16)$, hence $d^{\star}=T^{(\mathrm{LDPC})}_{\mathrm{surg}}=16$
(Equation~\eqref{eq:Tsurg-ldpc}), for the reasons given in the logical
error budget below. Equation~\eqref{eq:Tcad-ldpc} gives $T_{\mathrm{cad}} = 21$, set by
the $\ket{\CXCXX}$ factory, and Equation~\eqref{eq:factory_counts} one $\ket{\CXCXX}$
factory and two $\ket{i}$ factories. As in Section~\ref{ssec:physical_resource_count},
the factory zone has length $l_{\mathrm{fac}}=50$ and contains
$n_{\mathrm{fac}}=1872$ physical qubits, and the factory routing zone has
height $h_{\mathrm{routing,fac}}=10$, hence
$2h_{\mathrm{routing,fac}}l_{\mathrm{fac}} - (l_{\mathrm{fac}} - l_{\mathrm{buf}}) = 1000 - 23 = 977$.
Equation~\eqref{eq:Tbuf-ldpc} gives $T_{\mathrm{buf}}=49$ cycles, hence
$n_{\mathrm{buf}}^{(\Adder)}=3$. The code of Table~\ref{tab:adder_ldpc_codes} with
$d_{\mathrm{LDPC}} = 16$ has $H=7$ and $\Delta_{\mathrm{LDPC}}=26$, hence $n_L = 5045$,
$k_L = 1434$ and $\Gamma_{\mathrm{LDPC}} = 8656$
(Equation~\eqref{eq:Gamma-ldpc}).

Since all logical data qubits lie in the LDPC layer, each slice $s$ adds
to the $\Gamma_{\mathrm{LDPC}}$ qubits of the LDPC layer and to the
factories only the $2d_{\mathrm{rep}}\,n_{\mathrm{col}}(s)$ physical qubits of
its buffer positions (Buffer zone column of Table~\ref{tab:ldpc-layout}).
The maximum, $8656 + 27\times26 + 1872 = 11\,230$ physical qubits, is
reached by every adder slice.

Equation~\eqref{eq:ntotal-ldpc-num} gives
$n^{(\mathrm{LDPC})}_{\mathrm{total}} = 22\,593$ physical qubits. The
global routing of Equation~\eqref{eq:Gamma-global-ldpc} is
$\Gamma^{(\mathrm{rep})}_{\mathrm{global}} = 5603$, whose terms are 977
(factory routing zone), $\Gamma_{\mathrm{addr}} = 2560$ and
$\Gamma_{\mathrm{work}} = 2066$ (Equation~\eqref{eq:Gamma-work}): 1001
for the access zone under the work zone, 158 for the proxy row, 26 for the
left staircase ($\Delta_{\mathrm{L}} = 13$), 65 for the rectangle of width
$e_R = 5$ at the right end, and 816 for the routing zone (buffer). With $\Gamma_{\mathrm{QFT}} = 5760$
(Equation~\eqref{eq:Gamma-qft}), these values satisfy the identity of
Equation~\eqref{eq:identity-ldpc}: $11\,230 + 5603 + 5760 = 22\,593$. The LDPC layer accounts for 38\% of
the chip, and $\Gamma_{\mathrm{QFT}}$, idle until the final transform,
for 25\%. Each exponent logical qubit costs $12.5$ physical qubits:
$\Gamma_{\mathrm{LDPC}}/k_L = 6.0$ in the LDPC layer and $(2H-1)/2 = 6.5$
in the access zone below it, whose $2H$ qubits per column, minus the exposed ancilla row, serve the two logical qubits of the column.
The zone $\Gamma_{\mathrm{QFT}}$ holds a quarter of the physical qubits
but only serves the $n_e$ Z-basis measurements of the final inverse
quantum Fourier transform. A destructive Z-basis measurement of all the
physical qubits of the LDPC layer would give $Z_L$ of every exponent qubit
in a single cycle, without any access zone. The semi-classical transform,
however, measures the exponent qubits one after the other, the rotation of
each qubit depending on the outcomes of the qubits already measured, so
that it cannot use this simultaneous measurement as such. Avoiding
$\Gamma_{\mathrm{QFT}}$, or using its qubits for other purposes during the
computation, therefore requires another way of accessing $Z_L$ of the
exponent qubits, or a transform compatible with a simultaneous
measurement. We leave this investigation for future work.

\paragraph{Duration}
\label{sec:ldpc-cat-b-duration}
Table~\ref{tab:ldpc-cycles} gives the duration of each slice, in the
format of Table~\ref{tab:physical_cycle_breakdown}. For RSA-2048, $N_{\mathrm{PC}X} = 262|P|$
and $N_{\mathrm{prep}} = 5|P|$ (Section~\ref{sssec:global_layout_routing}), so that
Equation~\eqref{eq:Tother-ldpc} gives $T_{\mathrm{other}} = 267|P|\times16
\approx 9.7\times10^{7}$ cycles, 0.34\% of $T_{\mathrm{shot}}$, and the $N_{M_Z}
= 400|P| + 214$ Z-basis measurements of Table~\ref{tab:ldpc-cycles} take
$1.34\times10^{9}$ cycles. Equation~\eqref{eq:Tshot-ldpc} then gives
$T_{\mathrm{shot}} = 2.82\times10^{10}$ cycles per shot, about 7.84 hours. Adders and comparisons take
43.53\% of the duration, \Lookup{}s 44.63\%, \Phaseup{}s 1.65\%, Z-basis
measurements 4.73\% and register moves within subroutines 5.11\%
($1.44\times10^{9}$ cycles). The $X$-parity term of
Equation~\eqref{eq:TReshuffle} accounts for $1.4\times10^{8}$ of these
cycles, 0.50\% of $T_{\mathrm{shot}}$. For example, measuring the accumulation
register takes $T_{M_Z}(\kappa,3) = 160$ cycles
(Equation~\eqref{eq:TMZ}). The Z-basis measurements and the moves, which
are negligible on the repetition-cat architecture, thus take 9.8\% of a
shot. The moves between subroutines take $2560$ cycles per prime, that is
$5.8\times10^{7}$ cycles (Equation~\eqref{eq:T-transit}), 0.21\% of
$T_{\mathrm{shot}}$, and
are neglected, as on the repetition-cat architecture.
The final inverse
transform takes $T_{\mathrm{QFT}} = 1280\times1108 \approx 1.42\times10^{6}$
cycles for $n_\varphi=25$ (Equation~\eqref{eq:T-qft-ldpc}), that is
$5\times10^{-5}$ of $T_{\mathrm{shot}}$, and $3.1\times10^{4}$ Toffoli gates, and is
neglected as well.

\paragraph{Logical error budget, number of shots and total runtime}
Tables~\ref{tab:ldpc-layout} and~\ref{tab:ldpc-cycles} with
Equations~\eqref{eq:vol-phys-rep-ldpc} to~\eqref{eq:vol-eff-ldpc} give
$\mathrm{Vol}^{(\mathrm{eff,rep})}_{\mathrm{logical}} = 5.4\times10^{11}$ and
$\mathrm{Vol}^{(\mathrm{eff,LDPC})}_{\mathrm{logical}} = 2.5\times10^{12}$.
The global routing $\Gamma^{(\mathrm{rep})}_{\mathrm{global}}$ accounts for
91\% of the volume of the repetition layer, the address-routing allowance
$\Gamma_{\mathrm{addr}}$ alone for 41\%, and the repetition-code columns
of the operations for 9\%. The logical error rate
per logical cycle of the repetition layer is
$p^{(\mathrm{rep})}_{\mathrm{logical}} = 5.67\times10^{-14}$
(Equation~\eqref{eq:plogical} at $d=13$, Table~\ref{tab:rep_cat_recap}),
and that of the LDPC layer is
$p^{(\mathrm{LDPC})}_{\mathrm{logical}} = 0.1\times16\times
(29.8\times5\times10^{-4})^{8} = 3.9\times10^{-15}$
(Equation~\eqref{eq:plogical-ldpc}).
Equations~\eqref{eq:eps-layers} and~\eqref{eq:eps-total-ldpc} then give
$\epsilon^{(\mathrm{rep})}_{\mathrm{logical}} = 3.0\times10^{-2}$,
$\epsilon^{(\mathrm{LDPC})}_{\mathrm{logical}} = 9.8\times10^{-3}$ and
$\epsilon_{\mathrm{total}} = 4.0\times10^{-2}$.
With the factories of Table~\ref{tab:adder_reference_factories}, and counting Toffoli
gates and $\ket{i}$ states as in Section~\ref{ssec:physical_resource_count}, the
$\ket{\CXCXX}$ and $\ket{i}$ factories cause a failure during a shot with
probabilities $1.5\times10^{-4}$ and $0.20\%$, as on the repetition-cat
architecture. The per-shot failure probability is therefore
$\epsilon = 1 - (1-\epsilon_{\mathrm{total}})
(1-\epsilon_{\ket{\CXCXX}})(1-\epsilon_{\ket{i}}) = 4.2\%$, dominated by the
repetition layer.

We apply the criterion of Section~\ref{sec:C7-budget} to the pairs $(d_{\mathrm{rep}},d_{\mathrm{LDPC}})$: among the pairs for which $\epsilon$ is below $10\%$, we
choose the one with the fewest physical qubits
$n^{(\mathrm{LDPC})}_{\mathrm{total}}$ (Equation~\eqref{eq:ntotal-ldpc}).
Table~\ref{tab:ldpc-scan} gives the volumes and the logical error
probabilities per shot for several pairs, at $p_Z=10^{-3}$ and
$p_Z=5\times10^{-4}$. At both values of $p_Z$, with $d_{\mathrm{LDPC}}=12$, at least
one logical error occurs in the LDPC layer in almost every shot, and with
$d_{\mathrm{rep}}=11$ in the repetition layer. At $p_Z = 5\times10^{-4}$, the criterion selects
$(13,16)$, with $22\,593$ physical qubits, against $22\,701$ for the next
pair with $\epsilon<10\%$, $(15,16)$. The pair $(15,16)$ would divide $\epsilon$ by about 3.4, to
1.2\%, for $2\times18\,n_{\mathrm{buf}}^{(\Adder)} = 108$ more physical qubits
(Equation~\eqref{eq:ntotal-ldpc-num}). Since its surgeries also last
$T^{(\mathrm{LDPC})}_{\mathrm{surg}} = d_{\mathrm{LDPC}} = 16$ cycles (Equation~\eqref{eq:Tsurg-ldpc}), its shots
have the same duration, and its 3.0\% fewer shots would shorten the total
runtime from 3.14 to 3.04 days.
At $p_Z=10^{-3}$, the factories add about 1.8\% to
$\epsilon_{\mathrm{total}}$ (Section~\ref{ssec:physical_resource_count}), and the
pair $(15,22)$ gives $\epsilon = 10.9\%$, just above the threshold, since
its surgeries last $T^{(\mathrm{LDPC})}_{\mathrm{surg}} = 22$ cycles, which also sets
$T_{\mathrm{cad}} = 22$. Only $(17,22)$ and $(19,22)$ then have $\epsilon<10\%$,
and the criterion selects $(17,22)$, with $\epsilon = 2.3\%$, whose repetition distance coincides with the distance
$d=17$ selected at this $p_Z$ in Section~\ref{ssec:physical_resource_count}. It uses
$25\,673$ physical qubits, against $25\,781$ for $(19,22)$, and its shots last $3.32\times10^{10}$ cycles,
about 9.24 hours, for a total runtime of 3.63 days.

For $s=8$, the average number of shots is 9.2
(Equation~\eqref{eq:Rshots}). Accounting for the per-shot failure probability
$\epsilon$, it becomes $9.2/(1-\epsilon) = 9.60$ at $p_Z = 5\times10^{-4}$,
which brings the total duration to $9.60\times7.84~\mathrm{h} \approx
75.3~\mathrm{h} \approx 3.14$ days.

\paragraph{Comparison with the repetition-cat architecture}
Table~\ref{tab:arch-comparison} compares the two architectures at
$p_Z=5\times10^{-4}$ and $p_Z=10^{-3}$. At $p_Z=5\times10^{-4}$, the LDPC-cat architecture reduces the physical qubit
count by a factor of 2.2, from $49\,649$ to $22\,593$. The exponent
register accounts for 88\% of this reduction: an exponent logical qubit
costs $12.5$ physical qubits on the LDPC-cat architecture, against 31.1 on
the repetition-cat architecture, that is $2d-1 = 29$ for its repetition
code and 2.1 for its share of the routing zone between the two exponent
zones (Equation~\eqref{eq:Gglobal}). In exchange, one shot is 13\% longer, $2.82\times10^{10}$ against $2.493\times10^{10}$ cycles. The Z-basis measurements and register moves of Section~\ref{sec:ldpc-cat-zaccess} add $2.8\times10^{9}$ cycles, whereas the adders last $2nT_{\mathrm{cad}}$ on both architectures (Equations~\eqref{eq:Tadder} and~\eqref{eq:T-adder-ldpc}). The \Lookup{}s follow the same schedule on both architectures (Equations~\eqref{eq:lookup_cost_rep} and~\eqref{eq:T-lookup-ldpc}), but their $2^w - 1$ surgeries last 16 cycles instead of 15, so that a \Lookup$(6)$ lasts 1554 against 1491 cycles, which adds $5.1\times10^{8}$ cycles. The per-shot failure probability rises from 0.43\% to 4.2\%, which adds 3.9\% to the number of shots, and the total runtime grows by 18\%, from 2.67 to 3.14 days. The space-time cost of the computation, in physical qubit-days, is therefore divided by about 1.9. At $p_Z=10^{-3}$, with $(17,22)$ against $d=17$, the physical qubit count is again divided by 2.2, from $55\,497$ to $25\,673$. One shot, however, is 28\% longer, $3.32\times10^{10}$ against $2.598\times10^{10}$ cycles, since $T^{(\mathrm{LDPC})}_{\mathrm{surg}} = d^{\star} = 22$ sets both the surgery duration and the cadence $T_{\mathrm{cad}} = 22$. The total runtime grows by 27\%, from 2.85 to 3.63 days, so that the space-time cost is divided by 1.7.


\begin{table*}[t]
\centering
\begin{tabular}{llrrrr}
\hline\hline
Subroutine & Operation & Calls & Cycles/call & Cycles & Share\\
\hline
\ComputeExponent{} & \Lookup$(6,31)$ & 4\,849\,026 & 1554 & $7.54\times10^{9}$ & 26.70\%\\
 & Q-Q \Adder$(31)$ & 4\,849\,026 & 1302 & $6.31\times10^{9}$ & 22.37\%\\
 & $M_Z(31,3)$ & 4\,849\,026 & 160 & $7.76\times10^{8}$ & 2.75\%\\
 & \Reshuffle$(31,3)$ & 4\,849\,026 & 176 & $8.53\times10^{8}$ & 3.02\%\\
\hline
\ReduceExponent{} & Q-C \Adder$(21)$ & 249\,238 & 882 & $2.20\times10^{8}$ & 0.78\%\\
 & Q-Q \Adder$(20)$ & 249\,238 & 840 & $2.09\times10^{8}$ & 0.74\%\\
 & \Reshuffle$(21,3)$ & 249\,238 & 176 & $4.39\times10^{7}$ & 0.16\%\\
 & \Reshuffle$(2,1)$ & 249\,238 & 48 & $1.20\times10^{7}$ & 0.04\%\\
\hline
\ComputeResidue{} & \Lookup$(6,20)$ (loading) & 22\,658 & 1554 & $3.52\times10^{7}$ & 0.12\%\\
 & \Lookup$(6,21)$ & 793\,030 & 1554 & $1.23\times10^{9}$ & 4.37\%\\
 & Q-Q \Adder$^\dagger(21)$ & 793\,030 & 882 & $6.99\times10^{8}$ & 2.48\%\\
 & $M_Z(21,3)$ & 793\,030 & 160 & $1.27\times10^{8}$ & 0.45\%\\
 & Q-Q \Adder$(20)$ & 793\,030 & 840 & $6.66\times10^{8}$ & 2.36\%\\
 & $M_Z(1)$ & 793\,030 & 16 & $1.27\times10^{7}$ & 0.04\%\\
 & \Reshuffle$(20,3)$ & 679\,740 & 176 & $1.20\times10^{8}$ & 0.42\%\\
 & \Reshuffle$(20,3)$ (swap) & 113\,290 & 192 & $2.18\times10^{7}$ & 0.08\%\\
 & $M_Z(20,3)$ & 113\,290 & 160 & $1.81\times10^{7}$ & 0.06\%\\
\hline
\AccumulateResidue{} & \Lookup$(5,34)$ & 90\,632 & 727 & $6.59\times10^{7}$ & 0.23\%\\
 & Q-Q \Adder$^\dagger(34)$ & 90\,632 & 1428 & $1.29\times10^{8}$ & 0.46\%\\
 & $M_Z(34,3)$ & 90\,632 & 160 & $1.45\times10^{7}$ & 0.05\%\\
 & Q-Q \Adder$(33)$ & 90\,632 & 1386 & $1.26\times10^{8}$ & 0.45\%\\
 & \Lookup$(5,33)$ & 45\,316 & 727 & $3.29\times10^{7}$ & 0.12\%\\
 & Q-Q \Comparison$(33)$ & 45\,316 & 1428 & $6.47\times10^{7}$ & 0.23\%\\
 & $M_Z(33,3)$ & 45\,316 & 160 & $7.25\times10^{6}$ & 0.03\%\\
 & \Phaseup$(5)$ & 90\,632 & 153 & $1.39\times10^{7}$ & 0.05\%\\
 & \Reshuffle$(33,3)$ & 90\,632 & 176 & $1.60\times10^{7}$ & 0.06\%\\
\hline
\ComputeResidue$^\dagger$ & \Lookup$(6,21)$ & 1\,586\,060 & 1554 & $2.46\times10^{9}$ & 8.73\%\\
 & Q-Q \Adder$^\dagger(21)$ & 1\,586\,060 & 882 & $1.40\times10^{9}$ & 4.96\%\\
 & $M_Z(21,3)$ & 1\,586\,060 & 160 & $2.54\times10^{8}$ & 0.90\%\\
 & Q-Q \Adder$(20)$ & 1\,586\,060 & 840 & $1.33\times10^{9}$ & 4.72\%\\
 & \Lookup$(6,20)$ & 793\,030 & 1554 & $1.23\times10^{9}$ & 4.37\%\\
 & Q-Q \Comparison$(20)$ & 793\,030 & 882 & $6.99\times10^{8}$ & 2.48\%\\
 & $M_Z(20,3)$ & 793\,030 & 160 & $1.27\times10^{8}$ & 0.45\%\\
 & \Phaseup$(6)$ & 1\,586\,060 & 280 & $4.44\times10^{8}$ & 1.57\%\\
 & \Reshuffle$(20,1)$ (swap) & 113\,290 & 352 & $3.99\times10^{7}$ & 0.14\%\\
 & \Reshuffle$(20,3)$ & 1\,472\,770 & 176 & $2.59\times10^{8}$ & 0.92\%\\
 & \Reshuffle$(20,3)$ (swap) & 113\,290 & 192 & $2.18\times10^{7}$ & 0.08\%\\
 & \Phaseup$(6)$ (unload) & 22\,658 & 280 & $6.34\times10^{6}$ & 0.02\%\\
\hline
\ReduceExponent$^\dagger$ & Q-C \Adder$(21)$ & 249\,238 & 882 & $2.20\times10^{8}$ & 0.78\%\\
 & Q-Q \Adder$(20)$ & 249\,238 & 840 & $2.09\times10^{8}$ & 0.74\%\\
 & \Reshuffle$(21,3)$ & 249\,238 & 176 & $4.39\times10^{7}$ & 0.16\%\\
 & \Reshuffle$(2,1)$ & 249\,238 & 48 & $1.20\times10^{7}$ & 0.04\%\\
\hline
Last \Phaseup{} & \Phaseup$(6)$ & 214 & 280 & $5.99\times10^{4}$ & 0.00\%\\
\hline\hline
\end{tabular}
\caption{Breakdown of cycle counts by operation for the factorization of
an RSA-2048 number on the LDPC-cat architecture, for
$p_Z = 5\times10^{-4}$ and $(d_{\mathrm{rep}},d_{\mathrm{LDPC}})=(13,16)$, in the format of
Table~\ref{tab:physical_cycle_breakdown}. The total duration is $T_{\mathrm{shot}} =
2.82\times10^{10}$ cycles. Lattice surgeries involving the LDPC layer last
$T^{(\mathrm{LDPC})}_{\mathrm{surg}} = 16$ cycles, loadings of
factory outputs $d_{\mathrm{rep}} = 13$ cycles, and adders run at the cadence
$T_{\mathrm{cad}} = 21$
(Equations~\eqref{eq:T-adder-ldpc} to~\eqref{eq:T-phaseup-ldpc}).
$M_Z(x,k)$ is the Z-basis measurement of $x$ logical qubits of the
LDPC layer, and $\Reshuffle(x,k)$ a move of $x$ logical qubits
within the LDPC layer: drift compensation after the adders, role swap of
the two $\ell$-qubit registers, and moves of the underflow qubit and of the
next bit of $S_i$ in \ReduceExponent. In both cases, the qubits occupy one
column out of $k$, and $M_Z(x)$ stands for $M_Z(x,1)$.
$M_Z(x,k)$ lasts $T_{M_Z}(x,k)$ cycles and
$\Reshuffle(x,k)$ lasts $T_{\Reshuffle}(x,k)$ cycles
(Section~\ref{sec:ldpc-cat-zaccess}), with all $X$-parities in parallel
for a drift compensation and $n_{\mathrm{lane}} = 10$ lanes for a move
between zones, marked ``swap''. For clarity, the table omits the
$\mathrm{PC}X$ and $P_{\ket{+}}$ slices, which contribute 0.34\% of
$T_{\mathrm{shot}}$ and are included in the total duration and in the volumes. The
moves between subroutines (0.21\% of $T_{\mathrm{shot}}$) and the final
inverse quantum Fourier transform are not included. Cycles/call refers to
the number of cycles of one call of an operation, and Cycles to the
total number of cycles of all its calls.}
\label{tab:ldpc-cycles}
\end{table*}

\begin{table*}[t]
\centering
\begin{tabular}{rrccccccc}
\hline\hline
$d_{\mathrm{rep}}$ & $d_{\mathrm{LDPC}}$ &
$\mathrm{Vol}^{(\mathrm{eff,rep})}_{\mathrm{logical}}$ &
$\mathrm{Vol}^{(\mathrm{eff,LDPC})}_{\mathrm{logical}}$ &
$\epsilon^{(\mathrm{rep})}_{\mathrm{logical}}$ &
$\epsilon^{(\mathrm{LDPC})}_{\mathrm{logical}}$ &
$\epsilon_{\mathrm{total}}$ & $\epsilon$ & \\
\hline
\multicolumn{9}{c}{$p_Z = 10^{-3}$}\\
11 & 12 & $6.4\times10^{11}$ & $3.0\times10^{12}$ & 1.00 & 1.00 & 1.00 & 1.00 &  \\
   & 16 & $7.4\times10^{11}$ & $2.5\times10^{12}$ & 1.00 & $9.19\times10^{-1}$ & 1.00 & 1.00 &  \\
   & 22 & $9.0\times10^{11}$ & $2.2\times10^{12}$ & 1.00 & $3.89\times10^{-3}$ & 1.00 & 1.00 &  \\
13 & 12 & $4.7\times10^{11}$ & $3.1\times10^{12}$ & $9.62\times10^{-1}$ & 1.00 & 1.00 & 1.00 &  \\
   & 16 & $5.4\times10^{11}$ & $2.5\times10^{12}$ & $9.76\times10^{-1}$ & $9.19\times10^{-1}$ & 1.00 & 1.00 &  \\
   & 22 & $6.5\times10^{11}$ & $2.2\times10^{12}$ & $9.89\times10^{-1}$ & $3.89\times10^{-3}$ & $9.89\times10^{-1}$ & $9.89\times10^{-1}$ &  \\
15 & 12 & $3.8\times10^{11}$ & $3.2\times10^{12}$ & $6.95\times10^{-2}$ & 1.00 & 1.00 & 1.00 &  \\
   & 16 & $4.1\times10^{11}$ & $2.5\times10^{12}$ & $7.47\times10^{-2}$ & $9.19\times10^{-1}$ & $9.25\times10^{-1}$ & $9.27\times10^{-1}$ &  \\
   & 22 & $4.9\times10^{11}$ & $2.2\times10^{12}$ & $8.92\times10^{-2}$ & $3.89\times10^{-3}$ & $9.27\times10^{-2}$ & $1.09\times10^{-1}$ &  \\
17 & 12 & $3.1\times10^{11}$ & $3.4\times10^{12}$ & $1.61\times10^{-3}$ & 1.00 & 1.00 & 1.00 &  \\
   & 16 & $3.3\times10^{11}$ & $2.6\times10^{12}$ & $1.69\times10^{-3}$ & $9.24\times10^{-1}$ & $9.24\times10^{-1}$ & $9.26\times10^{-1}$ &  \\
   & 22 & $3.9\times10^{11}$ & $2.2\times10^{12}$ & $1.99\times10^{-3}$ & $3.89\times10^{-3}$ & $5.87\times10^{-3}$ & $2.34\times10^{-2}$ & $\blacktriangleleft$ \\
19 & 12 & $2.6\times10^{11}$ & $3.6\times10^{12}$ & $3.64\times10^{-5}$ & 1.00 & 1.00 & 1.00 &  \\
   & 16 & $2.8\times10^{11}$ & $2.7\times10^{12}$ & $3.83\times10^{-5}$ & $9.33\times10^{-1}$ & $9.33\times10^{-1}$ & $9.34\times10^{-1}$ &  \\
   & 22 & $3.1\times10^{11}$ & $2.2\times10^{12}$ & $4.29\times10^{-5}$ & $3.89\times10^{-3}$ & $3.93\times10^{-3}$ & $2.15\times10^{-2}$ &  \\
\hline
\multicolumn{9}{c}{$p_Z = 5\times10^{-4}$}\\
11 & 12 & $6.4\times10^{11}$ & $3.0\times10^{12}$ & $9.22\times10^{-1}$ & 1.00 & 1.00 & 1.00 &  \\
   & 16 & $7.4\times10^{11}$ & $2.5\times10^{12}$ & $9.49\times10^{-1}$ & $9.79\times10^{-3}$ & $9.49\times10^{-1}$ & $9.49\times10^{-1}$ &  \\
   & 22 & $9.0\times10^{11}$ & $2.2\times10^{12}$ & $9.72\times10^{-1}$ & $1.90\times10^{-6}$ & $9.72\times10^{-1}$ & $9.72\times10^{-1}$ &  \\
13 & 12 & $4.7\times10^{11}$ & $3.1\times10^{12}$ & $2.65\times10^{-2}$ & 1.00 & 1.00 & 1.00 &  \\
   & 16 & $5.4\times10^{11}$ & $2.5\times10^{12}$ & $3.00\times10^{-2}$ & $9.79\times10^{-3}$ & $3.95\times10^{-2}$ & $4.16\times10^{-2}$ & $\blacktriangleleft$ \\
   & 22 & $6.5\times10^{11}$ & $2.2\times10^{12}$ & $3.60\times10^{-2}$ & $1.90\times10^{-6}$ & $3.60\times10^{-2}$ & $3.81\times10^{-2}$ &  \\
15 & 12 & $3.8\times10^{11}$ & $3.2\times10^{12}$ & $2.98\times10^{-4}$ & 1.00 & 1.00 & 1.00 &  \\
   & 16 & $4.1\times10^{11}$ & $2.5\times10^{12}$ & $3.20\times10^{-4}$ & $9.79\times10^{-3}$ & $1.01\times10^{-2}$ & $1.22\times10^{-2}$ &  \\
   & 22 & $4.9\times10^{11}$ & $2.2\times10^{12}$ & $3.86\times10^{-4}$ & $1.90\times10^{-6}$ & $3.87\times10^{-4}$ & $2.54\times10^{-3}$ &  \\
17 & 12 & $3.1\times10^{11}$ & $3.4\times10^{12}$ & $3.35\times10^{-6}$ & 1.00 & 1.00 & 1.00 &  \\
   & 16 & $3.3\times10^{11}$ & $2.6\times10^{12}$ & $3.52\times10^{-6}$ & $1.00\times10^{-2}$ & $1.00\times10^{-2}$ & $1.22\times10^{-2}$ &  \\
   & 22 & $3.9\times10^{11}$ & $2.2\times10^{12}$ & $4.13\times10^{-6}$ & $1.90\times10^{-6}$ & $6.03\times10^{-6}$ & $2.16\times10^{-3}$ &  \\
19 & 12 & $2.6\times10^{11}$ & $3.6\times10^{12}$ & $3.81\times10^{-8}$ & 1.00 & 1.00 & 1.00 &  \\
   & 16 & $2.8\times10^{11}$ & $2.7\times10^{12}$ & $4.01\times10^{-8}$ & $1.05\times10^{-2}$ & $1.05\times10^{-2}$ & $1.26\times10^{-2}$ &  \\
   & 22 & $3.1\times10^{11}$ & $2.2\times10^{12}$ & $4.50\times10^{-8}$ & $1.90\times10^{-6}$ & $1.95\times10^{-6}$ & $2.15\times10^{-3}$ &  \\
\hline\hline
\end{tabular}
\caption{Effective logical volumes and logical error probabilities per
shot for the factorization of an RSA-2048 number on the LDPC-cat
architecture, as a function of the distances $d_{\mathrm{rep}}$ and $d_{\mathrm{LDPC}}$, for
$p_Z = 10^{-3}$ and $p_Z = 5\times10^{-4}$.
$\mathrm{Vol}^{(\mathrm{eff,rep})}_{\mathrm{logical}}$ and
$\mathrm{Vol}^{(\mathrm{eff,LDPC})}_{\mathrm{logical}}$ are given by
Equations~\eqref{eq:vol-eff-rep-ldpc} and~\eqref{eq:vol-eff-ldpc}, in
logical cycles of $d_{\mathrm{rep}}$ and $d_{\mathrm{LDPC}}$ physical cycles respectively, which
is why $\mathrm{Vol}^{(\mathrm{eff,LDPC})}_{\mathrm{logical}}$ decreases as
$d_{\mathrm{LDPC}}$ grows. $\epsilon^{(\mathrm{rep})}_{\mathrm{logical}}$,
$\epsilon^{(\mathrm{LDPC})}_{\mathrm{logical}}$ and
$\epsilon_{\mathrm{total}}$ are given by
Equations~\eqref{eq:eps-layers} and~\eqref{eq:eps-total-ldpc}, with
Equations~\eqref{eq:plogical} and~\eqref{eq:plogical-ldpc}, and exclude
the factories, which $\epsilon$ includes, with the factory failure
probabilities of Section~\ref{ssec:physical_resource_count}. All pairs use the
surgery durations of Equation~\eqref{eq:Tsurg-ldpc},
$T_{\mathrm{cad}} = \max(T^{(\mathrm{LDPC})}_{\mathrm{surg}},21)$, that is 21 for $d_{\mathrm{LDPC}}\le16$ and 22 for
$d_{\mathrm{LDPC}}=22$ (Equation~\eqref{eq:Tcad-ldpc}), and $n_{\mathrm{buf}}^{(\Adder)} = 3$
(Equation~\eqref{eq:Tbuf-ldpc}). An entry of 1.00 denotes a probability above
0.995. The pairs marked by $\blacktriangleleft$ are selected as the pairs
with the fewest physical qubits among those with $\epsilon < 10\%$
(criterion of Section~\ref{sec:C7-budget}): $(17,22)$ at $p_Z=10^{-3}$ and $(13,16)$
at $p_Z=5\times10^{-4}$.}
\label{tab:ldpc-scan}
\end{table*}

\begin{table*}[t]
\centering
\begin{tabular}{lcccc}
\hline\hline
 & \multicolumn{2}{c}{$p_Z = 5\times10^{-4}$} & \multicolumn{2}{c}{$p_Z = 10^{-3}$}\\
 & Repetition-cat & LDPC-cat & Repetition-cat & LDPC-cat\\
\hline
Distances & $d=15$ & $(13,16)$ & $d=17$ & $(17,22)$\\
Physical qubits & $49\,649$ & $22\,593$ & $55\,497$ & $25\,673$\\
$T_{\mathrm{shot}}$ (error-correction cycles) & $2.493\times10^{10}$ & $2.82\times10^{10}$ & $2.598\times10^{10}$ & $3.32\times10^{10}$\\
$T_{\mathrm{shot}}$ (hours) & 6.92 & 7.84 & 7.22 & 9.24\\
Physical volume & $1.17\times10^{15}$ & $4.2\times10^{14}$ & $1.37\times10^{15}$ & $5.5\times10^{14}$\\
(qubits $\times$ error-correction cycles) & & $1.7\times10^{14}$ (rep) & & $2.2\times10^{14}$ (rep)\\
 & & $2.4\times10^{14}$ (LDPC) & & $3.4\times10^{14}$ (LDPC)\\
Eff.\ logical volume, without factories & $2.68\times10^{12}$ & $5.4\times10^{11}$ (rep) & $2.43\times10^{12}$ & $3.9\times10^{11}$ (rep)\\
 & & $2.5\times10^{12}$ (LDPC) & & $2.2\times10^{12}$ (LDPC)\\
Eff.\ logical volume, with factories & $2.79\times10^{12}$ & $7.0\times10^{11}$ (rep) & $2.52\times10^{12}$ & $5.0\times10^{11}$ (rep)\\
 & & $2.5\times10^{12}$ (LDPC) & & $2.2\times10^{12}$ (LDPC)\\
Eff.\ logical qubits, without factories & $1615$ & $1681$ & $1592$ & $1631$\\
 & & ($247$ rep $+$ $1434$ LDPC) & & ($197$ rep $+$ $1434$ LDPC)\\
Eff.\ logical qubits, with factories & $1680$ & $1756$ & $1649$ & $1688$\\
 & & ($322$ rep $+$ $1434$ LDPC) & & ($254$ rep $+$ $1434$ LDPC)\\
Logical error/shot & 0.211\% & 4.0\% & 1.25\% & 0.59\%\\
Failure/shot, $\epsilon$ & 0.43\% & 4.2\% & 3.0\% & 2.3\%\\
Number of shots & 9.24 & 9.60 & 9.48 & 9.42\\
Total runtime (days) & 2.67 & 3.14 & 2.85 & 3.63\\
\hline\hline
\end{tabular}
\caption{Physical resources of one RSA-2048 factorization at
$p_Z = 5\times10^{-4}$ and $p_Z = 10^{-3}$ on the repetition-cat
architecture (Section~\ref{ssec:physical_resource_count}, Table~\ref{tab:rep_cat_recap})
and on the LDPC-cat architecture (Section~\ref{sec:ldpc-cat-b}), at the
distances selected at each $p_Z$. The failure probability per shot
$\epsilon$ includes the factories, and the number of shots is
$9.2/(1-\epsilon)$. Physical qubits include the
factories. The physical volume of one shot, in physical qubits times error-correction cycles, excludes the factories (Equations~\eqref{eq:Volphys}, \eqref{eq:vol-phys-rep-ldpc} and~\eqref{eq:vol-phys-ldpc}); on the LDPC-cat architecture, it is the sum over the two layers. The effective logical volume without factories is the one that enters the logical error probability, the factories being counted through their output infidelities; the volume with factories counts them as $n_{\mathrm{fac}}/(2d-1)$ effective logical qubits during the whole shot, with $d = d_{\mathrm{rep}}$ on the LDPC-cat architecture, where they belong to the repetition layer. The effective number of logical qubits is the effective logical volume divided by the number of logical cycles $T_{\mathrm{shot}}/d$ (Equation~\eqref{eq:neff}); on the LDPC-cat architecture, it is the sum over the two layers, each counted in its own logical cycles of $d_{\mathrm{rep}}$ and $d_{\mathrm{LDPC}}$ error-correction cycles, the LDPC layer contributing $k_L = 1434$ by construction (Equation~\eqref{eq:vol-eff-ldpc}).}
\label{tab:arch-comparison}
\end{table*}

\subsection{Physical cat qubits and bit-flip assumptions}\label{ssec:cat_qubits_bitflips}

\subsubsection{Cat qubits: encoding, stabilization, operations, and bit flips}\label{sssec:cat_qubits_physics}

In a cat qubit, the information is stored in a superconducting microwave resonator~\cite{mirrahimi2014dynamically}.
The computational states of the cat qubit are $\ket{0}\simeq\ket{\alpha}$ and $\ket{1}\simeq\ket{-\alpha}$, where the coherent states $\ket{\pm\alpha}$ are eigenstates of the annihilation operator $\hat a$ of the resonator mode, $\hat a\ket{\pm\alpha}=\pm\alpha\ket{\pm\alpha}$, and $\bar n=|\alpha|^2$ is their mean photon number.
The dominant error channel of the resonator is single-photon loss, with jump operator $\hat a$ and rate $\kappa_1$.\footnote{Elsewhere in this paper, $\kappa$ without subscript is the truncation length of Appendix~\ref{sec:algo_overview}, that is, the size of the accumulation register.}
In the basis $\{\ket{\alpha},\ket{-\alpha}\}$, $\hat a$ acts as a phase flip, $\hat a\,(a_0\ket{\alpha}+a_1\ket{-\alpha})=\alpha\,(a_0\ket{\alpha}-a_1\ket{-\alpha})$, while a bit flip requires a transition between two states separated by $2|\alpha|$ in phase space.
Once the manifold spanned by $\ket{\pm\alpha}$ is stabilized (see below), the phase-flip and bit-flip rates scale as
\begin{equation}\label{eq:cat_rates}
    \Gamma_X \simeq 2\kappa_1 \bar n, \qquad \Gamma_Z \propto e^{-c\,\bar n},
\end{equation}
where $c>0$.
Here, the subscript of $\Gamma$ refers to the observable whose expectation value decays, $\langle X\rangle$ for phase flips and $\langle Z\rangle$ for bit flips.  
In contrast, for the Pauli error probabilities used in the rest of this paper, the subscript denotes the Pauli error itself, and $p_X$ and $p_Z$ refer to bit flips and phase flips, respectively.
The factor $2$ in $\Gamma_X$ arises because photon losses occur at the rate $\kappa_1\bar n$ and each of them changes the sign of $\langle X\rangle$, which then decays as $e^{-2\kappa_1\bar n t}$.
We call $T_Z=1/\Gamma_Z$ the bit-flip time.
After a time $t$, a qubit prepared in $\ket{0}$ is found in $\ket{1}$ with probability $(1-e^{-t/T_Z})/2$, about $t/(2T_Z)$ for $t\ll T_Z$.
The bit-flip rate is thus exponentially suppressed with $\bar n$, while the phase-flip rate increases only linearly~\cite{mirrahimi2014dynamically,lescanne2020exponential,guillaud2019repetition}.
With such a noise bias, only phase flips need to be corrected actively, which we do with repetition codes~\cite{guillaud2019repetition} and LDPC-cat codes~\cite{Ruiz_2025}.

The cat qubit manifold is stabilized by an engineered two-photon dissipation~\cite{mirrahimi2014dynamically,leghtas2015confining,lescanne2020exponential}.
The resonator is coupled, through a nonlinear element, to a strongly dissipative mode, the buffer.
This nonlinear element converts pairs of resonator photons into single buffer photons, and conversely.
After adiabatic elimination of the buffer, the resonator undergoes the two-photon dissipation $\mathcal D[\hat a^2-\alpha^2]$ at a rate $\kappa_2$~\cite{mirrahimi2014dynamically,leghtas2015confining,guillaud2023quantum}, where the $\alpha^2$ term comes from a resonant drive on the buffer~\cite{lescanne2020exponential,reglade2024quantum}.
The steady states of this dissipation are the coherent states $\ket{\pm\alpha}$ and their superpositions, and any state of the resonator relaxes to the cat qubit manifold and remains in it.
The mean photon number $\bar n$, and with it the noise bias, is set by the amplitude of the buffer drive and can therefore be tuned in situ.

The physical operations of Section~\ref{sssec:layout_connectivity} have bias-preserving implementations~\cite{guillaud2019repetition,GouzienCatQubits23}.
These operations are the preparations in $\ket{0}$ and $\ket{+}$, the \textsc{Cnot} gates between neighboring qubits, and the measurements in the $X$ and $Z$ bases.
All the logical operations of this paper are built from these physical operations, including the lattice surgery on repetition codes~\cite{guillaud2019repetition,GouzienCatQubits23} and the computing layer of the LDPC-cat codes~\cite{Ruiz_2025}.
The only exceptions are the factories that prepare the resource states $\ket{\CXCXX}$ and $\ket{i}$~\cite{FeMoco2026prep,Ruiz_2026}, which also require rotations by $\pi/2$ and $\pi/4$ around the $X$ axis of the Bloch sphere.
These rotations do not need to be bias-preserving, since the bit flips they induce can be detected in the factory~\cite{Ruiz_2026}.
A rotation by $\pi/2$ around the $X$ axis has been realized experimentally with a holonomic gate~\cite{albert2016holonomic}, as part of the tomography protocol of~\cite{reglade2024quantum}.

As the physical realizations of the gates on cat qubits evolve quickly, it is simpler to use a generic phase-flip noise model, against which a specific physical implementation can then be compared.
Indeed, many gate designs have been proposed over the years~\cite{mirrahimi2014dynamically,guillaud2019repetition,puri2020bias,xu2022engineering,chamberland2022building,gautier2022combined,gautier2023designing,leregent2023high}, and recent proposals include \textsc{Cnot} gates between a cat qubit and a transmon~\cite{hann2025hybrid} and between two cat qubits~\cite{ye2026bias}.
The \textsc{Cnot} gates of the syndrome extraction can even be replaced by quantum non-demolition measurements of multi-qubit $Z$ operators~\cite{vuillot2026quantum,ruiz2026biased}.
We use the circuit-level model of~\cite[Supplementary Table~1(b)]{Ruiz_2025}, recalled in Table~\ref{tab:generic_phaseflip_model}, in which every noisy operation has the same phase-flip probability $p_Z$.
The correspondence between $p_Z$ and the hardware parameters is discussed in~\cite[Methods and Supplementary Table~1(c)]{Ruiz_2025}.

\begin{table}[ht]
\centering
\begin{tabular}{@{}lcc@{}}
\hline\hline
Operation & Error & Probability \\
\hline
Preparation in $\ket{+}$ & $Z$ & $p_Z$ \\
$X$-basis measurement & outcome flipped & $p_Z$ \\
\textsc{Cnot} & $Z_c$, $Z_t$ or $Z_cZ_t$ & $p_Z/3$ each \\
Idle, during any operation & $Z$ & $p_Z$ \\
\hline\hline
\end{tabular}
\caption{Generic circuit-level phase-flip noise model used throughout this work, with the single parameter $p_Z$ (Section~\ref{sssec:layout_connectivity}).
Every noisy operation is modeled as a perfect operation followed by a stochastic error, with the probability given in the third column.
For the $X$-basis measurement, this error flips the measurement outcome.
Every qubit that idles while an operation is performed undergoes a $Z$ error with probability $p_Z$.
Preparations in $\ket{0}$ and $Z$-basis measurements are insensitive to phase flips.}
\label{tab:generic_phaseflip_model}
\end{table}

In theory, the bit-flip time $T_Z$ of an idle cat qubit grows exponentially with $\bar n$ (Equation~\eqref{eq:cat_rates}).
During a computation, however, the circuit-level models of~\cite{leregent2023high,GouzienCatQubits23,Ruiz_2025} attribute the dominant bit flips to the \textsc{Cnot} gate. 
The probability per gate of a bit flip on either qubit is
\begin{equation}\label{eq:cnot_bitflip}
    p_X^{\mathrm{CX}} = 0.5\,e^{-2\bar n},
\end{equation}
obtained by fitting numerical simulations of the gate~\cite{leregent2023high,GouzienCatQubits23} and extrapolated to larger $\bar n$~\cite{GouzienCatQubits23}.
What matters for the estimates of this appendix is therefore not just the idle bit-flip time.
The gates, measurements and preparations must also preserve this bit-flip time, on every qubit of a large chip.

\subsubsection{Experimental state of the art}\label{sssec:cat_qubits_experiments}

In this section, we briefly review the measurements of the bit-flip time $T_Z$ of dissipative cat qubits.
Two-photon dissipation was first demonstrated in~\cite{leghtas2015confining}, where it confined the state of the resonator to the cat qubit manifold.
The exponential suppression of bit flips was then observed in~\cite{lescanne2020exponential}, up to a bit-flip time of about \SI{1}{\milli\second}, $300$ times the lifetime of the undriven resonator.
Above $\bar n\approx5$, this bit-flip time saturated, which the authors attributed to thermal excitations of the transmon used for the tomography.
In a later device without a transmon~\cite{berdou2023hundred}, the bit flips were detected by monitoring the field radiated by the resonator.
Without quantum control, the mean time between bit flips in this device reached about \SI{100}{\second} at $\bar n\approx40$, or a bit-flip time $T_Z$ of about \SI{50}{\second}.
With quantum control, the bit-flip time reached \SI{0.3}{\second} with an autoparametric stabilization~\cite{marquet2024autoparametric}.
In a device with a transmon-free tomography, it reached \SI{15}{\second} at rest at $\bar n=11.3$, and remained above \SI{10}{\second} under a continuous single-qubit drive~\cite{reglade2024quantum}.
Squeezed cat states reached a bit-flip time of \SI{22}{\second} at $\bar n=4.1$~\cite{rousseau2025squeezing}.
Note that all these bit-flip times were measured on a single cat qubit.

Three properties remain to be demonstrated.
First, the bit-flip time must keep growing exponentially with $\bar n$, with quantum control, up to about $18$ photons (see Section~\ref{sssec:bitflip_assumptions}).
This exponential growth was followed up to $\bar n\approx40$ only without quantum control~\cite{berdou2023hundred}, and the bit-flip time saturated in~\cite{lescanne2020exponential,rousseau2025squeezing} and, by $\bar n\approx20$, in~\cite{marquet2024autoparametric}.
Second, all the operations, and in particular the \textsc{Cnot} gates, must preserve this bit-flip time.
Bit-flip suppression has been observed during single-qubit $Z$ rotations~\cite{reglade2024quantum,rousseau2025squeezing} and, at small $\bar n$, during \textsc{Cnot} gates between a transmon and a cat qubit~\cite{putterman2025hardware}.
Since the operations add new bit-flip mechanisms, such as the one of Equation~\eqref{eq:cnot_bitflip}, they can only shorten the bit-flip time measured at rest.
Last, these two properties must hold on a chip with many cat qubits.
We discuss the requirement on this bit-flip time in Section~\ref{sssec:bitflip_assumptions}, and we present a mitigation strategy in Section~\ref{ssec:elevator_code}.

\subsubsection{Comparison with the bit-flip assumptions of the main text}\label{sssec:bitflip_assumptions}

\begin{table*}[t]
\centering
\setlength{\tabcolsep}{5pt}
\begin{tabular}{@{}llllcccccc@{}}
\hline\hline
Computation & $p_Z$ & Architecture & Distances & $n_{\mathrm{total}}$ & $T_{\mathrm{shot}}$ & $Q$ & $p_{\mathrm{bf}}^{\max}$ & $T_Z^{\min}$ & $\bar n_{\min}$ \\
\hline
RSA-2048 & $5\times10^{-4}$ & Repetition-cat & $d = 15$ & $49\,649$ & \SI{6.92}{\hour} & $1.24\times10^{15}$ & $8.08\times10^{-17}$ & $196$~years & $18.2$ \\
 & & LDPC-cat & $(13,16)$ & $22\,593$ & \SI{7.84}{\hour} & $6.38\times10^{14}$ & $1.57\times10^{-16}$ & $101$~years & $17.8$ \\
RSA-2048 & $10^{-3}$ & Repetition-cat & $d = 17$ & $55\,497$ & \SI{7.22}{\hour} & $1.44\times10^{15}$ & $6.93\times10^{-17}$ & $228$~years & $18.3$ \\
 & & LDPC-cat & $(17,22)$ & $25\,673$ & \SI{9.24}{\hour} & $8.54\times10^{14}$ & $1.17\times10^{-16}$ & $135$~years & $18.0$ \\
ECDLP-256 & $5\times10^{-4}$ & LDPC-cat & $(13,16)$ & $29\,588$ & \SI{60}{\minute} & $1.07\times10^{14}$ & $9.39\times10^{-16}$ & $16.9$~years & $17.0$ \\
FFDLP-512 & $5\times10^{-4}$ & Repetition-cat & $d = 11$ & $9\,336$ & \SI{2.28}{\minute} & $1.28\times10^{12}$ & $7.84\times10^{-14}$ & $2.4$~months & $14.7$ \\
 & & LDPC-cat & $(11,12)$ & $6\,656$ & \SI{2.51}{\minute} & $1.00\times10^{12}$ & $9.98\times10^{-14}$ & $1.9$~months & $14.6$ \\
\hline\hline
\end{tabular}
\caption{Requirement on bit flips of Equation~\eqref{eq:bitflip_requirement}, for a budget $\epsilon_{\mathrm{bf}} = \SI{10}{\percent}$ per shot, assuming that any bit flip on any physical qubit makes the shot fail.
The configurations are those selected in Sections~\ref{ssec:physical_resource_count}, \ref{sec:ldpc-cat-b}, \ref{ssec:FFDLP-512} and~\ref{ssec:ECDLP_tradeoffs}, with the distance $d$ on the repetition-cat architecture and the distances $(d_{\mathrm{rep}},d_{\mathrm{LDPC}})$ on the LDPC-cat architecture.
$n_{\mathrm{total}}$ is the number of physical qubits, including the factories and the routing qubits, and $T_{\mathrm{shot}}$ is the duration of one shot, rounded up to \SI{60}{\minute} for ECDLP-256.
$Q = n_{\mathrm{total}}\,T_{\mathrm{shot}}$ is the number of physical qubit-cycles of one shot, with $T_{\mathrm{shot}}$ in error-correction cycles of $T_{\mathrm{round}} = \SI{1}{\micro\second}$.
$p_{\mathrm{bf}}^{\max}$ is the maximum bit-flip probability per qubit and per cycle, and $T_Z^{\min}$ is the minimum bit-flip time of an idle qubit.
$\bar n_{\min}$ is the minimum mean photon number given by Equation~\eqref{eq:cnot_bitflip} with the equivalent of one \textsc{Cnot} gate per qubit and per cycle.}
\label{tab:bitflip_requirements}
\end{table*}

The results of the main text assume that no bit flip occurs during the computation.
In practice, this means that the bit-flip error probability is negligible.
In this section, we quantify what negligible means, and we estimate the bit-flip time required for each computation of this paper.

The codes of this work do not correct bit flips, and $X_L$ has weight one on the repetition codes and on the LDPC-cat codes (Section~\ref{sssec:error_correcting_codes}).
Therefore, a single physical bit-flip error corresponds to a logical bit-flip error.
Assuming pessimistically that any bit flip causes a logical error (some bit flips are harmless, for instance on a qubit prepared in an eigenstate of $X$), we count any bit flip on any physical qubit during a shot as a failure of this shot.

We denote by $p_{\mathrm{bf}}$ the probability of a bit flip, that is, of an $X$ or $Y$ error, per physical qubit and per error-correction cycle.
It is related to the bit-flip time by $p_{\mathrm{bf}}\simeq T_{\mathrm{round}}/(2T_Z)$ (Section~\ref{sssec:cat_qubits_physics}), where $T_{\mathrm{round}}=\SI{1}{\micro\second}$ is the duration of an error-correction cycle.
One shot involves $Q=n_{\mathrm{total}}\,T_{\mathrm{shot}}$ physical qubit-cycles, where $n_{\mathrm{total}}$ is the number of physical qubits, including the factories and the routing qubits, and $T_{\mathrm{shot}}$ is the duration of a shot in error-correction cycles.
A shot thus fails because of bit flips with probability $1-(1-p_{\mathrm{bf}})^{Q}\simeq p_{\mathrm{bf}}\,Q$.

As for logical phase-flip errors, a shot that fails because of bit flips is detected and retried (Section~\ref{par:composition-physical-error}).
A failure probability $\epsilon_{\mathrm{bf}}$ per shot thus multiplies the expected number of shots by $1/(1-\epsilon_{\mathrm{bf}})$.
We choose $\epsilon_{\mathrm{bf}}=\SI{10}{\percent}$, the threshold that we also impose on the per-shot failure probability due to phase flips (Section~\ref{ssec:physical_resource_count}).
Indeed, increasing $\bar n$ suppresses bit flips at the cost of more phase flips (Equation~\eqref{eq:cat_rates}), and we expect $\bar n$ to be chosen in practice such that logical phase flips and logical bit flips are balanced.
This budget lengthens the runtime by \SI{11}{\percent}, and a bit-flip time ten times longer than required would bring this overhead down to about \SI{1}{\percent}.
The code distances selected in Section~\ref{ssec:physical_resource_count} remain unchanged, since this threshold concerns only the failures due to phase flips, while bit flips multiply the number of shots by a separate factor.
The budget $\epsilon_{\mathrm{bf}}$ then gives the requirement
\begin{equation}\label{eq:bitflip_requirement}
  \begin{gathered}
    p_{\mathrm{bf}} \le p_{\mathrm{bf}}^{\max} = \frac{\epsilon_{\mathrm{bf}}}{n_{\mathrm{total}}\,T_{\mathrm{shot}}} , \\
    T_Z \ge T_Z^{\min} = \frac{n_{\mathrm{total}}\,T_{\mathrm{shot}}\,T_{\mathrm{round}}}{2\,\epsilon_{\mathrm{bf}}} .
  \end{gathered}
\end{equation}
This requirement concerns the bit-flip time $T_Z$ of an idle cat qubit, and it assumes that the gates, measurements and preparations preserve this bit-flip time with $T_{shot}$ expressed as the number of error-correction cycles per shot.
As a rough guide, we translate it into a minimum mean photon number $\bar n_{\min}=\frac{1}{2}\ln[1/(2p_{\mathrm{bf}}^{\max})]$ with Equation~\eqref{eq:cnot_bitflip}, assuming the equivalent of one \textsc{Cnot} gate per qubit and per cycle.
For RSA-2048, requiring the same budget for the whole computation instead of per shot would make this requirement only nine to ten times stricter, and would increase $\bar n_{\min}$ by about one photon.

For RSA-2048 on the LDPC-cat architecture at $p_Z=5\times10^{-4}$, with $22\,593$ physical qubits, the bit-flip probability per qubit and per cycle must remain below about $1.6\times10^{-16}$ (see Table~\ref{tab:bitflip_requirements} for the other configurations).
In other words, a bit flip should occur on the whole chip less than once every ten shots on average.
The bit-flip time $T_Z$ must then exceed about a hundred years, which requires about $18$ photons with the rough guide above.
Since $\bar n_{\min}$ depends only logarithmically on $Q$, it changes little between computations.
For instance, FFDLP-512, whose shots last only a few minutes, still requires a bit-flip time of about two months and about $15$ photons.

Experiments on dissipative cat qubits have already reached a noise bias of more than seven orders of magnitude, with bit-flip and phase-flip times of \SI{15}{\second} and \SI{490}{\nano\second} at $\bar n=11.3$~\cite{reglade2024quantum}, and of \SI{22}{\second} and \SI{1.3}{\micro\second} at $\bar n=4.1$~\cite{rousseau2025squeezing}.
The bit-flip time of about a hundred years required for RSA-2048 on the LDPC-cat architecture is still more than eight orders of magnitude longer than the bit-flip times measured with quantum control (Section~\ref{sssec:cat_qubits_experiments}).
Moreover, the bit-flip time eventually saturates in current experiments, and the mechanisms that limit it are not yet established.
Candidate mechanisms include the excitation of an ancillary transmon~\cite{lescanne2020exponential,marquet2024autoparametric}, the dephasing of the resonator and the thermalization of the buffer~\cite{marquet2024autoparametric}, and high-energy particle impacts~\cite{berdou2023hundred,rousseau2025squeezing}.
It is therefore not certain that the bit-flip time can be extended by another eight orders of magnitude.
In case it cannot, we discuss and quantify a mitigation strategy in the next section.

\subsubsection{Additional layer of bit-flip protection}\label{ssec:elevator_code}

In all the resource estimates of this paper, we assumed that bit flips are rare enough to require no active error correction, and we relied only on classical codes that correct phase flips, the repetition codes and the LDPC-cat codes~\cite{Ruiz_2025}.
As shown in Section~\ref{sssec:bitflip_assumptions}, this requires a bit-flip probability below about $8\times10^{-17}$ per qubit and per cycle on the repetition-cat architecture at $p_Z=5\times10^{-4}$, for a failure probability of \SI{10}{\percent} per shot.
In this section, we estimate the number of physical qubits required when the bit-flip probability is small but not zero.
More precisely, we assume a bit-flip probability $p_X=10^{-11}$ per fault location in the circuit-level noise model of~\cite[Table~I]{shanahan2026elevator}, whose phase flips follow the model of Table~\ref{tab:generic_phaseflip_model}.

To estimate the overhead of correcting these bit flips, we use the Elevator codes of~\cite{shanahan2026elevator}, which have asymmetric distances.
These codes concatenate inner repetition codes of distance $d_Z$, which correct phase flips, into classical outer codes that correct bit flips, using native transversal operations.
We use the $[15,9,3]$ outer code, in which $15$ inner repetition codes encode $9$ logical qubits, with $d_X=3$ and $d_Z$ the distance of the chosen inner repetition codes.
The $[15,9,3]$ Elevator code has been simulated as a quantum memory to obtain a logical error model.
We take the phase-flip model from~\cite{shanahan2026elevator}.
The bit-flip model comes from new memory simulations performed with the Tesseract decoder~\cite{beni2025tesseract}.
These simulations use the same noise model as~\cite{shanahan2026elevator}, with $d_Z=9$, $11$, $13$ and $15$, $10^{6}$ shots per point and $10^{-6}<p_X<5\times10^{-5}$.
The resulting logical error models, per logical qubit and per $d_Z$ rounds of syndrome extraction of the repetition codes, are
\begin{equation}\label{eq:Elevator_models}
    \begin{gathered}
    p_{Z_L}(p_Z,d_Z) = 0.12\,d_Z\,(34.4\,p_Z)^{0.94\,(d_Z+1)/2},\\
    p_{X_L}(p_X,d_Z) = d_Z^{4.10}\,(17.78\,p_X)^{2.06}.
    \end{gathered}
\end{equation}

These simulations only concern a quantum memory, since no logical gate set has been designed for Elevator codes yet.
We therefore assume a logical cycle of $d_Z$ rounds of syndrome extraction of the repetition codes.
This assumption is conservative, since more efficient schemes, for instance based on transversal gates, may yet be developed.
Under this assumption, we estimate the number of physical qubits by converting the effective logical volume of the repetition-cat estimate (Section~\ref{ssec:physical_resource_count}) to the Elevator code model.

At $p_Z=5\times10^{-4}$, the repetition-cat architecture has an effective logical volume $\mathrm{Vol}^{(\mathrm{eff})}_{\mathrm{logical}}=2.79\times10^{12}$ and $n^{(\mathrm{eff})}_{\mathrm{logical}}=1680$ effective logical qubits (Table~\ref{tab:arch-comparison}).
We choose $d_Z=17$.
With Equations~\eqref{eq:eps} and~\eqref{eq:Elevator_models}, and $p_X=10^{-11}$, we obtain the per-shot logical error probabilities $\epsilon^{(Z)}_{\mathrm{logical}}=6.7\times10^{-3}$ and $\epsilon^{(X)}_{\mathrm{logical}}=2.5\times10^{-3}$, due to phase flips and to bit flips, respectively.
The probability of no logical error during a shot is thus above \SI{99}{\percent}.
The total qubit overhead of the $[15,9,3]$ Elevator code is
\begin{equation}\label{eq:elevator_overhead}
    n_{\mathrm{over}} = \frac{16}{9}\,(2d_Z-1).
\end{equation}
Using this and $n^{(\mathrm{eff})}_{\mathrm{logical}}=1680$, we find that the overall number of physical qubits required is $98\,560$.

At $p_Z=10^{-3}$, the repetition-cat architecture has $\mathrm{Vol}^{(\mathrm{eff})}_{\mathrm{logical}}=2.52\times10^{12}$ and $n^{(\mathrm{eff})}_{\mathrm{logical}}=1649$.
If a $d_Z=21$ repetition code is used, the per-shot logical error probabilities are $\epsilon^{(Z)}_{\mathrm{logical}}=4.7\times10^{-3}$ and $\epsilon^{(X)}_{\mathrm{logical}}=5.5\times10^{-3}$, and the probability of no logical error during a shot is above \SI{98}{\percent}.
With Equation~\eqref{eq:elevator_overhead}, the overall number of physical qubits required is $120\,194$.

Using LDPC-cat codes~\cite{Ruiz_2025} as inner codes should further reduce this overhead~\cite{ConcatenatedLDPC2026prep}.
Until this construction is available, we estimate the overhead of such a concatenated LDPC-cat code by assuming the same proportional increase as from the repetition-cat architecture to the Elevator code.
At $p_Z=5\times10^{-4}$, this increase is \SI{99}{\percent}, and the number of physical qubits of the LDPC-cat architecture goes from $22\,593$ to $44\,850$.
At $p_Z=10^{-3}$, the increase is \SI{117}{\percent}, and the number of physical qubits goes from $25\,673$ to $55\,602$.

Conversely, with the budget $\epsilon^{(X)}_{\mathrm{logical}}=\epsilon_{\mathrm{bf}}=\SI{10}{\percent}$ of Section~\ref{sssec:bitflip_assumptions}, the Elevator code tolerates a bit-flip probability of order $10^{-11}$ per fault location.
Since a repetition code holds on average about two fault locations per qubit per cycle, this corresponds to a bit-flip time of about one hour, which requires about $11$ photons with the rough guide of Section~\ref{sssec:bitflip_assumptions}.
Without an outer code, the same budget requires a bit-flip time of about $200$ years and about $18$ photons (Table~\ref{tab:bitflip_requirements}).
Table~\ref{tab:elevator_requirements} summarizes these numbers for both values of $p_Z$, together with a rough estimate for the LDPC-cat architecture.
The outer code thus relaxes the requirement on the bit-flip time by a factor of more than $10^{6}$, or about seven photons, at the cost of about twice as many physical qubits.

\begin{table*}[t]
\centering
\setlength{\tabcolsep}{3pt}
\begin{tabular}{@{}llllcccccc@{}}
\hline\hline
Computation & $p_Z$ & Architecture & Distances & $n_{\mathrm{total}}$ & Overhead & $T_Z^{\min}$ without & $T_Z^{\min}$ with & Reduction & $\bar n_{\min}$ \\
\hline
RSA-2048 & $5\times10^{-4}$ & Repetition-cat & $d_Z = 17$ & $98\,560$ & $+99\%$ (from $49\,649$) & $196$~years & $1.2$~hours & $1.5\times10^{6}$ & $11.1$ \\
 & & LDPC-cat & estimate & $44\,850$ & $+99\%$ (from $22\,593$) & $101$~years & $1.3$~hours & $7.1\times10^{5}$ & $11.1$ \\
RSA-2048 & $10^{-3}$ & Repetition-cat & $d_Z = 21$ & $120\,194$ & $+117\%$ (from $55\,497$) & $228$~years & $1.7$~hours & $1.2\times10^{6}$ & $11.3$ \\
 & & LDPC-cat & estimate & $55\,602$ & $+117\%$ (from $25\,673$) & $135$~years & $1.8$~hours & $6.8\times10^{5}$ & $11.3$ \\
\hline\hline
\end{tabular}
\caption{Physical qubit count and requirement on bit flips with the additional Elevator layer, to be compared with Table~\ref{tab:bitflip_requirements}, for a logical bit-flip failure probability $\epsilon^{(X)}_{\mathrm{logical}} = \SI{10}{\percent}$ per shot.
$n_{\mathrm{total}}$ is the number of physical qubits with the outer code, and the overhead is the increase of $n_{\mathrm{total}}$ relative to the number of physical qubits without it.
$T_Z^{\min}$ is the minimum bit-flip time of an idle qubit, without the outer code (from Table~\ref{tab:bitflip_requirements}) and with it, and the reduction is the ratio of these two values of $T_Z^{\min}$.
$\bar n_{\min}$ is obtained as in Table~\ref{tab:bitflip_requirements}.
The LDPC-cat rows are rough estimates: the number of physical qubits assumes the same proportional increase as the repetition-cat rows, and $T_Z^{\min}$ applies the logical bit-flip model of Equation~\eqref{eq:Elevator_models}, per logical qubit and with the same $d_Z$, to the effective logical volume of the LDPC-cat architecture including the factories (Table~\ref{tab:arch-comparison}).}
\label{tab:elevator_requirements}
\end{table*}

These estimates rest on assumptions that remain to be checked.
First, the bit-flip model, fitted for $p_X>10^{-6}$ and $d_Z\le15$, is used at $p_X=10^{-11}$, five orders of magnitude lower, and at $d_Z=17$ and $21$.
Similarly, the phase-flip model is used at values of $p_Z$ five to ten times lower and at larger distances than in the simulations of~\cite{shanahan2026elevator}.
Second, we assume that the factories, the routing qubits and the logical operations are protected against bit flips at the logical rate of the memory, although their Elevator constructions remain to be designed.

    \clearpage

\section{Estimation for the other variations and space-time trade-offs}\label{sec:variant_and_tradeoff}

\subsection{RSA} \label{ssec:spacetime_tradeoff}

\subsubsection{Space-efficient configuration} \label{sssec:rsa_smaller}
Reference~\cite{FeMoco2026prep} proposes three layouts of the two-level $\ket{\CXCXX}$ factory, and Section~\ref{ssec:factories} uses the third, which is the fastest. 
The space-efficient configuration replaces the third layout with the second~\cite[Table~4]{FeMoco2026prep}, which takes 560 instead of 1566 physical qubits and a length of 8 instead of 29 columns. The second layout produces one state every 61.5 instead of 20.75 cycles, hence $T^{\ket{\CXCXX}}_{\mathrm{prod}} = 62$. Its output infidelity is slightly lower, $1.81\times10^{-13}$ instead of $2.02\times10^{-13}$ at $p_Z = 5\times10^{-4}$ ($1.32\times10^{-11}$ instead of $1.48\times10^{-11}$ at $p_Z = 10^{-3}$). Everything else is as in the balanced configuration of Section~\ref{sec:ldpc-cat-b}: the LDPC-cat architecture, $p_Z = 5\times10^{-4}$, the algorithmic parameters of Table~\ref{tab:algo_parameter_value}, the $\ket{i}$ factories and the selection criterion of Section~\ref{sec:C7-budget}.

Equation~\eqref{eq:Tcad-ldpc} gives $T_{\mathrm{cad}} = 62$ at every pair of distances of Table~\ref{tab:ldpc-scan}, and Equation~\eqref{eq:factory_counts} still gives one $\ket{\CXCXX}$ factory and two $\ket{i}$ factories. 
\Adder{}s last $2nT_{\mathrm{cad}}$ (Equation~\eqref{eq:T-adder-ldpc}) and the \textsc{And} gates of \Phaseup{}s are issued at the cadence (Equation~\eqref{eq:T-phaseup-ldpc}), so that both durations grow with $T_{\mathrm{cad}}$. 
In a \Lookup, the condition $T^{\ket{\CXCXX}}_{\mathrm{prod}} \le 2T_{\mathrm{surg}}$ of Section~\ref{ssec:lookup_and_phaseup} no longer holds, so that the $2^{w-1}-1$ \textsc{And} gates performed together with a parity-\textsc{Cnot} gate also wait for the factory. 
The duration of a \Lookup{} then takes the general form
\begin{equation}
\begin{split}
  T_{\Lookup}(w) ={}& (2^{w-1}-w)\,T_{\mathrm{cad}} + T_{\mathrm{surg}}\\
  &+ (2^{w-1}-1)\max(2T_{\mathrm{surg}}, T_{\mathrm{cad}}),
\end{split}
\label{eq:T-lookup-general}
\end{equation}
with $T_{\mathrm{surg}} = d$ on the repetition-cat architecture and $T_{\mathrm{surg}} = T^{(\mathrm{LDPC})}_{\mathrm{surg}}$ on the LDPC-cat architecture. 
For $T_{\mathrm{cad}} \le 2T_{\mathrm{surg}}$, Equation~\eqref{eq:T-lookup-general} coincides with Equations~\eqref{eq:Tlookup} and~\eqref{eq:T-lookup-ldpc}. 
Here, $T_{\mathrm{cad}} > 2T^{(\mathrm{LDPC})}_{\mathrm{surg}}$ at every pair, and Equation~\eqref{eq:T-lookup-general} reduces to $(2^w-w-1)\,T_{\mathrm{cad}} + T^{(\mathrm{LDPC})}_{\mathrm{surg}}$. A \Lookup$(6)$ then lasts 3550 cycles instead of 1554, and a \Lookup$(5)$ 1628 instead of 727.

At $T_{\mathrm{cad}} = 62$, Equation~\eqref{eq:Tbuf-ldpc} would give a single adder buffer. 
The buffer zone, however, also holds the auxiliary qubits of \Lookup{}s and the workspace of \Phaseup{}s, which use up to 20 of its $9n_{\mathrm{buf}}^{(\Adder)}$ positions (Table~\ref{tab:ldpc-layout}), hence $n_{\mathrm{buf}}^{(\Adder)} \ge 3$.
The buffers are therefore kept as in the balanced configuration, $n_{\mathrm{buf}}^{(\Adder)} = 3$ and $n_{\mathrm{buf}}^{(\Lookup)} = 2$. \Adder{}s then hold three buffers where one suffices, a slight overestimate of their volume and of their $\ket{i}$ count.
The factory zone has length $l_{\mathrm{fac}} = 8 + 2\times10.5 = 29$ and holds $n_{\mathrm{fac}} = 560 + 2\times153 = 866$ physical qubits, so that the factory routing zone of Equation~\eqref{eq:ntotal-ldpc} holds $2h_{\mathrm{routing,fac}}\,l_{\mathrm{fac}} - (l_{\mathrm{fac}} - l_{\mathrm{buf}}) = 578$ physical qubits instead of 977. 
The factories and their routing thus take 1444 instead of 2849 physical qubits.

Table~\ref{tab:rsa-space-efficient} compares the two configurations. 
The criterion again selects $(d_{\mathrm{rep}},d_{\mathrm{LDPC}}) = (13,16)$, with $21\,188$ physical qubits. 
One shot lasts $6.89\times10^{10}$ cycles, about \SI{19.15}{\hour}. The longer shot raises the logical error probabilities to $\epsilon^{(\mathrm{rep})}_{\mathrm{logical}} = 6.7\times10^{-2}$ and $\epsilon^{(\mathrm{LDPC})}_{\mathrm{logical}} = 2.4\times10^{-2}$. The $\ket{\CXCXX}$ and $\ket{i}$ factories add $1.4\times10^{-4}$ and $2.0\times10^{-3}$, hence $\epsilon = 9.1\%$, $9.2/(1-\epsilon) = 10.13$ shots and a total runtime of \SI{8.08}{\day}.
Compared with the balanced configuration, the physical qubit count decreases by 6.2\% and the total runtime grows by a factor of 2.58. 
The factories and their routing hold only 13\% of the physical qubits of the balanced configuration, whereas the cadence they set governs most of the duration of a shot.

\begin{table*}[t]
\centering
\begin{tabular}{@{}lccccccccccc@{}}
\toprule
 & $\ket{\CXCXX}$ & & & Physical & & \multicolumn{2}{c}{$T_{\mathrm{shot}}$} & & & Runtime & \\
\cmidrule(lr){7-8}
Configuration & layout & $T_{\mathrm{cad}}$ & $(d_{\mathrm{rep}},d_{\mathrm{LDPC}})$ & qubits & Factories & cycles & hours & $\epsilon$ & Shots & (days) & Qubit-days \\
\midrule
Balanced & third & 21 & $(13,16)$ & $22\,593$ & $2849$ & $2.82\times10^{10}$ & 7.84 & 4.2\% & 9.60 & 3.14 & $7.09\times10^{4}$ \\
Space-efficient & second & 62 & $(13,16)$ & $21\,188$ & $1444$ & $6.89\times10^{10}$ & 19.15 & 9.1\% & 10.13 & 8.08 & $1.71\times10^{5}$ \\
\bottomrule
\end{tabular}
\caption{RSA-2048 on the LDPC-cat architecture at $p_Z = 5\times10^{-4}$, with the third (balanced, Section~\ref{sec:ldpc-cat-b}) and the second (space-efficient) layouts of the $\ket{\CXCXX}$ factory~\cite[Table~4]{FeMoco2026prep}. The Factories column includes the factory routing zone. The number of shots is $9.2/(1-\epsilon)$, and the qubit-days are the physical qubit count times the total runtime.}
\label{tab:rsa-space-efficient}
\end{table*}

\subsubsection{Time-efficient configuration} \label{sssec:rsa_faster}

\ifdefined\CrepNe\else\newcommand{\CrepNe}{1280}          
\newcommand{\CrepNprimes}{22658}    
\newcommand{\CrepWone}{6}           
\newcommand{\CrepKappa}{33}         
\newcommand{\CrepLbuf}{27}          
\newcommand{\CrepLcomp}{103}        
\newcommand{\CrepLmem}{51}          
\newcommand{\CrepNfac}{1872}        
\newcommand{\CrepHroutingFac}{10}   
\newcommand{\CrepLfac}{50}          
\newcommand{\CrepHexp}{2}           
\newcommand{\CrepHrouting}{4}       
\newcommand{\CrepTcadMin}{21}       
\newcommand{\CrepNcolOther}{1460}   
\newcommand{\CrepParityPerPrime}{262} 
\newcommand{\CrepMzPerPrime}{400}   
\newcommand{\CrepPrepPerPrime}{5}   
\newcommand{\CrepAccOtherD}{8}      
\newcommand{\CrepAccOtherC}{6}      
\newcommand{\CrepDsel}{15}          
\newcommand{\CrepNtotalSel}{49649}  
\newcommand{\CrepTshotSel}{24926790112} 
\newcommand{\CrepGammaSel}{4678}    
\newcommand{\CrepPT}{2.02e-13}      
\newcommand{\CrepPi}{9.43e-13}      
\newcommand{\CpZ}{5e-4}             
\newcommand{\CpLogA}{0.12}          
\newcommand{\CpLogB}{23}            
\newcommand{\CpLogC}{0.99}          
\newcommand{\CepsCrit}{0.10}        
\newcommand{\CcycleUs}{1}           

\newcommand{\AshotsNeeded}{9}       
\newcommand{\APdeviant}{0.0125}     
\newcommand{\AlatticeSuccess}{0.99} 

\newcommand{\ClDrep}{13}            
\newcommand{\ClDldpc}{16}           
\newcommand{\ClTsurg}{16}           
\newcommand{\ClH}{7}                
\newcommand{\ClDeltaLDPC}{26}       
\newcommand{\ClER}{5}               
\newcommand{\ClLexp}{640}           
\newcommand{\ClLmem}{26}            
\newcommand{\ClLcomp}{51}           
\newcommand{\ClNbuf}{3}             
\newcommand{\ClHroutingBuf}{4}      
\newcommand{\ClFacRouting}{977}     
\newcommand{\ClNlane}{10}           
\newcommand{\ClParityPrepPerPrime}{267} 
\newcommand{\ClAccParity}{8}        
\newcommand{\ClNtotalSel}{22593}    
\newcommand{\ClGammaGlobalSel}{5603}
\newcommand{\ClTshotSel}{2.82e10}   
\newcommand{\ClEpsSel}{0.042}       
\newcommand{\ClpLogA}{0.1}          
\newcommand{\ClpLogB}{29.8}         
\newcommand{\ClpLogC}{1}            

\newcommand{\CfShotsNeeded}{8}      
\newcommand{\CfPdeviant}{0.019}     
\newcommand{\CfRepNtotal}{9336}     
\newcommand{\CfRepTshot}{1.37e8}    
\newcommand{\CfRepEps}{0.0162}      
\newcommand{\CfRepD}{11}            
\newcommand{\CfLdpcNtotal}{6656}    
\newcommand{\CfLdpcTshot}{1.51e8}   
\newcommand{\CfLdpcEps}{0.0234}     
\newcommand{\CfLdpcDrep}{11}
\newcommand{\CfLdpcDldpc}{12}
\newcommand{\CfNe}{104}             
\fi
\ifdefined\DrhoMax\else
\providecommand{\DrhoMax}{}\renewcommand{\DrhoMax}{7}
\providecommand{\DdLow}{}\renewcommand{\DdLow}{13}
\providecommand{\DrhoDdrop}{}\renewcommand{\DrhoDdrop}{2}
\providecommand{\DepsTwoAtDlow}{}\renewcommand{\DepsTwoAtDlow}{9.8}
\providecommand{\DepsMainOne}{}\renewcommand{\DepsMainOne}{0.43}
\providecommand{\DepsMainMax}{}\renewcommand{\DepsMainMax}{0.28}
\providecommand{\DqubitsOne}{}\renewcommand{\DqubitsOne}{49649}
\providecommand{\DshotOne}{}\renewcommand{\DshotOne}{6.92}
\providecommand{\DtotalOne}{}\renewcommand{\DtotalOne}{64.0}
\providecommand{\DqubitsThree}{}\renewcommand{\DqubitsThree}{68244}
\providecommand{\DtotalThree}{}\renewcommand{\DtotalThree}{21.9}
\providecommand{\DoverheadThree}{}\renewcommand{\DoverheadThree}{37}
\providecommand{\DspeedupThree}{}\renewcommand{\DspeedupThree}{2.9}
\providecommand{\DqubitsMax}{}\renewcommand{\DqubitsMax}{105434}
\providecommand{\DtotalMax}{}\renewcommand{\DtotalMax}{9.6}
\providecommand{\DoverheadMax}{}\renewcommand{\DoverheadMax}{112}
\providecommand{\DspeedupMax}{}\renewcommand{\DspeedupMax}{6.7}
\providecommand{\DwallMax}{}\renewcommand{\DwallMax}{1.06}
\providecommand{\DqubitsMaxMachines}{}\renewcommand{\DqubitsMaxMachines}{1054340}
\providecommand{\Dmachines}{}\renewcommand{\Dmachines}{10}
\providecommand{\DroundsOne}{}\renewcommand{\DroundsOne}{1.02}
\providecommand{\DroundsThree}{}\renewcommand{\DroundsThree}{1.02}
\providecommand{\DroundsThreeLow}{}\renewcommand{\DroundsThreeLow}{1.23}
\providecommand{\DwallThree}{}\renewcommand{\DwallThree}{2.42}
\providecommand{\DwallThreeLow}{}\renewcommand{\DwallThreeLow}{2.79}
\providecommand{\DqubitsThreeLow}{}\renewcommand{\DqubitsThreeLow}{60948}
\providecommand{\DsavingLow}{}\renewcommand{\DsavingLow}{11}
\providecommand{\DsavingLowLo}{}\renewcommand{\DsavingLowLo}{10}
\providecommand{\DsavingLowHi}{}\renewcommand{\DsavingLowHi}{11}
\providecommand{\DtotalMaxLow}{}\renewcommand{\DtotalMaxLow}{9.7}
\providecommand{\DwallMaxLow}{}\renewcommand{\DwallMaxLow}{1.13}
\providecommand{\DqubitsMaxLow}{}\renewcommand{\DqubitsMaxLow}{95242}
\providecommand{\DoneRoundThreeNine}{}\renewcommand{\DoneRoundThreeNine}{0.86}
\providecommand{\DoneRoundThreeTen}{}\renewcommand{\DoneRoundThreeTen}{0.98}
\providecommand{\DexpQubitsSel}{}\renewcommand{\DexpQubitsSel}{38400}
\providecommand{\DepsCritPct}{}\renewcommand{\DepsCritPct}{10}
\providecommand{\DmergeCycles}{}\renewcommand{\DmergeCycles}{6417}
\providecommand{\DeraseCycles}{}\renewcommand{\DeraseCycles}{19014}
\providecommand{\DtEndCycles}{}\renewcommand{\DtEndCycles}{597916}
\providecommand{\DmergeToff}{}\renewcommand{\DmergeToff}{132}
\providecommand{\DmergeIstates}{}\renewcommand{\DmergeIstates}{540}
\providecommand{\DmergeShareTwo}{}\renewcommand{\DmergeShareTwo}{2.3}
\providecommand{\DmergeShareMax}{}\renewcommand{\DmergeShareMax}{5.2}
\providecommand{\DfacToffMax}{}\renewcommand{\DfacToffMax}{1.64}
\providecommand{\DfacIMax}{}\renewcommand{\DfacIMax}{2.07}
\providecommand{\DfacEpsOne}{}\renewcommand{\DfacEpsOne}{0.219}
\providecommand{\DfacEpsMax}{}\renewcommand{\DfacEpsMax}{0.224}
\providecommand{\DvolRatioThree}{}\renewcommand{\DvolRatioThree}{0.44}
\providecommand{\DvolRatioMax}{}\renewcommand{\DvolRatioMax}{0.28}
\fi
\ifdefined\DLqubitsOne\else
\providecommand{\DLqubitsOne}{}\renewcommand{\DLqubitsOne}{22593}
\providecommand{\DLshotOne}{}\renewcommand{\DLshotOne}{7.84}
\providecommand{\DLtotalOne}{}\renewcommand{\DLtotalOne}{75.3}
\providecommand{\DLepsOne}{}\renewcommand{\DLepsOne}{4.2}
\providecommand{\DLqubitsThree}{}\renewcommand{\DLqubitsThree}{35413}
\providecommand{\DLtotalThree}{}\renewcommand{\DLtotalThree}{25.4}
\providecommand{\DLoverheadThree}{}\renewcommand{\DLoverheadThree}{57}
\providecommand{\DLspeedupThree}{}\renewcommand{\DLspeedupThree}{3.0}
\providecommand{\DLqubitsMax}{}\renewcommand{\DLqubitsMax}{61053}
\providecommand{\DLtotalMax}{}\renewcommand{\DLtotalMax}{11.1}
\providecommand{\DLoverheadMax}{}\renewcommand{\DLoverheadMax}{170}
\providecommand{\DLspeedupMax}{}\renewcommand{\DLspeedupMax}{6.8}
\providecommand{\DLepsMax}{}\renewcommand{\DLepsMax}{2.4}
\providecommand{\DLwallMax}{}\renewcommand{\DLwallMax}{1.25}
\providecommand{\DLqubitsMaxMachines}{}\renewcommand{\DLqubitsMaxMachines}{610530}
\providecommand{\DLmergeCycles}{}\renewcommand{\DLmergeCycles}{6656}
\providecommand{\DLmoveCycles}{}\renewcommand{\DLmoveCycles}{224}
\providecommand{\DLeraseCycles}{}\renewcommand{\DLeraseCycles}{20878}
\providecommand{\DLtEndCycles}{}\renewcommand{\DLtEndCycles}{683088}
\providecommand{\DLmergeShareTwo}{}\renewcommand{\DLmergeShareTwo}{2.2}
\providecommand{\DLmergeShareMax}{}\renewcommand{\DLmergeShareMax}{4.9}
\providecommand{\DLexpQubits}{}\renewcommand{\DLexpQubits}{16000}
\providecommand{\DLexpShare}{}\renewcommand{\DLexpShare}{71}
\providecommand{\DLengQubits}{}\renewcommand{\DLengQubits}{6410}
\providecommand{\DLvsRepMax}{}\renewcommand{\DLvsRepMax}{61053}
\providecommand{\DrepExpShare}{}\renewcommand{\DrepExpShare}{77}
\providecommand{\DFrepShot}{}\renewcommand{\DFrepShot}{2.28}
\providecommand{\DFldpcShot}{}\renewcommand{\DFldpcShot}{2.52}
\providecommand{\DFrepTotal}{}\renewcommand{\DFrepTotal}{19.1}
\providecommand{\DFldpcTotal}{}\renewcommand{\DFldpcTotal}{21.2}
\providecommand{\DFrepWallEight}{}\renewcommand{\DFrepWallEight}{2.9}
\providecommand{\DFldpcWallEight}{}\renewcommand{\DFldpcWallEight}{3.3}
\providecommand{\DFrepRoundsEight}{}\renewcommand{\DFrepRoundsEight}{1.26}
\providecommand{\DFldpcRoundsEight}{}\renewcommand{\DFldpcRoundsEight}{1.30}
\providecommand{\DFrepWallTen}{}\renewcommand{\DFrepWallTen}{2.3}
\providecommand{\DFldpcWallTen}{}\renewcommand{\DFldpcWallTen}{2.6}
\providecommand{\DFrepQubitsTen}{}\renewcommand{\DFrepQubitsTen}{93360}
\providecommand{\DFldpcQubitsTen}{}\renewcommand{\DFldpcQubitsTen}{66560}
\providecommand{\DFrepExpShare}{}\renewcommand{\DFrepExpShare}{25}
\providecommand{\DFldpcExpShare}{}\renewcommand{\DFldpcExpShare}{16}
\fi
\providecommand{\ket}[1]{\left|#1\right\rangle}

The estimate of Section~\ref{ssec:physical_resource_count} runs one shot after the other on one chip. Two levers shorten the clock time at the price of more qubits. The first runs several shots at once on several identical chips. The second runs several residue engines at once on one chip, sharing a single exponent register, as proposed in Ref.~\cite[Sec.~VI.B.1]{Webster2026Pinnacle}. We keep every algorithmic parameter fixed ($s$, $\kappa$, $S$, the window sizes and the prime set), as well as $p_Z = \num{\CpZ}$, and we treat RSA-2048 on the repetition-cat chip, then on the LDPC-cat chip.

\paragraph{Parallel shots.}
Shots are independent once $g$ is fixed, so $k$ identical chips can run them in rounds of $k$ 
simultaneous shots, on $k\, n_{\mathrm{total}}$ physical qubits. The expected clock time is
\begin{equation}
  T_{\mathrm{wall}} = T_{\mathrm{shot}}\; \mathbb{E}\!\left[N_{\mathrm{rounds}}(k)\right],
  \label{eq:twall}
\end{equation}
where $N_{\mathrm{rounds}}(k)$ is the number of rounds needed to collect $s+1$ consistent shots 
that pass the lattice step. Each shot keeps the statistics of the serial run (Section~\ref{sec:full-modern}), with a fresh
$g$ after a failed lattice step. For $k = 1$, $N_{\mathrm{rounds}}$ is the number of shots $R$, 
with mean $\mathbb{E}[R]$ given by Equation~\eqref{eq:Rshots} divided by $1-\epsilon$.

\paragraph{Parallel residue engines.}
Three properties of the oracle make the second lever cheap. First, $\tilde V$ is a sum of contributions, one per prime of the RNS, and each contribution only needs read access to the exponent register. Second, the exponent register is only used as a \Lookup{} address and as a control, and all phase corrections on it are deferred and merged into one final \Phaseup{} per window. Third, almost all of a shot is spent inside the loop over the primes (Table~\ref{tab:physical_cycle_breakdown}), while the exponent register holds most of the qubits. At $d = \CrepDsel$, the exponent register takes \num{\DexpQubitsSel} of the \num{\CrepNtotalSel} physical qubits, counting one ancilla per data qubit as in Table~\ref{tab:rep-code-columns}.

\begin{figure*}[t]
\centering
\includegraphics[width=\textwidth]{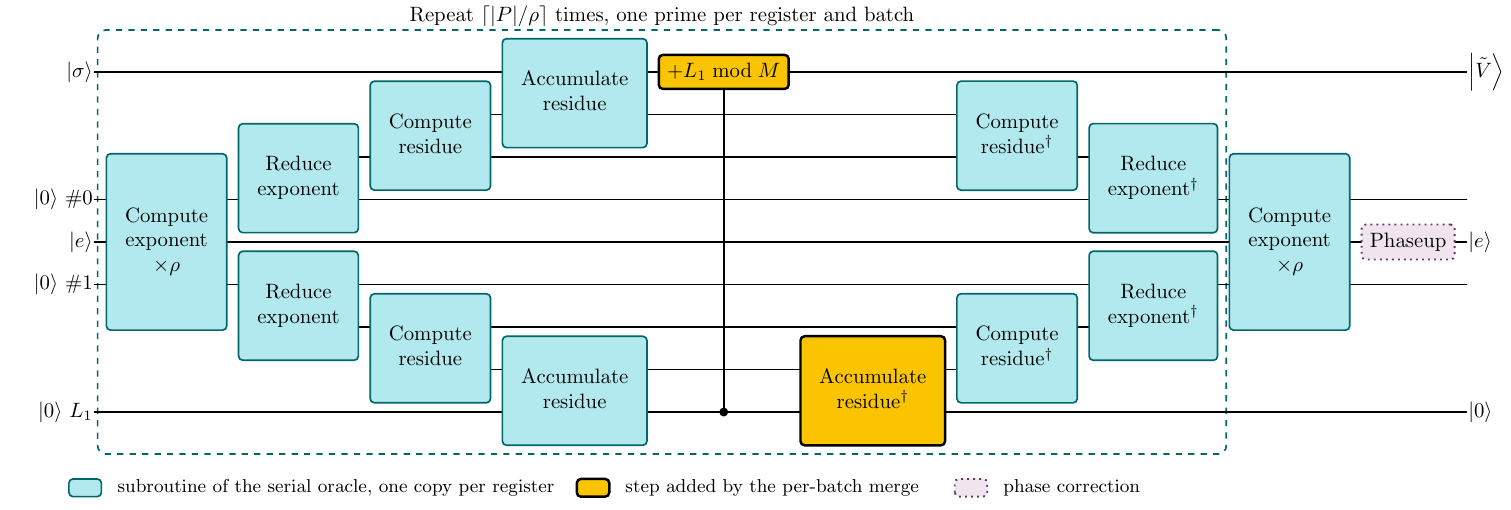}
\caption{The oracle of Figure~\ref{fig:full_shor_circuit} with $\rho = 2$ working registers, following the per-batch merge of Ref.~\cite[Sec.~VI.B.1]{Webster2026Pinnacle}. Register \#0 is drawn above $\ket{e}$ and register \#1 mirrored below it. Each register runs the subroutines of Figure~\ref{fig:full_shor_circuit} on its own primes $p_j, p_{j+\rho}, p_{j+2\rho}, \dots$, and both read $\ket{e}$ in staggered window orders. In each batch, register \#0 accumulates into $\ket{\sigma}$ and register \#1 into a local accumulator $L_1$ initialized to $\ket{0}$. $L_1$ is then added into $\ket{\sigma}$ modulo $M$, Equation~\eqref{eq:merge}, and returned to $\ket{0}$ by uncomputing its accumulation while the residue is still available. For larger $\rho$, the local accumulators $L_1, \dots, L_{\rho-1}$ are added into $\ket{\sigma}$ one after the other, and each is erased right after its own merge. After the loop, each register uncomputes its last exponent, and the deferred phase corrections of all registers are applied by the single final \Phaseup.}
\label{fig:oracle-parallel}
\end{figure*}

We follow the construction of Ref.~\cite[Sec.~VI.B.1]{Webster2026Pinnacle}. The chip holds $\rho$ working registers, each with its own residue registers, workspace and local accumulator, and the $|P|$ primes are processed in $B = \lceil |P|/\rho \rceil$ batches. Batch $b$ holds $m_b = \min(\rho, |P| - (b-1)\rho)$ primes. In each batch, every register runs the subroutines of Figure~\ref{fig:full_shor_circuit} on its own prime. Register \#0 accumulates its contribution into $\ket{\sigma}$, and register \#$j \geq 1$ into a local accumulator $L_j$ initialized to $\ket{0}$. The local accumulators are then added into $\ket{\sigma}$ one after the other, and each $L_j$ is returned to $\ket{0}$ by uncomputing its accumulation while its residue is still available. Figure~\ref{fig:oracle-parallel} shows the circuit for $\rho = 2$.

Three details of Figure~\ref{fig:full_shor_circuit} change. Register \#$j$ handles the primes $p_j, p_{j+\rho}, p_{j+2\rho}, \dots$, so the merged uncomputation of \ComputeExponent{} uses the differences between consecutive primes of that sequence. The sums still telescope, so each register starts from $\ket{0}$ and, after the loop, uncomputes its own last exponent. The deferred phase corrections of all registers act on the same exponent windows and are diagonal. These corrections therefore commute with the operations in between, and the single final \Phaseup{} applies them together, merging their phase tables. The registers read the exponent windows in staggered orders, register \#$j$ starting at window $j$, so that no two of them address the same window at the same time. Operations controlled by the exponent register that act on different registers commute, so the registers can share the exponent register without copying it. Unlike in Ref.~\cite{Webster2026Pinnacle}, no Clifford frame cleaning is needed, since the lattice-surgery circuits of Section~\ref{sec:resource_estim}execute their Clifford gates physically.

\paragraph{Modular merge.}\label{par:modular-merge}
Every accumulator, $\ket{\sigma}$ as well as each $L_j$, holds a value reduced modulo $M = N \gg \bar\kappa$ (Equation~\eqref{eq:def_hat_v}). The merge must therefore add modulo $M$,
\begin{equation}
  \ket{A}\ket{L_j} \longmapsto \ket{(A + L_j) \bmod M}\ket{L_j},
  \label{eq:merge}
\end{equation}
and an ordinary $(\kappa+1)$-qubit Q-Q adder does not suffice. 
We use four additions of $\kappa+1$ bits. 
The first computes $s = A + L_j$. 
The second subtracts $M$ with a Q-C adder and copies the sign bit to a no-wrap flag $f = [s < M]$. 
This subtraction is safe in $\kappa+1$ bits, since $-M \leq s - M \leq M - 2$ and $M \leq 2^{\kappa}$. 
The third adds $fM$, prepared in the operand slots by a fan-out of $f$, which leaves $(A + L_j) \bmod M$ in the accumulator. 
The fan-out is then reversed, which returns the operand slots to $\ket{0}$ before $f$ is cleared. 
The fourth, a Q-Q \Comparison$(\kappa)$, adds $[A \geq L_j]$ to $f$, where $A$ now holds the reduced sum. 
Without a wrap, this sum is $s \geq L_j$. With a wrap, it is $s - M < L_j$, since the original $A$ is below $M$.
Hence $[A \geq L_j] = f$, including at the boundary $s = M$, and the comparison returns $f$ to $\ket{0}$. 
The three parity-\textsc{Cnot} surgeries, which copy the sign bit, fan out $f$ and reverse the fan-out, account for the term $3d$ below. 
$L_j$ makes four register moves. 
$L_j$ enters the compute zone of register \#0 for the first step and returns to its engine during the second and third steps, which need the operand slots for $M$ and $fM$. 
$L_j$ then comes back for the comparison and returns to its engine for its erasure.
The flag $f$ uses the underflow qubit of \AccumulateResidue. 
We found no capacity conflict in that compute zone, but a detailed spatial allocation of the merge remains to be done. 
With the durations of Equation~\eqref{eq:Tadder} and register moves of three qubits per $d$ cycles (Section~\ref{sssec:global_layout_routing}), the merge lasts
\begin{equation}
  T_{\mathrm{merge}}(d) = 4\, T_{\Adder}(\kappa+1) + 3d + 4 \left\lceil \kappa/3 \right\rceil d ,
  \label{eq:tmerge}
\end{equation}
that is \num{\DmergeCycles} cycles at $d = \CrepDsel$, and consumes \DmergeToff{} Toffoli gates and \DmergeIstates{} $\ket{i}$ states.

The local accumulators are merged into $\ket{\sigma}$ one after the other, never into each other. Each $L_j$ therefore holds exactly its own contribution until it is erased, and the inverse of its \AccumulateResidue{} returns it to $\ket{0}$. A reduction tree would shorten the merge, but it turns some local accumulators into sums of several contributions, and these partial sums must be uncomputed before the erasures, as accounted for in Ref.~\cite[Eq.~(26)]{Webster2026Pinnacle}. For $\rho \leq \DrhoMax$, the sequential merge takes at most six merges per batch, against up to five merge layers for a tree with its cleanup, so we keep the simpler sequential merge. Each erasure starts as soon as its own merge ends and overlaps the later merges.

\paragraph{Layout on the repetition-cat chip.}
Each extra engine brings its own buffer, compute and memory zones, $l_{\mathrm{eng}} = l_{\mathrm{buf}} + l_{\mathrm{comp}} + l_{\mathrm{mem}} = \the\numexpr\CrepLbuf+\CrepLcomp+\CrepLmem\relax$ repetition-code columns, its own factory and the routing above its zones. The exponent register is shared. We keep the two-rectangle layout of Section~\ref{sssec:global_layout_routing} and place the $\rho$ engines in the bottom rectangle. The chip length and the bottom exponent zone become
\begin{equation}
  l_{\mathrm{tot}}(\rho) = \left\lceil \frac{n_e + \rho\, l_{\mathrm{eng}}}{2} \right\rceil,
  \qquad
  l_{\mathrm{exp},1}(\rho) = l_{\mathrm{tot}}(\rho) - \rho\, l_{\mathrm{eng}} .
  \label{eq:ltot-rho}
\end{equation}
The layout requires $l_{\mathrm{exp},1} \geq 0$, that is $\rho\, l_{\mathrm{eng}} \leq n_e + 1$ for integer lengths, which allows up to $\rho = \DrhoMax$ engines for RSA-2048. Larger $\rho$ needs engines in both rectangles and is not considered here.

The routing zone between the two exponent zones gives the engines access to the exponent qubits. Its $h_{\mathrm{exp}} = \CrepHexp$ rows suffice for one engine because the accesses are sequential. We consider two options for $\rho$ engines: a single shared lane, $h_{\mathrm{exp}} = \CrepHexp$, whose accesses are scheduled, and one lane per engine, $h_{\mathrm{exp}} = \CrepHexp\rho$. With the counting convention of Equations~\eqref{eq:ntotal}--\eqref{eq:ntotal_d}, the qubit count becomes
\begin{equation}
\begin{split}
  &n_{\mathrm{total}}(\rho, d) = (4d-1)\, l_{\mathrm{tot}}(\rho) + 2h_{\mathrm{exp}}\, l_{\mathrm{exp},1}(\rho)\\
  &\quad + \rho \bigl[ 2h_{\mathrm{routing}}\, l_{\mathrm{eng}} + 2h_{\mathrm{routing,fac}}\, l_{\mathrm{fac}} + n_{\mathrm{fac}} \bigr],
\end{split}
\label{eq:ntotal-rho}
\end{equation}
with $h_{\mathrm{routing}} = \CrepHrouting$, and the global routing of Equation~\eqref{eq:Gglobal_d} becomes $\Gamma_{\mathrm{global}}(\rho, d) = 2d\rho + 2h_{\mathrm{exp}}\, l_{\mathrm{exp},1}(\rho) + \rho\,(2h_{\mathrm{routing}}\, l_{\mathrm{eng}} + 2h_{\mathrm{routing,fac}}\, l_{\mathrm{fac}})$. Both reduce to Equations~\eqref{eq:Gglobal_d} and~\eqref{eq:ntotal_d} at $\rho = 1$ for every $d$.

The runtimes below assume that the engines never stall each other on their way to the exponent qubits or to $\ket{\sigma}$. 
Distinct windows at each time step are necessary for stall-free access but do not prove that the physical routes are disjoint. 
Staggering the engines in time also helps achieve stall-free access. 
The shared lane keeps the layout of Section~\ref{sec:resource_estim}, and it is the configuration we retain. 
Its runtime assumes that up to $\rho$ concurrent accesses fit in two rows, which remains to be established. 
One lane per engine would relax this requirement, at the cost of a new global layout.

\paragraph{Duration.}
We split the serial shot of Equation~\eqref{eq:Tshot} into the loop over the primes and two pieces of fixed work. 
As in Section~\ref{ssec:repetition_cat_architecture}, this shot excludes the final QFT. 
The first piece is the final uncomputation of the exponent, $T_{\mathrm{end}}(d) = \lceil n_e/w_1 \rceil \bigl(T_{\Lookup}(w_1, d) + T_{\Adder}(\ell + \log n_e) + 1\bigr)$, that is \num{\DtEndCycles} cycles at $d = \CrepDsel$. 
The second is the final \Phaseup, $T_{\mathrm{last}}(d)$.
The serial time per prime is $t_{\mathrm{loop}}(d) = \bigl(T_{\mathrm{shot}}(d) - T_{\mathrm{end}}(d) - T_{\mathrm{last}}(d)\bigr)/|P|$. With $\rho$ engines, the final uncomputations run in parallel, so $T_{\mathrm{end}}$ is paid once. 
A batch of $m_b > 1$ primes adds $m_b - 1$ merges and the erasure of the last local accumulator. 
One erasure costs one \AccumulateResidue, $t_{\mathrm{acc}}(d)$, including its slices of $T_{\mathrm{other}}$, that is \num{\DeraseCycles} cycles at $d = \CrepDsel$. 
Hence
\begin{equation}
\begin{split}
  &T_{\mathrm{shot}}(\rho, d) = B\, t_{\mathrm{loop}}(d) + T_{\mathrm{end}}(d) + T_{\mathrm{last}}(d)\\
  &\quad + \sum_{b=1}^{B} \mathbf{1}_{m_b > 1} \bigl[ (m_b - 1)\, T_{\mathrm{merge}}(d) + t_{\mathrm{acc}}(d) \bigr] .
\end{split}
\label{eq:tshot-rho}
\end{equation}
Relative to an ideal division of the loop by $\rho$, the merges and erasures add \SI{\DmergeShareTwo}{\percent} to the shot at $\rho = 2$ and \SI{\DmergeShareMax}{\percent} at $\rho = \DrhoMax$. 
We do not overlap the merges with the next batch.

\paragraph{Volume and factory errors.}
We evaluate the effective logical volume of Equation~\eqref{eq:Voleff} on the parallel schedule. During the loop, the $\rho$ registers run the same sequence of operations, so each slice $s$ of Table~\ref{tab:rep-code-columns} holds $n_e + \rho\,(n_{\mathrm{col}}(s) - n_e)$ repetition-code columns, the shared exponent register being counted once. The slices of $T_{\mathrm{other}}$, as in Section~\ref{ssec:repetition_cat_architecture}, and the merges and erasures are counted at the peak \begin{equation}   n_{\mathrm{peak}}(\rho) = n_e + \rho\,(\CrepNcolOther - n_e) = \CrepNe + \the\numexpr\CrepNcolOther-\CrepNe\relax\,\rho ,   \label{eq:npeak} \end{equation} an upper bound, since \CrepNcolOther{} columns include the exponent register of the serial run. The final uncomputation holds the columns of \ComputeExponent{} in every register. The global routing and the exposed tops follow the layout of Equation~\eqref{eq:ntotal-rho}, with $n_{e,2} = n_e - l_{\mathrm{exp},1}(\rho)$.

The merges, the erasures and the extra final uncomputations consume additional factory outputs. With $|P| - B$ local contributions,
\begin{equation}
\begin{split}
  &N_T(\rho) = N_T + (|P| - B)\bigl(N_T^{\mathrm{acc}} + N_T^{\mathrm{merge}}\bigr)\\
  &\qquad + (\rho - 1) \lceil n_e/w_1 \rceil N_T^{\mathrm{exp}},\\
  &\epsilon_{\mathrm{fac}}(\rho) = 1 - (1 - p_T)^{N_T(\rho)}\, (1 - p_i)^{N_i(\rho)},
\end{split}
\label{eq:efac}
\end{equation}
and the same expression for $N_i(\rho)$ with the $\ket{i}$ counts. Here, $N_T^{\mathrm{acc}}$, $N_T^{\mathrm{merge}}$ and $N_T^{\mathrm{exp}}$ count the Toffoli gates of one \AccumulateResidue, one merge and one window of \ComputeExponent. The output infidelities $p_T$ and $p_i$ are those of Table~\ref{tab:adder_reference_factories}. At $\rho = \DrhoMax$, the Toffoli count grows by \SI{\DfacToffMax}{\percent} and the $\ket{i}$ count by \SI{\DfacIMax}{\percent}, which raises the factory failure probability from \SI{\DfacEpsOne}{\percent} to \SI{\DfacEpsMax}{\percent} per shot.

The per-shot failure probability combines the logical errors of the data and routing qubits, Equation~\eqref{eq:eps}, with $\epsilon_{\mathrm{fac}}(\rho)$. We evaluate it for the shared lane.

\paragraph{Scope of the estimate.}\label{par:rsa-scope}
We base our estimate on the configurations with $d = \CrepDsel$ and a shared exponent lane, because they stay closest to the analysis of Section~\ref{ssec:physical_resource_count}. 
The distance $d = \CrepDsel$ is the one selected and evaluated there. 
The shared lane keeps the routing zone between the two exponent zones as in that section, so that only the engine zones, their factories and their routing are replicated. 
The qubit count then follows directly from Equations~\eqref{eq:ntotal}--\eqref{eq:ntotal_d}.
Keeping $d = \CrepDsel$ for every $\rho$ satisfies the failure-probability threshold of the criterion of Section~\ref{ssec:physical_resource_count}, since $\epsilon$ then stays below \SI{0.5}{\percent}. 
The full criterion also minimizes the number of qubits, whereas we deliberately keep $d = \CrepDsel$ fixed. 
From $\rho = \DrhoDdrop$ on, $d = \DdLow$ also meets the threshold and saves \DsavingLowLo{} to \SI{\DsavingLowHi}{\percent} of the qubits. 
The configurations at $d = \DdLow$ are extrapolations, and so are the lane-per-engine counts. 
At $d = \DdLow$, the per-shot failure probability reaches several percent. 
There, the logical error model and the upper bounds used for the volume of the merges weigh most on the result, and the choice of distance becomes sensitive to them. 
One lane per engine changes the global layout of Section~\ref{sec:resource_estim}, and its routing would need its own analysis. 
For every configuration, including the shared lane at $d = \CrepDsel$, the runtimes also rest on two scheduling assumptions that Section~\ref{sec:resource_estim} does not cover. 
The first is that the engines never stall each other on the exponent qubits or on $\ket{\sigma}$. 
The second is that the merges do not overlap the next batch. 
A proper evaluation of these configurations requires the same layout, schedule and volume analysis as in Section~\ref{ssec:repetition_cat_architecture}, which we leave to future work.

\begin{figure*}[t]
\centering
\includegraphics[width=\textwidth]{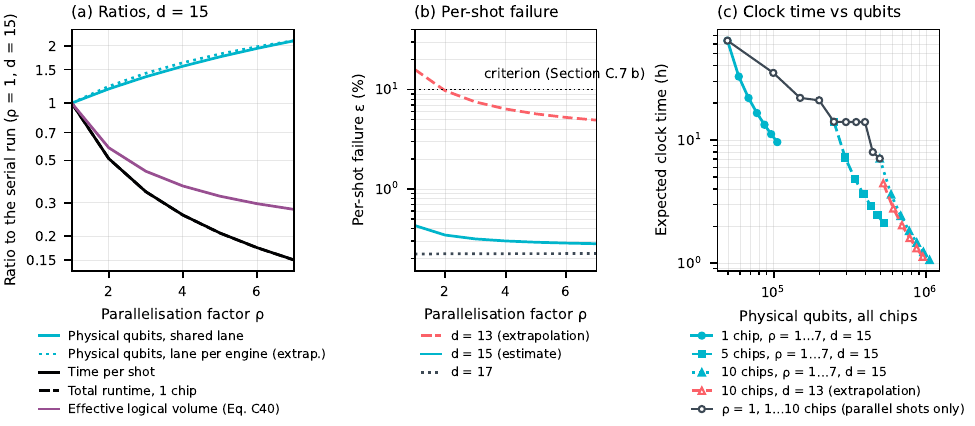}
\caption{Space-time trade-off for RSA-2048 on the repetition-cat chip at $p_Z = \num{\CpZ}$, with $\rho \leq \DrhoMax$ residue engines sharing one exponent register after Ref.~\cite[Sec.~VI.B.1]{Webster2026Pinnacle}, the limit of the layout of Equation~\eqref{eq:ltot-rho}. (a) Ratios to the serial run at $d = \CrepDsel$, for the shared lane (solid) and one lane per engine (dotted, extrapolation) of Equation~\eqref{eq:ntotal-rho}. The effective logical volume is that of Equation~\eqref{eq:Voleff} on the parallel schedule. (b) Per-shot failure probability with the shared lane for three distances. $d = \CrepDsel$ is our estimate, and $d = \DdLow$ an extrapolation. (c) Expected clock time against the total number of physical qubits at $d = \CrepDsel$, for one, five and ten chips, compared with $d = \DdLow$ on ten chips (extrapolation) and with parallel shots alone, with the shot model of Equation~\eqref{eq:twall}. The runtimes assume no routing stalls.}
\label{fig:time-efficient}
\end{figure*}

\paragraph{Results.}
Figure~\ref{fig:time-efficient} and Table~\ref{tab:time-efficient} give the results, with our estimate at $d = \CrepDsel$ in the top rows. Three observations stand out. First, the speed-up is close to $\rho$. With the shared lane, $\rho = 3$ costs \SI{\DoverheadThree}{\percent} more qubits than the serial run (\num{\DqubitsThree} against \num{\DqubitsOne}) and shortens the expected total runtime by a factor \DspeedupThree, from \SI{\DtotalOne}{\hour} to \SI{\DtotalThree}{\hour}. At the layout limit, $\rho = \DrhoMax$ costs \SI{\DoverheadMax}{\percent} more qubits and is \DspeedupMax{} times faster, \SI{\DtotalMax}{\hour} in total. Second, the effective logical volume falls with $\rho$, to \DvolRatioThree{} times the serial one at $\rho = 3$ and \DvolRatioMax{} times at $\rho = \DrhoMax$, but more and more slowly. Parallelism removes the idle volume of the exponent register but not the volume of the engines themselves. Third, the two levers combine. On \Dmachines{} chips with $\rho = \DrhoMax$, the expected clock time is \SI{\DwallMax}{\hour} on \num{\DqubitsMaxMachines} physical qubits.

Since the volume falls, the per-shot failure probability at $d = \CrepDsel$ drops from \SI{\DepsMainOne}{\percent} at $\rho = 1$ to \SI{\DepsMainMax}{\percent} at $\rho = \DrhoMax$, and $d = \DdLow$ meets the threshold from $\rho = \DrhoDdrop$ on. At $\rho = 2$, the margin is thin, with $\epsilon = \SI{\DepsTwoAtDlow}{\percent}$. These extrapolated configurations, in the bottom rows of Table~\ref{tab:time-efficient}, save \DsavingLowLo{} to \SI{\DsavingLowHi}{\percent} of the qubits but do not shorten the runtime, because the higher $\epsilon$ calls for more shots.

The last column of Table~\ref{tab:time-efficient} exceeds the time per shot because of the extra rounds of Equation~\eqref{eq:twall}. At $d = \CrepDsel$ the excess is small, with \DroundsOne{} rounds on average at $\rho = 1$ and \DroundsThree{} at $\rho = 3$. It grows at $d = \DdLow$, where $\epsilon$ reaches several percent, with \DroundsThreeLow{} rounds at $\rho = 3$. On parallel chips, the configurations at $d = \CrepDsel$ are therefore also faster. At $\rho = 3$ on \Dmachines{} chips, $d = \CrepDsel$ gives \SI{\DwallThree}{\hour} against \SI{\DwallThreeLow}{\hour} at $d = \DdLow$, for \num{\DqubitsThree} instead of \num{\DqubitsThreeLow} qubits per chip. All these numbers rest on the scheduling assumptions of Section~\ref{par:rsa-scope} and on the heuristic shot model of Equation~\eqref{eq:twall}.

\begin{table*}[t]
\centering
\begin{tabular}{@{}ccccccccc@{}}
\toprule
 & & & \multicolumn{2}{c}{Physical qubits per chip} & & & \multicolumn{2}{c}{Expected total runtime} \\
\cmidrule(lr){4-5}\cmidrule(l){8-9}
$\rho$ & $d$ & $\epsilon$ & shared lane & lane per engine & Time per shot & $\mathbb{E}[R]$ & 1 chip & 10 chips \\
\midrule
\multicolumn{9}{@{}l}{\emph{Estimate, $d = \CrepDsel$}} \\
1 & 15 & \SI{0.43}{\percent} & \num{49.6}\,k & \num{49.6}\,k & \SI{6.92}{\hour} & \num{9.25} & \SI{64.0}{\hour} & \SI{7.07}{\hour} \\
2 & 15 & \SI{0.34}{\percent} & \num{58.9}\,k & \num{60.8}\,k & \SI{3.54}{\hour} & \num{9.24} & \SI{32.7}{\hour} & \SI{3.62}{\hour} \\
3 & 15 & \SI{0.32}{\percent} & \num{68.2}\,k & \num{71.2}\,k & \SI{2.38}{\hour} & \num{9.24} & \SI{21.9}{\hour} & \SI{2.42}{\hour} \\
5 & 15 & \SI{0.29}{\percent} & \num{86.8}\,k & \num{89.8}\,k & \SI{1.44}{\hour} & \num{9.23} & \SI{13.3}{\hour} & \SI{1.47}{\hour} \\
7 & 15 & \SI{0.28}{\percent} & \num{105.4}\,k & \num{105.6}\,k & \SI{1.04}{\hour} & \num{9.23} & \SI{9.6}{\hour} & \SI{1.06}{\hour} \\
\midrule
\multicolumn{9}{@{}l}{\emph{Extrapolation, $d = \DdLow$}} \\
2 & 13 & \SI{9.75}{\percent} & \num{52.3}\,k & \num{54.2}\,k & \SI{3.39}{\hour} & \num{10.20} & \SI{34.6}{\hour} & \SI{4.45}{\hour} \\
3 & 13 & \SI{7.51}{\percent} & \num{60.9}\,k & \num{63.9}\,k & \SI{2.28}{\hour} & \num{9.95} & \SI{22.7}{\hour} & \SI{2.79}{\hour} \\
5 & 13 & \SI{5.69}{\percent} & \num{78.1}\,k & \num{81.1}\,k & \SI{1.38}{\hour} & \num{9.76} & \SI{13.5}{\hour} & \SI{1.60}{\hour} \\
7 & 13 & \SI{4.92}{\percent} & \num{95.2}\,k & \num{95.4}\,k & \SI{1.00}{\hour} & \num{9.68} & \SI{9.7}{\hour} & \SI{1.13}{\hour} \\
\bottomrule
\end{tabular}
\caption{RSA-2048 on the repetition-cat chip with $\rho$ residue engines sharing one exponent register, after Ref.~\cite[Sec.~VI.B.1]{Webster2026Pinnacle}, with the sequential modular merge of Section~\ref{par:modular-merge}, at $p_Z = \num{\CpZ}$. The top rows are our estimate, at the distance $d = \CrepDsel$ of Section~\ref{ssec:physical_resource_count} for every $\rho$. The bottom rows are extrapolations at $d = \DdLow$, shown only where $d = \DdLow$ also meets the criterion $\epsilon < \SI{\DepsCritPct}{\percent}$. The per-shot failure probability $\epsilon$ and the times are those of the shared lane. Qubits are given for both lane options of Equation~\eqref{eq:ntotal-rho}, the lane-per-engine counts being extrapolations. The times assume that the engines never stall each other. $\mathbb{E}[R]$ and the total runtimes follow the shot model of Equation~\eqref{eq:twall}, and the last column runs the shots in rounds of 10 simultaneous shots on 10 identical chips. Classical pre- and post-processing are not included.}
\label{tab:time-efficient}
\end{table*}

\paragraph{LDPC-cat chip.}\label{par:ldpc-engines}
The construction carries over to the LDPC-cat chip of Section~\ref{sec:ldpc-cat}. We keep the pair $(d_{\mathrm{rep}}, d_{\mathrm{LDPC}}) = (\ClDrep, \ClDldpc)$ selected in Section~\ref{sec:ldpc-cat-b} for every $\rho$, as we keep $d = \CrepDsel$ on the repetition-cat chip. The exponent register again holds most of the chip. Its $n_e$ logical qubits take \num{\DLexpQubits} of the \num{\DLqubitsOne} physical qubits (\SI{\DLexpShare}{\percent}), counting the LDPC layer and the access zone below it. Each extra engine appends its work zone, $l_{\mathrm{work}} = l_{\mathrm{mem}} + l_{\mathrm{comp}}$ columns, to the LDPC block, so that
\begin{equation}
\begin{split}
  L(\rho) &= l_{\mathrm{exp}} + \rho\, l_{\mathrm{work}}, \qquad k_L(\rho) = 2L(\rho),\\
  \Gamma_{\mathrm{LDPC}}(\rho) &= 2(H-1)\,L(\rho) + 2\Delta_{\mathrm{LDPC}} .
\end{split}
  \label{eq:ldpc-rho}
\end{equation}
In the repetition layer, each engine brings the access zone and the proxy row under its work zone, its routing zone (buffer) of length $l_{\mathrm{work}} + 2 + 9(n_{\mathrm{buf}}^{(\Adder)} - 1)$, its $n_{\mathrm{buf}}^{(\Adder)}$ adder buffers, and its factories with their routing. The access zone under the exponent zone, the left staircase and the right end of the block are shared. Equation~\eqref{eq:ntotal-ldpc-num} becomes
\begin{equation}
\begin{split}
  &n_{\mathrm{total}}^{(\mathrm{LDPC})}(\rho) = \Gamma_{\mathrm{LDPC}}(\rho) + (2H-1)\bigl(L(\rho) + e_R\bigr)\\
  &\quad + 2\Delta_L + 2h_{\mathrm{routing,buf}}\, e_R + \rho \bigl[ 2(l_{\mathrm{work}} + 2)\\
  &\qquad + 2h_{\mathrm{routing,buf}} \bigl(l_{\mathrm{work}} + 2 + 9(n_{\mathrm{buf}}^{(\Adder)} - 1)\bigr)\\
  &\qquad + 18\, n_{\mathrm{buf}}^{(\Adder)}\, d_{\mathrm{rep}} + 2h_{\mathrm{routing,fac}}\, l_{\mathrm{fac}}\\
  &\qquad - \max(l_{\mathrm{fac}} - l_{\mathrm{buf}}, 0) + n_{\mathrm{fac}} \bigr],
\end{split}
\label{eq:ntotal-ldpc-rho}
\end{equation}
which reduces to Equation~\eqref{eq:ntotal-ldpc-num} at $\rho = 1$ and adds \num{\DLengQubits} physical qubits per engine. The block only grows longer, so the layout does not limit $\rho$, but we stop at $\rho = 7$ for comparison with the repetition-cat chip.

The address-routing allowance $\Gamma_{\mathrm{addr}} = 4 l_{\mathrm{exp}}$ is shared. The engines read the exponent tips through the same two rows of the access zone, which Section~\ref{sec:ldpc-cat-a} sizes for sequential accesses. As for the shared lane of the repetition-cat chip, we assume that the concurrent accesses of the $\rho$ engines fit in these rows. Dedicated rows for up to three engines would fit in the rows of $\Gamma_{\mathrm{QFT}}$, which stay idle until the final transform.

The modular merge keeps its four additions. Its three parity-\textsc{Cnot} gates last $T_{\mathrm{surg}}^{(\mathrm{LDPC})}$ cycles, and the four moves of $L_j$ are moves between zones, each lasting $T_{\Reshuffle}^{\mathrm{swap}}(\kappa, 3) = \DLmoveCycles$ cycles (Equation~\eqref{eq:TReshuffle}). Hence
\begin{equation}
  T_{\mathrm{merge}}^{(\mathrm{LDPC})} = 4\, T_{\Adder}^{(\mathrm{LDPC})}(\kappa+1) + 3\, T_{\mathrm{surg}}^{(\mathrm{LDPC})} + 4\, T_{\Reshuffle}^{\mathrm{swap}}(\kappa, 3),
  \label{eq:tmerge-ldpc}
\end{equation}
that is \num{\DLmergeCycles} cycles. Equation~\eqref{eq:tshot-rho} applies with the durations of Table~\ref{tab:ldpc-cycles}. One erasure lasts $t_{\mathrm{acc}} = \num{\DLeraseCycles}$ cycles, including its parity-\textsc{Cnot} gates, and the final uncomputation of the exponent lasts $T_{\mathrm{end}} = \num{\DLtEndCycles}$ cycles. The merges and erasures add \SI{\DLmergeShareTwo}{\percent} to the shot at $\rho = 2$ and \SI{\DLmergeShareMax}{\percent} at $\rho = 7$. We evaluate the volumes of Equations~\eqref{eq:vol-phys-rep-ldpc} to~\eqref{eq:vol-eff-ldpc} on the parallel schedule. In the repetition layer, every engine contributes the buffer positions of its current slice, counted at the peak $9 n_{\mathrm{buf}}^{(\Adder)}$ during the merges, the erasures and the slices of $T_{\mathrm{other}}$. The global routing $\Gamma_{\mathrm{global}}^{(\mathrm{rep})}(\rho)$ follows the layout above. The LDPC layer counts its $k_L(\rho)$ logical qubits during the whole shot. The factory counts are those of Equation~\eqref{eq:efac}.

\begin{figure*}[t]
\centering
\includegraphics[width=\textwidth]{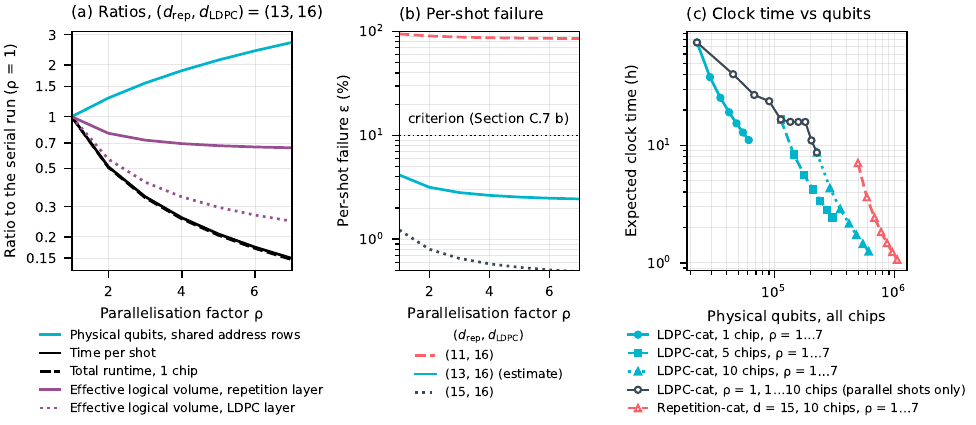}
\caption{Space-time trade-off for RSA-2048 on the LDPC-cat chip at $p_Z = \num{\CpZ}$, with $\rho \leq 7$ residue engines sharing one exponent register after Ref.~\cite[Sec.~VI.B.1]{Webster2026Pinnacle}, as in Figure~\ref{fig:time-efficient}. (a) Ratios to the serial run at $(d_{\mathrm{rep}}, d_{\mathrm{LDPC}}) = (\ClDrep, \ClDldpc)$, with shared address rows, Equation~\eqref{eq:ntotal-ldpc-rho}. The effective logical volumes are those of Equations~\eqref{eq:vol-eff-rep-ldpc} and~\eqref{eq:vol-eff-ldpc} on the parallel schedule. (b) Per-shot failure probability for three repetition distances at $d_{\mathrm{LDPC}} = \ClDldpc$, which keep the durations of Table~\ref{tab:ldpc-cycles}. $(\ClDrep, \ClDldpc)$ is our estimate. (c) Expected clock time against the total number of physical qubits for one, five and ten LDPC-cat chips, compared with parallel shots alone and with ten repetition-cat chips at $d = \CrepDsel$, with the shot model of Equation~\eqref{eq:twall}. The runtimes assume no routing stalls.}
\label{fig:time-efficient-ldpc}
\end{figure*}

Figure~\ref{fig:time-efficient-ldpc} and Table~\ref{tab:time-efficient-ldpc} give the results.
The per-shot failure probability falls from \SI{\DLepsOne}{\percent} at $\rho = 1$ to \SI{\DLepsMax}{\percent} at $\rho = 7$, so the pair of Section~\ref{sec:ldpc-cat-b} stays below the criterion.
$\rho = 3$ costs \SI{\DLoverheadThree}{\percent} more qubits than the serial run (\num{\DLqubitsThree} against \num{\DLqubitsOne}) and divides the expected total runtime by \DLspeedupThree, from \SI{\DLtotalOne}{\hour} to \SI{\DLtotalThree}{\hour}.
$\rho = 7$ costs \SI{\DLoverheadMax}{\percent} more qubits and is \DLspeedupMax{} times faster, \SI{\DLtotalMax}{\hour} in total.
An engine costs relatively more than on the repetition-cat chip, since the LDPC layer stores the exponent register more compactly. The LDPC-cat chip nevertheless stays smaller, with \num{\DLqubitsMax} physical qubits at $\rho = 7$ against \num{\DqubitsMax} on the repetition-cat chip.
On \Dmachines{} chips with $\rho = 7$, the expected clock time is \SI{\DLwallMax}{\hour} on \num{\DLqubitsMaxMachines} physical qubits.
The scheduling assumptions of Section~\ref{par:rsa-scope} apply to the shared address rows as they apply to the shared lane.

\begin{table*}[t]
\centering
\begin{tabular}{@{}ccccccc@{}}
\toprule
 & & & & & \multicolumn{2}{c}{Expected total runtime} \\
\cmidrule(l){6-7}
$\rho$ & $\epsilon$ & Qubits & Time per shot & $\mathbb{E}[R]$ & 1 chip & 10 chips \\
\midrule
1 & \SI{4.16}{\percent} & \num{22.6}\,k & \SI{7.84}{\hour} & \num{9.61} & \SI{75.3}{\hour} & \SI{8.69}{\hour} \\
2 & \SI{3.17}{\percent} & \num{29.0}\,k & \SI{4.01}{\hour} & \num{9.51} & \SI{38.1}{\hour} & \SI{4.32}{\hour} \\
3 & \SI{2.82}{\percent} & \num{35.4}\,k & \SI{2.69}{\hour} & \num{9.47} & \SI{25.4}{\hour} & \SI{2.87}{\hour} \\
5 & \SI{2.55}{\percent} & \num{48.2}\,k & \SI{1.63}{\hour} & \num{9.45} & \SI{15.4}{\hour} & \SI{1.73}{\hour} \\
7 & \SI{2.44}{\percent} & \num{61.1}\,k & \SI{1.18}{\hour} & \num{9.44} & \SI{11.1}{\hour} & \SI{1.25}{\hour} \\
\bottomrule
\end{tabular}
\caption{RSA-2048 on the LDPC-cat chip with $\rho$ residue engines sharing one exponent register, after Ref.~\cite[Sec.~VI.B.1]{Webster2026Pinnacle}, with the sequential modular merge of Section~\ref{par:modular-merge}, at $p_Z = \num{\CpZ}$ and $(d_{\mathrm{rep}}, d_{\mathrm{LDPC}}) = (\ClDrep, \ClDldpc)$ of Section~C~8~b for every $\rho$. Qubits follow Equation~\eqref{eq:ntotal-ldpc-rho}, with shared address rows. The times assume that the engines never stall each other. $\mathbb{E}[R]$ and the total runtimes follow Equation~\eqref{eq:twall}, and the last column uses 10 identical chips.}
\label{tab:time-efficient-ldpc}
\end{table*}

\subsection{FFDLP} \label{ssec:FFDLP-512}

\providecommand{\TBD}[1]{[TBD: #1]}

\newcommand{\ffdlpToffoli}{\ensuremath{4.70\times10^{6}}} 
\newcommand{\ffdlpEpsCCX}{\ensuremath{9.5\times10^{-7}}} 
\newcommand{\ffdlpEpsI}{\ensuremath{1.2\times10^{-5}}}   
\newcommand{\ffdlpPdeviant}{1.9\%}                       
\newcommand{\ffdlpER}{8.24}                              

\newcommand{\ffdlpRepD}{11}
\newcommand{\ffdlpRepNtotal}{9\,336}   
\newcommand{\ffdlpRepEps}{\ensuremath{1.61\times10^{-2}}}     
\newcommand{\ffdlpRepEpsFail}{\ensuremath{1.62\times10^{-2}}} 
\newcommand{\ffdlpRepTshot}{\ensuremath{1.37\times10^{8}}} 
\newcommand{\ffdlpRepTshotTime}{2.28~min}
\newcommand{\ffdlpRepShots}{8.37}      
\newcommand{\ffdlpRepTotalTime}{19.1~min}

\newcommand{\ffdlpLdpcNtotal}{6\,656}  
\newcommand{\ffdlpLdpcEpsRep}{\ensuremath{8.21\times10^{-3}}}  
\newcommand{\ffdlpLdpcEpsLdpc}{\ensuremath{1.53\times10^{-2}}} 
\newcommand{\ffdlpLdpcEpsTotal}{\ensuremath{2.34\times10^{-2}}}
\newcommand{\ffdlpLdpcEps}{\ensuremath{2.34\times10^{-2}}}     
\newcommand{\ffdlpLdpcTshot}{\ensuremath{1.51\times10^{8}}} 
\newcommand{\ffdlpLdpcTshotTime}{2.51~min}
\newcommand{\ffdlpLdpcShots}{8.44}
\newcommand{\ffdlpLdpcTotalTime}{21.2~min}

\subsubsection{Balanced configuration} \label{sssec:ffdlp_balanced}

\paragraph{Parameters.}
This section estimates the physical resources of FFDLP-512 on the repetition-cat and LDPC-cat architectures.
FFDLP uses the per-shot circuit of RSA-2048 with different register sizes.
The FFDLP-512 row of Table~\ref{tab:instances} gives $n = 512$, $s = 7$, $|P| = 591$, $\ell = 14$ and $\kappa = 26$.
The discrete logarithm has $m = 80$ bits (column $d$ of Table~\ref{tab:instances}).
The windows are $w_1 = 7$, $w_3 = 3$ and $w_4 = 5$.
Section~\ref{sec:input-register} gives an exponent register of $n_e = m + 2\lceil m/s\rceil = 80 + 2\times 12 = 104$ qubits.
The factories are those of Sections~\ref{ssec:repetition_cat_architecture} and~\ref{sec:ldpc-cat}, one $\ket{\CXCXX}$ factory and two $\ket{i}$ factories, which set $T_\mathrm{cad} = 21$ at the selected distances.
These factories are sized for the error budget of RSA-2048, and we keep them for simplicity.
Counting Toffoli gates as in Section~\ref{sec:C7-budget}, FFDLP-512 executes \ffdlpToffoli{} Toffoli gates per shot, about 160 times fewer than RSA-2048, so that the factories cause a failure during a shot with probabilities $\epsilon_{\ket{\CXCXX}} = \ffdlpEpsCCX$ and $\epsilon_{\ket{i}} = \ffdlpEpsI$ only.
We also still assume a cycle time of \SI{1}{\micro\second}, a reaction time of \SI{10}{\micro\second}, $p_Z = 5\times10^{-4}$ and no bit flips.

\paragraph{Layout.}
On the repetition-cat architecture, the layout keeps the zones of Figure~\ref{fig:global_layout}, with $l_\mathrm{mem} = 2\ell + \lceil \log_2 n_e \rceil = 28 + 7 = 35$, $l_\mathrm{comp} = 3(\kappa+1) + 1 = 82$ and $l_\mathrm{buf} = 9 n_{\mathrm{buf}}^{(\Adder)} = 27$.
At $d = 11$, Equation~\eqref{eq:adder_shared_buffer_count} gives $T_\mathrm{buf} = 2d + 20 = 42$ and would allow two adder buffers.
The buffer zone, however, also holds the \Lookup{} buffers and the \Lookup{} and \Phaseup{} auxiliary qubits (Section~\ref{sssec:global_layout_routing}).
For this reason, we keep $n_{\mathrm{buf}}^{(\Adder)} = 3$.
The sum $l_\mathrm{buf} + l_\mathrm{comp} + l_\mathrm{mem} = 144$ exceeds $n_e$.
The bottom rectangle therefore holds only the memory, compute and buffer zones.
The top rectangle holds the $n_e$ exponent qubits, above the memory and compute zones.
Counting the routing zones as in Equation~\eqref{eq:Gglobal} gives $\Gamma_\mathrm{global} = 2h_\text{routing,fac}\, l_\text{fac} + 2\times4\times 144 - 40 + 2d$, that is $1000 + 1152 - 40 + 2d = 2d + 2112$. The $40$ subtracted ancillae lie above the exposed top of the routing zone, in the $144 - n_e = 40$ columns not covered by the top rectangle (Section~\ref{sssec:global_layout_routing}). The $n_e = 104$ ancillae above the top rectangle are exposed as well.
Adding the $247$ repetition-code columns of the two rectangles other than the routing column (Equation~\eqref{eq:ntotal}) gives

\begin{equation}
  \begin{split}
  n_\mathrm{total}(d) &= 247\times 2d + \Gamma_\mathrm{global} + n_\mathrm{fac} - 104 \\
  &= 496\, d + 3880 .
  \end{split}
  \label{eq:ffdlp-ntotal-rep}
\end{equation}

On the LDPC-cat architecture, the layout of Figure~\ref{fig:ldpc-layout} applies unchanged.
Equation~\eqref{eq:ldpc-zones} gives $l_{\exp} = 52$, $l_\mathrm{mem} = 18$ and $l_\mathrm{comp} = 41$ (Table~\ref{tab:ffdlp-ldpc-layout}), hence $l_\mathrm{work} = 59$ and $L = 111$.
The distances are $d_\mathrm{rep} = 11$ and $d_\mathrm{LDPC} = 12$, hence $T^{(\mathrm{LDPC})}_\mathrm{surg} = d^\star = 12$ (Equation~\eqref{eq:Tsurg-ldpc}).
The code $[6L+17, 2L, 12]$ of Table~\ref{tab:adder_ldpc_codes}, i.e.\ $[683, 222, 12]$, has $H = 6$, $\Delta_\mathrm{LDPC} = 17$ and a right staircase of width $e_R = p_r = 4$.
Hence $n_L = 683$, $k_L = 222$ and $\Gamma_{\mathrm{LDPC}} = 1144$ (Equation~\eqref{eq:Gamma-ldpc}).
Since $\Delta_\mathrm{LDPC}$ is odd, the left staircase is rounded up to $\Delta_\mathrm{L} = \lceil \Delta_\mathrm{LDPC}/2 \rceil = 9$ qubits.
Equation~\eqref{eq:ntotal-ldpc} then gives \ffdlpLdpcNtotal{} physical qubits. The LDPC layer holds $1144$ of them, the access zone $1405$, the routing zone (buffer) $664$, the buffer zone $594$, the factory routing zone $1000 - 23 = 977$ and the factories $1872$. The routing zone (buffer) has length $l_\mathrm{routing,buf} = 83$.
The global routing of Equation~\eqref{eq:Gamma-global-ldpc} is $\Gamma^{(\mathrm{rep})}_\mathrm{global} = 977 + \Gamma_\mathrm{addr} + \Gamma_\mathrm{work} = 977 + 208 + 1497 = 2682$ and $\Gamma_\mathrm{QFT} = 364$ (Equation~\eqref{eq:Gamma-qft}), so that the identity of Equation~\eqref{eq:identity-ldpc} reads $3610 + 2682 + 364 = 6\,656$.

\begin{table*}[t]
\centering
\begin{tabular}{llrrrr}
\hline\hline
& & \multicolumn{3}{c}{LDPC-layer columns} & Buffer zone\\
Subroutine & Operation & Memory & Compute & Block & positions\\
\hline
\ComputeExponent{} & \Lookup$(7,21)$ & 13 & 32 & 97 & 18\\
 & Q-Q \Adder$(21)$ & 13 & 32 & 97 & 27\\
\hline
\ReduceExponent{} & Q-C \Adder$(15)$ & 17 & 26 & 95 & 27\\
 & Q-Q \Adder$(14)$ & 17 & 25 & 94 & 27\\
\hline
\ComputeResidue{} & \Lookup$(6,14)$ (loading) & 11 & 34 & 97 & 17\\
 & \Lookup$(6,15)$ & 18 & 36 & 106 & 17\\
 & Q-Q \Adder$^\dagger(15)$ & 18 & 36 & 106 & 27\\
 & Q-Q \Adder$(14)$ & 18 & 34 & 104 & 27\\
\hline
\AccumulateResidue{} & \Lookup$(5,27)$ & 18 & 41 & 111 & 16\\
 & Q-Q \Adder$^\dagger(27)$ & 18 & 41 & 111 & 27\\
 & Q-Q \Adder$(26)$ & 18 & 41 & 111 & 27\\
 & \Lookup$(5,26)$ & 18 & 41 & 111 & 16\\
 & Q-Q \Comparison$(26)$ & 18 & 41 & 111 & 27\\
 & \Phaseup$(5)$ & 18 & 41 & 111 & 17\\
\hline
\ComputeResidue{}$^\dagger$ & \Lookup$(6,15)$ & 18 & 36 & 106 & 17\\
 & Q-Q \Adder$^\dagger(15)$ & 18 & 36 & 106 & 27\\
 & Q-Q \Adder$(14)$ & 18 & 34 & 104 & 27\\
 & \Lookup$(6,14)$ & 18 & 34 & 104 & 17\\
 & Q-Q \Comparison$(14)$ & 18 & 36 & 106 & 27\\
 & \Phaseup$(6)$ & 18 & 34 & 104 & 20\\
 & \Phaseup$(6)$ (unload) & 11 & 13 & 76 & 20\\
\hline
Last \Phaseup{} & \Phaseup$(7)$ & 13 & 0 & 65 & 27\\
\hline\hline
\end{tabular}
\caption{Layout of each slice on the LDPC-cat architecture for FFDLP-512, with parameters ($\ell = 14$, $\kappa = 26$, $n_e = 104$, $w_1 = 7$), $n_{\mathrm{buf}}^{(\Adder)} = 3$ and $n_{\mathrm{buf}}^{(\Lookup)} = 2$.
Block adds the $l_{\mathrm{exp}} = 52$ columns of the exponent zone, used during the whole computation. 
As for RSA-2048, the accumulation register, packed in 13 columns, is parked in the memory zone during \ComputeExponent{} and \ReduceExponent{}, since $13 + 32 = 45 > 41$. 
The maxima, 18 memory columns, 41 compute columns and $L = 111$ columns of
the block, are reached in \AccumulateResidue{} and fix $l_{\mathrm{mem}}$, $l_{\mathrm{comp}}$ and $L$ in the global layout. 
The 27 buffer positions are used by every adder and by the last \Phaseup$(7)$, which fills the buffer zone. 
The table omits the $\mathrm{MZ}$, $\mathrm{Reshuffle}$, parity-\textsc{Cnot} and $\mathrm{P}_{\ket{+}}$ slices, which never exceed the adjacent \Adder{} or \Lookup{} slices.}
\label{tab:ffdlp-ldpc-layout}
\end{table*}

\paragraph{Results.}
On the repetition-cat architecture, the criterion of Section~\ref{sec:C7-budget} selects $d = \ffdlpRepD$, with \ffdlpRepNtotal{} physical qubits (Equation~\eqref{eq:ffdlp-ntotal-rep}).
With $p_\mathrm{logical} = 3.99\times10^{-12}$ (Equation~\eqref{eq:plogical}) and $\mathrm{Vol}^{(\mathrm{eff})}_\mathrm{logical} = 4.1\times10^{9}$, the exact form of Equation~\eqref{eq:eps} gives $\epsilon_\mathrm{logical} = \ffdlpRepEps$, and the factories bring the per-shot failure probability to $\epsilon = \ffdlpRepEpsFail$.
At $d = 9$, with $8\,344$ physical qubits, $\epsilon \approx 75\%$.

On the LDPC-cat architecture, $p^{(\mathrm{rep})}_\mathrm{logical} = 3.99\times10^{-12}$ and $p^{(\mathrm{LDPC})}_\mathrm{logical} = 0.1\times12\times(25.8\times5\times10^{-4})^{6} = 5.53\times10^{-12}$ (Equation~\eqref{eq:plogical-ldpc}), with $\mathrm{Vol}^{(\mathrm{eff,rep})}_\mathrm{logical} = 2.1\times10^{9}$ and $\mathrm{Vol}^{(\mathrm{eff,LDPC})}_\mathrm{logical} = 2.8\times10^{9}$.
Equations~\eqref{eq:eps-layers} and~\eqref{eq:eps-total-ldpc} then give $\epsilon_\text{logical}^{(\mathrm{rep})} = \ffdlpLdpcEpsRep$, $\epsilon_\text{logical}^{(\mathrm{LDPC})} = \ffdlpLdpcEpsLdpc$ and $\epsilon_\text{total} = \ffdlpLdpcEpsTotal$, hence $\epsilon = \ffdlpLdpcEps$ with the factories.
The criterion selects $(d_\mathrm{rep}, d_\mathrm{LDPC}) = (11, 12)$ (Table~\ref{tab:ffdlp-ldpc-scan}), with \ffdlpLdpcNtotal{} physical qubits. 
The pairs with fewer physical qubits, such as $(9, 12)$, give $\epsilon > 50\%$.

\begin{table*}[t]
\centering
\begin{tabular}{rrccccccc}
\hline\hline
$d_{\mathrm{rep}}$ & $d_{\mathrm{LDPC}}$ &
$\mathrm{Vol}^{(\mathrm{eff,rep})}_{\mathrm{logical}}$ &
$\mathrm{Vol}^{(\mathrm{eff,LDPC})}_{\mathrm{logical}}$ &
$\epsilon^{(\mathrm{rep})}_{\mathrm{logical}}$ &
$\epsilon^{(\mathrm{LDPC})}_{\mathrm{logical}}$ &
$\epsilon_{\mathrm{total}}$ & $\epsilon$ & \\
\hline
\multicolumn{9}{c}{$p_Z = 10^{-3}$}\\
9 & 12 & $3.0\times10^{9}$ & $2.8\times10^{9}$ & 1.00 & $6.27\times10^{-1}$ & 1.00 & 1.00 &  \\
   & 16 & $3.6\times10^{9}$ & $2.4\times10^{9}$ & 1.00 & $2.38\times10^{-3}$ & 1.00 & 1.00 &  \\
   & 22 & $4.5\times10^{9}$ & $2.1\times10^{9}$ & 1.00 & $3.73\times10^{-6}$ & 1.00 & 1.00 &  \\
11 & 12 & $2.1\times10^{9}$ & $2.8\times10^{9}$ & $3.97\times10^{-1}$ & $6.27\times10^{-1}$ & $7.75\times10^{-1}$ & $7.75\times10^{-1}$ &  \\
   & 16 & $2.5\times10^{9}$ & $2.4\times10^{9}$ & $4.55\times10^{-1}$ & $2.38\times10^{-3}$ & $4.56\times10^{-1}$ & $4.56\times10^{-1}$ &  \\
   & 22 & $3.0\times10^{9}$ & $2.1\times10^{9}$ & $5.26\times10^{-1}$ & $3.73\times10^{-6}$ & $5.26\times10^{-1}$ & $5.26\times10^{-1}$ &  \\
13 & 12 & $1.6\times10^{9}$ & $2.9\times10^{9}$ & $1.07\times10^{-2}$ & $6.38\times10^{-1}$ & $6.42\times10^{-1}$ & $6.42\times10^{-1}$ &  \\
   & 16 & $1.8\times10^{9}$ & $2.4\times10^{9}$ & $1.24\times10^{-2}$ & $2.38\times10^{-3}$ & $1.48\times10^{-2}$ & $1.49\times10^{-2}$ & $\blacktriangleleft$ \\
   & 22 & $2.2\times10^{9}$ & $2.1\times10^{9}$ & $1.53\times10^{-2}$ & $3.73\times10^{-6}$ & $1.53\times10^{-2}$ & $1.54\times10^{-2}$ &  \\
15 & 12 & $1.3\times10^{9}$ & $3.0\times10^{9}$ & $2.41\times10^{-4}$ & $6.59\times10^{-1}$ & $6.59\times10^{-1}$ & $6.60\times10^{-1}$ &  \\
   & 16 & $1.4\times10^{9}$ & $2.4\times10^{9}$ & $2.64\times10^{-4}$ & $2.38\times10^{-3}$ & $2.64\times10^{-3}$ & $2.75\times10^{-3}$ &  \\
   & 22 & $1.7\times10^{9}$ & $2.1\times10^{9}$ & $3.24\times10^{-4}$ & $3.73\times10^{-6}$ & $3.28\times10^{-4}$ & $4.38\times10^{-4}$ &  \\
\hline
\multicolumn{9}{c}{$p_Z = 5\times10^{-4}$}\\
9 & 12 & $3.0\times10^{9}$ & $2.8\times10^{9}$ & $5.61\times10^{-1}$ & $1.53\times10^{-2}$ & $5.68\times10^{-1}$ & $5.68\times10^{-1}$ &  \\
   & 16 & $3.6\times10^{9}$ & $2.4\times10^{9}$ & $6.28\times10^{-1}$ & $9.29\times10^{-6}$ & $6.28\times10^{-1}$ & $6.28\times10^{-1}$ &  \\
   & 22 & $4.5\times10^{9}$ & $2.1\times10^{9}$ & $7.04\times10^{-1}$ & $1.82\times10^{-9}$ & $7.04\times10^{-1}$ & $7.04\times10^{-1}$ &  \\
11 & 12 & $2.1\times10^{9}$ & $2.8\times10^{9}$ & $8.21\times10^{-3}$ & $1.53\times10^{-2}$ & $2.34\times10^{-2}$ & $2.34\times10^{-2}$ & $\blacktriangleleft$ \\
   & 16 & $2.5\times10^{9}$ & $2.4\times10^{9}$ & $9.83\times10^{-3}$ & $9.29\times10^{-6}$ & $9.84\times10^{-3}$ & $9.85\times10^{-3}$ &  \\
   & 22 & $3.0\times10^{9}$ & $2.1\times10^{9}$ & $1.21\times10^{-2}$ & $1.82\times10^{-9}$ & $1.21\times10^{-2}$ & $1.21\times10^{-2}$ &  \\
13 & 12 & $1.6\times10^{9}$ & $2.9\times10^{9}$ & $8.82\times10^{-5}$ & $1.58\times10^{-2}$ & $1.58\times10^{-2}$ & $1.59\times10^{-2}$ &  \\
   & 16 & $1.8\times10^{9}$ & $2.4\times10^{9}$ & $1.02\times10^{-4}$ & $9.29\times10^{-6}$ & $1.12\times10^{-4}$ & $1.25\times10^{-4}$ &  \\
   & 22 & $2.2\times10^{9}$ & $2.1\times10^{9}$ & $1.26\times10^{-4}$ & $1.82\times10^{-9}$ & $1.26\times10^{-4}$ & $1.39\times10^{-4}$ &  \\
15 & 12 & $1.3\times10^{9}$ & $3.0\times10^{9}$ & $9.94\times10^{-7}$ & $1.67\times10^{-2}$ & $1.67\times10^{-2}$ & $1.67\times10^{-2}$ &  \\
   & 16 & $1.4\times10^{9}$ & $2.4\times10^{9}$ & $1.09\times10^{-6}$ & $9.29\times10^{-6}$ & $1.04\times10^{-5}$ & $2.35\times10^{-5}$ &  \\
   & 22 & $1.7\times10^{9}$ & $2.1\times10^{9}$ & $1.34\times10^{-6}$ & $1.82\times10^{-9}$ & $1.34\times10^{-6}$ & $1.45\times10^{-5}$ &  \\
\hline\hline
\end{tabular}
\caption{Effective logical volumes and logical error probabilities per shot for FFDLP-512 on the LDPC-cat architecture, as a function of the distances $d_{\mathrm{rep}}$ and $d_{\mathrm{LDPC}}$, for $p_Z = 10^{-3}$ and $p_Z = 5\times10^{-4}$.
$T_{\mathrm{shot}}$ includes the moves between subroutines.
$\epsilon$ includes the factories, which cause a failure with probabilities
$\epsilon_{\ket{\CXCXX}} = 7.0\times10^{-5}$ and $\epsilon_{\ket{i}} = 4.0\times10^{-5}$ at $p_Z = 10^{-3}$, and $9.5\times10^{-7}$ and $1.2\times10^{-5}$ at $p_Z = 5\times10^{-4}$. 
All pairs use the surgery durations of Equation~\eqref{eq:Tsurg-ldpc}, $T_{\mathrm{cad}} = \max(T^{(\mathrm{LDPC})}_{\mathrm{surg}},21)$, that is 21 for $d_{\mathrm{LDPC}}\le16$ and 22 for $d_{\mathrm{LDPC}}=22$ (Equation~\eqref{eq:Tcad-ldpc}), and $n_{\mathrm{buf}}^{(\Adder)} = 3$. 
An entry of 1.00 denotes a probability above 0.995. 
The pairs marked by $\blacktriangleleft$ are selected as the pairs with the fewest physical qubits among those with $\epsilon < 10\%$ (criterion of Section~\ref{sec:C7-budget}): $(13,16)$ at $p_Z=10^{-3}$, with $7\,263$ physical qubits, and $(11,12)$ at $p_Z=5\times10^{-4}$, with $6\,656$ physical qubits.}
\label{tab:ffdlp-ldpc-scan}
\end{table*}

Durations follow from the call counts of Table~\ref{tab:logical_qubit_count} applied with the parameters above.
The $\ket{\CXCXX}$ factory sets $T_\mathrm{cad} = 21$ at every distance considered.
The register moves that Sections~\ref{sssec:global_layout_routing} and~\ref{ssec:ldpc_cat_layout} leave out of $T_\mathrm{shot}$ are included here.
All register moves take $0.67\%$ of a shot on the repetition-cat architecture, and moves between subroutines take $0.60\%$ on the LDPC-cat architecture.
One shot takes \ffdlpRepTshot{} cycles (\ffdlpRepTshotTime) on the repetition-cat architecture and \ffdlpLdpcTshot{} cycles (\ffdlpLdpcTshotTime) on the LDPC-cat architecture (Table~\ref{tab:ffdlp-ldpc-cycles}).

\begin{table*}[t]
\centering
\begin{tabular}{llrrrr}
\hline\hline
Subroutine & Operation & Calls & Cycles/call & Cycles & Share\\
\hline
\ComputeExponent{} & \Lookup$(7,21)$ & 8\,880 & 2721 & $2.42\times10^{7}$ & 16.04\%\\
 & Q-Q \Adder$(21)$ & 8\,880 & 882 & $7.83\times10^{6}$ & 5.20\%\\
 & MZ$(21,3)$ & 8\,880 & 96 & $8.52\times10^{5}$ & 0.57\%\\
 & Reshuffle$(21,3)$ & 8\,880 & 108 & $9.59\times10^{5}$ & 0.64\%\\
\hline
\ReduceExponent{} & Q-C \Adder$(15)$ & 4\,137 & 630 & $2.61\times10^{6}$ & 1.73\%\\
 & Q-Q \Adder$(14)$ & 4\,137 & 588 & $2.43\times10^{6}$ & 1.62\%\\
 & Reshuffle$(15,3)$ & 4\,137 & 108 & $4.47\times10^{5}$ & 0.30\%\\
 & Reshuffle$(2,1)$ & 4\,137 & 36 & $1.49\times10^{5}$ & 0.10\%\\
\hline
\ComputeResidue{} & \Lookup$(6,14)$ (loading) & 591 & 1302 & $7.69\times10^{5}$ & 0.51\%\\
 & \Lookup$(6,15)$ & 8\,865 & 1302 & $1.15\times10^{7}$ & 7.66\%\\
 & Q-Q \Adder$^\dagger(15)$ & 8\,865 & 630 & $5.58\times10^{6}$ & 3.71\%\\
 & MZ$(15,3)$ & 8\,865 & 96 & $8.51\times10^{5}$ & 0.57\%\\
 & Q-Q \Adder$(14)$ & 8\,865 & 588 & $5.21\times10^{6}$ & 3.46\%\\
 & MZ$(1)$ & 8\,865 & 12 & $1.06\times10^{5}$ & 0.07\%\\
 & Reshuffle$(14,3)$ & 7\,092 & 108 & $7.66\times10^{5}$ & 0.51\%\\
 & Reshuffle$(14,3)$ (swap) & 1\,773 & 120 & $2.13\times10^{5}$ & 0.14\%\\
 & MZ$(14,3)$ & 1\,773 & 96 & $1.70\times10^{5}$ & 0.11\%\\
\hline
\AccumulateResidue{} & \Lookup$(5,27)$ & 1\,773 & 603 & $1.07\times10^{6}$ & 0.71\%\\
 & Q-Q \Adder$^\dagger(27)$ & 1\,773 & 1134 & $2.01\times10^{6}$ & 1.33\%\\
 & MZ$(27,3)$ & 1\,773 & 96 & $1.70\times10^{5}$ & 0.11\%\\
 & Q-Q \Adder$(26)$ & 1\,773 & 1092 & $1.94\times10^{6}$ & 1.29\%\\
 & \Lookup$(5,26)$ & 886.5 & 603 & $5.35\times10^{5}$ & 0.35\%\\
 & Q-Q \Comparison$(26)$ & 886.5 & 1134 & $1.01\times10^{6}$ & 0.67\%\\
 & MZ$(26,3)$ & 886.5 & 96 & $8.51\times10^{4}$ & 0.06\%\\
 & \Phaseup$(5)$ & 1\,773 & 141 & $2.50\times10^{5}$ & 0.17\%\\
 & Reshuffle$(26,3)$ & 1\,773 & 108 & $1.91\times10^{5}$ & 0.13\%\\
\hline
\ComputeResidue{}$^\dagger$ & \Lookup$(6,15)$ & 17\,730 & 1302 & $2.31\times10^{7}$ & 15.33\%\\
 & Q-Q \Adder$^\dagger(15)$ & 17\,730 & 630 & $1.12\times10^{7}$ & 7.42\%\\
 & MZ$(15,3)$ & 17\,730 & 96 & $1.70\times10^{6}$ & 1.13\%\\
 & Q-Q \Adder$(14)$ & 17\,730 & 588 & $1.04\times10^{7}$ & 6.92\%\\
 & \Lookup$(6,14)$ & 8\,865 & 1302 & $1.15\times10^{7}$ & 7.66\%\\
 & Q-Q \Comparison$(14)$ & 8\,865 & 630 & $5.58\times10^{6}$ & 3.71\%\\
 & MZ$(14,3)$ & 8\,865 & 96 & $8.51\times10^{5}$ & 0.57\%\\
 & \Phaseup$(6)$ & 17\,730 & 252 & $4.47\times10^{6}$ & 2.97\%\\
 & Reshuffle$(14,1)$ (swap) & 1\,773 & 192 & $3.40\times10^{5}$ & 0.23\%\\
 & Reshuffle$(14,3)$ & 15\,957 & 108 & $1.72\times10^{6}$ & 1.14\%\\
 & Reshuffle$(14,3)$ (swap) & 1\,773 & 120 & $2.13\times10^{5}$ & 0.14\%\\
 & \Phaseup$(6)$ (unload) & 591 & 252 & $1.49\times10^{5}$ & 0.10\%\\
\hline
\ReduceExponent{}$^\dagger$ & Q-C \Adder$(15)$ & 4\,137 & 630 & $2.61\times10^{6}$ & 1.73\%\\
 & Q-Q \Adder$(14)$ & 4\,137 & 588 & $2.43\times10^{6}$ & 1.62\%\\
 & Reshuffle$(15,3)$ & 4\,137 & 108 & $4.47\times10^{5}$ & 0.30\%\\
 & Reshuffle$(2,1)$ & 4\,137 & 36 & $1.49\times10^{5}$ & 0.10\%\\
\hline
Last \Phaseup{} & \Phaseup$(7)$ & 15 & 399 & $5.98\times10^{3}$ & 0.00\%\\
\hline\hline
\end{tabular}
\caption{Breakdown of cycle counts by operation for FFDLP-512 on the LDPC-cat architecture, for $p_Z = 5\times10^{-4}$ and $(d_{\mathrm{rep}},d_{\mathrm{LDPC}})=(11,12)$. 
The total duration is $T_{\mathrm{shot}} = 1.51\times10^{8}$ cycles.
Lattice surgeries involving the LDPC layer last $T^{(\mathrm{LDPC})}_{\mathrm{surg}} = 12$ cycles, loadings of factory outputs $d_{\mathrm{rep}} = 11$ cycles, and adders run at the cadence $T_{\mathrm{cad}} = 21$(Equations~\eqref{eq:T-adder-ldpc} to~\eqref{eq:T-phaseup-ldpc}).
The \Lookup$(5,26)$, Q-Q \Comparison$(26)$ and $\mathrm{MZ}(26,3)$ of \AccumulateResidue{} are performed half of the time on average (Table~\ref{tab:logical_qubit_count}), hence their non-integer number of calls. 
For clarity, the table omits the parity-\textsc{Cnot} and $\mathrm{P}_{\ket{+}}$ slices, which contribute 0.60\% of $T_{\mathrm{shot}}$. 
Unlike in Table~\ref{tab:ldpc-cycles}, $T_{\mathrm{shot}}$ includes the moves between subroutines, which are not listed. 
The final inverse quantum Fourier transform ($1.1\times 10^{5}$ cycles, 0.08\% of $T_{\mathrm{shot}}$) is not included.}
\label{tab:ffdlp-ldpc-cycles}
\end{table*}

For $s = 7$, Equation~\eqref{eq:Rshots} with $P_\mathrm{deviant} \approx \ffdlpPdeviant$, estimated with Equation~\eqref{eq:kappa-budget} as for the FFDLP rows of Table~\ref{tab:instances}, gives $\mathbb{E}[R] \approx \ffdlpER$.
Accounting for $\epsilon$, the expected number of shots is \ffdlpRepShots{} on the repetition-cat architecture and \ffdlpLdpcShots{} on the LDPC-cat architecture, hence a total runtime of \ffdlpRepTotalTime{} and \ffdlpLdpcTotalTime{}, respectively.
The LDPC-cat architecture thus divides the physical qubit count by $1.4$ for an $11\%$ longer runtime.
One shot involves about $10^{12}$ physical qubit-cycles on either architecture, so that neglecting bit flips assumes a bit-flip probability well below $10^{-12}$ per qubit and cycle, against $10^{-15}$ for RSA-2048 (Section~\ref{sssec:layout_connectivity}).

\subsubsection{Space-efficient configuration} \label{sssec:ffdlp_smaller}
The method of Section~\ref{sssec:rsa_smaller} applies to the balanced configuration of Section~\ref{sssec:ffdlp_balanced}, on both architectures. 
The slower layout of the $\ket{\CXCXX}$ factory sets $T_{\mathrm{cad}} = 62$, and \Lookup{} durations follow Equation~\eqref{eq:T-lookup-general}. 
The factories and their routing take 1444 instead of 2849 physical qubits on the LDPC-cat architecture, and 1446 instead of 2872 on the repetition-cat architecture.

Table~\ref{tab:ffdlp-space-efficient} compares the two configurations. 
The criterion again selects $(d_{\mathrm{rep}},d_{\mathrm{LDPC}}) = (11,12)$ on the LDPC-cat architecture and $d = 11$ on the repetition-cat architecture.
On the LDPC-cat architecture, the physical qubit count decreases by 21\%, to $5\,251$, and one shot lasts \SI{6.69}{\minute}. 
With $\epsilon^{(\mathrm{rep})}_{\mathrm{logical}} = 1.9\times10^{-2}$ and $\epsilon^{(\mathrm{LDPC})}_{\mathrm{logical}} = 4.0\times10^{-2}$, $\epsilon = 5.8\%$. 
The expected 8.75 shots give a total runtime of \SI{58.5}{\minute}, 2.76 times longer than the balanced configuration. 
On the repetition-cat architecture, the configuration uses $7\,910$ physical qubits, 15\% fewer, and runs for \SI{56.1}{\minute}, 2.94 times longer. 
In the balanced configuration, the factories and their routing hold 43\% of the physical qubits on the LDPC-cat architecture and 31\% on the repetition-cat architecture, against 13\% for RSA-2048.
The qubit reduction is therefore larger than the 6.2\% of RSA-2048, for a similar increase in runtime.

\begin{table*}[t]
\centering
\begin{tabular}{@{}lccccccccccc@{}}
\toprule
 & $\ket{\CXCXX}$ & & & Physical & & \multicolumn{2}{c}{$T_{\mathrm{shot}}$} & & & Runtime & \\
\cmidrule(lr){7-8}
Configuration & layout & $T_{\mathrm{cad}}$ & Distances & qubits & Factories & cycles & min & $\epsilon$ & Shots & (min) & Qubit-hours \\
\midrule
\multicolumn{12}{@{}l}{\emph{LDPC-cat architecture}} \\
Balanced & third & 21 & $(11,12)$ & $6\,656$ & $2849$ & $1.51\times10^{8}$ & 2.51 & 2.3\% & 8.44 & 21.2 & $2.35\times10^{3}$ \\
Space-efficient & second & 62 & $(11,12)$ & $5\,251$ & $1444$ & $4.01\times10^{8}$ & 6.69 & 5.8\% & 8.75 & 58.5 & $5.12\times10^{3}$ \\
\midrule
\multicolumn{12}{@{}l}{\emph{Repetition-cat architecture}} \\
Balanced & third & 21 & $d = 11$ & $9\,336$ & $2872$ & $1.37\times10^{8}$ & 2.28 & 1.6\% & 8.37 & 19.1 & $2.97\times10^{3}$ \\
Space-efficient & second & 62 & $d = 11$ & $7\,910$ & $1446$ & $3.91\times10^{8}$ & 6.51 & 4.3\% & 8.61 & 56.1 & $7.39\times10^{3}$ \\
\bottomrule
\end{tabular}
\caption{FFDLP-512 at $p_Z = 5\times10^{-4}$ on the two architectures, with the third (balanced, Section~\ref{sssec:ffdlp_balanced}) and the second (space-efficient) layouts of the $\ket{\CXCXX}$ factory~\cite[Table~4]{FeMoco2026prep}. The Factories column includes the factory routing zone. The number of shots is $8.24/(1-\epsilon)$ (Equation~\eqref{eq:Rshots}), and the qubit-hours are the physical qubit count times the total runtime.}
\label{tab:ffdlp-space-efficient}
\end{table*}

\subsubsection{Time-efficient configuration} \label{sssec:ffdlp_faster}

\ifdefined\CfNe\else\fi
\ifdefined\DFrepTotal\else\fi
As for RSA-2048 (Section~\ref{sssec:rsa_faster}), the shots of FFDLP-512 can run in parallel on $k$ identical chips, with the expected clock time of Equation~\eqref{eq:twall} and $s + 1 = \CfShotsNeeded$ consistent shots.
Table~\ref{tab:time-efficient-ffdlp} gives the results. 
With $k = \CfShotsNeeded$ chips, the average duration drops from \SI{\DFrepTotal}{\minute} to \SI{\DFrepWallEight}{\minute} on the repetition-cat chip, and from \SI{\DFldpcTotal}{\minute} to \SI{\DFldpcWallEight}{\minute} on the LDPC-cat chip, with \DFrepRoundsEight{} and \DFldpcRoundsEight{} rounds on average. 
With 10 chips, the expected clock time approaches the duration of a single shot. The clock time reaches \SI{\DFrepWallTen}{\minute} on \num{\DFrepQubitsTen} physical qubits using the repetition-cat architecture, and \SI{\DFldpcWallTen}{\minute} on \num{\DFldpcQubitsTen} physical qubits using the LDPC-cat architecture.

We do not use parallel residue engines for FFDLP-512. 
Their benefit comes from sharing the exponent register, which holds only \SI{\DFrepExpShare}{\percent} of the physical qubits on the repetition-cat chip and \SI{\DFldpcExpShare}{\percent} on the LDPC-cat chip, against \SI{\DrepExpShare}{\percent} and \SI{\DLexpShare}{\percent} for RSA-2048. 
An extra engine would thus cost almost as much as an extra chip, and parallel shots reach the same clock times without the scheduling assumptions of Section~\ref{sssec:rsa_faster}.

\begin{table*}[t]
\centering

\begin{tabular}{@{}rrrrrrr@{}}
\toprule
 & \multicolumn{3}{c}{Repetition-cat} & \multicolumn{3}{c}{LDPC-cat} \\
\cmidrule(lr){2-4}\cmidrule(l){5-7}
$k$ & Qubits & Rounds & Clock time & Qubits & Rounds & Clock time \\
\midrule
1 & \num{9.3}\,k & \num{8.37} & \SI{19.1}{\minute} & \num{6.7}\,k & \num{8.44} & \SI{21.2}{\minute} \\
2 & \num{18.7}\,k & \num{4.29} & \SI{9.8}{\minute} & \num{13.3}\,k & \num{4.34} & \SI{10.9}{\minute} \\
4 & \num{37.3}\,k & \num{2.27} & \SI{5.2}{\minute} & \num{26.6}\,k & \num{2.31} & \SI{5.8}{\minute} \\
8 & \num{74.7}\,k & \num{1.26} & \SI{2.9}{\minute} & \num{53.2}\,k & \num{1.30} & \SI{3.3}{\minute} \\
10 & \num{93.4}\,k & \num{1.01} & \SI{2.3}{\minute} & \num{66.6}\,k & \num{1.02} & \SI{2.6}{\minute} \\
\bottomrule
\end{tabular}
\caption{FFDLP-512 on $k$ identical chips running shots in parallel, at $p_Z = \num{\CpZ}$, with the distances of Section~D~2~a: $d = \CfRepD$ on the repetition-cat chip and $(d_{\mathrm{rep}}, d_{\mathrm{LDPC}}) = (\CfLdpcDrep, \CfLdpcDldpc)$ on the LDPC-cat chip. Rounds and clock times follow Equation~\eqref{eq:twall} with $s + 1 = \CfShotsNeeded$ consistent shots. Classical pre- and post-processing are not included.}
\label{tab:time-efficient-ffdlp}
\end{table*}

\subsection{ECDLP} \label{ssec:ECDLP_tradeoffs}
\subsubsection{Balanced configuration} \label{sssec:ecdlp_balanced}

This section estimates the physical resources of ECDLP on secp256k1 ($n = 256$), on the LDPC-cat architecture only.
The circuit is the gate-optimized windowed point addition of~\cite{Schrottenloher26PointAdd} (Section~\ref{ssec:ecdlp-logical}).
Table~\ref{tab:ecdlp-params} gives the circuit parameters, and Tables~\ref{tab:ecdlp-qubits-a} and~\ref{tab:ecdlp-qubits-b} the operation counts.
One successful run suffices to recover $k$ (Section~\ref{sec:ecdlp} and Table~\ref{tab:ecdlp-relaxation}).
The window register of $w$ qubits is recycled (row ``Window recycling'' of Table~\ref{tab:ecdlp-qubits-b}), so the layout has no exponent zone.
The \Adder, the \Lookup, the Z-basis measurements, the parity-\textsc{Cnot} gates and the preparations take the costs compiled in Section~\ref{ssec:ldpc_cat_layout}.
The other operations are decomposed into these compiled operations, with a duration estimate but no dedicated layout or schedule.
The volume counts both layers, excluding the factories but including their routing zone, as allocated during the whole shot.
The resulting estimates are therefore approximate.
As in Section~\ref{ssec:FFDLP-512}, we assume a cycle time of \SI{1}{\micro\second}, a reaction time of \SI{10}{\micro\second}, $p_Z = 5\times10^{-4}$ and no bit flips.

\paragraph{Operations without a dedicated compilation.}
The compiled operations take the LDPC-cat durations of Section~\ref{sec:ldpc-cat-duration}.
All the surgeries of this section involve the LDPC layer, so $T_\mathrm{surg}$ denotes $T^{(\mathrm{LDPC})}_{\mathrm{surg}} = d^{\star}$ (Equation~\eqref{eq:Tsurg-ldpc}) throughout.
As in Section~\ref{sec:ldpc-cat}, operations are written in the color-changed convention of Section~\ref{ssec:colour_changed}.
The $\CX(x)$, $M_X(x)$ and $P_{\ket{0}}(x)$ rows of Tables~\ref{tab:ecdlp-qubits-a} and~\ref{tab:ecdlp-qubits-b} thus become parity-\textsc{Cnot} gates $\mathrm{PC}X(x)$, Z-basis measurements $M_Z(x)$ and $\ket{+}$ preparations $P_{\ket{+}}(x)$.
An \Adder{} and a \Lookup{} last $T_{\Adder}^{(\mathrm{LDPC})}$ and $T_{\Lookup}^{(\mathrm{LDPC})}$ (Equations~\eqref{eq:ldpc-tadder} and~\eqref{eq:T-lookup-ldpc}), and $\mathrm{PC}X(x)$ and $P_{\ket{+}}(x)$ last $T_\mathrm{surg}$.
A Q-C adder is costed as a Q-Q adder of the same size, and a comparison of size $m$ as an \Adder$(m+1)$.
Every $M_Z(x)$ is costed $T_{M_Z}(x,1)$, an upper bound since $T_{M_Z}(x,k) \le T_{M_Z}(x,1)$ (Equation~\eqref{eq:TMZ}).
We assume that each Toffoli or \textsc{And} gate outside the compiled operations is limited by the production of its resource states, and lasts one cadence step
\begin{equation*}
  T_{\mathrm{cad}} = \max\!\left(T_{\mathrm{surg}},\, T_{\mathrm{prod}}^{\ket{\CXCXX}},\, T_{\mathrm{prod}}^{\ket{i}}\right)
\end{equation*}
(Equation~\eqref{eq:Tcad-ldpc}).
This duration includes the \textsc{Cnot} gate that copies the \textsc{And} output into an existing target.
We assume that such a gate consumes at most one $\ket{\CXCXX}$ state and four $\ket{i}$ states, as in one adder step (Section~\ref{sssec:resource_states}).
We neglect the catalyst seeds of isolated gates.
An \textsc{And} gate is uncomputed by measurement. Its target is measured in the $X$ basis and, after the reaction time, a $\CZ$ correction is applied to its controls.
Each uncomputation therefore lasts about $2T_\mathrm{surg} + T_\mathrm{react}$.

\emph{Controlled Q-Q adder.}
A controlled adder maps $\ket{c}\ket{x}\ket{y}$ to $\ket{c}\ket{x}\ket{y + c\,x \bmod 2^n}$.
We implement the controlled adder with two uncontrolled adders and two multi-target
\textsc{Cnot} gates, and no other non-Clifford gate.
The construction is the single-term case of the controlled add-subtract method of Ref.~\cite{litinski2024schoolbook}.
Two qubits in $\ket{0}$ extend both registers to $n+1$ qubits.
The first qubit lies below the least significant qubit of $y$, so that the extended
register holds $Y = 2y$.
The second qubit lies above the most significant qubit of $x$.
A first adder computes $Y \leftarrow Y + x$.
A multi-target \textsc{Cnot} gate $\CX(n+1)$ (Section~\ref{sssec:multitarget_cnot}) controlled by $c$
then complements $Y$ when $c = 1$.
An $\Adder^\dagger$ subtracts $x$, and a second $\CX(n+1)$ undoes the
complement.
The bitwise complement maps $Y$ to $-Y-1 \bmod 2^{n+1}$, so the conjugated subtraction
acts as $Y \leftarrow Y - x$ if $c = 0$ and as $Y \leftarrow Y + x$ if $c = 1$.
The extended register therefore ends in $Y = 2(y + c\,x) \bmod 2^{n+1}$.
Its $n$ most significant qubits hold $y + c\,x \bmod 2^n$, and both added qubits return to $\ket{0}$.
A controlled $n$-bit adder therefore runs on the layout, buffers and factories of an
$(n+1)$-bit adder, and consumes twice the resource states of an $(n+1)$-bit adder.
On the LDPC-cat architecture, the controlled adder occupies the $3(n+1)$ slots of an
$(n+1)$-bit adder (Section~\ref{sssec:adder_physical_implementation_LDPC}) and the control qubit.
Executing the four operations sequentially bounds its duration by
\begin{equation}
  T_{\Adder}^{\mathrm{ctrl}}(n) \le 2\,T_{\Adder}(n+1) + 2\,T_{\mathrm{surg}},
  \label{eq:ctrl-adder}
\end{equation}
where $T_{\Adder}$ is given by Equation~\eqref{eq:ldpc-tadder}.
The second term corresponds to the duration of the two multi-target \textsc{Cnot} gates.
Controlled Q-C adders are costed as controlled Q-Q adders of the same size, which is an overestimate.

\emph{Controlled comparison.}
A controlled comparison maps $\ket{c}\ket{x}\ket{y}\ket{0}$ to $\ket{c}\ket{x}\ket{y}\ket{c\,b}$, with $b$ the comparison bit of $x$ and $y$.
The comparison circuit computes $b$ as the last carry of a subtraction.
An \textsc{And} gate controlled by $c$ and $b$ then copies $c\,b$ into an auxiliary logical qubit prepared in $\ket{0}$.
The comparison circuit finally uncomputes its carries, $b$ included, so only $c\,b$ remains.
Since we cost a comparison of size $m$ as an \Adder$(m+1)$, adding the duration of the \textsc{And} gate gives
\begin{equation}
  T^{\mathrm{ctrl}}_{\mathrm{Comp}}(m) \simeq T_{\Adder}(m+1) + T_{\mathrm{cad}}.
  \label{eq:ctrl-comp}
\end{equation}
The controlled comparison occupies the $3(m+1)$ slots of an \Adder$(m+1)$, and two slots, for the control and the auxiliary qubit.

\emph{CSWAP.}
A CSWAP on one pair of qubits uses two \textsc{Cnot} gates and one Toffoli gate.
With all gates and corrections executed sequentially, the CSWAP of two registers of size $n$ lasts
\begin{equation}
    T_\mathrm{CSWAP}(n) \simeq n\big[T_{\mathrm{cad}} + 2T_{\mathrm{surg}} + (2T_\mathrm{surg} + T_\mathrm{react})\big].
    \label{eq:ecdlp-cswap}
\end{equation}

\emph{MCX.}
A multi-controlled \textsc{Not} gate with $n$ controls, $\mathrm{MC}X(n)$, chains $n-1$ \textsc{And} gates using $n-2$ auxiliary qubits.
The auxiliary qubits are then uncomputed by measurement from the last one to the first, since each correction acts on the previous auxiliary qubit.
The duration is then about $T_{\mathrm{MC}X}(n) \simeq (n-1)\,T_\mathrm{cad} + (n-2)(2T_\mathrm{surg} + T_\mathrm{react})$.

\emph{\textsc{Absorber} and \textsc{Swapper}.} 
The \textsc{Absorber} is the in-place compression circuit of~\cite[Sec.~3.1 and Fig.~1]{Schrottenloher26PointAdd}.
The \textsc{Swapper} uses that circuit twice sequentially.
The circuit is composed of five Toffoli gates and six \textsc{Cnot} gates.
We execute all gates and corrections sequentially, which yields
\begin{equation}
    \begin{aligned}
        T_{\textsc{Absorber}} & \simeq 5\big[T_{\mathrm{cad}} + (2T_\mathrm{surg} + T_\mathrm{react})\big] + 6T_{\mathrm{surg}},\\
        T_{\textsc{Swapper}} & \simeq 2T_{\textsc{Absorber}}.
    \end{aligned}
    \label{eq:ecdlp-absorber}
\end{equation}

\emph{\Phaseup.} 
The \Phaseup{}s in ECDLP work on a window of size $w=16$, whose power products need more storage than the buffer zone offers.
Therefore, we store the $N_{\textsc{And}} = 2^h+2^l-w-2$ auxiliary qubits of the power products in free logical slots of the LDPC layer instead of repetition-code columns.
The layout below has room for these auxiliary qubits.
The duration of the \Phaseup{} remains the approximate one of Equation~\eqref{eq:T-phaseup-ldpc}.

\paragraph{Layout.}
An adder of size $m$ occupies $3m$ slots of the LDPC layer, for its inputs and its output (Section~\ref{sssec:adder_physical_implementation_LDPC}).
The width at a row of Tables~\ref{tab:ecdlp-qubits-a} and~\ref{tab:ecdlp-qubits-b} is therefore the subroutine scope, plus the out-of-scope qubits, plus the physical footprint of the operation.
This footprint counts $3m$ slots for an adder instead of its logical count.
As in Ref.~\cite[Sec.~3.2]{Schrottenloher26PointAdd}, where the Bézout reconstruction (replay) step dominates the space,
the widest point is the Q-Q \Adder$(n+1)$ controlled on $b_0$ of the replay in in-place div and in-place mul, with
\begin{equation}
  \begin{split}
  N_{\mathrm{max}} &= (G + a - 3) + w + 1 + 3(n+2) \\
  &= G + a + w + 3n + 4 = 1468
  \end{split}
  \label{eq:ecdlp-nmax}
\end{equation}
logical qubits.
The terms are the subroutine scope $G+a-3$, the $w$ out-of-scope qubits, the control $b_0$ and the $3(n+2)$ slots of the controlled Q-Q \Adder$(n+1)$ of Equation~\eqref{eq:ctrl-adder}.
The residual ancilla count $a$ is approximate (Table~\ref{tab:ecdlp-params}).
All other rows stay below $N_\mathrm{max}$.
In particular, the widest controlled comparison, the \Comparison$(\mu_R)$ of the replay, needs $(G+2n+a-2\mu_R-2) + w + 1 + 3(\mu_R+1) + 1 = G+2n+a+w+\mu_R+3$ logical qubits, $n-\mu_R+1 = 217$ below $N_\mathrm{max}$.
The $\C\SWAP(n)$ in the replay needs $G+a-1 + w + 1 + 2n$ logical qubits, $n+4=260$ below $N_\mathrm{max}$.
The closest are the controlled Q-C \Adder$(\varphi_R)$ of the replay, $111$ below $N_\mathrm{max}$, the coordinate \Adder$(n+1)$, $171$ below, and the controlled \Adder$^{\dagger}(\nu_j)$ at $\nu_j = n$, $187$ below.
The \Phaseup{} rows need $2n+w$ logical qubits plus the $N_{\textsc{And}}$ product slots.
For the parameters of Table~\ref{tab:ecdlp-params}, the \Phaseup{} rows need $1022$ logical qubits, which stays below $N_\mathrm{max}$.

With two logical qubits per column, the LDPC code has $L = \lceil N_\mathrm{max}/2 \rceil$, $k_L = 2L$ and $n_L = HL + \Delta_\mathrm{LDPC}$.
Table~\ref{tab:adder_ldpc_codes} gives $H$ and $\Delta_\mathrm{LDPC}$ at the chosen $d_\mathrm{LDPC}$.
The layout is that of Figure~\ref{fig:global_layout_ldpc} without the exponent zone, so all logical qubits lie in one work zone of length $L$.
The routing zone (buffer), of height $h_\text{routing,buf} = 4$, and its proxy row span this whole length, so \Adder{}s, \Lookup{}s and CSWAPs reach every logical qubit.
The routing zone (buffer) thus has length $l_\text{routing,buf} = L + 2 + e_R + 9(n_{\mathrm{buf}}^{(\Adder)}-1)$.
The factory zone and the routing zone (factory) are those of Section~\ref{sec:ldpc-factories}.

The buffer zone holds $9n_{\mathrm{buf}}^{(\Adder)}$ repetition-code columns of distance $d_\mathrm{rep}$, with $n_{\mathrm{buf}}^{(\Adder)}$ from Equation~\eqref{eq:Tbuf-ldpc}.
The buffer zone hosts, in turn, the adder buffers and the \Lookup{} buffers with their auxiliary qubits.
We assume that the other operations that need resource states are not limited by the space in the buffer zone.
The \Lookup{} needs $6n_{\mathrm{buf}}^{(\Lookup)}$ buffer columns and $w-1$ auxiliary qubits.
The buffer zone therefore suffices if $6n_{\mathrm{buf}}^{(\Lookup)} + w - 1 \le 9n_{\mathrm{buf}}^{(\Adder)}$.
At the distances selected below, $n_{\mathrm{buf}}^{(\Adder)} = 3$, and the condition holds with equality.

Equation~\eqref{eq:ntotal-ldpc}, with $l_\mathrm{work} = L$ and no $\Gamma_\mathrm{addr}$ or $\Gamma_\mathrm{QFT}$, gives
\begin{align}
  n^{(\mathrm{LDPC})}_{\mathrm{total}}
  &= \Gamma_{\mathrm{LDPC}} + \big[(2H-1)L + 2(L+2) + 2\Delta_{\mathrm{L}}\nonumber\\
  &\qquad + (2H-1)\,e_R\big] + 2h_\text{routing,buf}\,l_\text{routing,buf} \nonumber\\
  &\quad + 2d_{\mathrm{rep}}\,l_\text{buf} + 2h_\text{routing,fac}\,l_\text{fac} \nonumber\\
  &\quad - \max(l_\text{fac} - l_\text{buf}, 0) + n_\text{fac} \nonumber\\
  &= (4H+7)L + 3\Delta_{\mathrm{LDPC}} + (2H+7)\,e_R \nonumber\\
  &\quad + 18\,n_{\mathrm{buf}}^{(\Adder)} d_{\mathrm{rep}} + 81\,n_{\mathrm{buf}}^{(\Adder)} - 52 \nonumber\\
  &\quad + (2h_\text{routing,fac}-1)\,l_\text{fac} + n_\text{fac}.
  \label{eq:ecdlp-ntotal}
\end{align}
The second equality uses $\Gamma_\mathrm{LDPC} = 2(H-1)L + 2\Delta_\mathrm{LDPC}$ (Equation~\eqref{eq:Gamma-ldpc}), $\Delta_\mathrm{L} = \Delta_\mathrm{LDPC}/2$, $h_\text{routing,buf} = 4$, $l_\text{buf} = 9n_{\mathrm{buf}}^{(\Adder)}$ and $l_\text{fac} \ge l_\text{buf}$.

\paragraph{Duration, distances and factories.}
One shot lasts
\begin{equation}
  T_{\mathrm{shot}} = \sum_{o} N_o\,T_o ,
  \label{eq:ecdlp-tshot}
\end{equation}
where the sum runs over the rows of Tables~\ref{tab:ecdlp-qubits-a} and~\ref{tab:ecdlp-qubits-b}.
Row $o$ has $N_o$ calls of duration $T_o$, from Section~\ref{sec:ldpc-cat-duration} or the equations of this section, at $d^\star$.
The registers $u$ and $v$ of the dialog are truncated at iteration $j$ to
\begin{equation}
  \nu_j = \min\!\left(n,\ \big\lceil n - \tfrac{1}{2}\log_2(8/3)\,j \big\rceil + p\right)
  \label{eq:ecdlp-nuj}
\end{equation}
qubits~\cite[Sec.~3.1]{Schrottenloher26PointAdd}, with $p$ the register padding of Table~\ref{tab:ecdlp-params}.
For secp256k1, $\nu_j = n$ for the first $54$ iterations, then $\nu_j$ decreases to $10$.
The durations of Equations~\eqref{eq:ldpc-tadder}, \eqref{eq:ctrl-adder} and~\eqref{eq:ecdlp-cswap} are affine in the size.
The $N_\mathrm{it}$ iterations of the dialog rows $\C\SWAP(\nu_j)$ and controlled \Adder$^{\dagger}(\nu_j)$ therefore last as long as $N_\mathrm{it}$ calls at the mean width
\begin{equation*}
  \bar\nu = \frac{1}{N_\mathrm{it}} \sum_{j=0}^{N_\mathrm{it}-1} \nu_j \simeq 149,
\end{equation*}
which we use for these rows.
The semi-classical inverse quantum Fourier transform on the window registers is not included.
Its $N_A w = 448$ rotations, applied by phase kickback as in Equation~\eqref{eq:T-qft-ldpc}, take about $5\times10^{5}$ cycles and $1.1\times10^{4}$ Toffoli gates, less than $2\times10^{-4}$ of $T_{\mathrm{shot}}$ and of the Toffoli count.
Equation~\eqref{eq:ecdlp-tshot} then yields $3.26\times10^{9}$ cycles, that is \SI{54}{\minute}, which we round up to $T_{\mathrm{shot}} = 60$~min to leave slack for the register moves and the QFT.
The volumes and error probabilities below use the rounded $T_{\mathrm{shot}}$.

The LDPC layer contributes $\text{Vol}^{(\mathrm{eff,LDPC})}_\text{logical} = k_L T_\text{shot}/d_\mathrm{LDPC}$ (Equation~\eqref{eq:vol-eff-ldpc}).
The repetition layer, factories excepted, contributes $\text{Vol}^{(\mathrm{rep})}_\text{physical} \le (n^{(\mathrm{LDPC})}_\mathrm{total} - \Gamma_{\mathrm{LDPC}} - n_\text{fac})\,T_\text{shot}$, converted by Equation~\eqref{eq:vol-eff-rep-ldpc}.
Since the routing zone (buffer) spans the whole length $L$, this bound stays close to the actual allocation.
Equations~\eqref{eq:error_rep_layer} to~\eqref{eq:total_error_ldpc} then give the error probabilities.

The factories of Table~\ref{tab:adder_reference_factories} run at the cadence of Equation~\eqref{eq:Tcad-ldpc}, as in Section~\ref{sec:ldpc-factories}.
The state counts $N_{\ket{\CXCXX}}$ and $N_{\ket{i}}$ are not summed over the rows of Tables~\ref{tab:ecdlp-qubits-a} and~\ref{tab:ecdlp-qubits-b}.
We take instead the Toffoli count of the gate-optimized variant of~\cite[Table~2]{Schrottenloher26PointAdd} for the whole run, $N_{\ket{\CXCXX}} = 2^{25.78} \simeq 5.76\times10^{7}$, one $\ket{\CXCXX}$ state per Toffoli gate.
We assume four $\ket{i}$ states per Toffoli gate, as in one adder step (Section~\ref{sssec:adder_scheduling}), hence $N_{\ket{i}} = 4N_{\ket{\CXCXX}} \simeq 2.30\times10^{8}$.
RSA-2048 executes $7.56\times10^{8}$ Toffoli gates per shot (Section~\ref{sec:C7-budget}), about $13$ times more, so the factories sized for RSA-2048 suffice for ECDLP-256.
As in Section~\ref{sec:C7-budget}, the factories cause a failure with probability
\begin{equation}
  p_{\mathrm{fac}} = 1 - \big(1 - N_{\ket{\CXCXX}}\,p^{\ket{\CXCXX}}_{Z,\mathrm{out}}\big)\big(1 - N_{\ket{i}}\,p^{\ket{i}}_{Z,\mathrm{out}}\big),
  \label{eq:ecdlp-pfac}
\end{equation}
with the output infidelities $p^{j}_{Z,\mathrm{out}}$ of Table~\ref{tab:adder_reference_factories}.
As in Section~\ref{sec:ldpc-factories}, the per-shot failure probability is $\epsilon = 1 - (1-\epsilon_\text{total})(1-p_\mathrm{fac})$.
We apply the criterion of Section~\ref{sec:C7-budget} to the pairs $(d_\mathrm{rep}, d_\mathrm{LDPC})$ with $d_\mathrm{LDPC}$ from Table~\ref{tab:adder_ldpc_codes} and odd $d_\mathrm{rep}$ from $9$ to $19$.
Among the pairs for which $\epsilon$ is below $10\%$, the criterion chooses the one with the fewest physical qubits $n^{(\mathrm{LDPC})}_{\mathrm{total}}$ (Equation~\eqref{eq:ecdlp-ntotal}).

A run fails with probability at most $p_\mathrm{fail} = \epsilon_\text{total} + p_\mathrm{fac} + \delta_\times$, with $\delta_\times$ the corruption bound of Table~\ref{tab:ecdlp-relaxation}.
A failed run is detected classically by checking $Q = [k]P$.
The expected number of runs is then $1/(1-p_\mathrm{fail})$, and the total runtime is $T_\text{shot}/(1-p_\mathrm{fail})$.

\paragraph{Results.}
The criterion selects $(d_\mathrm{rep}, d_\mathrm{LDPC}) = (13, 16)$, with $H = 7$ and $\Delta_\mathrm{LDPC} = 26$.
Hence $L = 734$, $k_L = 1468$, $n_L = 5164$ and $n^{(\mathrm{LDPC})}_\mathrm{total} = 29\,588$ physical qubits, against $29\,696$ for the next pair with $\epsilon < 10\%$, $(15,16)$.
One shot takes $3.26\times10^{9}$~cycles (\SI{54}{\minute}), rounded up to $T_\mathrm{shot} = 3.6\times10^{9}$~cycles (\SI{60}{\minute}).
Then $\epsilon_\text{logical}^{(\mathrm{rep})} = 1.2\times10^{-2}$, $\epsilon_\text{logical}^{(\mathrm{LDPC})} = 1.3\times10^{-3}$, $\epsilon_\text{total} = 1.3\times10^{-2}$, $p_\mathrm{fac} = 2.3\times10^{-4}$, $\epsilon = 1.3\times10^{-2}$ and $p_\mathrm{fail} = 2.2\times10^{-2}$.
The expected number of runs is $1.03$, and the total runtime \SI{61.8}{\minute}.

\subsubsection{Space-efficient configuration} \label{sssec:ecdlp_smaller}
The method of Section~\ref{sssec:rsa_smaller} applies to the balanced configuration of Section~\ref{sssec:ecdlp_balanced}.
The slower layout of the $\ket{\CXCXX}$ factory sets $T_{\mathrm{cad}} = 62$, and the factories and their routing take 1444 instead of 2849 physical qubits.

The criterion keeps $(d_{\mathrm{rep}},d_{\mathrm{LDPC}}) = (13,16)$, with $28\,183$ physical qubits, 4.7\% fewer than the balanced configuration. 
Equation~\eqref{eq:ecdlp-tshot} gives $7.61\times10^{9}$ cycles (\SI{127}{\minute}) instead of $3.26\times10^{9}$.
As in Section~\ref{sssec:ecdlp_balanced}, a margin of about 10\% on the duration brings $T_{\mathrm{shot}}$ to $8.41\times10^{9}$ cycles (\SI{140}{\minute}). 
Then $\epsilon_\text{logical}^{(\mathrm{rep})} = 2.7\times10^{-2}$, $\epsilon_\text{logical}^{(\mathrm{LDPC})} = 3.0\times10^{-3}$, $p_\mathrm{fac} = 2.3\times10^{-4}$, $\epsilon = 3.0 \times 10^{-2}$ and $p_\mathrm{fail} = 3.8 \times 10^{-2}$. 
The expected number of runs is $1.04$, and the total runtime is \SI{146}{\minute}, 2.36 times longer than the balanced configuration.

\subsubsection{Time-efficient configuration} \label{sssec:ecdlp_faster}

The time-efficient variants of RSA-2048 and FFDLP-512 do not apply directly to ECDLP, which needs a single successful run and does not make use of RNS-type techniques.

\clearpage
\bibliographystyle{apsrev4-2}
\bibliography{biblio/add}

\end{document}